\documentclass[11pt,fleqn]{book}
\usepackage[utf8]{inputenc}
\usepackage[polish,english]{babel}
\usepackage[numbers,sort&compress]{natbib}

\usepackage[top=3cm,bottom=3cm,left=3cm,right=3cm,headsep=10pt,a4paper]{geometry} 

\usepackage{graphicx} 

\usepackage{lipsum} 

\usepackage{tikz} 

\usepackage[english]{babel} 

\usepackage{enumitem} 
\setlist{nolistsep} 

\usepackage{booktabs} 

\usepackage{xcolor} 
\definecolor{ocre}{RGB}{243,102,25} 

\usepackage{avant} 
\usepackage{mathptmx} 

\usepackage{microtype} 
\usepackage[utf8]{inputenc} 
\usepackage[T1]{fontenc} 

\usepackage{calc} 
\usepackage{makeidx} 
\makeindex 

\usepackage{titletoc} 

\contentsmargin{0cm} 

\titlecontents{part}[0cm]
{\addvspace{20pt}\centering\large\bfseries}
{}
{}
{}

\titlecontents{chapter}[1.25cm] 
{\addvspace{12pt}\large\sffamily\bfseries} 
{\color{ocre!60}\contentslabel[\Large\thecontentslabel]{1.25cm}\color{ocre}} 
{\color{ocre}}  
{\color{ocre!60}\normalsize\;\titlerule*[.5pc]{.}\;\color{ocre}\thecontentspage} 

\titlecontents{section}[1.25cm] 
{\addvspace{3pt}\sffamily\bfseries} 
{\contentslabel[\thecontentslabel]{1.25cm}} 
{}
{\color{gray}\ \titlerule*[.5pc]{.}\;\color{black}\thecontentspage} 
[]

\titlecontents{subsection}[1.25cm] 
{\addvspace{1pt}\sffamily\small} 
{\contentslabel[\thecontentslabel]{1.25cm}} 
{}
{\ \titlerule*[.5pc]{.}\;\thecontentspage} 
[]

\titlecontents{figure}[0em]
{\addvspace{0pt}\sffamily}
{\thecontentslabel\hspace*{1em}}
{}
{\ \titlerule*[.5pc]{.}\;\thecontentspage}
[]

\titlecontents{table}[0em]
{\addvspace{-5pt}\sffamily}
{\thecontentslabel\hspace*{1em}}
{}
{\ \titlerule*[.5pc]{.}\;\thecontentspage}
[]

\titlecontents{lchapter}[0em] 
{\addvspace{15pt}\large\sffamily\bfseries} 
{\color{ocre}\contentslabel[\Large\thecontentslabel]{1.25cm}\color{ocre}} 
{}  
{\color{ocre}\normalsize\sffamily\bfseries\;\titlerule*[.5pc]{.}\;\thecontentspage} 

\titlecontents{lsection}[0em] 
{\sffamily\small} 
{\contentslabel[\thecontentslabel]{1.25cm}} 
{}
{\small\sffamily\;\titlerule*[.5pc]{.}\;\thecontentspage}

\titlecontents{lsubsection}[.5em] 
{\normalfont\footnotesize\sffamily} 
{}
{}
{}

\usepackage{fancyhdr} 

\renewcommand{\chaptermark}[1]{\markboth{\sffamily\normalsize\bfseries\chaptername\ \thechapter.\ #1}{}} 
\renewcommand{\sectionmark}[1]{\markright{\sffamily\normalsize\thesection\hspace{5pt}#1}{}} 
\fancypagestyle{plain}{\fancyhead{}} 

\makeatletter
\renewcommand{\cleardoublepage}{
\clearpage\ifodd\c@page\else
\hbox{}
\vspace*{\fill}
\thispagestyle{empty}
\newpage
\fi}

\usepackage{amsmath,amsfonts,amssymb,amsthm} 

\newcommand{\ud}{\mathop{\mathrm{{}d}}\mathopen{}}

\newtheoremstyle{ocrenumbox}
{0pt}
{0pt}
{\normalfont}
{}
{\small\bf\sffamily\color{ocre}}
{\;}
{0.25em}
{\small\sffamily\color{ocre}\thmname{#1}\nobreakspace\thmnumber{\@ifnotempty{#1}{}\@upn{#2}}
\thmnote{\nobreakspace\the\thm@notefont\sffamily\bfseries\color{black}---\nobreakspace#3.}} 

\newtheoremstyle{blacknumex}
{5pt}
{5pt}
{\normalfont}
{} 
{\small\bf\sffamily}
{\;}
{0.25em}
{\small\sffamily{\tiny\ensuremath{\blacksquare}}\nobreakspace\thmname{#1}\nobreakspace\thmnumber{\@ifnotempty{#1}{}\@upn{#2}}
\thmnote{\nobreakspace\the\thm@notefont\sffamily\bfseries---\nobreakspace#3.}}

\newtheoremstyle{blacknumbox} 
{0pt}
{0pt}
{\normalfont}
{}
{\small\bf\sffamily}
{\;}
{0.25em}
{\small\sffamily\thmname{#1}\nobreakspace\thmnumber{\@ifnotempty{#1}{}\@upn{#2}}
\thmnote{\nobreakspace\the\thm@notefont\sffamily\bfseries---\nobreakspace#3.}}

\newtheoremstyle{ocrenum}
{5pt}
{5pt}
{\normalfont}
{}
{\small\bf\sffamily\color{ocre}}
{\;}
{0.25em}
{\small\sffamily\color{ocre}\thmname{#1}\nobreakspace\thmnumber{\@ifnotempty{#1}{}\@upn{#2}}
\thmnote{\nobreakspace\the\thm@notefont\sffamily\bfseries\color{black}---\nobreakspace#3.}} 
\makeatother

\newcounter{dummy} 
\numberwithin{dummy}{section}
\theoremstyle{ocrenumbox}
\newtheorem{theoremeT}[dummy]{Theorem}

\newtheorem{exerciseT}{Exercise}[chapter]
\theoremstyle{blacknumex}
\newtheorem{exampleT}{Example}[chapter]
\theoremstyle{blacknumbox}

\newtheorem{definitionT}{Definition}[section]
\newtheorem{corollaryT}[dummy]{Corollary}
\theoremstyle{ocrenum}

\RequirePackage[framemethod=default]{mdframed} 

\newmdenv[skipabove=7pt,
skipbelow=7pt,
backgroundcolor=black!5,
linecolor=ocre,
innerleftmargin=5pt,
innerrightmargin=5pt,
innertopmargin=5pt,
leftmargin=0cm,
rightmargin=0cm,
innerbottommargin=5pt]{tBox}

\newmdenv[skipabove=7pt,
skipbelow=7pt,
rightline=false,
leftline=true,
topline=false,
bottomline=false,
backgroundcolor=ocre!10,
linecolor=ocre,
innerleftmargin=5pt,
innerrightmargin=5pt,
innertopmargin=5pt,
innerbottommargin=5pt,
leftmargin=0cm,
rightmargin=0cm,
linewidth=4pt]{eBox}	

\newmdenv[skipabove=7pt,
skipbelow=7pt,
rightline=false,
leftline=true,
topline=false,
bottomline=false,
linecolor=ocre,
innerleftmargin=5pt,
innerrightmargin=5pt,
innertopmargin=0pt,
leftmargin=0cm,
rightmargin=0cm,
linewidth=4pt,
innerbottommargin=0pt]{dBox}	

\newmdenv[skipabove=7pt,
skipbelow=7pt,
rightline=false,
leftline=true,
topline=false,
bottomline=false,
linecolor=gray,
backgroundcolor=black!5,
innerleftmargin=5pt,
innerrightmargin=5pt,
innertopmargin=5pt,
leftmargin=0cm,
rightmargin=0cm,
linewidth=4pt,
innerbottommargin=5pt]{cBox}

\makeatletter
\renewcommand{\@seccntformat}[1]{\llap{\textcolor{ocre}{\csname the#1\endcsname}\hspace{1em}}}                    
\renewcommand{\section}{\@startsection{section}{1}{\z@}
{-4ex \@plus -1ex \@minus -.4ex}
{1ex \@plus.2ex }
{\normalfont\large\sffamily\bfseries}}
\renewcommand{\subsection}{\@startsection {subsection}{2}{\z@}
{-3ex \@plus -0.1ex \@minus -.4ex}
{0.5ex \@plus.2ex }
{\normalfont\sffamily\bfseries\setcounter{paragraph}{0}}}
\renewcommand{\subsubsection}{\@startsection {subsubsection}{3}{\z@}
{-2ex \@plus -0.1ex \@minus -.2ex}
{.2ex \@plus.2ex }
{\normalfont\small\sffamily\bfseries\setcounter{paragraph}{0}}}
\renewcommand\paragraph{\@startsection{paragraph}{4}{3.5ex}
{0.4ex \@plus-.2ex \@minus .2ex}
{\z@}
{\normalfont\small\sffamily\bfseries\stepcounter{paragraph}}
{\arabic{paragraph}.~}}

\newcommand{\@mypartnumtocformat}[2]{%
\setlength\fboxsep{0pt}%
\noindent\colorbox{ocre!20}{\strut\parbox[c][.7cm]{\ecart}{\color{ocre!70}\Large\sffamily\bfseries\centering#1}}\hskip\esp\colorbox{ocre!40}{\strut\parbox[c][.7cm]{\linewidth-\ecart-\esp}{\Large\sffamily\centering#2}}}%
\newcommand{\@myparttocformat}[1]{%
\setlength\fboxsep{0pt}%
\noindent\colorbox{ocre!40}{\strut\parbox[c][.7cm]{\linewidth}{\Large\sffamily\centering#1}}}%
\newlength\esp
\newlength\ecart
\def\@part[#1]#2{%
\ifnum \c@secnumdepth >-2\relax%
\refstepcounter{part}%
\addcontentsline{toc}{part}{\texorpdfstring{\protect\@mypartnumtocformat{\thepart}{#1}}{\partname~\thepart\ ---\ #1}}
\else%
\addcontentsline{toc}{part}{\texorpdfstring{\protect\@myparttocformat{#1}}{#1}}%
\fi%
\startcontents%
\markboth{}{}%
{\thispagestyle{empty}%
\begin{tikzpicture}[remember picture,overlay]%
\node at (current page.north west){\begin{tikzpicture}[remember picture,overlay]%
\fill[ocre!20](0cm,0cm) rectangle (\paperwidth,-\paperheight);
\node[anchor=north west] at (3cm,-3.25cm){\color{ocre!40}\fontsize{220}{100}\sffamily\bfseries\@Roman\c@part}; 
\node[anchor=south east] at (\paperwidth-1cm,-\paperheight+1cm){\parbox[t][][t]{11.5cm}{ 
\printcontents{l}{0}{\setcounter{tocdepth}{0}}%
}};
\node[anchor=north east] at (\paperwidth-1.5cm,-3.25cm){\parbox[t][][t]{15cm}{\strut\raggedleft\color{white}\fontsize{40}{40}\sffamily\bfseries#2}};
\end{tikzpicture}};
\end{tikzpicture}}%
\@endpart}
\def\@spart#1{%
\startcontents%
\phantomsection
{\thispagestyle{empty}%
\begin{tikzpicture}[remember picture,overlay]%
\node at (current page.north west){\begin{tikzpicture}[remember picture,overlay]%
\fill[ocre!20](0cm,0cm) rectangle (\paperwidth,-\paperheight);
\node[anchor=north east] at (\paperwidth-1.5cm,-3.25cm){\parbox[t][][t]{15cm}{\strut\raggedleft\color{white}\fontsize{30}{30}\sffamily\bfseries#1}};
\end{tikzpicture}};
\end{tikzpicture}}
\addcontentsline{toc}{part}{\texorpdfstring{%
\setlength\fboxsep{0pt}%
\noindent\protect\colorbox{ocre!40}{\strut\protect\parbox[c][.7cm]{\linewidth}{\Large\sffamily\protect\centering #1\quad\mbox{}}}}{#1}}%
\@endpart}
\def\@endpart{\vfil\newpage
\if@twoside
\if@openright
\null
\thispagestyle{empty}%
\newpage
\fi
\fi
\if@tempswa
\twocolumn
\fi}

\newif\ifusechapterimage
\usechapterimagetrue
\newcommand{\thechapterimage}{}%
\newcommand{\chapterimage}[1]{\ifusechapterimage\renewcommand{\thechapterimage}{#1}\fi}%
\def\@makechapterhead#1{%
{\parindent \z@ \raggedright \normalfont
\ifnum \c@secnumdepth >\m@ne
\if@mainmatter
\begin{tikzpicture}[remember picture,overlay]
\node at (current page.north west)
{\begin{tikzpicture}[remember picture,overlay]
\node[anchor=north west,inner sep=0pt] at (0,0) {\ifusechapterimage\includegraphics[width=\paperwidth]{\thechapterimage}\fi};
\draw[anchor=west] (\Gm@lmargin,-9cm) node [line width=2pt,rounded corners=15pt,draw=ocre,fill=white,fill opacity=0.7,inner sep=50pt]{\strut\makebox[22cm]{}};
\draw[anchor=west] (\Gm@lmargin+.3cm,-9.1) node {\fontsize{78}{98}\sffamily\bfseries\color{black}\thechapter.\ \huge #1\strut};
\end{tikzpicture}};
\end{tikzpicture}
\else
\begin{tikzpicture}[remember picture,overlay]
\node at (current page.north west)
{\begin{tikzpicture}[remember picture,overlay]
\node[anchor=north west,inner sep=0pt] at (0,0) {\ifusechapterimage\includegraphics[width=\paperwidth]{\thechapterimage}\fi};
\draw[anchor=west] (\Gm@lmargin,-9cm) node [line width=2pt,rounded corners=15pt,draw=ocre,fill=white,fill opacity=0.7,inner sep=15pt]{\strut\makebox[22cm]{}};
\draw[anchor=west] (\Gm@lmargin+.3cm,-9.1cm) node {\fontsize{80}{100}\sffamily\bfseries\color{black}\thechapter.\ \huge #1\strut};
\end{tikzpicture}};
\end{tikzpicture}
\fi\fi\par\vspace*{270\p@}}}

\def\@makeschapterhead#1{%
\begin{tikzpicture}[remember picture,overlay]
\node at (current page.north west)
{\begin{tikzpicture}[remember picture,overlay]
\node[anchor=north west,inner sep=0pt] at (0,0) {\ifusechapterimage\includegraphics[width=\paperwidth]{\thechapterimage}\fi};
\draw[anchor=west] (\Gm@lmargin,-9cm) node [line width=2pt,rounded corners=15pt,draw=ocre,fill=white,fill opacity=0.7,inner sep=15pt]{\strut\makebox[22cm]{}};
\draw[anchor=west] (\Gm@lmargin+.35cm,-9.05cm) node {\huge\sffamily\bfseries\color{black}#1\strut};
\end{tikzpicture}};
\end{tikzpicture}
\par\vspace*{270\p@}}
\makeatother

\usepackage{hyperref}
\hypersetup{hidelinks,colorlinks=false,breaklinks=true,urlcolor=ocre,bookmarksopen=false,pdftitle={Noise suppression and long-range exchange coupling for gallium arsenide spin qubits},pdfauthor={Filip K. Malinowski}}
\usepackage{bookmark}
\bookmarksetup{
open,
numbered,
addtohook={%
\ifnum\bookmarkget{level}=0 
\bookmarksetup{bold}%
\fi
\ifnum\bookmarkget{level}=-1 
\bookmarksetup{color=ocre,bold}%
\fi
}
}

\usepackage{braket}
\usepackage{multicol}
\usepackage{multirow}
\usepackage{rotating}
\usepackage[notbib]{tocbibind}
\usepackage[breakable]{tcolorbox}

\renewcommand{\ud}{\uparrow\downarrow}
\newcommand{\du}{\downarrow\uparrow}

\newcommand{\uud}{\uparrow\uparrow\downarrow}
\newcommand{\udu}{\uparrow\downarrow\uparrow}
\newcommand{\duu}{\downarrow\uparrow\uparrow}
\newcommand{\uuu}{\uparrow\uparrow\uparrow}

\author{Filip Kazimierz Malinowski}
\title{Noise suppression and long-range exchange coupling for gallium arsenide spin qubits}
\begin{document}

\newcommand{\rot}[1]{\begin{rotate}{90}#1\end{rotate}}


\begingroup
\thispagestyle{empty}
\begin{tikzpicture}[remember picture,overlay]
\coordinate [below=23cm] (midpoint) at (current page.north);
\node at (current page.north west)
{\begin{tikzpicture}[remember picture,overlay]
\node[anchor=north west,inner sep=0pt] at (0,0) {\includegraphics[width=\paperwidth]{Pictures/
sweet_eye_2_A4.png}}; 
\draw[anchor=north] (midpoint) node [fill=white,fill opacity=0.8,text opacity=1,inner sep=1.2cm]{\Huge\centering\bfseries\sffamily\parbox[c][][t]{\paperwidth}{\centering Noise suppression  \\[6pt] 
{\Huge and long-range exchange coupling}\\[6pt] 
{\Huge for gallium arsenide spin qubits}\\[10pt] 
{\LARGE Filip Kazimierz Malinowski}}}; 
\end{tikzpicture}};
\end{tikzpicture}
\vfill
\endgroup

\cleardoublepage
\thispagestyle{empty}
\frontmatter

\begin{center}
	~\vspace{0.17\textheight}\\
	{\LARGE\centering\bfseries\sffamily\color{ocre} Noise suppression \\[3pt] and long-range exchange coupling \\[6.5pt] for gallium arsenide spin qubits
	}\vspace{0.03\textheight}\\
	\textsc{Filip Kazimierz Malinowski}
	\vspace{0.32\textheight}\\
	Ph.D. Thesis \\
	Center for Quantum Devices \\
	Niels Bohr Institute \\
	University of Copenhagen \vspace{0.04\textheight} \\
	Academic advisors: \\
	Charles M. Marcus \& Ferdinand Kuemmeth \vspace{0.04\textheight} \\
	{This thesis has been submitted \\
	to the PhD School of the Faculty of Science, \\
	University of Copenhagen} \vspace{0.04\textheight} \\
	
	{\em June 2017}
\end{center}

\cleardoublepage
\thispagestyle{empty}
\begin{flushright}
\begin{minipage}{0.5\textwidth}
\begin{flushleft}
	~\vspace{0.69\textheight} \\
	{\em Cover:} \\
	The fingerprint of the exchange interaction mediated by the multielectron quantum dot for relatively small tunnel couplings. \\
	Chapter~\ref{ch:jb-mediated}
	\vspace{0.04\textheight}\\
	The artwork in this thesis is the visualization of the experimental data or is adopted from the documentation of the experiment.
	\vspace{0.04\textheight}
	
	The thesis layout is adopted from \\
	\href{https://www.latextemplates.com/template/the-legrand-orange-book}{``The Legrand Orange Book''} template \\
	by Mathias Legrand with modifications by Vel, \\
	licensed under \href{https://creativecommons.org/licenses/by-nc-sa/3.0/}{CC BY-NC-SA 3.0}
\end{flushleft}
\end{minipage}
\end{flushright}


\chapterimage{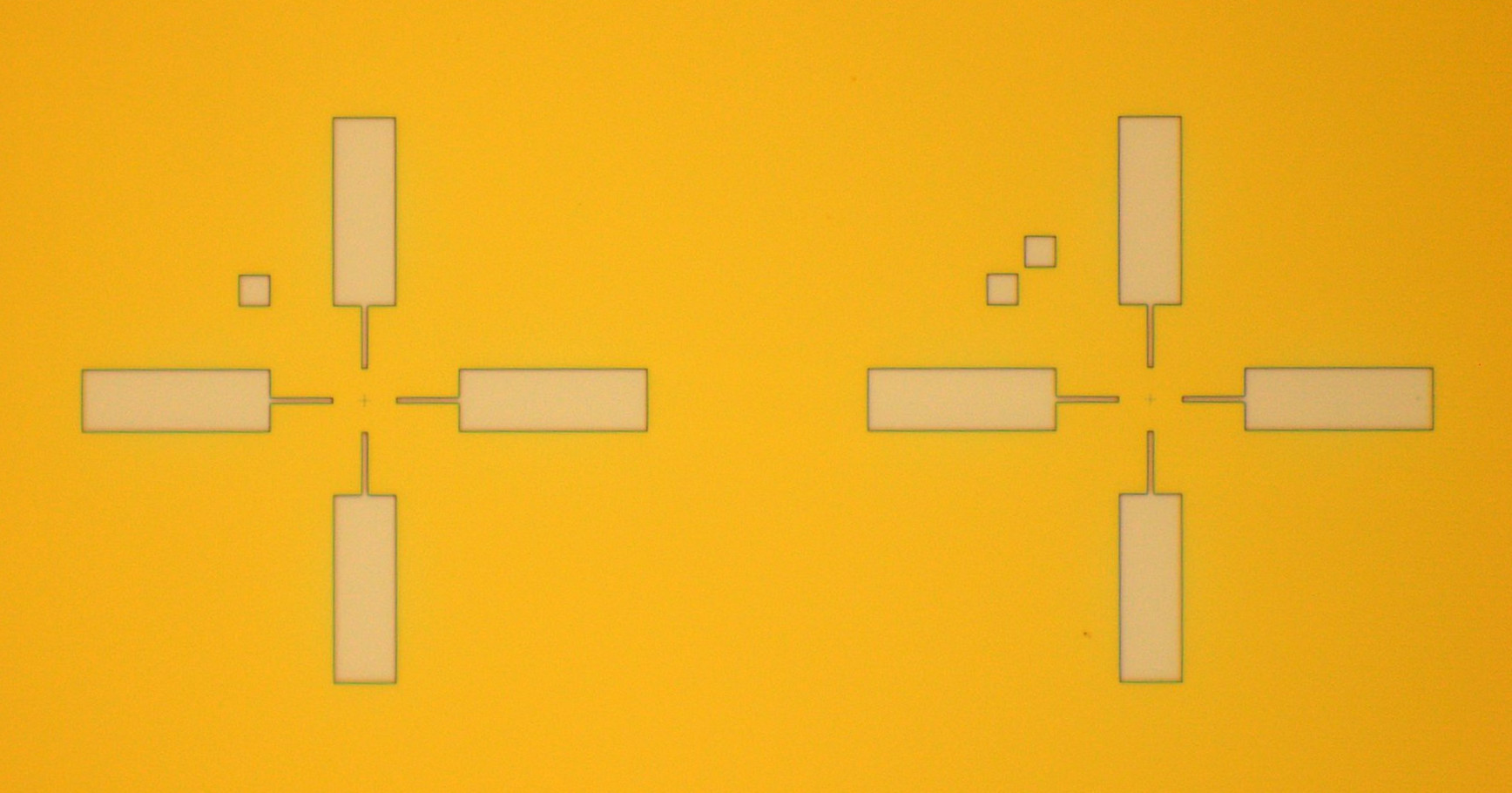}
\chapter*{Acknowledgements}
\addcontentsline{toc}{chapter}{Acknowledgements}
\vspace{-5pt}

Only looking back at the last three years of my studies I am realizing how short time it was, and how unrealistic it would be to conclude this period with a thesis if not for a huge help of people with from whom and with whom I learned.

In the first place I have to thank Frederico Martins with whom I have spent two and a half year at the experimental setup, where we were constantly measuring, analyzing the data and writing. I can hardly imagine more enjoyable and fruitful cooperation than this time spent together at the fridge.

I am hugely grateful to my supervisors, Charlie Marcus and Ferdinand Kuemmeth. Charlie taught me to see the broader context of the experiment, how to separate interesting from boring and important from irrelevant while Ferdinand was the ultimate teacher and guru regarding all experimental techniques.

Peter Nissen, his help with the experiments and, especially, his amazing fab were a key to make my PhD a successful one -- his devices provide enough research material not for one thesis but a dozen. Christian Volk was always there to help with the most tedious fixes. Rasmus Eriksen was a great person to learn from and with in the first months of the PhD.

The articles in this thesis would be half as plentiful and half as valuable if not for all the theoreticians we have spent hours and hours discussing with -- Ed Barnes, \L{}ukasz Cywi\'nski, Mark Rudner, Stephen Bartlett, Andrew Doherty and Tom Smith. I especially loved to work with \L{}ukasz who would answer to each short question with pages and pages of detailed notes, and who was a great link with Warsaw.

Thorvald Larsen and Lucas Casparis from the fellow qubit team (or the \emph{only} qubit team, as they would certainly argue) were great friends never refused to help out and rarely to go out for a cocktail. And so did many present and former QDevers (and not only) including (but not limited to) Rob, Karl, Fabrizio, Fabio, Federico, Anton, Nastasia, Jana, Natalie, Filip, Henri, Giulio, Sole, Shiv, Anders, Mingtang, several Mortens and many others. And Gerbold M\`enard who read and corrected huge parts of this thesis and enjoyed pointing out piles of my errors.

Last but not least (and rather the most) I want to thank my parents, Ma\l{}gosia and Szymon. It was them who directed me towards science, showed how to enjoy physics and love sharing what I have learned.

\chapterimage{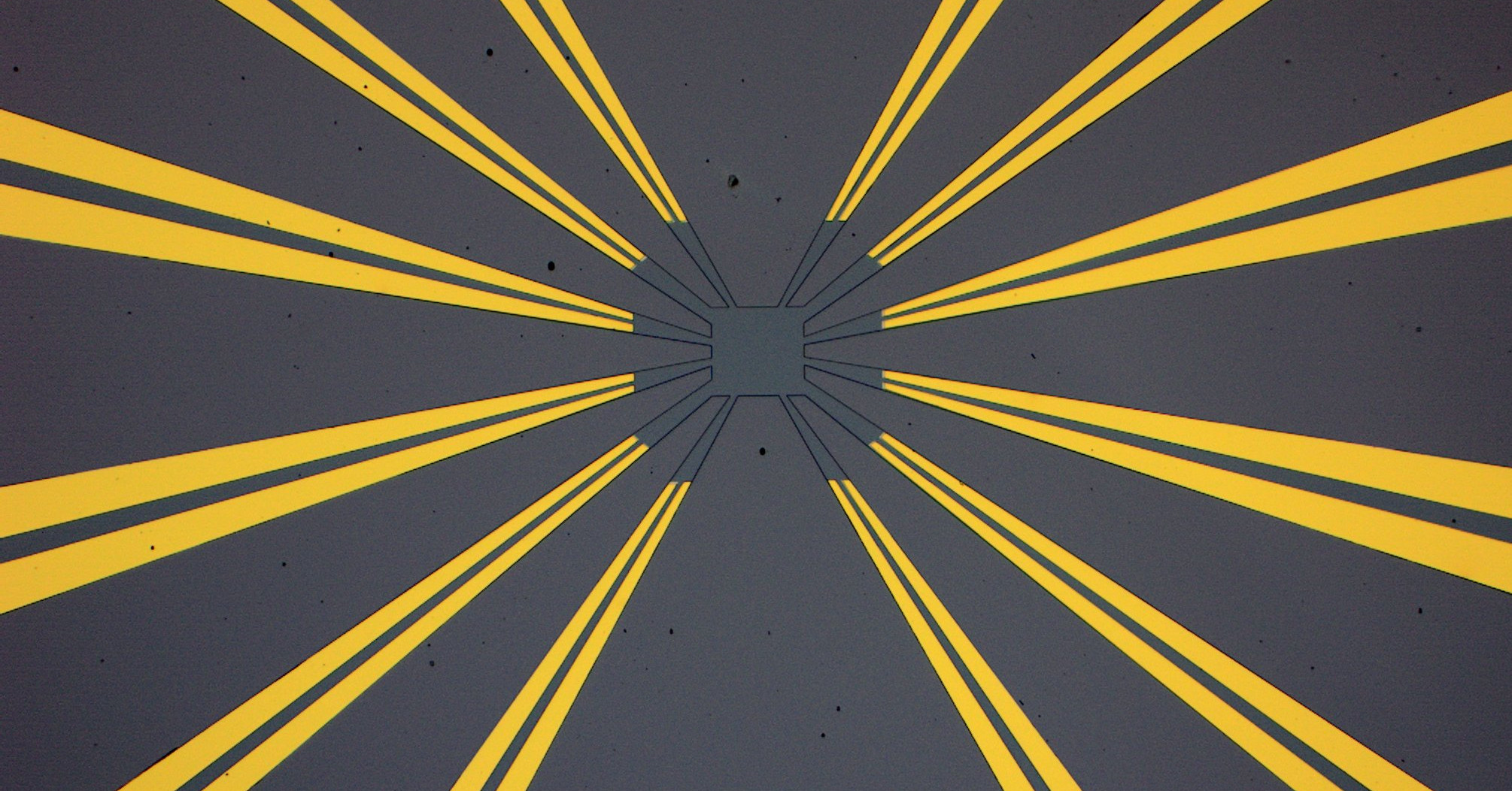}
\chapter*{Abstract in English}
\addcontentsline{toc}{chapter}{Abstract in English}

This thesis presents the results of the experimental study performed on spin qubits realized in gate-defined gallium arsenide quantum dots, with the focus on noise suppression and long-distance coupling.

First, we show that the susceptibility to charge noise can be reduced by reducing the gradient of the qubit splitting with respect to gate voltages. We show that for singlet-triplet and resonant exchange qubit this can be achieved by operating a quantum dot array in a highly symmetric configuration. The symmetrization approach results in a factor-of-six improvement of the double dot singlet-triplet exchange oscillations quality factor while the dephasing times for the three-electron resonant exchange qubit are marginally longer.

Second, we present the study of the Overhauser field noise arising due to interaction with the nuclear spin bath. We show that the Overhauser field noise conforms to classical spin diffusion model in range from 1~mHz to 1~kHz. Meanwhile the megahertz-scale noise spectrum is focused in three narrow bands related to relative Larmor precession of the three nuclear species. Application of the dynamical decoupling sequence designed to notch-filter the narrowband noise enables us to put the highest, up to date, lower bound on the electron spin coherence time in gallium arsenide: 870~$\mu$s.

Later, we study the perspectives of exploiting a multielectron quantum dot as a mediator of the exchange interaction. We investigate interaction between a single spin and the multelectron quantum dot in nine different charge occupancies and identify ground state spin in all cases. For even-occupied spin-1/2 multielectron quantum dot a variation of the gate voltage by a few milivolts in the vicinity of the charge transition leads to sign change of the exchange interaction with a single neighboring electron.

Finally, we demonstrate the exchange coupling between distant electrons mediated by the even-occupied spin-0 multielectron quantum dot. The exchange interaction strength can be controlled up to several gigahertz frequencies. Small level spacing and many body effects give raise to the positions in the gate voltage space that are characterized by decreased susceptibility to charge noise which can be used to implement high fidelity, long-range two-qubit gates.

\chapterimage{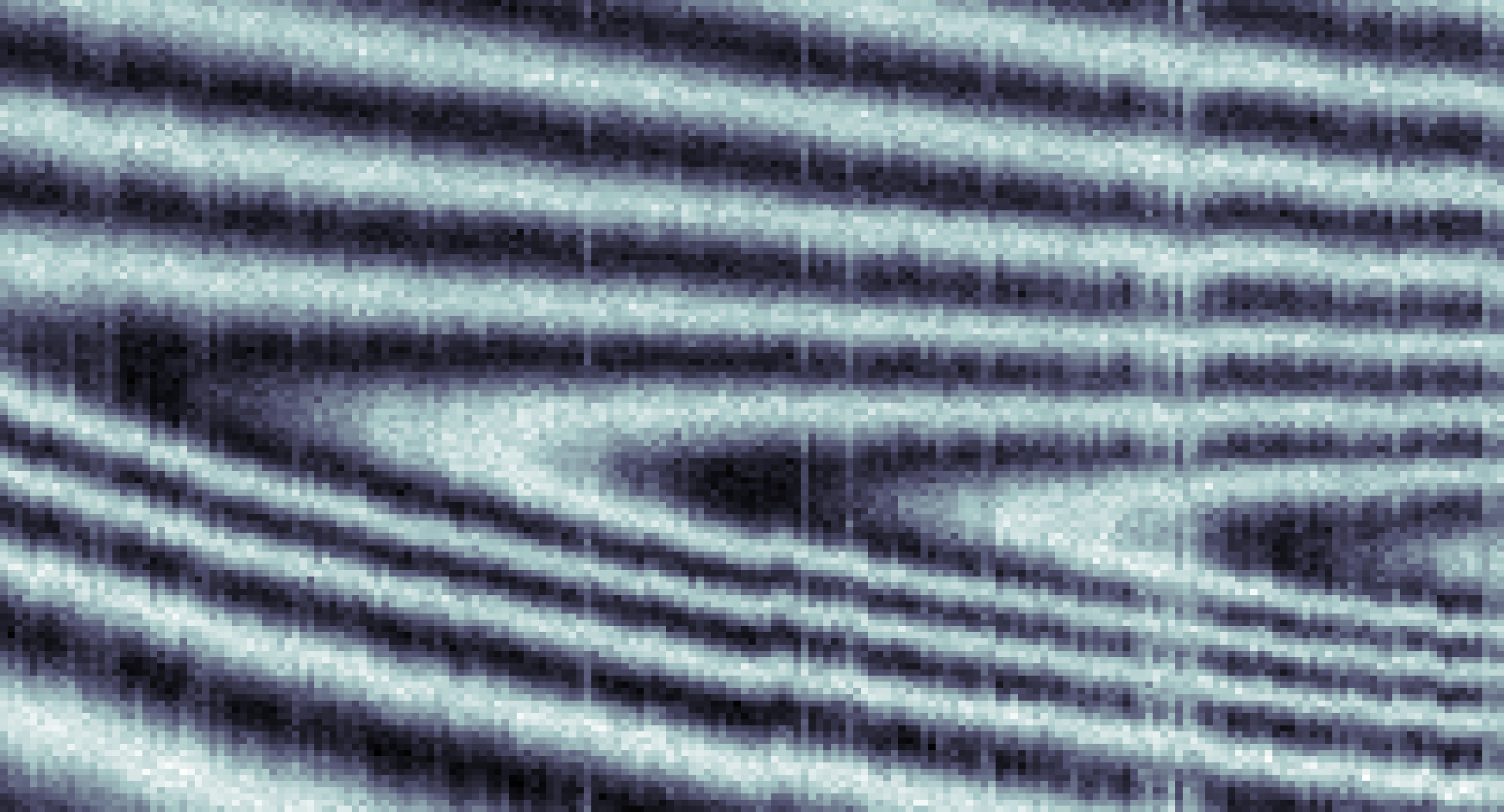}
\chapter*{Streszczenie w języku polskim}
\addcontentsline{toc}{chapter}{Streszczenie w języku polskim}

W niniejszej pracy demonstruję wyniki eksperymentów przeprowadzonych na kubitach spinowych zrealizowanych w elektrostatycznie zdefiniowanych kropkach kwantowych z arsenku galu. Szczególny nacisk położony jest na tłumienie szumu oraz oddziaływanie długozasięgowe.

W pierwszej części pokazujemy że podatność na szum ładunkowy może być ograniczona przez minimalizację gradientu rozszczepienia qubitu względem napięć na bramkach. Dla kubitu syngletowo-trypletowego i kubitu rezonansowej wymiany gradient jest zminimalizowany w symetryczej konfiguracji szeregu kropek kwantowych. Dzięki symetryzacji uzyskujemy sześciokrotną poprawę dobroci oscylacji wymiany pomiędzy dwoma elektronami, natomiast kubit rezonansowej wymiany wykazuje jedynie nieznaczną poprawę czasu koherencji.

Druga część pracy dotyczy szumu Overhausera wynikającego z oddziaływania nadsubtelnego pomiędzy elektronem i kompielą spinową. Spektrum szumu Overhausera jest zgodne z klasycznym modelem dyfuzji w zakresie od mili- do kilohertza. Tymczasem szum w wysokoczęstotliwościowy jest skoncentrowany w trzech wąskich pasmach odpowiadających względnej precesji Larmora jąder trzech izotopów występujących w arsenku galu. Wykorzystując specjalnie zaprojektowaną dynamicznie odsprzęgającą sekwencę, działającą jako filtr środkowozaporowy, jesteśmy w stanie pokazać że czas koherencji spinu elektronu w arsenku galu przekracza 870~$\mu$s.

Następnie badamy perspektywę wykorzystania wieloelektronowej kropki kwantowej jako pośrednika oddziaływania wymiany. Dla dziewięciu kolejnych obsadzeń jesteśmy w stanie określić spin stanu podstawowego przez sprzęganie jej z pojedyńczym sąsiadującym elektronem. Dla obsadzeń nieparzystych obserwujemy zmianę znaku oddziaływania wymiany przy strojeniu napięcia na bramkach w zakresie kilku miliwoltów w pobliżu przejścia ładunkowego.

W ostatnim eksperymencie demonstrujemy długozasiegowe oddziaływanie wymiany któremu pośredniczy wieloelektronowa kropka kwantowa z bezspinowym stanem podstawowym. Siła oddziaływania wymiany może być kontrolowana w zakresie sięgającym kilku gigaherców. Niewielkie rozszczepienie poziomów energetycznych i efekty wielociałowe prowadzą do pojawienia się punktów w przestrzeni napięć na bramkach charakteryzujących się niską wrażliwością na szum ładunkowy, które mogą być wykorzystane do realizacji długozasięgowych bramek dwukubitowych o wysokiej dobroci.

\chapterimage{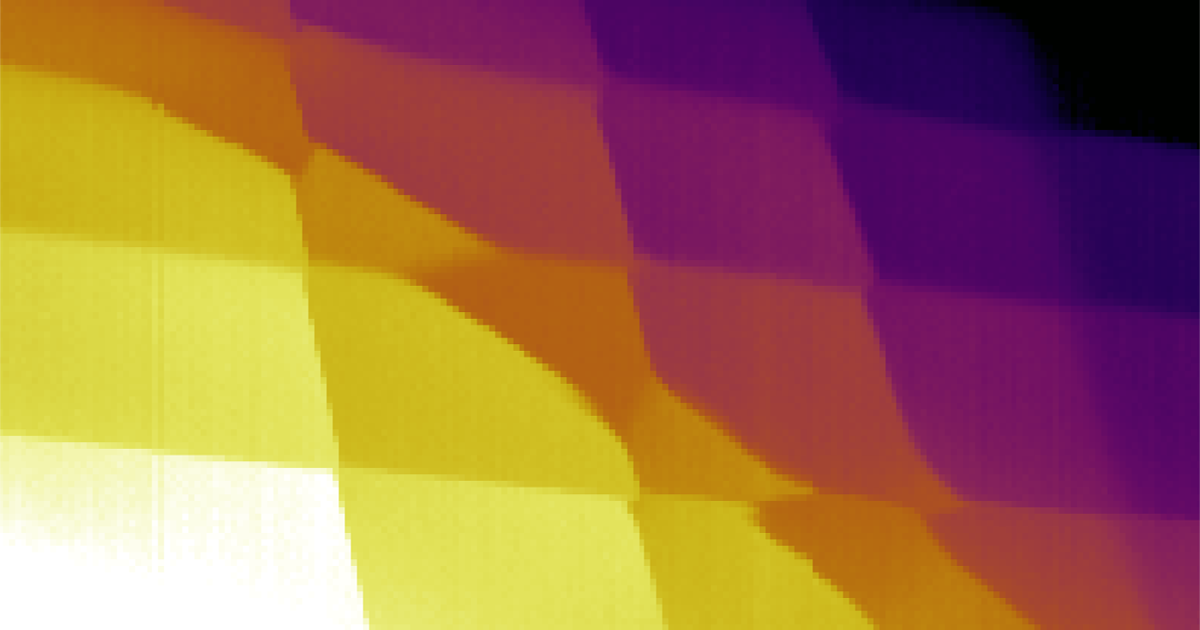}
\chapter*{R\'esum\'e i dansk}
\addcontentsline{toc}{chapter}{R\'esum\'e i dansk}

Denne afhandling præsenterer resultater fra et ekperimentelt studie af spin kvantebit fremstillet i gatedefinerede galliumarsenid kvantepunkter med fokus på støjdæmpning og langdistancekopling.

For det første viser vi, at følsomheden til ladningsstøj kan reduceres ved at reducere gradienten af kvantebit energien med hensyn til  gatespændinger. Vi viser, at dette for singlet-triplet- og resonant udvekslingskvantebit kan opnås ved at drive en kvantepunktsrække i en yderst symmetrisk konfiguration. Symmetriseringstilgangen resulterer i en faktor seks forbedring af kvalitetsfaktoren af singlet-tripletkvantebittens elektronudvekslingsoscillationer, mens coherenstiden for resonant udevekslingskvantebitten er marginalt længere.

For det andet præsenterer vi en undersøgelse af Overhauser-feltstøj, der opstår som følge af interaktion med atomspinbadet. Vi viser, at Overhauser-feltstøj følger en klassisk spindiffusionsmodel i området fra 1~mHz til 1~kHz, mens megahertz-skalaens støjspektrum er fokuseret i tre smalle bånd relateret til relativ Larmor præcession af de tre nukleare arter. Anvendelse af en dynamisk afkoblingssekvens, som er designet til at notch-filtrere smalbåndsstøjen, gør det muligt at sætte den højeste, up-to-date, nedre grænse på elektronens spinconherenstid i galliumarsenid: 870~$\mu$s.

Senere studerer vi perspektiverne i at udnytte et multielektronkvantepunkt som mediator for udvekslingsinteraktionen.
Vi undersøger samspillet mellem et enkelt spin og et multielektronkvantepunkt  i ni forskellige ladningsbesættelser og identificerer spingrundtilstanden i alle tilfælde. For et lige besat, spin-1/2 multielektronkvantepunkt medfører variationer af gatespændingen med et par milivolt i nærheden af ladningstransitionen et fortegnsskift af udvekslingsinteraktionen med en enkelt naboelektron.

Endelig demonstrerer vi udvekslingskoblingen mellem fjerne elektroner formidlet af et lige besat, spin-0 multielektronkvantepunkt. Udvekslingsinteraktionsstyrken kan styres op til nogle-gigahertz frekvenser. Lille energiniveauafstand og mangelegemeeffekter giver anledning til positioner i gatespændingsrummet, der er karakteriseret af nedsat følsomhed til ladningsstøj, som kan bruges til at implementere nøgagtige, langdistanse to-kvantebitoperationer.

%

\chapterimage{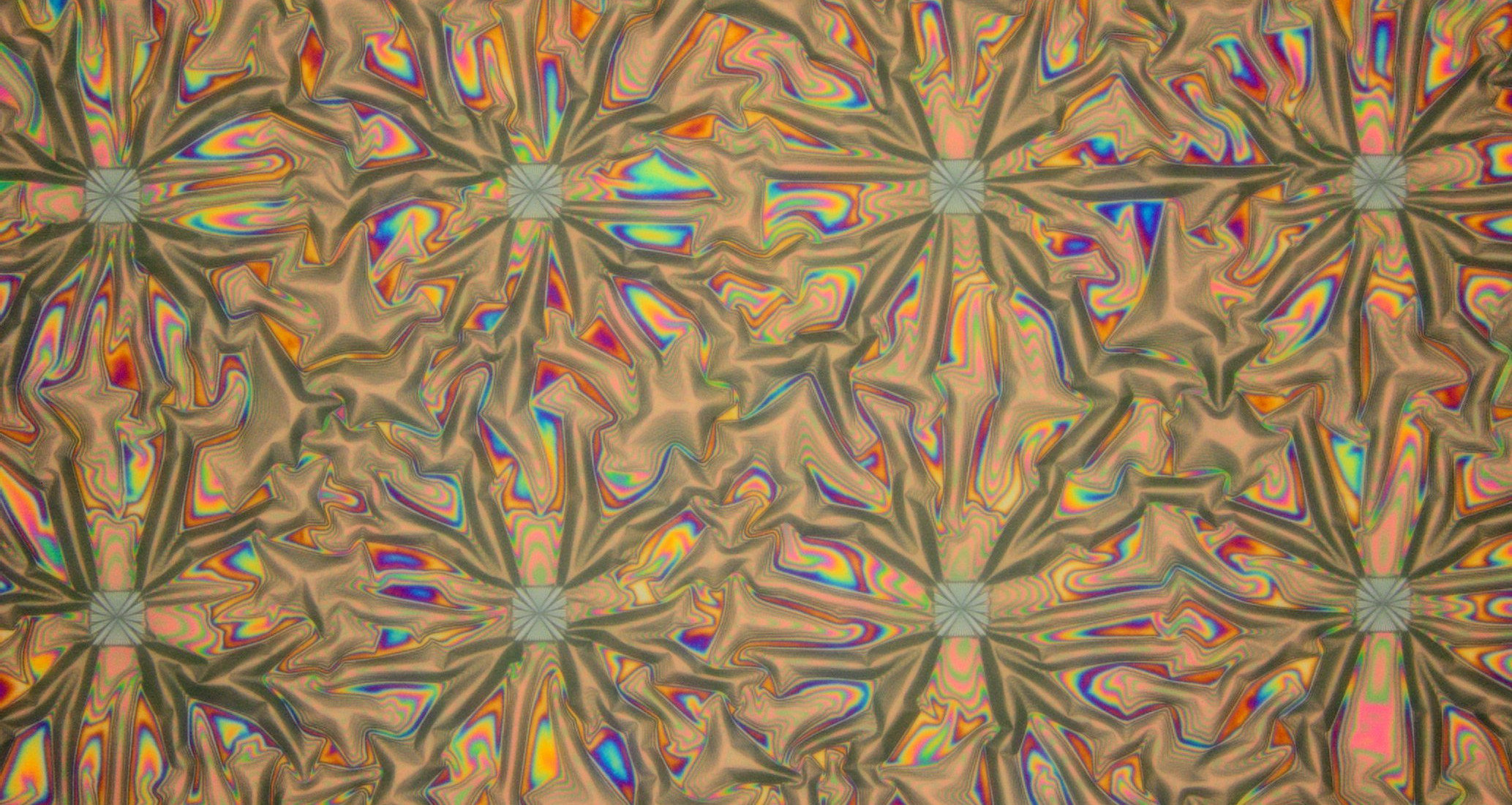}
\pagestyle{fancy}
\setcounter{tocdepth}{1}
\tableofcontents

\chapterimage{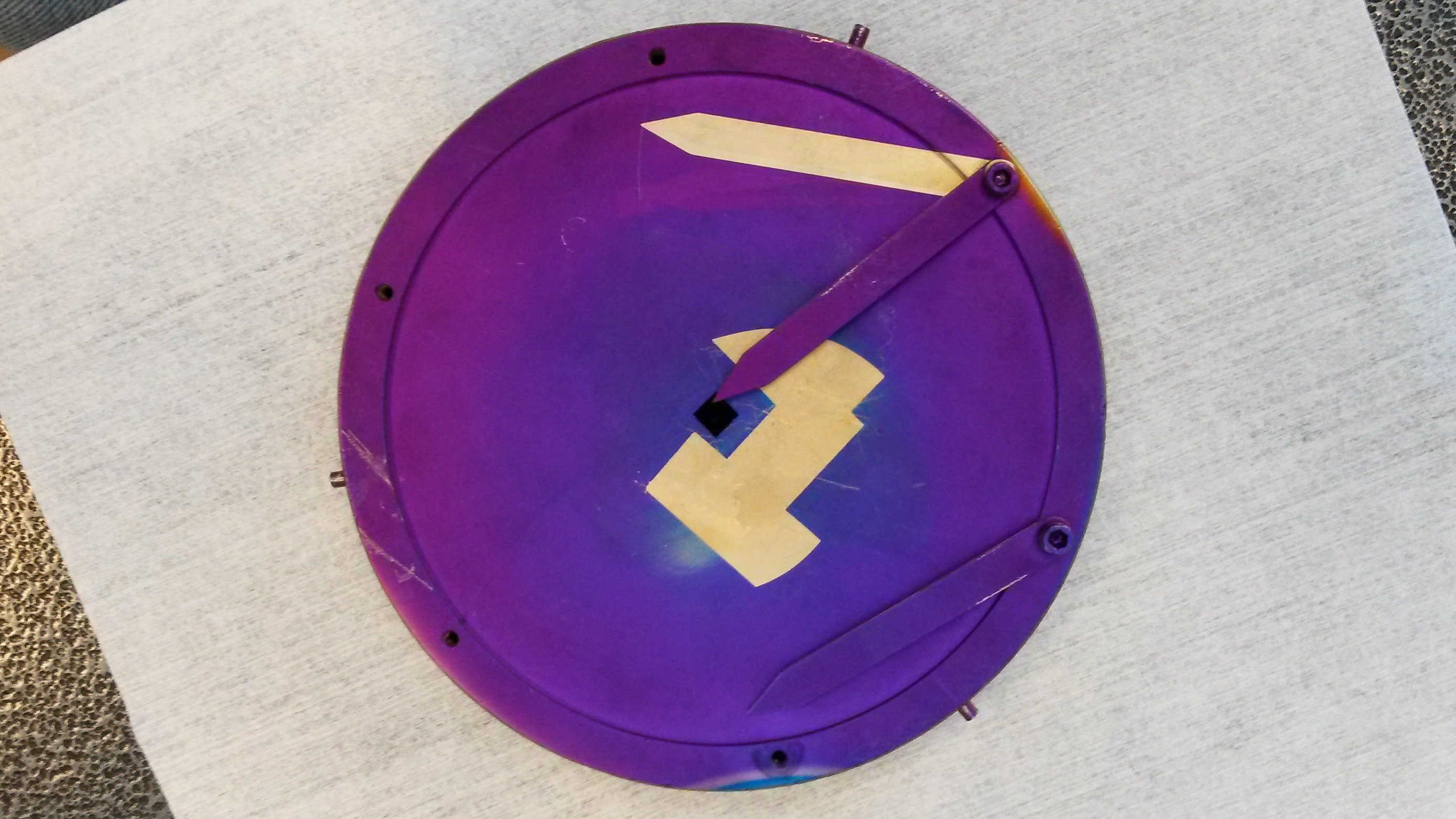}
\listoffigures




\mainmatter

\part{Introduction}
\label{part:introduction}
\setcounter{page}{3}

\chapterimage{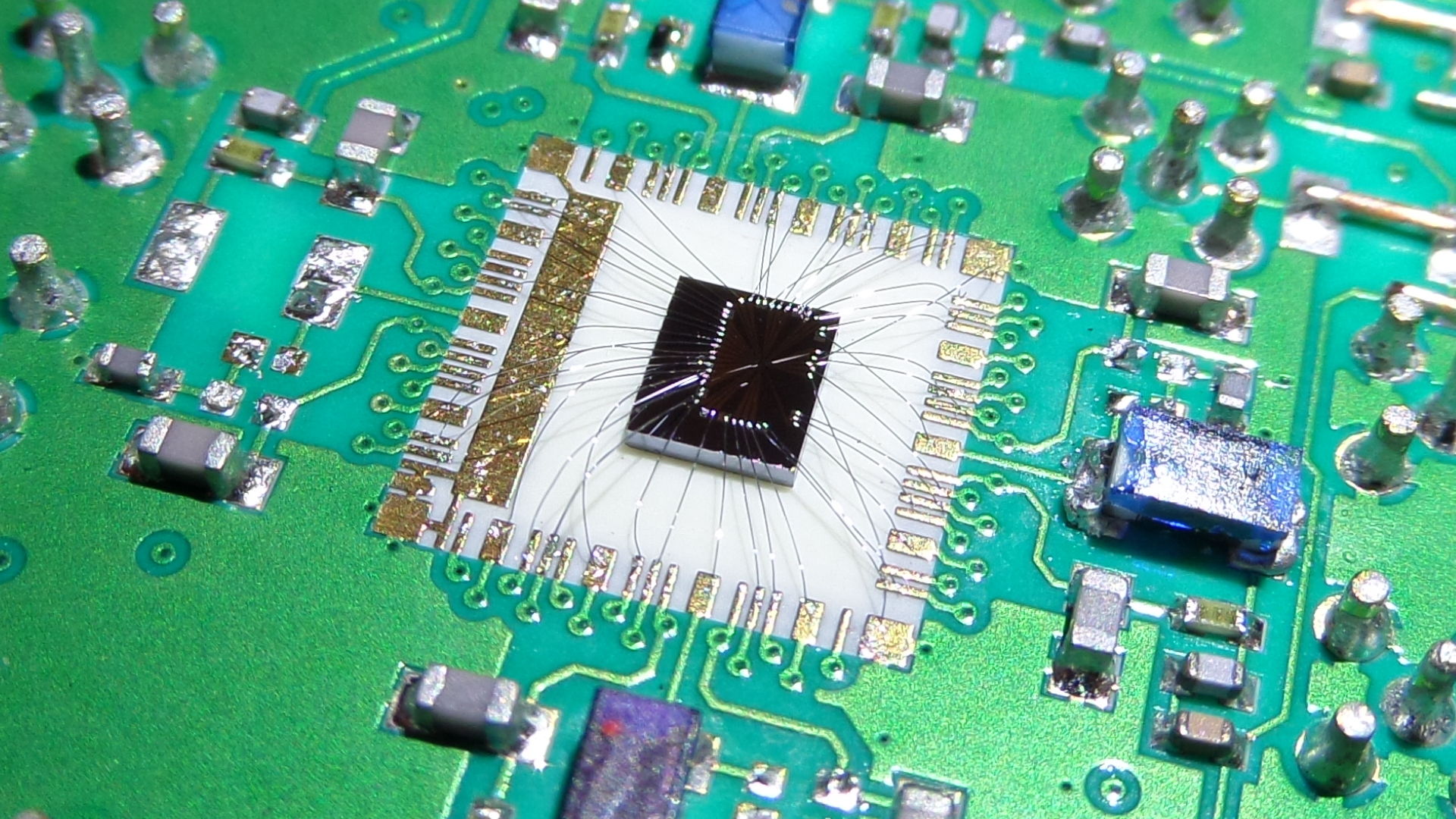}
\chapter[Overview]{\protect\parbox{0.9\textwidth}{Overview}}
\label{ch:intro}

\section{Organization of the thesis}

Part~\ref{part:introduction} is the overview of the field of spin qubits in gate-defined quantum dots. In the next section of this chapter I draw a broad picture of the field to present the state of the art and the main challenges the community is facing. Chapter~\ref{ch:qubits} includes a proper introduction to the field of spin qubits realized in gate-defined quantum dots. It provides a description of the techniques that can be employed to manipulate the electronic spins and the list of qubit implementations in quantum dot arrays.

The main results of this thesis will be contained in three parts, corresponding to three topics which I studied. Each part consists of the reproductions of published results and is extended with the description of the essential concepts, or puts the published results in a broader context. In several places I supplement the published results with several additional datasets and elements of the analysis that did not find their place in articles.

Part~\ref{part:symmetrization} is dedicated to the mitigation of the charge noise which affects the fidelity of the exchange gates. Chapter~\ref{ch:symmetrization} describes the principle of symetrization of spin qubits. Chapter~\ref{ch:symm} presents how to apply this idea to the two-electron exchange gate in a double quantum dot. Chapter~\ref{ch:SRX} describes the symmetrization applied to the three-electron resonant exchange qubit.

Part~\ref{part:nuclei} focuses on the other source of the spin qubits dephasing, i.e. the interaction with the spinful nuclei of the crystal lattice that hosts the quantum dot. Chapter~\ref{ch:nuclei} reviews the most important properties of the nuclear spin bath and briefly introduces the dynamical decoupling techniques. Chapter~\ref{ch:ovh} presents nuclear noise spectrum study over a frequency range of 9 orders of magnitude. Chapter~\ref{ch:notch} explains how the dynamical decoupling sequence can be viewed as a filter in a frequency domain It also demonstrates that the sequence can be adjusted to suppress the component of the nuclear noise related to the Larmor precession of the different nuclear species.

Part~\ref{part:multielectron} is dedicated to multielectron quantum dots. Chapter~\ref{ch:many_charge_states} contains an article that provides the effective model of the multielectron dot. It also presents the study of 9 subsequent multielectron dot occupancies, by coupling it to the two-electron double quantum dot. Chapter \ref{ch:negative-j} focuses on one particular occupancy of the multielectron quantum dot where the negative exchange interaction was observed. The following chapter~\ref{ch:jb-mediated} describes the experiment in which two distant electronic spins are exchange-coupled via a multielectron quantum dot, which serves as an interaction mediator. The final chapter \ref{ch:outlook} contrasts the long-range exchange coupling with other proposals for long range coupling and suggests several relatively simple, high-impact follow-up experiments.

The last part~\ref{part:experiment} consists of appendices. In chapter~\ref{ch:setup} I include the schematics of the experimental setup. Chapter~\ref{ch:samples} contains the recipe for fabrication of GaAs quantum dot devices with a single layer of confining gates.
Chapter~\ref{ch:RT} introduces the technique that has been a key to fast tuning of the dot arrays: real time measurements of the charge diagrams. The final chapter~\ref{ch:AWGs} is a hybrid between a manual and documentation of the Igor Pro procedures that were used to design, upload and apply the sequences of voltage pulses using the Tektronix 5014c arbitrary waveform generator.

\section{Gate-defined quantum dots as qubits (DiVincenzo's criteria)}

Quantum computation is one of the holy grails of modern physics. As such, it is a topic of numerous books and articles with which this brief introduction has no chance to compete. For this reason I restrain myself from describing the ultimate goal to which this work contributes. 
Instead, I will give a brief introduction to the techniques used for spin manipulation and readout and describe the state of the art. To maintain the logical structure I will organize these considerations according to the DiVincenzo's criteria for quantum computing~\cite{DiVincenzo2000} (in a slightly rearranged order). In this manner I attempt to put the results presented in this thesis in their broader context.

\subsection{A (scalable) physical system with \emph{well characterized qubits}}

A single spin 1/2 is conceivably the most natural realization of the qubit, since it forms a natural 2-level quantum-mechanical system. This is the reason why the first implementation of the quantum algorithms were performed on systems of spin-1/2 nuclei~\cite{Jones1998,Chuang1998}. However it was realized that small strength of two qubit coupling and lack of tunability prevents this kind of quantum processor to be scalable~\cite{Warren1997}. An alternative solution~\cite{Loss1998} proposed by Daniel Loss and David DiVincenzo was to employ single electronic spins in gate tunable quantum dots. Compared to NMR implementation this proposal has numerous advantages: the electron magnetic moment is larger than the nuclear magnetic moment by a factor of $\sim$1000 enabling faster manipulation. Moreover, the artificial molecule consisting of multiple quantum dots can be arbitrarily designed, the strength of couplings -- voltage controlled and the qubits themselves could be moved within the quantum processor.

\subsection{The ability to initialize the state of the qubits to a simple fiducial state, such as $\ket{000}$}

Using a higher magnetic moment (or g-factor) becomes a clear advantage of electron spin qubits when considering the state initialization. A g-factor of the order of 1 leads to a Zeeman splitting $g\mu_B B$ of tens of microelectronvolts at 1~T. For comparison the thermal fluctuations at 100~mK have a characteristic energy of $k_BT=8.6$~$\mu$eV. The ratio of these two quantities leads to the conclusion that for moderate magnetic fields and at dilution-refrigerator temperature the electronic spin can be prepared in the ground state, simply by letting it relax~\cite{Elzerman2004}.

Freedom in control of the quantum dot occupation provides also a mean for the initialization of a more complex multielectron state, by exploiting the Pauli exclusion principle~\cite{Petta2005,Koppens2006}. The exclusion principle forces two electrons with parallel spins (i.e. in one of the triplet states), located in the same quantum dot to occupy different orbitals. On the contrary for the antiparallel configuration of spins (i.e. singlet state) the electrons can occupy the same, lowest orbital. The arising energy splitting between the two-electron states enables the initialization of the electron pair in a singlet configuration simply by letting the system relax to its ground state. Once the relaxation happens one electron can be moved to a different quantum dot resulting in a one-step initialization of the entangled state. In fact this very method is exploited in all  the original experiments presented in this thesis. The idea could in principle be employed to initialize more complex highly entangled multielectron states, by preparing a multielectron quantum dot in the ground state, followed by shuttling of the electrons to the designated single-electron quantum dots.

\subsection{A qubit-specific measurement capability}

For gate-defined quantum dots the ability to initialize the electronic spins turns out to be closely related to our ability to read them out. In most works up to date the readout has been performed in two steps. The first step is to convert the spin state to charge state, i.e. induce a difference of charge distribution between different spin states in an incoherent manner. In turn, the difference in charge distribution affects the conductance of neighboring transistor (in practice a quantum point contact or a single electron transistor). In the second step the conductance is readout by means of lock-in or reflectometry measurement.

The two techniques for spin-to-charge conversion are dual to the two initialization techniques, and both were shown to provide a single-shot readout fidelity of individual spins~\cite{Elzerman2004,Barthel2010a}. In the first technique the spin-up and spin-down states are tuned in such a way that one is located above the Fermi level of neighboring lead, while other one remains below~\cite{Elzerman2004}. Whenever the electron has high energy it will eventually tunnel out of the lead, resulting in a change of the charge distribution, and then the electron will tunnel back in onto a dot. The observation of this process relies, similarly to initialization, on adjusting the Zeeman splitting to be larger the temperature of the electrons in the lead. The advantage of this technique is that towards the end of the readout phase the electronic spin is guaranteed to be initialized in the ground state. On the other hand, the limitation is that the readout speed is set by the tunneling rates which have to be small for high fidelity qubit manipulations.

The counterpart of singlet initialization provides a mean for measurement of the relative orientation of the two spins. This method exploits the fact that, as explained before, the triplet states with two electrons occupying the same dot have significantly larger energy than a singlet state. This implies that for a carefully tuned double quantum dot the spin triplet state of the lowest energy will consist of the two electrons residing on different dots. Meanwhile for the singlet state with the lowest energy two electrons will reside on a single dot. Once more, the resulting difference in the charge distribution can be picked up by the neighboring charge sensor. This technique, used in all experiments described in this thesis, has three significant advantages over the alternative described above: in principle it is non-destructive (excluding decoherence between triplet states and avoidable relaxation) and does not require a neighboring lead.

\subsection{Long relevant decoherence times, much longer than the gate operation time}

Another DiVincenzo criterium refers to the preservation of the multiqubit quantum state. In this context the exploitation of the spin degree of freedom yields significant advantages. In the simplest view the advantage is related to the relative weakness of magnetic interactions compared to electrostatic interactions. This implies that the external disturbances that can lead to decoherence are either weak\footnote{typical hyperfine interaction with between an electron and a single nuclei of the host lattice is of the order $10^{-9}$~eV, for the electronic wavefunction which typically spreads over a million of crystal unit cells~\cite{Cywinski2009a}} or indirect\footnote{related to mixing of a spin with other degrees of freedom due to spin-orbit~\cite{Scarlino2014} or exchange interaction~\cite{Dial2013}}. Nonetheless these decoherence mechanisms remain a significant challenge. This thesis yields two results related to the decoherence problem.

Part~\ref{part:nuclei} is dedicated to the study of the interaction between the electronic spin and the collection of millions of nuclear spin in the gallium arsenide crystal. In these studies we address the decoherence due to the collective influence of multiple nuclear spins. In particular the use of elaborated decoupling sequences enables us to increase the spin coherence time by over five orders of magnitude. This method can be expected to find application in areas closely related to quantum computing, but is not an ultimate solution to the issue of the electron spin decoherence due to interaction with the nuclear bath. In a long run, more promising path is either to change the semiconducting material hosting the quantum dots to one where nuclei with a non-zero spin are sparse (e.g. silicon-based quantum dots)~\cite{Pla2012,Maune2012,Maurand2016}, or to use hole spins~\cite{Higginbotham2014c,Li2015} which enjoy much weaker hyperfine interaction.

Reduction of the indirect influence of the charge noise is the topic of Part~\ref{part:symmetrization} of this thesis. I describe there how introduction of the additional symmetries reduces admixture of the charge degree of freedom to the multielectron spin state. On the fundamental level our study shows that the spin qubits are at their best when the difference between the spin states does not entail the difference in the charge distribution.

\subsection{A ``universal'' set of quantum gates}

The number of methods for manipulation electronic spins in gate-defined quantum dots is numerous. The difficulty lies in the choice of the qubit definition and manipulation method that provides a set of universal gates that do not entail to an increased susceptibility to noise. The almost complete list of available variants will be included in the introduction to spin qubits in chapter~\ref{ch:qubits}, and therefore I will here limit the discussion to presenting the state of the art in the manipulation of electronic spins.

In terms of single qubit gates the most promising candidates are simple single-electron qubits defined in Si/SiGe quantum wells equipped with a micromagnet. The two recent experiments~\cite{Kawakami2016,Takeda2016} showed that electron rotation induced by shaking a single electron in the gradient of an external magnetic field can reach the limit of 99\% fidelity which is also a threshold for the application of quantum error correction. These results are very promising, especially because they did not take advantage of the isotopically enriched silicon which would result in further improvements in coherence by removing the residual atoms of spinful isotope $^{29}$Si.

A similar result was obtained for a single electronic spin in MOS structure that uses isotopically purified silicon~\cite{Veldhorst2014}. Moreover, this result was followed by the  presentation of the two qubit gate~\cite{Veldhorst2015}, although it fidelity was not quantified. A significant limitation in this experiment was the insufficient tunability of the tunnel coupling, due to which the exchange interaction between electrons was turned on and off adiabatically, rather than rapidly with respect to difference of Zeeman splittings of the two electrons. Nevertheless, the realization of single- and two-qubit gates in these two kinds of silicon structures is a signature that exploiting a spin-free nuclear lattices is, in my opinion, the most promising path towards spin-qubit based quantum computation.

Even higher degree of control over two-qubit system was obtained with singlet-triplet qubits in GaAs nanostructures. Reference~\cite{Nichol2017} presents a two-qubit gate with 90\% fidelity and single-qubit gates of 99\% fidelity. That achievement was reached by harnessing the nuclear spin bath by means of dynamical nuclear polarization and reduction of the charge noise by operating the qubits in a rotating frame. As of today realization of such a complex operation is not possible in silicon-based qubits, since it is much more difficult to fabricate Si/SiGe or MOS quantum dots.

The amount of overhead related to dynamical nuclear polarization makes the electron spin qubits systems in GaAs unsuitable for scalability. However due to easier fabrication and greater tunability this system serves as a playground for increasing the size of electron array. In that field arrays of five, tunnel-coupled quantum dots equipped with two independent charge sensors (chapter~\ref{ch:jb-mediated}) are, to my knowledge, the largest system of coherently controlled electronic spins. The result presented there also addresses another problem that needs to be solved in order to begin scaling quantum dot based systems: the gate crowding

\subsection{A \emph{scalable} physical system (with well characterized qubits)}

The gate crowding is a consequence of the small dimensions of the quantum dots. Even though small size may ultimately be an advantage, once the quantum dot array reaches a phase of industrialization, at the stage of developing few-qubit devices it is a significant obstacle. This is because the typical gate size required to tune a potential sufficiently precisely must have a width of tens of nanometers. These dimensions are at the limit of state of the art electron beam lithography techniques, and almost certainly exclude possibility of building large, two-dimensional arrays of directly tunnel-coupled quantum dots within the next few years. This concern leads to an extensive search for a potential long-range coupling mechanism.

In chapter~\ref{ch:jb-mediated}, I present a proof-of-principle experiment for performing a long-range (micrometer scale) exchange gate between electronic spins. This approach involves a multielectron quantum dot as the interaction mediator and avoids exploiting degrees of freedom different than spin.

Another extensively studied long range coupling mechanism is a dipole-dipole coupling between hybridized spin-charge qubits mediated by a superconducting cavity~\cite{Mi2016,Stockklauser2017}. This concept is based on the enormous success of the superconducting qubits and may provide a method of qubit coupling over enormous distances (of several milimeters). The difficulty this approach poses is hybridization with the charge degree of freedom which increases the susceptibility to charge noise. Therefore an improved understanding of its origin may be a key for scaling quantum dot systems.

\chapterimage{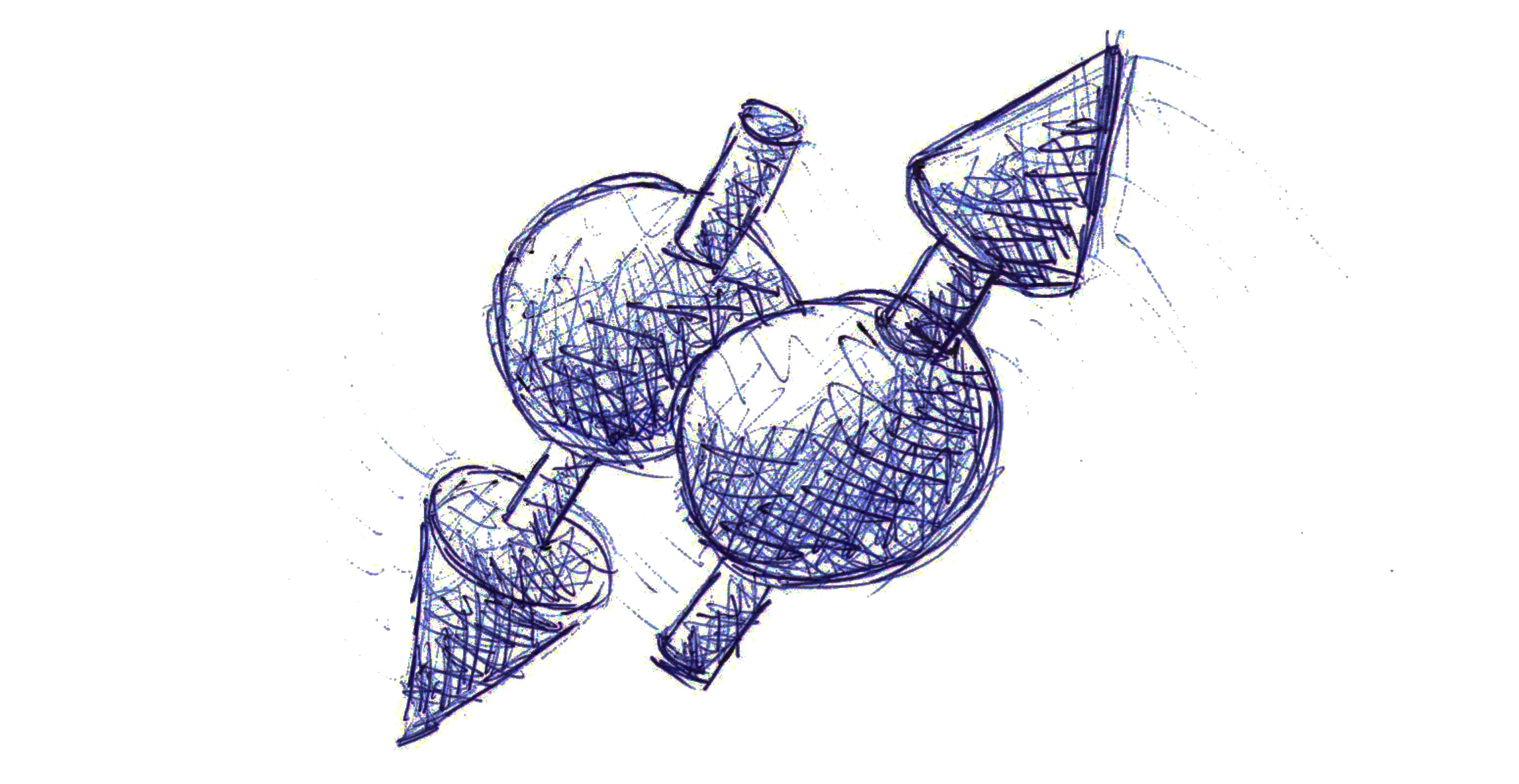}
\chapter[Spin qubits]{\protect\parbox{0.9\textwidth}{Spin qubits}}
\label{ch:qubits}

Along the lines of the overview chapter~\ref{ch:intro} I will devote this chapter to spin qubits and minimize the introduction to quantum mechanics and quantum computing. Section~\ref{qubits:qubit} provides an elementary background to the description of the two level system with the main purpose of introducing the Bloch sphere representation. Section~\ref{qubits:confining} describes how single electrons can be confined and isolated in a solid state system, in particular in GaAs gate-defined quantum dots. Section~\ref{qubits:implementations} introduces the implementations of the qubit in a single-electron quantum dots that were demonstrated experimentally.

\section{Qubit}
\label{qubits:qubit}

\subsection{Definition}

A qubit is the basic unit of quantum information, in analogy to a (classical) bit being the unit of (classical) information. Both qubits and bits can be in one of two states, usually labeled ``0'' and ``1''. Of these two a qubit is however infinitely more complex\footnote{In a sense that an infinite (countable) number of bits is necessary to fully describe state of a single qubit.} since by definition it may be in a superposition of the two states.

Formally the qubit state $\ket{\Psi}$ can be represented as a linear combination of the two orthogonal vectors, $\ket{0}$ and $\ket{1}$ (representing 0 and 1 states)
\begin{equation}
	\ket{\Psi} = \alpha \ket{0} + \beta \ket{1},
\end{equation}
where complex coefficients $\alpha$ and $\beta$ are the probability amplitudes and are related to probabilities $P_i$ of finding the qubit in state $i=0, 1$:
\begin{equation}
	P_0 = |\alpha|^2, \qquad P_1 = |\beta|^2.
\end{equation}
The probabilistic interpretation of $\alpha$ and $\beta$ implies normalization of the vector $\ket{\Psi}$ representing the state of the qubit
\begin{equation}
	|\alpha|^2 + |\beta|^2 = 1.
\end{equation}
Moreover it turns out that the relative phase between the two coefficients has no physically observable consequences.

\subsection{Most handy representation}

\begin{figure}
\begin{center}
\includegraphics[scale=1]{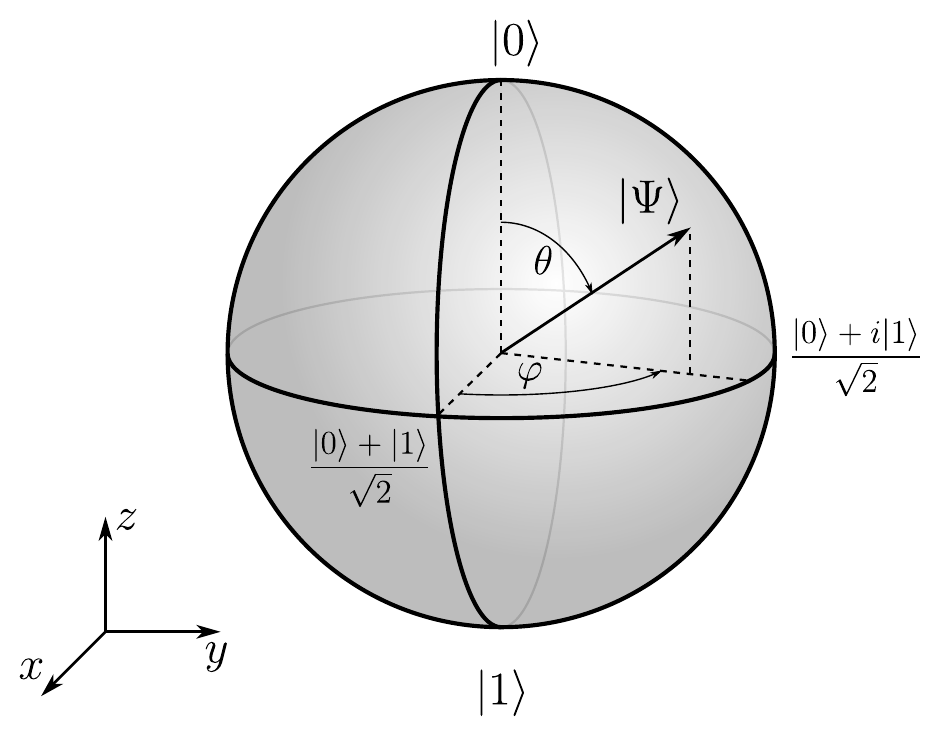}
\caption[Representation of the qubit state on a Bloch sphere]{
Representation of the qubit state on a Bloch sphere
}
\label{qubits:bloch}
\end{center}
\end{figure}

Normalization condition and irrelevance of the global phase allows to represent the qubit state in a most helpful manner, i.e. as a point of a sphere (called Bloch sphere). For that purpose the coefficients $\alpha$ and $\beta$ must be rewritten, so that the qubit state is represented by
\begin{equation}
	\ket{\Psi} = \cos(\theta/2) \ket{0} + e^{i\varphi} \sin(\theta/2) \ket{1}
\end{equation}
where $\theta \in [0,\pi]$ and $\varphi \in [0, 2\pi)$.
In this form of $\theta$ and $\varphi$ can be interpreted as coordinates of a point on the Bloch sphere (Fig.~\ref{qubits:bloch}). 

The operations on the qubit state also have a natural representation on the Bloch sphere, as rotations around some axis. Some of the most common single qubit gates are even named after their graphical representation, e.g. $\pi/2$-rotation around $y$ axis.

\subsection{Imperfect qubit states and operations}

In experiments one never finds the qubit to be perfectly prepared or manipulated. Even though the above notation very often suffices to explain the experimental results, one often needs to quantify the errors of manipulations. One can derive mathematical formalism to describe such situation by considering that the qubit might be in one of several pure states $\ket{\Psi_i}$ with classical probability $p_i$. Such representation turns out to be an excessive one, since many of such states are physically indistinguishable. The density matrix representation turns out to be the simplest sufficient description of imperfect states.

For the physical state in one of several pure states $\ket{\Psi_i}$ with classical probability $p_i$ one defines the density matrix by
\begin{equation}
	\hat{\rho} = \sum_i p_i \ket{\Psi_i}\bra{\Psi_i}.
\end{equation}
Such positive definite matrix has trace 1.
In case of a single qubit $\hat{\rho}$ is a $2\times2$ matrix, which can be fully characterized by only three parameters $x$, $y$ an $z$
\begin{equation}
	\hat{\rho} = \frac{1}{2} \left( \mathbb{I} + x\hat{\sigma}_x + y\hat{\sigma}_y + z\hat{\sigma}_z \right) = \frac{1}{2} \left( \mathbb{I} + \vec{r}\vec{\sigma} \right),
\end{equation}
where $\mathbb{I}$ is the identity and $\sigma_i$ are the Pauli matrices.
Parameters $x$, $y$ an $z$, represented as a single vector $\vec{r}$ must fulfill $|\vec{r}| \leq 1$ for the matrix to represent a physical state.

Drawing $\vec{r}$ on the Bloch sphere (with its center at the origin of the coordinate system and radius 1) is an accurate representation of the imperfect, mixed state. This means that the rotations representing the operations on the pure qubit state will act in an identical way on $\vec{r}$. Finally, imperfect qubit manipulations can be represented as combinations of the sphere rotation, shrinking or squeezing along some directions.

\subsection{Multi-qubit (and multi-spin) states}

Mathematically the multi-qubit states are represented by the vectors in the tensor-product space of several two-level systems. Such a general state can be written as
\begin{equation}
	\ket{\Psi} = \sum_{\lbrace{s_i\rbrace = 0,1}} \alpha_{\lbrace s_i \rbrace} \ket{ \lbrace s_i \rbrace }
\end{equation}
where $\lbrace s_i \rbrace$ indicates a list of 0's and 1's. The number of $2^N-2$ real parameters necessary to fully characterize such state is a manifestation of increased computing power of a quantum system over a classical one.

\subsection{Universal set of gates}

Even though the number of states available in multi-qubit system is huge the set of operations necessary to perform an arbitrary computation is rather limited~\cite{Nielsen2010}.

First part of the universal set of gates are the arbitrary single qubit rotations. The standard way to achieve these is to combine the rotations around two orthogonal qubit axes (usually $x$ and $y$). Then an arbitrary rotation can by realized by rotation by three Euler angles.

One of the spin-qubit realizations, exchange-only qubit~(Subcection~\ref{qubits:XO&RX}), gives a possibility of performing rotations about two axes which are tilted by 120$^\circ$ with respect to each other. Although these rotations are sufficient to construct an arbitrary single-qubit gate. Such twist to Euler angles increases difficulty of designing and optimization of a quantum algorithm. On the other hand, it is beneficial to supplement the $x$ and $y$ rotations with the $z$ rotations to reduce depth of the final quantum circuit.

In principle the set of single qubit operations can be minimized to only two $\pi/2$ rotations and $T$-gate (which is a rotation by $\pi/8$ angle). This can be proven by showing that the set of states achievable by combining these three rotations is dense on a Bloch sphere, and therefore arbitrary rotations can be realized with arbitrary precision. However one must remember that the number of primary gates that must be composed to perform such arbitrary rotation can introduce a huge overhead in execution of a quantum algorithm. Therefore one must remember that in the set of available single qubit rotations affects the length and of the algorithm and ultimately has an impact on resources required to implement quantum error correction.

The second part of the required gate set is an two-qubit entangling gate. Usually three kinds of such gates are considered. 

A CNOT gate is a quantummechanical equivalent of the classical gate with the same name It changes the state on 

CPHASE gate is the most commonly realized two qubit gate. The name indicates that changing the phase between $\ket{1}$ and $\ket{0}$ state of the target qubit conditioned by the control qubit. This operation is equivalent to adding a phase factor to $\ket{11}$ state with respect to $\ket{00}$, $\ket{10}$ and $\ket{01}$.

The final possibility, most natural for exchange-coupled spins, is a $\sqrt{\mathrm{SWAP}}$ gate. This gate results from interrupting the flip-flop process halfway through. For example, it can take the two spin state between different states in a cycle
\begin{equation}
	\ket{10} \ \xrightarrow{\sqrt{\mathrm{SWAP}}} \ \frac{\ket{10}+i\ket{01}}{\sqrt{2}} \ \xrightarrow{\sqrt{\mathrm{SWAP}}} \ \ket{01} \ \xrightarrow{\sqrt{\mathrm{SWAP}}} \ \frac{\ket{10}-i\ket{01}}{\sqrt{2}} \ \xrightarrow{\sqrt{\mathrm{SWAP}}} \ \ket{10}.
\end{equation}

\section{Confining electronic spins}
\label{qubits:confining}

\subsection{Semiconductor platforms for confining electronic spins}

A single electronic spin 1/2 is such a natural realization of the qubit that virtually any structure capable of confining an electron was at some point in the history considered to be a possible platform for quantum computing. In particular, semiconductor nanostructures allow to benefit from the wide range of the available fabrication techniques. Roughly speaking, there are two types of such structures in which coherent spin manipulations were realized: electrons bound on the crystal impurities or confined by combination of band-gap engineering and gating (i.e. quantum dots). On the other hand, these systems can be divided into two groups according to the method of manipulation and detection: all-electrical and mixed optical and electrical. The summary of semiconductor systems in which coherent spin manipulations of single electrons was performed up to date is presented in Table~\ref{tab:systems}.

\begin{table}
	\caption{A choice of the semiconductor systems for which coherent operation of single electronic spins was demonstrated up to date.}
	\label{tab:systems}
	\vspace{3pt}
	\centering
	\begin{tabular}{cc||c|c}
	 &  & \multicolumn{2}{c}{\bf Manipulation method} \\ 
	 &  & All-electrical & Optical and electrical \\ 
	\hline 
	\hline 
	\multirow{6}{*}{\rotatebox{90}{\bf Structure}} &  & GaAs, Si/Ge quantum wells &  \\ 
	 & Quantum & MOS structures & \multirow{2}{*}{In(Ga)As/GaAs} \\ 
	 & dot & IsAs nanowires &  \\ 
	 &  & carbon nanotubes & \\ 
	\cline{2-4}
	 & \multirow{2}{*}{Impurity} & \multirow{2}{*}{P donor in Si} & NV in diamond \\ 
	 &  &  & vacancy in SiC \\ 
	\end{tabular} 
\end{table}

In a nutshell, the electrons in optically active structures are expected to serve as an interface between a solid state quantum processor and photonic quantum communication network. In particular the depth of the binding on such impurities as NV centers in diamond~\cite{Childress2006,Dolde2013,Hensen2015} and vacancy in silicon carbide~\cite{Koehl2011} enables their coherent control at room temperature. On the other hand, self-assembled quantum dots~\cite{Press2008,Berezovsky2008} can be embedded into carefully designed optical cavities enabling very efficient extraction of emitted photons~\cite{Dousse2010}.

The electronically manipulated spins are rather viewed as the platform for quantum processor, as suggested in the seminal articles by Loss \& DiVincenzo~\cite{Loss1998} and Kane~\cite{Kane1998}. The leading impurity-based implementation of electrically controlled qubit is based on individual phosphorus donors in silicon~\cite{Kane1998,Pla2012,Laucht2015}. Meanwhile there are multiple competitive quantum dot realizations. Among these, the qubits based on nanowires~\cite{Nadj-Perge2010} or carbon nanotubes~\cite{Laird2013} are less promising, due to scaling difficulties related to arranging multiple nanowires or nanotubes in a deterministic manner. The most complex qubits are currently realized in GaAs heterostructures~\cite{Petta2005,Shulman2012}, however the scaling perspectives are limited by the decoherence related to interaction of the electron with the nuclear spin bath. Finally, in silicon quantum wells~\cite{Maune2012,Kawakami2014} and MOS structures~\cite{Veldhorst2015} this problem is eliminated. However so far the reliability of device fabrication in this systems is limited. A huge hope is that small adjustments to CMOS technology will be sufficient to create spin qubit devices with industrial processes~\cite{Maurand2016}.

In what follows I will focus on the description of techniques used to control GaAs spin qubits, however most of them can be readily applied to Si/SiGe and MOS qubits with the added value of improved coherence.

\subsection{Electron confinement GaAs nanostructures}

\begin{figure}
\begin{center}
\includegraphics[width=\textwidth]{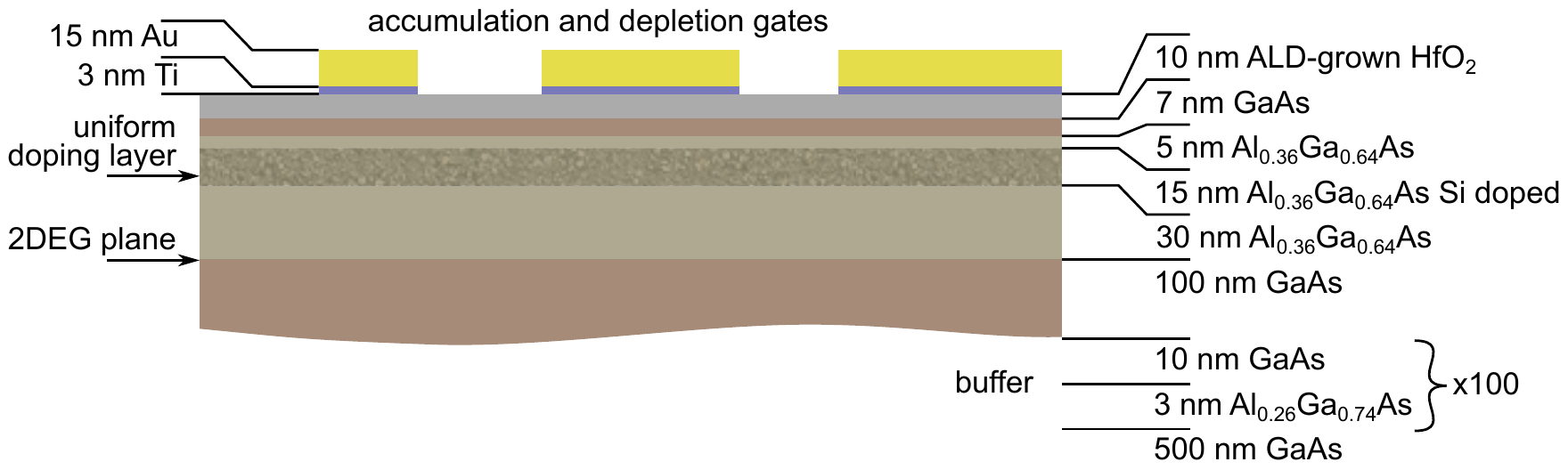}
\caption[Schematic of the uniformly doped GaAs/AlGaAs heterostructure]{
Schematic of the uniformly doped GaAs/AlGaAs heterostructure forming a 2-dimensional electron gas (2DEG) 57~nm below the surface, used in all experiments presented in this thesis. ALD-deposited HfO$_2$ insulating layer and and Ti/Au metallic gates are deposited in the further fabrication steps. Dimensions in the schematics are drawn to scale.
}
\label{qubits:heterostructure}
\end{center}
\end{figure}

Confinement of the electrons in the GaAs quantum dot is achieved by the combination of the band gap engineering and electrostatic gating. The 2-dimensional electron gas in this structure is formed at the interface between GaAs and AlGaAs crystals, on the GaAs side. It appears there as a result of attraction to the ionized silicon dopants located in the uniform layer above~\cite{Manfra2014} (Fig.~\ref{qubits:heterostructure}). The three key characteristics of the heterostructure essential for spin qubits applications are the following.

First, the 2DEG density should be such that the typical area per electron matches closely the expected size of the quantum dot. For a circular single electron quantum dot of 25~nm radius we obtain that 2DEG density should be approximately $2\times10^{-15}$~m$^{-2}$. Second, the distance to the surface should be as small as possible so the shape of potential created by the metallic gates deposited on top of the heterostructure was not excessively smoothened. In particular, the created potential variations should be much stronger than the intrinsic ones due to built-in inhomogeneity of the doping layer. Otherwise the positions of the dots will be accidental, rather that intentional. For the dot diameter of 50~nm the depth of the 2DEG smaller that $\sim$100~nm is desirable. Third, the 2DEG should have high mobility. In principle this condition is not necessary for quantum dot creation. However high mobility increases the dots tunability, is likely related to the sample stability. Moreover, the mobility may affect the performance of the parts of the device other than spin qubit itself (e.g. leads, especially the ones attached to radio-frequency sensor dots).

Further confinement of the electrons is performed by means of an adjustable electrostatic potential controlled by metallic gates located on top of the device. Due to small dimensions the gates are defined by electron beam lithography\footnote{Except for CMOS structures. The idea behind studying them is to use conventional industrial optical lithography techniques.}. Typically the gates are made of gold with thin sticking layer (e.g. titanium), however in the structures that require multilayer gating, aluminum is a common choice~\cite{Zajac2016} since native oxide that forms in contact with the air can serve as a naturally insulating layer.

\begin{figure}
\begin{center}
\includegraphics[scale=1]{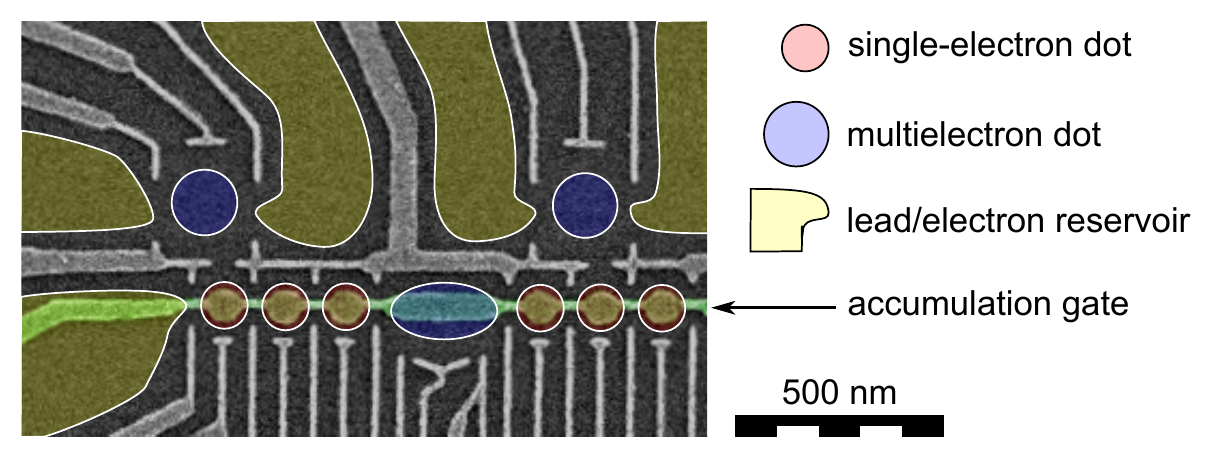}
\caption[Scanning electron micrograph of the device designed to create a linear array of quantum dots]{
Scanning electron micrograph of the device designed to create a linear array of quantum dots. Colors indicate intended locations of single- and multielectron dots as well as leads. Light gray structures are Ti/Au gates used to deplete the two-dimensional electron gas and confine the electrons. Light green structure is the accumulation gate attracting the electrons. From the fabrication perspective this gate is identical to all other gates.
}
\label{qubits:SEM}
\end{center}
\end{figure}

The gates are arranged so as to surround a small area of the 100-250~nm diameter\footnote{For n-type GaAs devices operated in single electron regime. Dimensions in different materials are chosen differently due to different effective mass of the carriers, to maintain tunability of the tunnel couplings and control the energy of the lowest excited orbitals. The actual dot size is smaller than gate to gate distance, since 2DEG is depleted within a certain radius around the gate.} as shown in figure~\ref{qubits:SEM}. The optional gate in the center, operated at positive voltage helps to create a deep potential well, but requires deposition of the insulating layer between the semiconductor and the gates to prevent leakage through the Schottky diode. The insulating layer has the positive side effect of blocking the tunneling events that give rise to the sample instabilities~\cite{Buizert2008}.

\section{Qubit implementations}
\label{qubits:implementations}

\emph{Note: the description of various qubit implementations is written under the assumption that they are realized in GaAs or other material with negative g-factor. This implies that states with spin up $\uparrow$ have lower energy than states with spin down $\downarrow$. The description of the implementations applies also to realization of the qubits in materials with positive g-factor, in which case one must replace every $\uparrow$ symbol with $\downarrow$ and vice versa.}

\subsection{Single spin (Loss-DiVincenzo qubit)}

\begin{figure}
\begin{center}
\includegraphics[scale=1.1]{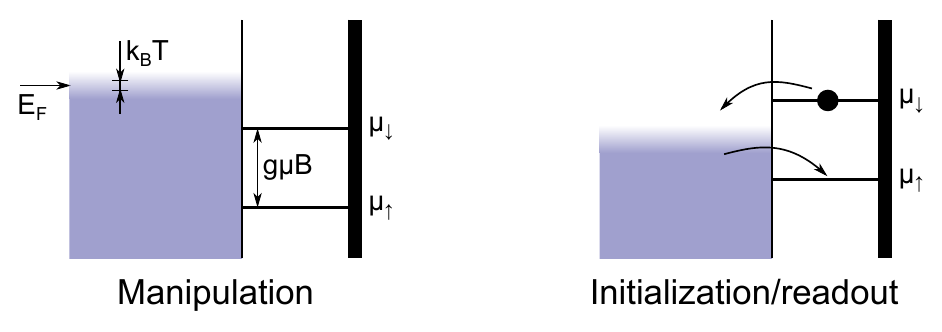}
\caption[Arrangement of chemical potentials during manipulation, initialization and readout of Loss-DiVincenzo qubit]{
Arrangement of chemical potentials during manipulation, initialization and readout of Loss-DiVincenzo qubit.
}
\label{qubits:LD}
\end{center}
\end{figure}

It is most natural to use the spin projections $\ket{\uparrow}$ and $\ket{\downarrow}$ as the two qubit states. However to name a system a qubit one needs to indicate methods of initialization, coherent manipulation and readout. For that reason it is necessary to apply an external magnetic field that Zeeman-splits the two states.

If the Zeeman splitting is larger than the typical energy of thermal fluctuations $k_BT$, the Loss-DiVincenzo qubit initialization can be realized by letting the system reach the thermal equilibrium. At this point one has to remember that relaxation at any other stage of computation is an undesired effect. Therefore one must design a scheme in which relaxation rate can be manipulated. It is usually achieved by adjusting the chemical potential of $\ket{\uparrow}$ and $\ket{\downarrow}$ relative to the Fermi level of the neighboring lead (Fig.~\ref{qubits:LD}).

In the manipulation stage the quantum dot is Coulomb blocked. In the initialization stage the chemical potential of $\ket{\downarrow}$ is moved above the Fermi energy of the leads~\cite{Elzerman2004}. In the latter case the electron in $\ket{\downarrow}$ state can tunnel out, and later an electron in $\ket{\uparrow}$ state can take its place. If the tunneling rates are sufficiently high the relaxation rates are significantly increased. Moreover, the tunneling events (or their absence) can be detected by the neighboring charge sensor which yields a readout mechanism.

Manipulation of the Loss-DiVincenzo qubit can be done either with oscillating magnetic fields~\cite{Koppens2006} (electron spin resonance, ESR) or with electric fields (electron dipole spin resonance, EDSR). To achieve the latter, there must be a mechanism which couples charge and spin degrees of freedom, i.e. spin-orbit interaction~\cite{Nowack2007} or a magnetic field gradient~\cite{Pioro-Ladriere2008}. The advantage of EDSR is that is easier to address specific, individual spins, by applying the radio-frequency excitation only to the gate located closely to one of the quantum dots. AC magnetic field is much more difficult to localize -- the excitation is created with an oscillating current through the macroscopic stripline. Addressability is achieved by detuning the resonant frequencies of different qubits - either with variations in local magnetic field created by micromagnets~\cite{Takeda2016} or by locally adjusting the electronic g-factor~\cite{Veldhorst2014}.

The two qubit gate between Loss-DiVincenzo qubit can be most easily realized with exchange interaction. Conventionally the exchange interaction is induced by tilting a potential of the double quantum dot~\cite{Petta2005}. This procedure admixes (2,0) to (1,1) charge occupations (where the two numbers indicate the number of electrons on the two dots). Due to Pauli exclusion principle the admixing process differs between singlet $\ket{S}=(\ket{\ud}-\ket{\du})/\sqrt{2}$ and any of the triplet states $\ket{T_0}=(\ket{\ud}+\ket{\du})/\sqrt{2}$, $\ket{T_+}=\ket{\uparrow\uparrow}$  and $\ket{T_-}=\ket{\downarrow\downarrow}$. In singlet configuration the two electrons are allowed to occupy the same orbital, while in triplet configuration added electron must occupy first excited state. This results in Heisenberg type interaction whose Hamiltonian has the form
\begin{equation}
	J \hat{S}_1 \cdot \hat{S}_2
\end{equation}
where $J$ indicates the exchange strength and $\hat{S}_i$ are the spin operators corresponding to the two electrons. Depending on whether the exchange interaction is turned on and off diabatically or adiabatically with respect to magnetic field gradients the interaction if this character can be used to perform a SWAP-like~\cite{Petta2005} of CPHASE-like gate~\cite{Veldhorst2015}.

Multiple alternative methods of inducing exchange interaction were considered up to date, with the goal if decreasing susceptibility of the system to charge noise during exchange gate and increasing the range of the interaction. These are discussed in detail in Parts~\ref{part:symmetrization} and~\ref{part:multielectron} of this thesis.

\subsection{Singlet-triplet qubit (S-T$_0$)}

\begin{figure}
\begin{center}
\includegraphics[width=\textwidth]{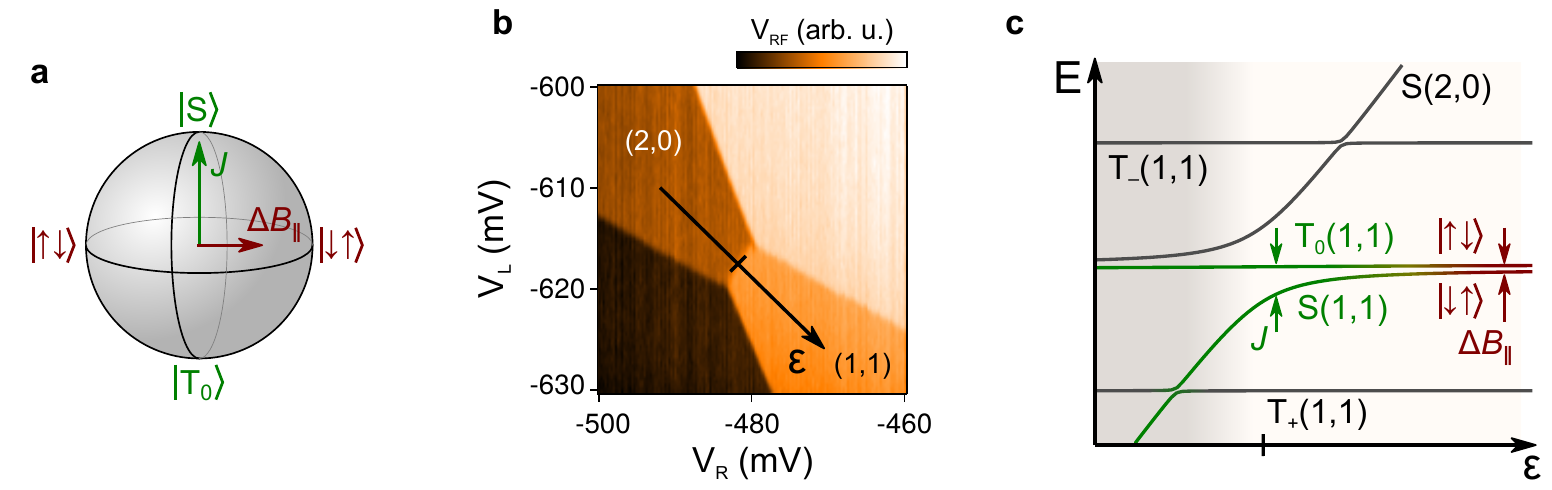}
\caption[Double dot as a S-T$_0$ qubit]{
Double dot as a S-T$_0$ qubit. (a) The Bloch sphere representation of the singlet-triplet qubit. The two rotation axis are defined by exchange splitting (green arrow) and gradient of the effective magnetic field (or more general -- difference of the Zeeman splitting; red arrow). (b) Representation of the detuning axis $\varepsilon$ in the double quantum dot charge diagram.
}
\label{qubits:S-T0}
\end{center}
\end{figure}

The singlet-triplet S-T$_0$ qubit is defined in a $S_z=0$ subspace of the two-electron double quantum dot, where $S_z$ is the total spin projection on the direction of the magnetic field. The two level system is therefore spanned by $\ket{S}=(\ket{\ud}-\ket{\du})/\sqrt{2}$ and $\ket{T_0}=(\ket{\ud}+\ket{\du})/\sqrt{2}$ states or, equivalently, by $\ket{\ud}$ and $\ket{\du}$ states (Fig.~\ref{qubits:S-T0}a).
In absence of the exchange energy the ifference in the Zeeman splitting between the two electrons sets $\ket{\ud}$ and $\ket{\du}$ to be the eigenstates and provides a first rotation axis (red colored arrow labeled $\Delta B_\parallel$). When the exchange interaction dominates over the difference between the Zeemans splitting the $\ket{S}$ and $\ket{T_0}$ are the system eigenstates, which defines the second rotation axis (green colored arrow labeled $J$).
Experimentally, the difference of the Zeeman splittings is usually fixed, set by the gradient of the external~\cite{Wu2014} or effective~\cite{Foletti2009} magnetic field, or the difference between g-factors in the two dots. Meanwhile the exchange strength can be controlled in a range of several orders of magnitude by switching between (2,0) and (1,1) charge configurations. For the experimentalist this is equivalent to moving along the detuning $\varepsilon$ axis showed in the charge diagram in Fig.~\ref{qubits:S-T0}b.

Figure~\ref{qubits:S-T0}c shows a simplified schematic of the spin states along the detuning $\varepsilon$ axis, in the vicinity of the charge transition. In the left side of the diagram, in (2,0) charge state, the exchange interaction dominates over the Zeeman splitting difference and the eigenstates of the system are common eigenstates of the total spin operator $\hat{S}$ and the operator of the total spin projection on the direction of the magnetic field $\hat{S}_z$. In the right side, in (1,1) charge state, the electrons are decoupled from each other so the eigenstates are the tensor products of individual spin states $\ket{\uparrow}$ and $\ket{\downarrow}$. In the middle region the eigenstates continuously change, which is used to incoherently convert $\ket{S}$ and $\ket{T_0}$ states into $\ket{\ud}$ and $\ket{\du}$, or vice versa.

Using the two spins to define a single qubit has the disadvantage of leaving several states unused, in this case the fully polarized triplet states $\ket{T_+}=\ket{\uparrow\uparrow}$ and $\ket{T_-}=\ket{\downarrow\downarrow}$. This introduces the threat of leakage error. Due to typical long relaxation times and large energy separation the leakage risk when performing single qubit gates appears only at the crossing between $\ket{S}$ and $\ket{T_+}$ states. This is usually avoided by performing subnanosecond, diabatic ``jumps'' between points on the $\epsilon$ axis at a significant separation from the crossing. 

The leakage errors become a significant problem when considering implementation of the two-qubit gate with the exchange interaction. This is because such interaction enables transfer of the energy between the two qubits and lifts the protection provided by energy conservation. In fact, the exchange interaction between the two electrons assigned to different qubits is a perfect leakage mechanism since it connects $\ket{\ud}\ket{\ud}$ and $\ket{T_+}\ket{T_-}$ states. This leakage can be suppressed by performing CPHASE-like gate in which the exchange between the electrons belonging to different qubits is turned on and off adiabatically with respect to Zeeman splitting of the correspoinding spins, in analogy to Ref.~\cite{Veldhorst2015}. In principle~\footnote{which is extremely hard to realize experimentally} it can be also suppressed by keeping the exchange interaction within the qubits finite.

Exploiting the dipole moment between $\ket{S}$ and $\ket{T_0}$ at the (2,0)-(1,1) charge transition provides alternative methods of performing two qubit gates. One has to take these ideas with a grain of salt, since use of the dipole moment entails increased susceptibility to the charge noise. The simplest way is to use direct dipole-dipole interaction between neighboring double dots~\cite{Shulman2012}. Although this scheme allows to achieve 90\% gate fidelities~\cite{Nichol2017}, such gates may turn out to be too slow to be practical in the long perspective.

Dipole moment also provides a mean for coupling the S-T$_0$ qubit to the microwave superconducting resonator~\cite{Mi2016,Stockklauser2017,Bruhat2017}, which can mediate interaction between qubits separated by milimiter-scale distances. In recent experiments the electron-cavity coupling was shown to be as strong as 240~MHz~\cite{Stockklauser2017} which can be translated to a potential gate time of about 4~ns.

\subsection{Exchange-only qubit and Resonant exchange qubit}
\label{qubits:XO&RX}

Both, Loss-DiVincenzo and singlet-triplet qubits require several different manipulation techniques to achieve a full single qubit control. Using a three-electron systems allows to perform all manipulations using only exchange interaction~\cite{DiVincenzo2000a}. This can be easily understood when one realizes that exchange interaction conserves the total spin. For single- and two- electron system all spin states differ either by amplitude of the total spin or its projection. Only addition of the third electron creates a subspace with two states having both of these quantum numbers identical.

\begin{figure}
\begin{center}
\includegraphics[width=\textwidth]{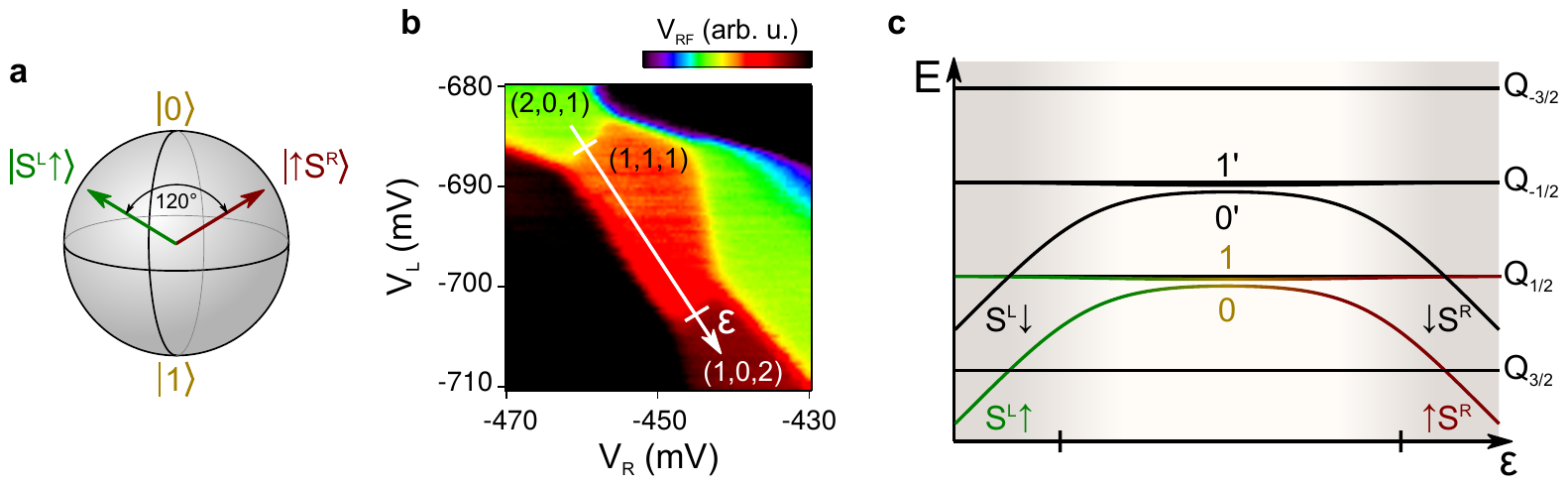}
\caption[Exchange-only qubit]{
Exchange-only qubit. (a) Bloch sphere representation showing the two rotation axis corresponding to only left or right exchange being non-zero.
(b) The triple quantum dot charge diagram showing a single detuning axis $\varepsilon$ connecting (2,0,1), (1,1,1) and (1,0,2) charge states.
(c) Energy diagram of the triple quantum dot with the two qubit states indicated with colored lines.
}
\label{qubits:XO}
\end{center}
\end{figure}

\begin{figure}
\begin{center}
\includegraphics[width=0.65\textwidth]{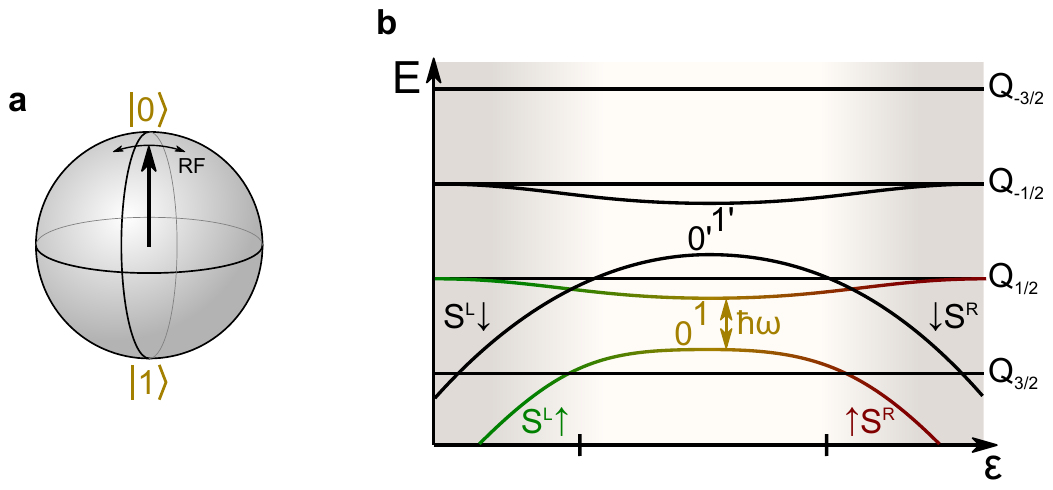}
\caption[Resonant exchange qubit]{
Resonant exchange qubit. (a) Bloch sphere representation showing the two rotation axes corresponding to only left or right exchange being non-zero.
(b) Energy diagram of the triple quantum dot with the two qubit states indicated with colored lines. Note that much larger range of $\varepsilon$ for which both, $J_L$ and $J_R$ are non-zero.
}
\label{qubits:RX}
\end{center}
\end{figure}

Exchange-only qubit and resonant exchange qubit are both defined in the three-electron triple quantum dot~\cite{Laird2010,Taylor2013}. The two qubit states are $\ket{0} = (\ket{\duu}-2\ket{\udu}+\ket{\uud})/\sqrt{6}$ and $\ket{1} = (\ket{\uud}-\ket{\duu})/\sqrt{2}$, which both belong to $S=1/2$, $S_z=-1/2$ subspace of the three-electron system. These are the eigenstates when electron in the middle quantum dot interacts equally strongly with the electrons located in the left and right dot. By changing ratio between the two exchange interactions it is possible to change quantization (or rotation) axis by up to 120$^\circ$. This is illustrated in Fig.~\ref{qubits:XO}a with the two arrows showing rotation axes in the extreme case of only left or right exchange being non-zero.

Readout and initialization of the triple dot qubits can be performed using spin-blockade, in a similar manner to singlet-triplet qubit. For extreme negative values of $\varepsilon$ two qubit eigenstates are $\ket{S^L \uparrow}$ and $\ket{T_0^L \uparrow}/\sqrt{3} - 2\ket{T_+^L \downarrow}/\sqrt{6}$. Here the first symbol indicates spin state of the electrons located on the middle and the left quantum dot, while the arrow represents the spin of the electron in the right dot. One can notice that the middle-left electron pair is either in a pure singlet or pure triplet configuration, giving rise to Pauli blockade and enabling spin-to-charge conversion, as well as initialization via relaxation to the ground state.

Exchange-only qubit is usually operated by adjusting gate voltages along a single axis~\cite{Medford2013,Eng2015} (called detuning $\varepsilon$), between (2,0,1), (1,1,1) and (1,0,2) electron occupancies (Fig.~\ref{qubits:XO}b). To better understand the methods of operation it is helpful to analyze the energy diagram of spin states along $\varepsilon$ (Fig.~\ref{qubits:XO}c). As visualized with the two colored lines, one can control the rotation axis by changing $\varepsilon$.

Resonant exchange is a variation of the exchange-only qubit. The qubit subspace is the same, however resonant exchange is operated in the regime where both $J_L$ and $J_R$ are large (on the order of hundreds of megahertz), while operations are performed by means of RF excitation applied to one of the gates~\cite{Medford2013a} (Fig.~\ref{qubits:RX}). The RF drive causes a small tilt of the quantization axis which leads to coherent precession between $\ket{0}$ and $\ket{1}$, if the RF frequency is adjusted to match the qubit splitting. Finally, adjusting the phase of the excitation enables the universal two-axis control in the rotating frame.

Larger number of electrons used to define a single qubit results in in increased number of leakage states. Similarly to singlet-triplet qubit, leakage to the states with different spin projection on the external magnetic field direction does not pose a significant threat. However, the state with $S=3/2$, $S_z=1/2$ (labeled $Q_{-1/2}$ in Figs.~\ref{qubits:XO} and~\ref{qubits:RX}) is energetically split from the qubit states only when \emph{both} $J_L$, $J_R>0$, while the leakage to this state is driven by the gradients of the Overhauser field.

The leakage becomes even more limiting when discussing the two qubit gates realized by means of the exchange interaction. In principle it is possible\footnote{experimentally it is virtually impossible} to exploit the exchange interaction for the two qubit gates but the this requires very tuning of multiple exchanges at the same time~\cite{Doherty2013} (while not being able to measure them separately) or performing very long sequences to suppress leakage~\cite{Setiawan2014}.

In principle it is also possible to use a dipole moment to couple such qubits using dipole-dipole interaction or via superconducting cavity~\cite{Srinivasa2016} but the required coherence time are currently far beyond the experimental state of the art.

\subsection{Hybrid (three-electron double dot) qubit}

\begin{figure}
\begin{center}
\includegraphics[width=0.65\textwidth]{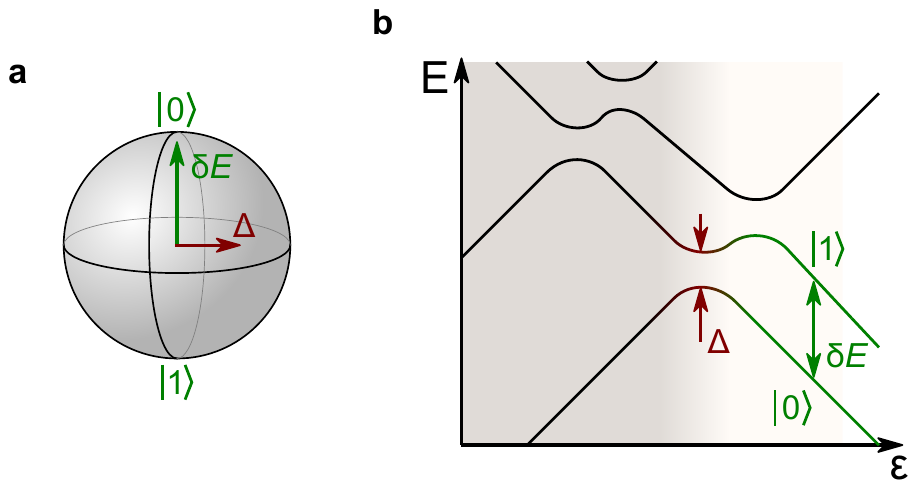}
\caption[Hybrid (three-electron double dot) qubit]{
Hybrid (three-electron double dot) qubit. (a) Bloch sphere representation showing the two rotation axis. The green colored rotation axis corresponds to the level spacing of one of the dots.
(b) Energy diagram of the $S=1/2$, $S_z=1/2$ subspace of the three-electron double quantum dot, where one of the dots has the level spacing significantly different than another.
}
\label{qubits:hybrid}
\end{center}
\end{figure}

The final implementation of the quantum dot based spin qubit is the so-called hybrid qubit, realized in a three-electron double quantum dot. Similarly to exchange-only and resonant exchange qubit it is defined in $S=1/2$, $S_z=1/2$ subspace of the three-spin system~\cite{Shi2012}. However it is operated in the regime of where two of the electrons are located on the first dot (with smaller level spacing) while the third one resides on another dot (with larger level spacing). This results in the qubit eigenstates $\ket{0} = \ket{S^L \uparrow}$ and $\ket{1} = 1/\sqrt{3}\ket{T_0^L \uparrow} -\sqrt{2/3} \ket{T_+^L \downarrow}$ being split by the level spacing in the dot containing two electrons.

The qubit benefits from the fact that the level spacing that defining the qubit splitting in the idle configuration (green in Fig.~\ref{qubits:hybrid}) is virtually insensitive to small changes in gate voltages, in particular is insensitive to charge noise\footnote{To the best of my knowledge the literature~\cite{Koh2012,Kim2014,Shi2012,Kim2015,Cao2016} does not mention the possibility of the leakage from the qubit space to the single state with $S=3/2$ and $S_z=1/2$, i.e. $\ket{Q_{1/2}} = \sqrt{2/3}\ket{T_0^L \uparrow} + 1/\sqrt{3} \ket{T_+^L \downarrow}$. The leakage to that state is driven by the gradient of the Overhauser field which is not suppressed by any mechanism. One can observe that the operation position of the hybrid qubit is analogous to the extreme (1,0,2) configuration of exchange-only qubit, where $\ket{1}$ and $\ket{Q_{1/2}}$ state are degenerate (Fig.~\ref{qubits:XO}). The coherence time measured in Refs.~\cite{Kim2014,Kim2015,Cao2016} is below the timescale related to the leakage and therefore does not provide an experimental counterevidence to the leakage.}. On the other hand, inducing the exchange interaction with the third electron results in a change of the eigenstates and enables the universal qubit control~\cite{Kim2014}. At the same time, large level spacing in the dot that is usually occupied by a single electron enables initialization and readout in the manner analogous to the exchange-only qubit.

The fact that the hybrid qubit consists of the three electrons allows to employ most of the two qubit exchange-based coupling schemes from exchange-only qubit. Additionally, the qubit splitting given by the level spacing, which can effectively suppress some of the leakage paths~\cite{Shi2012}. At the same time the definition of this qubit in the double dot also enable the use of a dipole moment in a manner analogous to the singlet-triplet qubit.

\part{Symmetric operation of spin qubits}
\label{part:symmetrization}

\chapterimage{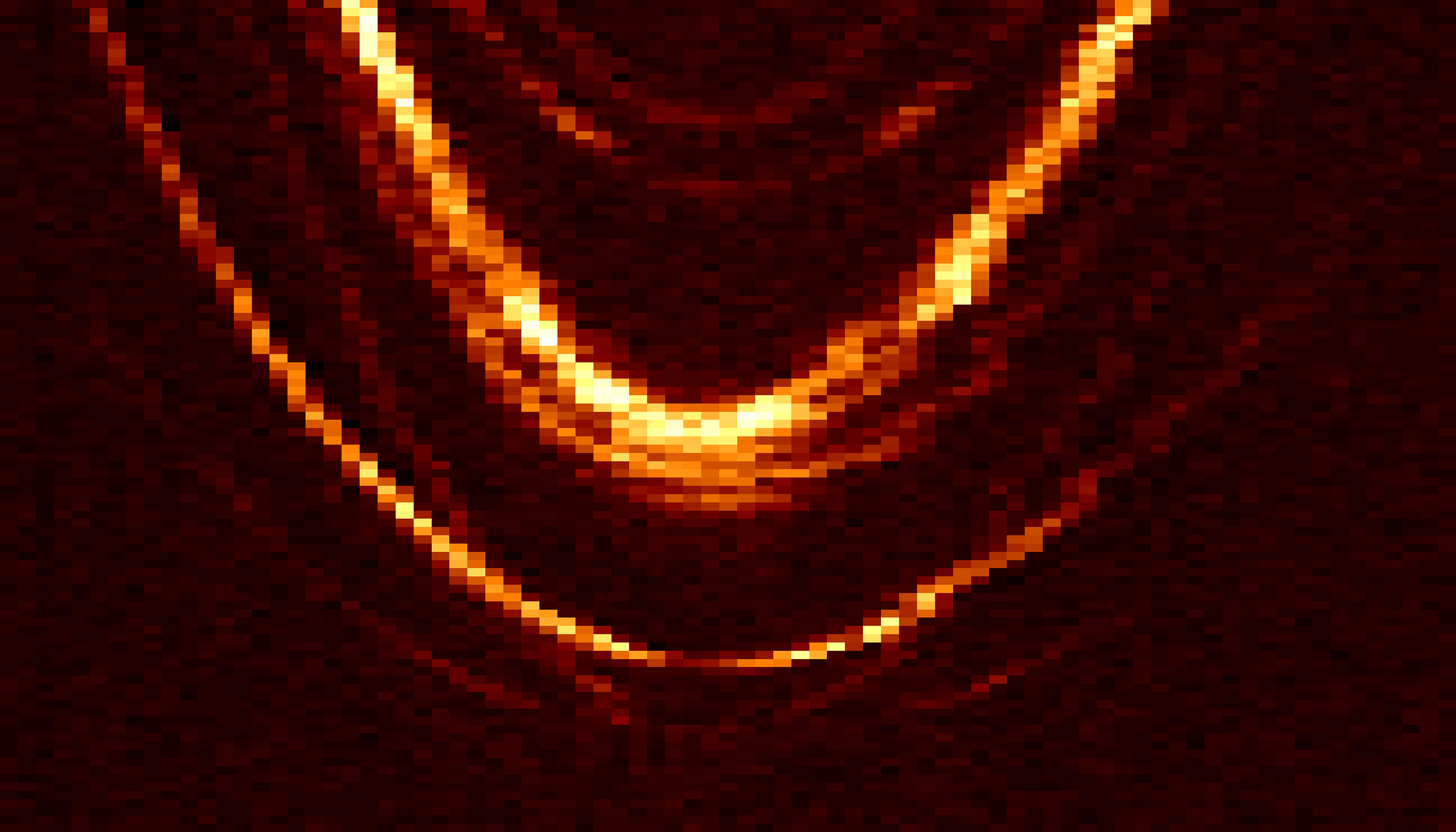}
\chapter[The principle of symmetrization]{\protect\parbox{0.9\textwidth}{The principle of symmetrization}}
\label{ch:symmetrization}

Charge noise poses a huge obstacle in the development of spin qubits. Depending on the qubit implementation it can affect about every parameter that is used to tune the qubit frequency. For example it affects the Zeeman splitting and the EDSR drive strength by moving the electrons wavefunction in the gradient of the external magnetic field~\cite{Takeda2016,Kawakami2016} or effectively modifying the g-factor~\cite{Muhonen2014,Maurand2016}. The charge noise effects are most pronounced whenever the exchange interaction is induced between the two neighboring electrons. The exchange interaction arises from the combination of the Pauli exclusion principle and the Coulomb interaction or confinement of the electrons. Crucially, in the presence of exchange different spin states of the two electrons are characterized by different charge distributions which opens the channel for the electric noise to affect the spins.

\section{Coupling of the charge noise to the tilted double quantum dot}

\begin{figure}[t]
\begin{center}
\includegraphics[width=0.9\textwidth]{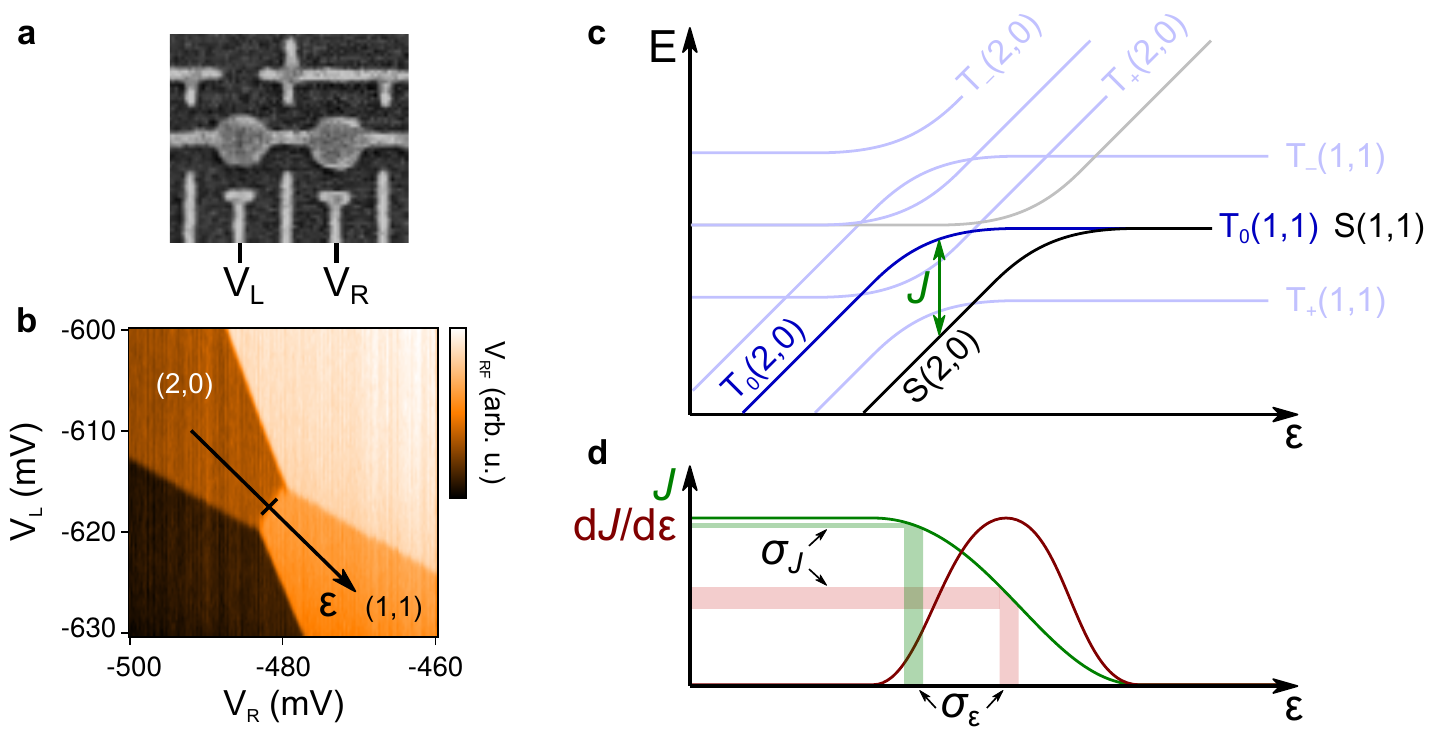}
\caption[Charge diagram and the energy diagram of the spin states of the double quantum dot in the vicinity of the (2,0)-(1,1) charge transition]{
(a) SEM of the double quantum dot device
Charge diagram (b) and the energy diagram of the spin states (c) of the double quantum dot in the vicinity of the (2,0)-(1,1) charge transition.
(d) Exchange interaction strength and the susceptibility to the detuning noise.
}
\label{symmetrization:tilt}
\end{center}
\end{figure}

The canonical example of how the charge noise affects the exchange interaction strength is the two-electron double quantum dot (Fig.~\ref{symmetrization:tilt}a), where the exchange $J$ is induced by the potential tilt. The tilt is realized by changing the difference between the gate voltages controlling the occupancies of the two dots ($\varepsilon$ axis in the charge diagram presented in Fig.~\ref{symmetrization:tilt}b). Figure~\ref{symmetrization:tilt}c shows a schematic energy diagram of the two-electron charge-spin states in the vicinity of the charge transition between (2,0) and (1,1) occupancies~\cite{Hanson2007}, where $(N,M)$ indicates the number of the electrons in the left and the right dot. The lines indicated with the strong contrast correspond to the low energy singlet $S$ and unpolarized triplet $T_0$ states which are to be controlled. For large positive values of the detuning $\varepsilon$ the two electrons reside on different dots and are virtually decoupled. In particular the exchange splitting $J$ between $S$ and $T_0$ states is equal to zero and completely insensitive to charge noise. In these conditions other decoherence mechanisms, such as Overhauser noise (Part.~\ref{part:nuclei}), dominate the dephasing.

In the large negative $\varepsilon$ limit (the two electrons residing on a single quantum dot) $S$ and $T_0$ states are split by (approximately) the spacing between the ground and a first excited orbital of the quantum dot. That is the case since in a singlet configuration both electrons occupy the ground orbital, while in the triplet configuration one of the electrons is forced to occupy excited orbital due to Pauli exclusion principle. This splitting, typically of the order of several gigahertz (depending on the material and the size of the dot) is only weakly susceptible to charge noise. The insensitivity is a consequence of the fact that the difference in the charge distribution corresponding to different orbitals in the same dot is very small. The residual susceptibility to noise, which is not a current concern of the community and is beyond the scope of this thesis, arises due to the dependence of the level spacing on the changes in the confining potential.

For the intermediate values of $\varepsilon$ the exchange interaction strength gradually increases. Starting from large $\varepsilon$, first the singlet state changes the character from (1,1) to (2,0). This leads to an increase of both, the singlet-triplet splitting (Fig.~\ref{symmetrization:tilt}d) and the the difference in the charge distribution between these two states. The latter implies an increase of the susceptibility to the charge noise (Fig.~\ref{symmetrization:tilt}d), since the local variation of the confining potential will affect the two spin states to a different extent. As $\varepsilon$ becomes more negative, the triplet states also change character from (1,1) to (2,0). Consequently the difference in charge distribution between singlet and triplet states is reduced and susceptibility to noise decreases.

\section{Detuning noise}

The quantitative analysis of the decoherence usually assumes that the characteristic length scale of the charge noise is larger than the physical size of the double quantum dot. Since the level spacing within dots and the tunnel coupling are invariant with respect to global shifts of the potential, it follows that the tilts of the potential are the dominant kind of noise affecting the singlet triplet splitting. For that reason the noise can be effectively represented as an uncertainty in the detuning control parameter $\varepsilon$.

Considering the effective quasistatic detuning noise, characterized by the rms value $\sigma_\varepsilon$, we observe that the susceptibility to the charge noise is given by the derivative of the singlet-triplet splitting with respect to detuning $\mathrm{d}J/\mathrm{d}\varepsilon$ (Fig.~\ref{symmetrization:tilt}d). The resulting formula for the decoherence time due to charge noise is
\begin{equation}
	T_2^* = \frac{1}{\sqrt{2}\pi\sigma_J} = \frac{1}{\sqrt{2}\pi\frac{\mathrm{d}J}{\mathrm{d}\varepsilon} \sigma_\varepsilon},
\end{equation}
where $\sigma_J$ is the rms exchange noise, turns out to be very accurate whenever the exchange interaction in a double quantum dot is induced by the tilt of the potential~\cite{Dial2013,Higginbotham2014,Martins2016}.

Such a basic model provides a simple recipe for reducing the dephasing while performing the exchange oscillations -- to reduce the dependence of the exchange $J$ on the detuning at the operating position. The energy diagram and the plot of exchange strength with respect to detuning $\varepsilon$ (Fig.~\ref{symmetrization:tilt}c,d) reveals two such regions, mentioned in the previous section. These are deep Coulomb blockade in (1,1) and (2,0) charge configuration. However both of them present significant disadvantages.

When operating in the (2,0) configuration the oscillations have such a large frequency ($\gg~1$~GHz) that conventionally used arbitrary waveform generators (with $\approx~1$~GHz bandwidth) do not provide sufficient control to perform rotations by a small angle and take advantage of high fidelity oscillations. On the other hand in the (1,1) charge configuration the exchange interaction is virtually 0, and no exchange oscillations take place.

The symmetrization principle is the variation to the latter of this approache, where the reduction of the exchange strength is compensated by an increase in the tunnel coupling to reach required strength of the exchange interaction.

\section{The symmetrization principle}
\label{symmetrization:principle}

We will begin by proving that for a double quantum dot there exists a value of the detuning $\varepsilon$ for which the susceptibility to detuning noise $\mathrm{d}J/\mathrm{d}\varepsilon = 0$. For that purpose we observe that, starting in the (1,1) electron configuration, exchange interaction can be increased by tilting the potential either towards the (2,0) or (0,2) electron configuration. In the first of these cases the exchange increases for decreasing $\varepsilon$, that is $\frac{\mathrm{d}J}{\mathrm{d}\varepsilon}<0$, while in the second case $\frac{\mathrm{d}J}{\mathrm{d}\varepsilon}>0$. By continuity of $\frac{\mathrm{d}J}{\mathrm{d}\varepsilon}$ there must be exist a value of $\varepsilon$ for which the susceptibility to the detuning noise vanishes.

Interestingly, one can also show easily that the point with vanishing susceptibility lies exactly in the center of the (1,1) region, provided that the level spacing of each of the dots is much larger than the tunnel coupling. This can be most easily seen when the tunnel coupling $t$ is much smaller than the energy difference between singlet states consisting of two singly occupied dots $E_{S(1,1)}$ and singlet states consisting of the one doubly occupied dots: $E_{S(2,0)}$ and $E_{S(0,2)}$. In this approximation the energy of the triplet state $T(1,1)$ is 0, while the energy of the singlet state $S(1,1)$ is slightly reduced due to admixtures of singlets with doubly occupied dots
\begin{equation}
	\tilde{E}_{S(1,1)} = -t^2 \left( \frac{1}{E_{S(2,0)} - E_{S(1,1)}} + \frac{1}{E_{S(0,2)} - E_{S(1,1)}}\right)
\end{equation}
where the tilde symbol indicates the energy that includes the first order correction due to the tunneling $t$ between the dots. Energy differences can be expressed in terms of the detuning between single particle levels in the two dots\footnote{note that here we use the theoretical definition of detuning $\varepsilon$ (the difference of the energies of the two single particle orbitals) rather than experimental definition in terms of gate voltages} $\varepsilon$ and charging energies of the two dots $U_{L/R}$
\begin{equation}
	\tilde{E}_{S(1,1)} = -t^2 \left( \frac{1}{U_L+\varepsilon} + \frac{1}{U_R-\varepsilon}\right).
\end{equation}
The symmetry of this formula implies that $\tilde{E}_{S(1,1)}$ has an extremum at $\varepsilon=(U_R-U_L)/2$, which is exactly in the middle between the (2,0)-(1,1) charge transition (which lies at $\varepsilon=-U_L$) and the (1,1)-(0,2) charge transition (at $\varepsilon=U_R$). Finally, within the validity range of our approximations the splitting between singlet and triplet states at the symmetry point can be tuned by adjusting the tunnel coupling.

\begin{figure}[t]
\begin{center}
\includegraphics[width=0.6\textwidth]{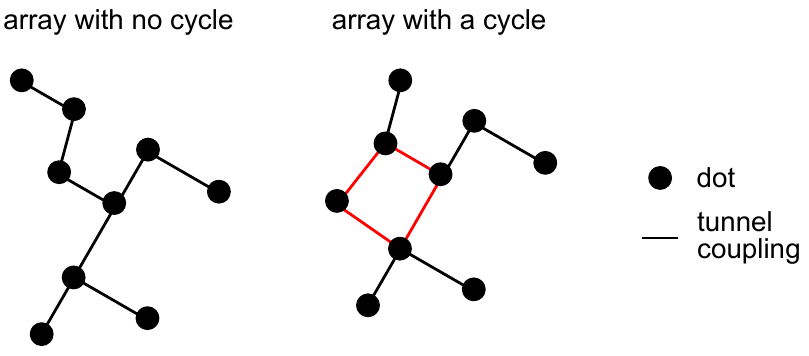}
\caption[Illustration of quantum dots array with and without any cycles]{
Illustration of quantum dots array with and without any cycles. In an array without cycles one can always find a tuning with every exchange splitting independent on detuning.
}
\label{symmetrization:cycles}
\end{center}
\end{figure}

This result can be extended to any array of singly-occupied tunnel-coupled quantum dots which has no cycles (a cycle is a group of dots between which the electron can tunnel in a circle, as opposed to the tree-like structure; Fig.~\ref{symmetrization:cycles}). To prove that we start with a double quantum dot and a new dots, one by one, to the existing array. In an array without a cycle the new dot is always tunnel coupled to only one of the old dots. Therefore we can adjust the detuning (and find the noise-insensitive point for the newly introduced exchange interaction) by setting correctly the single particle energy in the new dot. In the limit of small tunnel couplings this does not influence the fine tuning of all other exchange interactions in the array. One can also observe that in such case the noise-insensitive point lies in the geometrical center of the region with each dot being singly occupied. Notably it is irrelevant whether the tunnel couplings between pairs of dots are identical.

The two particular examples of the dot array without a cycle are: (a) the three-electron triple quantum dot, in which one can implement the resonant exchange or exchange-only qubit~\cite{Russ2015,Shim2016a}, and (b) a central quantum dot surrounded by three tunnel-coupled dots, which is a basis of the proposed exchange-only singlet-only qubit~\cite{Sala2017}.

This theoretical result has two limitations for practical applications. First, the control of the quantum dots is performed by means of gate voltages that couple not only to the chemical potential of the closest dot, but also to other dots, and to the tunnel couplings. Second, the charge noise couples to the tunnel couplings.

\section{Effective gate voltage noise}

The tunnel couplings, controlled by dedicated gates, are the second knob the experimentalist has to tune the array of the quantum dot, next to the chemical potential of the dots (in particular -- detuning). It is also the second most significant channel through which the charge noise affects the splitting between the singlet and triplet configuration of the electron pairs, and the one that dominates when the detuning noise is suppressed by means of the symmetrization.

To include the tunnel coupling noise in an experimentally accessible manner it is convenient to use the effective gate-voltage noise. It relies on the assumption that the charge noise experienced by the qubit is either an actual gate voltage noise, or that the potential fluctuations have similar length scales to the changes in the potential introduced by modifications of the gate voltages. In this model the dephasing time due to the charge noise is given by
\begin{equation}
	T_2^* = \frac{1}{\sqrt{2}\pi\sigma_J} = \frac{1}{\sqrt{2}\pi | \mathrm{d}J /\mathrm{d}\vec{V}| \sigma_V}.
\end{equation}
Here the exchange splitting is considered to be a function of the vector of the gate voltages applied on individual gates $J(\vec{V})$ and $\sigma_V$ is the rms effective gate voltage noise. The susceptibility to the charge noise is given by the gradient of the amplitude of the exchange at the operating point $| \mathrm{d}J /\mathrm{d}\vec{V}|$.

Within this model a reduced sensitivity of the exchange splitting to the charge noise is an experimental observation concerning tunability of the chemical potentials and the tunnel coupling. For the typical GaAs quantum dot devices the lever arm between the dedicated gate and the chemical potential of the dot is much larger than the lever arm between the gate voltage and the tunnel coupling. This observation implies that the symmetric operation of the spin qubits is beneficial~\cite{Martins2016,Reed2016}.

\section{Experimentally finding the optimal working point}

The ultimate question about the utility of the optimal symmetric configurations concerns the experimental methods used to locate them. This task is relatively easy in the case of the double quantum dot, but its complexity increases very quickly for larger quantum dot arrays.

In the double dot there is only a single exchange splitting and its value corresponds to the frequency of the exchange oscillations which can be measured directly~\cite{Petta2005}. This enables, for example, to apply the iterative gradient descent algorithm to converge to the optimal tuning. Moreover, the dimensionality of the gate voltage space that needs to be searched is relatively small (3) as the only relevant parameters that can be adjusted are the chemical potentials of the two quantum dots and the tunnel coupling between them. This enables a mapping out of the full topography of the exchange splitting with respect to gate voltages with only a few fingerprint measurements~\cite{Reed2016} (Sec.~\ref{symm_sup:mapping}).

Neither of these approaches is directly applicable to the triple quantum dot case (Ch.~\ref{ch:SRX}). The gradient descend approach does not work, because one must simultaneously optimize two parameters (susceptibility to the charge noise of the two exchange splittings) while being able to measure only a combination of the two (the qubit splitting). On the other hand the dimension of the parameter space in this problem increases to 5 (3 chemical potentials and two tunnel couplings) which makes it impossible to fully map out the dependence of the qubit splitting with respect to all gate voltages within a reasonable time.

In the study presented in chapter~\ref{ch:SRX} we assumed that the experimentally controlled $\varepsilon$ and $\delta$ parameters (defined by Eq.~\ref{SRX:parameters1}) couple only to the chemical potential of the dots. Moreover we decided to equalize the two exchange splittings only approximately. With these two simplifications the optimization procedure reduces to setting the gradient of the qubit splitting with respect to $\varepsilon$ and $\delta$ to zero, while ensuring that the qubit splitting is nearly symmetric with respect to flipping the sign of $\varepsilon$. This turned out to be sufficient to reduce the susceptibility to the effective gate voltage noise by one order of magnitude. Finding the approximate optimal points in larger quantum dot arrays is most likely possible as well, but it would require a much larger number of practical assumptions.

\chapterimage{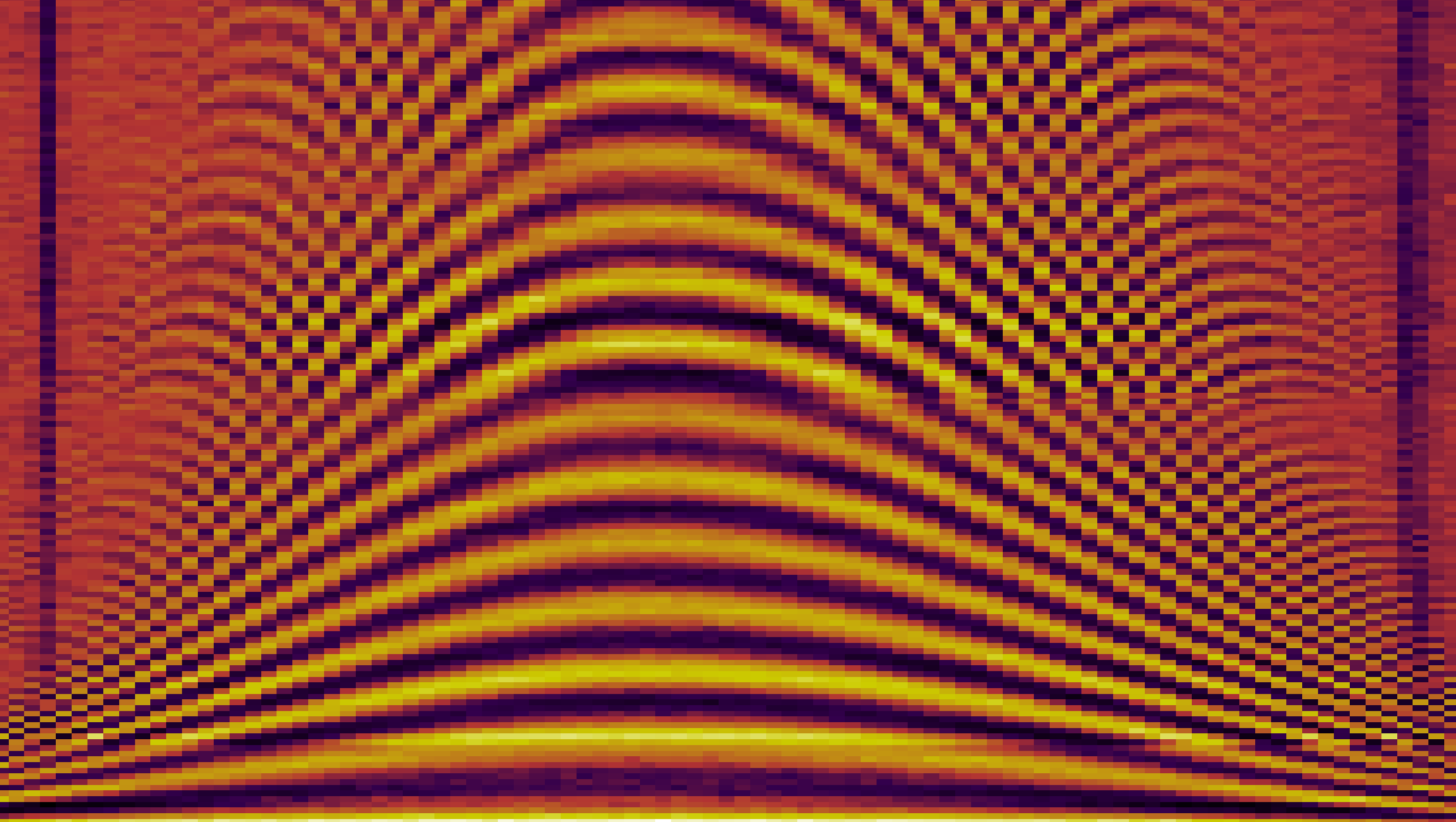}
\chapter[Noise suppression using symmetric exchange gates in spin qubits]{\protect\parbox{0.9\textwidth}{Noise suppression using symmetric\\ exchange gates in spin qubits}}
\label{ch:symm}

{\let\thefootnote \relax\footnote{This chapter and chapter \ref{ch:symm_sup} are adapted from Ref. \cite{Martins2016}. \copyright~(2016) by the American Physical Society.}}
\addtocounter{footnote}{-1}

\begin{center}
Frederico Martins$^{1,*}$, Filip K. Malinowski$^{1,*}$, Peter D. Nissen$^{1}$, Edwin Barnes$^{2,3}$, \\
Saeed Fallahi$^{4}$, Geoffrey C. Gardner$^{5}$, Michael J. Manfra$^{4,5,6}$, \\
Charles M. Marcus$^{1}$, Ferdinand Kuemmeth$^{1}$
\end{center}

\begin{center}
	\scriptsize
	$^{1}$ Center for Quantum Devices, Niels Bohr Institute, University of Copenhagen, 2100 Copenhagen, Denmark\\
	$^{2}$ Department of Physics, Virginia Tech, Blacksburg, Virginia 24061, USA\\
	$^{3}$ Condensed Matter Theory Center and Joint Quantum Institute, Department of Physics, \\
	University of Maryland, College Park, Maryland 20742-4111, USA\\
	$^{4}$ Department of Physics and Astronomy and Birck Nanotechnology Center, Purdue University, West Lafayette, Indiana 47907, USA\\
	$^{5}$ School of Materials Engineering and Birck Nanotechnology Center, Purdue University, West Lafayette, Indiana 47907, USA\\
	$^{6}$ School of Electrical and Computer Engineering, Purdue University, West Lafayette, Indiana 47907, USA\\
	$^{*}$ These authors contributed equally to this work
\end{center}

\begin{center}
\begin{tcolorbox}[width=0.8\textwidth, breakable, size=minimal, colback=white]
	\small
	We demonstrate a substantial improvement in the spin-exchange gate using symmetric control instead of conventional detuning in GaAs spin qubits, up to a factor-of-six increase in the quality factor of the gate. For symmetric operation, nanosecond voltage pulses are applied to the barrier that controls the interdot potential between quantum dots, modulating the exchange interaction while maintaining symmetry between the dots. Excellent agreement is found with a model that separately includes electrical and nuclear noise sources for both detuning and symmetric gating schemes. Unlike exchange control via detuning, the decoherence of symmetric exchange rotations is dominated by rotation-axis fluctuations due to nuclear field noise rather than direct exchange noise. 
\end{tcolorbox}
\end{center}

\section{Introduction}

Spin qubits, basic units of quantum information built from the spin states of electrons in solid-state systems, are one of the most promising realizations of a qubit~\cite{Kloeffel2013}. This is due to their potential for minituarization, scalability and fault tolerance~\cite{Taylor2005, Awschalom2013}. In fact, experiments in recent years have demonstrated remarkable progress in the coherent manipulation of single- and multi-spin devices~\cite{Foletti2009,Bluhm2010,Petta2010,Shulman2012}.
Nevertheless, one of the main difficulties with spin qubits, and more generally with solid-state qubits, is the decoherence due to interactions with the environment.
In the case of electron spins confined in semiconductor quantum dots, two main types of environmental noise limit coherence: electrical noise and hyperfine interactions with nuclear spins in the surrounding lattice~\cite{Bluhm2011,Dial2013,Barthel2012}. 
To reach the high control fidelities necessary for quantum computing, the coupling between a quantum dot spin qubit and its environment can be reduced by the use of sweet spots~\cite{Wong2015, Hiltunen2015, Fei2015}, and pulse errors can be reduced by bootstrap tomography~\cite{Dobrovitski2010, Cerfontaine2014}.

A crucial component of any spin-based quantum computing platform is strong spin-spin interaction.
In their seminal article, Loss and DiVincenzo proposed that exchange interactions between electron spins could be controlled by the height of the tunnel barrier between neighboring quantum dots~\cite{Loss1998}.
However, until recently this proposal was not implemented in the laboratory, and instead exchange interactions were induced by raising or lowering the potential of one dot relative to the other, an approach referred to as tilt or detuning control~\cite{Petta2005}.  Unlike the dot-symmetric tunnel barrier control method, tilt control affects the two dots asymmetrically and hybridizes the (1,1) and (0,2) charge states. 
Here numbers within each parenthesis denote occupation number of the left dot and right dot. 
In Fig.~\ref{symm:fig1}(a) we illustrate the difference between the two methods. Firstly, a singlet state (0,2)S is prepared (P). 
Thereafter the electrons are adiabatically separated to the $\ket{\uparrow\downarrow}$ state in the (1,1) charge configuration.  At the exchange point (X), a pulse is performed. For the tilt case, during this pulse the wavefunctions of the electrons are brought together by asymmetrically deforming the confining potential of the dots. In the case of the symmetric mode of operation, the exchange interaction is increased by lowering the potential barrier between the two dots. Finally, reversing the slow adiabatic passage first projects the final two-spin state onto $\ket{\uparrow\downarrow}$ and then maps it onto (0,2)S, which is then read out at the measurement point (M).

In this Letter, we demonstrate rapid, high-quality exchange oscillations implemented by pulsing the barrier between two dots, as envisioned in the original Loss-DiVincenzo proposal. 
We also show that, unlike tilt-induced qubit rotations, the coherence of barrier-induced rotations is not limited by electrical detuning noise, but rather by nuclear spin fluctuations parallel to the applied magnetic field. 
We quantify the improvements by studying exchange oscillations within a singlet-triplet qubit, corresponding to $\sqrt{\mathrm{SWAP}}$ operations between the two spins. 
Alternatively benchmarking of single-qubit gate fidelities is in principle possible but requires nuclear programming~\cite{Foletti2009}. 
Recent work on surface acoustic waves and silicon triple quantum dots showed results consistent with some of our observations~\cite{Bertrand2015, Reed2016}, indicating that symmetric exchange finds applications beyond GaAs qubits. 

\section{The device and pulse sequences}

\begin{figure}[tbh]
\begin{center}
\includegraphics[width=96 mm]{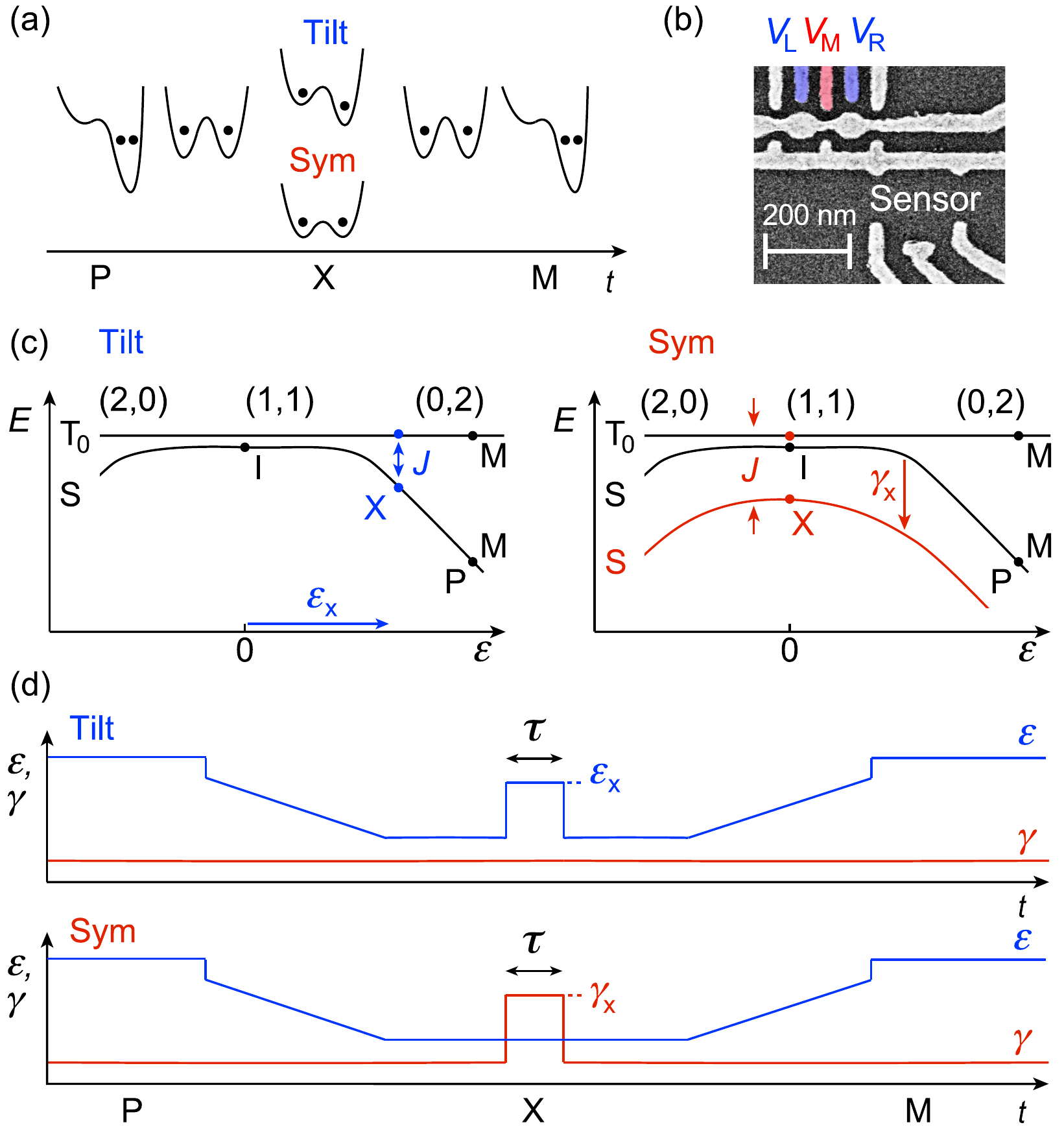}
\caption[Schematic comparison of detuning (tilt) and symmetric exchange pulse sequences]{
(a) Schematic comparison of detuning (tilt) and symmetric exchange pulse sequences, showing double-dot potentials and dot occupancies.
Tilt: wave function overlap controlled by detuning the confining potential; Symmetric: wave function overlap controlled by lowering the potential barrier between dots.
(b) Electron micrograph of the device consisting of a double dot and charge sensor. Note the gate that runs through the center of the dots. A 10 nm HfO$_2$ layer is deposited below the gates to allow positive and negative gating. 
High-bandwidth lines are connected to left and right plungers gates, $V_{\mathrm{L}}$, $V_{\mathrm{R}}$ (blue), and the middle barrier gate,  $V_{\mathrm{M}}$ (red).
(c) Energy diagrams of the two-electron spin singlet, S, and spin-zero triplet, T$_0$,  as a function of detuning $\varepsilon$.
(Left) Tilt mode:  Exchange, $J$, is controlled by detuning $\varepsilon$, set by $V_{\mathrm{L}}$ and $V_{\mathrm{R}}$; (Right) Symmetric mode: $J$ is controlled interdot coupling, $\gamma$, set by $V_{\mathrm{M}}$ (red curve).
(d) Pulse sequences for tilt and symmetric modes, with amplitudes $\varepsilon_{\mathrm{x}}$ and  $\gamma_{\mathrm{x}}$ during the exchange pulse, respectively.
}
\label{symm:fig1}
\end{center}
\end{figure}

The double quantum dot device with integrated charge sensor \cite{Barthel2009} is shown in Fig.~\ref{symm:fig1}(b). 
The device was fabricated on a GaAs/AlGaAs heterostructure 57~nm below the surface, producing a two-dimensional electron gas with bulk  density $n$~=~$2.5\times10^{15}$~m$^{-2}$ and mobility  $\mu$~=~230~m$^{2}$/Vs. 
To minimize stray capacitance a mesa was patterned using electron-beam lithography and wet etching. Metallic gates (Ti/Au) were deposited after atomic layer deposition of 10~nm HfO$_2$, which allows both positive and negative gating, and obviates gate-bias cooling~\cite{Buizert2008}.
 All measurements were conducted in a dilution refrigerator with mixing chamber temperature below 50~mK  and in-plane magnetic field $B = 300$~mT applied perpendicular to the axis between dots.

Voltages pulses were applied via high-bandwidth coaxial lines to the left and right plunger gates, $V_{\mathrm{L}}$, $V_{\mathrm{R}}$, and the barrier between the dots, $V_{\mathrm{M}}$. In practice, to account for the small coupling asymmetries, all three gates are involved in applying detuning $\varepsilon$ and symmetric barrier control $\gamma$:
\begin{align}
	\label{definition}
	\begin{split}
	\varepsilon &=k_{\mathrm{0}}[(V_{\mathrm{R}}-V_{\mathrm{R}}^0)-(V_{\mathrm{L}}-V_{\mathrm{L}}^0)]+k_{\mathrm{1}}(V_{\mathrm{M}}-V_{\mathrm{M}}^0),
	\\
	\gamma &=V_{\mathrm{M}}-V_{\mathrm{M}}^0, 
	\end{split}
\end{align}
where $V_{\mathrm{R}}^0$, $V_{\mathrm{L}}^0$ and  $V_{\mathrm{M}}^0$ are DC offset voltages (see Supplementary Material). Parameters $k_{\mathrm{0}} = 0.5$ and $k_{\mathrm{1}}=-0.075$ were determined  experimentally by mapping out the charge stability diagram. The value of $k_{\mathrm{0}}$ is consistent with previous experiments and sets the difference between left and right dot electrochemical potential, whereas $k_{\mathrm{1}}$ keeps other charge states  energetically unaccessible during $\gamma$ pulses.

Energy levels for the two-electron singlet S and triplet T$_0$ states as a function of detuning, $\varepsilon$,  are shown in Figs.~1(c), along with the pulse sequences for the tilt and symmetric operation modes in Fig.~1(d). For both, tilt and symmetric operation, two electrons are prepared (P) in a singlet (0,2)S state and, by slowly ramping $\varepsilon$ to (1,1), the system is initialized (I) into the ground state  of the nuclear Overhauser field, either $\ket{\uparrow\downarrow}$ or $\ket{\downarrow\uparrow}$.  For tilt operation, the exchange pulse, $J$, is applied by detuning to the exchange (X) point $\varepsilon_{\mathrm{x}}$ for a duration $\tau$, inducing rotations between $\ket{\uparrow\downarrow}$ and $\ket{\downarrow\uparrow}$. For symmetric operation, the exchange pulse is applied by pulsing the middle gate to $\gamma_{\mathrm{x}}$.

\section{Exchange oscillations}

\begin{figure}[tb]
\begin{center}
\includegraphics[width=96 mm]{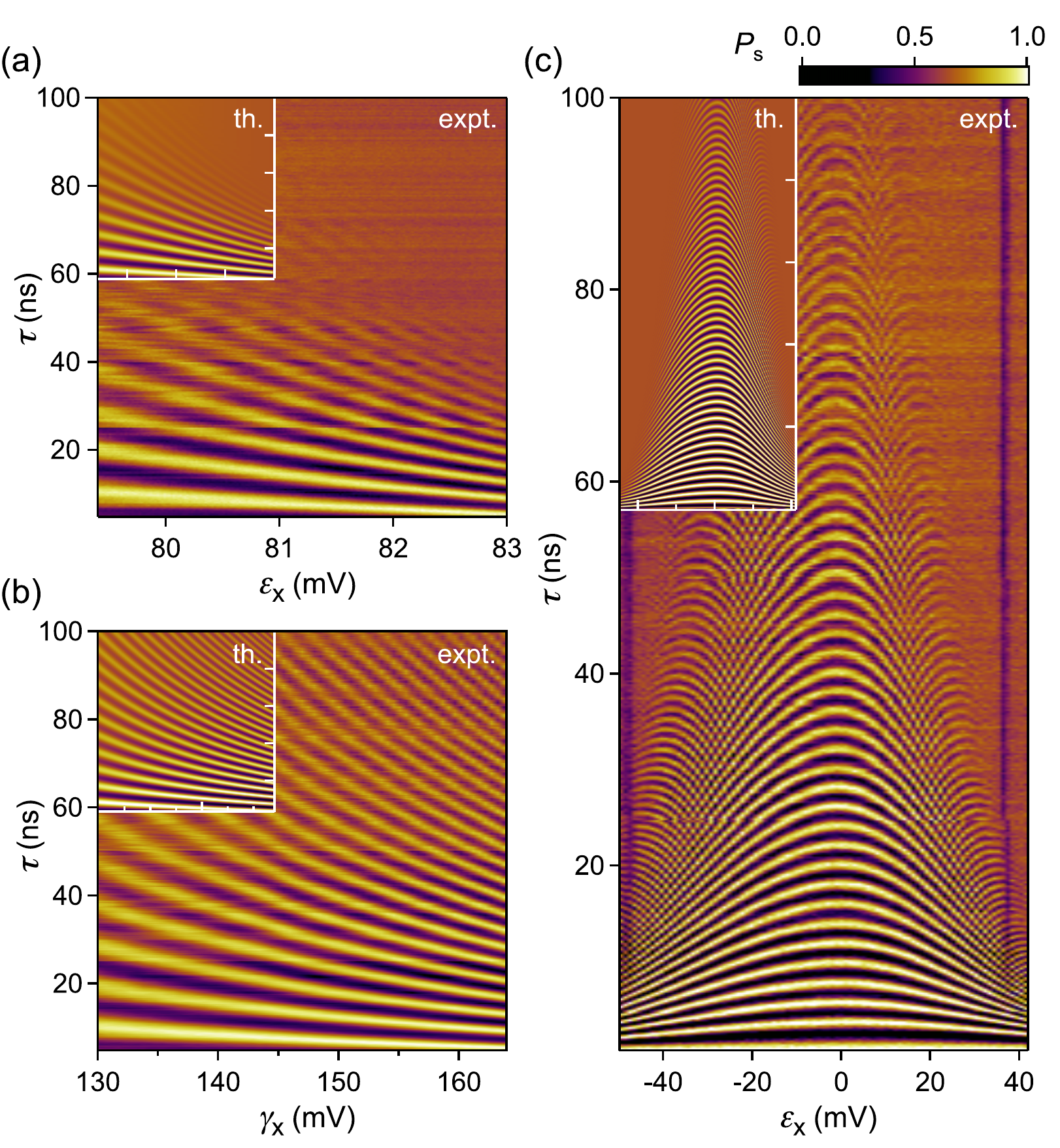}
\caption[Map of exchange oscillations for tilt and symmetric operation]{
(a) Probability of detecting a singlet, $P_{\mathrm{s}}$, as a function of $\varepsilon_{\mathrm{x}}$ and
exchange time $\tau$ for tilt-induced oscillations ($\gamma_{\mathrm{x}}$ = 0 mV).
(b) $P_{\mathrm{s}}$ as a function of $\gamma_{\mathrm{x}}$ and exchange time $\tau$ obtained for barrier-induced oscillations near the symmetry point ($\varepsilon_{\mathrm{x}}$ = 13.5 mV).
(c) Same as (a) with barrier pulse activated, $\gamma_{\mathrm{x}}$ = 190 mV, revealing the sweet spot of the symmetric operation. 
The dark vertical features near 39~mV and -44 mV are due to leakage from the singlet state to the spin-polarized triplet state.
Insets show theoretical simulations for each experimental situation.
}
\label{symm:fig2}
\end{center}
\end{figure}

Two-dimensional images of exchange oscillations, controlled by either tilt [Fig.~\ref{symm:fig2}(a)] or symmetric operation near the midpoint of (1,1) [Fig.~\ref{symm:fig2}(b)], show a striking difference in quality. 
In both images, each pixel represents the singlet return probability, $P_{\mathrm{S}}$, measured from an ensemble of $\sim 10^{3}$ single-shot measurements. Each single-shot measurement is assigned a binary value by comparing the reflectometery signal at the measurement (M) point, integrated for $T_\mathrm{M}=10$~$\mu$s, to a fixed threshold  \cite{Barthel2009,Barthel2010}. 
Figure \ref{symm:fig2}(c) shows exchange oscillations using both tilt and exchange. This image is generated by applying a tilt pulse of amplitude $\varepsilon_{\mathrm{x}}$ (of either sign) along with a fixed symmetric pulse $\gamma_{\mathrm{x}} = 190$~mV for a duration $\tau$. As $|\varepsilon_{\mathrm{x}}|$ is increased $J$ also increases, producing a chevron-like pattern centered around the sweet spot $J(\varepsilon_{\mathrm{x}} = 0)$ that occurs in the middle of the (1,1) charge state. 
Defining a quality factor, $Q$, to be the number of oscillations before the amplitude decays to $1/e$ of its initial value, we measure $Q\sim 35$ at the symmetry point, $\varepsilon_{\mathrm{x}} = 0$ ~\footnote{Fig.~\ref{symm:fig2}(c) shows that barrier-induced exchange oscillations have high quality factors for a wide range of operating points $\varepsilon_{\mathrm{x}}$ near the sweet spot. For example, data shown in Fig.~\ref{symm:fig2}(b) was obtained at $\varepsilon_{\mathrm{x}}=13.5$ mV, which for practical purposes we also classify as symmetric operation.}.

\begin{figure}[tb]
\begin{center}
\includegraphics[width=96 mm]{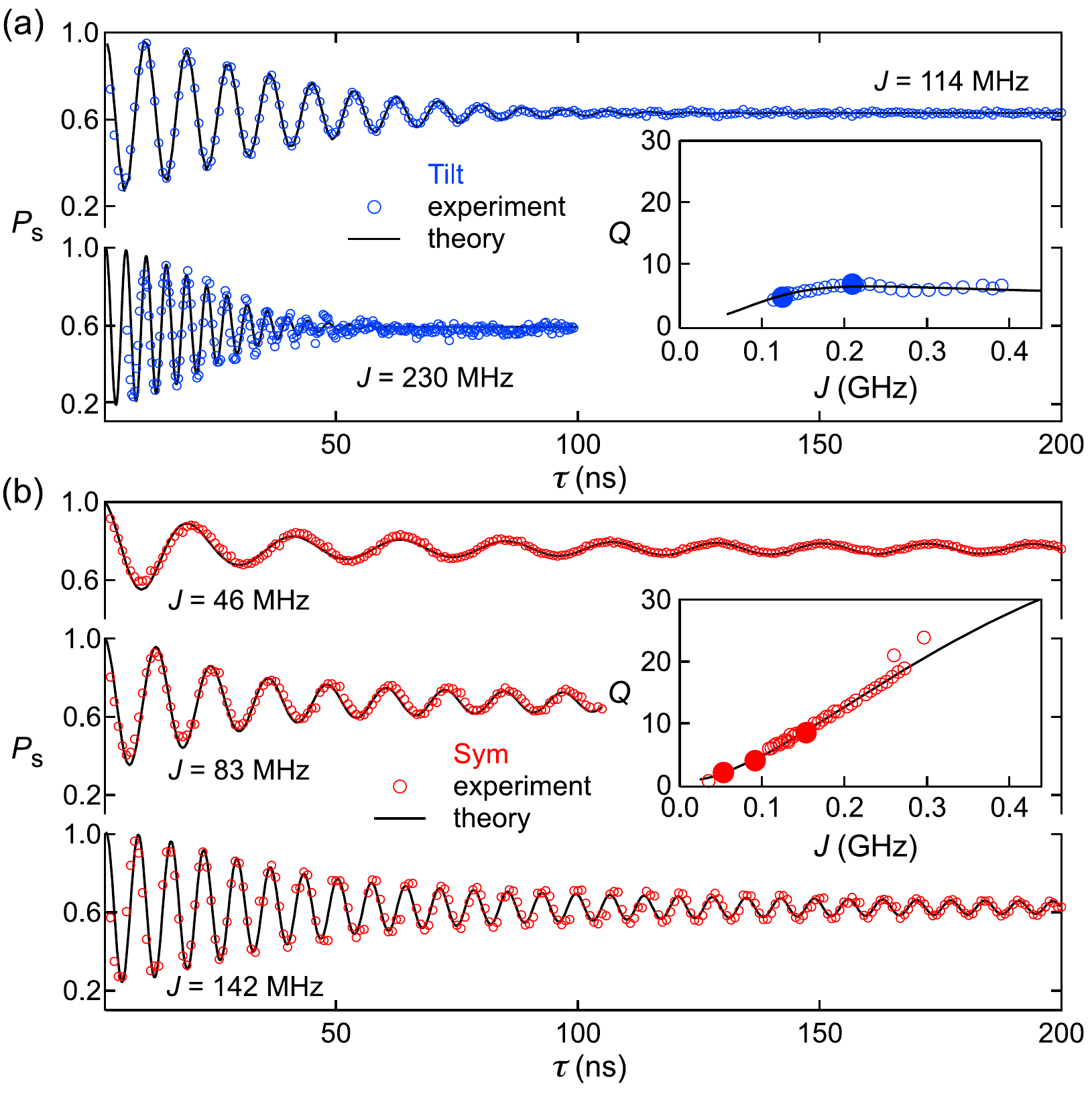}
\caption[Exchange oscillations for tilt and symmetric operation]{
(a) Tilt-induced exchange oscillations ($i.e.$ $\gamma_{\mathrm{x}}=$~0~mV) for  $\varepsilon_{\mathrm{x}} = $~79.5~mV and 82~mV, generating oscillation frequencies indicated by $J$.
(b) Same as (a) but for the symmetric mode of operation ($\varepsilon_{\mathrm{x}}=$~13.5~mV), with $\gamma_{\mathrm{x}}=$~100~mV, 120 mV and 140 mV. 
Open circles are experimental data. 
Solid lines correspond to the theoretical model in Eq.~\eqref{Ed_Model}, with $J$ and a horizontal offset being the only adjustable parameters. 
Insets show the quality factor $Q$, defined as the number of oscillations before the amplitude damps by a factor of $e$, as a function of $J$ for both tilt and symmetric operation modes. Solid circles correspond to data in the main panel, and solid lines are theoretical predictions.
}
\label{symm:fig3}
\end{center}
\end{figure}

The oscillation frequency of $P_{\mathrm{S}}(\tau)$ gives a direct measure of $J$ at the exchange point X. Interestingly, the frequency does not depend on the Overhauser field, even when it is comparable in size to $J$~\cite{Barnes2016}. 
Figures~\ref{symm:fig3}(a) and (b) show a set of experimental exchange oscillations representative of the tilt and symmetric operation mode, respectively. 
$Q$ extracted from such oscillations is shown in the insets. 
Consistent with previous observations ~\cite{Petta2005,Higginbotham2014}, tilt-induced exchange oscillations result in $Q\sim6$ independent of $J$. On the other hand, for the symmetric mode, $Q$ increases with $J$ for the range measured of 40~MHz~$<J<$~700~MHz. This is in agreement with recent results in singlet-triplet qubits fabricated in the Si/SiGe heterostructures~\cite{Reed2016}. Much higher values of Q can be obtained by tilting the double dot potential so far that both S and T$_0$ states share the same (0,2) charge state~\cite{Dial2013}. However, it is unclear if qubit operations at frequencies of tens of GHz are practical.
 
\section{Noise quantification}

To quantify the noise sensitivity of the symmetric exchange gate as well as gain insight into why it outperforms exchange by detuning, we compare both methods to a simple model that includes both nuclear Overhauser gradient noise and voltage noise on the detuning and barrier gates. Noise is assumed gaussian and quasistatic on the timescale of the exchange oscillations. Nuclear noise is characterized by a mean longitudinal Overhauser gradient energy $h_0$ between dots, with standard deviation $\sigma_h$. Exchange noise is assumed to result from voltage noise on left and right plungers and the barrier, with mean exchange energy $J$ with standard deviation $\sigma_J$.  The model also accounts for triplet-to-singlet relaxation at the measurement point, with a relaxation time $T_{\mathrm{RM}}$ during the measurement interval of length $T_\mathrm{M}$. Within this model, the singlet return probability $\langle\langle P_\mathrm{s}\rangle\rangle$ over both noise ensembles is given by~\cite{Barnes2016}:
\begin{equation}
	\begin{split}
	\langle\langle P_\mathrm{s}\rangle\rangle =&1-\frac{T_{\mathrm{RM}}}{T_\mathrm{M}}\left(1-e^{-\frac{T_\mathrm{M}}{T_{\mathrm{RM}}}}\right)
	\frac{e^{-\frac{h_0^2}{2\sigma_h^2}}e^{-\frac{J^2}{2\sigma_J^2}}}{\sqrt{\pi}\sigma_h\sigma_J}
	\\
	&
	\times\int_{-\pi/2}^{\pi/2} d\chi
	\Bigg\{\frac{b\left(\chi\right)}{a\left(\chi\right)^{3/2}} e^\frac{b\left(\chi\right)^2}{a\left(\chi\right)}
	\\
	&-\mathrm{Re}\Bigg[ \frac{b\left(\chi\right)+i\tau\mathrm{sec}(\chi)}{a\left(\chi\right)^{3/2}}
	e^\frac{[b\left(\chi\right)+i\tau\mathrm{sec}(\chi)]^2}{a\left(\chi\right)} \Bigg]\Bigg\}
	\end{split},
	\label{Ed_Model}
\end{equation}
where $\chi$ is the tilt of the qubit rotation axis during an exchange pulse due to the Overhauser field gradient~\cite{Barnes2016}, $a\left(\chi\right) \equiv 2\mathrm{tan}^2\chi/\sigma_h^2+2/\sigma_J^2$ and $b\left(\chi\right)\equiv h_0\mathrm{tan}\chi/\sigma_h^2+J/\sigma_J^2$.

The black solid lines in Fig.~\ref{symm:fig3}, together with the insets in Figs.~\ref{symm:fig2}(a), (b) and (c), are generated by evaluating Eq.~\eqref{Ed_Model} numerically. Two fit parameters per curve are the oscillation frequency $J$ and a horizontal offset associated with the rise time of the waveform generator. All other parameters were obtained from independent measurements:
The Overhauser energy gradient fluctuations, $\sigma_h= 23$~MHz, was obtained by measuring the distribution of free induction decay frequencies \cite{Barthel2012} over a 30 min.~interval and fitting the distribution to a gaussian.

\begin{figure}[tbh]
\begin{center}
\includegraphics[width=96 mm]{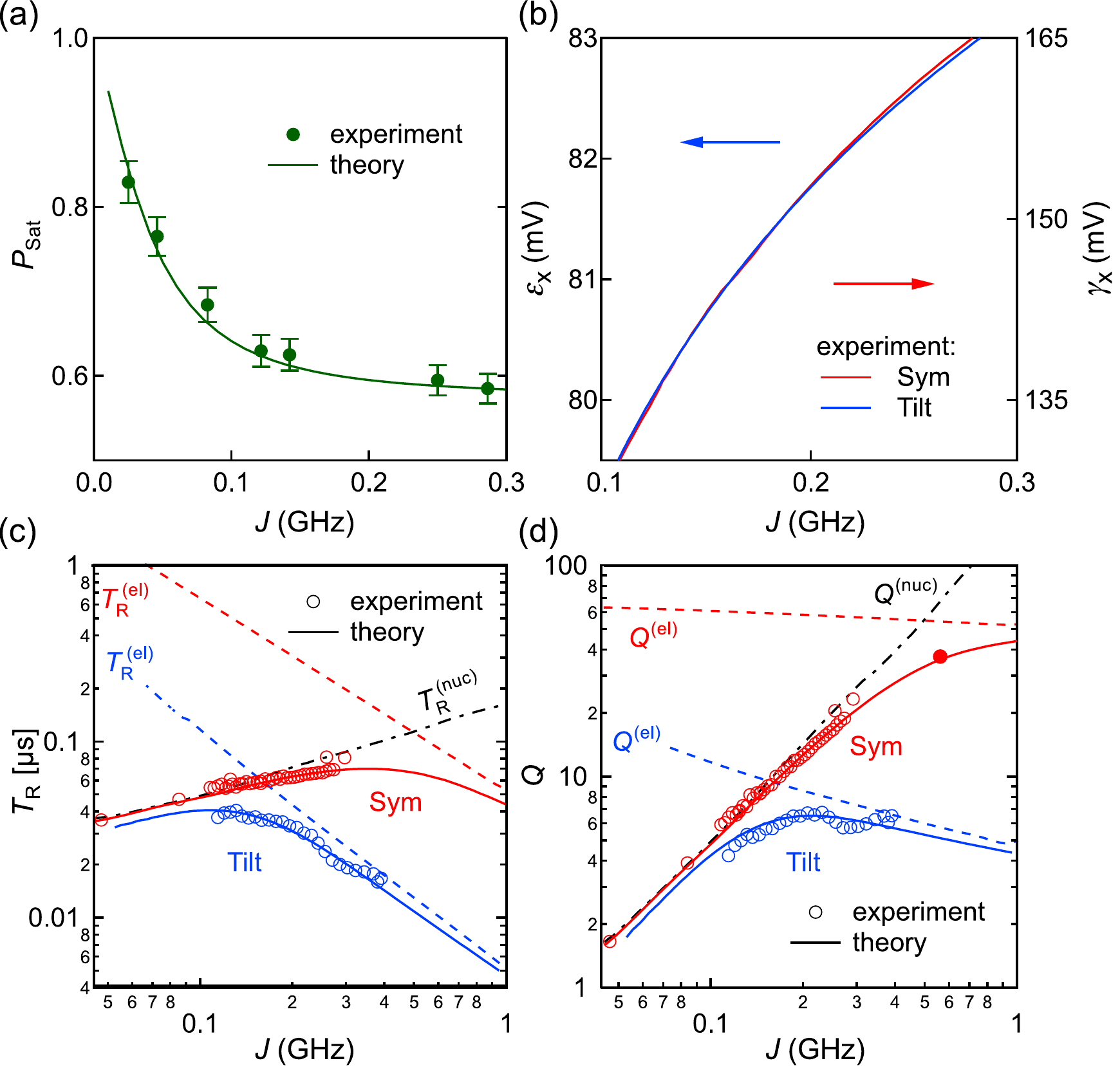}
\caption[Detailed analysis of noise sources and quality factors]{
(a) Saturation probability of the symmetric mode of operation, $P_{\mathrm{Sat}}$, as a function of $J$ (symbols). Comparison with theory (solid line) determines $T_\mathrm{RM}$ and $h_{\mathrm{0}}$.
(b) Plot of $\varepsilon_{\mathrm{x}}$ for the tilt  and  $\gamma_{\mathrm{x}}$ for the symmetric mode of operation, as functions of the exchange coupling extracted experimentally.
(c) Decoherence time $T_{\mathrm{R}}$, $i.e.$ time before the amplitude of oscillations is reduced by a factor of $e$, as a function of $J$ for both tilt and symmetric modes.
(d) Quality of the exchange rotations, defined as $Q = J T_{\mathrm{R}}$, for different $J$.
In (c) and (d) the open circles are obtained experimentally and solid lines correspond to a model that includes dephasing due to electrical and nuclear noise. 
Black dashed lines are the same model if we only consider nuclear noise contributions ($T_{\mathrm{R}}^{\mathrm{(nuc)}}$, $Q^{\mathrm{(nuc)}}$). Blue and red dashed lines correspond to the electrical noise contributions ($T_{\mathrm{R}}^{\mathrm{(el)}}$, $Q^{\mathrm{(el)}}$) for the tilt and symmetric modes of operation, respectively.
Solid circle indicates the maximum $Q$ value observed in Fig.~\ref{symm:fig2}(c).
}
\label{symm:fig4}
\end{center}
\end{figure}

The saturation of the singlet return probability, $P_S$, at long $\tau$, denoted $P_\mathrm{sat}$, will deviate from $P_\mathrm{sat}$~=~0.5 in the presence of a nonzero mean Overhauser field gradient, $h_0$, or finite relaxation time, $T_{\mathrm{RM}}$. Fitting the $J$ dependence of $P_\mathrm{sat}$ [Fig.~\ref{symm:fig4}(a)], yields fit values $T_{\mathrm{RM}}=30$~$ \mu$s and $h_0/h=40$~MHz. 

Exchange noise $\sigma_J$ is obtained by assuming (i) all noise is gate noise, (ii) noise on different gates is independent:
$\sigma_J^2=\sigma_\mathrm{el}^2[\left(dJ/dV_\mathrm{L}\right)^2+\left(dJ/dV_\mathrm{M}\right)^2+\left(dJ/dV_\mathrm{R}\right)^2]$.
In giving all three components equal weight, we have further assumed that all three gates are equally noisy as quantified by the parameter $\sigma_\mathrm{el}$. Taking into account the definitions in Eq.~\eqref{definition} we obtain:
\begin{equation}
	\begin{split}
	\sigma_J=\sigma_\mathrm{el}\sqrt{2k_0^2\left(\frac{dJ}{d\varepsilon_\mathrm{x}}\right)^2+\left(\frac{dJ}{d\gamma_\mathrm{x}}+k_1\frac{dJ}{d\varepsilon_\mathrm{x}}\right)^2}
	\end{split}
	\label{Ed_Model_complement}
\end{equation}

The derivatives are calculated from a phenomenological smooth exchange profile $J(\varepsilon_{\mathrm{x}},\gamma_{\mathrm{x}})$ fitted to a discrete map of $J$ measured at various operating points (see chapter~\ref{ch:symm_sup}). 
The effective gate noise $\sigma_\mathrm{el}$ is extracted from tilt exchange oscillations measured in a regime where effective detuning noise dominates, giving $\sigma_\mathrm{el}=0.18$~mV (see Supplementary Material). 
This value, together with Eq.~\ref{Ed_Model_complement},  determines $\sigma_{J}(\varepsilon_{\mathrm{x}},\gamma_{\mathrm{x}})$ used in all simulations, and yields excellent agreement with data. 

The origin of the improved electrical performance becomes apparent when comparing the required pulse amplitudes for symmetric and tilted operation for a given $J$ [Fig.~\ref{symm:fig4}(b)].  Although the dependences of $\varepsilon_\mathrm{x}$ and $\gamma_\mathrm{x}$ on $J$ are  similar, the range of $\varepsilon_\mathrm{x}$ is significantly smaller than $\gamma_\mathrm{x}$. Note in Fig.~\ref{symm:fig4}(b) that $J$ changes from 0.1 to 0.3 GHz
for a $\sim$ 3 mV change in $\varepsilon_\mathrm{x}$, or a $\sim$ 30 mV change in $\gamma_\mathrm{x}$ [see Fig.~\ref{symm:fig4}(b)]. Because of this difference in derivatives of $J$ with respect to $\varepsilon_\mathrm{x}$ and $\gamma_\mathrm{x}$, the symmetric operation has much less noise for a given noise in the gate voltages.

The contributions of nuclear and electrical noise to limiting the quality factor $Q$ of and dephasing time, $T_\mathrm{R} = Q/J$, comparing experiment and model, is shown in Figs.~\ref{symm:fig4}(c) and (d).  
Note that for detuning (tilt) operation, electrical noise dominates above $\sim 0.2$ GHz, so that going any faster (using larger $J$) just makes the exchange noise greater in proportion, limiting the number of oscillations to $Q \sim 6$. For symmetric exchange, on the other hand, electrical noise doesn't dominate until above $J\sim 0.6$~GHz, resulting in a monotonically increasing quality factor up to $\sim 1$ GHz.
From the model, we find $Q$ as high as 50, 8 times larger than in the conventional tilt operation mode. 
Finally, we note that the origin of the effective electrical noise may be within the sample and not in the instrumentation. To distinguish actual voltage fluctuations on the gate electrodes (due to instrumentation) from intrinsic noise source (e.g. two-phonon processes~\cite{Kornich2014}), further studies including temperature dependence are needed.       

\section{Summary}

In summary, we have investigated experimentally and modeled the application of an exchange gate applied by opening the middle barrier at a symmetry point of a two-electron spin qubit system instead of the conventional method, which is to detune the potential. 
The model allows the influences of nuclear and electrical noise to be disentangled for both symmetric and detuning exchange control, and is in excellent agreement with experimental data.
We find that symmetric mode of control is significantly less sensitive to electrical noise due to the symmetric arrangement, making exchange only quadratically sensitive to detuning gate voltage noise.  
With this new symmetric control method, we were able to increase the quality factor of coherent oscillations from around 6 to 35, and expect that improvements beyond $Q\sim 50$ are possible by further increasing $J$. 
The corresponding enhancement of coherence times by nearly an order of magnitude will also benefit other single- and multi-qubit implementations that rely on exchange interactions~\cite{Veldhorst2015}.  

\section*{Acknowledgements}

We thank Rasmus Eriksen for help in fast data acquisition and Daniel Loss and Mark S. Rudner for helpful discussions. 
This work was supported by IARPA-MQCO, LPS-MPO-CMTC, the EC FP7- ICT project SiSPIN no. 323841, the Army Research Office and the Danish National Research Foundation.  

\section*{Authors contributions}
S.F., G.C.G. and M.J.M. grew the heterostructure. P.D.N. fabricated the device. F.M., P.D.N., F.K. and F.K.M. prepared the experimental setup. F.K.M., F.M. and F.K. performed the experiment. E.B. developed the theoretical model. E.B. and F.M. performed the simulations. F.M., F.K., F.K.M., E.B. and C.M.M. analysed the data and prepared the manuscript.

\chapterimage{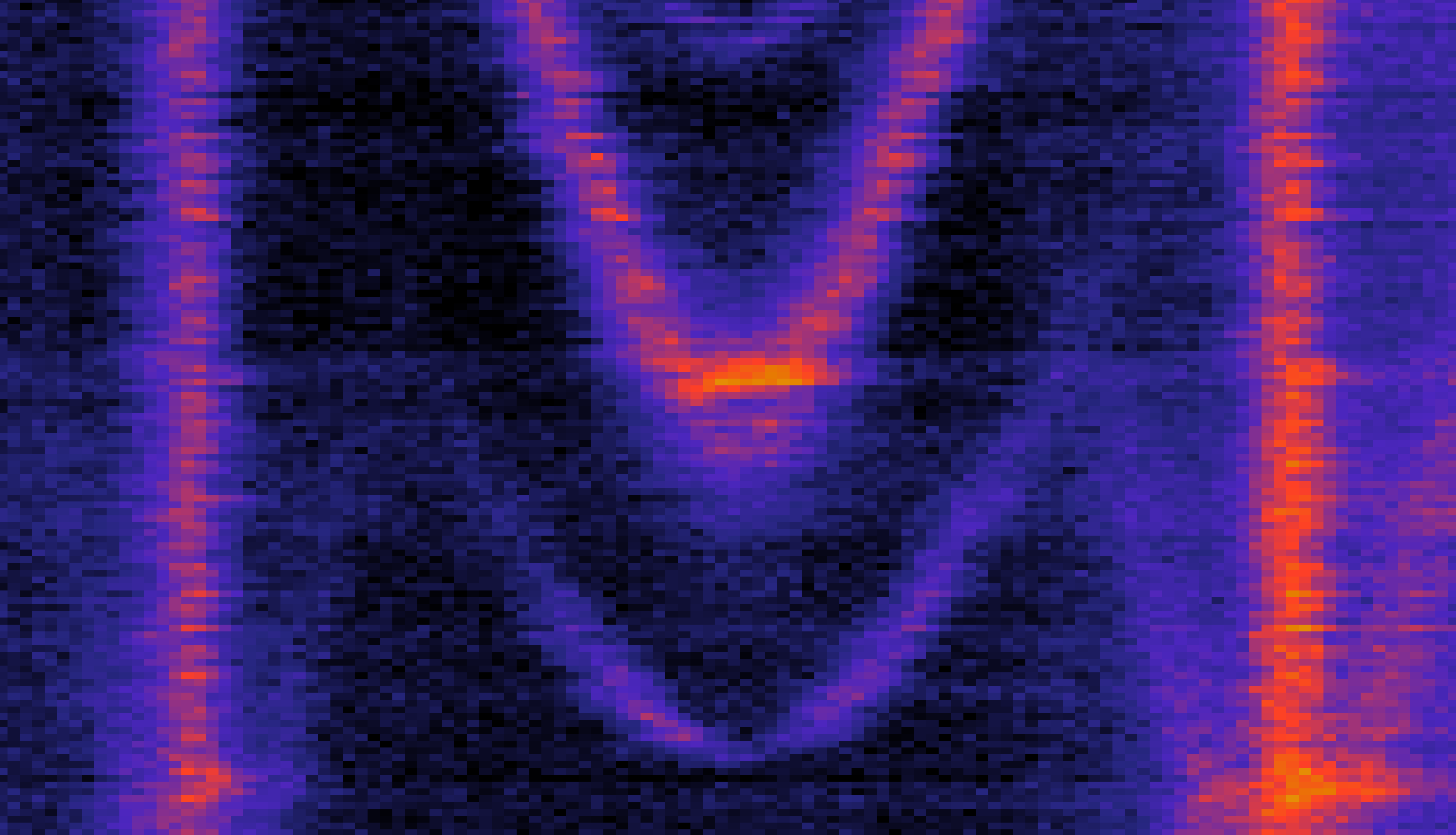}
\chapter[Symmetric operation of the resonant exchange qubit]{\protect\parbox{0.9\textwidth}{Symmetric operation\\of the resonant exchange qubit}}
\label{ch:SRX}

\newcommand{\VLP}{V_\mathrm{LP}}
\newcommand{\VLB}{V_\mathrm{LB}}
\newcommand{\VMP}{V_\mathrm{MP}}
\newcommand{\VRB}{V_\mathrm{RB}}
\newcommand{\VRP}{V_\mathrm{RP}}

\newcommand{\TR}{T_\mathrm{R}}
\newcommand{\fR}{f_\mathrm{R}}
\newcommand{\TCPMG}{T_2^\mathrm{CPMG}}
\renewcommand{\vec}[1]{{\bf #1}}

{\let\thefootnote \relax\footnote{This chapter is adapted from Ref. \cite{Malinowski2017b}.}}
\addtocounter{footnote}{-1}

\begin{center}
Frederico Martins$^{1,*}$, Filip K. Malinowski$^{1,*}$, Peter D. Nissen$^{1}$, \\
Saeed Fallahi$^{2}$, Geoffrey C. Gardner$^{2,3}$, Michael J. Manfra$^{4,5}$, \\
Charles M. Marcus$^{6}$, Ferdinand Kuemmeth$^{1}$
\end{center}

\begin{center}
	\scriptsize
	$^{1}$ Center for Quantum Devices, Niels Bohr Institute, University of Copenhagen, 2100 Copenhagen, Denmark\\
	$^{2}$ Department of Physics and Astronomy, Birck Nanotechnology Center, Purdue University, West Lafayette, Indiana 47907, USA \\
	$^{3}$ School of Materials Engineering and School of Electrical and Computer Engineering, \\ Purdue University, West Lafayette, Indiana 47907, USA \\
	$^{4}$ Department of Physics and Astronomy, Birck Nanotechnology Center, and Station Q Purdue, \\ Purdue University, West Lafayette, Indiana 47907, USA \\
	$^{5}$ School of Materials Engineering, Purdue University, West Lafayette, Indiana 47907, USA \\
	$^{6}$ Center for Quantum Devices and Station Q Copenhagen, Niels Bohr Institute, \\ University of Copenhagen, 2100 Copenhagen, Denmark \\
	$^{*}$ These authors contributed equally to this work
\end{center}

\begin{center}
\begin{tcolorbox}[width=0.8\textwidth, breakable, size=minimal, colback=white]
	\small
	We operate a resonant exchange qubit in a highly symmetric triple-dot configuration using IQ-modulated RF pulses. 
At the resulting three-dimensional sweet spot the qubit splitting is an order of magnitude less sensitive to all relevant control voltages, compared to the conventional operating point, but we observe no significant improvement in the quality of Rabi oscillations. 
For weak driving this is consistent with Overhauser field fluctuations modulating the qubit splitting. 
For strong driving we infer that effective voltage noise modulates the coupling strength between RF drive and the qubit, thereby quickening Rabi decay. 
Application of CPMG dynamical decoupling sequences consisting of up to $n=32$ $\pi$ pulses significantly prolongs qubit coherence, leading to marginally longer dephasing times in the symmetric configuration. 
This is consistent with dynamical decoupling from low frequency noise, but quantitatively cannot be explained by effective gate voltage noise and Overhauser field fluctuations alone.  
Our results inform recent strategies for the utilization of partial sweet spots in the operation and long-distance coupling of triple-dot qubits. 
	
	The version of the article presented in the thesis is supplemented with the derivation of the formula for the Rabi decay time (Sec.~\ref{SRX:derivation}).
\end{tcolorbox}
\end{center}

\section{Introduction}

Spin qubits are widely investigated for applications in quantum computation~\cite{Loss1998,Petta2005,Shulman2012,Koppens2006,Nowack2011,Veldhorst2015,Nichol2017}, with several operational choices depending on whether the qubit is encoded in the spin state of one~\cite{Koppens2006,Nowack2011,Veldhorst2015,Kawakami2016,Maurand2016,Takeda2016}, two~\cite{Petta2005,Shulman2012,Foletti2009,Nichol2017} or three electrons~\cite{Laird2010,Gaudreau2011,Medford2013,Medford2013a,Kim2014,Eng2015,Russ2015}. 
In particular, spin qubits encoded in three-electron triple quantum dots allow universal electrical control with voltage pulses, and enable integration with superconducting cavities~\cite{Petersson2012,Stockklauser2015,Liu2015,Russ2015a,Srinivasa2016,Mi2016}. 
Multi-qubit coupling via superconducting cavities, however, is challenging due to the effects of environmental noise on resonant exchange (RX) qubits \cite{Medford2013a,Srinivasa2016}.
A recent approach to improve coherence times is the operation at sweet spots, where the qubit splitting is to first order insensitive to most noisy parameters~\cite{Martins2016,Reed2016,Russ2015,Shim2016a}. 
Here, we operate a symmetric resonant exchange (SRX) qubit in which the qubit splitting is highly insensitive to all three single-particle energies~\cite{Shim2016a}, and compare its performance to its conventional configuration as a RX qubit~\cite{Medford2013a,Taylor2013}.

\begin{figure}[tbh]
	\centering
	\includegraphics[width=0.6\textwidth]{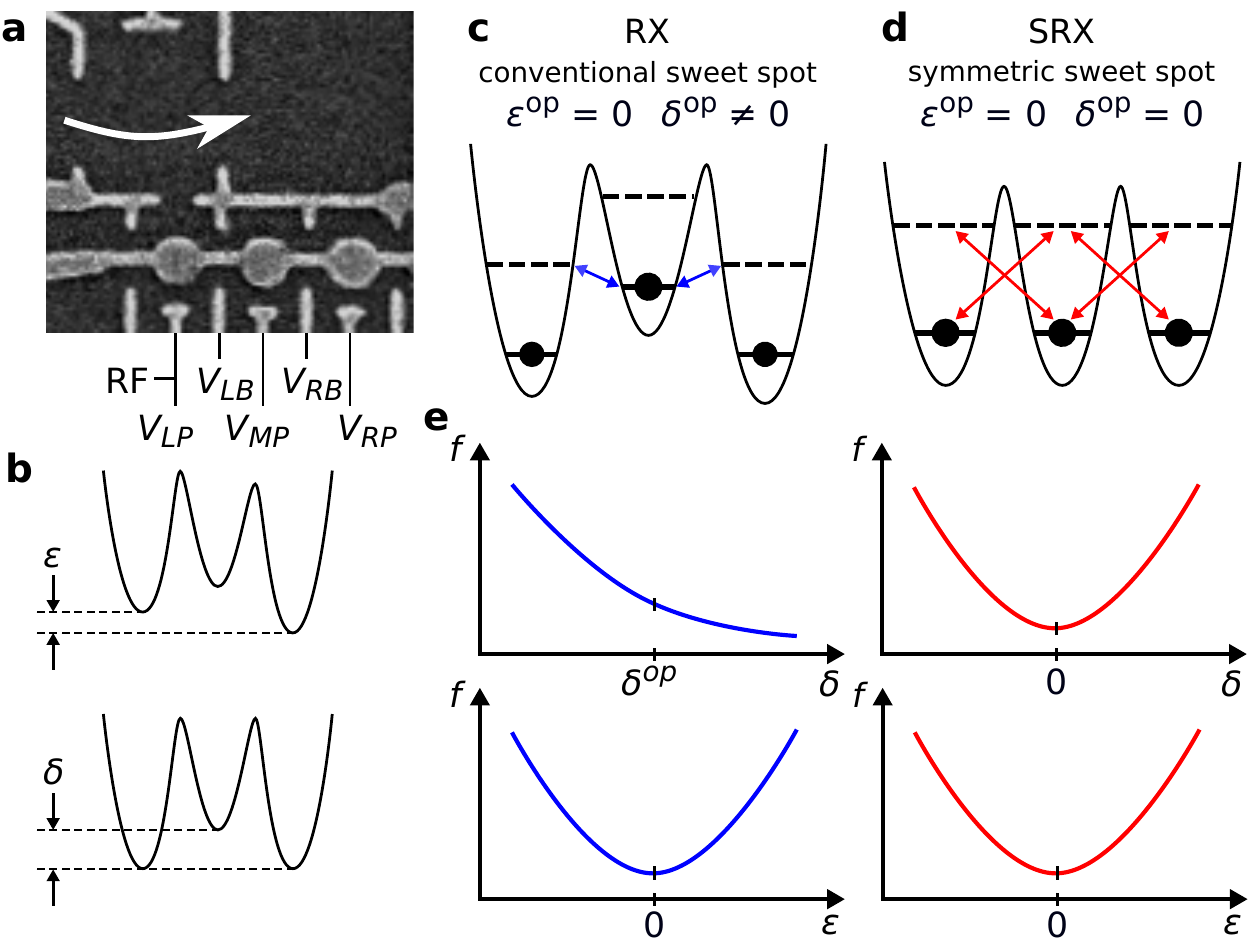}
	\caption[Schematic illustration of the RX and SRX qubit]{(a) Scanning electron micrograph of a GaAs triple quantum dot, formed under the rounded accumulation gate, and a proximal sensor dot (white arrow), formed by depletion gates. The five depletion gates used for qubit manipulation are labeled.  
(b) Schematic illustration of two control parameters, $\delta$ and $\varepsilon$, resulting in energy shifts $\delta|e|$ and $\varepsilon|e|$.
(c) Potential along the RX qubit. The qubit splitting arises from virtual tunneling of the central electron to the outer dots (blue arrows), and is therefore sensitive to potential fluctuations of each dot. 
(d) Potential along the SRX qubit. Tunneling of the outer electrons to the central dot contributes to charge hybridization equally strongly as tunneling of the central electron to the outer dots (red arrows), making the qubit splitting insensitive to potential fluctuations of all three dots. 
(e) Schematic dependence of the qubit frequency $f$ on $\varepsilon$ and $\delta$ around the operating point of the RX and SRX qubit.
	}
	\label{SRX:fig1}
\end{figure}

\section{RX and SRX qubit}

We configure a triple-quantum-dot device either as a SRX or RX qubit by appropriate choice of gate voltages.
Gate electrodes are fabricated on a doped, high-mobility GaAs/AlGaAs quantum well,
and the triple dot is located $\sim 70$ nm below three circular portions of the accumulation gate (Fig.~\ref{SRX:fig1}a). 
The occupation of the dots is controlled on nanosecond timescales by voltage pulses on gates $V_i$, where $i$ refers to the left/middle/right plunger gate (LP/MP/RP) or left/right barrier gate (LB/RB). 
Radio frequency (RF) bursts for resonant qubit control are applied to the left plunger gate. 
The conductance through the proximal sensor dot is sensitive to the charge occupation of the triple quantum dot, allowing qubit readout (see below). 

In the presence of an in-plane magnetic field, $B=400$~mT in this experiment, the triple-dot qubit is defined by the two three-electron spin states with total spin $S=1/2$ and spin projection $S_z=1/2$
~\cite{Laird2010,Medford2013a,Taylor2013,Russ2015}.
Ignoring normalization, these spin states can be represented by $\ket{0} \propto (\ket{\duu}-\ket{\udu})+(\ket{\uud}-\ket{\udu})$ and $\ket{1} \propto (\ket{\uud}-\ket{\duu})$. 
Here, arrows indicate the spin of the electron located in the left, middle and right quantum dot.
Note that the spin state of  $\ket{0}$ and $\ket{1}$ is, respectively, symmetric and antisymmetric under exchange of the outer two electrons.
In the presence of interdot tunneling this exchange symmetry affects hybridization of the associated orbital wavefunctions, splitting $\ket{0}$ and $\ket{1}$ by $h f$ (where $h$ is Planck's constant and $f$ sets the frequency of the qubit's rotating frame).
Similarily, an additional triple-dot state with $S=3/2$ and $S_z = 1/2$ is split from the qubit states due to interdot tunneling. 
All other triple-dot states have different $S_z$ and are energetically separated from the qubit states due to the Zeeman effect. 

\begin{figure}[tb]
	\centering
	\includegraphics[width=0.6\textwidth]{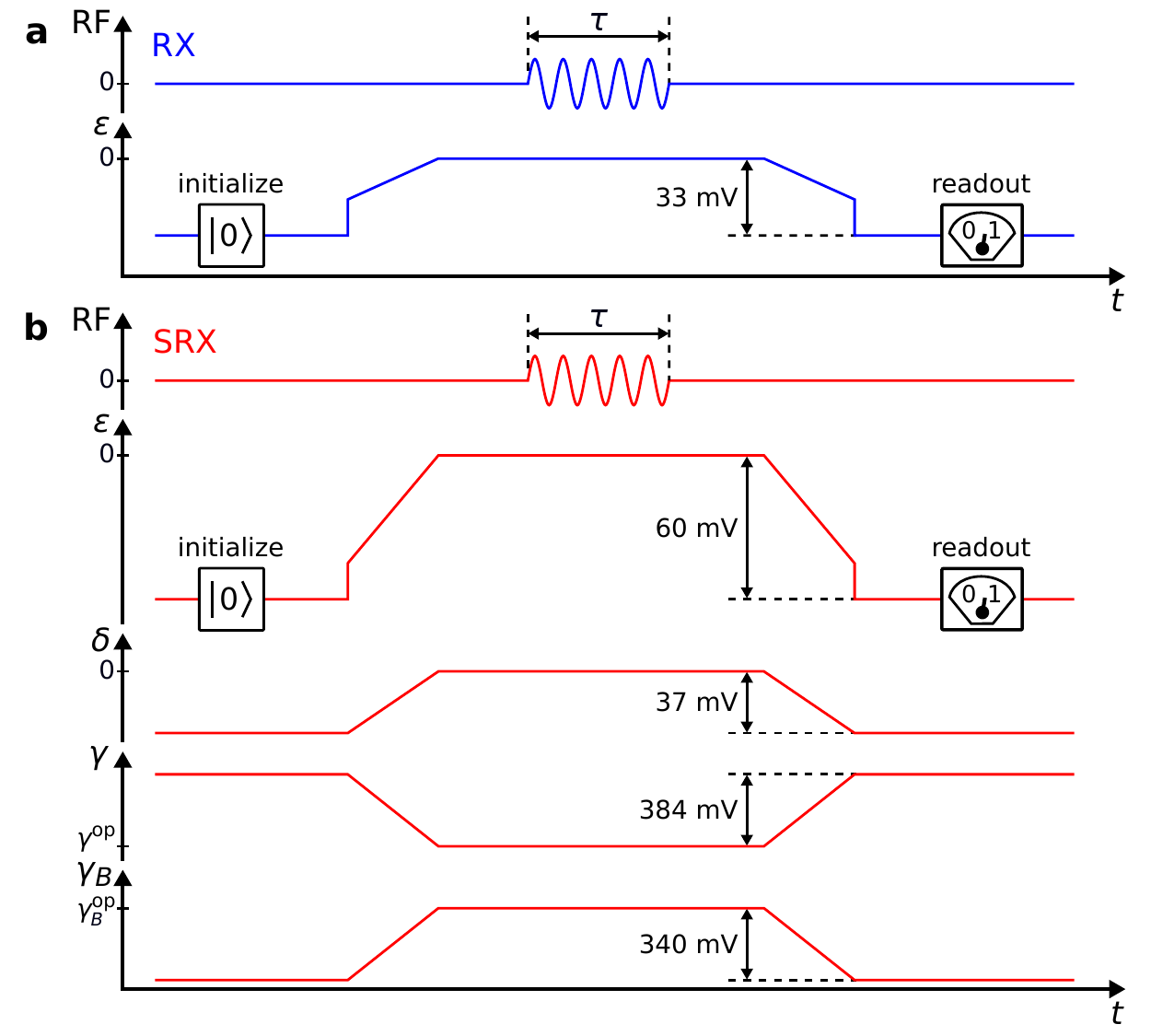}
	\caption[Schematic pulse cycle for measuring Rabi oscillations of the RX and SRX qubit]{
Schematic pulse cycle for measuring Rabi oscillations of the RX (a) and SRX (b) qubit. 
An IQ-modulated RF burst is applied on resonance with the qubit splitting for duration $\tau$. 
Linear detuning ramps, with typical amplitudes indicated, implement spin-to-charge conversion needed for qubit initialization and readout. 
For qubit spectroscopy and CMPG measurements the RF burst is replaced by a continuous RF tone or a sequence of calibrated RF pulses, respectively.
	}
	\label{SRX:fig2}
\end{figure}

In the conventional operating regime of the RX qubit (Fig.~\ref{SRX:fig1}c) the (111) charge state of the triple dot is hybridized weakly with charge states (201) and (102) (here number triplets denote the charge occupancy of the triple dot). 
This lowers the energy of $\ket{0}$ with respect to $\ket{1}$ and makes the resulting qubit splitting sensitive to detuning of the central dot, $\delta$ (cf. Fig.~\ref{SRX:fig1}b,e)~\cite{Medford2013a}. The qubit splitting is, however, to first order insensitive to detuning between the outer dots, $\varepsilon$, \cite{Taylor2013}, reflecting that tunneling across left and right barrier contribute equally to the qubit splitting (Fig.~\ref{SRX:fig1}c,e).  
Qubit rotations in the rotating frame are implemented by applying RF bursts to gate $\VLP$, such that the operating point oscillates around $\varepsilon=0$. When the RF frequency matches the qubit splitting, the qubit nutates between $\ket{0}$ and $\ket{1}$, allowing universal control using IQ modulation \cite{Medford2013a}. 
When the detuning of the outer dots is ramped towards (201), $\ket{0}$ maps to a singlet state of the left pair ($\ket{S_L}\propto (\ket{\duu}-\ket{\udu})$, see first terms in $\ket{0}$), whereas $\ket{1}$ remains in the (111) charge state due to the Pauli exclusion principle \cite{Laird2010,Medford2013,Medford2013a}. 
This spin-to-charge conversion allows us to perform single-shot readout on microsecond timescales, by monitoring a proximal sensor dot using high-bandwidth reflectometry~\cite{Barthel2010a}. In this work we estimate the fraction of singlet outcomes, $P_\mathrm{S}$, by averaging 1000-10000 single-shot readouts.

In the case of the SRX qubit, however, all three single-particle levels are aligned, and the (111) state hybridizes with the charge states (201), (102), (120) and (021)~\cite{Shim2016a}. This introduces additional symmetries between the tunneling of the electron from the central dot to the outer dots and tunneling of the outer electrons to the central dot (Fig.~\ref{SRX:fig1}d). 
As a consequence, the qubit splitting is expected to be insensitive to first order to both $\varepsilon$ and $\delta$ (Fig.~\ref{SRX:fig1}e) as well as to the barrier detuning, $\varepsilon_B$ (introduced below).
Due to the required alignment of single-particle levels, hybridization is suppressed by the charging energy within each dot (indicated by the large energy spacing between solid and dashed lines in Fig.~\ref{SRX:fig1}c,d).
Accordingly, we find that much larger tunnel couplings have to be tuned up to maintain a significant qubit splitting. 
In practice, the gate voltage configuration needed to achieve a SRX qubit splitting of a few hundred megahertz does no longer allow spin-to-charge conversion solely by a ramp of $\varepsilon$. Therefore, we also apply voltage pulses to the barrier gates when ramping the qubit between the operation configuration (indicated by superscript $\mathrm{op}$) and readout configuration (see below).

Figure ~\ref{SRX:fig2}a (\ref{SRX:fig2}b) defines the pulse cycle used for spectroscopy and operation of the RX (SRX) qubit.
Taking into account the physical symmetries of the device (cf. Fig.~\ref{SRX:fig1}), control parameters $ \varepsilon, \gamma, \delta, \varepsilon_B, \gamma_B $ are specified in terms of gate voltages $V_i$,

\begin{equation}
	\left(
	\begin{array}{c}
		\varepsilon \\
		\delta \\
		\gamma
	\end{array}
	\right) = \frac{1}{\sqrt{6}}\left(
	\begin{array}{ccccc}
		-\sqrt{3} & 0 & \sqrt{3} \\
		-1 & 2 & -1 \\
		\sqrt{2} & \sqrt{2} & \sqrt{2}
	\end{array}
	\right) \left(
	\begin{array}{c}
		\VLP-\VLP^\mathrm{sym} \\
		\VMP-\VMP^\mathrm{sym} \\
		\VRP-\VRP^\mathrm{sym}
	\end{array}
	\right)
	\label{SRX:parameters1}
\end{equation}
\begin{equation}
	\left(
	\begin{array}{c}
		\varepsilon_B \\
		\gamma_B
	\end{array}
	\right) = \frac{1}{\sqrt{2}}\left(
	\begin{array}{ccccc}
		-1 & 1 \\
		1 & 1
	\end{array}
	\right) \left(
	\begin{array}{c}
		\VLB-\VLB^\mathrm{sym} \\
		\VRB-\VRB^\mathrm{sym}
	\end{array}
	\right),
	\label{SRX:parameters2}
\end{equation}
and the power ($P_\mathrm{RF}$), duration ($\tau$), frequency ($f_\mathrm{RF}$) and phase of the IQ-modulated RF burst. 
The operating point of the SRX qubit, defined by $V_i=V_i^\mathrm{sym}$, was chosen to yield a qubit frequency of 530~MHz
~\footnote{($ \VLP^\mathrm{sym}, \VLB^\mathrm{sym}, \VMP^\mathrm{sym}, \VRB^\mathrm{sym}, \VRP^\mathrm{sym} $) = (-1.53, -0.14, -1.05, -0.15, -0.83) V.}. 
The operating point of the RX qubit, located at  $\{ \delta^\mathrm{op}>0, \gamma^\mathrm{op}>0, \gamma_B^\mathrm{op}<0\}$, was chosen to yield a comparable qubit frequency of 510~MHz.
The linear ramps before (after) the RF burst facilitate initialization (readout) of the qubit state via an adiabatic conversion of a two-electron spin singlet state in the left dot.
For the RX qubit $\{ \delta-\delta^\mathrm{op}, \gamma-\gamma^\mathrm{op}, \varepsilon_B, \gamma_B-\gamma_B^\mathrm{op} \}$ all remain zero throughout the pulse cycle, i.e. the operation and readout configuration differ only in detuning $\varepsilon$ (Fig.~\ref{SRX:fig2}a). In contrast, to adiabiatically connect the initialization/readout point of the SRX qubit to its operating point, we found it necessary to vary $\varepsilon$, $\delta$, $\gamma$ and $\gamma_B$ during the pulse cycle (Fig.~\ref{SRX:fig2}b), which involves voltage pulses on all five gates indicated in Fig.~\ref{SRX:fig1}a.

\begin{figure}[tbh]
	\centering
	\includegraphics[width=0.6\textwidth]{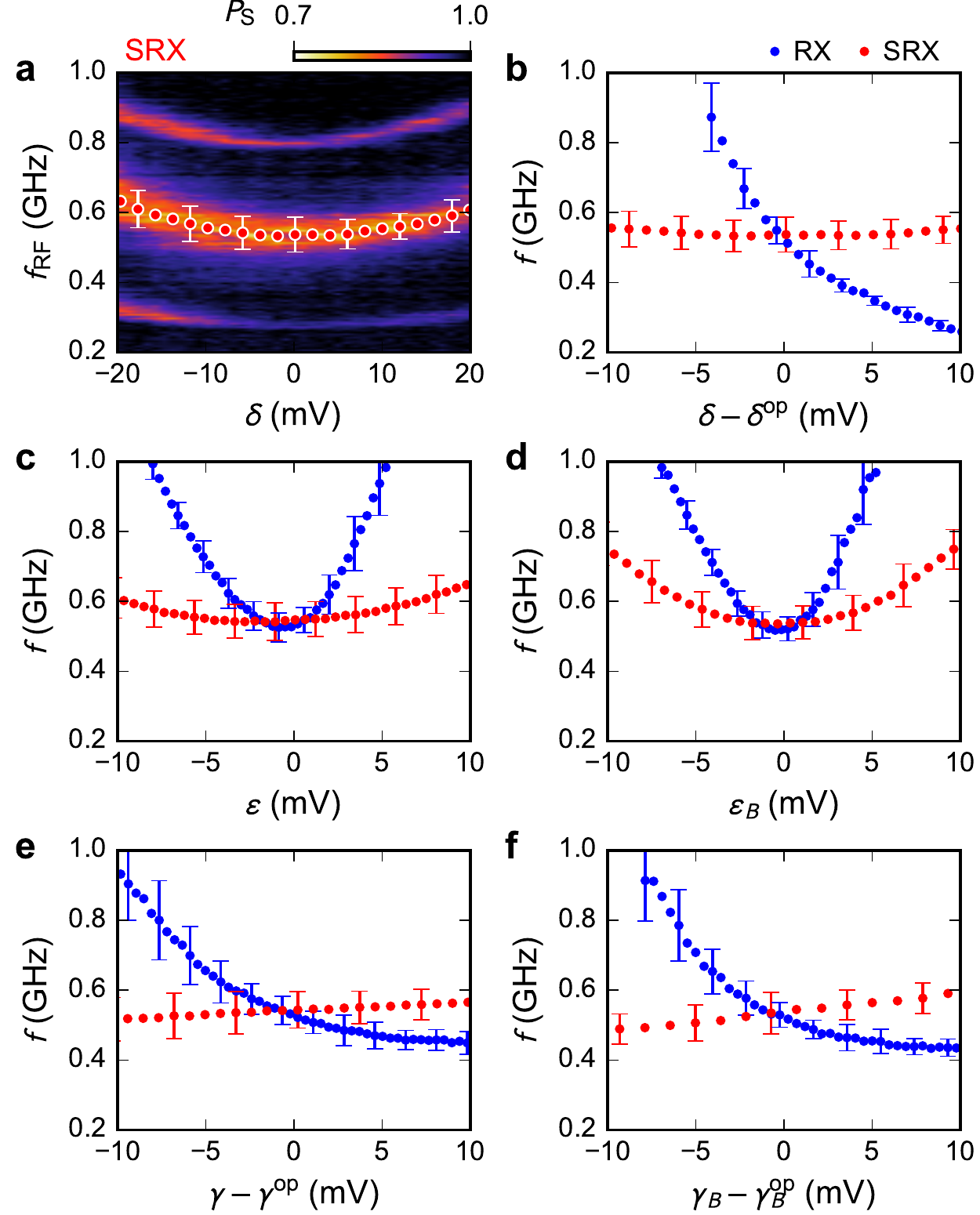}
	\caption[Qubit spectroscopy around the operating point]
	{(a) Qubit spectroscopy along $\delta$ around the SRX operating point (see text). The red circles indicate the extracted qubit splitting $f$. Additional resonances correspond to multiphoton excitations of the triple dot. (b-f) Extracted qubit splitting along $\delta$, $\varepsilon$, $\varepsilon_B$, $\gamma$ and $\gamma_B$ for the SRX (red) and RX (blue) configuration, around their corresponding operating points. 
	Error bars indicate the inhomogeneous line width of the resonance.
	}
	\label{SRX:fig3}
\end{figure}

\section{Qubit spectroscopy and Rabi oscillations}

Qubit spectroscopy performed in the vicinity of the operating point quantitatively reveals each qubit's symmetries and susceptibilities to gate voltage fluctuations. First, maps as in Fig~\ref{SRX:fig3}a are acquired by repeating a pulse cycle with $\tau=150$~ns fixed, and plotting the fraction of singlet readouts, $P_\mathrm{S}$, as a function of $f_\mathrm{RF}$, while stepping the control parameters along five orthogonal axes that intersect with the operating point. 
The qubit frequency $f$ is extracted from the center of the dominant $P_\mathrm{S}(f_\mathrm{RF})$ resonance (cf. red circles in Fig~\ref{SRX:fig3}a), and plotted as a function of $\varepsilon$, $\delta$, $\gamma$, $\varepsilon_B$ and $\gamma_B$ (Fig.~\ref{SRX:fig3}b-f).
Indeed, the dependence of $f$ on $\delta$ reveals that the SRX qubit splitting is to first order insensitive to $\delta$, in contrast to the conventional RX qubit (Fig.~\ref{SRX:fig3}b). Further, we observe that both qubits show a sweet spot with respect to $\varepsilon$ and $\varepsilon_B$ (Fig.~\ref{SRX:fig3}c,d), indicating that the symmetry breaking associated with $\varepsilon_B\neq0$ is analogues to the well-known symmetry breaking associated with $\varepsilon\neq0$~\cite{Taylor2013}. 
Interestingly, for both detuning parameters, the curvature of the qubit splitting is significantly smaller for the SRX configuration, compared to the RX configuration. 
Moreover, the SRX qubit frequency is also significantly less susceptible to changes in parameters $\gamma$ and $\gamma_B$, compared to the conventional RX qubit (Fig.~\ref{SRX:fig3}e,f), corroborating the potential use of this highly symmetric configuration for prolonging qubit coherence. 

The qubit spectra from Figures \ref{SRX:fig3} allow us to quantify the susceptibility of the qubit splitting to gate voltage fluctuations, by evaluating
\begin{equation}
	S = \sqrt{ \sum\limits_{\begin{subarray}{c}
	i \in \{\mathrm{LP, LB,} \\ \mathrm{MP, RB, RP}\}
	\end{subarray}}
	 \left( \frac{\partial f}{\partial V_i} \right)^2}
	 = \sqrt{ \sum\limits_{\begin{subarray}{c}
	\xi \in \{\varepsilon, \delta, \\ \gamma, \varepsilon_B, \gamma_B \}
	\end{subarray}}
	 \left( \frac{\partial f}{\partial \xi} \right)^2
	 }
	 \label{eq:S}
\end{equation}
for both operating points. 
For the SRX qubit we find a susceptibility to gate noise ($S = 6$~MHz/mV) that is one order of magnitude smaller compared to the RX qubit ($S = 66$~MHz/mV). 
For the linear coupling regime this means that voltage fluctuations on gate electrodes, including instrumentation noise propagating on the cryostats wideband transmission lines, are expected to be much less detrimental to the SRX qubit than to the RX qubit.

\begin{figure}[tbh]
	\centering
	\includegraphics[width=0.6\textwidth]{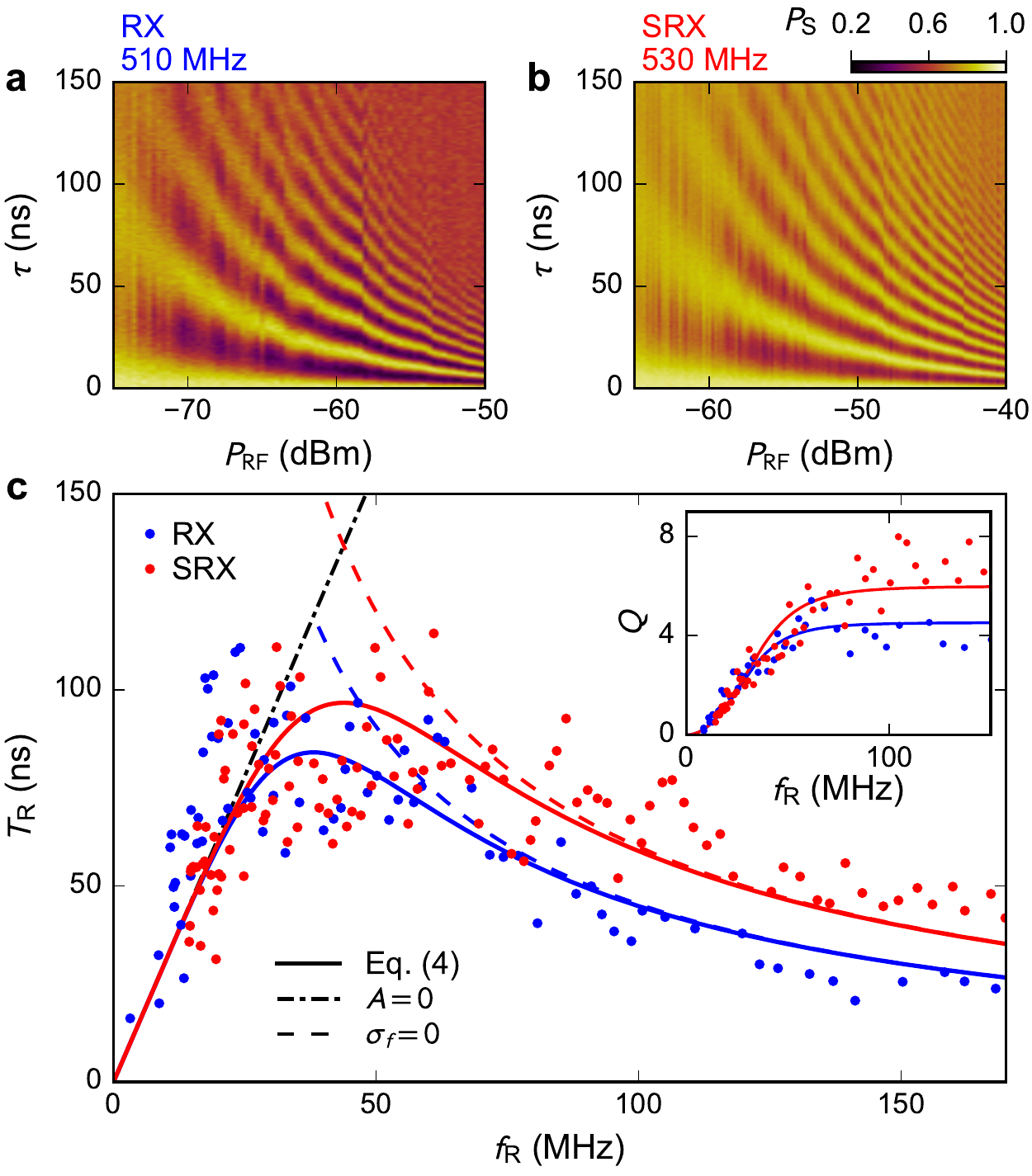}
	\caption[Rabi oscillations of the RX and SRX qubit]
	{(a,b) Rabi oscillations of the RX and SRX qubit as a function of RF burst time ($\tau$) and excitation power ($P_\mathrm{RF}$) obtained at nearly identical qubit splitting of 510 MHz (RX) and 530 MHz (SRX).
	(c) Parametric plot of Rabi decay time $\TR$ and quality factor $Q$ (inset) as a function of Rabi frequency $\fR$, extracted from vertical cuts of (a) and (b). Solid lines are theory fits based on Eq.~\ref{eq:TR} and $Q\equiv\TR \times \fR$. Broken lines indicate the limits imposed by solely detuning noise (black) or solely drive noise (red and blue).
	}
	\label{SRX:fig4}
\end{figure}

Next we investigate whether the reduced noise susceptibility of the SRX qubit results in improved Rabi oscillations (Fig.~\ref{SRX:fig4}). 
To achieve a comparable Rabi frequency, $\fR$,  we find that $P_\mathrm{RF}$ needs to be 10 dB larger for the SRX qubit compared to the RX qubit, consistent with the smaller curvatures observed in Fig.~\ref{SRX:fig3}. However, only for high $P_\mathrm{RF}$ do we observe improvement in SRX qubit performance relative to the RX qubit. For quantitative comparison we fit an exponentially damped cosine to $P_\mathrm{S}(\tau)$ for each RF power. Figure~\ref{SRX:fig4}c parametrically plots the extracted $1/e$ decay time ($\TR$) and quality factor ($Q=\TR \times \fR$) of Rabi oscillations as a function of $\fR$. For $\fR<50$~MHz the quality of SRX Rabi oscillations is comparable to the RX qubit, while for $\fR>50$~MHz $\TR$ and $Q$ are enhanced by approximately 50\%, relative to the RX qubit.

The marginal performance improvement observed for the SRX qubit can be analyzed quantitatively by extending theory from Ref.~\cite{Taylor2013} to include the dependence of the Rabi oscillations decay time $\TR$ on the Rabi frequency $\fR$. Assuming quasistatic gate-voltage noise and quasistatic nuclear spin noise, we derive
\begin{equation}
	\left( \frac{1}{\TR} \right)^2 =  \frac{\sigma_f^4}{4 \fR^2} + \fR^2 A^2,
	\label{eq:TR}
\end{equation}
where $\sigma_f$ quantifies the rms deviation of $f$ from $f_\mathrm{RF}$ due to effective voltage flucuations and Overhauser field fluctuations (discussed below). The quantity $A^2$ captures the effect of voltage fluctuations on the coupling strength of the RF drive
\begin{equation}
	\label{eq:A2}
	A^2 = \frac{8 \pi}{\eta^2}
	\sum\limits_{\begin{subarray}{c}
	\xi = \varepsilon, \delta, \\ \gamma, \varepsilon_B, \gamma_B
	\end{subarray}}
	\left(
	\frac{\partial \eta}{\partial \xi} \sigma_\xi
	\right)^2,
\end{equation}
with $\sigma_\xi$ being the standard deviation of the fluctuating paramater $\xi$ and $\eta$ being the lever arm between amplitude of the RF drive and the qubit nutation speed in the rotating frame. 
We find that the observed $\TR (f_\mathrm{R})$ is well fitted by our theoretical model, using $A= 0.17$ (0.22) for the SRX (RX) qubit and a common value $\sigma_f= 0.025$ (solid lines in Fig.~\ref{SRX:fig4}c). Although Ref.~\cite{Taylor2013} formally identified $\eta$ with 
\begin{equation}
	\label{eq:eta}
	\eta = \sqrt{ \left( \frac{\partial J}{\partial V_{LP}} \right)^2 + 3 \left( \frac{\partial j}{\partial V_{LP}} \right)^2 }
\end{equation}
(here $J = (J_\mathrm{L }+ J_\mathrm{R})/2$ and $j = (J_\mathrm{L} - J_\mathrm{R})/2$ are symmetry-adapted exchange energies arising from exchange $J_\mathrm{L/R}$ between central and left/right dot), its implications for the properties of the $A^2$ term and associated Rabi coherence were not considered. 
Equations~\eqref{eq:A2},\eqref{eq:eta} would in principle allow the extraction of voltage noise in more detail, but experimentally the partial derivatives are not easily accessible. 
However, by plotting the expected limit of $\TR$ if only detuning noise (black dash-dotted line) or only drive noise (red and blue dashed lines) is modeled, we deduce that the dominating contribution to $\sigma_f$ arises not from effective gate voltage noise, but from fluctuations of the Overhauser gradient between dots.    
Assigning $\sigma_f= 0.025$ entirely to Overhauser fluctuations, we estimate the rms Overhauser field in each dot to be approximately 4.2~mT, in good agreement with previous work on GaAs triple dots~\cite{Medford2013a,Delbecq2016}.

The detrimental effect of fluctuating Overhauser fields on qubit dephasing is not surprising, given that the qubit states are encoded in the $S_z=1/2$ spin texture:
For $\ket{0}$ the spin angular momentum resides in the outer two dots, whereas for  $\ket{1}$ it resides in the central dot. This makes the qubit splitting to first order sensitive to Overhauser gradients between the central and outer dots \cite{Taylor2013}. Equation~\ref{eq:eta} further suggests that the qubit drive strength depends on $j = (J_\mathrm{L} - J_\mathrm{R})/2$, which likely is first-order-sensitive to both $\varepsilon$ and $\varepsilon_B$, and hence we suspect that $f_\mathrm{R}$, unlike $f$, remains sensitive to the charge noise.
These conclusions suggest that triple-dot qubits will benefit from implementation in nuclear-spin-free semiconductors, and possibly from replacing IQ-control in the rotating frame by baseband voltage pulses. Recent theoretical work indicates that this may allow efficient two-qubit gates between neighboring qubits using exchange pulses~\cite{Shim2016a} and long-distance coupling via superconducting resonators~\cite{Russ2015a,Srinivasa2016}. 

\section{Coherence under CPMG sequence}

\begin{figure}[tbh]
	\centering
	\includegraphics[width=0.6\textwidth]{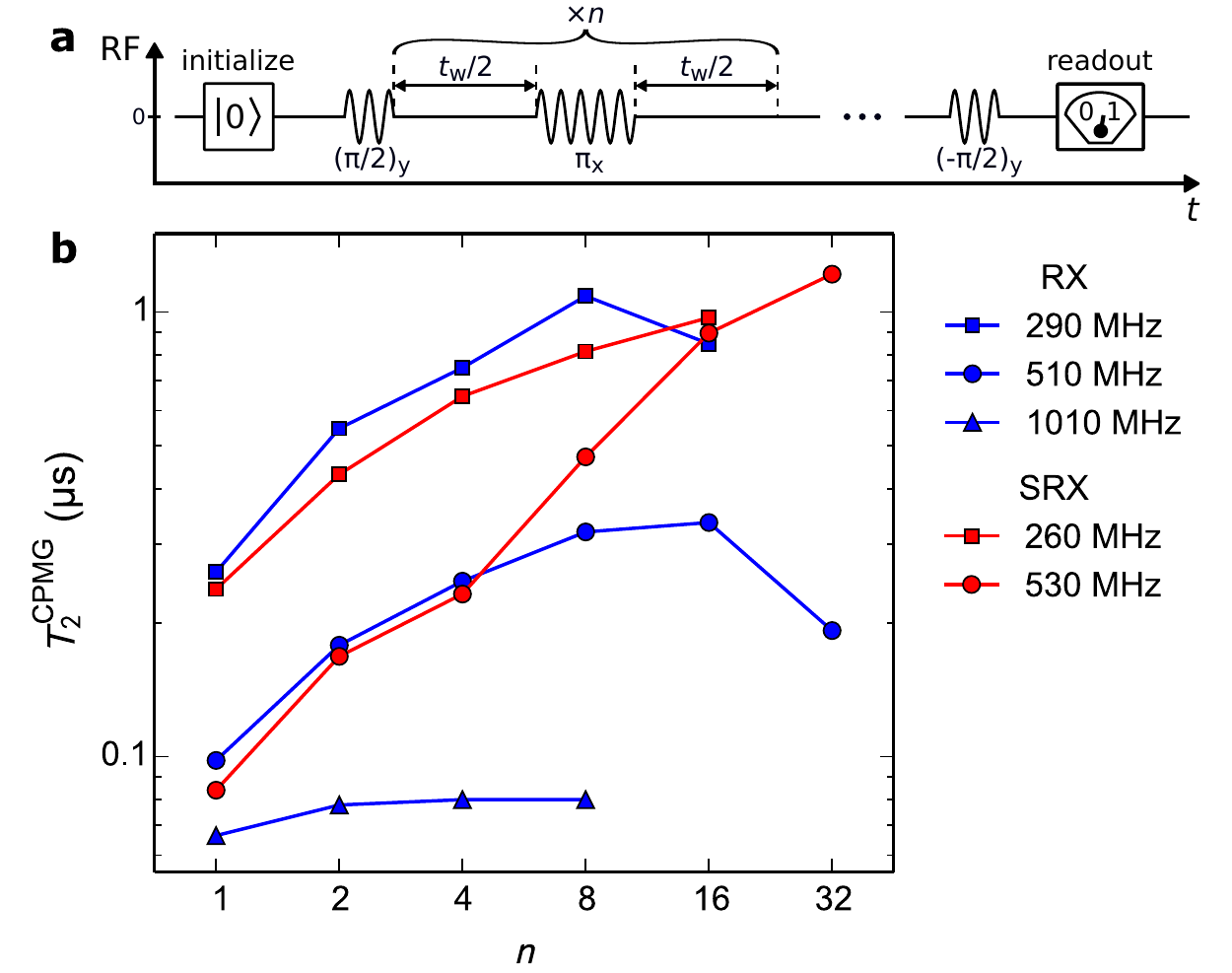}
	\caption[$T_2^\mathrm{CPMG}$ as a function of the number of $\pi$ pulses for various  SRX and RX qubit frequencies]
	{(a) CPMG dynamical decoupling sequence adapted from Ref.~\cite{Medford2013a}. 
	The $(\pi/2)_y$ pulse prepares the superposition state $(1/\sqrt{2})(\ket{0}+\ket{1})$. The segment consisting of a waiting time, $t_W/2$, a $\pi_x$ pulse, and another waiting time, $t_W/2$, is repeated $n$ times ($n=1$ for Hahn echo). The $(-\pi/2)_y$ pulse projects the resulting state onto $\ket{0}$ or $\ket{1}$. The fraction of $\ket{0}$ outcomes, for increasing waiting time and fixed $n$, is used to extract the coherence time $T_2^\mathrm{CPMG}$ (see main text). 
	(b) $T_2^\mathrm{CPMG}$ as a function of the number of $\pi$ pulses for various  SRX and RX qubit frequencies.
	}
	\label{SRX:fig5}
\end{figure}

Finally, we test the prospect of the SRX qubit as a quantum memory, using Hahn echo and CPMG sequences consisting of relatively strong ($\tau \lesssim 10$ ns) $\pi$-pulses (defined in Fig.~\ref{SRX:fig5}a). 
These dynamical decoupling sequences are particularly effective against nuclear noise~\cite{Bluhm2011,Malinowski2017}, which is known to display relative long correlation times~\cite{Reilly2008,Barthel2009,Malinowski2017a}.
Figure \ref{SRX:fig5} shows the resulting coherence time, $T_2^\mathrm{CPMG}$, for different qubit frequencies, for up to $n$=32 pulses. Values for $T_2^\mathrm{CPMG}$ were extracted from Gaussian fits to $P_\mathrm{S}(T)$, where $T=n \cdot t_w$ is the total dephasing time. 
For small number of $\pi$-pulses we see no difference in the performance of the RX and SRX qubit, indicating that effective voltage noise (incl. instrumentation noise on gate electrodes) is not limiting coherence. Qualitatively, this may point towards high-frequency Overhauser fluctuations playing a dominant role, although we find coherence times significantly shorter than expected from nuclear spin noise alone~\cite{Petta2005,Bluhm2011,Malinowski2017a} and values reported for RX qubits \cite{Medford2013a}.  
While $\TCPMG$ strongly depends on the qubit frequency, the ratio $f \times \TCPMG$ is roughly independent of $f$ (not shown). This is reminiscent of gate defined quantum dots that showed a nearly exponential dependence of the exchange splitting on relevant control voltages ~\cite{Higginbotham2014,Veldhorst2015,Dial2013,Medford2013a,Martins2016,Reed2016}. 

Although we do not know the exact origin of the effective noise observed here and in previous work \cite{Medford2013a, Medford2013}, we note that the overall noise levels need to be reduced by several orders of magnitude to allow high-fidelity entangling gates~\cite{Srinivasa2016}.
As a cautionary advice against the overuse of partial sweet spots, we note that for any qubit tuned smoothly by $N$ (in our work five) gate voltages one can always (i.e., for any operating point) define at least $N-1$ (in our work 4) independent control parameters that to first order do not influence the qubit splitting. This underlines the importance of careful analysis of noise sources and noise correlations~\cite{Szankowski2016} in determining optimal working points of qubits~\cite{Cywinski2014}.

For the 530~MHz tuning the SRX qubit appears to outperform the RX qubit for $n>8$, indicating that the spectral noise density at higher frequencies, filtered by the CMPG sequence~\cite{Martinis2003,Cywinski2008,Soare2014,Malinowski2017}, may indeed be reduced for the SRX qubit.
The scaling of $T_2^\mathrm{CPMG}$ with (even) number of pulses appears to follow a power law. Although the exponent (0.77$\pm$0.07) for the SRX data is consistent with values reported for RX qubits~\cite{Medford2013a}, a spectral interpretation may need to take into account unconventional decoherence processes that can occur at sweet spots, such as non-Gaussian noise arising from quadratic coupling to Gaussian distributed noise and the appearance of linear coupling to noise arising from low-frequency fluctuations around a sweet spot~\cite{Bergli2006,Cywinski2014}.

\section{Conclusions}

In conclusion, we have operated a triple-dot resonant exchange qubit in a highly symmetric configuration.
At the three-dimensional sweet spot the overall sensitivity of the qubit frequency to five control voltages is reduced by an order of magnitude, but resonant operation of the qubit is technically more demanding. 
For weak resonant driving the quality of Rabi oscillations show no significant improvement due to the dominant contributions of nuclear Overhauser gradients to fluctuations of the qubit splitting, motivating the future use of nuclear-spin-free semiconductors. 
For strongly driven Rabi oscillations and CPMG decoupling sequences the coherence times are significantly shorter than expected from instrumentation noise alone and Overhauser fluctuations, suggesting that recent theory must be extended to include the dependence of drive strength on control voltages.  
An optimization of gate lever arms and materials' charge noise may then allow non-resonant operation of multi-qubit structures that take advantage of highly symmetric configurations of triple-dot qubits.

\section*{Acknowledgements}

This work was supported by the Army Research Office, the Villum Foundation, the Innovation Fund Denmark, and the Danish National Research Foundation.

\section*{Authors contributions}
S.F., G.C.G. and M.J.M. grew the heterostructure. P.D.N. fabricated the device. F.M., F.K.M., P.D.N., and F.K. prepared the experimental setup. F.K.M. and F.M. performed the experiment. F.K.M., F.M., F.K. and C.M.M. analysed the data and prepared the manuscript.

\section{Unpublished: Derivation of the formula for the Rabi decay time}
\label{SRX:derivation}

To derive a formula for the Rabi decay time (Eq.~\eqref{eq:TR} and \eqref{eq:A2}) we start with the RX qubit splitting in the rotating frame
\begin{equation}
	2 \pi f_R = \Omega_R = \sqrt{(\Omega_d + \delta \Omega_d)^2 + \delta \Omega_Q^2}
\end{equation}
in units of angular frequency, where $\Omega_R$ is the Rabi angular frequency, $\Omega_d$ is the Rabi drive, $\delta \Omega_d$ is the Rabi drive noise and $\delta \Omega_Q$ the noise in the detuning of the qubit frequency. From here we can calculate noise of the Rabi angular frequency the lowest order
\begin{equation}
	\delta \Omega_R \equiv \Omega_R - \Omega_d = \delta \Omega_d + \frac{\left(\delta \Omega_Q\right)^2}{2\Omega_d}.
\end{equation}

Assuming that the noise affecting the qubit frequency and the drive strength are independent, and using the fact that $(\sigma_{X^2})^2= 2 (\sigma_X)^4$ we can convert the above identity into relation between variances
\begin{equation}
	\sigma_R^2 = \sigma_d^2 + \frac{\left(\sigma_Q\right)^4}{2\left( \Omega_d \right)^2} 
\end{equation}
where $\sigma_{R,d,Q}$ indicates, respectively, the variance of the Rabi angular frequency, drive and the qubit frequency. We neglect higher order moments of $\left( \Omega_d \right)^2$ distribution. Knowing $\sigma_R$ we can write the formula for the Rabi decay time
\begin{equation}
	T_R = \frac{1}{\sqrt{2 \pi^2 \left( \sigma_d^2 + \frac{\left(\sigma_Q\right)^4}{2\left( \Omega_d \right)^2}  \right)}}.
	\label{eq:T_R_formula}
\end{equation}
What remains now is to analyze the origin of the noise in the drive strength ($\sigma_d$ term) and in the qubit frequency ($\sigma_Q$ term).

First we focus on the noise in the qubit frequency. The splitting between the qubit states is~\cite{Taylor2013}
\begin{equation}
	\Omega_Q = \sqrt{J^2 + 3 j^2} + \frac{2}{3}(B_L - 2 B_M + B_R)
\end{equation}
where $J = (J_L + J_R)/2$, $j = (J_L - J_R)/2$ and $B_{L/M/R}$ is the electron Zeeman splitting in the left/middle/right dot including contributions from the Overhauser field. We can safely assume that the electrical noise affecting the first term (captured by $\sigma_J$) and the nuclear noise affecting the second term (captured by $\sigma_B$) are independent and so
\begin{equation}
	\sigma_Q^2 = \sigma_J^2 + \sigma_B^2.
\end{equation}
We represent the value of $\sigma_J$ as a result of the effective gate voltage noise ($\sigma_V$), and so its' coupling to the qubit is determined by the gradient of the qubit splitting with respect to gate voltages
\begin{equation}
	\sigma_J^2 = \sigma_V^2 \sum\limits_{\begin{subarray}{c}
	\xi \in \{\varepsilon, \delta, \\ \gamma, \varepsilon_B, \gamma_B \}
	\end{subarray}}
	 \left( \frac{\partial \Omega_Q}{\partial \xi} \right)^2.
\end{equation}
The sum in this formula is the equivalent of the susceptibility $S$ defined in Eq.~\eqref{eq:S}, up to $2\pi$ factor between frequency and angular frequency.
To estimate $\sigma_B$ we assume that the Overhauser field in the three quantum dots is independent and characterized by the same variance $\sigma_{B,0}$. This leads to
\begin{equation}
	\sigma_B^2 = \frac{8}{3} \sigma_{B,0}^2.
\end{equation}
As described in the main text, taking $\sigma_{B,0}=4.2$~mT, which is consistent with direct measurements of the Overhauser field in device of the same geometry~\cite{Malinowski2017a} (Fig.~\ref{ovh_sup:distribution}c), and neglecting the charge noise contribution is sufficient to explain the Rabi decay time for weak driving.

To analyse the drive noise we first write down the Rabi drive as a product of rf exciting voltage amplitude applied to the left plunger gate $V_{LP}^0$ and the lever arm to the qubit drive $\eta$ (defined in Eq.~\eqref{eq:eta})~\cite{Taylor2013}:
\begin{equation}
	\Omega_d = \frac{V_{LP}^{rf}}{2} \eta = \frac{V_{LP}^{rf}}{2} \sqrt{ \left( \frac{\partial J}{\partial V_{LP}} \right)^2 + 3 \left( \frac{\partial j}{\partial V_{LP}} \right)^2 }.
	\label{eq:Omega_d}
\end{equation}
This indicates that the drive strength can be affected either by the noise in the driving rf voltage amplitude $V_{LP}^{rf}$ or in the lever arm $\eta$. We exclude the first possibility relying on two facts: stability of the rf electronics and small power of the charge noise at qubit frequency~\cite{Dial2013}. On the other hand the lever arm $\eta$ is a function of exchange splittings and therefore is susceptible to the effective gate voltage noise. Using these observation we can write down
\begin{equation}
	\sigma_{d} = \frac{V_{LP}^{rf}}{2} \sigma_V \sum\limits_{\begin{subarray}{c}
	\xi \in \{\varepsilon, \delta, \\ \gamma, \varepsilon_B, \gamma_B \}
	\end{subarray}}
	\left( \frac{\partial \eta}{\partial \xi} \right)^2.
\end{equation}
Inserting Eq.~\eqref{eq:Omega_d} we finally get that the variance of the drive noise is proportional to the drive
\begin{equation}
	\sigma_{d} = \frac{\Omega_d}{\eta} \sigma_V \sum\limits_{\begin{subarray}{c}
	\xi \in \{\varepsilon, \delta, \\ \gamma, \varepsilon_B, \gamma_B \}
	\end{subarray}}
	\left( \frac{\partial \eta}{\partial \xi} \right)^2.
	\label{eq:sigma_d}
\end{equation}
Substituting Eq.~\eqref{eq:sigma_d} to Eq.~\eqref{eq:T_R_formula} leads to Eq.~\eqref{eq:TR} and \eqref{eq:A2}.

\chapterimage{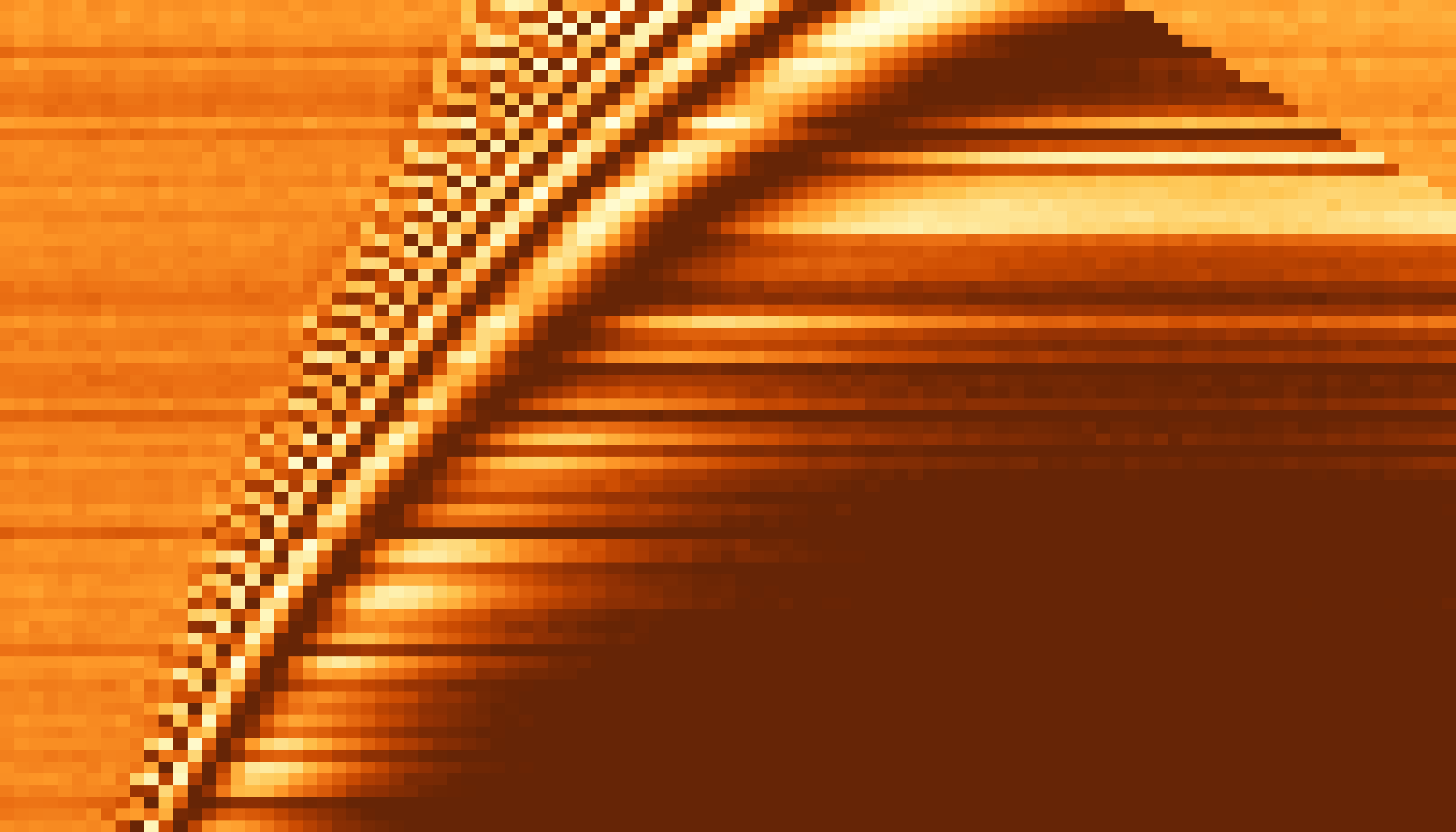}
\chapter[Supplementary Information for ``Noise suppression using symmetric exchange gates in spin qubits'']{\protect\parbox{0.9\textwidth}{Supplementary Information for \\ ``Noise suppression using symmetric \\ exchange gates in spin qubits''}}
\chaptermark{Supplementary Information for ``Noise suppression using...''}
\label{ch:symm_sup}

\begin{center}
\begin{tcolorbox}[width=0.8\textwidth, breakable, size=minimal, colback=white]
	\small
	Supplementary material for ``Noise suppression using symmetric exchange gates in spin qubits" is given on the following topics:
	\begin{enumerate}
		\item[\ref{symm_sup:epsilongamma}] Relationship between control parameters $\varepsilon$, $\gamma$ and gate voltages $V_\mathrm{L}$,  $V_\mathrm{M}$, and $V_\mathrm{R}$
		\item[\ref{symm_sup:fittingJ}] Extracting $J$, $T_\mathrm{R}$ and $Q$ from $P_\mathrm{s}(\tau)$
		\item[\ref{symm_sup:Jmodel}] Model of $J(\varepsilon_{\mathrm{x}},\gamma_{\mathrm{x}})$
		\item[\ref{symm_sup:sigmaJmodel}] Calculation of exchange noise $\sigma_J$, decoherence time $T_\mathrm{R}^\mathrm{(el)}$, and quality factor $Q^\mathrm{(el)}$ arising from quasistatic electrical noise $\sigma_\mathrm{el}$
		\item[\ref{symm_sup:extractsigmael}] Determination of $\sigma_\mathrm{el}$
		\item[\ref{symm_sup:sigmaJcomparison}] Comparison of electrical noise in tilt and symmetric operation
		\item[\ref{symm_sup:calcTQ}] Calculation of $T_\mathrm{R}^\mathrm{(nuc)}$, $T_\mathrm{R}$, $Q^\mathrm{(nuc)}$ and $Q$.
	\end{enumerate}
	Supplementary information appended to this thesis include additional unpublished results on:
	\begin{enumerate}[resume]
		\item[\ref{symm_sup:mapping}] Mapping of $J$ with exchange pulses of fixed time
		\item[\ref{symm_sup:samef}] Exchange oscillations in symmetric, tilt and semi-tilt mode at the same frequency
		\item[\ref{symm_sup:sweet}] Exchange oscillations at sweet spot for various $\gamma_X$
	\end{enumerate}
\end{tcolorbox}
\end{center}

\let\mysectionmark\sectionmark
\renewcommand\sectionmark[1]{}
\section{Relationship between control parameters $\varepsilon$, $\gamma$\\ and gate voltages $V_\mathrm{L}$,  $V_\mathrm{M}$ and $V_\mathrm{R}$}
\let\sectionmark\mysectionmark
\sectionmark{Relationship between control parameters $\varepsilon$, $\gamma$ and gate voltages...}
\label{symm_sup:epsilongamma}

Here, we explain in more detail the actual voltage pulses employed for qubit operation, and their relationship to control parameters for detuning and barrier height.

\begin{figure}[tb]
\begin{center}
\includegraphics[width=\textwidth]{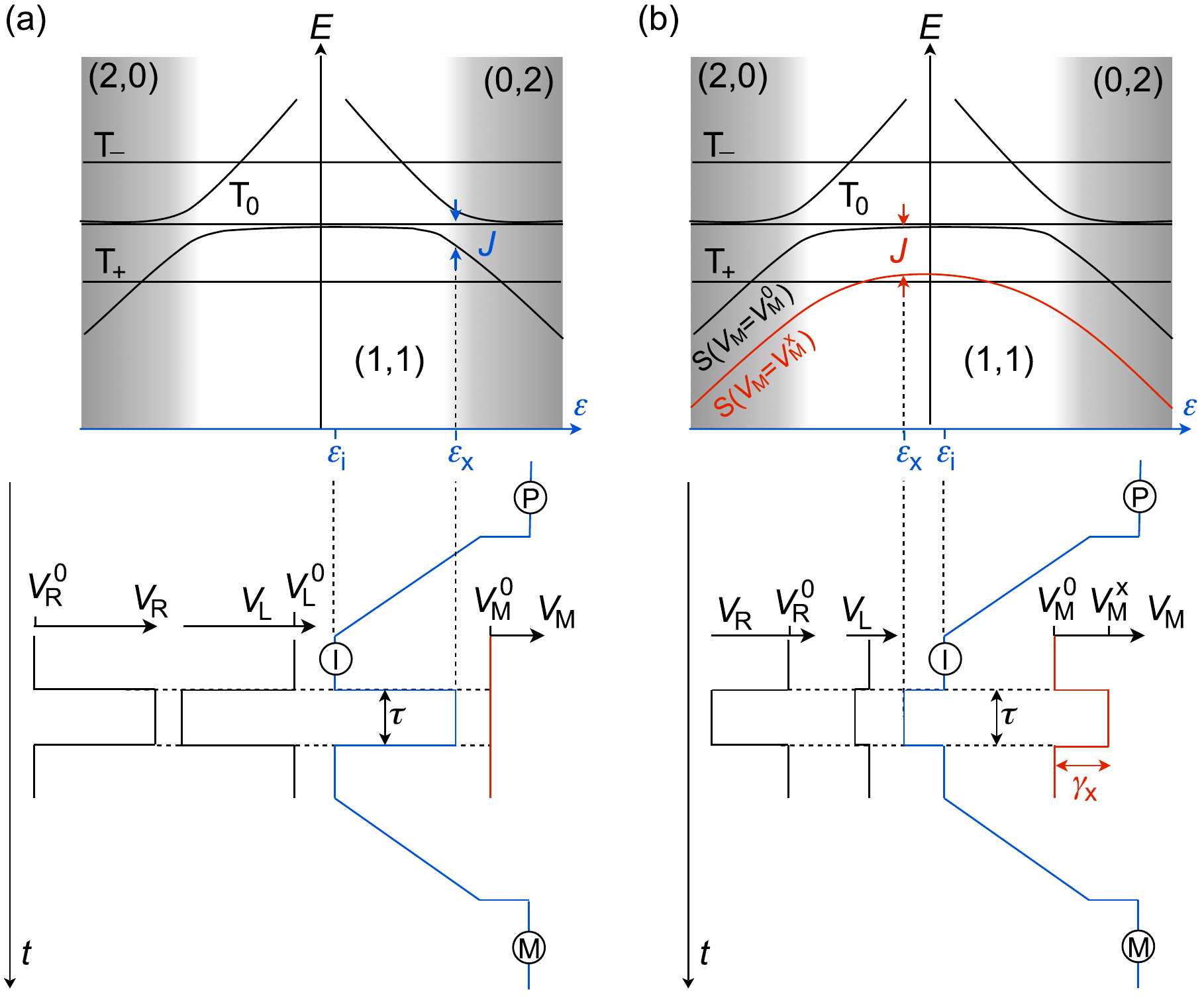}
\caption[Schematic of the energy levels of the two-electron double quantum dot]{Schematic of the energy levels of the two-electron double quantum dot together with voltage pulses $V_{\mathrm{L}}(t)$, $V_{\mathrm{M}}(t)$, $V_{\mathrm{R}}(t)$ that implement a tilted (a) and barrier-induced exchange gate (b). The blue trace indicates detuning of the double dot during preparation of the singlet (P), after initialization of the $\ket{\uparrow\downarrow}$ state (I), and during the measurement of the charge sensor (M).
}
\label{symm_sup:fig1}
\end{center}
\end{figure}

Fig.~\ref{symm_sup:fig1} illustrates the pulse sequence.
Before each exchange pulse the qubit is prepared (P) in the eigenstate of the nuclear gradient field (denoted $\ket{\uparrow\downarrow}$) using standard voltage pulses applied to the left and right gate similar to earlier experiments~\cite{Petta2005}. These pulses realize exchange of electrons with the reservoirs to reset a singlet (0,2) state, a fast crossing of the $S$-$T_+$ degeneracy to avoid leakage into the $T_+$ state, and an adiabatic ramp to the (1,1) charge state that maps the singlet state into the $\ket{\uparrow\downarrow}$ state (I). After each exchange pulse the qubit is read out using standard voltage pulses applied to the left and right gate electrodes. These involve an adiabatic ramp and a fast crossing of the  $S$-$T_+$ degeneracy that maps the $\ket{\uparrow\downarrow}$ qubit state into a (0,2) singlet state or the $\ket{\downarrow\uparrow}$ qubit state into a (1,1) $T_0$ state. In the resulting measurement configuration (M), the charge states (0,2) and (1,1) are discriminated using single shot readout of the sensor quantum dot based on rf reflectometry and thresholding of the demodulated rf voltage~\cite{Barthel2009}. 

The exchange pulse itself differs from conventional operating schemes as it involves fast voltage pulses applied to left, middle and right gate electrodes ($V_{\mathrm{L}}$,$V_{\mathrm{M}}$ and $V_{\mathrm{R}}$ in Fig.~\ref{symm_sup:fig1}). For practical reasons, low-frequency and high-frequency signals are transmitted to the sample holder using twisted pairs and coax transmission lines, respectively, and combine on the sample holder using home-built RC bias tees.
After each readout pulse we apply pulse compensation pulses such that the time average of each coax voltage signal is equal to the voltage of the coax signal just before the exchange pulse. This means that the idling configuration of the qubit, characterized by voltages $V_{\mathrm{L}}^0$, $V_{\mathrm{M}}^0$, $V_{\mathrm{R}}^0$, corresponds to the DC voltages of the twisted pairs connected to the low-frequency input of the bias tees. This ensures that the idling configuration of the qubit does not change within a data set, even when changing amplitude or duration of the exchange pulses. 

In order to turn on a well-defined exchange splitting for a certain amount of time it is convenient to construct voltage pulses ($V_{\mathrm{L}}(t)-V_{\mathrm{L}}^0$, $V_{\mathrm{M}}(t)-V_{\mathrm{M}}^0$, $V_{\mathrm{R}}(t)-V_{\mathrm{R}}^0$) based on three control parameters $\delta$, $\varepsilon$, and $\gamma$:
\begin{equation}
	\left\lbrace
	 \begin{array}{rcl}
	 	V_{\mathrm{L}}-V_{\mathrm{L}}^0 & = & \delta-\varepsilon-\alpha_1\gamma
		\\
		V_{\mathrm{R}}-V_{\mathrm{R}}^0 & = & \delta+\varepsilon-\alpha_2\gamma
		\\
		V_{\mathrm{M}}-V_{\mathrm{M}}^0 & = & \gamma
 	 \end{array}
 	 \right.
\end{equation}
\label{definition_0}
where, $\alpha_1 =0.675$ and $\alpha_2=0.525$.

These equations show that the three parameters $\delta$, $\varepsilon$, and $\gamma$ parameterize physically different manipulations of the (1,1) charge configuration. Namely, $\delta$ controls the common mode of the plunger gates (it appears with a plus sign in each equation) and brings the (1,1) charge state, which is in deep Coulomb blockade, toward the energy of the (2,2) or (0,0) charge states. In contrast, the detuning parameter $\varepsilon$ controls how much the double well potential is tilted towards the (0,2) charge state ($\varepsilon >0$) or the (2,0) charge state ($\varepsilon <0$). The barrier height in the double well potential is controlled by $\gamma$, which appears with a positive sign in the equation for $V_{\mathrm{M}}$ (i.e. positive $\gamma$ corresponds to lower barrier height/increased exchange splitting) and with a negative gain in $V_{\mathrm{L,R}}$ (in order to minimize its contribution to the common mode voltage).

For the symmetric operation of the exchange gate the choice of detuning and barrier during the time of exchange rotation, ($\varepsilon_{\mathrm{x}}$,$\gamma_{\mathrm{x}}$), are most important, as these parameters determine how much virtual tunneling to (0,2) and (2,0) can occur (setting the speed of the exchange gate), and how balanced these processes are (minimizing the sensitivity to $\varepsilon$ noise). The common mode voltage during the exchange pulse, $\delta_{\mathrm{x}}$, as well as the detuning voltage after initialization of the $\ket{\uparrow\downarrow}$ state, $\varepsilon_{\mathrm{i}}$, have a much weaker effect on the quality of observed exchange rotations, and therefore have not been studied systematically. For the measurement presented in the main text, we choose $\delta_{\mathrm{x}}=0$ and $\varepsilon_{\mathrm{i}}=13.5$~mV. 

\begin{figure}[tb]
\begin{center}
\includegraphics[width=\textwidth]{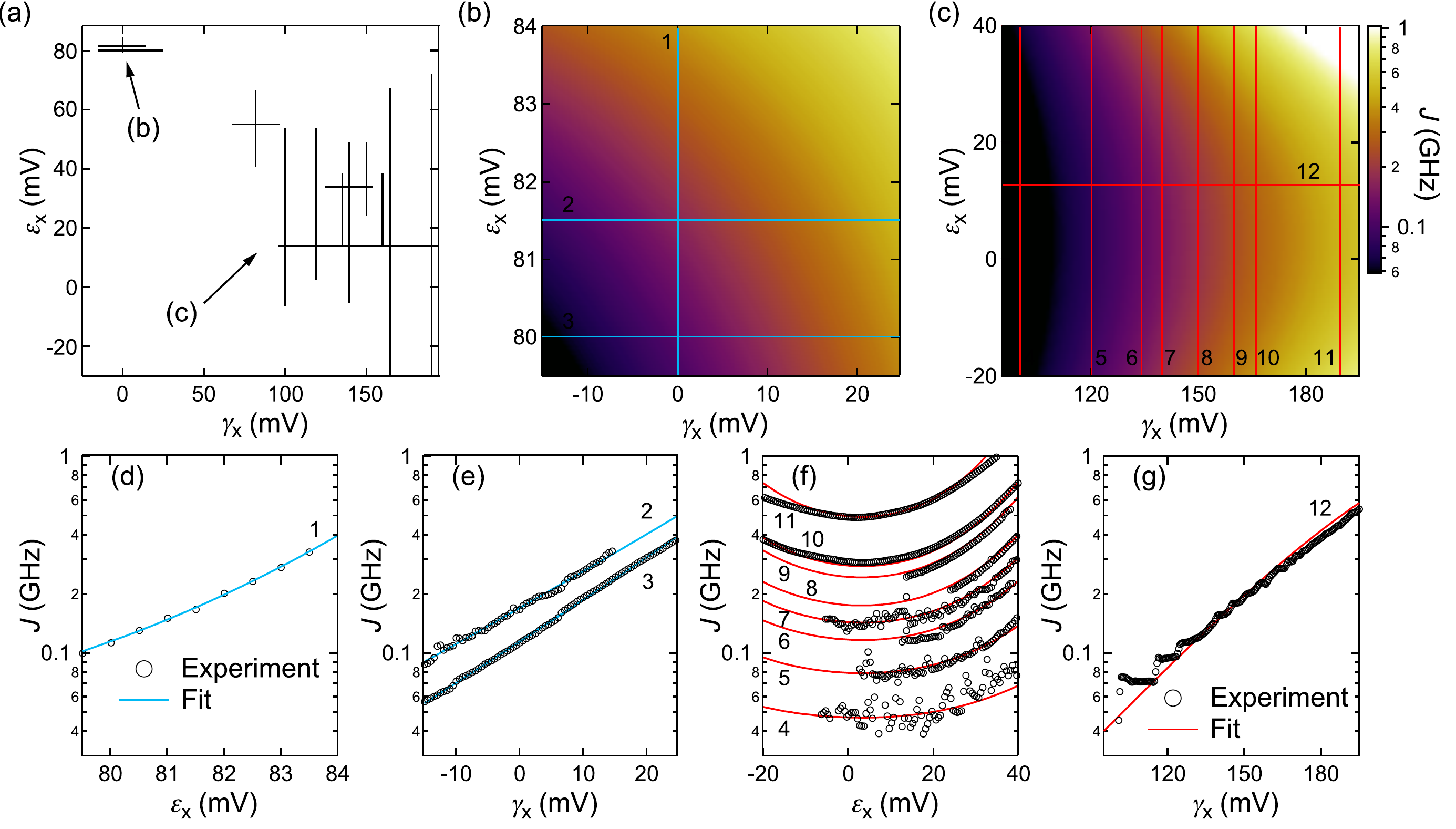}
\caption[Fit of phenomenological exchange model]{
(a) Range of control parameters $\varepsilon_{\mathrm{x}}$ and $\gamma_{\mathrm{x}}$ for which a numerical model of $J$ was developed (gray shaded regions). Each black line corresponds to operating points ($\varepsilon_{\mathrm{x}}$,$\gamma_{\mathrm{x}}$) where $P_\mathrm{s}(\tau)$ was measured. Extracting $J$ from $P_\mathrm{s}(\tau)$ on a subset of these lines yields the data points (symbols) plotted in panels (d-g). Fitting equation \eqref{fit} to the symbols in panels (d-g) yields a two-dimensional model $J(\varepsilon_{\mathrm{x}},\gamma_{\mathrm{x}})$ presented in panels (b) and (c). Cuts indicated by numbers in (b) and (c) correspond to the solid lines in panels (d-g).  
}
\label{symm_sup:fig2}
\end{center}
\end{figure}

\section{Extracting $J$, $T_\mathrm{R}$ and $Q$ from $P_\mathrm{s}(\tau)$}
\label{symm_sup:fittingJ}

For each operating point of the exchange oscillation, $(\varepsilon_{\mathrm{x}},\gamma_{\mathrm{x}})$, the exchange interaction $J(\varepsilon_{\mathrm{x}},\gamma_{\mathrm{x}})$ can be determined by measuring $P_\mathrm{s}(\tau)$ and extracting the oscillation frequency. This method is justified even for $h_0\neq0$ provided that $P_\mathrm{s}(\tau)$ represents an average over a sufficiently large (quasistatic) ensemble characterized by $h_0<\sigma_h$. For our data sets, which typically involve averaging times exceeding 10 minutes, this condition is satisfied, and hence we do not have to take into account nuclear contributions to the oscillation frequency of type $\sqrt{J^2+h_0^2}$~\cite{Barnes2016}.

Specifically, we extract the frequency of the exchange oscillations for selected operating points [black lines in Fig.~\ref{symm_sup:fig2}(a)] in a two step process. First we calculate the discrete Fourier transform of $P_\mathrm{s}(\tau)$ and identify the main peak. Then we use the frequency associated with the main peak as an initial guess for fitting a damped sine wave of frequency $J$ to $P_\mathrm{s}(\tau)$, with a decay of the form exp$\left[-\left(\tau/T_\mathrm{R}\right)^\alpha\right]$. The quality factor is obtained using the relation $Q=J T_\mathrm{R}$. Values of $J$ obtained by this method are plotted as symbols in Fig.~\ref{symm_sup:fig2}(d-g).

\section{Model of $J(\varepsilon_{\mathrm{x}},\gamma_{\mathrm{x}})$}
\label{symm_sup:Jmodel}

The operating points ($\varepsilon_{\mathrm{x}},\gamma_{\mathrm{x}}$) associated with data presented in panels Fig.~\ref{symm_sup:fig2} (d,e,f,g) fall onto a grid in the two-dimensional barrier-detuning space, as shown by black lines in  Fig.~\ref{symm_sup:fig2}(a). In order to inspect the sensitivity of $J$ to small fluctuations in  
$\varepsilon_{\mathrm{x}}$ and $\gamma_{\mathrm{x}}$ a two-dimensional model 
$J(\varepsilon_{\mathrm{x}},\gamma_{\mathrm{x}})$ is needed. 

Our phenomenological model of $J(\varepsilon_{\mathrm{x}},\gamma_{\mathrm{x}})$ is given by:
\begin{equation}
 	J(\varepsilon_{\mathrm{x}},\gamma_{\mathrm{x}}) =e^{\left[c+\left(y_0+y_1\varepsilon_{\mathrm{x}}+y_2 \varepsilon_{\mathrm{x}}^2+y_3 \varepsilon_{\mathrm{x}}^3+y_4 \varepsilon_{\mathrm{x}}^4+y_5 \varepsilon_{\mathrm{x}}^6\right)\times\left((\gamma_{\mathrm{x}}-x_0) s_0+(\gamma_{\mathrm{x}}-x_1)^2 s_1\right)\right]} \mathrm{GHz}
\end{equation}
\label{fit}

\begin{table}[t]
\centering
\begin{tabular}{ c||c|c }
 Parameters & Fig.~\ref{symm_sup:fig2} (b) & Fig.~\ref{symm_sup:fig2} (c)\\
  \hline  \hline  
$c$&  -2.62 &  -1.80\\
  \hline
$x_0$& 650  mV&-487  mV\\
  \hline 
$s_0$&0.351 mV$^{-1}$ &0.441 mV$^{-1}$\\
  \hline
$x_1$&-1695 mV& 1770 mV\\ 
  \hline
$s_1$  &8.01 $10^{-5}$ mV$^{-2}$&-7.21 $10^{-5}$ mV$^{-2}$\\
  \hline
$y_0$&0.205 & 0.0705\\
   \hline
$y_1$& 8.39 $10^{-6}$ mV$^{-1}$& -9.39 $10^{-5}$ mV$^{-1}$\\
  \hline 
$y_2$& 2.11 $10^{-5}$  mV$^{-2}$ & 1.42 $10^{-5}$  mV$^{-2}$\\
  \hline
$y_3$& -1.67 $10^{-7}$  mV$^{-3}$ &5.80 $10^{-12}$ mV$^{-3}$\\ 
  \hline
$y_4$  & 8.25 $10^{-9}$ mV$^{-4}$ & 2.70 $10^{-9}$ mV$^{-4}$\\
  \hline
$y_5$ &-3.46 $10^{-13}$  mV$^{-6}$ & -3.80 $10^{-13}$ mV$^{-6}$\\
\end{tabular}
\caption[Parameters for calculating the smooth exchange profile]
{Parameters for calculating the smooth exchange profile $J(\varepsilon_{\mathrm{x}},\gamma_{\mathrm{x}})$ from Eq.~\ref{fit}, shown in Figure~\ref{symm_sup:fig2}(b,c). These parameters were obtained by fitting Eq.~\ref{fit} to symbols in Fig.~\ref{symm_sup:fig2}(d-g).}
\label{symm_sup:table1}
\end{table}

Using parameters from Table~\ref{symm_sup:table1}, this model provides an excellent interpolation of $J$ in the gray shaded regions in Fig.~\ref{symm_sup:fig2}(a). 
These two regions have been selected based on the insight that they provide into the origin of the drastically different performance of tilted exchange gates and symmetric exchange gates (cf. section \ref{symm_sup:sigmaJmodel} below).
Comparing line cuts of the model $J(\varepsilon_{\mathrm{x}},\gamma_{\mathrm{x}})$ with observed values of $J$ shows that our numerical model of $J$ accurately captures the observed exchange profile of the device [cuts of panels Fig.~\ref{symm_sup:fig2}(b,c) are shown as solid lines in panels Fig.~\ref{symm_sup:fig2}(d,e,f,g)]. The next section uses partial derivatives of this model to calculate the effects of effective gate noise.

\let\mysectionmark\sectionmark
\renewcommand\sectionmark[1]{}
\section{Calculation of exchange noise $\sigma_J$, decoherence time $T_\mathrm{R}^\mathrm{(el)}$,\\ and quality factor $Q^\mathrm{(el)}$ arising from quasistatic electrical noise $\sigma_\mathrm{el}$}
\let\sectionmark\mysectionmark
\sectionmark{Calculation of exchange noise $\sigma_J$, decoherence time $T_\mathrm{R}^\mathrm{(el)}$...}
\label{symm_sup:sigmaJmodel}

\begin{figure}[tb]
\begin{center}
\includegraphics[width=\textwidth]{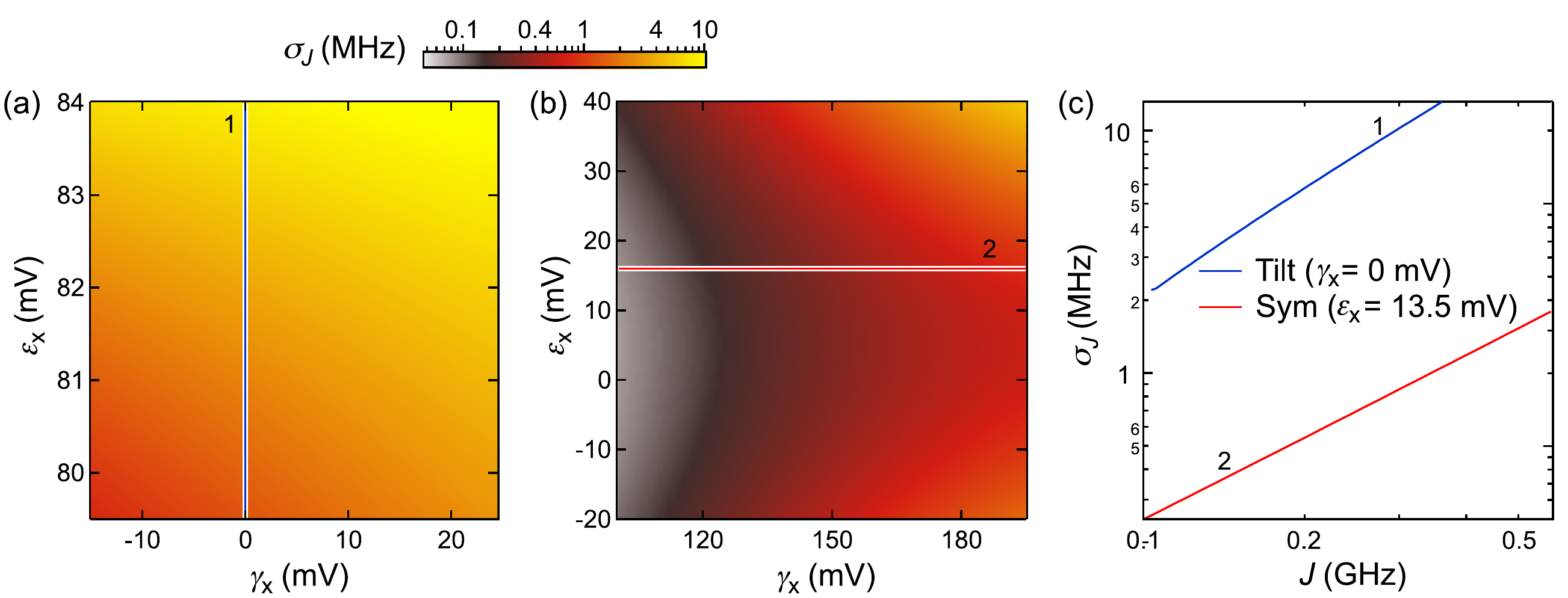}
\caption[$J$ noise $\sigma_J$ as a function of $\varepsilon$ and $\gamma$]{
(a) and (b) $J$ noise $\sigma_J$ as a function of $\varepsilon$ and $\gamma$ for the surfaces in Fig.~\ref{symm_sup:fig2}(b) and (c). Both panels assume the same amount of effective gate noise ($\sigma_\mathrm{el}$~=~0.18~mV).
(c) $\sigma_J$, extracted from cuts 1 and 2 indicated in panel (a) and (b), as a function of $J$ for tilt and symmetric mode of operation.
}
\label{symm_sup:fig3}
\end{center}
\end{figure}

In this section we describe how $T_\mathrm{R}^\mathrm{(el)}$, and quality factor $Q^\mathrm{(el)}$ in Fig.~4(c) and 4(d) in main text were calculated. 

We disregard nuclear fluctuations and consider decoherence caused by quasistatic effective gate noise only. Small fluctuations of control parameters $\varepsilon$ or $\gamma$ result in fluctuations of $J$ with an amplitude that is proportional to the partial derivative of $J$ with respect to $\varepsilon$ or $\gamma$. For comparison with other experiments, and in order to model the decoherence due to electrical noise, it is useful to express fluctuations of $J$ arising fo $\varepsilon$ or $\gamma$ noise in terms of effective gate noise on $V_\mathrm{L}$, $V_\mathrm{M}$, and $V_\mathrm{R}$.

From Eq.\ \ref{definition_0} we obtain the following relations between partial derivatives of $J$:
\begin{equation}
	\left\lbrace
	\begin{array}{rcl}
	\frac{dJ}{dV_\mathrm{L}} &=& -k_{\mathrm{0}} \frac{dJ}{d\varepsilon}
	\\
	\frac{dJ}{dV_\mathrm{R}} &=& k_{\mathrm{0}} \frac{dJ}{d\varepsilon}
	\\
	\frac{dJ}{dV_\mathrm{M}} &=& k_{\mathrm{1}} \frac{dJ}{d\varepsilon}+\frac{dJ}{d\gamma}
	\end{array}
	\right.
	\label{definition_2}
\end{equation}

Assuming that the effective gate noise associated with $V_{\mathrm{L}}$, $V_{\mathrm{M}}$ and $V_{\mathrm{R}}$ is quasistatic, independent, and Gaussian distributed with a common standard deviation $\sigma_\mathrm{el}$, i.e. $\sigma_\mathrm{L}=\sigma_\mathrm{M}=\sigma_\mathrm{R}=\sigma_\mathrm{el}$, we can write the expected fluctuations of $J$:
\begin{equation}
	\begin{array}{c}
	\sigma_J=\sqrt{\left(\frac{dJ}{dV_\mathrm{L}}\sigma_L\right)^2+\left(\frac{dJ}{dV_\mathrm{M}}\sigma_M\right)^2+\left(\frac{dJ}{dV_\mathrm{R}}\sigma_R\right)^2}=\sigma_\mathrm{el}\sqrt{2k_0^2\left(\frac{dJ}{d\varepsilon}\right)^2+\left(\frac{dJ}{d\gamma}+k_1\frac{dJ}{d\varepsilon}\right)^2} 
	\end{array}
	\label{suppl_Ed_Model_complement}
\end{equation}

Averaging over a quasistatic, Gaussian ensemble of $J$ with standard deviation $\sigma_J$ yields a Gaussian decay envelope $\exp\left[-(\tau/T_{\mathrm{R}})^2\right]$, a decoherence time given by $T_{\mathrm{R}}^\mathrm{(el)}=1 /\left(\sqrt{2}\pi\sigma_J\right)$, and a quality factor given by $Q^\mathrm{(el)}= JT_{\mathrm{R}}^\mathrm{(el)}$ \cite{Barnes2016,Dial2013}. To generate the associated curves in Fig.~\ref{symm:fig4}(c) and \ref{symm:fig4}(d) in the main text we use $\sigma_\mathrm{el}$~=~0.18~mV, determined as described in the next section. 

\section{Determination of $\sigma_\mathrm{el}$}
\label{symm_sup:extractsigmael}

We determine the effective gate noise $\sigma_\mathrm{el}$ by measuring tilt-induced exchange oscillations in a regime where effective detuning noise $\sigma_\varepsilon$ dominates, i.e. for $\varepsilon_{\mathrm{x}}$~=~84~mV. 
Fitting a sinusoid with a Gaussian envelope, $ \exp\left[-(\tau/T_{\mathrm{R}})^2\right]$, yields $T_{\mathrm{R}}= 14$ ns. Using $T_{\mathrm{R}}=1/\left(\sqrt{2}\pi\sigma_J\right)$ yields $\sigma_J~=~1.6$ MHz. 
Taking in account Equations \ref{fit},\ref{definition_2} and \ref{suppl_Ed_Model_complement} and Table \ref{symm_sup:table1} this value corresponds to an effective gate noise $\sigma_\mathrm{el}$~=~0.18~mV. 

\section{Comparison of electrical noise in tilt and symmetric operation}
\label{symm_sup:sigmaJcomparison}

Application of Eq.~\ref{suppl_Ed_Model_complement} to the model $J(\varepsilon_{\mathrm{x}},\gamma_{\mathrm{x}})$ shown in Fig.~\ref{symm_sup:fig2}(b) and (c) allows us to calculate  $\sigma_J$ [Fig.~\ref{symm_sup:fig3} (a) and (b)].
To highlight the difference in magnitude of $\sigma_J$ between tilt and symmetric mode of operation we use the same color scale for panels (a) and (b), and compare two cuts plotted against $J$ in panel (c). 
This analysis demonstrates that the same amount of effective gate noise ($\sigma_\mathrm{el}$~=~0.18~mV) results in exchange noise ($\sigma_J$) that is more than one order of magnitude larger in the tilt mode of operation than the symmetric mode of operation, for  a given $J$.

\section{Calculation of $T_\mathrm{R}^\mathrm{(nuc)}$, $Q^\mathrm{(nuc)}$}
\label{symm_sup:calcTQ}

Here we describe how theoretical curves $T_\mathrm{R}^\mathrm{(nuc)}$ and $Q^\mathrm{(nuc)}$ in Figures 4(c,d) of the main text were calculated. 
These quantities represent the contribution of nuclear noise to the total noise.
First, exchange oscillations were simulated using Eqs. 3 of the main text for both tilt and symmetric mode of operations, similar to simulating insets shown in Fig. 2(a) and (b) in the main text, but using $\sigma_\mathrm{el}=0$ while keeping all other parameters unchanged. 
From these simulations, $T_\mathrm{R}^\mathrm{(nuc)}$ and $Q^\mathrm{(nuc)}$ were extracted in the same way as $T_\mathrm{R}$ and $Q$ were extracted as described in section~\ref{symm_sup:fittingJ}.

\begin{figure}[tb]
\begin{center}
\includegraphics[width=\textwidth]{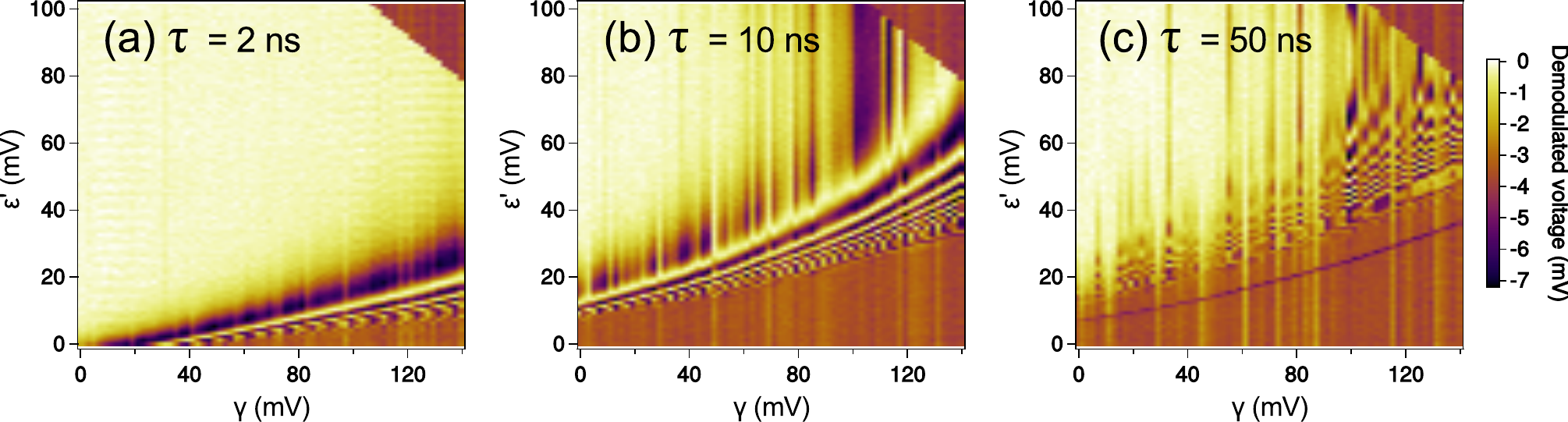}
\caption[Exchange maps with constant $\tau$]{
Exchange maps with constant $\tau$ = 2, 10, 50 ns ((a), (b), (c) respectively). Change of color in vertical lines noise is related to random changes in Overhauser field, that become more relevant for longer exchange pulses. Black line in Fig.~(c) ($\tau=50$ ns) shows the position of $S$-$T_+$ crossing. Data in top right corner of each plot is missing because required voltage for exchange pulse exceeded amplitude of Tektronix 5014c AWG.
}
\label{symm_sup:fig4}
\end{center}
\end{figure}

\section{Unpublished: Mapping of $J$ with exchange pulses of fixed time}
\label{symm_sup:mapping}

Alternatively to extracting $J$ profile from oscillations $P_S(\tau)$ for various values of $\gamma$ and $\delta$ (Sec.~\ref{symm_sup:fittingJ} and Fig.~\ref{symm_sup:fig2}) $\tau$ can be kept fixed while both $\gamma$ and $\delta$ are sweeped. In the resulting map (Fig.~\ref{symm_sup:fig4}) lines of constant $P_S(\gamma,\delta)$ indicate lines of constant $J$. And so the first minimum (black stripe) will correspond to $J \tau / \hbar = \pi$, first maximum (bright stripe): $J \tau / \hbar = 2 \pi$ and so on. Density of oscillations represents therefore the gradient of $J$ and susceptibility to charge noise.

In Fig.~\ref{symm_sup:fig4} we can see that for large $\varepsilon=-\varepsilon '+83\approx0$ the density of the oscillations decreases indicating the symmetry point. Unfortunately, we couldn't observed the full symmetry of the pattern~\cite{Reed2016} because of limited voltage range of the waveform generator and high attenuation of the fast lines.

Fast decay of exchange oscillations requires usage of various times of exchange time $\tau$ to probe various regimes of $J$. Choice of $\tau = 50$ ns will correspond to probing $J\sim 50$ MHz which is comparable to the gradient of Overhauser field. As a result the measurement becomes more noisy. 50~ns is also sufficiently long time enable leakage to fully polarized $T_+$ state. The crossing between $S$ and $T_+$ states is visible as a single dark line in Fig.~\ref{symm_sup:fig4}(c).

\let\mysectionmark\sectionmark
\renewcommand\sectionmark[1]{}
\section{Unpublished: Exchange oscillations in symmetric, tilt\\ and semi-tilt mode at the same frequency}
\label{symm_sup:samef}
\let\sectionmark\mysectionmark
\sectionmark{Exchange oscillations in symmetric, tilt and semi-tilt mode...}

\begin{figure}[bt]
\begin{center}
\includegraphics[width=0.7\textwidth]{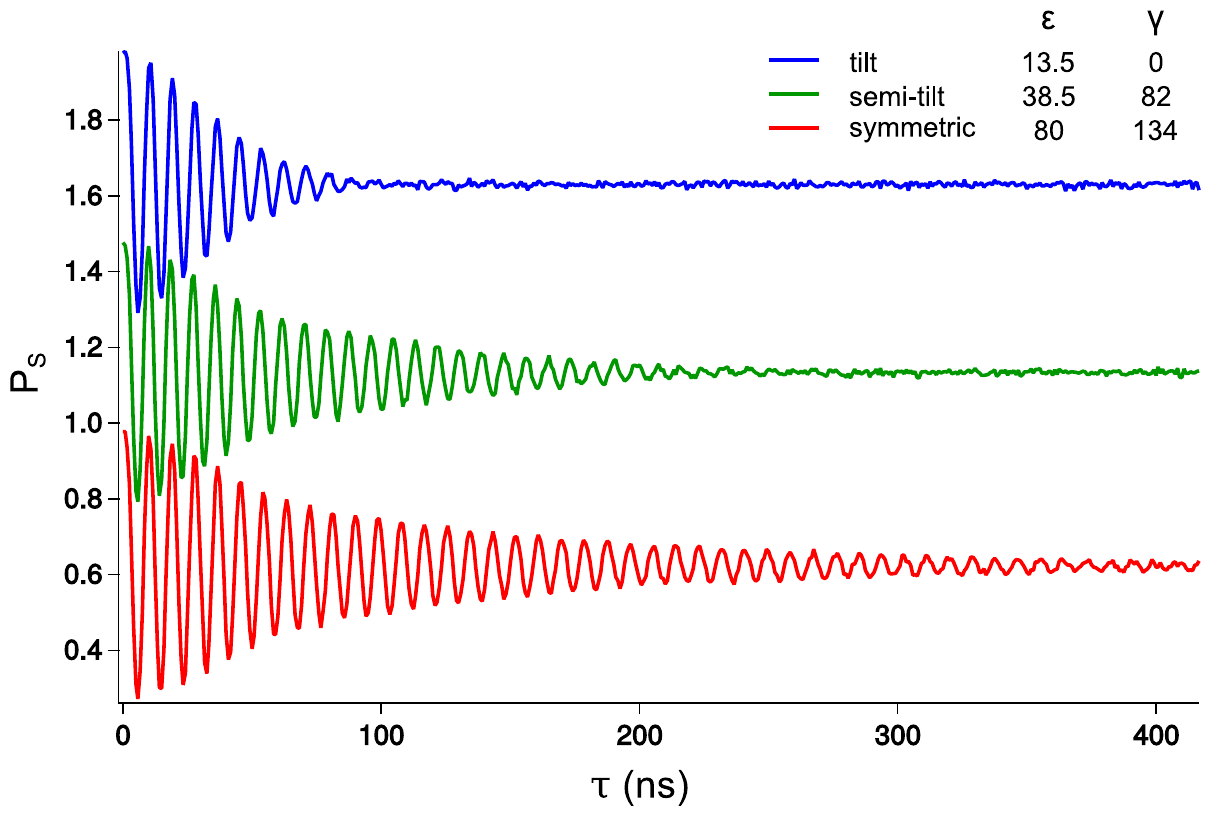}
\caption[Exchange oscillations in tilt, semi-tilt and symmetric mode at the same frequency]{
Exchange oscillations in tilt (blue), semi-tilt (green) and symmetric (red) mode at the same frequency. In the legend values of $\varepsilon$ and $\gamma$ at the exchange point are given.
}
\label{symm_sup:fig5}
\end{center}
\end{figure}

The additional set of exchange oscillations presented here (Fig.~\ref{symm_sup:fig5}) shows how $Q$ of exchange oscillations for fixed frequency changes between charge-noise-dominated (tilt) and nuclear-noise dominated (symmetric) regime. The top curve shows Gaussian decay for the pure tilt mode. The middle curve probes intermediate regime. This exchange rotations involved pulse along both $\varepsilon$ and $\gamma$ axis. Non-Gaussian tail indicates reduced influence of a quasistatic charge noise. The bottom curve was acquired in the pure symmetric mode. The decay throughout approximately first 50 ns remains unchanged compared to the first two traces, but the extremely long power-law tail that it is the Overhauser field fluctuations that limit the fidelity of oscillations.

\let\mysectionmark\sectionmark
\renewcommand\sectionmark[1]{}
\section{Unpublished: Exchange oscillations at sweet spot for various $\gamma_X$}
\label{symm_sup:sweet}
\let\sectionmark\mysectionmark
\sectionmark{Exchange oscillations at sweet spot...}

\begin{figure}[bt]
\begin{center}
\includegraphics[width=0.7\textwidth]{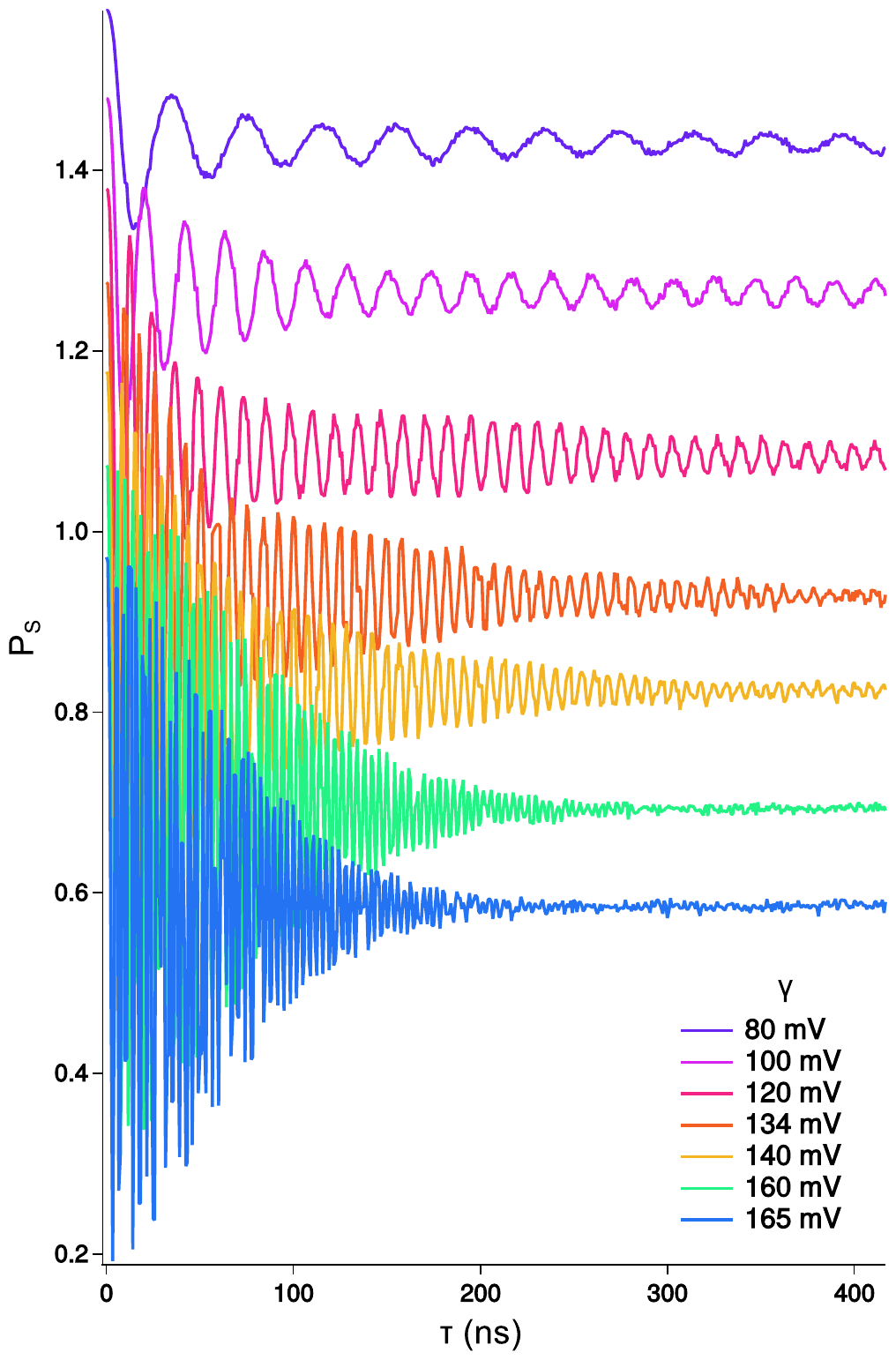}
\caption[Exchange oscillations at sweet spot for various $\gamma_X$]{
Exchange oscillations at sweet spot for various $\gamma_X$. Traces are offset vertically for clarity. Irregularity of oscillations at highest frequencies is caused by aliasing.
}
\label{symm_sup:fig6}
\end{center}
\end{figure}

The final dataset (Fig.~\ref{symm_sup:fig6}) presents crossover from nuclei-dominated decay to charge-noise-dominated decay. At smallest $\gamma_X$ decay has a power-law tail, revealing influence of the nuclei. Notably up to $\gamma_X\approx140$ mV the characteristic decay time remains roughly unchanged. Meanwhile the frequency of oscillations dramatically increases, indicating substantial increase of $Q$. For $\gamma_X>140$ mV the envelope shortens and becomes Gaussian. This is a result of dominant role of charge noise. At this point $Q$ is close to the saturation (Fig.~\ref{symm:fig4}d).

\part{Nuclear dynamics and dynamical decoupling}
\label{part:nuclei}
\providecommand{\VL}{}
\renewcommand{\VL}{V_\mathrm{L}}
\providecommand{\VM}{}
\renewcommand{\VM}{V_\mathrm{M}}
\providecommand{\VR}{}
\renewcommand{\VR}{V_\mathrm{R}}

\providecommand{\Btot}{}
\renewcommand{\Btot}{B^\mathrm{tot}}
\providecommand{\Bext}{}
\renewcommand{\Bext}{B^\mathrm{ext}}
\providecommand{\Bznuc}{}
\renewcommand{\Bznuc}{B_\mathrm{z}^\mathrm{nuc}}
\providecommand{\Bpnuc}{}
\renewcommand{\Bpnuc}{B_\perp^\mathrm{nuc}}

\providecommand{\ud}{}
\renewcommand{\ud}{\uparrow\downarrow}
\providecommand{\du}{}
\renewcommand{\du}{\downarrow\uparrow}

\providecommand{\drv}{}
\renewcommand{\drv}{\mathrm{d}}

\providecommand{\FHahn}{}
\renewcommand{\FHahn}{F_\mathrm{Hahn}}
\providecommand{\FCPMG}{}
\renewcommand{\FCPMG}{F_{\mathrm{CPMG},n}}
\providecommand{\FFID}{}
\renewcommand{\FFID}{F_\mathrm{FID}}
\providecommand{\Fenv}{}
\renewcommand{\Fenv}{F_\mathrm{env}}

\providecommand{\Ga}{}
\renewcommand{\Ga}{^{69}\mathrm{Ga}}
\providecommand{\Gb}{}
\renewcommand{\Gb}{^{71}\mathrm{Ga}}
\providecommand{\As}{}
\renewcommand{\As}{^{75}\mathrm{As}}
\providecommand{\fGa}{}
\renewcommand{\fGa}{f_{^{69}{\rm Ga}}}
\providecommand{\fGb}{}
\renewcommand{\fGb}{f_{^{71}{\rm Ga}}}
\providecommand{\fAs}{}
\renewcommand{\fAs}{f_{^{75}{\rm As}}}

\providecommand{\TCPMG}{}
\renewcommand{\TCPMG}{T_2 ^\mathrm{CPMG}}
\renewcommand{\vec}[1]{{\bf #1}}

\chapterimage{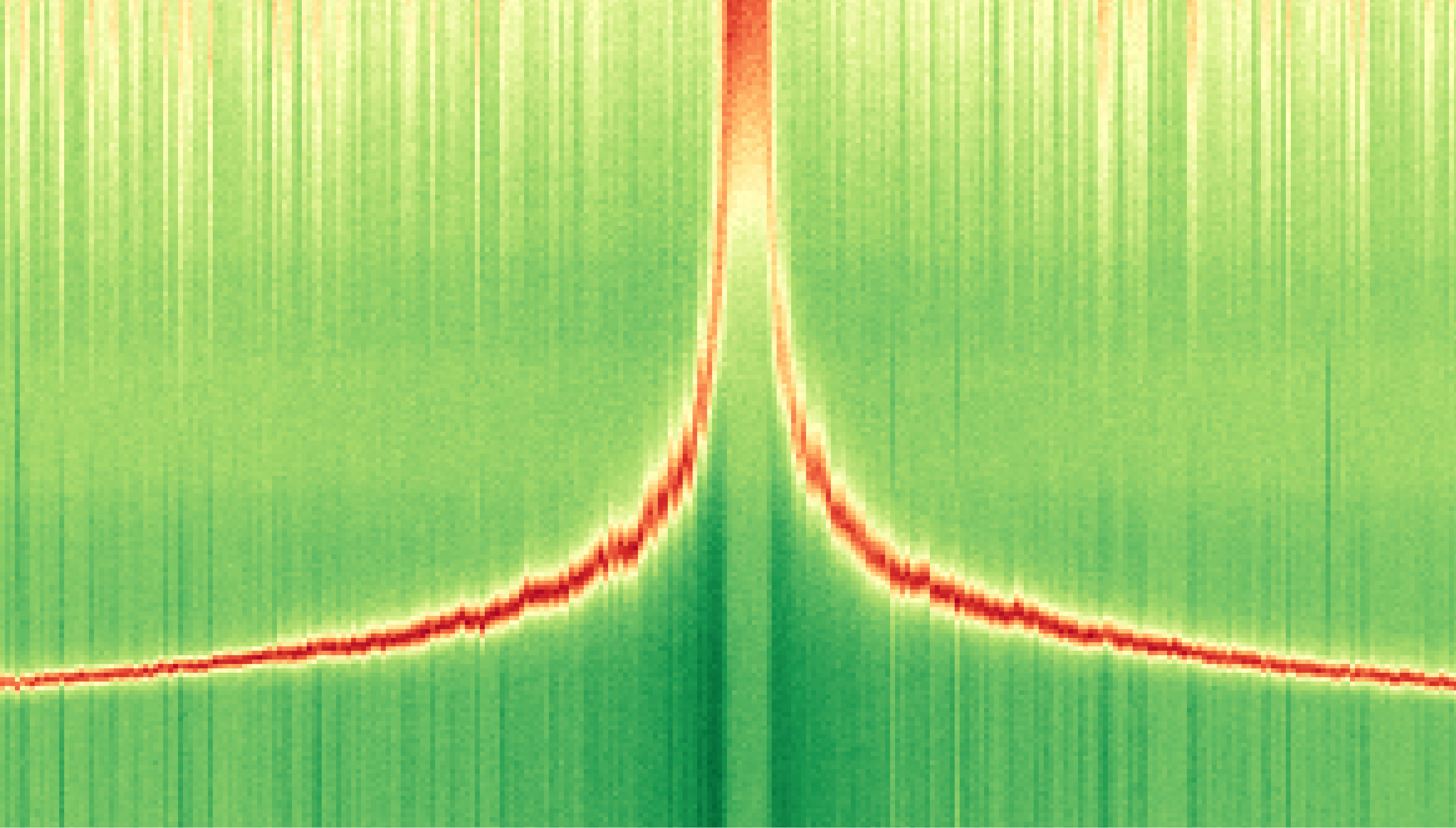}
\chapter[Overhauser noise and dynamical decoupling]{\protect\parbox{0.9\textwidth}{Overhauser noise \\ and dynamical decoupling}}
\label{ch:nuclei}

The noise limiting utility of GaAs spin qubits arises due to hyperfine interaction between each electronic spin and the nuclear spin bath. In striking contrast to the electrical noise the nuclear noise has been researched in depth in just about every system that is a potential candidate for a spin qubit implementation~\cite{Chekhovich2013,Reilly2008,DeLange2010a,Seo2016,Stockill2016}. These studies covered a wide range of topics including the free dynamics of the nuclear spin bath, dynamical decoupling of the qubit and controlling the bath by means of the dynamical nuclear polarization. In this chapter I will focus on the elementary properties of the GaAs spin bath dynamics interacting with a single electron and dynamical decoupling techniques that are most relevant for the novel results presented in chapters~\ref{ch:ovh} and~\ref{ch:notch}.

\section{Hamiltonian governing the dynamics of an electron and nuclei}

The dynamics of the coupled electron-nuclear bath system is ruled by three main mechanisms~\cite{Cywinski2009a} -- the Zeeman effect, the hyperfine interaction and  dipole-dipole interaction between nuclei.

Typically the largest energies in the system are related to the Zeeman splitting due to external magnetic field which can be described by the Hamiltonian
\begin{equation}
	\hat{H}_Z = g\mu_B B \hat{S}_z + \sum_{i} g_{\alpha(i)} \mu_B B \hat{I}_i
\end{equation}
where $g$ is the g-factor of the electron (or the nuclei of species $\alpha$), $B$ is the magnitude of the external magnetic field (assumed to be pointing along the $z$ direction) while $\hat{S}$ and $\hat{I}$ are the electron and nuclear spin operators. The summation is performed over the nuclei of several species $\alpha$ ($^{69}$Ga, $^{71}$Ga and $^{75}$As in case of GaAs quantum dot). For the purpose of this chapter we will assume that both the electron and nuclear Zeeman splitting are much larger than other energy scales.

The hyperfine interaction between the electron and nuclei is given by
\begin{equation}
	\hat{H}_\mathrm{hf} = \sum_i A_i \hat{\vec{S}} \hat{\vec{I}}_i = \sum_i A_i \hat{S}_z \hat{I}_{i,z} + \sum_i \frac{A_i}{2} \left(\hat{S}_+ \hat{I}_{i,-} + \hat{S}_- \hat{I}_{i,+} \right),
	\label{nuclei:hyperfine}
\end{equation}
where $A_i$ is the hyperfine coupling between $i$-th nuclei and the electron, which depends on the nuclear species and the amplitude of the electronic wavefunction at the nuclear site. Due to anisotropy introduced by the external magnetic field it is convenient to rewrite this Hamiltonian as two separate terms. The $z$ component gives an effective magnetic field experienced by the electron due to nuclei and vice versa. The $+/-$ terms describe a second-order flip-flop processes between nuclei, mediated by the electron.

The dipole-dipole interaction between the nuclei can be described by
\begin{equation}
	\hat{H}_\mathrm{dip} = \sum_{i \neq j} b_{i,j} \hat{\vec{I}}_i \hat{\vec{I}}_j = \sum_{i \neq j} b_{i,j} \left( \hat{I}_{i,z} \hat{I}_{j,z} + \hat{I}_{i,+} \hat{I}_{j,-} \right),
	\label{nuclei:dipole}
\end{equation}
which I have already written down in a way that distinguishes between components parallel and perpendicular to an external magnetic field.

Finally, the quadrupolar coupling between electric field gradients and nuclear spin also has an influence on the system dynamics~\cite{Chekhovich2015,Botzem2016}, which I will mention in the relevant places of the discussion.

\section{Electron dephasing in a quasistatic nuclear field}

Due to the assumption about the dominant influence of the magnetic field we can, in the first approximation, neglect the flip-flop processes between nuclei\footnote{for magnetic field >100~mT and timescales up to a microsecond this is a very accurate approximation}. On the other hand we observe that even at the typical dilution refrigerator base temperature of 20~mK ($kT\approx 2$~$\mu$eV) the nuclear bath is in the infinite temperature limit ($g_\mathrm{nuc} \mu_B B$ is of the order of 0.01~$\mu$eV at $B=1$~T).

Large magnetic field and high temperature motivate treatment of the nuclear bath as a collection of classical spins with random projection on the direction of the magnetic field. Since all flip-flop processes are over each of the nuclear spins induces a small energy shift to the electron Zeeman splitting $A_i I_{i,z}$. Since the electron wavefunction overlaps with tens to millions of the nuclei it is convenient to threat their collective influence in terms of an effective magnetic field, called the Overhauser field
\begin{equation}
	B_\mathrm{nuc} = \frac{1}{g \mu_B} \frac{1}{2} \sum_i A_i I_{i,z}
\end{equation}
which points in the direction parallel to the external magnetic field. Since the flip-flop terms play a role on a longer timescale it is common to consider the orientation of each spin to be random each time the electron-spin manipulation experiment is performed. This leads to the probability distribution of Overhauser field amplitudes which are characterized by several parameters, and for most purposes can be considered Gaussian and centered on $B_\mathrm{nuc}=0$.

Firstly, the Overhauser field can be characterized by the value of the external magnetic field corresponding to the full polarization of the nuclear spins. Secondly, the crucial parameter is the number of nuclei of each spinful species within the electronic wavefunction. These two give a typical (rms) value of the Overhauser field experienced by the electron and an inhomogeneous dephasing time.

For a GaAs quantum dot the electronic wavefunction overlaps with about a million~\cite{Petta2005} nuclei of $^{69}$Ga, $^{71}$Ga and $^{71}$As. In the case of full polarization these nuclei would create an affective magnetic field of between 3 and 6~T~\cite{Taylor2007,Assali2011}. The actual rms value of the Overhauser field is reduced by a factor $\sqrt{N}$ compared with the full polarization, where $N$ is the number of nuclei, resulting in typical magnetic fields of several millitesla~\cite{Malinowski2017a,Delbecq2016}. Finally, this translates into a decoherence time of about~\cite{Petta2005,Taylor2007}
\begin{equation}
	T_2^* = \frac{\hbar}{g \mu \langle B_\mathrm{nuc} \rangle} \approx 10~\mathrm{ns}.
\end{equation}

Such a short coherence time is the ultimate obstacle to scaling of the GaAs spin qubits, since it puts a severe constraint on the maximum gate fidelity. Several methods, including stabilization of the spin bath by means of dynamical nuclear polarization~\cite{Petta2008,Foletti2009,Shulman2014,Cerfontaine2016,Nichol2017} and dynamical decoupling (Sec.~\ref{nuclei:DD}) allow relaxing the limitation arising from the short inhomogeneous dephasing time, but they both incur significant overhead.

The ultimate solution to this problem is to reduce the fraction of the spinful nuclei within the electronic wavefunction. Unfortunately this is impossible for spin qubits in III-V material, such as GaAs, since the odd number of protons in the atomic nucleus means a non-zero total spin. For that reason the focus many in the spin qubit community shifts to semiconductors of group IV, in particular Si and SiGe. In the case of both, silicon and germanium the spinful isotopes are in a minority -- the natural abundance of $^{29}$Si is 4.7\%, while $^{73}$Ge is 7.7\%~\cite{Audi1997}. It has been demonstrated that usage of natural silicon is sufficient to extend inhomogenuous dephasing times to more than perform a single spin manipulation at the quantum error correction threshold~\cite{Fowler2012,Takeda2016,Kawakami2016}. Further isotopical purification virtually eliminates the spinful nuclei from the lattice (<1000 ppm) and the related decoherence effects~\cite{Veldhorst2014}.

\section{Overhauser field fluctuations}

The degree to which a certain kind of noise is detrimental for the qubit coherence depends not only on the variance of fluctuations but also their rate. In case of the Overhauser field the fluctuation rate is related to the internuclei interactions leading to flip-flops.

The two relevant processes that have to be considered are the dipole-dipole interaction (Eq.~\eqref{nuclei:dipole}), and flip-flops mediated by virtual flipping of the electron (a second-order process resulting from the hyperfine interaction; Eq.~\eqref{nuclei:hyperfine})~\cite{Taylor2007,Cywinski2009a}. The exact rates for both of these depend on the mismatch between the Zeeman splitting of the interacting nuclei.

In the first place, different nuclear species are characterized by different g-factors, and therefore flip-flops between pairs of identical nuclei are most relevant.

Second, the nuclei experience the Knight field (effective magnetic field due to interaction with the electron, dual to the Overhauser field), that varies depending on the wavefunction amplitude at the nuclear site~\cite{Witzel2006}. Curiously, large differences of Knight field lead to suppression of the nuclear flip-flops, but each such event has more significant impact on the Overhauser field. Conversly, if the two electrons experience identical Knight fields the flip-flops are not suppressed, but they have no impact on the Overhauser field. This interplay makes the degree of flip-flop suppression strongly dependent on the (unknown) electronic wavefunction shape.

Third, the magnetic field affects the detuning between electronic and nuclear Zeeman splitting, influencing the rate of the electron-mediated flip-flops~\cite{Gong2011,Malinowski2017a}.

Finally, the strain-related gradients of electric field couple to the quadrupole moment of the nuclei~\cite{Sundfors1969}. This is most evident in the study of the Overhauser field performed in optically active self-assembled GaAs quantum dots where the diffusion is supressed by several orders of magnitude relative to gate defined GaAs dots~\cite{Chekhovich2015,Stockill2016,Reilly2008,Malinowski2017a}.

The resulting rate of Overhauser field fluctuations is sufficiently small to allow performing thousands of repetitions of the electron spin manipulation experiment within the nuclear spin bath coherence time (which is of the order of seconds)~\cite{Barthel2009,Delbecq2016,Malinowski2017a}. This enables either measuring the Overhauser fields, and adjusting the spin manipulating sequence accordingly~\cite{Shulman2014} or exploiting spin transfer between electron and nuclei to manipulate the spin bath~\cite{Petta2008,Foletti2009,Bluhm2010,Nichol2015}. Quantification of the Overhauser field fluctuation timescales is the topic of chapter~\ref{ch:ovh}.

\section{Larmor precession of the nuclei}
\label{nuclei:larmor}

Except for diffusion-like low frequency dynamics, the Overhauser field has nontrivial high frequency behaviour, related to the Larmor precession of the nuclei. The amplitude of related fluctuations is very small and can be revealed only when the electron decoherence due to slow Overhauser field dynamics is suppressed by means of dynamical decoupling, which will be the topic of the next section. Nevertheless the particular power spectrum makes these fluctuations an interesting topic for experimental and theoretical study.

\begin{figure}
	\centering
	\includegraphics[width=0.5\textwidth]{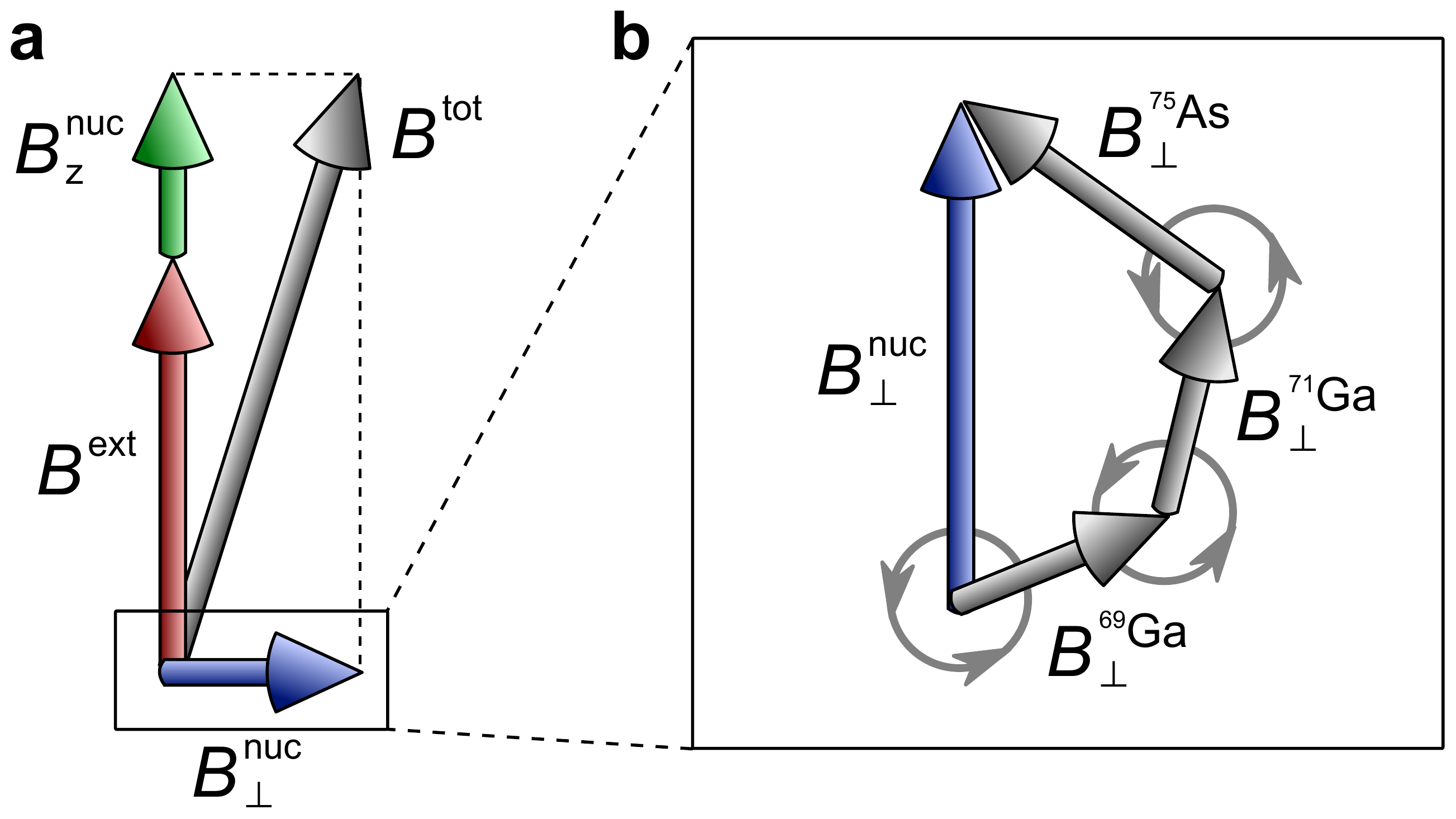}
	\caption[External and Overhauser contributions to the total magnetic field experienced by the electron]{
	{\bf a} The total effective magnetic field experienced by the electron is composed of an external magnetic field as well as two components of the Overhauser field -- parallel and perpendicular to the external magnetic field.
	{\bf b} The transverse component of the Overhauser field is composed of three ``partial'' fields related to three nuclear species, undergoing Larmor precession at different frequencies, as represented by the circular arrows.
	}
	\label{nuclei:sum_B}
\end{figure}

To understand the Overhauser field dynamics related to the Larmor precession one needs to include in the considerations the Overhauser field perpendicular to the external magnetic field. Within this extended framework the effective magnetic field experienced by the electron has three contributions: external magnetic field, Overhauser field both parallel and perpendicular to the external magnetic field (Fig.~\ref{nuclei:sum_B}). Since we assume that the external magnetic field is much larger than the other fields we can write the amplitude of the total effective magnetic field (to the lowest nonvanishing order in $\mathbf{B}_{\perp}^{\rm nuc}$) as
\begin{equation}
	B^{\rm tot} =  B^{\rm ext} + B_{z}^{\rm nuc}(t) + \frac{|\mathbf{B}_{\perp}^{\rm nuc}(t)|^2}{2|\mathbf{B}^{\rm ext}|}.
\end{equation}
This formula shows that a relatively small influence of the transverse field on the electron coherence is a consequence of the suppression by a factor of $|\mathbf{B}_{\perp}^{\rm nuc}(t)|/|\mathbf{B}^{\rm ext}|$, which is typically somewhere between 0.2 and 0.01, depending on the external magnetic field, dot size etc. Nevertheless it is the transverse Overhauser field that gives raise to the high frequency nuclear noise.

The curious high frequency dynamics of the Overhauser field can be understood by treating the transverse field as consisting of several ``partial'' fields related to each nuclear species~\cite{Bluhm2011,Botzem2016,Malinowski2017} (Fig.~\ref{nuclei:sum_B}b). Each of this species Larmor precesses at a different frequency (megahertz-scale for typical magnetic field of a few hundred militesla), giving rise to the noise focused at very specific narrow bands. The quadratic coupling to the qubit splitting implies that these are mostly the Larmor difference frequencies. In chapter~\ref{ch:notch} we demonstrate how to deterministically suppress a noise at these frequencies with dynamical decoupling. But first it is relevant to introduce the principles of dynamical decoupling.

\section{Dynamical decoupling from a low frequency noise}
\label{nuclei:DD}

\begin{figure}
	\centering
	\includegraphics[width=0.7\textwidth]{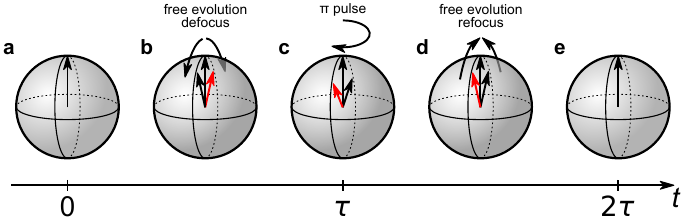}
	\caption[Schematic of state evolution under Hahn-echo sequence in the Bloch sphere representation]{
	Schematic of state evolution under Hahn-echo sequence in the Bloch sphere representation. The ensemble of states is initialized {\bf a}, and then evolves freely {\bf b}. After time $\tau$ a $\pi$-pulse {\bf c} swaps the  two eigenstates. Continued free evolution {\bf d} leads to the recovery of coherence after total time of $2\tau$ {\bf e}.
	}
	\label{nuclei:hahn}
\end{figure}

The principle of dynamical decoupling can be most easily understood in the case of quasistatic noise. When considering a superposition of $\ket{\uparrow}$ and $\ket{\downarrow}$, the two eigenstates will be acquiring a predictable phase due to magnetic-field-induced Zeeman splitting and a random phase due to the Overhauser field $\phi(t) = t \times g\mu B^\mathrm{nuc} /\hbar$. Let's consider the Overhauser field fixed on the timescale we are interested in. Swapping $\ket{\uparrow}$ and $\ket{\downarrow}$ states with a $\pi$-pulse after certain time $\tau$ (Fig.~\ref{nuclei:hahn}) results in the random phase being perfectly cancelled out after another time $\tau$. And so the coherence will be restored after the total time $2\tau$. This technique, called spin echo or Hahn echo~\cite{Hahn1950} was first demonstrated in nuclear magnetic resonance. The decay time of the echo signal is a simple measure of the non-quasistatic noise and is one of the measures of the qubit performance~\cite{Vandersypen2005}.

\begin{figure}
	\centering
	\includegraphics[width=0.75\textwidth]{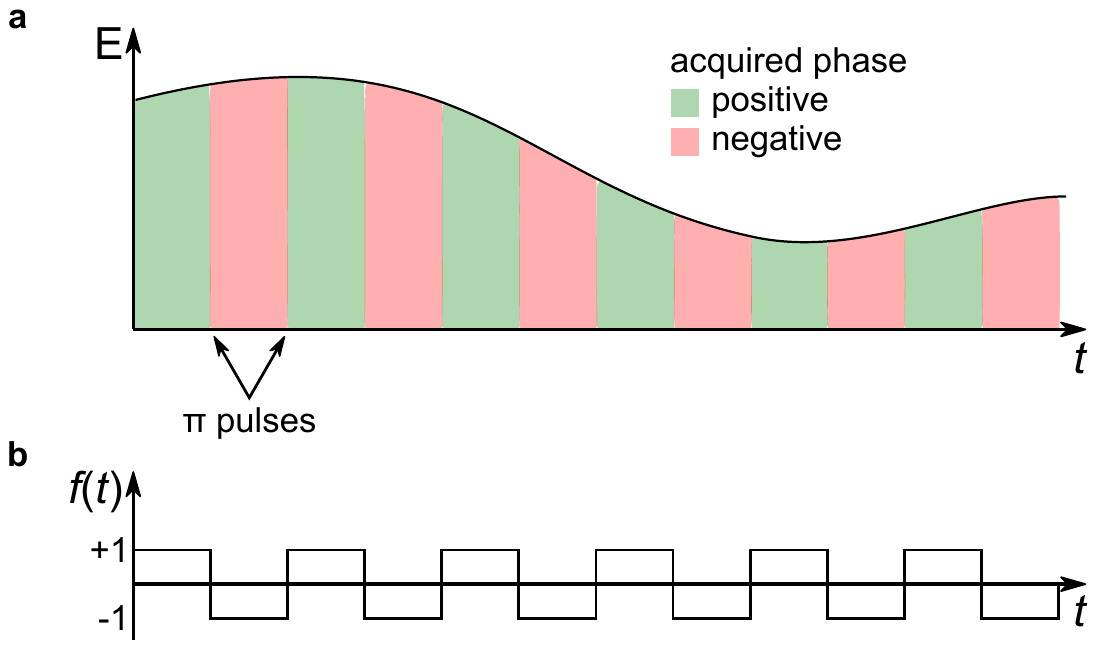}
	\caption[An illustration of noise suppression with a train of $\pi$-pulses]{
	{\bf a} An illustration of noise suppression with a train of $\pi$-pulses. The black curve represents a stochastic time dependence of the qubit energy splitting. The colors indicate whether the phase acquired by the two qubit eigenstates was positive or negative. Application of sufficiently dense, evenly spaced train of $\pi$-pulses results in the red and green regions having similar area, indicating more efficient decoupling from random fluctuations in the qubit splitting $\Delta E$.
	{\bf b} Plot of the $f(t)$ function corresponding to the periodic train of $\pi$ pulses, which leads to the filter function according to Eq.~\eqref{nuclei:ff}
	}
	\label{nuclei:slow}
\end{figure}

An echoing technique can be extended to minimize the effect of the slow fluctuations in the qubit splitting by implementing a train of evenly spaced $\pi$-pulses, called Carr-Purcell (CP) sequence~\cite{Carr1954}. As illustrated in Fig.~\ref{nuclei:slow}a the phase acquired between $n$th and $(n+1)$th $\pi$-pulse will almost perfectly cancel the phase acquired between $(n-1)$th and $n$th $\pi$-pulse. Effectively, the CP sequence will suppress any components of the noise fluctuation that do not change on the $2\tau$ timescale.

\section{Dynamical decoupling as a filter design}

The difference between Hahn echo and CP sequences in the efficiency of decoupling from low frequency noise can be demonstrated in the frequency domain. For that purpose one represents the environmental noise as a power spectrum\footnote{Throughout this chapter we assume the noise to be Gaussian. This assumption implies that the noise is fully quantified by its power spectrum.}, $S(\omega)$, and assigns a filter function, $F(\omega)$, to the sequence of $\pi$-pulses~\cite{Martinis2003,Cywinski2008}. The filter function for a sequence of $\pi$-pulses is given by
\begin{align}
	F(\omega) & = \frac{\omega^2}{2} \left| \tilde{f}(\omega) \right|^2, \\
	f(t) & = \sum\limits_{k=0}^n (-1)^k \Theta(t_{k+1}-t) \Theta(t - t_k),
	\label{nuclei:ff}
\end{align}
where $\omega$ is the angular frequency, $\tilde{\cdot}$ indicates a Fourier transform, $f(t)$ represents the decoupling sequence, $\Theta$ is the Heaviside step function and $t_k$ is the time at which the $k$th $\pi$-pulse is applied (in particular $t_0=0$ and $t_n=T$ are the beginning and the end of the sequence). The function $f(t)$ has a value of $\pm1$ and changes sign at the moment of $\pi$-pulse application, and is 0 beyond the duration in the sequence (Fig.~\ref{nuclei:slow}b). In most of the studies of dynamical decoupling the sequences are defined by specifying the fractions of the total sequence duration $T$ at which $\pi$-pulses are applied. For example for Hahn echo $t_0=0$, $t_1=T/2$, $t_2=T$.

With these one can quantify coherence at the end of the pulse sequence by calculating
\begin{align}
	W(T) & = \exp(-\chi(T)), \\
	\chi(T) & = \int\limits_0^\infty \frac{\mathrm{d}\omega}{\pi} S(\omega) \frac{F(\omega)}{\omega^2}.
\end{align}

A formulation of the dynamical decoupling has led to tremendous amount of work oriented at suppression of the low frequency noise~\cite{Khodjasteh2007,Uhrig2007,Biercuk2009}. These works usually aim at setting the $F(0)=0$ along with one or several derivatives, or at minimization of $\chi(T)$ given certain noise spectrum $S(\omega)$.

The suppression of the now frequency noise achieved with dynamical decoupling sequences is sufficient for the high frequency noise related to the Larmor precession of the nuclei (Sec.~\ref{nuclei:larmor}) to leave a fingerprint on the decoherence pattern~\cite{Malinowski2017,Bluhm2011,Botzem2016,Seo2016}. To gain additional insight into high-frequency noise one can focus on the filter function shape at frequencies comparable to the typical spacing between the $\pi$-pulses in a sequence. It is at these frequencies that the filter functions exhibit a pattern of minima and maxima. The minima can be used either to address a narrow-band high-frequency noise~\cite{Soare2014,Malinowski2017} while the maxima enhance a specific band enabling the noise spectroscopy~\cite{Dial2013,Bylander2011,Alvarez2011,Yuge2011,Kotler2011,Szankowski2016,Ramon2017}.

\chapterimage{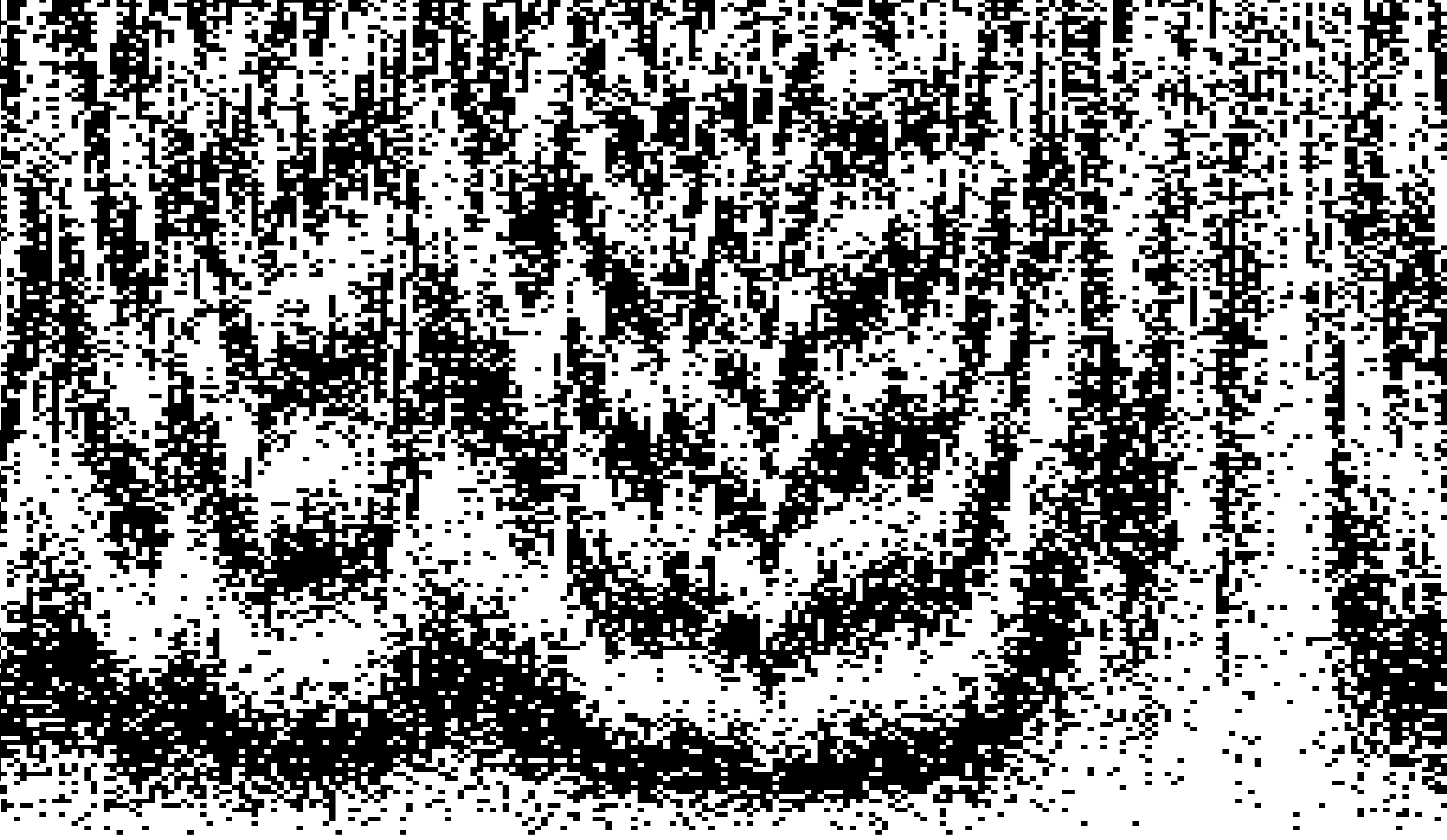}
\chapter[Spectrum of the nuclear environment for GaAs spin qubits]{\protect\parbox{0.9\textwidth}{ Spectrum of the nuclear \\ environment for GaAs spin qubits}}
\label{ch:ovh}

{\let\thefootnote \relax\footnote{This chapter and chapter \ref{ch:ovh_sup} are adapted from Ref. \cite{Malinowski2017a}. \copyright~(2017) by the American Physical Society.}}
\addtocounter{footnote}{-1}

\newcommand{\Tn}{T_{2,n}}

\begin{center}
 Filip K. Malinowski$^{1}$, Frederico Martins$^{1}$, Peter D. Nissen$^{1}$, \L{}ukasz~Cywi\'nski$^{2}$, Mark~S.~Rudner$^{1,3}$, \\
 Saeed Fallahi$^{4}$, Geoffrey C. Gardner$^{5}$, Michael~J.~Manfra$^{4,5,6}$, \\
 Charles M. Marcus$^{1}$, Ferdinand~Kuemmeth$^{1}$
\end{center}

\begin{center}
	\scriptsize
	$^{1}$ Center for Quantum Devices, Niels Bohr Institute, University of Copenhagen, 2100 Copenhagen, Denmark\\
	$^{2}$ Institute of Physics, Polish Academy of Sciences, al.~Lotnik{\'o}w 32/46, PL 02-668 Warsaw, Poland \\
	$^{3}$ Niels Bohr International Academy, Niels Bohr Institute, 2100 Copenhagen, Denmark\\
	$^{4}$ Department of Physics and Astronomy and Birck Nanotechnology Center, Purdue University, West Lafayette, Indiana 47907, USA\\
	$^{5}$ School of Materials Engineering and Birck Nanotechnology Center, Purdue University, West Lafayette, Indiana~47907, USA\\
	$^{6}$ School of Electrical and Computer Engineering, Purdue University, West Lafayette, Indiana 47907, USA
\end{center}

\begin{center}
\begin{tcolorbox}[width=0.8\textwidth, breakable, size=minimal, colback=white]
	\small
	Using a singlet-triplet spin qubit as a sensitive spectrometer of the GaAs nuclear spin bath, we demonstrate that the  spectrum of Overhauser noise agrees with a classical spin diffusion model over six orders of magnitude in frequency, from 1~mHz to 1~kHz, is flat below 10~mHz, and falls as $1/f^2$ for frequency $f \! \gtrsim \! 1$~Hz. Increasing the applied magnetic field from 0.1~T to 0.75~T suppresses electron-mediated spin diffusion, which decreases spectral content in the $1/f^2$ region and lowers the saturation frequency, each by an order of magnitude, consistent with a numerical model. Spectral content at megahertz frequencies is accessed using dynamical decoupling, which shows a crossover from the few-pulse regime ($\lesssim \! 16~\pi$-pulses), where transverse Overhauser fluctuations dominate dephasing, to the many-pulse regime ($\gtrsim \! 32$~$\pi$-pulses), where  longitudinal Overhauser fluctuations with a $1/f$ spectrum dominate.
\end{tcolorbox}
\end{center}

\section{Introduction}

Precise control of single electron spins in gate-defined quantum dots makes them a promising platform for quantum computation~\cite{Loss1998,Veldhorst2015,Petta2005,Nowack2011,Shulman2012}.
In particular, GaAs spin qubits benefit from unmatched reliability in fabrication and tuning.
However, being a III-V semiconductor, the GaAs lattice hosts spinful nuclei that couple to electron spins via the hyperfine interaction~\cite{Petta2005,Shulman2012,Malinowski2017,Bluhm2011,Foletti2009}.
Nuclear dynamics lead to fluctuations of the Overhauser field, which affect the coherent evolution of spin qubits. 
In turn, advances in qubit operation, including single-shot readout \cite{Barthel2009} and long dynamical decoupling sequences \cite{Malinowski2017}, allow spin qubits to serve as sensitive probes of the electron-plus-nuclear-environment system, an interesting coupled nonlinear many-body system.

\begin{figure}[tb]
	\centering
	\includegraphics[scale=.9]{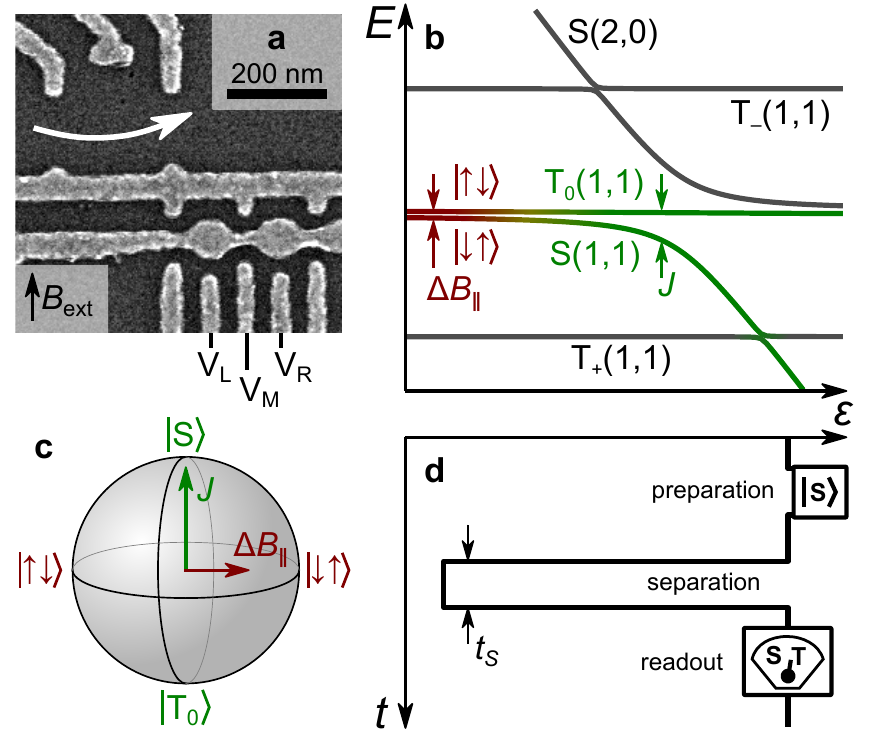}
	\caption[Device, qubit and pulse]
	{(a) Electron micrograph of the device. Gate voltages  $V_i$ control the double dot state on ns timescales. Reflectance from the RF resonant circuit incorporating a sensor dot (white arrow) measures the charge state of the double dot located below the round accumulation gates.
	(b) Energy levels of the two-electron double dot as a function of detuning $\varepsilon \! = \! V_L-V_R$ at the (1,1)-(2,0) charge transition. Red-green lines indicate the qubit states.
	(c) Bloch sphere representation of the qubit. Rotation axes correspond to exchange interaction $J$ (green) and gradient of the Overhauser field $\Delta B_\parallel$ (red).
	(d) Pulse cycle used to probe the qubit precession in the gradient of the Overhauser field. The qubit is initialized in the S(2,0) state by exchanging electrons with the lead. Next, one electron is moved to the right dot, and the qubit evolves for the time $t_S$ in the gradient of the Overhauser field. Finally, $\varepsilon$ is pulsed back to the readout point, projecting $\ket{S}$ into a (2,0) charge state, whereas $\ket{T_0}$ remains in (1,1). 
	}
	\label{ovh:fig1}
\end{figure}

\section{Measuring Overhauser field gradient with $S$-$T_0$ qubit}

In this Letter, we use a singlet-triplet (S-T$_0$) qubit as a probe to reveal the dynamics and magnetic field dependence of the GaAs nuclear spin bath over a wide range of frequencies, without the use of nuclear pumping~\cite{Shulman2014,Bechtold2015,Nichol2015}  or postselection~\cite{Delbecq2016} techniques. 
The qubit is defined in a two-electron double quantum dot (Fig.~\ref{ovh:fig1}a). The external magnetic field $\Bext$ separates the qubit states singlet, $\ket{\mathrm{S}} \! = \! \tfrac{1}{\sqrt{2}}(\ket{\ud} \! - \! \ket{\du})$, and the unpolarized triplet, $\ket{\mathrm{T_0}} \! = \! \tfrac{1}{\sqrt{2}}(\ket{\ud} \! + \! \ket{\du})$, from the fully polarized triplet states, $\ket{T_+} \! = \! \ket{\uparrow\uparrow}$ and $\ket{T_-} \! = \! \ket{\downarrow\downarrow}$. In this notation, the first (second) arrow indicates the spin in the left (right) dot. 
The resulting energy diagram of the spin states at the transition between (1,1) and (2,0) charge states is presented in Fig.~\ref{ovh:fig1}b.
Here ($N$,$M$) indicates the number of electrons in the left ($N$) and the right ($M$) dot. 
The Bloch sphere representation of the qubit is shown in Fig.~\ref{ovh:fig1}c.

\begin{figure}[tb]
	\centering
	\includegraphics[scale=0.9]{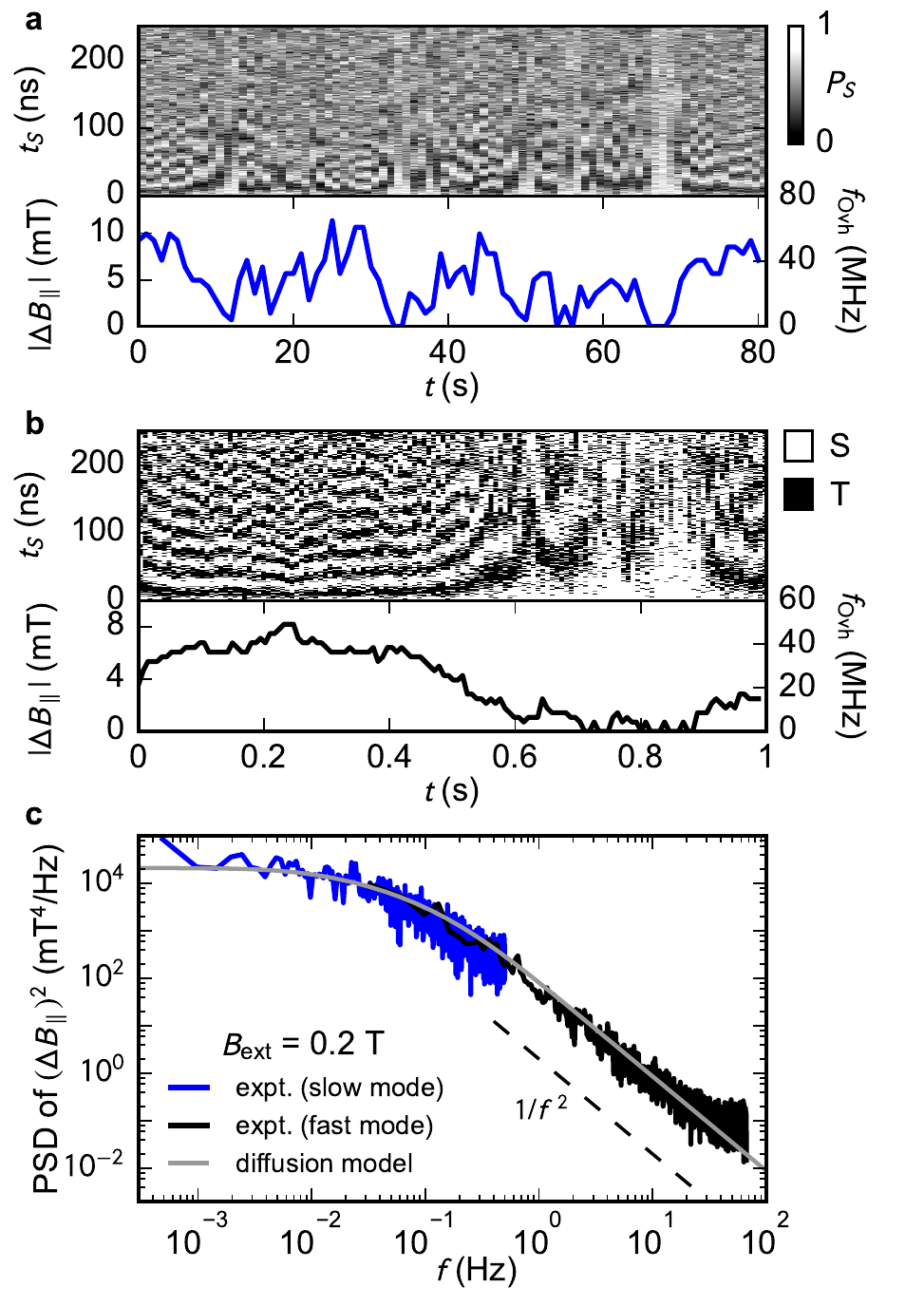}
	\caption[Overhauser field low frequency dynamics]
	{(a,b) Top panels present S-T$_0$ oscillations resulting from the relative precession of the two electron spins in the Overhauser field gradient, as a function of laboratory time at $\Bext \! = \! 0.2$~T (see main text). In the bottom panels we show the extracted frequency of oscillations, $f_\mathrm{Ovh}$, converted to $|\Delta B_\parallel|$.
	(c) Power spectral density of $(\Delta B_\parallel)^2$ at $\Bext \! = \! 0.2$~T obtained from traces such as in (a) (blue) and (b) (black). Transition from white spectrum at low frequencies to $1/f^2$ at high frequencies is reproduced by the nuclear spin diffusion model (gray). A deviation from this dependence at the highest frequencies is a numerical artifact caused by the discreteness of $|\Delta B_\parallel|$ values obtained from the Fourier analysis.
}
	\label{ovh:fig2}
\end{figure}

Dynamics of the S-T$_{0}$ qubit in the well-separated (1,1) charge state, i.e., for vanishing exchange, $J$, between the two electrons, is governed by the static external magnetic field $\Bext$ and dynamic Overhauser fields. For large $\Bext$, we can model the qubit evolution using the Hamiltonian~\cite{Malinowski2017,Bluhm2011,Neder2011}
\begin{equation}
	\label{H}
	\hat{H}(t) \! = \! g \mu_B \sum\limits_{i=L,R} \left( B_\parallel^i(t) \! + \! \frac{|\vec{B}_\perp^i(t)|^2}{2|\Bext|}\right) \hat{S}^i_z,
\end{equation}
where $g \sim -0.4$ is the electronic $g$-factor, $\mu_B$ is a Bohr magneton, $\hat{S}_z^i$ is the spin operator of the electron in left or right dot $i \! = \! L,R$, and $B_\parallel^i$ is the Overhauser field component parallel to $\Bext$. The influence of the transverse Overhauser field component $\vec{B}_\perp^i$ on the qubit is strongly suppressed when $\Bext$ is much larger than the typical Overhauser field. 
Hence the transverse Overhauser field fluctuations play a significant role in the qubit evolution only when the influence of the fluctuating longitudinal Overhauser field $B_\parallel^i$ is eliminated by dynamical decoupling~\cite{Bluhm2011, Malinowski2017}. The splitting between qubit states $\ket{\du}$ and $\ket{\ud}$ for $J=0$ is thus proportional to the longitudinal component of the Overhauser field gradient, $\Delta B_\parallel \! = \! B_\parallel^L \! - \! B_\parallel^R$, and can be measured by monitoring the qubit precession between $\ket{S}$ and $\ket{T_0}$~\cite{Foletti2009,Barthel2009,Barthel2012}. 

To measure this precession, we apply a cyclic pulse sequence that first prepares the singlet, then separates the two electrons to allow free precession in the Overhauser field for time $t_S$, and finally performs a projective readout of the qubit in the S-T$_{0}$ basis (Fig.~\ref{ovh:fig1}d). 
The total length of the pulse sequence is approximately 30 $\mu$s, including 10 $\mu$s of readout time. 
For each $t_S$ we use 16 single-shot readouts of this sequence to estimate the singlet return probability, $P_S$.
By repeatedly sweeping $t_S$ from 0 to 250~ns in 300 steps allows the precession of the qubit in the evolving Overhauser field to be measured with roughly 1~s temporal resolution (slow mode). 
A time trace showing 80~s  of slow-mode probability data is shown in Fig.~\ref{ovh:fig2}a. 
To increase the temporal resolution from 1 s to 12 ms we omit the probability estimation and record one single-shot outcome for each $t_S$ (fast mode). A time trace showing 1 s of fast-mode single-shot data is shown in Fig.~\ref{ovh:fig2}b. The time evolution of the qubit precession frequency, $f_\mathrm{Ovh}(t)$, is then extracted from these data as described in Sec.~\ref{ovh_sup:extracting}. The frequency corresponds to the absolute value of the Overhauser field gradient $|\Delta B_\parallel(t)| \! = \! h f_\mathrm{Ovh}(t)/|g| \mu_B$.
Examples of $|\Delta B_\parallel(t)|$ for $\Bext \! = \! 0.2$~T are shown in Figs.~\ref{ovh:fig2}a,b. In contrast to experiments performing dynamic nuclear polarization~\cite{Danon2009,Bluhm2010,Forster2015} the observed distributions of $\Delta B_\parallel$ reveal no sign of multistable behaviour (see Sec.~\ref{ovh_sup:distribution}).

\section{$\Delta B_z$ low frequency dynamics}

Next, we focus on the power spectral density (PSD) of $\Delta B_\parallel$ for $\Bext \! = \! 0.2$~T. Since taking the absolute value of $\Delta B_\parallel$ introduces kinks in $|\Delta B_\parallel|$ traces, adding spurious high-frequency content, we instead extract the PSD of $(\Delta B_\parallel)^2$ (Fig.~\ref{ovh:fig2}c). The resulting spectrum is flat below $10^{-2}$~Hz and falls off as $1/f^2$ above 1~Hz, indicating a correlation time of $\Delta B_\parallel$ of a few seconds.

A classical model of Overhauser field fluctuations due to nuclear spin diffusion is used to fit the experimental data in Fig.~\ref{ovh:fig2}c~\cite{Reilly2008} (Sec.~\ref{ovh_sup:classical}).
In the model we use the double dot geometry estimated from the lithographic dimensions of the device and the heterostructure growth parameters (distance between the dots $d \! = \! 150$~nm, dot diameter $\sigma_\perp \! = \! 40$~nm and width of the electron wave function in the crystal growth direction $\sigma_z \! = \! 7.5$~nm). We fit the effective diffusion constant $D \! = \! 33$~nm$^2$/s and the equilibrium width of the $\Delta B_\parallel$ distribution $\sigma_{\Delta B} \! = \! 6.0$~mT.
This model yields the power spectrum of $\Delta B_\parallel$, which has the same qualitative behavior as the spectrum of $(\Delta B_\parallel)^2$ -- it is flat at low frequencies ($<10^{-2}$~Hz) and falls off as $1/f^2$ at high frequencies ($> \! 1$~Hz).
Such a relation between the PSD of a Gaussian distributed variable and that of its square is expected whenever the PSD has a $1/f^\beta$ dependence over a wide frequency range~\cite{Cywinski2014}.


\begin{figure}[tb]
	\centering
	\includegraphics[scale=.9]{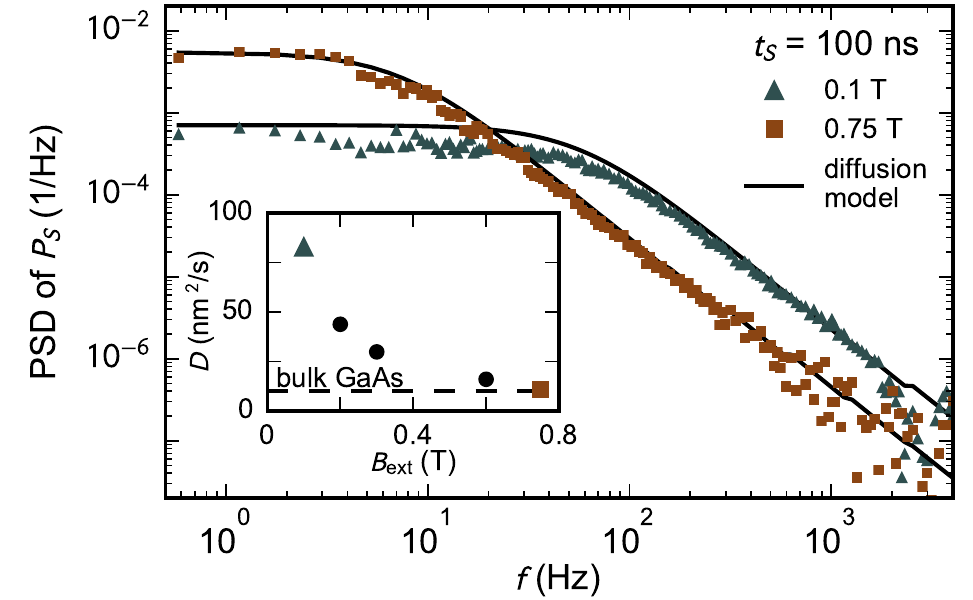}
	\caption[Dependence of $P_S$ power spectral density on external magnetic field]
	{Magnetic field dependence of the power spectral density of $P_S$, keeping $t_S=100$~ns fixed. Increasing $\Bext$ from 0.1 to 0.75~T suppresses the $1/f^2$ noise by an order of magnitude. 
	Solid lines are fits of the diffusion model with the effective diffusion constant $D$ being the only free parameter.
	Inset: $D$ as a function of magnetic field $\Bext$. Dashed line indicates the spin diffusion constant for bulk GaAs, $D \! = \! 10$~$\mathrm{nm^2/s}$~\cite{Paget1982}.
	}
	\label{ovh:fig3}
\end{figure}

In order to extend the spectral range to higher frequencies we apply the pulse cycle with a fixed separation time $t_S \! = \! 100$~ns, acquiring a single-shot measurement every 30~$\mu$s. This can be visualized as a horizontal cut through the data in Fig.~\ref{ovh:fig2}b (top) at 100 ns, though, of course, now without taking the rest of the data at other values of $t_{S}$. Although the series of single-shot outcomes at fixed $t_{S}$ does not allow a direct measure of $\Delta B_\parallel$ from temporal oscillations, it does give statistical spectral information \cite{Reilly2008}. In particular, the Fourier transform of the windowed autocorrelation of single-shot outcomes (Sec.~\ref{ovh_sup:autocorrelation}) yields a PSD of the singlet return probability $P_S$, now extended to 4~kHz.

Power spectra of $P_{S}$ for the lowest and highest applied fields studied, $\Bext \! = \! 0.1$ and 0.75~T are shown in Fig.~\ref{ovh:fig3}. We observe that the spectrum for $\Bext \! = \! 0.75$~T is reduced by an order of magnitude in the $1/f^2$ regime, compared to the spectrum at $\Bext \! = \! 0.1$~T. To quantify the observed magnetic field dependence of the PSD of $P_S$ we fit the nuclear spin diffusion constant $D$ of the classical diffusion model~(Sec.~\ref{ovh_sup:classical}) to data, using fixed $\sigma_{\Delta B} \! = \! 6.0$~mT (obtained from the fit in Fig.~\ref{ovh:fig2}) and the same geometrical parameters as above.
The observed agreement with experimental data suggests that the effects of the nuclear spin bath are well described by classical evolution up to at least 1~kHz.

At low $\Bext$ we observe a strong enhancement of the effective spin diffusion constant compared to the literature value for bulk GaAs in the absence of free electrons, $D \! \sim \! 10$~$\mathrm{nm^2/s}$~\cite{Paget1982} (Fig.~\ref{ovh:fig3}, inset).
Qualitatively, this increase may be attributed to electron-mediated nuclear flip-flop processes~\cite{Klauser2008,Latta2011,Gong2011, Reilly2008,Reilly2010}, which dominate over nuclear dipole-dipole mediated diffusion.
At 0.75~T the effective diffusion constant drops down to the value for bulk GaAs. Despite this agreement, we note that our values for $D$ are not corrected for possible changes of electronic wavefunctions with increasing magnetic field. A quantitative statement about the underlying bare diffusion constant is difficult, as the fitting results for D are sensitive to assumptions about the spatial extent of the quantum dots (in particular $\sigma_\perp$) and the fraction of time spent in (1,1) and (2,0). 
Since spin diffusion due to nuclear dipole-dipole interaction is strongly suppressed by the Knight field gradient~\cite{Deng2005} and quadrupolar splittings, we expect further suppression of $D$ at higher magnetic fields~\cite{Gong2011}, and saturation below the bulk GaAs value. Indeed, this is observed in self-assembled quantum dots, where quadrupolar splittings are significantly stronger due to strain~\cite{Nikolaenko2009,Latta2011,Chekhovich2015}.

\section{High frequency dynamics of the Overhauser field}

Overhauser field fluctuations above 100 kHz are too fast to be observed as oscillation between $\ket{S}$ and $\ket{T_0}$ with the present setup. However, we can infer spectral features from the decoherence of $\ket{\ud}$ and $\ket{\du}$ states using Hahn echo and Carr-Purcell-Meiboom-Gill (CPMG) dynamical decoupling sequences~\cite{Medford2012,Malinowski2017}. Since these decoupling sequences act as filters in frequency domain, we can relate the Overhauser spectrum to the decay of qubit coherence \cite{Malinowski2017,Martinis2003,Cywinski2008,Biercuk2011}. In particular, Hahn echo and CPMG sequences suppress the low frequency fluctuations, making the coherence decay a sensitive probe of high-frequency Overhauser fields. 

\begin{figure}[tb]
	\centering
	\includegraphics[scale=.9]{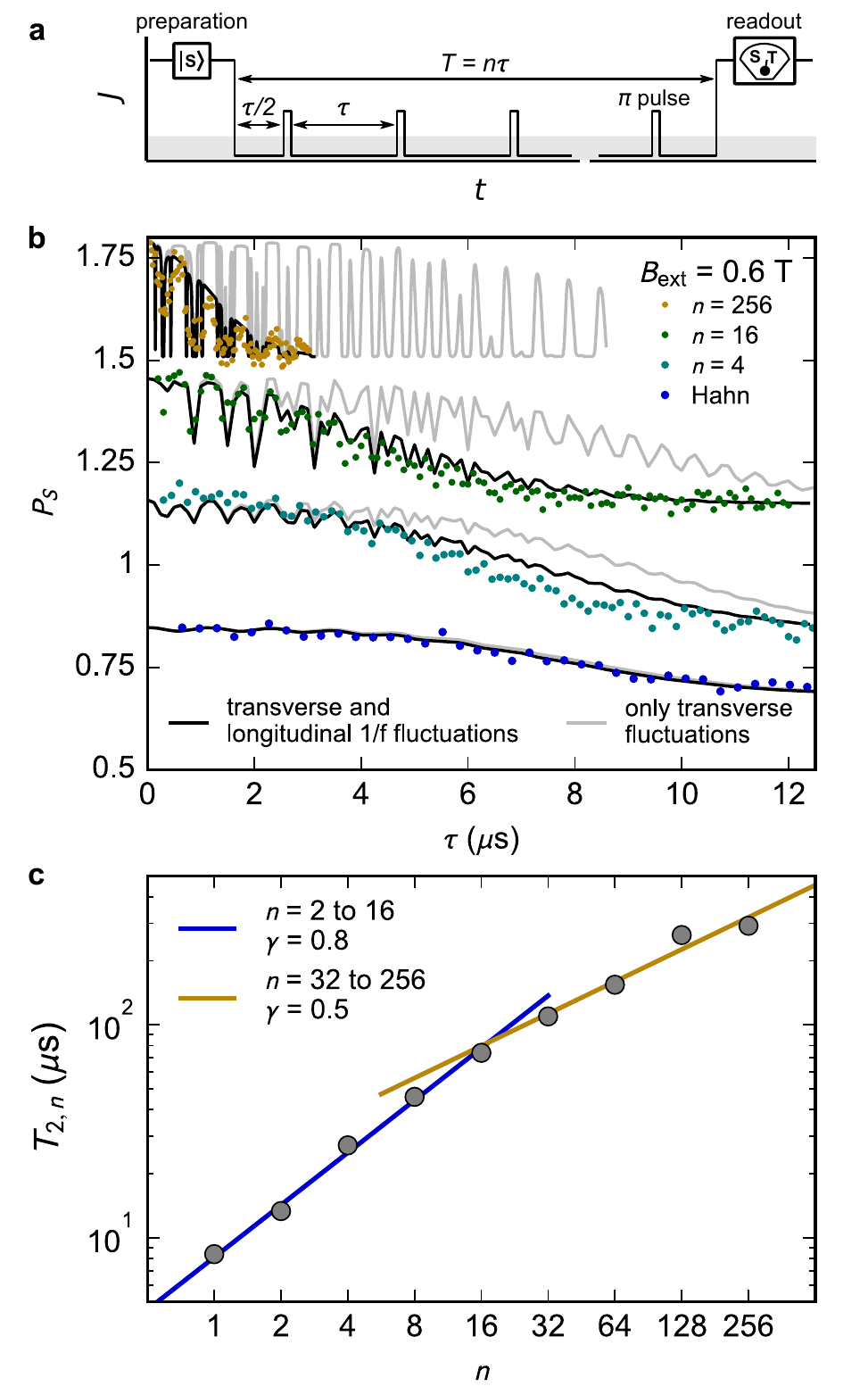}
	\caption[CPMG decay and $T_2^\mathrm{CPMG}$ scaling]
	{(a) Schematic of a CPMG dynamical decoupling sequence applied to a $S$-$T_0$ qubit, presented as a time dependent exchange energy $J$ (see text).
		(b) Coherence of the S-T$_0$ qubit after Hahn echo and CPMG sequences with number of $\pi$ pulses $n$. $\tau \! = \! T/n$ is the repetition period between pulses. Black curves present simulations including longitudinal $1/f$ noise and transverse fluctuations due to Larmor precession of the nuclei. Gray curves assume transverse Overhauser field fluctuations only. Data and curves are offset for clarity.
	(c) Scaling of the extracted coherence decay envelope $\Tn$ with $n$. Solid blue and yellow lines indicate fits of the power law $\propto \! n^\gamma$ to data in the indicated range.
	A large value of $\gamma \! = \! 0.8$ for small number of $\pi$ pulses indicates that decay is dominated by the transverse noise. $\gamma \! = \! 0.5$ for large $n$ is consistent with decay due to longitudinal $1/f$ noise.}
	\label{ovh:fig4}
\end{figure}

The decoupling sequence in Fig.~\ref{ovh:fig4}a uses symmetric exchange pulses \cite{Martins2016}, but is otherwise standard \cite{Medford2012}: initialize in S(2,0), evolve for time $\tau/2$ in (1,1), apply symmetric exchange $\pi$-pulse, evolve for another $\tau/2$, repeat the $\tau/2 \! - \! \pi \! - \! \tau/2$ segment a total of $n$ times. After the total evolution time $T \! = \! n\tau$, project onto S-T$_{0}$ by pulsing to (2,0) and perform single-shot readout.  Averaging $\sim$1000 such single-shot readouts then yields the singlet return probability. 
For such a sequence the resulting singlet return probability is related to the qubit coherence by $P_{S} = \frac{1}{2} + \frac{1}{2}\text{Re}[W_{L}(n\tau)W^{*}_{R}(n\tau) ]$, where $W_{i}(t)$ is the normalized coherence of the spin in dot $i$ at time $t$. 

Figure  \ref{ovh:fig4}b shows the singlet return probability for Hahn echo and CPMG sequences with various numbers of $\pi$ pulses, $n$, as a function of the interpulse time $\tau \! = \! T/n$.
For sequences with small $n$, coherence decreases smoothly with $\tau$, while for sequences with large $n$ the decay is strongly modulated. It was previously shown~\cite{Bluhm2011,Malinowski2017} that the coherence modulations are due to narrowband spectral content at megahertz frequencies in the transverse Overhauser field $\vec{B}^{i}_\perp$, arising from the relative Larmor precession of the three nuclear species.

The influence of  transverse Overhauser fluctuations, $\vec{B}^{i}_\perp$, on the CPMG signal decay was simulated using a semiclassical theory \cite{Cywinski2009,Cywinski2009a,Neder2011} that previously gave good agreement with echo \cite{Bluhm2010,Botzem2016} and CPMG \cite{Malinowski2017} experiments (see Sec.~\ref{ovh_sup:transverse} for details). 
Comparisons of experimental data with numerical simulations are shown in Fig.~\ref{ovh:fig4}b.
First, we include only narrowband transverse fields (gray curves), assuming two identical dots each containing $N = 9\times 10^5$ nuclei and a spread of effective fields experienced by the nuclei of $\delta B \! = \! 1$~mT, arising, for example, from quadrupolar splittings~\cite{Bluhm2011,Stockill2016,Botzem2016,Stockill2016}. This simulation reproduces the coherence decay for Hahn echo and the coherence modulations.
The decay envelopes for the simulated CPMG, however, do not agree well with experiment, especially for large $n$. In order to gain additional insight into the source of decoherence we extract the envelope decay time, $\Tn$, from the experimental data and plot it as a function of $n$ (Fig.~\ref{ovh:fig4}c and Fig.\ref{ovh_sup:figS4})~\cite{Medford2012}. 
We observe an initial scaling of $\TCPMG \! \propto \! n^\gamma$ with $\gamma \! \sim \! 0.8$, and a crossover to $\gamma \! \sim \! 0.5$ for large $n$.

We ascribe the change in the observed $\TCPMG$ scaling to a crossover between decoherence limited by transverse to longitudinal Overhauser field dynamics. For small $n$ the fluctuations of $\vec{B}^{i}_\perp$ dominate the decoherence, leading to scaling with large $\gamma$; purely transverse low-frequency fluctuations are expected to yield $\TCPMG \! \propto n^\gamma$ with $\gamma \! = \! 1$ (see Sec.~\ref{ovh_sup:transverse}).
With increasing $n$ other decoherence sources start playing a dominant role. 
The intermediate-frequency fluctuations of $\Delta B_\parallel$ cause additional superexponential decay, which for large $n$ is given by $\exp[-4TS_{\parallel}(1/2\tau)/\pi^2]$, where $S_{\parallel}(f)$ is the PSD of $\Delta B_\parallel$~\cite{Alvarez2011,Yuge2011,Bylander2011}. 
Assuming that this PSD has a $1/f^{\beta}$ power-law behavior in the relevant frequency range, the CPMG decay for fixed $n$ and varying $\tau$ is then  $\exp[-(T/\Tn)^{\beta+1}]$, with $\Tn \! \propto \! n^{\gamma}$ and $\gamma \! = \! \beta/(\beta \! + \! 1)$~\cite{Medford2012}.
The observed scaling with $\gamma \! \sim \! 0.5$ is therefore consistent with $1/f$ noise and a Gaussian decay. 

As shown in Fig.~\ref{ovh:fig4}b (black lines), adding the $\beta = 1$ envelope function, $\exp[-(T/\Tn)^2]$ and $\Tn \! = \! n^{1/2} \! \times \! 25$~$\mu$s, appropriate for $\beta = 1$, gives good agreement with  experimental results. From the agreement between the simulations and the measurements we estimate that for $f \! > \! 100$~kHz the PSD $S_{\parallel}(f) \! \sim \! A^{2}/(2\pi f)$ with  $A^{-1} \! \sim \! 9$~$\mu$s.
For comparison with results presented in Ref.~\cite{Malinowski2017} we extrapolate this frequency dependence to $667$~kHz. Using the extrapolated value we estimate the CPMG decay time in an experiment in which $\tau$ is fixed but $n$ is varied, $\TCPMG \! = \! \pi^2/4S_{\parallel}(1/2\tau)$. Such estimate yields $\approx\!0.83$~ms for $\tau \! = \! 750$~ns, which is close to $\TCPMG \! = \! 0.87 \! \pm \! 0.13$~ms measured in Ref.~\cite{Malinowski2017}.

The $1/f$ power law found for $f \! > \! 100$~kHz differs from the $1/f^2$ spectrum observed below 1~kHz. This is not surprising, since for frequencies higher than the strength of intra-nuclear interactions ($\sim$1~kHz) the diffusion model is no longer applicable.
Whether the high-frequency $\Delta B_\parallel$ fluctuations have the same physical origin (i.e.~flip-flops of nuclei due to dipolar and hyperfine-mediated interactions) as the low-frequency ones is an open question.

Theory for CPMG decay caused by spectral diffusion due to dipolar interactions predicts a coherence decay of the form $\exp[-(T/\Tn)^6]$,  with $\Tn \! \propto \! n^{2/3}$ for small and even $n$ \cite{Witzel2007}.
This decay form (and scaling) is in disagreement with our observations.
In particular for large $n$, existing spectral diffusion theories based on cluster expansion~\cite{Witzel2006,Yao2006,Yang2009a} may need to be refined, for example taking into account realistic shapes of the electronic wave functions. Based on our findings, such theories can be tested experimentally at $\Bext>1$~T, where bare dipole-dipole coupling is the dominant internuclear interaction. 

Finally, it is possible that the $\Delta B_\parallel$ fluctuations are not of intrinsic origin (nuclear dynamics), but of extrinsic origin.
For example, charge noise, which generically has a $1/f^\beta$ spectrum with $\beta \! \sim \! 1$~\cite{Dial2013}, can shift the electron wavefunction and effectively result in Overhauser field fluctuations~\cite{Neder2011}.

\section{Conclusion}

In conclusion, we have experimentally investigated the spectrum of the GaAs nuclear environment for spin qubits and find it consistent with classical diffusion over six orders of magnitude in frequency, from millihertz to kilohertz. For applied fields below $\sim\!0.75$~T, nuclear diffusion is dominated by the electron-mediated flip-flop, enhancing diffusion by a factor of 8. Decoherence of the S-T$_0$ qubit is dominated by fluctuations of the transverse Overhauser field for short CPMG sequences, and by longitudinal Overhauser field for CPMG sequences with more than 32 $\pi$ pulses.

\section*{Acknowledgements}

This work was supported by the Army Research Office, the Polish National Science Centre (NCN) under Grants No.~DEC-2012/07/B/ST3/03616 and DEC-2015/19/B/ST3/03152, the Innovation Fund Denmark, the Villum Foundation and the Danish National Research Foundation. 

\section*{Authors contributions}
S.F., G.C.G. and M.J.M. grew the heterostructure. P.D.N. fabricated the device. F.M., P.D.N., F.K. and F.K.M. prepared the experimental setup. F.K.M., F.M. performed the experiment. \L{}.C. and M.S.R. developed the theoretical model. F.K.M., F.K., \L{}.C., M.S.R., F.M. and C.M.M. analysed the data and prepared the manuscript.

\chapterimage{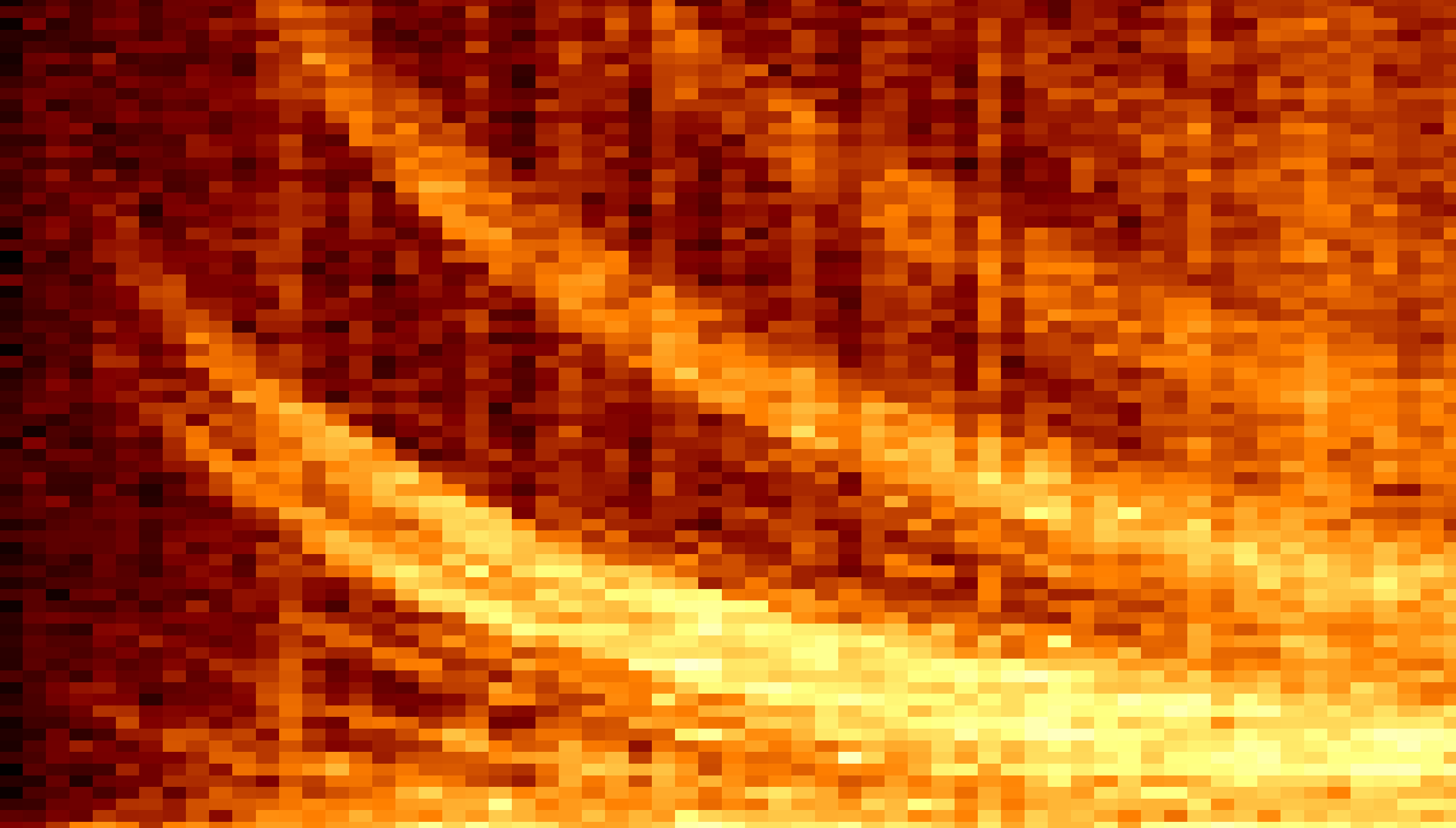}
\chapter[Notch filtering the nuclear environment of a spin qubit]{\protect\parbox{0.9\textwidth}{Notch filtering the nuclear\\ environment of a spin qubit}}
\label{ch:notch}

{\let\thefootnote \relax\footnote{This chapter and chapter \ref{ch:notch_sup} are adapted from Ref.~\cite{Malinowski2017}. \copyright~(2017) by the Nature Publishing Group.}}
\addtocounter{footnote}{-1}

\begin{center}
 Filip K. Malinowski$^{1,*}$, Frederico Martins$^{1,*}$, Peter D. Nissen$^{1}$, \\
 Edwin Barnes$^{2,3}$, \L ukasz~Cywi\'nski$^{4}$, Mark~S.~Rudner$^{1,5}$, Saeed Fallahi$^{6}$, Geoffrey C. Gardner$^{7}$, \\
 Michael~J.~Manfra$^{6,7,8}$, Charles M. Marcus$^{1}$, Ferdinand~Kuemmeth$^{1}$
\end{center}

\begin{center}
	\scriptsize
	$^{1}$ Center for Quantum Devices, Niels Bohr Institute, University of Copenhagen, 2100 Copenhagen, Denmark\\
	$^{2}$ Department of Physics, Virginia Tech, Blacksburg, Virginia 24061, USA\\
	$^{3}$ Condensed Matter Theory Center and Joint Quantum Institute, Department of Physics, University of Maryland, College Park, Maryland 20742-4111, USA\\
	$^{4}$ Institute of Physics, Polish Academy of Sciences, al.~Lotnik{\'o}w 32/46, PL 02-668 Warsaw, Poland \\
	$^{5}$ Niels Bohr International Academy, Niels Bohr Institute, 2100 Copenhagen, Denmark\\
	$^{6}$ Department of Physics and Astronomy and Birck Nanotechnology Center, Purdue University, West Lafayette, Indiana 47907, USA\\
	$^{7}$ School of Materials Engineering and Birck Nanotechnology Center, Purdue University, West Lafayette, Indiana~47907, USA\\
	$^{8}$ School of Electrical and Computer Engineering, Purdue University, West Lafayette, Indiana 47907, USA\\
	$^{*}$ These authors contributed equally to this work
\end{center}

\begin{figure}[t]
	\centering
	\includegraphics[scale=1]{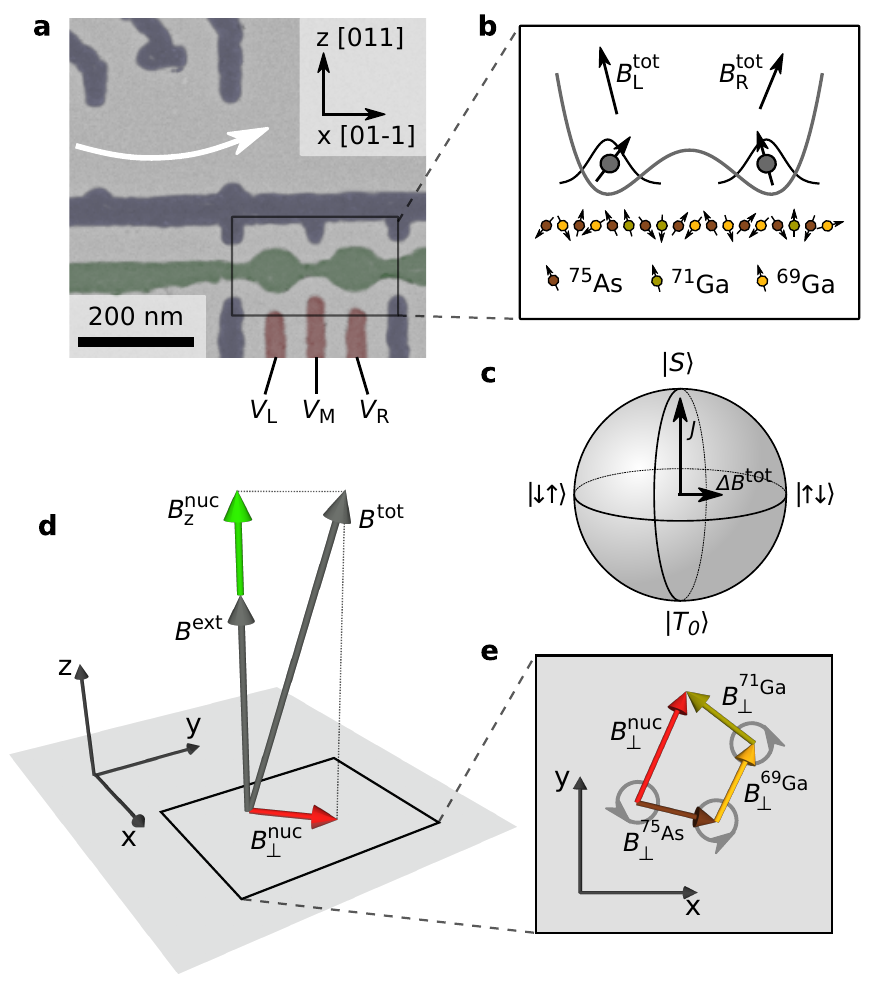}
	\caption[Singlet-triplet qubit interacting with a nuclear spin bath]
	{\textbf{Singlet-triplet qubit interacting with a nuclear spin bath.}
	\textbf{a}, False-color scanning electron micrograph of a device similar to the one measured, consisting of a double dot (surrounded by black rectangle) and a proximal readout dot (indicated by white arrow).
	\textbf{b}, Double-well potential occupied by two electrons. Within the left (right) dot an effective magnetic field $\Btot_\mathrm{L(R)}$ splits the electron spin states due to the Zeeman effect and hyperfine interaction with spinful nuclei of $^{69}$Ga, $^{71}$Ga, and $^{75}$As.
	\textbf{c}, Bloch sphere representation of the qubit with corresponding two-electron spin states indicated. Two rotation axes are defined by the exchange interaction, $J$, and 
the total field gradient between the dots, $\Delta \Btot=\Btot_\mathrm{L}-\Btot_\mathrm{R}$.
	\textbf{d}, The effective magnetic field $\Btot$ acting on each spin is set by 
the external magnetic field $\vec{B}^{\rm ext}$ (nominally aligned with the [011] crystal axis), the slowly fluctuating Overhauser field component $\vec{B}^\mathrm{nuc}$ parallel to $\vec{\Bext}$, and the rapidly changing transverse Overhauser field $\vec{B}_\perp^\mathrm{nuc}$. 
Here we suppress the dot label indices for brevity.
	\textbf{e}, The transverse Overhauser field $\vec{B}^{\rm nuc}_\perp = \vec{B}_\perp^{^{69}\mathrm{Ga}} + \vec{B}_\perp^{^{71}\mathrm{Ga}} + \vec{B}_\perp^{^{75}\mathrm{As}}$ is a sum of fields
of the three nuclear species, each precessing at its Larmor frequency.
	}
	\label{notch:fig1}
\end{figure}

\section{Introduction}

Electron spins in gate-defined quantum dots provide a promising platform for quantum computation \cite{Loss1998,Veldhorst2015,Nowack2011,Bluhm2011,Petta2005,Nowack2007,Maune2012}.
In particular, spin-based quantum computing in gallium arsenide takes advantage of the high quality of semiconducting materials, reliability in fabricating arrays of quantum dots, and accurate qubit operations \cite{Nowack2007,Petta2005,Foletti2009,Dial2013,Maune2012,Martins2016}.
However, the effective magnetic noise arising from the hyperfine interaction with uncontrolled nuclear spins in the host lattice constitutes a major source of decoherence \cite{Petta2005,Bluhm2011,Botzem2016,Martins2016}. 
Low frequency nuclear noise, responsible for fast (10 ns) inhomogeneous dephasing \cite{Petta2005}, can be removed by echo techniques \cite{Petta2005,Bluhm2011,Viola1998,Barthel2010,Medford2012,Botzem2016}. High frequency nuclear noise, recently studied via echo revivals \cite{Bluhm2011,Botzem2016}, occurs in narrow frequency bands related to differences in Larmor precession of the three isotopes $^{69}$Ga, $^{71}$Ga,  and $^{75}$As \cite{Cywinski2009,Cywinski2009a, Neder2011}.
Here we show that both low and high frequency nuclear noise can be filtered by appropriate dynamical decoupling sequences, resulting in a substantial enhancement of spin qubit coherence times. Using nuclear notch filtering, we demonstrate a spin coherence time (${T_{2}}$) of 0.87 ms, five orders of magnitude longer than typical exchange gate times, and exceeding the longest coherence times reported to date in Si/SiGe gate-defined quantum dots~\cite{Eng2015,Kawakami2016}.

\section{Qubit and nuclear spin bath}

The qubit under study is implemented in a gate-defined double dot, with a potential that can be manipulated via nanosecond voltage pulses applied to gate electrodes $\VL$, $\VM$ and $\VR$ (Fig.~\ref{notch:fig1}a and Methods). 
The qubit states are encoded in the two-electron spin singlet state, $\ket{S} = \frac{1}{\sqrt{2}}(\ket{\ud} - \ket{\du})$, and the spin triplet state,  $\ket{T_0} = \frac{1}{\sqrt{2}}(\ket{\ud} + \ket{\du})$, where the arrows indicate the spin projections of the electrons in the left and right dots \cite{Petta2005,Foletti2009}. 
These qubit states are energetically separated from the spin-polarized two-electron states, $\ket{\uparrow\uparrow}$ and $\ket{\downarrow\downarrow}$, by an external magnetic field $\vec{B}^{\rm ext}$, ranging from 0.2 to 1 tesla in this experiment.
Single-shot readout of the qubit is accomplished using spin-to-charge conversion followed by readout of a proximal sensor dot \cite{Barthel2009, Barthel2010} (see Methods).

As illustrated in Fig.~\ref{notch:fig1}b,d, 
the local Zeeman energy in dot $d = {\rm L,R}$ is perturbed by the Overhauser field $\vec{B}^{{\rm nuc}}_d$ arising from the hyperfine interaction with the nuclear spin bath. 
In our device, each electron is in contact with $\sim 10^6$ nuclear spins, comprised of three species:
 $^{69}$Ga, $^{71}$Ga, and $^{75}$As \cite{Petta2005,Bluhm2011,Neder2011}.

\begin{figure}[tb]
	\centering
	\includegraphics[scale=1]{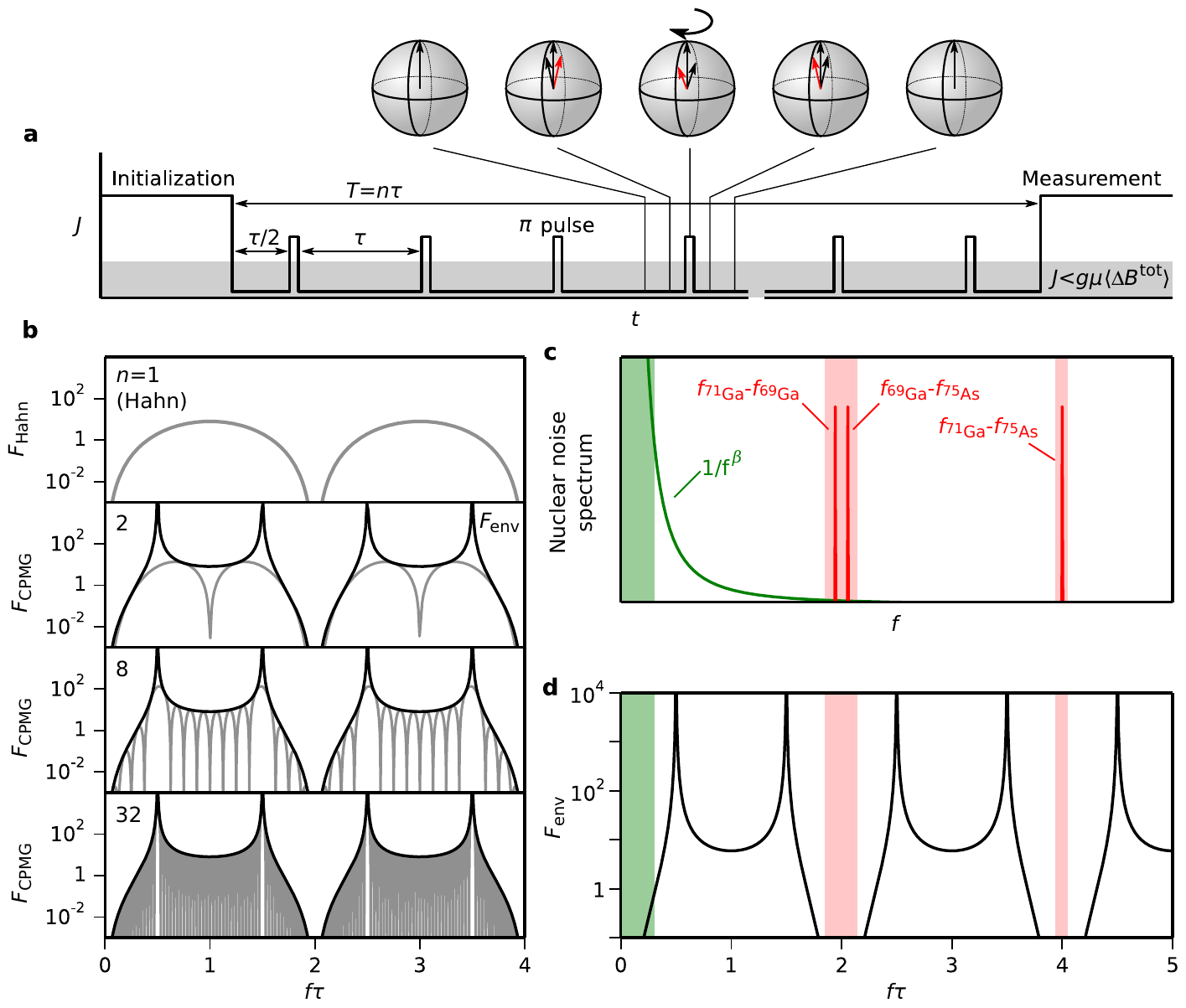}
	\caption[Frequency-selective dynamical decoupling]
	{\textbf{Frequency-selective dynamical decoupling.}
	\textbf{a}, CPMG pulse sequence consisting of $n$ pulses separated by time $\tau$. At the beginning, two electrons prepared in a singlet state $|S\rangle$ (Initialization) are rapidly separated into two dots with negligible exchange splitting (shaded region of $J$). After a total separation time $T=n \tau$ the preserved qubit state is detected by the readout dot via spin-to-charge conversion (Measurement). During the separation time the two-electron state evolves in the fluctuating gradient of total magnetic field $\Delta \Btot$. For slow fluctuations, the phases acquired before and after each $\pi$ pulse cancel each other, due to the sign reversal of the acquired phase by the $\pi$ pulse.
This is exemplified for three different values of $\Delta \Btot$ by arrows in the Bloch sphere. 
	\textbf{b}, Filter functions of Hahn echo ($n$ = 1) and CPMG sequence with $n$ = 2, 8 and 32 $\pi$ pulses (gray). Envelope of the filter function $\Fenv$ reveals a frequency selectivity that is independent of $n$ (black).
	\textbf{c}, Schematic spectral density of nuclear noise. The linear low frequency part (green), described by a power law, is dominated by fluctuations associated with diffusion of the longitudinal component of the nuclear spin. The quadratic high frequency noise (red) results from fluctuations of $\Btot$ at differences of nuclear Larmor frequencies.
	\textbf{d}, By adjusting the time between $\pi$ pulses, the minima of the filter function envelope $\Fenv$ (black) can be aligned with the nuclear noise spectrum (green and red shading), thereby decoupling the qubit from both linear low frequency and quadratic high frequency noise. 
	}
	\label{notch:fig2}
\end{figure}

The Bloch sphere of the $S$-$T_0$ qubit is shown in Fig.~\ref{notch:fig1}c.  Bold arrows indicate the rotation axes associated with the exchange interaction, $J$, and the gradient of the effective 
field between the dots, $\Delta \Btot = \Btot_{\rm L} - \Btot_{\rm R}$, where $\Btot_d = \sqrt{|\vec{B}^{\rm ext} + \vec{B}^{{\rm nuc}}_d|^2}$ is the magnitude of the total effective field in dot ${d}$~\cite{Foletti2009}.
Note that transverse nuclear field gradients tilt the quantization axes in the two dots relative to each other.
For large external fields this primarily leads to a minor redefinition of the qubit subspace~\cite{Neder2011}; for simplicity throughout this work we refer to the states in the qubit subspace by the conventional labels $S$ and $T_0$.

Overhauser field fluctuations in each dot are non-Markovian, with low frequency (power-law) spectral content parallel to the external field, denoted $\Bznuc$ (suppressing the dot index), and narrow-band spectral components at the nuclear Larmor frequency scale perpendicular to the external field, denoted $\vec{B}^{\rm nuc}_\perp$. Low frequency fluctuations arise primarily from nuclear spin diffusion~\cite{Reilly2008}, driven by dipole-dipole interactions between neighboring nuclei, and nonlocal electron-mediated flip-flops \cite{DeSousa2003,Yao2006,Cywinski2009,Cywinski2009a}.
High frequency fluctuations of $\vec{B}^{\rm nuc}_\perp$ arise primarily due to the megahertz-scale relative Larmor precession of different nuclear spins~\cite{Bluhm2011,Neder2011,Botzem2016}. 
The transverse Overhauser field $\vec{B}^{\rm nuc}_\perp$ is given by the sum of contributions $\vec{B}_\perp^{^{69}\mathrm{Ga}}$, $\vec{B}_\perp^{^{71}\mathrm{Ga}}$, and $\vec{B}_\perp^{^{75}\mathrm{As}}$ of the three isotopic species, each of which precesses at its own  Larmor frequency, see Fig.~\ref{notch:fig1}e.
This leads to modulations of the total field in each dot, $\Btot$, which are concentrated near the {\it differences} of the nuclear Larmor frequencies, and contribute quadratically to the qubit splitting.

\section{Dynamical decoupling and notch filtering}

To decouple the qubit from the multiscale nuclear noise, we employ the Carr-Purcell-Meiboom-Gill (CPMG) pulse sequence shown in Fig.~2a. 
We first initialize the double dot in a spin singlet by temporarily loading two electrons into the left dot.
Then we quickly separate the electrons in the double-well potential, thereby rapidly turning off the exchange interaction, $J$. In this configuration, the gradient of the total effective field, $\Delta \Btot$, causes uncontrolled qubit rotation around the horizontal axis of the Bloch sphere in Fig.~\ref{notch:fig1}c. 
After a time $\tau/2$, an exchange pulse is applied by temporarily lowering the barrier between dots with a voltage pulse on gate $\VM$ \cite{Martins2016}, implementing a $\pi$ rotation around the vertical axis of the Bloch sphere (see Section~\ref{notch_sup:calibration}).
We repeat this set of operations $n$ times (where $n$ is even) and, after a total evolution time $T = n\tau$, read out the state of the qubit. The fraction of singlet outcomes is denoted $P_S$. Setting $n=1$ implements a Hahn-echo sequence, and allows comparison to previous work \cite{Bluhm2011,Childress2006,Botzem2016}.

For quasistatic nuclear noise, the effective field acting on the qubit before and after the $\pi$ pulse is nearly the same, causing the qubit state to be refocused to the singlet after an interval $\tau/2$.
For nuclear noise with power spectrum $S(f)$, Hahn and CPMG sequences 
act as a filter of the noise in the frequency domain~\cite{Martinis2003,Cywinski2008,Biercuk2011,Soare2014,Kabytayev2014,Alvarez2011}.
For Gaussian noise, decoherence is described by a function
\begin{equation}
\label{notch:W}	W(\tau) = \exp \left( -\int_0^\infty \frac{\drv f}{2 \pi^2} S(f) \frac{F(f \tau)}{f^2} \right),
\end{equation}
corresponding to a singlet probability $P_S(\tau)=\frac12[W(\tau) + 1]$.
In this expression, $F(f \tau)$ is a filter function that depends on the particular pulse sequence.

Filter functions for Hahn echo ($\FHahn$) and several CPMG sequences ($\FCPMG$) for fixed $\tau$ are plotted in Fig.~\ref{notch:fig2}b (gray) for varying numbers of $\pi$ pulses.
We write the CPMG filter function as a product $\FCPMG = \tfrac{1}{2} \FFID \times \Fenv$, where $\FFID$ is the filter function corresponding to the free induction decay and $\Fenv$ is a slowly varying envelope (see Methods). $\Fenv$ is periodic with period $2/\tau$, with minima occurring at zero frequency and multiples of $2/\tau$ (Fig.~\ref{notch:fig2}(b), black), independent of $n$.
Specific features of the filter functions can be exploited to decouple the qubit from its characterisitic noise environment.
First, for fixed separation time $T = n\tau$, the filter minimum near zero frequency becomes wider for increasing $n$ (i.e., decreasing $\tau$, note that the horizontal axis in Fig.~\ref{notch:fig2}b is normalized frequency $f\tau$).
Thus for fixed $T$, decoupling from low frequency $1/f^\beta$-type noise ($\beta>0$) becomes more efficient as $n$ increases. 
Second, the minima that occur at multiples of $1/\tau$ indicate that noise at these frequencies  is notch-filtered, in the sense that specific narrow frequency windows are suppressed.

A schematic of the spectral density of nuclear noise for the $S$-$T_0$ qubit fabricated in a GaAs heterostructure is shown in Fig.~\ref{notch:fig2}c, distinguishing longitudinal low frequency noise (green) and transverse narrow-band noise (red). 
The low frequency longitudinal contribution is well described by a power-law spectrum~\cite{Reilly2008,Medford2012}, and can be removed efficiently by any CPMG sequence (Fig.~\ref{notch:fig2}d). 
The high frequency transverse contribution due to relative Larmor precession of nuclei is concentrated near the three Larmor frequency differences \cite{Cywinski2009a}, at megahertz frequencies for tesla-scale applied fields. Remarkably, two of the Larmor difference frequencies, $\fGb - \fGa$ and $\fGa - \fAs$, are nearly equal, independent of magnetic field, and hence the third frequency difference, $\fGb - \fAs$, occurs at twice that frequency. This coincidental property of the three nuclear species allows us to approximately align minima of the filter function with {\it all three} frequency differences  by correctly choosing the time between $\pi$ pulses, $\tau$, thereby decoupling the qubit from low and high frequency nuclear noise simultaneously.

\section{Revivals of coherence}

\begin{figure}[t]
	\centering
	\includegraphics[scale=1]{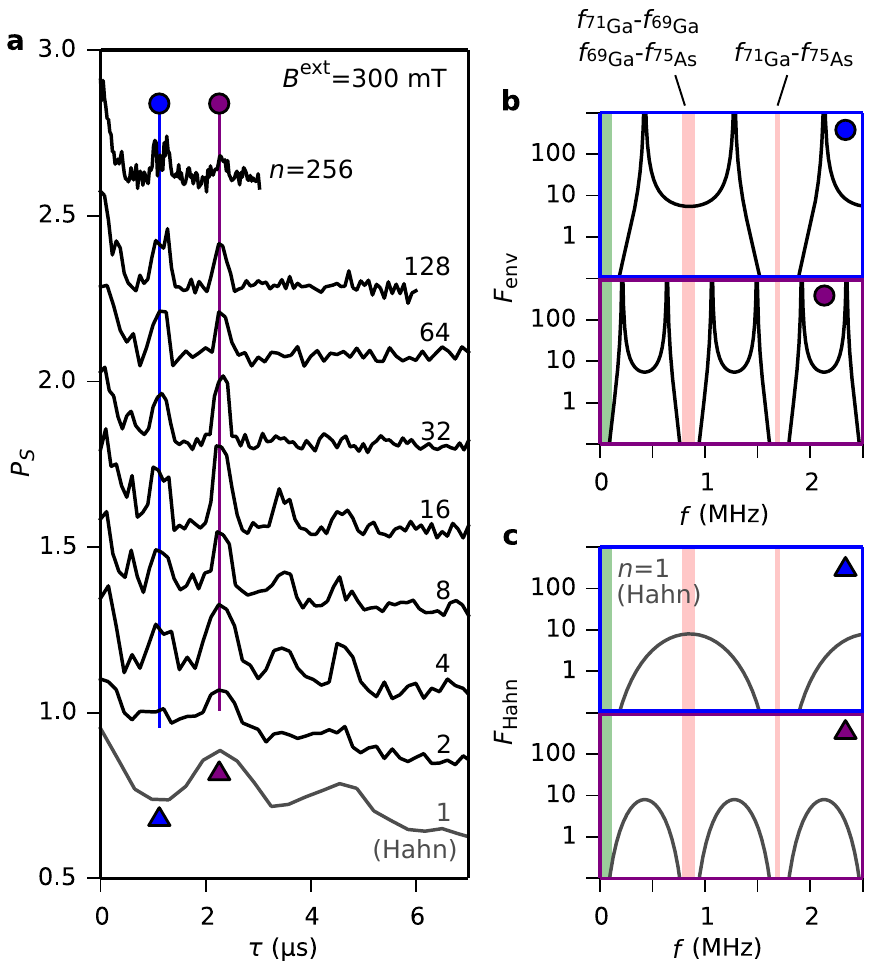}
	\caption[Revival of coherence due to decoupling from nuclear Larmor precession]
	{\textbf{Revival of coherence due to decoupling from nuclear Larmor precession.}
	\textbf{a}, Singlet return probability, $P_S$, as a function of the time between $\pi$ pulses, $\tau$, for various numbers of $\pi$ pulses, $n$. Curves are offset for clarity.
	\textbf{b}, Filter function envelope (black) and nuclear noise frequencies expected at 300 mT (shaded) for two choices of $\tau$. In both cases (marked by blue and purple lines in \textbf{a}) the revival in $P_S$ appears when minima of the filter function align with nuclear difference frequencies.  
	\textbf{c}, Filter function of Hahn-echo sequence for the same choices of $\tau$ as in \textbf{b}. The absence of the first revival (marked by a blue triangle in \textbf{a}) indicates that coherence is lost when the maximum of the filter function overlaps with the peaks in the nuclear noise spectrum (shaded). 
	The revival of $P_S$ for the second choice of $\tau$ (marked by the purple triangle in \textbf{a}) corroborates the destructive role of nuclear Larmor dynamics in qubit decoherence. 
	}
	\label{notch:fig3}
\end{figure}

We now demonstrate the efficacy of this notch filter strategy in our experimental setup.
The narrow-band character of the high frequency nuclear noise is revealed by plotting the observed singlet return probability $P_S$ as a function of $\pi$-pulse separation time $\tau$ (rather than total separation time $T$).
Independent of the choice of $n$, we observe  an initial loss of coherence followed by revivals at $\tau\approx$ 1.1, 2.2, 3.3, ... $\mu$s (Fig.~\ref{notch:fig3}a).
These values of $\tau$ correspond to decoupling conditions shown in Fig.~\ref{notch:fig3}b, namely the alignment of nuclear difference frequencies (shaded red) with minima of the filter function envelope. 
Qualitatively, the alternating depth of filter minima in Fig.~\ref{notch:fig3}b also explains the alternating heights of revivals, most pronounced for $n=4$ in Fig.~\ref{notch:fig3}a. 
With increasing $\tau$, the height of the revivals decreases. This is related to decoherence arising from low frequency noise (shaded green in Fig.~\ref{notch:fig3}b) \cite{Medford2012}.
Revivals observed for Hahn-echo sequences can be explained similarly, except that the filter function for $\tau\approx$ 1.1 $\mu$s has a maximum near  0.9 MHz (Fig.~\ref{notch:fig3}c), rather than a minimum. Accordingly,  $P_S$ shows a minimum near $\tau=1.1$ $\mu\mathrm{s}$ instead of a revival (cf. $n=1$ data in Fig.~\ref{notch:fig3}a).

\begin{figure*}
	\centering
	\includegraphics[width=\textwidth]{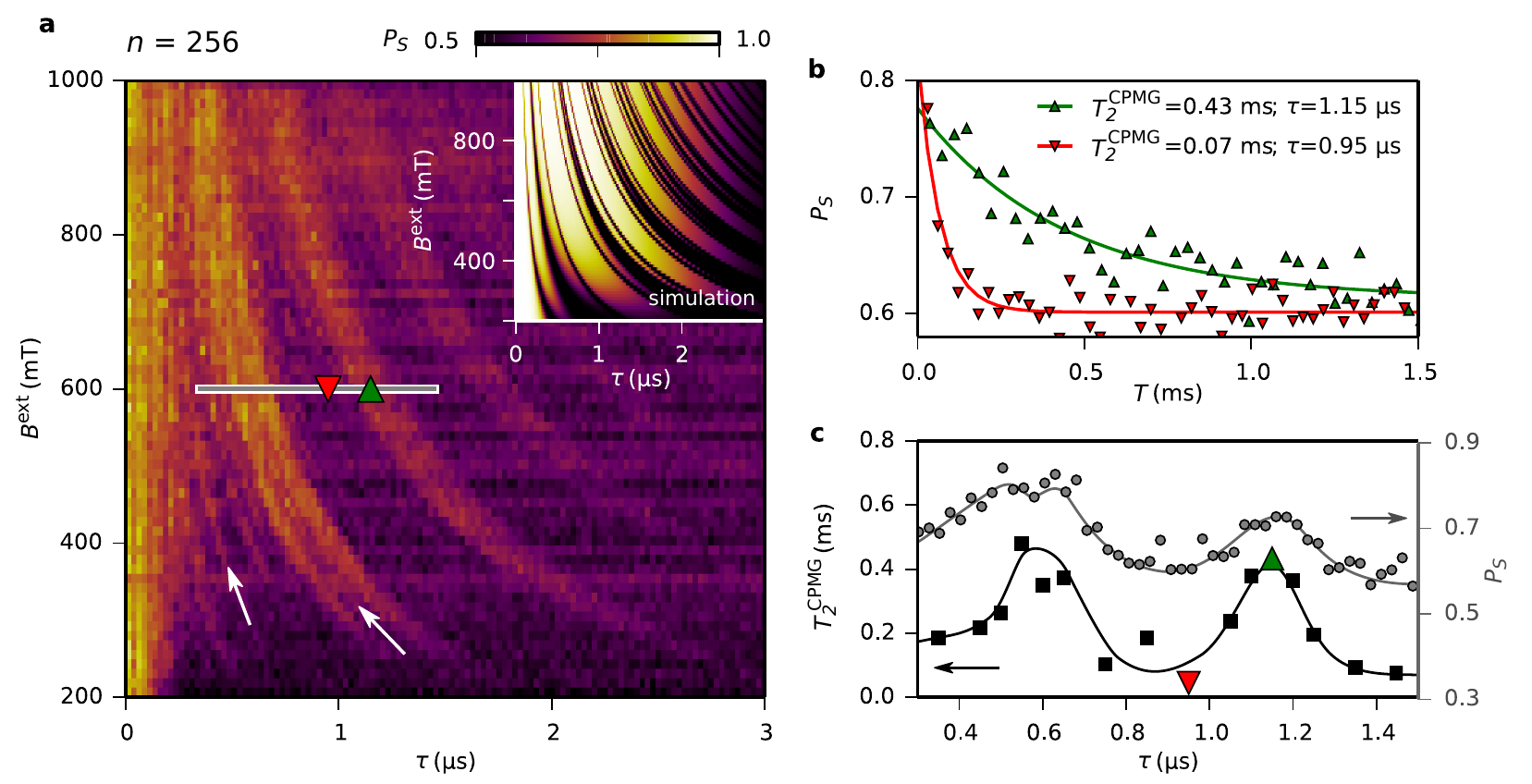}
	\caption[Effect of magnetic field and ${\tau}$ on qubit coherence]
	{\textbf{Effect of magnetic field and ${\tau}$ on qubit coherence.}
	\textbf{a}, Singlet return probability $P_S$ as a function of time between $\pi$ pulses, $\tau$, and external magnetic field, $\Bext$, for  a fixed number of $\pi$ pulses, $n=256$.
	Inset: A semiclassical model, generalizing the model of \cite{Neder2011} to the case of CPMG sequences, with no free parameters (see Section~\ref{notch_sup:model}). Arrows indicate fine features that the model fails to reproduce.
	\textbf{b}, Singlet return probability, $P_S$, measured as a function of total separation time $T=n\tau$, where $n$ is varied between 32 and 1536, for $\tau=0.95$ and $1.15$ $\mu$s at $\Bext=600$ mT, the values marked in \textbf{a}. Solid lines are fits to a decay law with exponential decay time $T_2^\mathrm{CPMG}$ (see Methods).
	\textbf{c}, Coherence time $T_2^\mathrm{CPMG}$, measured by increasing $n$ as in \textbf{b}, as a function of $\tau$ (squares). For comparison, $P_S$ for constant $n$, reproduced from the grey cut in \textbf{a}, is also shown (circles). Lines are guides to the eye. Triangles indicate the coherence times obtained from \textbf{b}.
	}
	\label{notch:fig4}
\end{figure*}

The dependence of the decoupling condition for $\tau$ on nuclear Larmor dynamics can be verified by changing the applied magnetic field. 
In Fig.~\ref{notch:fig4}a we fix $n=256$ and measure the decay of coherence as a function of $\Bext$.
As expected for a linear nuclear Zeeman splitting we find that the positions of the revival peaks  follow a $1/\Bext$ dependence.
We further observe that the peaks in $P_S(\tau)$ disappear at low magnetic fields.
This may arise from several effects.
First, the transverse Overhauser field in each dot, $\vec{B}^{{\rm nuc}}_{\perp}$, affects the total electronic Zeeman energy more strongly at low magnetic field (see Fig.~\ref{notch:fig1}d), thereby accelerating dephasing.
Second, the energy mismatch between nuclear and electron Zeeman splittings becomes smaller at low fields, increasing electron-mediated interactions between nuclear spins and the associated low frequency noise \cite{Yao2006,Deng2005,Cywinski2009a}.
Third, an increase in $\tau$, as needed to maintain the decoupling condition at lower fields, narrows the filter function minima and thus reduces decoupling from high frequency noise. 

\section{Extending coherence time}

Next we show that revivals in $P_S$ translate to prolonged qubit coherence times, by increasing $n$ while keeping $\tau$ and $\Bext$ fixed. 
This method, pioneered in NMR~\cite{Carr1954}, differs from other spin qubit experiments in which $n$ is held constant while $\tau$ is swept proportionally to $T$~\cite{Bluhm2011,Medford2012}.
Figure~\ref{notch:fig4}b plots decay curves $P_S(T=n\tau)$  obtained for $\tau=0.95$ and $1.15$ $\mu$s at $\Bext=600$ mT (the corresponding points are indicated in Fig.~\ref{notch:fig4}a). 
For large $n$ and Gaussian noise, an exponential decay of coherence is expected, independent of the power spectrum of the noise~\cite{Alvarez2011}. 
By fitting exponential decay curves~\cite{Carr1954,Alvarez2011} (see Methods) we extract drastically different coherence times $\TCPMG$, as indicated. 
Values of $\TCPMG$ for more choices of $\tau$ are plotted in Fig.~\ref{notch:fig4}c, along with $P_S(\tau)$ extracted from Fig.~\ref{notch:fig4}a. 
We observe a clear correlation between $\TCPMG$ and $P_S(\tau)$, indicating that qubit coherence is significantly prolonged whenever the decoupling condition is fulfilled. 
The exponential decay indicates that coherence is limited by either incompletely filtered longitudinal noise or pulse errors.

\begin{figure}
	\centering
	\includegraphics[scale=1]{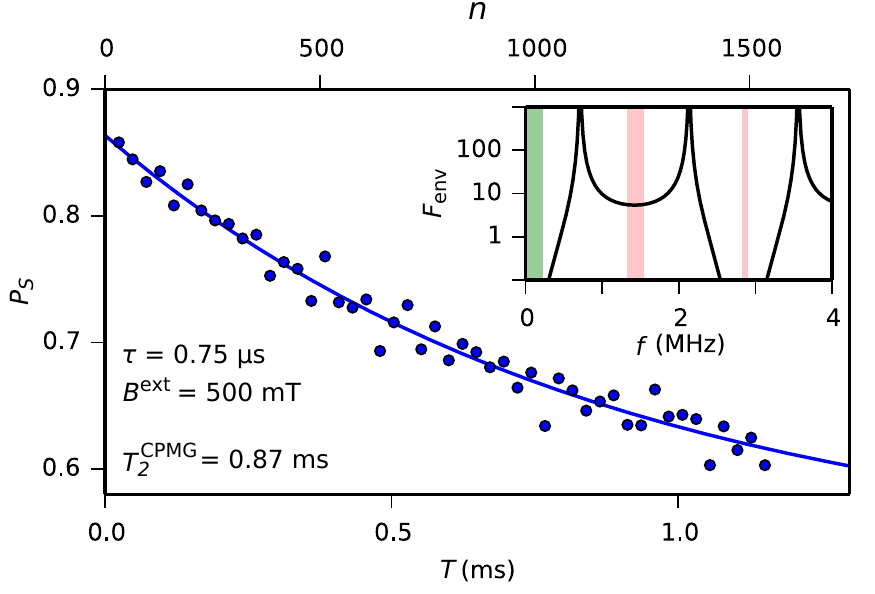}
	\caption[Singlet return probability, ${P_S}$, as a function of total separation time, ${T=n\tau}$, for optimized and fixed values of ${\tau}$ and ${\Bext}$.]
	{\textbf{Singlet return probability, $\mathbf{P_S}$, as a function of total separation time, $\mathbf{T=n\tau}$, for optimized and fixed values of $\mathbf{\tau}$ and $\mathbf{\Bext}$.} 
	An exponential fit to the data yields $T_2^\mathrm{CPMG} = 0.87 \pm 0.13$ ms. The data points correspond to the CPMG sequences with $n=32$ to $n=1536$ $\pi$ pulses. Inset: Alignment of the filter function envelope with nuclear noise spectrum for this choice of $\tau$ and $\Bext$.}
	\label{notch:fig5}
\end{figure}

Finally we comment on the limits of preserving qubit coherence. 
Most of the observed features in Fig.~\ref{notch:fig4}a are captured by a generalization of the semiclassical model of Ref.~\cite{Neder2011}, modified to include the details of the CPMG pulse sequence.
The model involves four device-specific parameters (Fig.~\ref{notch:fig4} inset): 
the effective number of nuclei interacting with each electron, $N=7\times10^5$, 
a phenomenological broadening, $\delta B=1.1$ mT,  of the effective magnetic field acting on nuclei (likely due to quadrupolar splitting arising from electric field gradients \cite{Bluhm2011,Neder2011,Botzem2016}),
the spectral diffusion time, $T_\mathrm{SD} = 600$ $\mu$s, 
and the exponent associated with the linear low frequency noise, $\beta=3$ (all determined by independent measurements as described in Sections~\ref{notch_sup:NandB} and \ref{notch_sup:TandBeta}). 
The model suggests that the longest coherence time may be achieved by choosing a decoupling condition corresponding to the second revival at high magnetic fields, consistent with a reduced contribution of $\vec{B}^\mathrm{nuc}_{\perp}$ to $\Btot$ in each dot (see Fig.~\ref{notch:fig1}d) and the decoupling condition depicted in Fig.~\ref{notch:fig2}d. 
We note that the model does not take pulse errors into account and does not show several fine features observed in experiment (see white arrows in Fig.~\ref{notch:fig4}a, see Section~\ref{notch_sup:splitting}). 

By exploring the parameter space between $\Bext=300$ and $1000$ mT with $\tau$ corresponding to the first revival peak, we observe coherence times around 0.7 ms for $\Bext=500$ to 600~mT, with the largest being $\TCPMG = 0.87 \pm 0.13$ ms (Fig.~\ref{notch:fig5}), measured at $\Bext=500$ mT and $\tau=0.75$ $\mu$s. However, the number of examined values of $\Bext$ and $\tau$ remains insufficient to resolve the fine structure apparent in the first revival peak.

We expect further improvements by using shorter $\pi$ pulses and nuclear programming \cite{Foletti2009}. This will improve the fidelity of $\pi$ pulses and suppress low frequency noise, allowing the advantageous use of the decoupling condition in Fig.~\ref{notch:fig2}d at high magnetic fields and high pulse rates.  

\section{Summary}

In summary, dynamical decoupling sequences were demonstrated to provide decoupling from narrow-band high frequency noise, acting as a notch filter for the nuclear environment. This technique was used to efficiently decouple a GaAs-based $S$-$T_0$ qubit from its nuclear environment. By synchronizing the repetition rate of $\pi$ pulses in CPMG sequences with differences of nuclear Larmor frequencies, the coherence of a $S$-$T_0$ qubit coupled to nuclear spin bath was extended the millisecond regime (0.87 ms), five orders of magnitude longer than the gate operation time.

\section{Methods}

\subsection{The sample}

The sample, identical to the one shown in Fig.~1a, is fabricated from a GaAs/AlGaAs quantum well grown by molecular beam epitaxy. 
Crystallographic axes are shown in Fig.~1a. 
A high-mobility 2D electron gas (2DEG) is formed 57 nm below the sample surface with carrier density $n_s$~=~$2.5\times10^{15}$~m$^{-2}$ and mobility $\mu$~=~230~m$^{2}$/Vs. Metallic gates, separated from the heterostructure by a 10 nm layer of HfO$_2$, are used to confine two electrons in the region indicated by a rectangle in Fig.~1a.  Gates indicated in blue and red are operated at negative voltages to deplete the 2DEG underneath, while gates colored in green are biased with positive voltages to accumulate electrons beneath. The charge state and tunnel coupling of the double dot can be controlled on a nanosecond timescale by applying voltage pulses to gates $\VL$, $\VM$, and $\VR$. 

\subsection{Initialization and readout of the qubit}

The sample is measured at a base temperature of 25~mK in a cryofree dilution refrigerator, with an external magnetic field $\Bext$ applied parallel to the $z$ direction indicated in Fig.~1a. 
The qubit is initialized in a singlet state by tilting its charge state into the (2,0) charge configuration and allowing the exchange of electrons with the left lead near the (1,0) charge transition \cite{Petta2005}.

After qubit manipulation the state of the qubit is measured by tilting the double well potential to favour the (2,0) charge state. If the two electrons are in the spin triplet configuration, Pauli blockade prevents reaching the (2,0) state, and the charge configuration remains (1,1).
The charge state of the double dot modifies the conductance through a proximity sensor dot operated as a single electron transistor. This sensor dot is embedded in a radio-frequency resonant circuit, enabling us to distinguish singlet and triplet states in 8 $\mu$s with a readout visibility of approximately 80\%, as defined in Ref. \cite{Barthel2010}.

\subsection{ Envelope of a filter function for CPMG sequence } 

Filter functions for Hahn echo and CPMG sequence (for even number of $\pi$ pulses, $n$) are given by \cite{Cywinski2008}
\begin{equation}
	\FHahn(f \tau) = 8 \sin^4 \left( \pi f \tau/2 \right);
\end{equation}

\begin{equation}
	\FCPMG(f \tau) = \frac{ 8 \sin^4 \left( \pi f \tau / 2 \right) \sin^2 \left(\pi f \tau n \right) }{ \cos^2 \left(\pi f \tau \right) }.
\end{equation}
\vspace{5pt}

To emphasize the qualitative difference between CPMG sequences and the Hahn echo sequence, and represent features of CPMG filter functions relevant for large number of $\pi$ pulses, $n$, we rewrite
\begin{equation}
	\FCPMG = \frac{1}{2} \Fenv \times \FFID
\end{equation}
using $n$-independent filter function envelope
\begin{equation}
	\Fenv(f \tau) = \frac{ 8 \sin^4 \left( \pi f \tau / 2 \right) }{ \cos^2 \left(\pi f \tau \right) }. 
\end{equation}
obtained by dividing
 $\FCPMG$ by the filter function corresponding to free induction decay
\begin{equation}
	\FFID(f T) = 2 \sin^2(\pi f T).
\end{equation}
Here $T=\tau n$ corresponds to a free induction decay time equal to the total duration as a CPMG sequence. This normalization removes a fine comb related to the total length of the sequence.

\subsection{Exponential fits to $P_S(T)$}

In contrast to many spin qubit experiments~\cite{Bluhm2011,Petta2005,Dial2013,Botzem2016,Barthel2010,
Medford2012,Cywinski2009,Cywinski2009a,Kawakami2016,DeSousa2003,Yao2006,Childress2006}
we measure coherence not by keeping $n$ constant and sweeping $\tau$, but by increasing $n$ while keeping $\tau$ constant. This method, which is standard in NMR experiments~\cite{Carr1954}, results in an exponential decay of coherence for large number of $\pi$ pulses $n$ and long evolution times $T=n\tau$, independent of the power spectrum of the Gaussian noise~\cite{Alvarez2011}. The rate of such a decay is determined by the noise spectrum at a frequency corresponding to the first peak of the filter function from Fig.~\ref{notch:fig2}d at $f=1/2\tau$.

Therefore, we perform an exponential fit of the form $A+B\exp(-T/\TCPMG)$ to the data, where $A$ and $B$ account for preparation and readout fidelity as well as rapid initial decay of the signal~\cite{Bluhm2011,Botzem2016}, and $\TCPMG$ is a coherence time of the qubit. Typical values of $A$ and $B$ obtained from fits used to extract values of $\TCPMG$, shown in Fig.~\ref{notch:fig4}c, are $A\sim 0.6$ and $B\sim 0.2$. Fit to the data presented in Fig.~5 yields $A=0.53$ and $B=0.34$. 

\section*{Acknowledgements}
We thank Rasmus Eriksen for help in preparation of the reflectometry setup.
This work was supported by IARPA-MQCO, LPS-MPO-CMTC, the Polish National Science Centre (NCN) under Grant No.~DEC-2012/07/B/ST3/03616, the Army Research Office, the Villum Foundation and the Danish National Research Foundation. 

\section*{Author contributions}
S.F., G.C.G. and M.J.M. grew the heterostructure. P.D.N. fabricated the device. F.M., P.D.N., F.K. and F.K.M. prepared the experimental setup. F.K.M., F.M. and F.K. performed the experiment. E.B., \L.C. and M.S.R. developed the theoretical model and performed the simulations. F.K.M., F.K., F.M., E.B., \L.C., M.S.R. and C.M.M. analysed data and prepared the manuscript.

\chapterimage{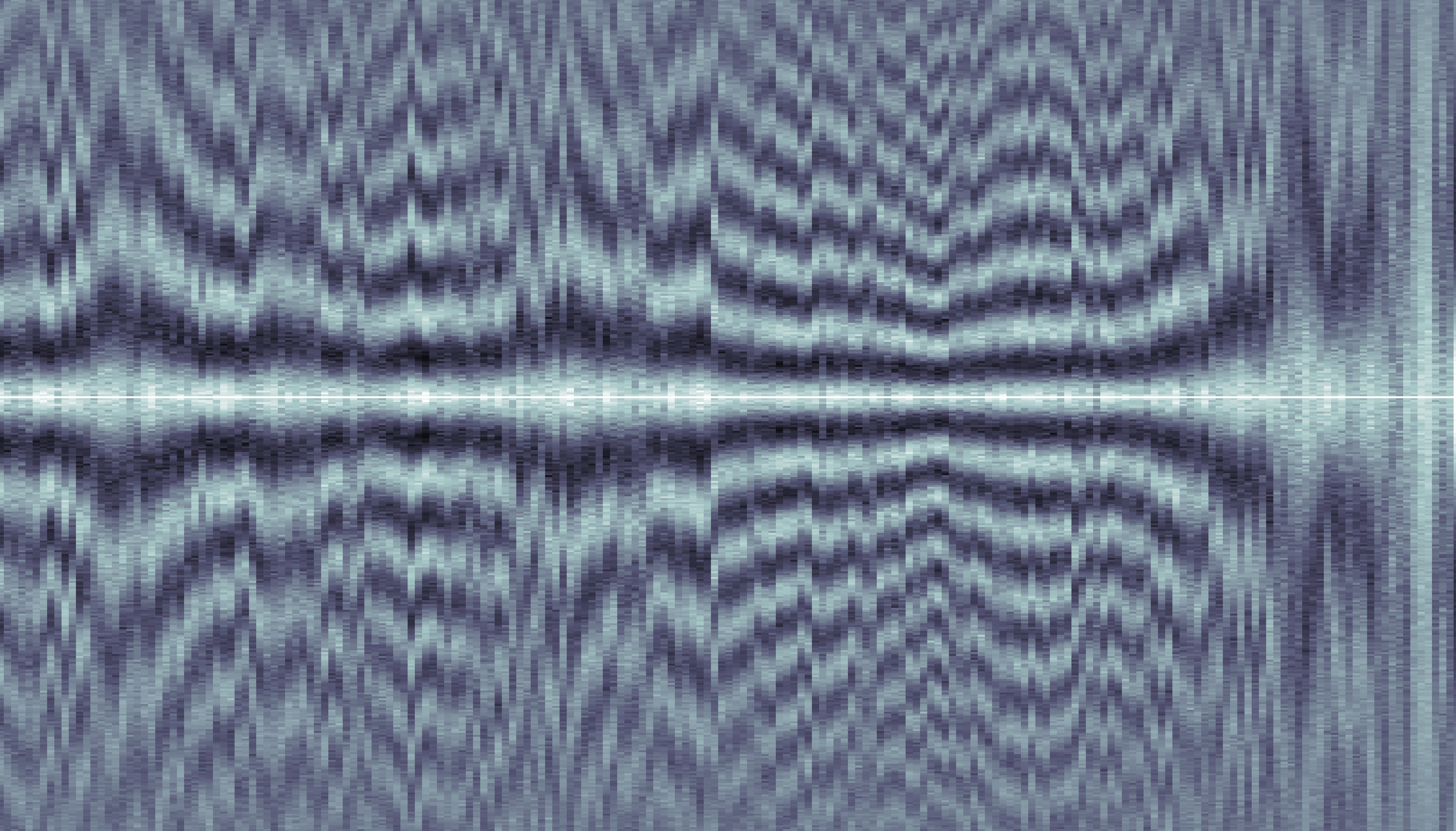}
\chapter[Supplementary information for ``Spectrum of the nuclear environment for GaAs spin qubits'']{\protect\parbox{0.9\textwidth}{Supplementary information for \\ ``Spectrum of the nuclear \\ environment for GaAs spin qubits''}}
\chaptermark{Supplementary Information for ``Spectrum of the nuclear...''}
\label{ch:ovh_sup}

\newcommand{\MatEl}[3]{\langle \, #1 \,\vert\,#2\,\vert\,#3\,\rangle}
\newcommand{\Amp}[2]{\langle \, #1\, \vert\,  #2 \, \rangle}
\newcommand{\mpar}[1]{\marginpar{\small \it #1}}
\newcommand{\Avg}[1]{\langle  #1  \rangle}
\newcommand{\la}{\langle}
\newcommand{\ra}{\rangle}
\newcommand{\lb}{\left[}
\newcommand{\rb}{\right]}
\newcommand{\lp}{\left(}
\newcommand{\rp}{\right)}
\newcommand{\E}{{\cal E}}
\newcommand{\HH}{{\cal H}}
\newcommand{\LL}{{\cal L}}
\newcommand{\tr}{{\rm tr}\,}
\newcommand{\p}{\partial}

\begin{center}	
\begin{tcolorbox}[width=0.8\textwidth, breakable, size=minimal, colback=white]
	\small This supplementary information discusses the following topics:
	\begin{enumerate}
		\item[\ref{ovh_sup:extracting}] Extracting the frequency of oscillations in Overhauser field from averaged and single-shot data
		\item[\ref{ovh_sup:distribution}] Gaussian distribution of $\Delta B_\parallel$
		\item[\ref{ovh_sup:autocorrelation}] Obtaining power spectral density of $P_S$ from truncated autocorrelation of single-shot measurements
		\item[\ref{ovh_sup:fitting}] Fitting procedures for PSD in Figs. \ref{ovh:fig2}c and \ref{ovh:fig3}
		\item[\ref{ovh_sup:classical}] Classical model of Overhauser field noise due to nuclear spin diffusion
		\item[\ref{ovh_sup:transverse}] Decoherence of the qubit subjected to the transverse Overhauser noise
	\end{enumerate}
\end{tcolorbox}
\end{center}

\let\mysectionmark\sectionmark
\renewcommand\sectionmark[1]{}
\section{Extracting the frequency of oscillations in Overhauser field from averaged and single-shot data}
\label{ovh_sup:extracting}
\let\sectionmark\mysectionmark
\sectionmark{Extracting the frequency of oscillations...}

\begin{figure}[ht]
	\centering
	\includegraphics[width=\textwidth]{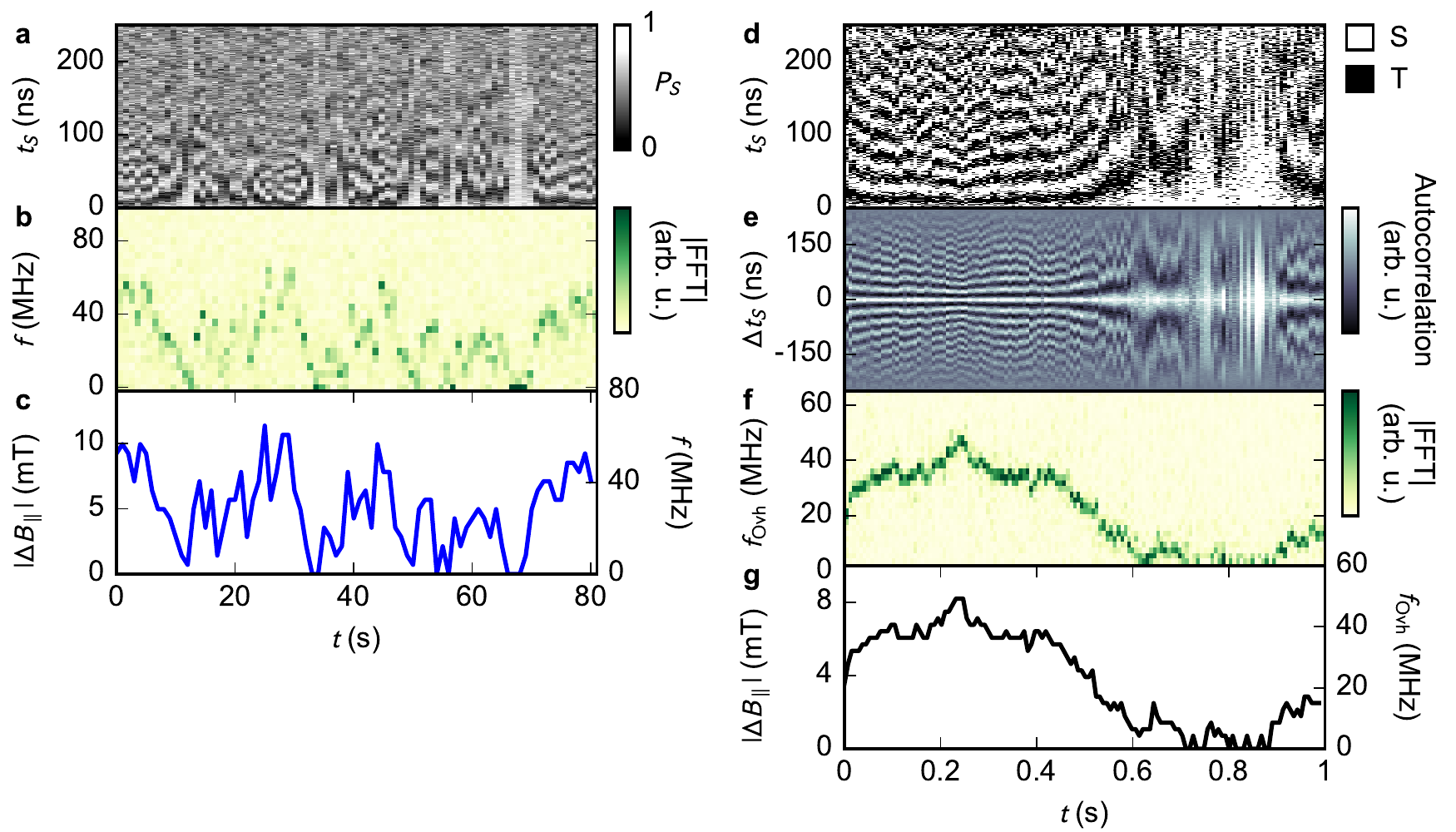}
	\caption[Intermediate steps of $\Delta B_z$ calculation from raw data]
	{Intermediate steps of extracting $\Delta B_\parallel$ from raw data.
	(a) Raw data is taken by repeating a 300-point sequence with $t_S$ from 0 to 250~ns 16 times, allowing us to estimate the probability $P_S(t_S)$. Such data forms a single column of the presented map. (b) Absolute value of FFT of probability oscillations. (c) Extracted position of the peak, $f$, and corresponding $|\Delta B_\parallel|$.
	(d) Raw single-shot data (one sequence run per column) and (e) its autocorrelation. Color scale in autocorrelation 2d map is chosen to show oscillations and hide the peak at $\Delta t_S = 0$. (f) Absolute value of FFT of autocorrelation. (g) Extracted position of the peak and corresponding $|\Delta B_\parallel|$.
	}
	\label{ovh_sup:figS1}
\end{figure}

The power spectral density of the gradient of Overhauser field squared $(\Delta B_\parallel)^2$ is obtained from two kinds of data sets. 

The first one consists of oscillations in the singlet probability $P_S$ as a function of electron separation time $t_S$, varying from 0 to 250~ns, measured with 1~s repetition rate. 
For each studied value of magnetic field, we measure a 1.5~hour-long data set, and a fragment of such a data set is presented in Fig.~\ref{ovh_sup:figS1}a. The frequencies of the oscillations are obtained by means of Fourier analysis. We calculate the Fast Fourier Transform (FFT) of each vertical column and inspect its absolute value (Fig.~\ref{ovh_sup:figS1}b). The position of the maximum indicates the frequency of oscillations, which is related to the gradient of the Overhauser field between the dots by $hf_\mathrm{Ovh} = g\mu |\Delta B_\parallel|$ (Fig.~\ref{ovh_sup:figS1}c).

The second data set consists of ten non-averaged measurements, each 30~s long.
A 1~s excerpt of one of them is shown in Fig.~\ref{ovh_sup:figS1}d. 
Extracting the underlying oscillation frequency from each column requires more careful treatment, since the probabilistic nature of binary measurements adds large amounts of shot noise. 
In our analysis we first assign S and T$_0$ outcomes to, respectively, $1$ and $-1$. 
Then we subtract from each column its mean, and calculate the autocorrelation (Fig.~\ref{ovh_sup:figS1}e).  
The obtained autocorrelation reveals oscillations at the same frequency as unprocessed data. However, in the autocorrelated data the shot noise is averaged out for all $\Delta t_S$ except for 0, where the shot noise accumulates. Next we replace the autocorrelation value at $\Delta t_S = 0$ with the value at the smallest $|\Delta t_S|\neq 0$. This minimizes the influence of shot noise without affecting the visibility of the oscillations. The absolute value of the FFT (Fig.~\ref{ovh_sup:figS1}f) of the autocorrelations processed in such way exhibits a clear peak, which we associate with the qubit oscillation frequency.  
Namely, the peak position in frequency, $f_\mathrm{Ovh}$, is used to extract $|\Delta B_\parallel|$ via $hf_\mathrm{Ovh} = g\mu |\Delta B_\parallel|$ (Fig.~\ref{ovh_sup:figS1}g).

\section{Gaussian distribution of $\Delta B_\parallel$}
\label{ovh_sup:distribution}

\begin{figure}[ht]
	\centering
	\includegraphics[scale=0.9]{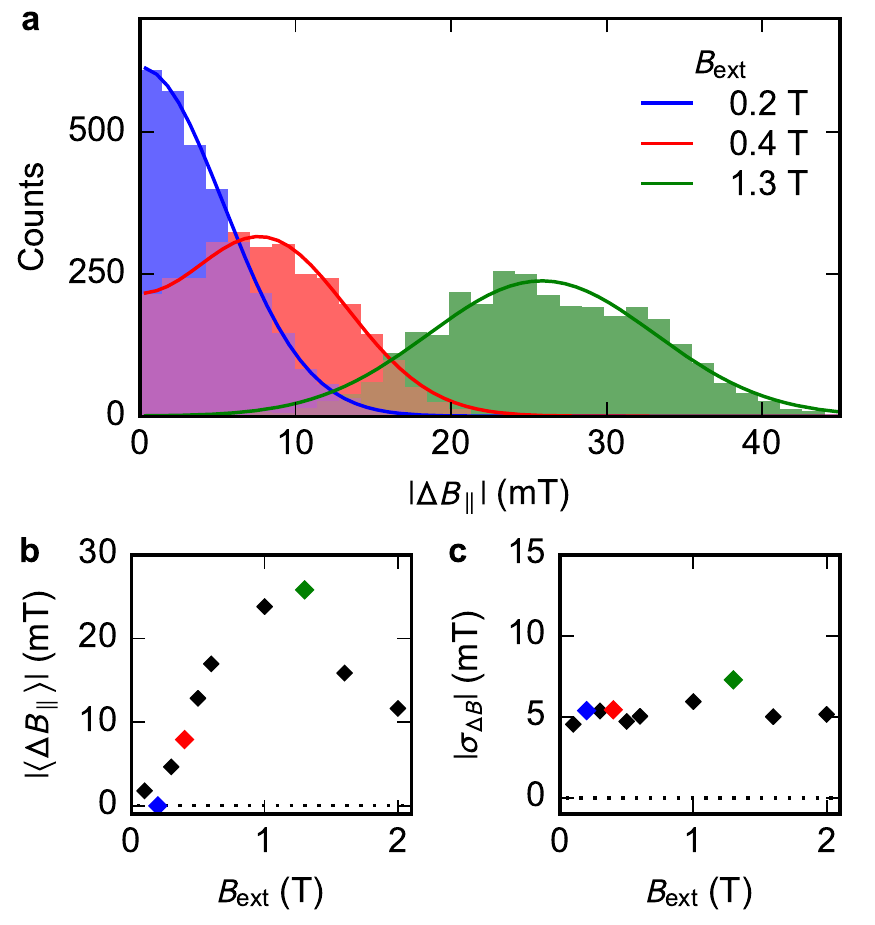}
	\caption[Measured distribution of $|\Delta B_\parallel|$ for various applied magnetic fields]
	{(a) Measured distribution of $|\Delta B_\parallel|$ for various applied magnetic fields $\Bext$. Solid lines are fits assuming Gaussian distributed $\Delta B_\parallel$ with mean $|\langle \Delta B_\parallel\rangle |$.
	(b,c) Fitted mean Overhauser field gradient $|\langle \Delta B_\parallel\rangle |$ and the Overhauser field gradient distribution width $\sigma_{\Delta B}$ as a function of applied magnetic field $\Bext$. Colored points correspond to $|\Delta B_\parallel|$ distributions presented in panel (a).}
	\label{ovh_sup:figS2}
\end{figure}

Several points of the analysis presented in the main text assume a Gaussian distribution of the Overhauser field gradients. To show that this assumption is justified we plot in Fig.~\ref{ovh_sup:figS2}a histograms of $|\Delta B_\parallel|$ for several values of the external magnetic field. The fits to the data confirm our assumption, indicating there spin bath does not have multiple stable points, in contract to experiments involving intentional dynamical nuclear polarization~\cite{Danon2009,Bluhm2010,Forster2015}. However we observe that the mean Overhauser field gradient is shfted away from zero for increasing external magnetic field, while the width of the $\Delta B_\parallel$ distribution remains unchanged (Fig.~\ref{ovh_sup:figS2}b,c). We suspect that this non-zero mean arises from unintentional nuclear polarization, as reported previously for this device~\cite{Martins2016,Malinowski2017} and for devices studied by other groups~\cite{Botzem2016}.

\let\mysectionmark\sectionmark
\renewcommand\sectionmark[1]{}
\section{Obtaining power spectral density of $P_S$ from truncated autocorrelation of single-shot measurements}
\label{ovh_sup:autocorrelation}
\let\sectionmark\mysectionmark
\sectionmark{Obtaining power spectral density of $P_S$...}

\begin{figure}[tb]
	\centering
	\includegraphics[width=\textwidth]{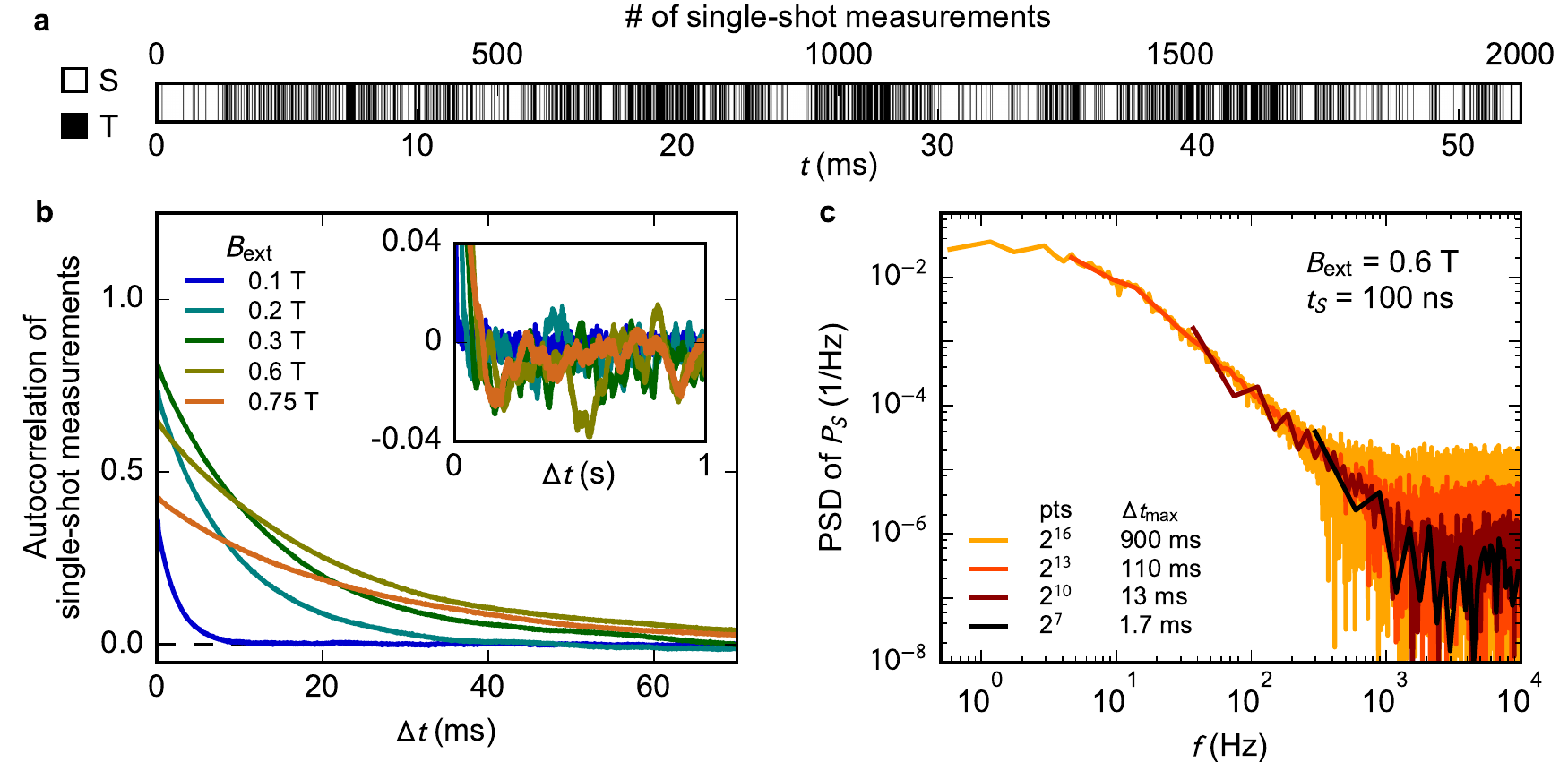}
	\caption[Calculating $P_S$ using truncated autocorrelation traces]
	{(a) Example trace of single-shot measurements obtained from pulses with $t_S=100$~ns at 0.6~T.
	(b) Autocorrelation of single-shot traces for $t_S=100$~ns. Inset shows long-time tail with oscillations caused by finite size of the sample.
	(c) Power spectral density of $P_S$ obtained by FFT of the autocorrelation truncated at $\pm \Delta t_\mathrm{max}$.}
	\label{ovh_sup:figS3}
\end{figure}

\begin{figure}[tb]
	\centering
	\includegraphics[scale=.9]{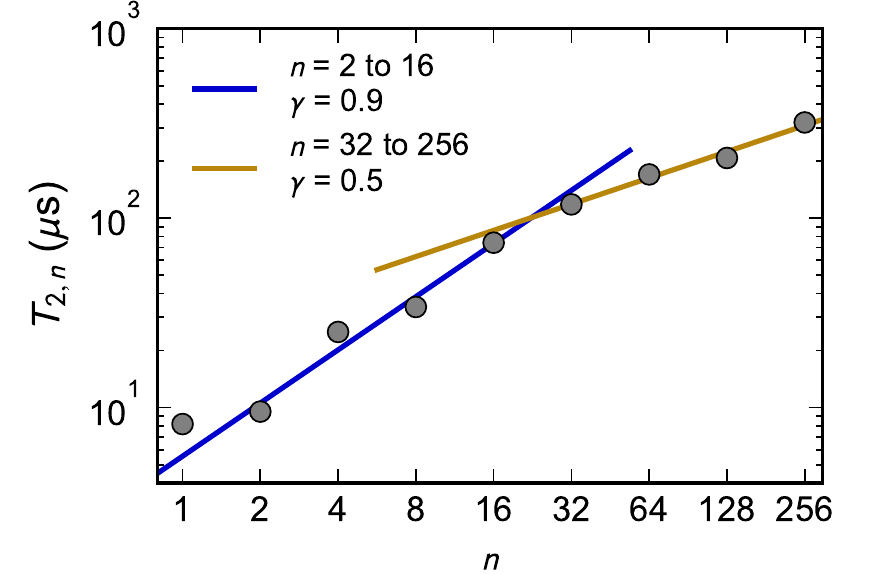}
	\caption[Scaling of the extracted coherence decay envelope $\Tn$ with $n$ at external magnetic field $B=0.75$~T]
	{Scaling of the extracted coherence decay envelope $\Tn$ with $n$ at external magnetic field $B=0.75$~T. Solid blue and yellow lines indicate fits of the power law $\propto n^\gamma$ to data in the indicated range.}
	\label{ovh_sup:figS4}
\end{figure}

As explained in the main text, to maximize the repetition rate at which $\Delta B_\parallel$ is probed, we fix the separation time $t_S=100$~ns and repeat the pulse cycle continuously. Then we map S and T$_0$ outcomes to, respectively, $1$ and $-1$. As a result we obtain binary traces of over 2 million points. A small piece of such a trace is presented in Fig.~\ref{ovh_sup:figS3}a, obtained for external magnetic field $\Bext = 0.6$~T and $t_S=100$~ns.

The sequence of single-shot outcomes is dominated by shot noise, which obscures the underlying oscillations when using conventional methods for calculating the PSD. To eliminate this noise contribution, we apply the same procedure mentioned in the previous section. That is, we find the autocorrelation and replace its value at $\Delta t=0$ with the value at the smallest $|\Delta t|\neq0$. The autocorrelation of single-shot measurements for $\Bext = 0.6$~T are plotted in Fig.~\ref{ovh_sup:figS3}b. Now we can take advantage of the fact that the Fourier transform of the autocorrelation is identical to the power spectral density of the original trace.

Even though the sample is huge, we observe artifacts related to its finite size. Namely, the autocorrelation has a long, irregular tail (Fig.~\ref{ovh_sup:figS3}b, inset). If we perform FFT over the entire available range of $\Delta t$, fluctuations in the tail dominate over the relevant features at $\Delta t \sim 1/f$.

In our further analysis we assume that relevant information about the nuclear noise at frequency $f$ is contained within the window $|\Delta t| \lesssim A/f = \Delta t_\mathrm{max}$ where $A$ is of the order of ten. In other words, to obtain an accurate value of the power spectral density of the noise at frequency $f$, it is sufficient to take the Fourier transform of the autocorrelation in the range $-\Delta t_\mathrm{max}<\Delta t<\Delta t_\mathrm{max}$. In our analysis we use $16\leq A\leq32$. 
To avoid the necessity of windowing we keep the range of $\Delta t$ such that the number of points within the $-\Delta t_\mathrm{max}<\Delta t<\Delta t_\mathrm{max}$ range is $2^n$, for integer $n$.

Power spectral densities for $\Bext = 0.6$~T, $t_S=100$~ns, and various choices of $\Delta t_\mathrm{max}$ are presented in Fig.~\ref{ovh_sup:figS3}c, visualizing the trade-off between frequency range and noise floor level due to finite sample size. 
Wide $\Delta t$ windows (i.e. large $t_\mathrm{max}$) give access to lower frequencies but raise the noise floor, while narrow $\Delta t$ windows sacrifice low-frequency information for a reduction in noise. By adjusting the window dynamically we are able to achieve a wide spectral range without suffering from the background noise.

\section{Fitting procedures for PSD in~Figs.~\ref{ovh:fig2}c and \ref{ovh:fig3}}
\label{ovh_sup:fitting}

The classical diffusion model used to describe the experimental results presented in Figs.~\ref{ovh:fig2}c~and~\ref{ovh:fig3}, provides analytical expressions for the autocorrelation of $(\Delta B_\parallel)^2$ and $P_S$, but no analytical expression for the PSD. Therefore the fitted expressions involve numerical Fourier transforms of the autocorrelation obtained from the analytical formulas.

To fit data in Fig.~\ref{ovh:fig2}c we simulate two sets of autocorrelation traces, such that after performing a FFT they produce PSD points at the same frequencies as in the experimental data. Simulated traces obtained in such way are suitable for optimization via the method of least squares.

To fit the data in Fig.~\ref{ovh:fig3} we simulate autocorrelation traces with identical time resolution as the experimental data. To these traces we then apply the same procedure as to experimental data (that is we perform FFT of data limited to a suitable range of $\Delta t$). The obtained set of frequencies is identical to those of the experimental PSD, making this method suitable for least squares fitting.

\section{Classical model of Overhauser field noise due to nuclear spin diffusion}
\label{ovh_sup:classical}

Here we construct a model for the dynamics of a $S$-$T_0$ qubit in a double quantum dot, arising from slow fluctuations of the longitudinal Overhauser (difference) field in the two dots.
The Overhauser field is produced by nuclear spins in the host crystal, which undergo their own dynamics due to their mutual dipole-dipole coupling.
These dynamics lead to fluctuations of the longitudinal Overhauser field, which in turn affect the evolution of the spin qubit, which plays the role of a sensor in this work. 
Our approach is similar to that employed previously to describe the results of the experiment in Ref.~\cite{Reilly2008}.

The electronic system is influenced by the nuclear spins through the hyperfine (hf) interaction,
\begin{equation}
H_{\rm HF} = A_0 \sum_j \delta(\hat{\vec{r}} - \vec{R}_j)\, \hat{\vec{S}} \cdot \hat{\vec{I}}_j,
\end{equation}
where $\hat{\vec{r}}$ is the electron position operator, $\vec{R}_j$ is the position of nucleus $j$, and $\hat{\vec{S}}$ and $\hat{\vec{I}}_j$ are the spin operators for the electron and nucleus $j$, respectively.
Here $A_0$ has units $[{\rm Energy\ }\times\ {\rm Volume}/\hbar^2]$, with a characteristic value in GaAs of $\hbar^2A_0/v_0 \approx 100\ \mu {\rm eV}$ \cite{Cywinski2009a}, where $v_0$ is the unit cell volume.
For simplicity we consider a single nuclear species.
On a coarse-grained scale encompassing many atomic sites, we describe the nuclear spin state in terms of a spin density field $I(\vec{x},t)$.
Throughout this treatment we focus only on the spin component parallel to the externally-applied magnetic field.
Here $I(\vec{x},t)$ has units $[\hbar/{\rm Volume}]$.

In a double dot, the $(1,1)$ singlet and triplet ($T_0$) states are coupled by the longitudinal Overhauser difference field
\begin{equation}
  \Delta B_z(t) = \frac{\hbar A_0}{g_*\mu_{\rm B}} \int d^3x\, \Delta\rho(\vec{x})  I(\vec{x},t),  
\end{equation}
where $g_*$ is the electronic effective $g$-factor, $\mu_{\rm B}$ is the Bohr magneton, and $\Delta\rho(\vec{x}) =|\psi_R(\vec{x})|^2 - |\psi_L(\vec{x})|^2$, with $|\psi_R(\vec{x})|^2$ and $|\psi_L(\vec{x})|^2$ the electronic density profiles in the right and left dot, respectively.
Later it will be convenient to work in Fourier space,
\begin{equation}
 \label{eq:dBzFourier} \Delta B_z(t) = \frac{\hbar A_0}{g_*\mu_{\rm B}} \int \frac{d^3q}{(2\pi)^3}\, \Delta\tilde\rho_{\vec{q}}\,  \tilde I_{-\vec{q}}(t),
\end{equation}
where $\tilde{f}_{\vec{q}} = \int d^3x\, e^{-i\vec{q}\cdot\vec{x}} f(\vec{x})$.
Notably, although the nuclear spin field $I(\vec{x},t)$ extends throughout the entire sample, the Overhauser field is only sensitive to the value of $I(\vec{x},t)$ in a limited region where the electrons are localized.

For simplicity, we take a model where the nuclear spin field $I(\vec{x},t)$ evolves under its own dynamics, unperturbed by the presence of the electronic system.
In the absence of nuclear spin relaxation (i.e., for infinite nuclear $T_1$), the dipolar interaction between nuclear spins leads to a diffusive-type dynamics of nuclear spin polarization:
\begin{equation}
\label{eq:Diff_realspace}\partial_t I(\vec{x},t) = D \nabla^2 I(\vec{x},t) + \xi(\vec{x},t),
\end{equation}
where $\xi(\vec{x},t)$ is a stochastic field 
that accounts for the randomness of dipole-dipole induced nuclear spin flips.
The units of $\xi(\vec{x},t)$ are [Energy / Volume].
Such a diffusive model is also expected to at least qualitatively describe the dynamics caused by electron-mediated nuclear flip-flops~\cite{Gong2011}.

The smooth diffusive dynamics, as described by the first term in Eq.~\eqref{eq:Diff_realspace}, are only manifested on timescales longer than that for a single nuclear spin flip due to its interaction with its neighbors (typically $\sim 10-100\ \mu$s for GaAs \cite{Cywinski2009a}).  
On times longer than this scale, where the diffusion model applies, the noise has zero average, $\Avg{\xi(t)} = 0$, and is essentially white: $\Avg{\xi(t)\xi(t')} \sim \delta(t-t')$.
Here the angle brackets indicate averaging over noise realizations.
The conservation of total nuclear spin is ensured by taking the noise to have the following spatial correlations on scales larger than the atomic lattice spacing:
\begin{equation}
\label{eq:Xi_stats}\Avg{\xi(\vec{x},t)\xi(\vec{x}',t')} = -\eta D\nabla^2\delta(\vec{x}-\vec{x}')\delta(t- t'), 
\end{equation}
where the proportionality constant $\eta$ will be fixed below to ensure the correct RMS value of the Overhauser difference field in equilibrium.
The units of $\eta$ are $[\hbar^2/\,{\rm Volume}]$.

Fourier transforming Eq.~\eqref{eq:Diff_realspace}, we obtain an independent differential equation for each nuclear spin mode $\tilde{I}_{\vec{q}}$, labeled by the 3D wave vector $\vec{q}$:
\begin{equation}
\label{eq:Diff_fourierspace} \partial_t \tilde{I}_{\vec{q}}(t) = -Dq^2 \tilde{I}_{\vec{q}} + \tilde{\xi}_{\vec{q}}(t).
\end{equation}
The Fourier modes $\tilde{\xi}_{\vec{q}}(t)$ of the noise field satisfy
\begin{equation}
\label{eq:Xi_stats_q} \Avg{\tilde\xi_{\vec{q}}(t)\tilde\xi_{\vec{q}'}(t')} =  \eta Dq^2 (2\pi)^3   \delta(\vec{q}+\vec{q}')\delta(t- t').
\end{equation}

The differential equation \eqref{eq:Diff_fourierspace} has the formal solution 
\begin{equation}
\tilde{I}_{\vec{q}}(t) = \tilde{I}_{\vec{q}}(t_0) e^{-Dq^2 (t-t_0)} + \int_{t_0}^t dt'\, e^{-Dq^2(t-t')}\tilde{\xi}_{\vec{q}}(t').
\end{equation}
Using the explicit form for $\tilde{I}_{\vec{q}}(t)$ above, along with Eq.~\eqref{eq:Xi_stats_q}, we obtain the correlation functions for the nuclear spin field:
\begin{equation}
 \Avg{\tilde{I}_{\vec{q}}(t)\tilde{I}_{\vec{q}'}(t')}
 = e^{-Dq^2 (t + t'- 2t_0)}\left[\Avg{\tilde{I}_{\vec{q}}(t_0)\tilde{I}_{\vec{q}'}(t_0)} - \eta (2\pi)^3\delta(\vec{q} + \vec{q}')\right] + \eta (2\pi)^3\delta(\vec{q} + \vec{q}')e^{-Dq^2|t -t'|},
\end{equation}
where in the second line we have used Eq.~\eqref{eq:Xi_stats_q}.
If the initial state $I(t_0)$ is drawn from the (stationary) equilibrium distribution, $\Avg{\tilde{I}_{\vec{q}}(t_0)\tilde{I}_{\vec{q}'}(t_0)} = \Avg{\tilde{I}_{\vec{q}}(t)\tilde{I}_{\vec{q}'}(t)}$, then we find
\begin{equation}
\label{eq:I_corr}\Avg{\tilde{I}_{\vec{q}}(t)\tilde{I}_{\vec{q}'}(t')} = \eta (2\pi)^3 \delta(\vec{q} + \vec{q}')e^{-Dq^2|t -t'|}.
\end{equation}
In position space, the equilibrium Overhauser field fluctuations are thus uncorrelated:
\begin{equation}
\label{eq:I_corr_x}\Avg{I(\vec{x},t)I(\vec{x}',t)} = \eta\, \delta(\vec{x}-\vec{x}').
\end{equation}

We now use the results above for the correlation function of the nuclear spin field to calculate the noise correlations of the Overhauser difference field, $\Delta B_z(t)$.
The correlation function $\Avg{\Delta B_z(t) \Delta B_z(t')}$ is straightforward to evaluate using Eqs.~\eqref{eq:dBzFourier} and \eqref{eq:I_corr}:
\begin{equation}
\Avg{\Delta B_z(t) \Delta B_z(t')} 
\label{eq:dBzCorr}= \eta \frac{\hbar^2 A^2_0}{(g_*\mu_{\rm B})^2} \int \frac{d^3q}{(2\pi)^3}\, |\Delta\tilde\rho_{\vec{q}}|^2\,e^{-Dq^2|t -t'|},
\end{equation}
where we have used $\tilde{\rho}_{-\vec{q}} = \tilde{\rho}^*_{\vec{q}}$.

We set the value of $\eta$ by demanding that the equilibrium RMS Overhauser field fluctuations should match the measured value:
\begin{equation}
 \label{eq:dB2Eq} \sigma^2_{\Delta B} \equiv \Avg{\Delta B^2_z}_{\rm eq} = \eta \frac{\hbar^2 A^2_0}{(g_*\mu_{\rm B})^2} \int \frac{d^3q}{(2\pi)^3}\, |\Delta\tilde\rho_{\vec{q}}|^2.
\end{equation}
To evaluate $|\Delta \tilde{\rho}_{\vec{q}}|^2$ in the integrand above, we must specify a particular form for the electron density profiles in the two dots.
For simplicity we take the densities in the two dots to be Gaussian, centered at positions $\vec{x}_L = (x_L, 0, 0)$ and $\vec{x}_R = (x_R, 0, 0)$:
\begin{equation}
\label{eq:wavefunction}|\psi_\alpha(\vec{x})|^2 = \frac{e^{-[(x-x_\alpha)^2 + y^2]/(2\sigma_\perp^2)}}{2\pi\sigma_\perp^2}\frac{e^{-z^2/(2\sigma_z^2)}}{\sqrt{2\pi\sigma^2_z}},
\end{equation}
where $\alpha = L,R$. 
Setting $x_L = -d/2,\ x_R = d/2$ and taking the Fourier transform of the electron density in Eq.~\eqref{eq:wavefunction}, we obtain
\begin{equation}
|\Delta\tilde\rho_{\vec{q}}|^2 = 4\sin^2 (q_x d/2)\,e^{- (\sigma_z^2q_z^2\, +\, \sigma_\perp^2 q_x^2\, +\, \sigma_\perp^2 q_y^2)}.
\end{equation}

Rewriting $4\sin^2 (q_x d/2) = 2(1 - \cos q_xd)$ and substituting into Eq.~\eqref{eq:dB2Eq} gives
\begin{eqnarray}
 \Avg{\Delta B^2_z}_{\rm eq}
&=&  \frac{\hbar^2 A^2_0}{(g_*\mu_{\rm B})^2}\frac{\eta}{4\pi^{3/2} \sigma_z\sigma_\perp^2}\left(1 - e^{-d^2/4\sigma_\perp^2}\right).
\end{eqnarray}
To cast the result into a more convenient form, we define the effective number of spins $N_\alpha$ in dot $\alpha = L,R$ via $N_\alpha^{-1} = v_0\int d^3x\, |\psi_\alpha(\vec{x})|^4$, where $v_0$ is the unit cell volume (see above).
For the wave functions in Eq.~\eqref{eq:wavefunction} we have $N_\alpha = 8\pi^{3/2}\sigma_z\sigma^2_\perp/v_0 \equiv N$.
Letting $E_N = (g_*\mu_{\rm B})\sqrt{\Avg{\Delta B^2_z}_{\rm eq}}$, we have
\begin{equation}
\label{eq:Eq_Fluctuations}
E^2_N = \frac{2}{N}\cdot\frac{v_0 \eta}{\hbar^2}\cdot\left(\frac{\hbar^2 A_0}{v_0}\right)^2\cdot\left(1 - e^{-d^2/4\sigma_\perp^2}\right).
\end{equation}
In this form, the $\sqrt{N}$ dependence of the RMS Overhauser field fluctuations is explicitly displayed. 

The oscillations observed in Fig.~\ref{ovh:fig2} of the main text reveal the {\it magnitude} of the gradient, but do not yield information about its {\it sign}.
Therefore the correlation function in Eq.~\eqref{eq:dBzCorr}, which depends on both the magnitude and sign of $\Delta B(t)$, is not of direct relevance.
Instead, we compute the experimentally-relevant noise correlations and power spectrum for $\Delta B^2_z(t)$. 
Assuming that the noise is Gaussian, which is justified by the fact that the continuous nuclear spin field $I(\vec{x},t)$ is produced by a large density of randomly polarized individual spins, the fourth-order correlators that appear in the expression for $\Avg{\Delta B^2_z(t)\Delta B^2_z(t')}_c = \Avg{\Delta B^2_z(t)\Delta B^2_z(t')}_c - \Avg{\Delta B^2_z(t)}^2$ can be factorized.
This gives:
\begin{equation}
\label{eq:dBz2Corr2} \Avg{\Delta B^2_z(t) \Delta B^2_z(t')}_c 
 =  \frac{2\eta^2\hbar^4 A^4_0}{(g_*\mu_{\rm B})^4} 
\left[\int\!\! \frac{d^3q}{(2\pi)^3}|\Delta\tilde\rho_{\vec{q}}|^2 e^{-Dq^2|t - t'|}\right]^2\!\!\!.
\end{equation}

The integral in Eq.~\eqref{eq:dBz2Corr2} is very similar to the one evaluated above to compute the RMS nuclear field.
Again using the electronic density profile, Eq.~\eqref{eq:wavefunction}, we get
\begin{equation}
\label{eq:CorrelationInt}
C(t-t') \equiv \int \frac{d^3q}{(2\pi)^3}|\Delta\tilde\rho_{\vec{q}}|^2 e^{-Dq^2|t - t'|} \ =\ 
\frac{1 - e^{-\frac14 d^2/(\sigma_\perp^2  + D|t-t'|)}}{4\pi^{3/2}(D|t-t'| + \sigma_\perp^2)\sqrt{D|t-t'| + \sigma_z^2}}.
\end{equation}
Substituting back into Eq.~\eqref{eq:dBz2Corr2}, we get
\begin{equation}
\label{eq:AutoCorrdB2}
\Avg{\Delta B^2_z(t) \Delta B^2_z(t')}_c = \frac{\eta^2}{8\pi^3} \frac{\hbar^4 A^4_0}{(g_*\mu_{\rm B})^4}\frac{\left(1 - e^{-\frac14 d^2/(\sigma_\perp^2  + D|t-t'|)}\right)^2}{(D|t-t'| + \sigma_\perp^2)^2\,(D|t-t'| + \sigma_z^2)}. 
\end{equation}

The autocorrelation function in Eq.~\eqref{eq:AutoCorrdB2} above was used for fitting the experimentally obtained power spectral densities for $(\Delta B_\parallel)^2$ in Fig.~\ref{ovh:fig2}c.
The geometric parameters were taken from the lithographic dimensions of the device, and known growth parameters of the heterostructure: $d = 150$ nm, $\sigma_\perp = 40$ nm, and $\sigma_z = 7.5$ nm.
The diffusion constant $D$ and the equilibrium nuclear field fluctuations $E_N = g^*\mu_B\sigma_{\Delta B}$, see Eqs.~\eqref{eq:dB2Eq} and \eqref{eq:Eq_Fluctuations}, were taken as fit parameters.
The extracted values were $D = 33$ nm$^2/s$ and $\sigma_{\Delta B} = 6.0$ mT. 

\subsection{Correlations for fixed separation time}

In Fig.~\ref{ovh:fig3} of the main text, power spectral densities for measurements of the singlet return probability with fixed separation time are shown.
This type of measurement was employed previously by Reilly and coworkers \cite{Reilly2008}.
For separation time $t_S$, the singlet return probability $P_S$ is given by
\begin{equation}
\label{eq:PSDef}P_S(t) = \frac12 {\rm Re}\left[1 + e^{i(g_*\mu_B/\hbar)\Delta B_z(t)t_S}\right].
\end{equation}
Here we assume that $B_z(t)$ is frozen on the timescale of one experiment, but its value may change from run to run.
Averaging over Gaussian fluctuations gives
\begin{equation}
\Avg{P_S(t)} = \frac12\left[1 + e^{-\frac12 (g_*\mu_B/\hbar)^2\Avg{\Delta B_z^2}t^2_S}\right].
\end{equation}

Using Eq.~\eqref{eq:PSDef}, the autocorrelation function $\Avg{P_S(t + \Delta t)P_S(t)} - \Avg{P_S(t)}^2$ then follows (as also found in Ref.~\cite{Reilly2008}):
\begin{equation}
\Avg{P_S(t + \Delta t)P_S(t)} - \Avg{P_S(t)}^2 = \frac14 e^{-(g_*\mu_B/\hbar)^2\Avg{\Delta B^2_z}t_S^2}\left[\cosh \left( (g_*\mu_B/\hbar)^2\Avg{\Delta B_z(t + \Delta t) B_z(t)}t_S^2 \right) - 1 \right].
\end{equation}
The quantity $\Avg{\Delta B_z^2}$ in the exponent was calculated above, see Eq.~\eqref{eq:dB2Eq} and below.
Furthermore, the correlation function $C(t - t') = \frac{(g_*\mu_B)^2}{\eta\hbar^2A_0^2} \Avg{\Delta B_z(t + \Delta t) B_z(t)}$ was calculated in Eq.~\eqref{eq:CorrelationInt}.
Thus we obtain the autocorrelation function for the singlet return probability in experiments with fixed separation time, used for fitting the data in Fig.~\ref{ovh:fig3} of the main text. 

\let\mysectionmark\sectionmark
\renewcommand\sectionmark[1]{}
\section{Decoherence of the qubit subjected to the transverse Overhauser noise}
\label{ovh_sup:transverse}
\let\sectionmark\mysectionmark
\sectionmark{Decoherence of the qubit subjected to...}

\begin{equation}
\hat{H}_{\mathrm{SN}} = \sum_{i=L,R}\left( \hat{h}^{i}_{z} + \frac{(\hat{h}^{i}_{x})^2 + (\hat{h}^{i}_{y})^2}{\Omega} \right )  \hat{S}^i_z \,\, \label{eq:HSN}
\end{equation}
where $\hat{h}^{i}_{a}$ with $a \! =\! x$, $y$, and $z$ are the operators of Overhauser field components, given by
$\hat{h}^{i}_{a} = \sum_{k} A^{i}_{k} \hat{I}^{a}_{k}$,
in which $\hat{I}^{a}_{k}$ are the spin operators of $k$-th nucleus and  $A^{i}_{k} = \mathcal{A}_{\alpha[k]} |\Psi_{i}(\mathbf{r}_{k})|^2$ (with $\Psi_{i}(\mathbf{r})$ being the envelope wavefunction of the electron in dot $i$, $\mathcal{A}_{\alpha[k]}$ the hyperfine interaction energy for nucleus of species $\alpha$, and $\mathbf{r}_{k}$ the position of $k$-th nucleus) are the hyperfine couplings of the $k$-th nucleus to the electron in dot $i$.

When the exchange interaction between the two electrons is strongly suppressed due to a large barrier height separating the two potential minima \cite{Martins2016}, the overlap between the $\Psi_{L}(\mathbf{r})$ and $\Psi_{R}(\mathbf{r})$ functions is negligible, and every contributing nucleus is coupled only to one electron, residing either in $L$ or $R$ dot. The Hamiltonian \eqref{eq:HSN} is then a sum of two commuting terms, each pertaining to another dot. We also assume that the nuclear density matrices in the two dots are uncorrelated, and that the $L$ and $R$ dots have the same size and shape. 
The singlet return probability $P_{S}(T)$ is then given by
\begin{equation}
P_{S}(T) = \frac{1}{2} + \frac{1}{2}|W(T)|^2 \,\, ,  \label{eq:PSW}
\end{equation}
where $W(T)$ is the coherence function of a single spin in one of the QDs (i.e.~an off-diagonal element of its density matrix normalized to unity), calculated for the respective dynamical decoupling sequence. 

Although the longitudinal and transverse Overhauser field operators do not strictly commute, their commutator is $\! \sim \sigma/N$, where $\sigma$ is the rms of the Overhauser field and $N$ is the number of nuclei appreciably interacting with the electron \cite{Neder2011}. In the following we use a semiclassical approach to dynamics of large nuclear bath, and neglect this commutator. The decoherence function can then be written as
\begin{equation}
W(T) \approx W_{z}(T)W_{\perp}(T) \,\, ,
\end{equation}
in which $W_{z}(T)$ is the contribution to decoherence that originates from the first (longitudinal) term in Eq.~\eqref{eq:HSN}, while $W_{\perp}(T)$ is the contribution due to the second term (quadratic in transverse Overhauser operators). 

The Hamiltonian of the nuclei is the sum of a Zeeman term, a quadrupolar splitting term, and a dipole-dipole interaction term:
\begin{align}
\hat{H}_{\mathbf{N}} & =  \hat{H}_{Z} +  \hat{H}_{Q} + \hat{H}_{D} = \sum_{k}\omega_{k}\hat{I}^{z}_{k} + \sum_{j}q_{k}(I^{z}_{k})^2  \nonumber \\
&  + \sum_{k>l}b_{kl}(\hat{I}^{+}_{k}\hat{I}^{-}_{l} + \hat{I}^{-}_{k}\hat{I}^{+}_{l} - 2 \hat{I}^{z}_{k}\hat{I}^{z}_{l}) \,\, ,
\end{align}
where $\omega_{k}$  and $q_{k}$ are the Zeeman and quadrupolar splittings of the $k$-th nucleus respectively, and $b_{kl}$ is the dipolar coupling between $k$ and $l$ nuclei. 

It is important to note that $W_{\perp}(T)$ calculated for echo or other dynamical decoupling sequences has noticeable time-dependence due to the presence of $\hat{H}_{Z}$ and $\hat{H}_{Q}$ terms involving only single nuclei. The characteristic oscillations of $W_{\perp}(T)$ for spin echo \cite{Cywinski2009,Cywinski2009a,Bluhm2011,Neder2011,Botzem2016} and for CPMG \cite{Malinowski2017} arise from the presence of three distinct nuclear Larmor precession frequencies in GaAs.
In contrast, the echo decay envelope was explained \cite{Bluhm2011,Neder2011,Malinowski2017,Botzem2016} by the presence of a quadrupolar-induced spread of effective fields $\delta B$, with $\delta B$ sufficiently large to dominate over broadening due to $\hat{I}^{z}_{k}\hat{I}^{z}_{l}$ dipolar interactions. Reported values are $\delta B \! \approx \! 0.3$ mT in  \cite{Bluhm2011} and  $\delta B \! \approx \! 1$ mT in \cite{Malinowski2017,Botzem2016}. On the other hand, the decay of $W_z(T)$ is solely due to the dipolar $\hat{I}^{-}_{k}\hat{I}^{+}_{l}$  flip-flop term, which does not commute with $\hat{h}^{z}$ and thus leads to dynamics of the longitudinal Overhauser field. 

We calculate $W_{\perp}(T)$ using a semiclassical theory \cite{Neder2011}, in which the averages of products of any number $\hat{h}^{2}_{x,y}$ are evaluated to the lowest order in $1/N$ expansion \cite{Cywinski2009,Cywinski2009a}. Following \cite{Cywinski2009a,Neder2011} we define a $\mathcal{T}$-matrix, the components of which are given by
\begin{align}
\mathcal{T}_{kl}(T) & = \frac{2}{3}I(I+1)\sqrt{N_{k}N_{l}} \frac{A_{k}A_{l}}{2\Omega}\int_{0}^{T} f(t') e^{i\omega_{kl}t'} \nonumber\\ 
& \times \cos \left( \int_{0}^{t'}f(t'')A_{kl}\mathrm{d}t'' \right )  \,\, \label{eq:Tkl}
\end{align}
where $k(l)$ labels the group of $N_{k(l)}$ nuclei having (approximately) common values of hf coupling $A_{k(l)}$ and Zeeman splitting $\omega_{k(l)}$, $f(t')$ is the time-domain filter function corresponding to the given pulse sequence ($f(t')$ is nonzero for $t' \in [0,T]$, and it changes between $1$ and $-1$ value at times at which $\pi$ pulses are applied), $I$ is the length of individual nuclear spins (assumed to be the same for all spins, as is the case for GaAs, for which $I\! =\! 3/2$), and $\omega_{kl}\! = \! \omega_{k}-\omega_{l}$, $A_{kl} \! =\! (A_{k}-A_{l})/2$. We have then \cite{Cywinski2009,Cywinski2009a,Neder2011}
\begin{equation}
W_{\perp}(T) \! =\! \frac{1}{\mathrm{det}[1+i\mathcal{T}(T)]} = \exp \left [ \sum_{k=1}^{\infty} \frac{(-i)^k}{k}R_{k}(T) \right ] \,\, ,
\end{equation}
where $R_{k}(T) = \mathrm{Tr}[\mathcal{T}^{k}(T)]$.

For a spin echo sequence of length $T\! = \! \tau$ the $\mathcal{T}$-matrix is given by \cite{Cywinski2009,Cywinski2009a,Neder2011}
\begin{align}
\mathcal{T}^{\mathrm{SE}}_{kl}(\tau) & = \frac{\bar{b}_{k}\bar{b}_{l}}{\Omega}  \frac{-i\omega_{kl}}{\omega^{2}_{kl}-A^{2}_{kl}}  
\left( \cos \frac{\omega_{kl}\tau}{2} - \cos \frac{A_{kl}\tau}{2} \right) e^{i\omega_{kl}\tau/2} \,\, , \label{eq:TklSE}
\end{align}
while for a CMPG sequence with even number of pulses $n$ and interpulse spacing $\tau$ we have \cite{Malinowski2017}
\begin{align}
\mathcal{T}^{\mathrm{CP,n}}_{kl}(T=n\tau) & = \frac{\bar{b}_{k}\bar{b}_{l}}{\Omega}  \frac{\omega_{kl}}{\omega^2_{kl}-A^{2}_{kl}} \nonumber 
\frac{\cos \frac{\omega_{kl}\tau}{2} - \cos \frac{A_{kl}\tau}{2}}{\cos \frac{\omega_{kl}\tau}{2}}  \nonumber \\
& \times \sin \frac{\omega_{kl}n\tau}{2} e^{i\omega_{kl}n\tau/2} \,\, . \label{eq:TklCP}
\end{align}
where $\bar{b}_{k} \! =\! \sqrt{\frac{2}{3}I(I+1)N_{k}}A_{k}$ is the rms strength of the Overhauser field arising from $N_{k}$ spins, all spins having Knight shift $A_{k}$. 

\begin{figure}
	\centering
	\includegraphics[scale=.9]{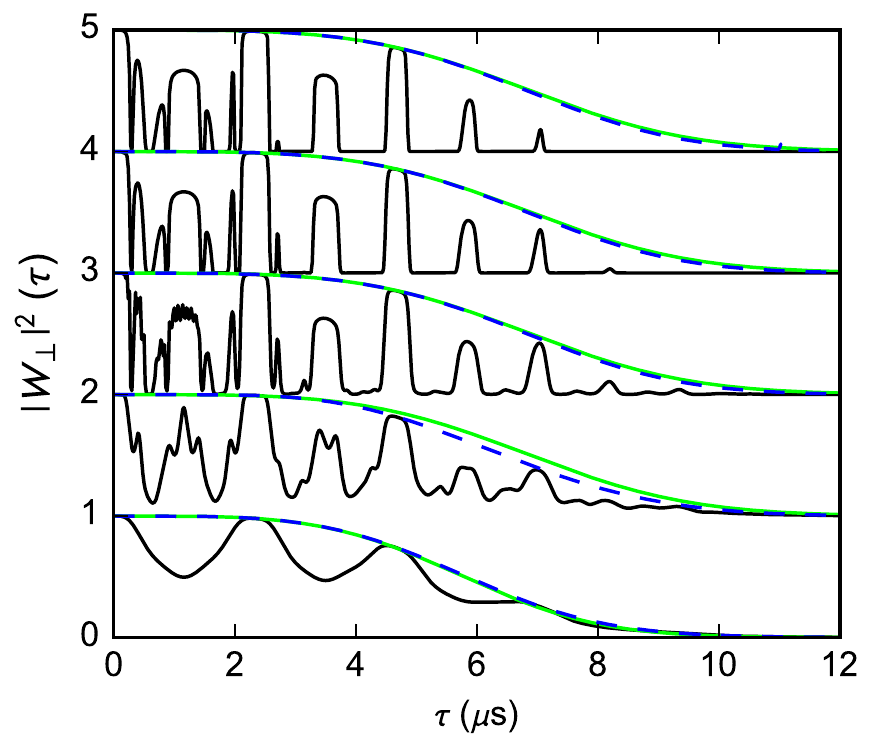}
	\caption[Two-spin decoherence function $|W_{\perp}(T=n\tau)|^2$ calculated using the $T$-matrix approach]
	{Two-spin decoherence function $|W_{\perp}(T=n\tau)|^2$ calculated using the $T$-matrix approach for $n\! = \! 1$, $4$, $16$, $64$, and $256$ (from bottom to top). Black solid lines are the exact calculation, green lines are the homonuclear-only result, and dashed lines correspond to analytical approximation to the homonuclear result from Eq.~\eqref{eq:Whoman}.}
	\label{ovh_sup:figS5}
\end{figure}

It is crucial now to recognize the distinct roles of two kinds of contributions to the $\mathcal{T}$-matrix: the heteronuclear ones, in which $\omega_{k}$ and $\omega_{l}$ correspond to distinct nuclear isotopes (i.e.~$^{69}$Ga, $^{71}$Ga, and $^{75}$As in the case of GaAs, labeled by $\alpha \! =\! 1,2,3$), so that $\omega_{kl} \! \approx \! \omega_{\alpha\beta}$, and the homonuclear ones, in which $\omega_{k}$ and $\omega_{l}$ correspond to groups of nuclei of the same isotope, so that $\omega_{kl} \! \ll \! \omega_{\alpha\beta}$. 
The former terms govern the presence of characteristic oscillations of $W_{\perp}(t)$ in echo \cite{Bluhm2011,Botzem2016,Malinowski2017} and CPMG case \cite{Malinowski2017}, while the latter smooth these oscillations in CPMG case and, more importantly,  lead to an irreversible decay of the signal. Note that the homonuclear terms are nonzero in the presence of intra-species spread of nuclear splittings, thereby contributing to low- and intermediate-frequency noise. 

Note that the dependence of $\mathcal{T}_{kl}$ on Knight shift differences $A_{kl}$ is negligible in experimentally relevant range of parameters for both homonuclear and heteronuclear terms. For QD with number of nuclei $N \! \approx \! 10^{6}$ we have $A_{kl}\tau \! \ll \! 1$ for $\tau \! \ll \! 20$ $\mu$s. For heteronuclear terms we can then put $A_{kl}\! =\! 0$ in Eqs.~\eqref{eq:TklSE} and \eqref{eq:TklCP} provided that $\omega_{\alpha\beta} \! \gg \! A_{kl}$ (which is fulfilled for $B_{\text{ext}} \! > \! 100$ mT in dots considered here), while for homonuclear terms for isotope $\alpha$ we have to assume that $\omega_{kl}\tau \! \ll \! 1$ (which is easily fulfilled for considered values of $\tau$ with $\delta B \! \approx \! 1$ mT used here). We can then perform the calculation by dividing the nuclei of each isotope into $K$ groups, each having the same $A_{k}$ (equal to the typical hf coupling for a give isotope), value of $\omega_{k}$ taken from $[\omega_{\alpha}-5\sigma_{\alpha},\omega_{\alpha}+5\sigma_{\alpha}]$ range, and $N_{k}$ taken from a Gaussian distribution:
\begin{equation}
N_{k\in \alpha} = N_{\alpha} \frac{1}{\sqrt{2\pi}\sigma_{\alpha}} \exp\left (-\frac{(\omega_{k}-\omega_{\alpha})^2}{2\sigma^{2}_{\alpha}} \right) \,\, ,
\end{equation}
in which $\sigma_{\alpha}$ is the rms value of the nuclear splitting due to a spread $\delta B$ of the effective field experienced by the nuclei. The calculations converge on timescales relevant for measurements presented in this paper for $K \! < \! 100$ for $n\! =\! 256$ pulses (and for smaller $K$ for lower numbers of pulses).

Examples of results for $n\! = 1$, $4$, $16$, $64$, and $256$ are shown in Fig.~\ref{ovh_sup:figS5}. In the calculations we have used the values of $A_{\alpha} \! = \! 2\mathcal{A}_{\alpha}/N$, with $\mathcal{A}_{\alpha} \! =\! 35.9$, $45.9$, and $42.9$ $\mu$eV, and $\omega_{\alpha} \! = \! -42.1$, $-53.6$, and $-30.1$ neV at $1$ T field corresponding to $^{69}$Ga, $^{71}$Ga, and $^{75}$As, respectively (note that $N$ here is the number of nuclei, while in \cite{Cywinski2009a} it denoted the number of unit cells, i.e.~twice the number of nuclei). The experimental results presented in the main text are best fit by $N \! =\! 9 \times 10^{5}$ in each dot, and effective broadening $\sigma_{\alpha}$ corresponding to $1$ mT field, and these are the values used in Fig.~\ref{ovh_sup:figS5}.

The key observation is that the calculated $W_{\perp}(T)$ can be very well approximated by a product of decoherence functions calculated while keeping {\it only} the homonuclear terms and {\it only} the heteronuclear ones:
\begin{equation}
W_{\perp}(T) \approx W_{\perp,\text{het}}(T) \times W_{\perp,\text{hom}}(T)   \,\, .
\end{equation}
The heteronuclear term is responsible for large-amplitude oscillations of the signal, while the homonuclear term gives a decay envelope of the coherence signal, see Fig.~\ref{ovh_sup:figS5}.
Furthermore, the homonuclear contribution can be approximated very well (at least on timescale of the signal decay) by a simple solution obtained using a bimodal approximation to the distribution of $\omega_{k}$ frequencies of nuclei of each species (first used for spin echo case in \cite{Neder2011}). In this approximation we have
\begin{equation}
W_{\perp,\text{hom}}(T) \approx \prod_{\alpha}\frac{1}{1+(T/t^{(n)}_{\alpha})^4} \,\, ,\label{eq:Whoman}
\end{equation}
where for echo we have 
\begin{equation}
t^{(1)}_{\alpha} = \frac{2\sqrt{2}}{\bar{b}_{\alpha}} \sqrt{\frac{\Omega}{\sigma_{\alpha}}} \,\, ,\label{eq:tSE}
\end{equation}
while for CPMG sequence with even $n$ we obtain
\begin{equation}
t^{(n)}_{\alpha} = 2^{1/4} n t^{(1)}_{\alpha} \,\, .
\end{equation}
This is the main result here: when decoherence due to tranvserse Overhauser field fluctuations is dominated by the homonuclear contribution, the characteristic coherence half-decay time $T_{2}$ (defined by $W_{\perp,\text{hom}}(T_{2})\! =\! 1/e$) scales {\textit {linearly}} with the number of pulses $n$, i.e.~we have $T_{2}\propto n^{\gamma_{\perp}}$ with $\gamma_{\perp}\!=\! 1$.

\chapterimage{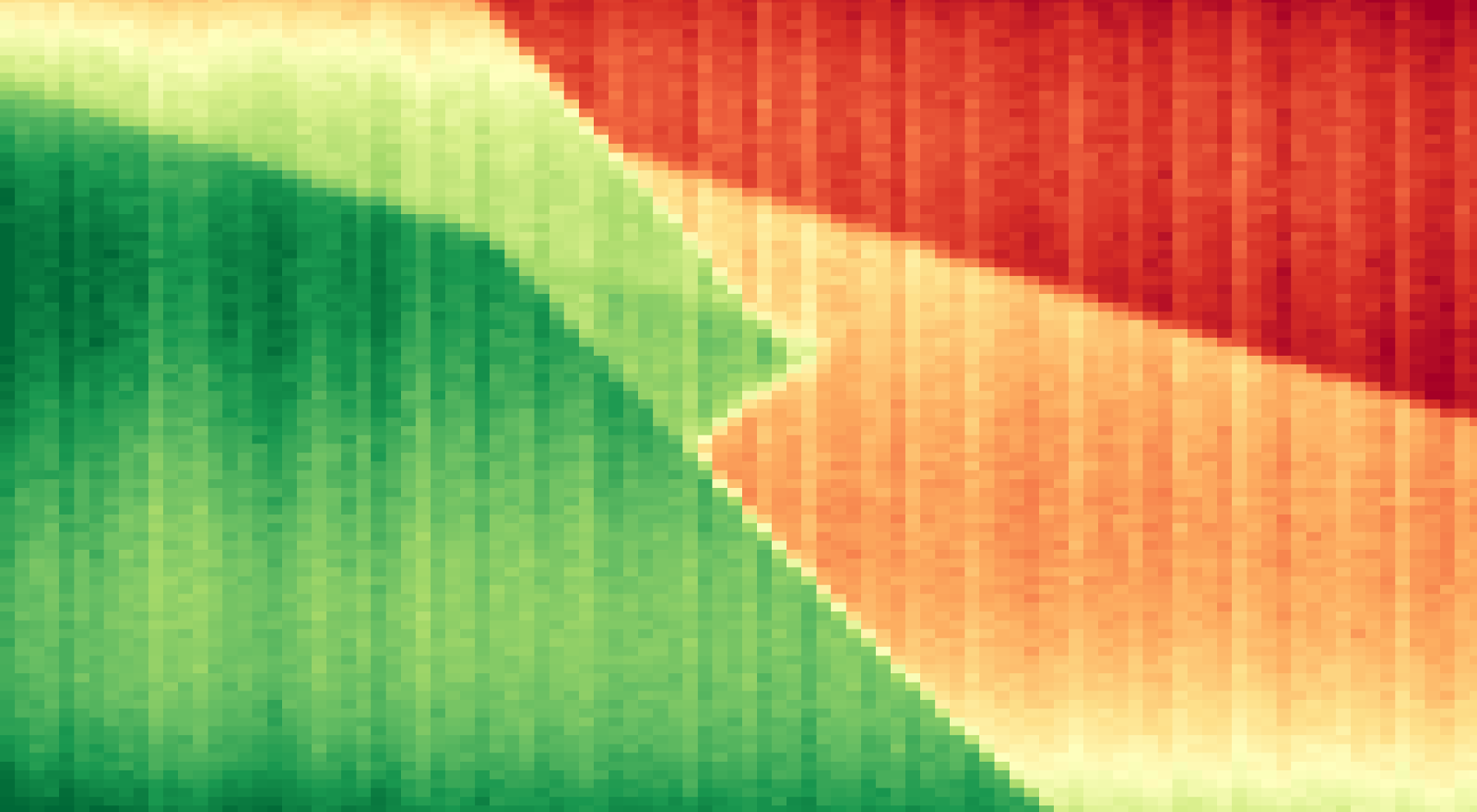}
\chapter[Supplementary Information for ``Notch filtering the nuclear environment of a spin qubit'']{\protect\parbox{0.9\textwidth}{Supplementary Information for\\ ``Notch filtering the nuclear\\ environment of a spin qubit''}}
\chaptermark{Supplementary Information for ``Notch filtering...''}
\label{ch:notch_sup}

\providecommand{\VL}{}
\renewcommand{\VL}{V_\mathrm{L}}
\providecommand{\VM}{}
\renewcommand{\VM}{V_\mathrm{M}}
\providecommand{\VR}{}
\renewcommand{\VR}{V_\mathrm{R}}

\providecommand{\Btot}{}
\renewcommand{\Btot}{B^\mathrm{tot}}
\providecommand{\Bext}{}
\renewcommand{\Bext}{B^\mathrm{ext}}
\providecommand{\Bznuc}{}
\renewcommand{\Bznuc}{B_\mathrm{z}^\mathrm{nuc}}
\providecommand{\Bpnuc}{}
\renewcommand{\Bpnuc}{B_\perp^\mathrm{nuc}}

\providecommand{\ud}{}
\renewcommand{\ud}{\uparrow\downarrow}
\providecommand{\du}{}
\renewcommand{\du}{\downarrow\uparrow}

\providecommand{\drv}{}
\renewcommand{\drv}{\mathrm{d}}

\providecommand{\FHahn}{}
\renewcommand{\FHahn}{F_\mathrm{Hahn}}
\providecommand{\FCPMG}{}
\renewcommand{\FCPMG}{F_{\mathrm{CPMG},n}}
\providecommand{\FFID}{}
\renewcommand{\FFID}{F_\mathrm{FID}}
\providecommand{\Fenv}{}
\renewcommand{\Fenv}{F_\mathrm{env}}

\providecommand{\Ga}{}
\renewcommand{\Ga}{^{69}\mathrm{Ga}}
\providecommand{\Gb}{}
\renewcommand{\Gb}{^{71}\mathrm{Ga}}
\providecommand{\As}{}
\renewcommand{\As}{^{75}\mathrm{As}}
\providecommand{\fGa}{}
\renewcommand{\fGa}{f_{^{69}{\rm Ga}}}
\providecommand{\fGb}{}
\renewcommand{\fGb}{f_{^{71}{\rm Ga}}}
\providecommand{\fAs}{}
\renewcommand{\fAs}{f_{^{75}{\rm As}}}

\providecommand{\TCPMG}{}
\renewcommand{\TCPMG}{T_2 ^\mathrm{CPMG}}
\renewcommand{\vec}[1]{{\bf #1}}

\begin{center}
\begin{tcolorbox}[width=0.8\textwidth, breakable, size=minimal, colback=white]
	\small
	The supplementary information is divided into sections, which discuss the following topics:
	\begin{enumerate}
		\item[\ref{notch_sup:calibration}] Calibration of $\pi$ pulses
		\item[\ref{notch_sup:model}] Semiclassical model of decoherence due to nuclear noise
		\item[\ref{notch_sup:NandB}] Estimating $N$ and $\delta B$ from Hahn echo signal
		\item[\ref{notch_sup:TandBeta}] Estimating $T_\mathrm{SD}$ and $\beta$ from scaling of coherence time
		\item[\ref{notch_sup:splitting}] Extension of the model to take into account anisotropy of electron $g$-factor. Discussion of the origin of the splitting of the first revival peak.
	\end{enumerate}
	Supplementary information appended to this thesis include additional unpublished results on:
	\begin{enumerate}[resume]
		\item[\ref{notch_sup:moremaps}] CPMG revival maps for 32, 64 and 128 $\pi$ pulses
	\end{enumerate}
\end{tcolorbox}
\end{center}

\section{Calibration of $\pi$ pulses}
\label{notch_sup:calibration}

To generate decoupling sequences consisting of as many as 1000 $\pi$ pulses we took advantage of charge-noise-insensitive symmetric exchange pulses. This new technique improves the quality factor of exchange oscillations by a factor of six relative to conventional method of tilting the double dot potential~\cite{Martins2016,Reed2016}. A detailed analysis of this technique, and results obtained in the preceding experiment from the same setup and same sample, can be found in Ref.~\cite{Martins2016,Barnes2016}.

\begin{figure}[tb]
	\centering
	\includegraphics[scale=1]{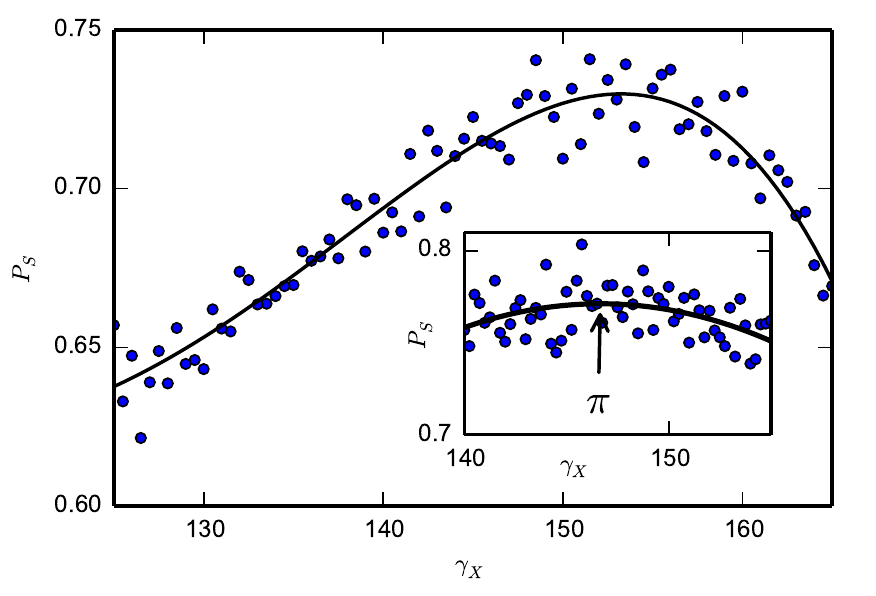}
	\caption[Calibration of $\pi$ pulses]{
	\textbf{Calibration of $\mathbf{\pi}$ pulses} Singlet return probability $P_S$ as a function of a symmetric exchange pulse amplitude, for exchange time of 4.167 ns in a Hahn echo experiment. The maximum probability indicates $\pi$ pulse. Solid line is a guide to the eye. Inset: around the maximum the parabola is fitted to the data. Symbol $\pi$ indicates value of $\gamma_X$ corresponding to the $\pi$ pulse.
	}
	\label{notch_sup:figS1}
\end{figure}

The optimization was performed by maximizing the Hahn echo signal by varying the amplitude of the symmetric exchange pulse, $\gamma_X$, while keeping detuning $\varepsilon_X=0$ mV, exchange time $t_X=4.167$ ns and total evolution time $\tau=0.75$ $\mu$s fixed (Fig.~\ref{notch_sup:figS1}). The experiment was performed on the same device and in identical tuning as Ref.~\cite{Martins2016}, where $\gamma_X$ and $\varepsilon_X$ are defined and discussed in detail.

We note that symmetric exchange pulses show a weaker dependence on gate voltages than ordinary tilt pulses.
Hence, symmetric $\pi$ pulses are more robust against fluctuations of pulse amplitudes. 
On the other hand, symmetric pulses require a larger amplitude, resulting in somewhat slower exchange gates compared to conventional tilted exchange gates. This limitation makes $\pi$ pulses more susceptible to errors induced by gradients of the Overhauser field, causing a tilted rotation axis of the qubit. In future experiments, larger pulse amplitudes can be achieved straightforwardly by decreasing the attenuation in the transmission lines in the cryostat.

Nevertheless, 
the CPMG sequence is particularly robust to
two kinds of $\pi$ pulse errors that affect exchange gates~\cite{Borneman2010}. The first is over or under rotation around the vertical axis of the Bloch sphere due to miscalibration of pulse amplitude and duration. 
The second is tilt of the rotation axis in the $\ket{S}$-$\ket{T_0}$--$\ket{\ud}$-$\ket{\du}$ plane due to uncontrolled gradients of the Overhauser field. 

\section{Semiclassical model for decoherence}
\label{notch_sup:model}

The inset of Fig.~4 shows theoretical results for coherence revivals.
The model is derived closely following the semiclassical approach developed in Ref.~\cite{Neder2011}.
The starting point is to express the Hamiltonian for the $\ket{\ud}$, $\ket{\du}$ subspace of the two-spin system as
\begin{equation}
\label{eq:NoiseHam}	\hat{H}(t) = g^* \mu_B \sum\limits_{d=L,R} \left( B_{z,d}^{\rm nuc}(t) + \frac{|\mathbf{B}_{\perp,d}^{\rm nuc}(t)|^2}{2|\mathbf{B}^{\rm ext}|}\right)  c(t) \hat{S}_d^z,
\end{equation}
where $g^*$ is the electron $g$-factor, $\mu_B$ is the Bohr magneton, $\mathbf{B}^{\rm ext}$ is the external magnetic field, $B_{z,d}^{\rm nuc}$  ($\mathbf{B}_{\perp,d}^{\rm nuc}$) is the Overhauser field component parallel (perpendicular) to external magnetic field, $\hat{S}_d^z$ is the electron spin operator, $d=L,R$ labels the left and right dots, and we assume $|\mathbf{B}^{\rm ext}| \gg |\mathbf{B}_{d}^{\rm nuc}|$.
Here, the sequence of $\pi$ pulses applied to the qubit is captured by
\begin{equation}
	c(t) = \sum\limits_{j=0}^n (-1)^j \theta(t_{j+1}-t) \theta(t-t_j),
\end{equation}
where $t_j$ is the time at which the $j$-th $\pi$ pulse of the CPMG sequence is applied (with $t_0=0$, $t_{n+1}=T$), and $\theta(t)$ is the Heaviside step function. 

Reference \cite{Neder2011} treated only the Hahn echo sequence.
This corresponds to $n = 1$ in the above.
Following the same sequence of steps, we obtain results for arbitrary $n$.
As in that case, the decoherence function $W(\tau) = W_z(\tau)W_\perp(\tau)$ is separated into a product of contributions from the longitudinal and transverse noise sources.
The low-frequency longitudinal noise contribution is of the form $W_z(\tau) = e^{-(\tau/T_\mathrm{SD})^{\alpha}}$, where $\alpha=\beta+1$ is related to the exponent in the power law $1/f^\beta$ describing the spectrum of this noise source~\cite{Medford2012}, and $T_\mathrm{SD}$ is the spectral diffusion time.
Because the transverse field enters the Hamiltonian \eqref{eq:NoiseHam} as a square, $|\mathbf{B}_{\perp,d}^{\rm nuc}(t)|^2$, this noise source is effectively non-Gaussian \cite{Cywinski2014,Neder2011}.
As a result, the decoherence function for dot $d$ is of the form $W_{\perp,d}(\tau) = 1/\det(1 + iT_d)$ with components of the matrix $T_d$ given by
\begin{equation}
	T_{kl,d} = \frac{5 A_{\xi(k)} A_{\xi(l)} \sqrt{N_k N_l}}{2 g^* \mu_B |\mathbf{B}^{\rm ext}|}
	\frac{\omega_{kl}}{\omega_{kl}^2 - A_{kl}^2}
	\left\lbrace 1- \frac{\cos\left( \frac{A_{kl} T}{2n} \right)}{\cos \left( \frac{\omega_{kl} T}{2n} \right)} \right\rbrace
	\sin\left( \frac{\omega_{kl} T + n\pi}{2} \right) e^{i \frac{\omega_{kl}T + n\pi}{2}}.
\end{equation}
Here $k,l$ labels groups of nuclei associated by isotope and local nuclear Zeeman coupling, $A_{\xi(k)}$ is the hyperfine coupling constant for nuclei in group $k$, $A_{kl}=A_{\xi(k)}-A_{\xi(l)}$, $N_k$ is number of nuclei in a group, $\omega_{kl} = \omega_k - \omega_l$ is a difference of Larmor frequencies between nuclei from two groups, and $T=n\tau$ is the total evolution time.
Specifically, the nuclei of each isotope are divided into $K$ groups using the relation $N_k=n_{\xi(k)}N/(2K)$, where $n_\xi$ is the number of nuclei of isotope $\xi$ per unit cell, and where all nuclei within each group have the same Larmor angular frequency $\omega_k$.  
The value of $\omega_k$ for each group is drawn from a Gaussian distribution centered at the bare Larmor frequency $\omega_\xi$ for the corresponding isotope, with standard deviation $\delta B$. 
The broadening $\delta B$ is introduced as a phenomenological parameter to take into account an effective spread in the Larmor frequencies due to inhomogeneous quadrupolar splittings and dipole-dipole interactions.
For the simulation shown in Fig.~4, differences between hyperfine couplings within the same isotope were neglected, and convergence was obtained with $K=4$ groups.

Larmor frequencies and hyperfine couplings used to perform the simulation shown in Fig.~4 were taken from Ref.~\cite{Cywinski2009a} (table \ref{notch_sup:tab1}). The remaining parameters are: the effective number, $N$, of nuclei interacting with each electron, inhomogeneity, $\delta B$, of the effective magnetic field acting on the nuclei, the spectral diffusion time, $T_\mathrm{SD}$, and the exponent, $\beta$, associated with the low-frequency noise. The following sections explain how these parameters are obtained.

\begin{table}[b]
	\caption{Bare Larmor angular frequencies, $\omega_\xi$, hyperfine constants, $A_\xi$, in units of angular frequency, and abundances, $n_\xi$, of $^{69}$Ga, $^{71}$Ga and $^{75}$As, taken from Ref. \cite{Cywinski2009a}.}
	\centering
	\renewcommand\tabcolsep{10pt}
	\begin{tabular}{c||ccc}
		& $\omega_\xi/B$ [s$^{-1}$T$^{-1}$] & $A_\xi$ [s$^{-1}$] & $n_\xi$\\ \hline \hline
		$^{69}$Ga & 64.2 & $5.47\times10^{10}$ & 0.604\\
		$^{71}$Ga & 81.6 & $6.99\times10^{10}$ & 0.396\\
		$^{75}$As & 45.8 & $6.53\times10^{10}$ & 1\\
	\end{tabular}
	\label{notch_sup:tab1}
\end{table}

\section{Estimating $N$ and $\delta B$ from Hahn echo signal}
\label{notch_sup:NandB}

\begin{figure}[tb]
	\centering
	\includegraphics[scale=1]{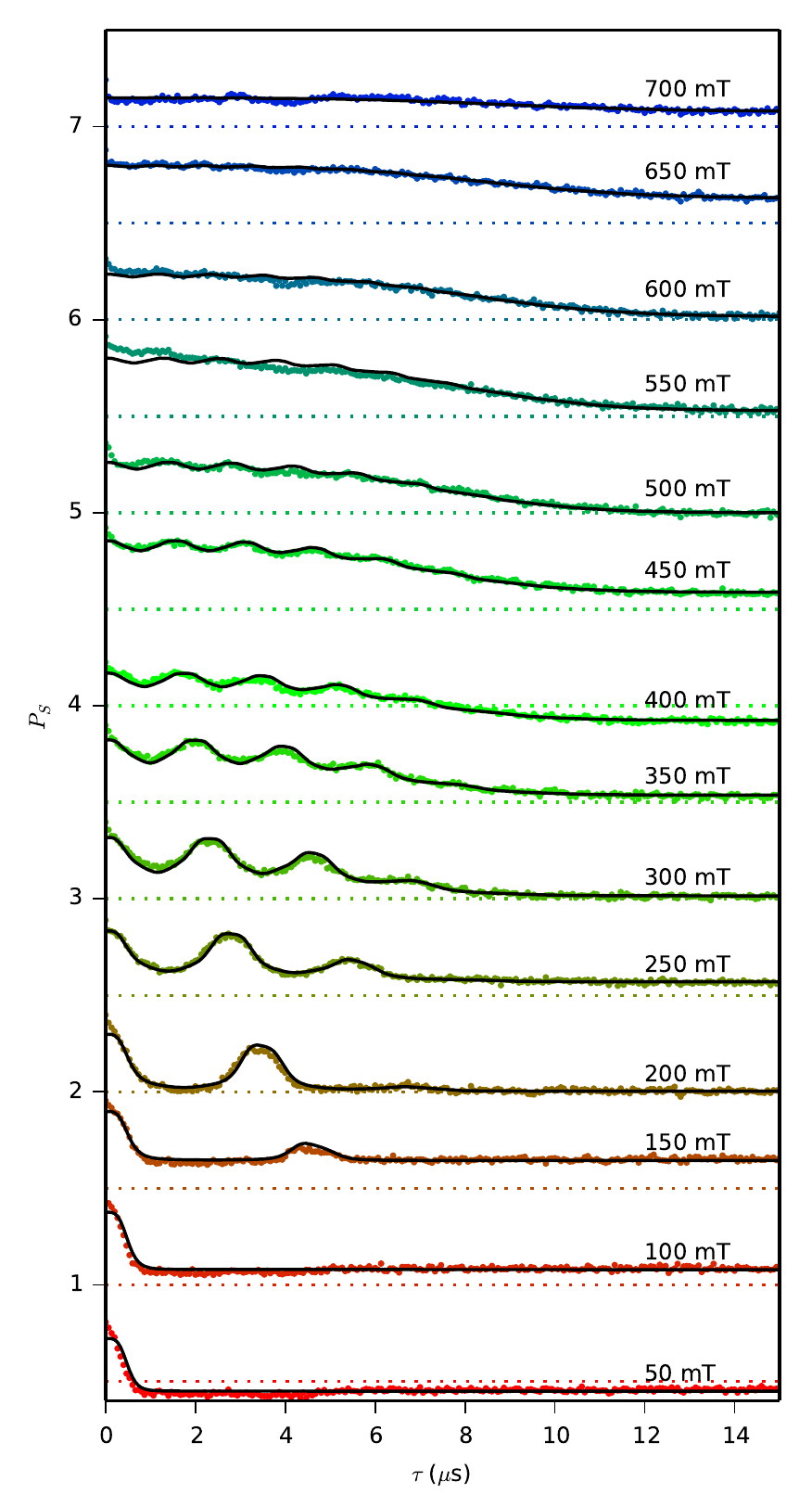}
	\caption[Revival of coherence under Hahn-echo sequence]
	{\textbf{Revival of coherence under Hahn-echo sequence.} Singlet return probability $P_S$ as a function of separation time $\tau$ for various magnetic fields. Datasets are offset for clarity. Dotted lines indicate $P_S=0.5$ for data plotted in corresponding color. Black lines are simulations with $\delta B = 1.1$ mT, $T_\mathrm{SD}^\mathrm{Hahn} = \infty$, $N = 7\times 10^5$. They are fitted to experimental data using offset and visibility, different for each curve.}
	\label{notch_sup:figS2}
\end{figure}

The simulation of revivals under CPMG sequences requires knowledge of four device-specific parameters, two of which, the effective number, $N$, of nuclei interacting with each electron and the inhomogeneity, $\delta B$, of the effective magnetic field acting on the nuclei, are extracted from Hahn echo data obtained at several magnetic fields (Fig.~\ref{notch_sup:figS2}). Following previous work \cite{Bluhm2011,Neder2011} we first fit theory to Hahn echo data at each magnetic field separately, keeping $\delta B$, $N$, the spectral diffusion time for Hahn echo, $T_\mathrm{SD}^\mathrm{Hahn}$, vertical offset, and vertical scaling as free parameters.
Setting $T_\mathrm{SD}^\mathrm{Hahn} \gg 1$ ms gives essentially equally good fits (i.e. $T_\mathrm{SD}^\mathrm{Hahn}$ cannot be determined accurately by this method) but values for $\delta B$ and $N$ obtained at various magnetic fields (150-350 mT) differ from each other by less than 20\%. Therefore we average these values and obtain  $\delta B = 1.1$ mT and $N = 7\times 10^5$. Fixing these values and $T_\mathrm{SD}^\mathrm{Hahn}=\infty$, we leave the vertical offset and vertical scaling as the only free parameters, and obtain excellent agreement for all magnetic fields, as seen in Fig.~\ref{notch_sup:figS2}. Visibility and offset are left as free parameters, independent for each curve, to accommodate a fluctuating readout visibility that is likely due to a buildup of the gradient of Overhauser field for large $\Bext$ \cite{Barthel2012}.

The only systematic deviation between the experimental results and the model is a slight, rapid, initial decay of the signal (first 3-5 data points of each data set). This effect was also observed in Refs.~\cite{Bluhm2011,Botzem2016}. 
The effect depended on the external magnetic field as well as the gradient of the Overhauser field~\cite{Botzem2016}, and was speculated to be related to the entanglement of the qubit with the nuclear bath or to $\pi$ pulse errors~\cite{Bluhm2011}.

\section{Estimating $T_\mathrm{SD}$ and $\beta$ scaling of coherence time}
\label{notch_sup:TandBeta}

To estimate the spectral diffusion time $T_\mathrm{SD}$ for the simulation in Fig.~4a we quantify the scaling of the CPMG coherence time with the number of $\pi$ pulses $n$ in a regime where revivals are not yet developed (i.e., for $n\leq 32$ at 750 mT)  \cite{Medford2012}.  
The coherence time is found to be proportional to $n^\gamma$, with $\gamma \sim 0.75$ \cite{Malinowski2017}.
Using this scaling behaviour we infer $T_\mathrm{SD} \approx 0.6$ ms for a CPMG sequence with 256 $\pi$-pulses. Using the relationship $\beta = \gamma / (1-\gamma)$ \cite{Medford2012} the exponent $\gamma=0.75$ corresponds to a power law of low-frequency noise governed by $1/f^\beta$ behaviour, with $\beta=3$, in reasonable agreement with previous measurements \cite{Medford2012}.

\section{Splitting of the first revival peak}
\label{notch_sup:splitting}

\begin{figure}[tb]
	\centering
	\includegraphics[width=\textwidth]{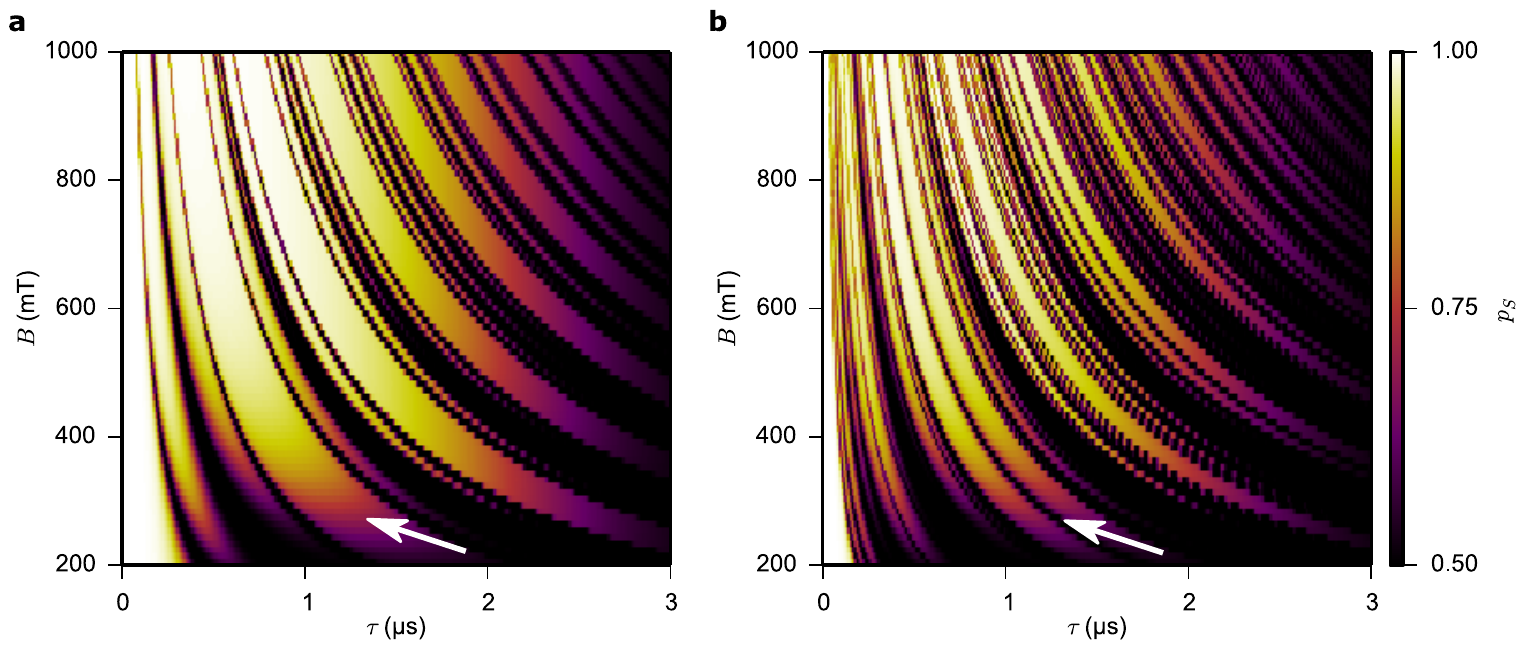}
	\caption[Simulation of revivals of coherence under CPMG sequence for 256 $\mathbf{\pi}$ pulses]
	{\textbf{Simulation of revivals of coherence under CPMG sequence for 256 $\mathbf{\pi}$ pulses.} \textbf{a}, Simulation omitting effects of g-factor anisotropy, identical to map in the inset of Fig.~4a, i.e., $g_\perp / g_\parallel = 0$. \textbf{b}, Simulation assuming $g_\perp / g_\parallel= 0.01$.}
	\label{notch_sup:figS3}
\end{figure}

A possible explanation for the observed splitting in the first revival peak (Fig.~4a) is based on the anisotropy of the electronic $g$-factor. The $g$-factor anisotropy between [011] and [01-1] primary axes can be as high as 15\% in asymmetric GaAs/AlGaAs quantum wells \cite{Nefyodov2011}. In Ref.~\cite{Botzem2016} it was shown that the anisotropy has a strong impact on $S$-$T_0$ qubit coherence when the magnetic field is not parallel to one of the main axes. The combination of the anisotropy and small misalignment of the external magnetic field with the [011] crystal axis changes the magnetic field term in the system Hamiltonian \eqref{eq:NoiseHam} to:
\begin{equation}
\label{eq:BAniso}
B_{z,d}^{\rm nuc}(t) + \frac{|\mathbf{B}_{\perp,d}^{\rm nuc}(t)|^2}{2|\mathbf{B}^{\rm ext}|} + \frac{g_\perp}{g_\parallel} \left[ B_{x,d}^{\rm nuc}(t) + B_{y,d}^{\rm nuc}(t) \right],
\end{equation}
where $g_\parallel$ ($g_\perp$) are diagonal (off-diagonal) elements of a $g$-tensor in the basis set by direction of the external magnetic field. The latter leads to the appearance of individual nuclear Larmor frequencies  in the nuclear noise spectrum in addition to nuclear difference frequencies \cite{Botzem2016}. As a result, the CPMG sequence will not be as efficient in suppressing nuclear noise even when the pulses are commensurate with all three difference frequencies.

In Fig.~\ref{notch_sup:figS3} we present simulations showing the consequences of $g$-factor anisotropy. Panel (a) shows the simulation presented in the inset of Fig.~4a, i.e. $g_\perp / g_\parallel = 0$. In panel (b) we show a simulation that assumes $g_\perp / g_\parallel = 0.01$. Although our external magnetic field was nominally aligned with the [011] crystal axis (cf. Fig.~\ref{notch:fig1}a), the choice of $g_\perp / g_\parallel = 0.01$ is consistent with the smallest value observed in ~\cite{Botzem2016} for the same direction of magnetic field as in our setup. 
In our simulation a splitting of the first revival peak appears (indicated by a white arrow) as well as more complex fine structure in other revival peaks. Such fine structure is beyond the resolution of the experimental data.

We note that a splitting of the first revival peak appears exactly when the frequency of $\pi$ pulses coincides with a difference of Larmor frequencies $\fGb - \fGa$ and $\fGa - \fAs$. Therefore other mechanisms might also lead to the appearance of the splitting. We speculate that weak driving of the nuclei by a periodic Knight field could enhance flip-flops between nuclei of different species and therefore increase spin diffusion, leading to faster decoherence.

\section{Unpublished: CPMG revival maps for 32, 64 and 128 $\pi$ pulses}
\label{notch_sup:moremaps}

\begin{figure}[tb]
	\centering
	\includegraphics[width=\textwidth]{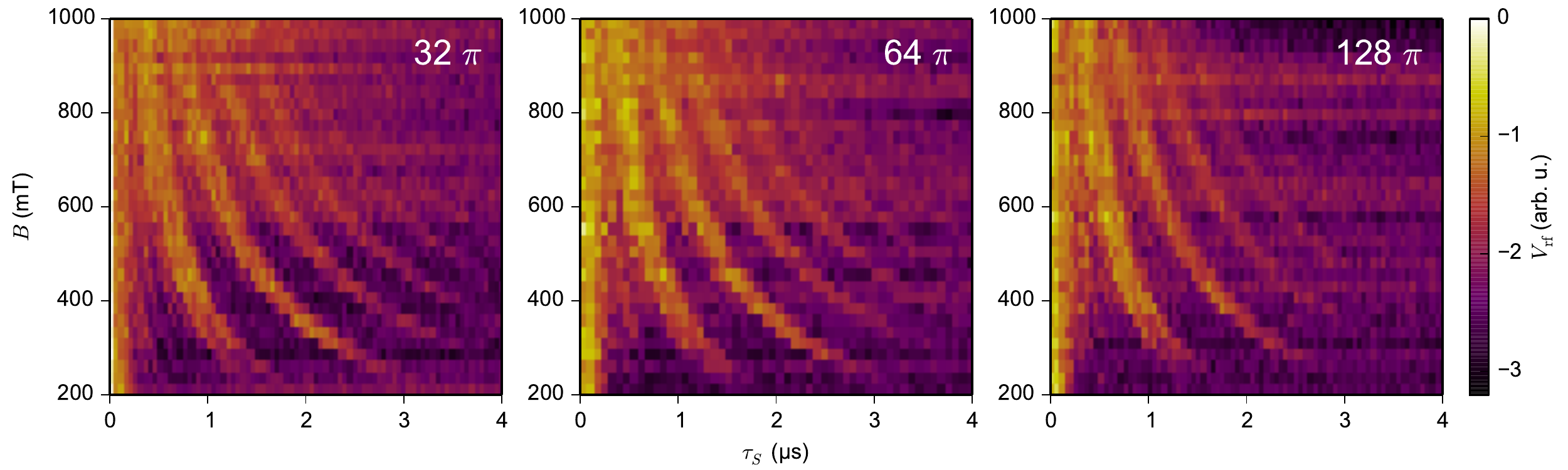}
	\caption[CPMG revival maps for 32, 64 and 128 $\pi$ pulses]
	{\textbf{CPMG revival maps for 32, 64 and 128 $\pi$ pulses.}}
	\label{notch_sup:figS4}
\end{figure}

The measurement of the CPMG decay as a function of the magnetic field, such as in Fig.~\ref{notch:fig4}a, was also repeated for smaller number of $\pi$ pulses (Fig.~\ref{notch_sup:figS4}). Here we can see that more and more ridges are visible for CPMG sequences with fewer $\pi$ pulses. For $n=32$ we can see that mysterious features before the first revival ridge and the splitting of the first revival ridge persist as well.

\part{Multielectron quantum dot}
\label{part:multielectron}

\chapterimage{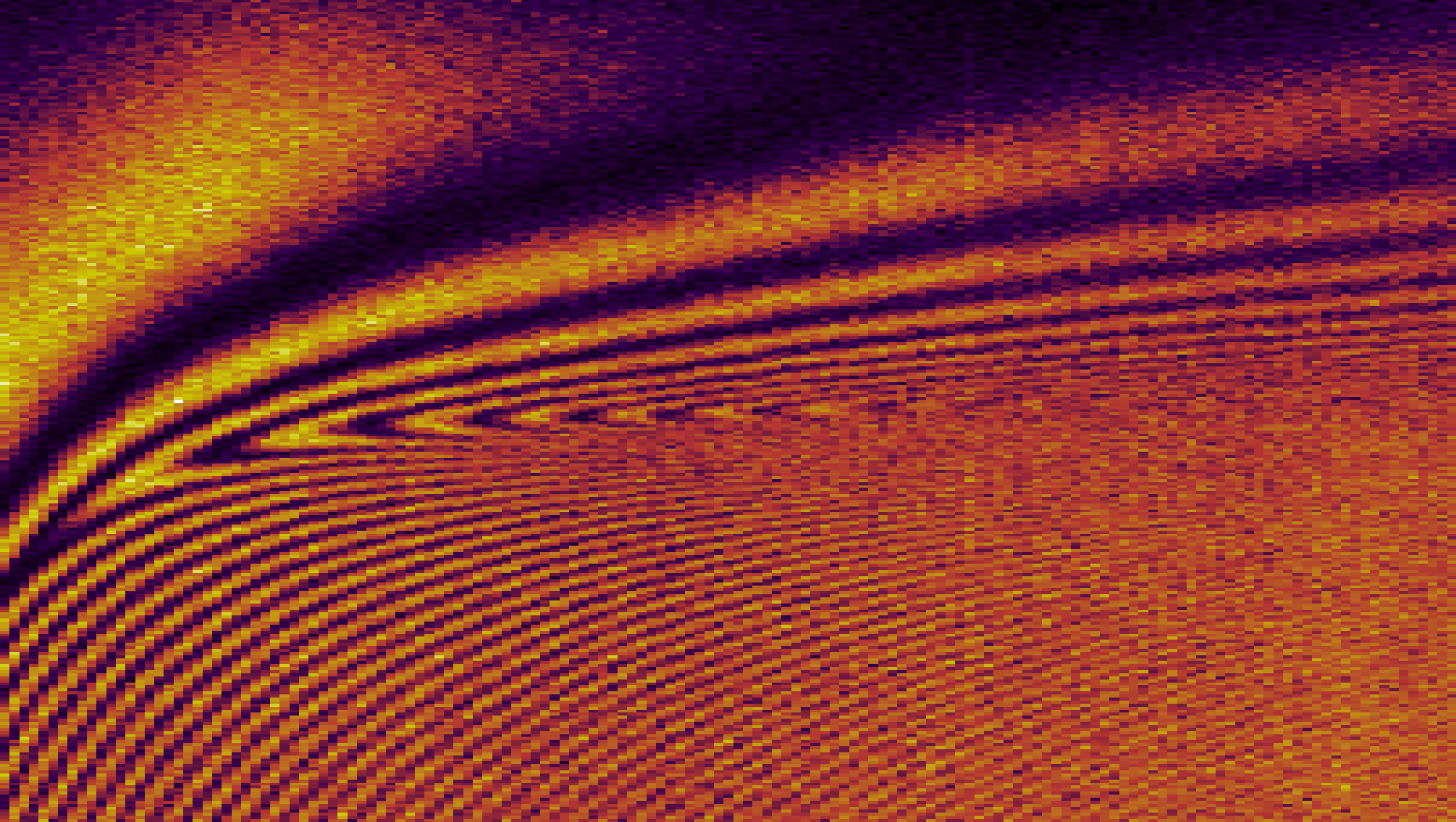}
\chapter[Ground state spin of a multielectron quantum dot and its interaction with a single neighboring spin]{\protect\parbox{0.9\textwidth}{Ground state spin of a multielectron \\ quantum dot and its interaction \\ with a single neighboring spin}}
\chaptermark{Ground state spin of a multielectron quantum dot...}
\label{ch:many_charge_states}

{\let\thefootnote \relax\footnote{This chapter is adapted from the manuscript in preparation.}}
\addtocounter{footnote}{-1}

\begin{center}
Filip K. Malinowski$^{1,*}$, Frederico Martins$^{1,*}$, Thomas Smith$^{2}$, Stephen D. Bartlett$^{2}$, \\
Andrew C. Doherty$^{2}$, Peter D. Nissen$^{1}$, Saeed Fallahi$^{3}$, Geoffrey C. Gardner$^{3,4}$, \\
Michael J. Manfra$^{5,6}$, Charles M. Marcus$^{7}$, Ferdinand Kuemmeth$^{1}$
\end{center}

\begin{center}
	\scriptsize
	$^{1}$ Center for Quantum Devices, Niels Bohr Institute, University of Copenhagen, 2100 Copenhagen, Denmark\\
	$^{2}$ Centre for Engineered Quantum Systems, School of Physics, The University of Sydney, Sydney NSW 2006, Australia  \\
	$^{3}$ Department of Physics and Astronomy, Birck Nanotechnology Center, Purdue University, West Lafayette, Indiana 47907, USA \\
	$^{4}$ School of Materials Engineering and School of Electrical and Computer Engineering, \\ Purdue University, West Lafayette, Indiana 47907, USA \\
	$^{5}$ Department of Physics and Astronomy, Birck Nanotechnology Center, and Station Q Purdue, \\ Purdue University, West Lafayette, Indiana 47907, USA \\
	$^{6}$ School of Materials Engineering, Purdue University, West Lafayette, Indiana 47907, USA \\
	$^{7}$ Center for Quantum Devices and Station Q Copenhagen, Niels Bohr Institute, \\ University of Copenhagen, 2100 Copenhagen, Denmark \\
	$^{*}$ These authors contributed equally to this work
\end{center}

\begin{center}
\begin{tcolorbox}[width=0.8\textwidth, breakable, size=minimal, colback=white]
	\small
	We study the ground state spin of nine subsequent occupancies of the multielectron GaAs quantum dot, by means of coupling it to the neighboring two-electron double quantum dot. For all nine occupancies we map the crossing of the spin states as a function of the external magnetic field, in the vicinity of the interdot charge transition between a single- and a multielectron quantum dot. We also perform a time-resolved measurement of the exchange oscillations in the same regime. These measurements enable us to identify even and odd occupancies of the multielectron quantum dot. For three even occupancies we observe no exchange interaction leading indicating a spin-0 ground state. For one even occupancy we observe the exchange interaction and assign a spin 1 to the multielectron quantum dot ground state. For all five odd occupancies we observe the exchange interaction and assign a spin 1/2 to the ground state. We demonstrate that for three odd occupancies the exchange interaction changes sign in the vicinity of the charge transition. For one of these the exchange interaction is negative (i.e. triplet-preferring) beyond the interdot charge transition, consistently with the observed spin-1 for the next occupancy. We develop the Hubbard model involving two orbitals of the multielectron quantum dot. Allowing for the spin correlation energy (i.e. including a term favoring the Hund's rules) and different tunnel coupling to different orbitals we qualitatively reproduce the measured exchange profiles.
\end{tcolorbox}
\end{center}

\section{Introduction}

Spins in semiconductornanostructures offer a wide variety of approaches to quantum computing. These include approaches based on gate-defined single-electron quantum dots realized in GaAs/AlGaAs heterostructures~\cite{Nowack2011,Shulman2012,Gaudreau2011,Studenikin2012,Cao2016,Malinowski2017,Bertrand2016}, Si/SiGe quantum wells~\cite{Maune2012,Kim2014,Eng2015,Kawakami2016,Takeda2016} or in MOS nanodevices~\cite{Veldhorst2014,Maurand2016}, as well as spins localized on crystal defects such as phosphorus donors in silicon~\cite{Pla2012,Muhonen2014}.  Along with the range of materials, spins trapped in quantum dots offer a myriad of qubit encodings, including single-~\cite{Veldhorst2014,Kawakami2016,Takeda2016}, double-~\cite{Foletti2009,Maune2012,Kim2014,Cerfontaine2016} and triple-dot~\cite{Medford2013a,Medford2013,Eng2015} schemes, each with distinct advantages.

In contrast to the vast range of spin qubit encodings, there are only a handful of demonstrations of two-qubit entangling operations~\cite{Nowack2011,Shulman2012,Nichol2017,Veldhorst2015} that are required for quantum computing.  Approaches to two-qubit entangling gates based on direct exchange interaction between neighboring tunnel-coupled quantum dots~\cite{Nowack2011,Veldhorst2015} offer fast, high-fidelity operation~\cite{Martins2016,Reed2016}.  Unfortunately, these approaches require the dots to be extremely close, which makes fabrication and cross-coupling between qubits a challenge for multi-qubit systems~\cite{Taylor2005a,Wang2015,Zajac2016}.  In contrast, approaches based on direct charge dipole-dipole interaction can offer longer ranges, but suffer from weak coupling (and thus slow gate times) and comparatively lower fidelities~\cite{Shulman2012,Nichol2017}.  This dipole-dipole interaction could be mediated by a superconducting cavity~\cite{Burkard2006,Liu2014,Viennot2015,Mi2016,Russ2015a,Srinivasa2016}, providing a mechanism to couple over even longer distances, which is commonly used for superconducting qubits~\cite{DiCarlo2009}.  However, the small dipole moments and susceptibility to charge noise make it unclear whether these approaches will lead to improvements in gate speed and fidelity.

An attractive alternative that has recently been proposed~\cite{Srinivasa2015,Mehl2014a} and demonstrated~\cite{Baart2017,JB-mediated_exchange} is to base two-qubit coupling on the fast exchange interaction, using an intermediate quantum system as a mediator.
This approach combines the fast coupling of the exchange interaction with the ability to arrange quantum dots at separations that are compatible with current fabrication techniques.
In particular, a mesoscopic multielectron quantum dot~\cite{Folk1996,Kouwenhoven1997,Stewart1997,Folk2001,Alhassid2000,Negative-J} could serve as both coupling mediator and spacer~\cite{JB-mediated_exchange}, providing a pathway for scalability to multi-qubit systems.

To serve as a mediator and spacer, a multielectron quantum dot needs to fulfill several requirements:
\begin{enumerate}
	\item Its physical size must space neighboring dots by at least a few hundred nanometers. This distance is necessary to facilitate the fabrication of all of the required control and readout gates for each qubit.  It would also be necessary if one wished to couple more than two qubits to the mediator.
	\item The spin of the multielectron quantum dot ground state must be well defined, to enable the interaction between qubits without entangling with the mediator~\cite{Hu2001}.  The multielectron quantum dot with a non-degenerate spinless ground state would be the easiest to exploit as a coupler~\cite{Mehl2014a,Srinivasa2015}.
	\item The level spacing of the multielectron quantum dot and the tunnel couplings must be larger than the energy of the thermal fluctuations ($k_BT\approx 10$~$\mu$eV for $T=100$~mK) and the excitation spectrum of the control voltage pulses ($\approx 20$~$\mu$eV for 5~GHz bandwidth). This condition is necessary to guarantee that the mediator will be prepared in the ground state and to avoid its accidental excitation.
	\item  The ground state spin, level spacing and tunnel coupling of the multielectron quantum dot must be tunable with a high yield.  These parameters depend on the mesoscopic details of the multielectron dot, which may not be easily controlled.
	\item The strength of the exchange interaction must provide a competitive timescale for two qubit gates. We estimate that 100~ns is the upper bound on a viable two-qubit gate. This puts a bound on the minimum coupling strength of roughly $0.01$~$\mu$eV.
\end{enumerate} 

\begin{figure}[tb]
	\includegraphics[width=\textwidth]{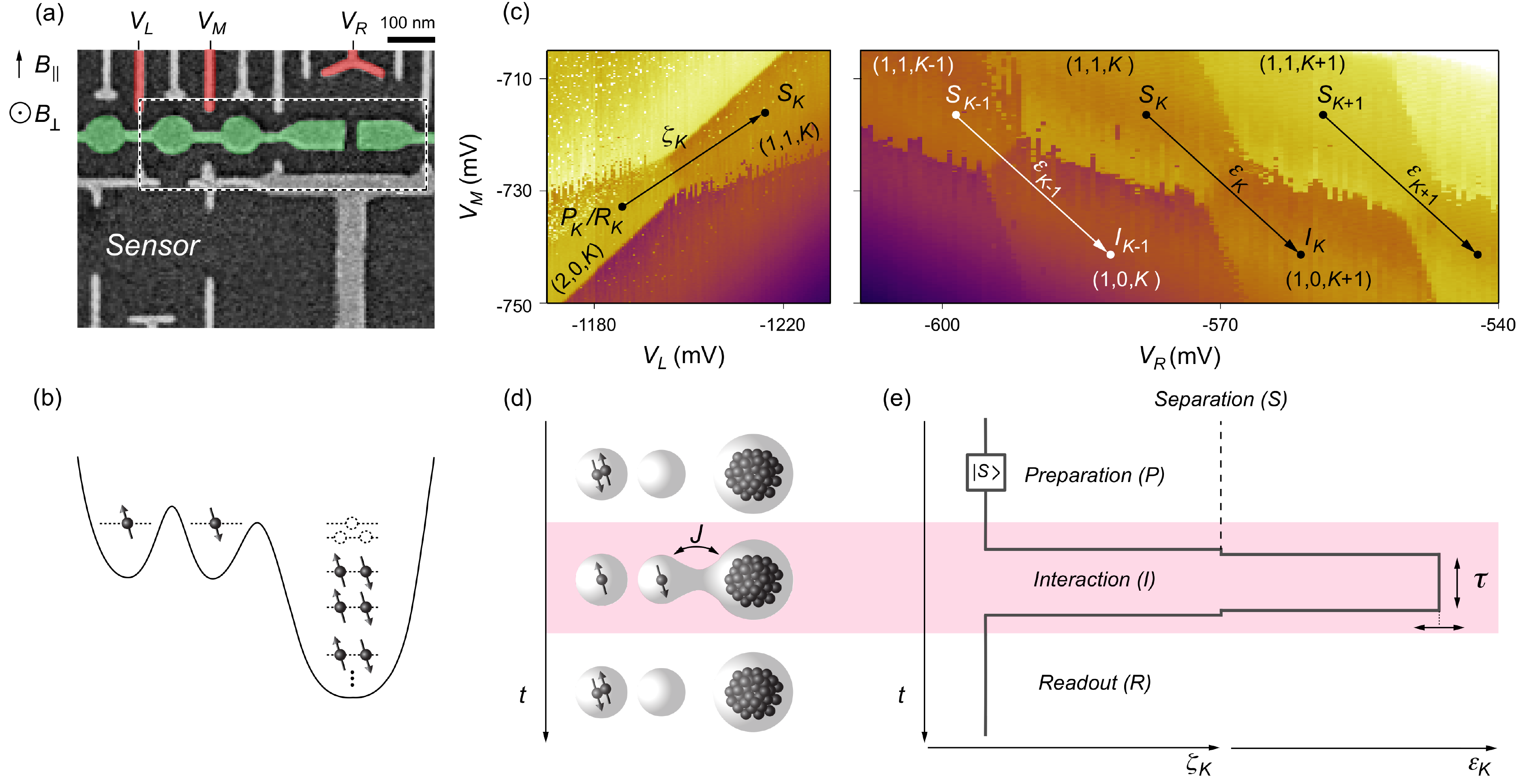}
	\caption[The device and performed pulse sequence]{
	(a) Micrograph of the device. Light-gray and red-colored gates metallic gates deplete the 2DEG underneath. Green-colored accumulation gate deepens the confining potential of the quantum dots formed underneath. Voltage pulses applied to the gates $V_{L,M,R}$ are used to control the electrons on nanosecond timescale.
	(b) Illustration of the electron configuration in a triple quantum dot. 
	(c) Charge diagrams of the triple quantum dot. The left panel shows the interdot charge transition of two-electron double quantum dot. The right panel presents the charge transition at which the transfer of the electron between the middle dot and multielectron quantum dot occurs. Labels $P_K$, $R_K$ and $S_K$ indicate positions in gate voltage space at which the electron pair is, respectively, prepared, readout and separated. Arrows, labeled $\zeta_K$ and $\varepsilon_K$ indicate axes in gate-voltage space used to define the voltage pulses. The voltages during the interaction step ($I$) are varied, but remain on $\zeta_K$ or $\varepsilon_K$ axis.
	(d) A cartoon illustrating the performed pulses. First, the pair of electrons is prepared in a singlet state on the leftmost dot. Next, one of electrons is transfered to the middle dot, and the interaction with the multielectron dot is induced. In the end, the state of the middle-dot electron spin is measured relative to the left-dot electron spin by means of Pauli blockade.
	(e) Performed voltage pulse illustrated in terms of $\zeta_K$ and $\varepsilon_K$ parameters.
	}
	\label{many_charge_states:fig1}
\end{figure}

In this article, we demonstrate that these requirements can be fulfilled by a multi-electron quantum dot (except for the final requirement, addressed elsewhere~\cite{JB-mediated_exchange}).  We implement a composite system consisting of the two-electron double quantum dot coupled to the multielectron quantum dot in GaAs device [Fig.~\ref{many_charge_states:fig1}(a),(b)]. Our approach is to study the interaction of one of the electrons in a two-electron double quantum dot with a multielectron quantum dot, in the vicinity of the charge transition at which the single electron tunnels into the multielectron quantum dot. The double quantum dot provides a mechanism to prepare the electronic spin state and for spin readout.

By these means, we study the properties of the multi-electron dot in nine different charge occupancies. We are able to identify even and odd occupancy of the dot. In all of the cases we are able to identify the total spin of the ground state of the multielectron quantum dot; in terms of the occupancy of the dot, the ground states form a sequence of alternating spin-0 (even occupancy) and spin-1/2 (odd occupancy) states interrupted once by single case of a spin-1 ground state in a dot with even occupancy. This progression of the ground state spin is consistent with previous studies of the ground state spin of a multielectron quantum dot~\cite{Folk2001,Lindemann2002}.

Moreover, we discover a peculiar behavior of the exchange interaction at the charge transition for the case of spin 1/2 multielectron dot ground state. Namely, the exchange interaction changes sign as a result of a few-millivolt change of a dot-defining gate voltages. A Hubbard model that includes two levels of the multielectron dot enables us to capture the energetics associated with the total spin of the multielectron quantum dot. From that model we derive a ``phase diagram'' that indicates four regimes with qualitatively distinct energy spectra and associated exchange interaction dependencies.

The article is organized as follows. In Sec.~\ref{many_charge_states:setup} we describe in detail the studied system and the sequences of voltage pulses used to induce the interaction between a single electron an the multielectron quantum dot. In Sec.~\ref{many_charge_states:progression} we present the sequence of the observed ground states as the occupancy of the multielectron quantum dot is increased one electron at the time. This enables us to propose a Hubbard model for description of the multielectron quantum dot. In Sec.~\ref{many_charge_states:0} we present the evidence for the spin 0 ground state for three of the studied electron occupancies. Sec.~\ref{many_charge_states:1/2} contains an in-depth study of the interaction between an electron and spin-1/2 state of the multielectron quantum dot for five different electron occupancies. In Sec.~\ref{many_charge_states:1} we present the data supporting the observation of a spin-1 ground state. In the final Sec.~\ref{many_charge_states:summary} we summarize the results.

\section{Experimental setup and techniques}
\label{many_charge_states:setup}

The quantum dots are defined in GaAs/AlGaAs two-dimensional electron gas (2DEG), with electron density $2.5 \times 10^{15}$ m$^{-2}$ and a mobility $230$ m$^2$/Vs. The 2DEG is located 57 nm below the heterostructure surface. A layer of HfO$_2$ of thickness of 10~nm is deposited on top of the heterostructure to isolate the gold gates defined on top by electron beam lithography. The oxide layer has a double purpose: first, it prevents the current leakage through the Schottky barrier that would appear under application of positive voltage to metallic gate deposited directly on top of GaAs; second, it blocks the tunneling events between the gates and the donor layer, which are the main sources of the samples instability~\cite{Buizert2008}. The experiment is performed in the dilution refrigerator with the mixing chamber at 20~mK.

The SEM of the innermost part of the device is presented in Fig.~\ref{many_charge_states:fig1}(a). The light gray and colored structures are the metallic gates that are used to define the quantum dot confining potential. The green-colored accumulation gate is operated at small positive voltage of +40~mV. The remaining gates are operated at negative voltages to deplete the 2DEG. The accumulation gate is introduced in this design to increase the depth the quantum dot potential and improve the tunability.  With the use of accumulation gates the typical center-to-center distance of the single-electron quantum dots is 150~nm, which is approximately 30\% less than in designs that do not employ the accumulation gate~\cite{Medford2013,Yoneda2014,Nichol2017}.

In our device we tune two single electron quantum dots and a multielectron quantum dot, under the accumulation gate, in the region indicated by a dashed rectangle. Based on the 2DEG density and the device geometry (dot size roughly 120$\times$250~nm) we estimate the electron occupancy of the multielectron quantum dot to be between 50 and 100. Splitting of the accumulation gate has no effect on the formation of the multielectron dot. In red we indicate the gates, labeled $V_{L,M,R}$ connected to the wide bandwidth coaxial lines in the dilution refrigerator. Voltage pulses on these gates are used to perform the sub-microsecond charge and spin manipulations, while all DC gate voltages are modified to explore various occupancies of the multielectron dot.

In Fig.~\ref{many_charge_states:fig1}(c) we present the typical charge diagrams of the double-dot -- multielectron-dot system. On the left we present the charge diagram with respect to gate voltages $V_L$ and $V_M$. These gate voltages are dedicated to control the state of the double quantum dot.  This diagram presents the interdot charge transition for the double dot, connecting the (2,0,$K$) and (1,1,$K$) charge states, where ($L$,$M$,$R$) indicates a number of electrons in the left, middle and multielectron dot, respectively. By adjusting the voltages $V_L$ and $V_M$, in such a way that we remain on the gate voltage axis labeled $\zeta_K$ [Fig.~\ref{many_charge_states:fig1}(c), left panel] we control the position of the electrons in the double dot while maintaining the fixed number of the electrons on the multielectron dot.

On the $\zeta_K$ axis we define a point $P_K/R_K$ that serves as the preparation and readout of the double quantum dot spin state. We also define a separation point $S_K$ at which the electrons in the double dot do not interact with the multielectron quantum dot and weakly exchange interact with each other. For symbols $P$, $R$ and $S$ we will use the subscript $K$ to indicate the initial occupancy of the multielectron dot. Having chosen the separation point $S_K$ we map out the dot occupancies as a function of voltages $V_M$ and $V_R$ as illustrated in the right panel of Fig.~\ref{many_charge_states:fig1}(c). In this charge diagram we identify the point $S_K$ in the gate voltage space. Once we identify that point we define $\varepsilon_K$ axis that contains the $S_K$ point and goes through the interdot charge transition connecting the (1,1,$K$) and (1,0,$K+1$) charge states. By controlling the position on this axis we can induce the interaction between a single electron and a multielectron quantum dot, while preserving the reference electronic spin in the leftmost quantum dot. By slightly changing the DC tuning of the quantum dots, we can change the occupancy of the multielectron dot and define analogous axis for different charge states. These are schematically illustrated with the axes labeled $\varepsilon_{K-1}$ and $\varepsilon_{K+1}$ in Fig.~\ref{many_charge_states:fig1}(c).

Having defined the points $P_K/R_K$, $S_K$ and axes $\zeta_K$ and $\varepsilon_K$ for each occupancy $K$ of the multielectron quantum dot, we can perform pulses of the gate voltage pulses that will take us between the (2,0,$K$), (1,1,$K$) and (1,0,$K+1$) charge states and allow to study the interaction of the single electron with the multielectron quantum dot [illustrated in Fig.~\ref{many_charge_states:fig1}(d),(e)].  The pulse initiates at point $P_K$, with a pair of electrons prepared in the singlet state $\ket{S}$ on the leftmost quantum dot. From there, we move to the point $S_K$ and separate the two electrons.  We wait a single clock cycle of the waveform generator at point $S_K$, which varies between 0.83 and 2.5~ns. This step is necessary to ensure that we transfer the electron through the middle dot to the multielectron, instead of ejecting it into a lead, followed by injection of the electron to the multielectron dot from the opposite side. The waiting time is shorter then the dephasing time due to interaction with the nuclear spins~\cite{Petta2005,Reilly2008,Bluhm2010}, $T_2^* \approx 10$~ns.  The next step of the pulse moves along the $\varepsilon_K$ axis for time $\tau$. Both parameters $\varepsilon_K$ and $\tau$ are varied within a sequence of pulses.  It is during this stage that the interaction between the electron and multielectron quantum dot occurs.  We then return to the point $S_K$ for another clock cycle of the waveform generator.  Finally, we pulse back to the (2,0,$K$) charge configuration at point $P_K/R_K$. This charge configuration is reached only if the pair of electrons on the double quantum dot forms a spin singlet state, otherwise the system is blocked in (1,1,$K$) charge state. The reflectometry readout of the conductance trough the neighboring sensor dot lasting between 5 and 20~$\mu$s allows us to distinguish these states yielding a single-shot spin readout.

\section{The multielectron quantum dot}
\label{many_charge_states:progression}

Before we discuss the experiments revealing the exchange interaction between a single electron and a multielectron quantum dot, we present and discuss a model that can be used to describe the multielectron quantum dot and its coupling to the double dot.

For the several subsequent occupancies we observe alternating sequence of spin-0 and spin-1/2 ground states as summarized in Tab.~\ref{tab1} (inferred from experiments described in sections~\ref{many_charge_states:0}-\ref{many_charge_states:1}). This sequence is interrupted once by a spin-1 ground state instead of spin-0. These are consistent with findings of Folk \emph{et al.}~\cite{Folk2001} and Lindemann \emph{et al.}~\cite{Lindemann2002} who identify the ground states spin by studying the change of the Coulomb peak spacing in magnetic field. Based on these observations we model the multielectron quantum dot with the Hamiltonian:
\begin{equation}
	\hat{H}_R = \frac{e^2}{2C_R} \hat{N}_R^2 + \sum\limits_{\substack{\lambda \in \mathbb{N} \\ \sigma=\uparrow,\downarrow}} E_\lambda \hat{a}_{\lambda,\sigma}^\dag \hat{a}_{\lambda,\sigma} - \frac{\xi}{2} \hat{S}^2,
	\label{eq:JB}
\end{equation}
where $e$ is the electron charge; $C_R$ is the dot self-capacitance; $\hat{N}_R$ is the operator counting the total number of electrons; $\hat{a}_{\lambda,\sigma}^{(\dag)}$ are the annihilation (creation) operators for electron on the single particle level $\lambda$ with spin $\sigma$; $E_\lambda$ are the energies of the single particle levels; $\hat{S}$ is the total spin operator and $\xi$ is the spin correlation energy. The subscript $R$ in this formula refers to the multielectron dot as right dot, as opposed to the left ($L$) and middle ($M$) single-electron dots.

\begin{table}[tb]
	\caption{ Summary of the inferred ground state spins for 9 subsequent charge occupancies of the multielectron quantum dot. This sequence of alternating 0 and 1/2 spin states is interrupted once with a spin 1 ground state. The even reference occupancy $2N$ is chosen arbitrarily to unambiguously label the multielectron dot occupancy and emphasize the electron parity.}
	\vspace{5pt}
	\label{tab1}
	\centering
\begin{tabular}{lll}
\hline \hline
\textbf{Multielectron dot} & \textbf{Inferred ground} & \textbf{Experimental} \\ 
\textbf{occupancy} & \textbf{state spin} & \textbf{evidence} \\ \hline
$2N \! - \! 5$ & 1/2 & Fig.~\ref{many_charge_states:fig10} \\ 
$2N \! - \! 4$ & 0 & Fig.~\ref{many_charge_states:fig4}(a) \\ 
$2N \! - \! 3$ & 1/2 & Fig.~\ref{many_charge_states:fig7} \\ 
$2N \! - \! 2$ & 0 & Fig.~\ref{many_charge_states:fig4}(b) \\ 
$2N \! - \! 1$ & 1/2 & Fig.~\ref{many_charge_states:fig8} \\ 
$2N$ & 0 & Fig.~\ref{many_charge_states:fig3} \\ 
$2N \! + \! 1$ & 1/2 & Fig.~\ref{many_charge_states:fig9} \\ 
$2N \! + \! 2$ & 1 & Fig.~\ref{many_charge_states:fig13} \\ 
$2N \! + \! 3$ & 1/2 & Fig.~\ref{many_charge_states:fig11} \\ \hline\hline
\end{tabular} 
\end{table}

The relative strength of the three terms present in this Hamiltonian determine the properties of the multielectron quantum dot. The charging energy of the multielectron quantum dot $e^2/2C_R \approx 1$~meV is estimated from the distance between the multielectron dot charge transitions [$\Delta V_R \approx 20$~mV; Fig.~\ref{many_charge_states:fig1}(c)] and typical lever arm between the gates and the dots in devices of similar design ($\approx 0.05~e$). The charging energy may vary slightly as a function of the dot occupancy, as with additional electrons the quantum dots increase in size. For the results presented here it is only relevant that the charging energy is much larger than all other energy scales.  

From the lithographic size of the device, we estimate the typical level spacing~\cite{Datta1997} to be $\langle \Delta E \rangle = \pi \hbar/m^* A \approx 120$~$\mu$eV, where $\hbar$ is the reduced Planck constant, $m^*$ is the effective electron mass in GaAs and $A$ is the area of the 2-dimensional quantum dot. However, the lack of symmetry causes the level spacings to vary.  It remains an open question of how the distribution of the level spacings $\Delta E$ and correlations between them are determined for a given mesoscopic quantum dot. These kind of distributions may be described using random matrix theory with the orthogonal ensemble~\cite{Folk1996,Folk2001,Brouwer1999,Kurland2000,Baranger2000}, which by itself neglects the interaction effect. The interaction effects are usually introduced by means of random-phase approximation~\cite{Blanter1997}, mean-field approximation\cite{Baranger2000}, using density functional theory~\cite{Hirose2002}, Anderson model~\cite{Sivan1996} or by an on-site Hubbard interaction term~\cite{Brouwer1999} (for review see Ref.~\cite{Alhassid2000}). For the results presented here it is most relevant that the level spacings distribution has the width $\sigma_{\Delta E}$ comparable to $\Delta E$ and $\xi$. Also, the excitation spectrum is highly correlated for a few subsequent charge states~\cite{Stewart1997}.

In general, the spin correlation energy $\xi$ is the most difficult quantity to estimate due to the lack of data in the literature.  We make the assumption that $\xi/2$ comparable, but smaller than, $\langle \Delta E\rangle$, based on two observations. On the one hand there is no macroscopic polarization of the electronic spins, which would originate from the Stoner instability that appears in case\cite{Andreev1998,Kurland2000,Alhassid2000} $\xi/2 > \langle \Delta E\rangle$. On the other hand, it is not uncommon that the ground state has a spin $>1/2$, which can occur only when energy of the first excited state is smaller than $\xi$. This excludes the possibility $\xi \ll \langle \Delta E\rangle$.

From the perspective of designing two-qubit entangling gates, configurations of the multielectron quantum dot with total spin 0 are desirable.  In our study, we observe a spin-0 ground state for 3 out of 4 even occupancies. We claim this provides sufficient reliability for use in a scalable quantum dot system -- in case of finding the spin-1 ground state of the coupler it is sufficient to change its occupancy by 2.

Starting from the multielectron-dot Hamiltonian (Eq.~\ref{eq:JB}), we construct a simplified model that is sufficient to describe a multielectron dot with a given occupancy to which a neighbouring double quantum dot is coupled. We include a gate voltage $V_g$ tuned such that the ground state charge configuration of the multielectron dot has $N_g = C_g V_g/e$ electrons, where $C_g$ is the gate capacitance.  For the spin-0 case we consider here, $N_g$ is even.  The Coulomb energy term in the Hamiltonian with this gate voltage is then given by $\frac{e^2}{2C_R} (\hat{N}_R - N_g)^2 = \frac{e^2}{2C_R} \hat{n}_R^2$, where $\hat{n}_R = \hat{N}_R - N_g$ describes the number of \emph{excess} electrons on the multielectron dot in addition to the electrons paired up as singlets and forming an effective spin-0 ``vacuum''.  We can now consider a restricted model of the multielectron dot, where we need only describe the multielectron dot level $\lambda$ corresponding to the lowest unoccupied level (or levels) of the dot above this effective ``vacuum'', and label this level by $R$ rather than $\lambda$.  (If we wish to consider more levels, they are labelled $R1$, $R2$, etc.)  The excess occupancy $\hat{n}_{R,\sigma}$ counts the electrons with spin $\sigma$ in this single level and is restricted to 0 or 1.  As the occupancy is restricted, we can combine the energies of the Couloumb term $\frac{e^2}{2C_R} \hat{n}_R^2 $ with the energy levels $E_\lambda \hat{a}_{\lambda,\sigma}^\dag \hat{a}_{\lambda,\sigma}$ into a single Hamiltonian term describing the energetics of the multielectron dot:  $\hat{H}_R = \sum\limits_{\substack{\lambda \in R1, R2 \\ \sigma=\uparrow,\downarrow}} E_\lambda \hat{n}_{\lambda} - \frac{\xi}{2} \hat{S}^2$.

Our study involves modeling of the interaction between the spin occupying the middle dot $M$ [Fig.~\ref{many_charge_states:fig1}(b)] and one of the aforementioned levels of the multielectron quantum dot. The occupancy of the multielectron quantum dot will determine the nature of this interaction.  We consider three cases, ordered by increasing complexity:
\begin{itemize}
	\item All levels of the multielectron quantum dot are empty or doubly occupied and the total spin is zero.  In this case, the interaction of the double quantum dot with the multielectron quantum dot can be modeled as an effective interaction of the spin of the middle dot $M$ tunnel coupled to an \emph{unoccupied} dot $R$ (Sec.~\ref{many_charge_states:0}).
	\item There is a single unpaired spin in the multielectron quantum dot, and so it has a total spin 1/2 (Sec.~\ref{many_charge_states:1/2}).  In this case, the interaction of the double quantum dot with the multielectron quantum dot can be modeled as an effective interaction of the spin of the middle dot $M$ tunnel coupled to a single spin in the right dot $R$.  Depending on the details of the spin interaction terms of the multielectron quantum dot, we must consider both the partially occupied level of the multielectron quantum dot as well as other low-energy unoccupied.
	\item Several unpaired spins in the multielectron quantum dot couple to total nonzero spin, e.g., spin 1 (Sec.~\ref{many_charge_states:1}).
\end{itemize}

\section{Spin 0}
\label{many_charge_states:0}

\begin{figure}
	\centering
	\includegraphics[width=0.57\textwidth]{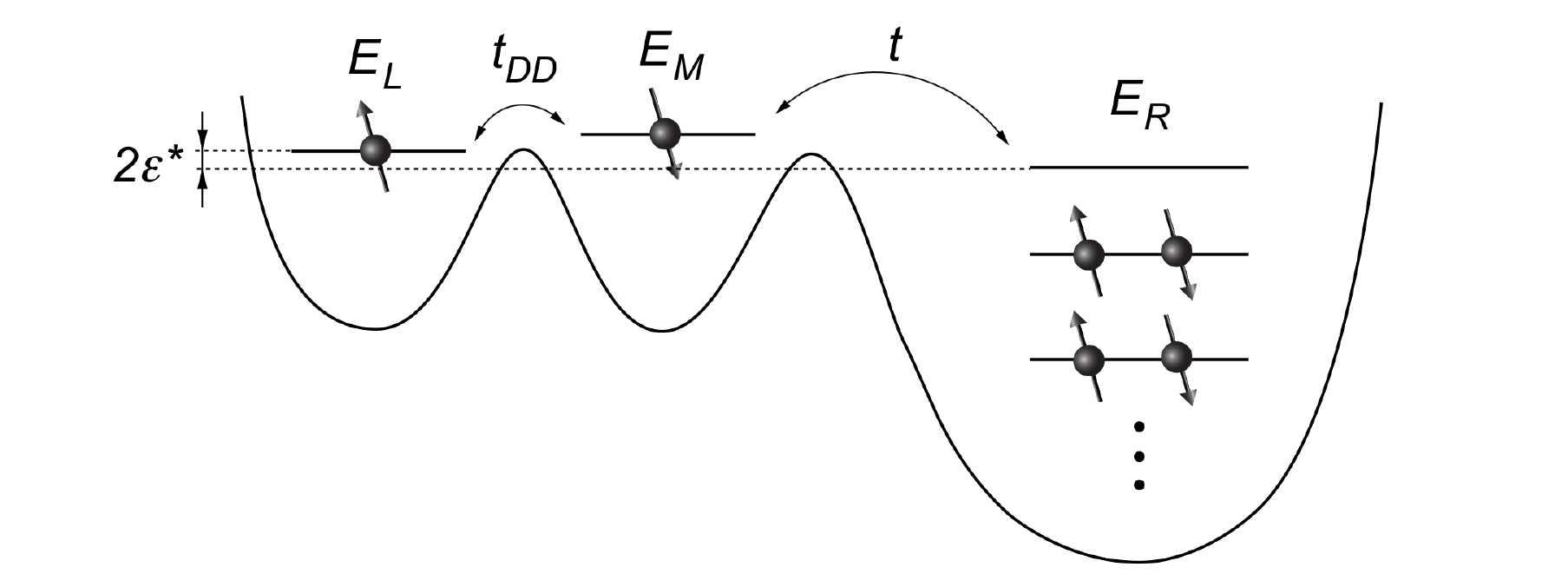}
	\caption[A schematics of the two-electron double quantum dot coupled to the even-occupied spinless multielectron dot]{
	A schematics of the two-electron double quantum dot coupled to the even-occupied spinless multielectron dot. Symbols $E_{L/M/R}$ indicate a single particle energy of the lowest orbitals in a double dot and a lowest unoccupied orbital in the multielectron dot. Arrows indicate tunnel couplings between two small dots $t_{DD}$ and between a middle and multielectron dot $t$. Detuning $\varepsilon^*$ is varied to obtain the leakage spectroscopy reconstruction presented in Fig.~\ref{many_charge_states:fig3}(d).
	}
	\label{many_charge_states:fig2}
\end{figure}

\begin{figure}[tb]
	\centering
	\includegraphics[width=0.57\textwidth]{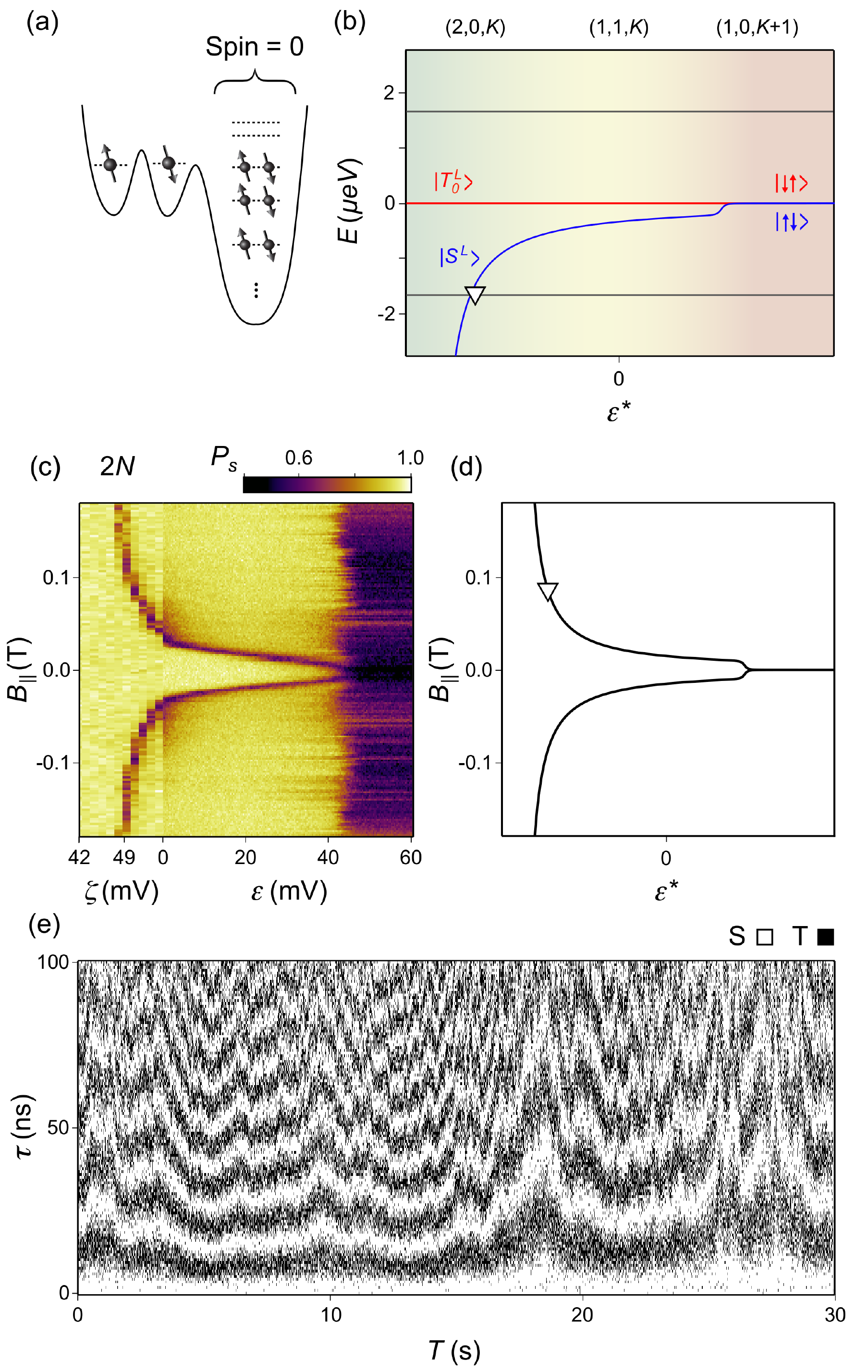}
	\caption[Evidence for a spin-0 ground state for the multielectron dot occupied by $2N$ electrons]{Evidence for a spin-0 ground state for the multielectron dot occupied by $2N$ electrons. (a) Schematic representation of the electron configuration for even-occupied multielectron dot with spin-0 ground state. The electrons in the multielectron dot are paired-up in singlets on the orbitals located below the Fermi level of the leads.
	(b) The energy diagram of the DD coupled to the multielectron dot in charge states $(2,0,2N)$, $(1,1,2N)$ and $(1,0,2N \! + \! 1)$. The white triangle indicate the features observed in (c).
	(c) The leakage spectroscopy revealing vanishing exchange interaction between two electrons. The line feature corresponds to the $S$-$T_+$ anticrossing presented in panel (b).
	(d) Reconstruction of the leakage spectrum using the simple microscopic model. 
	(e) The time-resolved measurement of the precession between $\ket{S}$ and $\ket{T_0}$ two-electron spin states in $(1,0,2N \! + \! 1)$ charge state due to Overhauser field gradient.
	}
	\label{many_charge_states:fig3}
\end{figure}

We first focus on the evidence that the even occupancies of the multielectron dot, specifically $2N \! - \! 4$, $2N \! - \! 2$ and $2N$, have a spin 0 ground state.  (Here, $2N$ indicates a specific, even, but unknown number of electrons).  The model for the multielectron dot introduced in Sec.~\ref{many_charge_states:progression} suggests that, in this case, the ground state is described by a configuration where all single-particle states below the Fermi energy are occupied by singlet pairs of electrons [Fig.~\ref{many_charge_states:fig2} and \ref{many_charge_states:fig3}(a)] and form an effective vacuum state, provided that the spin correlation term is smaller than the level spacing between the ground and the first excited state.  We therefore expect that the double dot will interact with the multielectron dot as if it were an unoccupied dot, and the spin of an electron tunneling into the multielectron dot would not change in the process.  In this sense, the double dot coupled to the multielectron dot with this even occupancy should be qualitatively similar to a two-electron triple-dot.

We describe this situation using a phenomenological model based on the Hamiltonian for the multielectron dot detailed in Sec.~\ref{many_charge_states:progression} and add terms describing the neighboring tunnel-coupled two-electron double quantum dot. Then, for spinless even-occupancy ground states we neglect the electron pairs singlet-paired on the orbitals below the Fermi energy. We also neglect all but the lowest unoccupied orbital, and finally we arrive at a Hubbard model of the three dots, each having a single orbital, labeled $L$, $M$, and $R$: 
\begin{align}
	\hat{H}_{\text{spin-0}} &= \sum_{i=L,M,R} E_i \hat{n}_i \nonumber \\
    &- t_{DD} \sum\limits_{\sigma=\uparrow,\downarrow}  (\hat{a}_{L,\sigma}^\dag \hat{a}_{M,\sigma} + \hat{a}_{M,\sigma}^\dag \hat{a}_{L,\sigma}) \nonumber \\
	&- t\sum\limits_{\sigma=\uparrow,\downarrow}  (\hat{a}_{M,\sigma}^\dag \hat{a}_{R,\sigma} + \hat{a}_{R,\sigma}^\dag \hat{a}_{M,\sigma}) \,.
	\label{eq:spin0Ham}
\end{align}
As illustrated in Fig.~\ref{many_charge_states:fig2} the first term describes the gate-tunable chemical potential of the left $L$, middle $M$ and the right multielectron $R$ dot. Second term describes the quantum dots charging energy and the third, capacitative coupling between the dots. The second and final lines incorporate the tunnel coupling $t_{DD}$ within a double dot and the tunnel coupling $t$ between the middle and right multielectron dot.

This effective Hamiltonian can be solved in the 2-electron configuration to yield a qualitative energy level diagram for the coupled system; a representative plot is shown in Fig.~\ref{many_charge_states:fig3}(b).  Recall that the ``unoccupied'' state of the multielectron dot is describing the effective ``vacuum'' state with $2N$ electrons in a spin-0 configuration.  We find that, in the (2,0,$2N$) charge state, the singlet and triplet states of the two-electron double dot are split by the exchange energy. In the (1,1,$2N$) state, the exchange splitting gradually decreases as the overlap of the electronic wavefunctions decreases. Finally, in the (1,0,$2N \! + \! 1$) configuration, the exchange splitting reduces to zero, as the two electrons occupy distant dots.

\begin{figure}[tb]
	\centering
	\includegraphics[width=0.57\textwidth]{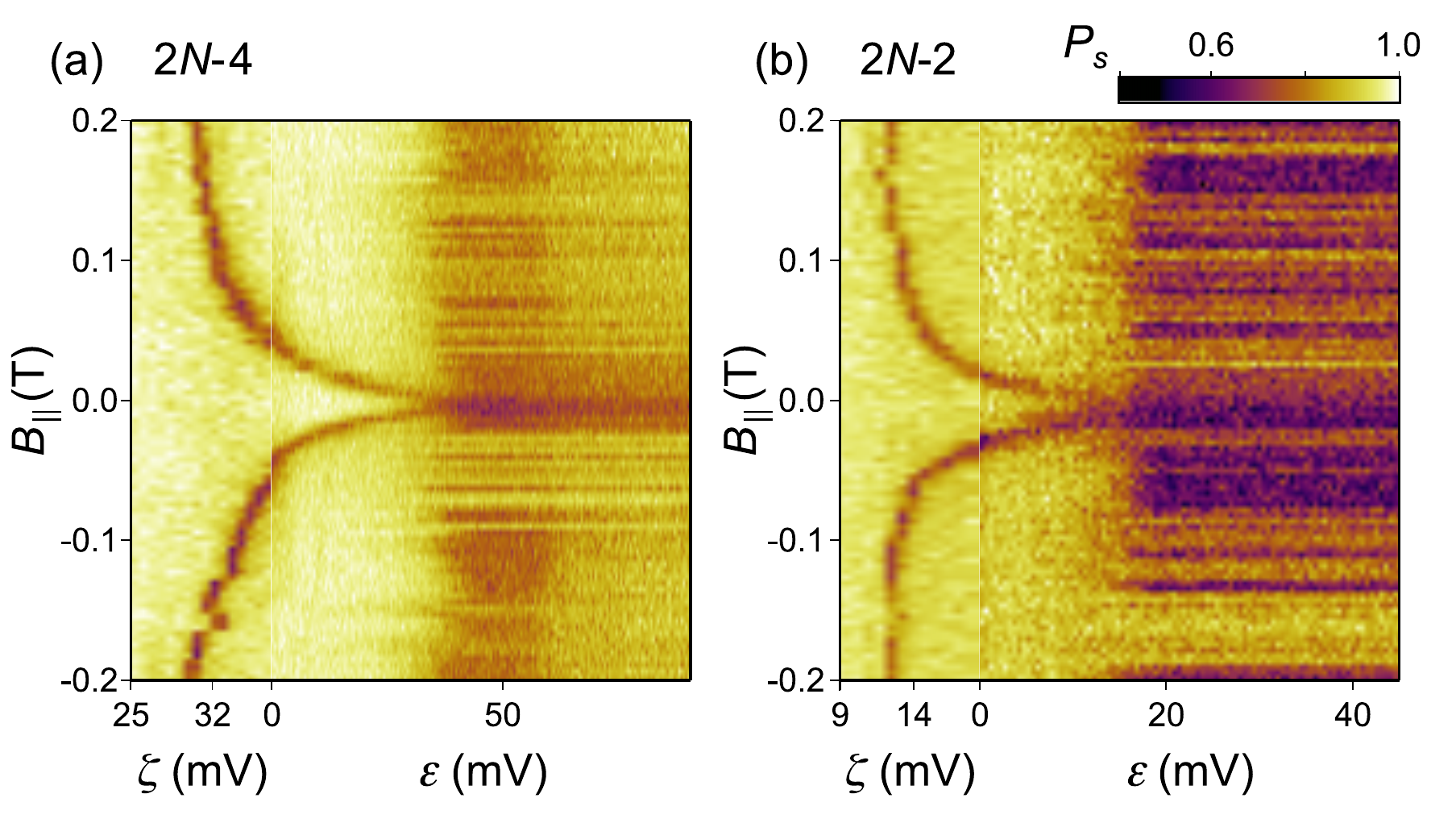}
	\caption[The leakage spectroscopy of the DD coupled to the spin-0 multielectron dot with $(2N \! - \! 4)$ and $(2N \! - \! 2)$]{
	The leakage spectroscopy of the DD coupled to the multielectron dot at the transition between 
	(a) $(2,0,2N \! - \! 4)$, $(1,1,2N \! - \! 4)$ and $(1,0,2N \! - \! 3)$ and
	(b) $(2,0,2N \! - \! 2)$, $(1,1,2N \! - \! 2)$ and $(1,0,2N \! - \! 1)$.
	}
	\label{many_charge_states:fig4}
\end{figure}

In Fig.~\ref{many_charge_states:fig3}(c) we map the exchange profile in these three regimes by means of leakage spectroscopy (an analogue of the ``spin funnel'' measurement of a double quantum dot~\cite{Petta2005,Maune2012}). In this measurement we prepare the double dot in a singlet state $\ket{S}$, and then pulse to the various interaction points $I_{2N}$ along the $\zeta_{2N}$ and $\varepsilon_{2N}$ axis for an interaction time of 150~ns. The decrease of singlet return probability measurement indicates leakage from the singlet state. We repeat this procedure for various values of the in-plane magnetic field $B_\parallel$ up to 200 mT. The result of such a measurement, for the multielectron dot occupancy of $2N$ is presented in Fig.~\ref{many_charge_states:fig3}(c). The observed line feature (white triangle) indicates the crossing of the singlet state $\ket{S}$ and the fully polarized triplet state $\ket{T_+} = \ket{\uparrow\uparrow}$. This line diverges to the high field in the (2,0,$2N$) configuration, indicating that the exchange interaction between the two electrons within the DD is very strong. The line gradually moves towards $B_\parallel = 0$ in the (1,1,2) configuration indicating a decrease of the exchange interaction strength. Finally, it converges to zero field in the (1,0,$2N \! + \! 1$) configuration when the two electrons are far separated. In this configuration we observe an increased triplet return probability, independent on the magnetic field. This increase is due to mixing of singlet $\ket{S}$ and unpolarized triplet $\ket{T_0}$ states the Overhauser field gradient between the left and the multielectron dot. The measurement yields very similar features in cases of $2N \! - \! 4$ and $2N \! - \! 2$ multielectron dot occupancy, as presented in Fig.~\ref{many_charge_states:fig4}.

We can reconstruct the qualitative features of the leakage spectrum using our microscopic model of the system. Using the spectrum obtained from the Hamiltonian of Eq.~\ref{eq:spin0Ham} with the addition of Zeeman energy due to an external perpendicular magnetic field $B_\parallel$, allows us to reproduce the expected leakage spectroscopy behaviour, as shown in Fig.~\ref{many_charge_states:fig3}(d).  We identify the ground state associated with preparation of the double dot in the singlet state, and then select the locus of points as a function of detuning $\varepsilon$ and in-plane magnetic field $B_\parallel$ where this singlet intersects with the fully polarized state $\ket{T_+}$.

To confirm the origin of the increased triplet outcome probability for (1,0,$2N \! + \! 1$) configuration, we perform a time-resolved measurement of the Overhauser field gradient~\cite{Barthel2009,Foletti2009,Barthel2012,Shulman2014,Delbecq2016,Malinowski2017a} between the leftmost dot and the multielectron dot. For that purpose we fix the interaction point in the (1,0,$2N \! + \! 1$) charge configuration and cyclically vary the waiting time $\tau$ from 0 to 100 ns. The cycle is run continuously and a result of single shot readout at the end of each pulse is recorded. A time trace showing 30~s reveals coherent oscillations between the singlet $\ket{S}$ and triplet $\ket{T_0}$ states [Fig.~\ref{many_charge_states:fig3}(e)]. The oscillation frequency varies on a timescale of seconds and within a range of tens of megahertz, consistent with the dynamics of the GaAs nuclear spin bath~\cite{Reilly2008,Barthel2009,Malinowski2017a,Delbecq2016}.

\section{Spin 1/2}
\label{many_charge_states:1/2}

When the multielectron dot has odd occupancy, the ground state is spin-1/2 and the resulting coupled system is much more complex than the  previous spin-0 case. As a reference, we begin this section by first presenting the set of results for a three-electron triple quantum dot (tuned in the same device). Having described the relevant physics of this coupled system, we present two scenarios observed in our experiments that result from replacing a single-electron quantum dot by a odd-occupied multi-electron quantum dot. This investigation involves a detailed study of the two distinct profiles of the exchange interaction together with their interpretation within a microscopic model that predicts two other possible profiles.

\subsection{Three-electron triple quantum dot}
\label{subsection:TQD}

\begin{figure}[tb]
	\centering
	\includegraphics[width=0.57\textwidth]{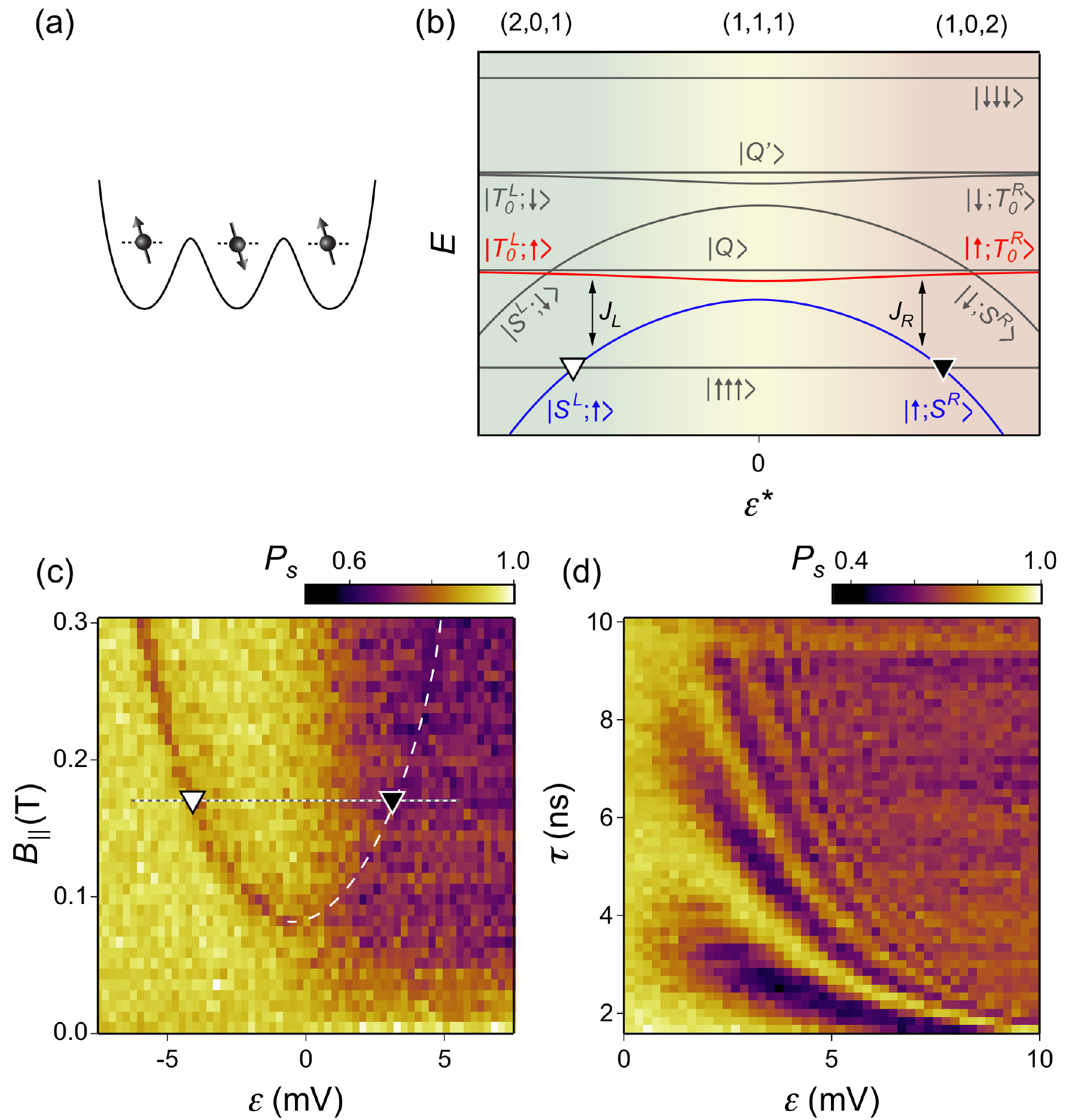}
	\caption[Leakage spectroscopy and exchange oscillations in a three-electron triple quantum dot]{
	(a) Schematic of the three-electron triple quantum dot which serve as a reference for the study of a multielectron quantum dot tunnel-coupled to the double quantum dot.
	(b) Energy diagram of the three-electron triple dot spin states for finite external magnetic field. Markers indicate crossings observed in the leakage spectroscopy measurement (c).
	(d) Exchange oscillations pattern revealing oscillations of monotonously increasing frequency.
	}
	\label{many_charge_states:fig5}
\end{figure}

The description of the three-electron triple quantum dot [Fig.~\ref{many_charge_states:fig5}(a)] requires understanding the spectrum of the possible spin states. The spectrum at the transition between the (2,0,1), (1,1,1) and (1,0,2) charge states for finite magnetic field is presented in the energy diagram~\cite{Laird2010,Gaudreau2011,Medford2013a,Poulin-Lamarre2015} in Fig.~\ref{many_charge_states:fig5}(b). In the leftmost part, in the vicinity of the (2,0,1)-(1,1,1) charge transition, only the exchange energy $J_L$ between left and the middle dot is significant, while the exchange energy $J_R$ between middle and the right quantum dot can be neglected. Therefore, near this charge transition, the eigenstates are approximated by a tensor product of the two-electron spin states of the left double quantum dot $\ket{S^L}$, $\ket{T_i^L}$ (where $i=0,+,-$) together with a state of the right single-electron quantum dot $\ket{\uparrow}$ or $\ket{\downarrow}$. These states will be Zeeman-split, according to the projection of the total spin on the external magnetic field direction. Moreover the two singlet-like states $\ket{S^L; \uparrow}$ and $\ket{S^L; \downarrow}$ will have lower energy than the triplet-like states $\ket{T_i^L; \uparrow}$ and $\ket{T_i^L; \downarrow}$ due to the exchange interaction.

Conversely, in the vicinity of the (1,1,1)-(1,0,2) charge transition, $J_L$ is negligible while $J_R>0$ is significant.  In this region, the eigenstates are approximated by tensor products of the left dot spin states $\ket{\uparrow}$ or  $\ket{\downarrow}$ and the two-electron spin states of the right double quantum dot $\ket{S^R}$ or $\ket{T_i^R}$. In the middle region with (1,1,1) charge occupancy, the eigenstates continuously vary between these two limiting cases. Their exact structure is not relevant for the analysis presented here. However the important observation is that simultaneous presence of the exchange $J_L$ and $J_R$ lifts all degeneracies.

The splitting between the lower singlet-like state and the triplet-like state with the total spin $3/2$ [red colored lines in Fig.~\ref{many_charge_states:fig5}(b)] can be mapped out using leakage spectroscopy~\cite{Poulin-Lamarre2015} [Fig.~\ref{many_charge_states:fig5}(c)] with a procedure similar to the one described in Sec.~\ref{many_charge_states:0}. In this case the system is prepared in the $\ket{S^L ; \uparrow}$ state, while the line feature  indicates leakage from the singlet-like state to the fully polarized $\ket{\uuu}$ state [white and black triangles in Fig.~\ref{many_charge_states:fig5}(b),(c)]. We observe that the line diverges to high magnetic field for large positive and negative values of $\varepsilon$, consistent with the decrease of the energy of the singlet-like state $\ket{S^L; \uparrow}$ or $\ket{\uparrow; S^R}$ in the (2,0,1) or (1,0,2) electron configuration, respectively.

Moreover, in this leakage spectroscopy experiment, we observe that the background probability of the singlet return $P_S$, which is independent of the magnetic field, decreases with increasing value of $\varepsilon$. 
This indicates that the eigenstates on the left side ($\varepsilon<0$) differ from the right side ($\varepsilon>0$) of the spectrum. In fact, preparing the three-electron system in the $\ket{S^L ; \uparrow}$ state and interacting it on the right side ($\varepsilon>0$) leads to coherent exchange oscillations~\cite{Laird2010,Medford2013}.
These kind of oscillations are presented in Fig.~\ref{many_charge_states:fig5}(d), in which the interaction time $\tau$ is varied on the few-nanosecond timescale. Additionally, the frequency of the oscillations becomes faster for larger values of $\varepsilon$ which reflects the increase of the exchange interaction $J_R$ between the middle and the right quantum dot.

This precession can be exploited for the operation of the exchange-only qubit~\cite{Laird2010,Medford2013,Eng2015}. Based on this description of the three-electron triple dot behavior as a reference, in the following sections we will present the results of the leakage spectroscopy and exchange oscillations measurements performed on the system consisting of the two-electron double quantum dot coupled to the third multielectron dot with an odd-occupancy spin-1/2 ground state. 

\subsection{Case study: Negative exchange interaction at the charge transition}
\label{subsection:positive-J}

\begin{figure}[tb]
	\centering
	\includegraphics[width=0.57\textwidth]{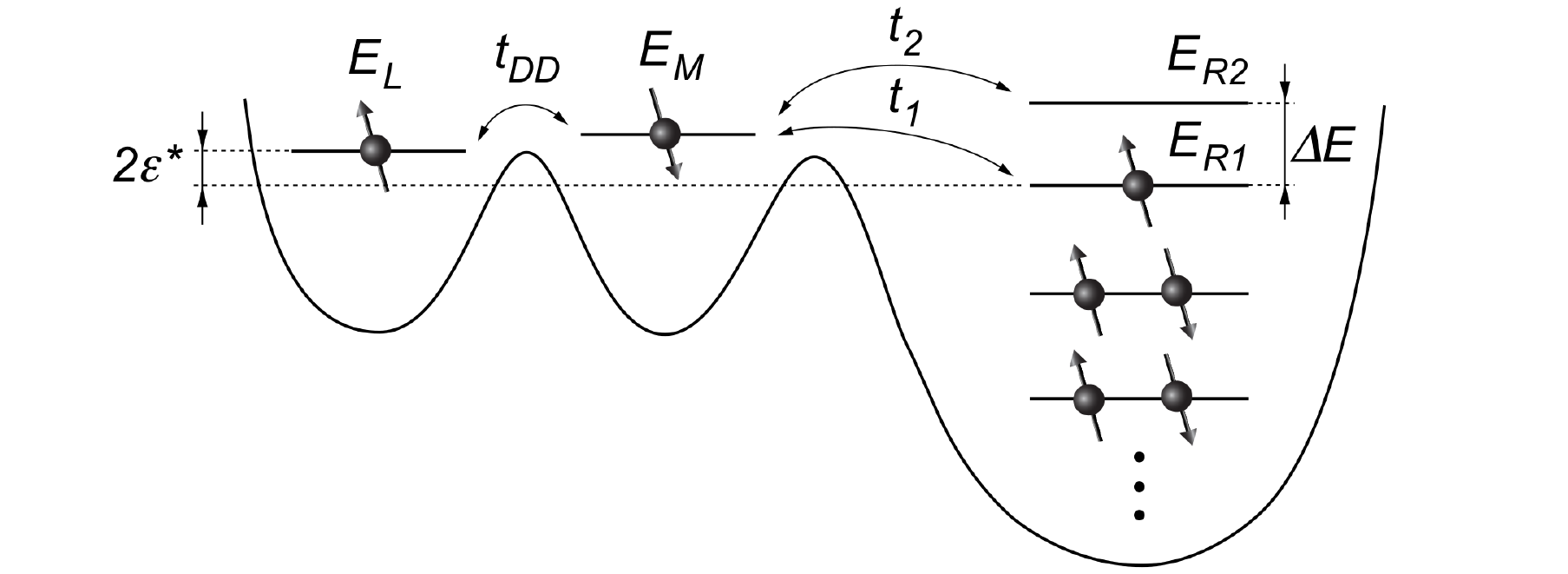}
	\caption[Schematics of a two-electron double quantum dot coupled to a odd-occupied spin-1/2 multielectron dot]{
	Schematics of a two-electron double quantum dot coupled to a odd-occupied spin-1/2 multielectron dot. Symbols $E_{L/M/R1/R2}$ indicate a single particle energy of the lowest orbitals in a double dot and two lowest orbitals above the effective vacuum in the multielectron dot. Arrows indicate tunnel couplings between two small dots $t_{DD}$ and between a middle and each of the orbitals in a multielectron dot $t_{1/2}$. Energy difference between the two orbitals on the multielectron dot is indicated by $\Delta E = E_{R2} - E_{R1}$. Detuning $\varepsilon^*$ is varied to obtain the leakage spectroscopy reconstruction presented in Fig.~\ref{many_charge_states:fig7}(e).
	}
	\label{many_charge_states:fig6}
\end{figure}

\begin{figure}[tb]
	\centering
	\includegraphics[width=0.57\textwidth]{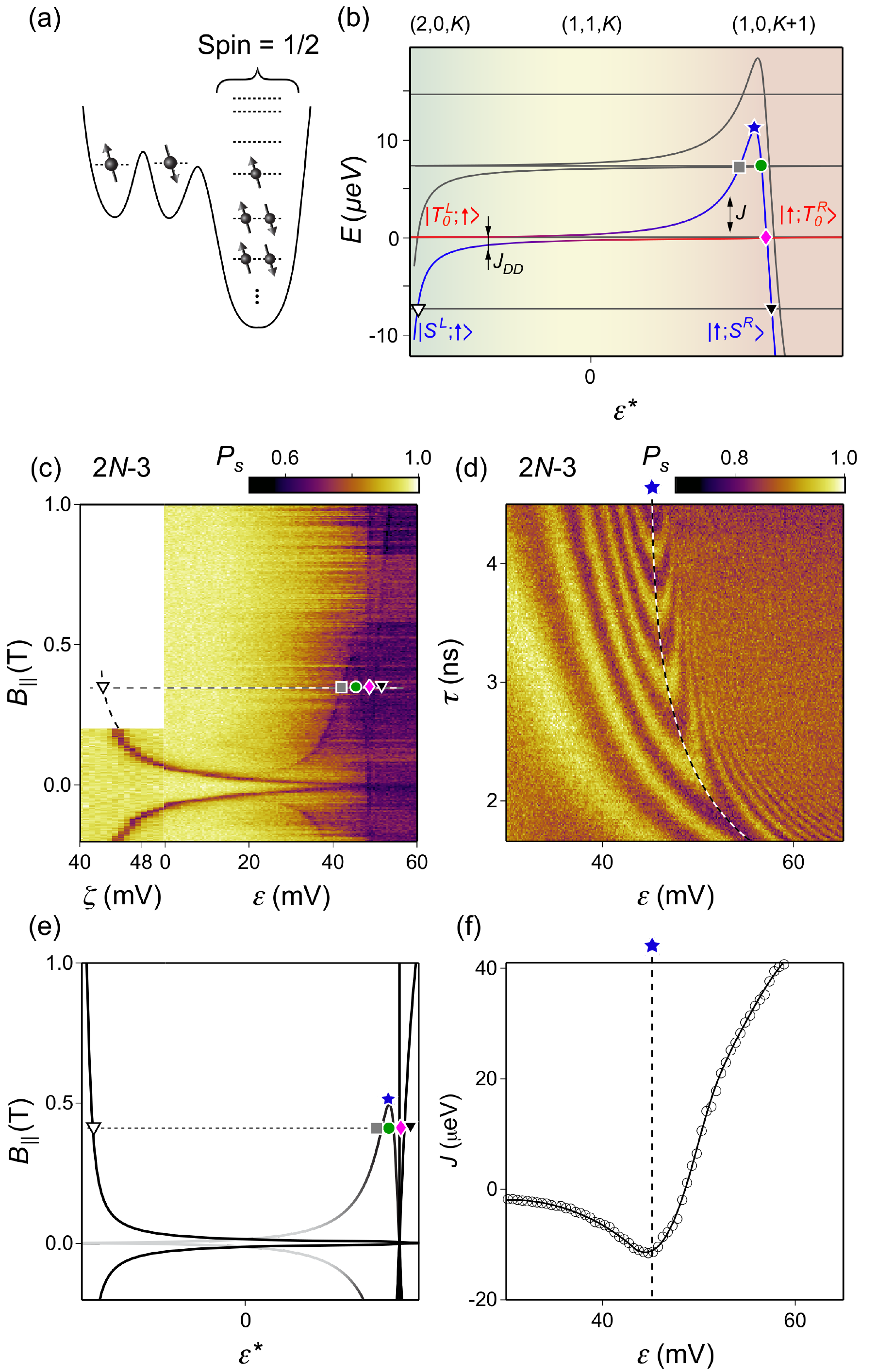}
	\caption[Leakage spectroscopy and exchange oscillations in a two-electron double quantum dot coupled to $(2N \! - \! 3)$-occupied multielectron quantum dot]{
	(a) Schematics of the odd-occupied multielectron quantum dot with spin 1/2 ground state tunnel coupled to the two electron double quantum dot.
	(b) The inferred energy diagram at the transition between (2,0,$2N \! - \! 3$), (1,1,$2N \! - \! 3$) and (1,0,$2N \! - \! 2$) electronic configurations, for a finite magnetic field. The markers indicate the crossings revealed by the leakage spectroscopy measurement presented in panel (c).	
	(d) Time resolved measurement of the exchange oscillations between the $(2N \! - \! 3)$-occupied spin-1/2 multielectron quantum dot and the neighboring electron.
	(e) Reconstruction of the leakage spectrum using the simple microscopic model.
	(f) Dependence of the exchange energy extracted from the pattern of exchange oscillations in panel (d).
	}
	\label{many_charge_states:fig7}
\end{figure}

Now we focus on the two odd occupancies of the multielectron quantum dot, $2N \! - \! 3$ and $2N \! - \! 1$. In these cases the multielectron quantum dot has a single unpaired spin on the highest occupied orbital, while remaining electrons are paired up in singlets, forming the effective vacuum state [Fig.~\ref{many_charge_states:fig6} and \ref{many_charge_states:fig7}(a)].

We first perform a leakage spectroscopy measurement for the multielectron quantum dot in $2N \! - \! 3$ occupancy [Fig.~\ref{many_charge_states:fig7}(c)]. The left part of the Fig.~\ref{many_charge_states:fig7}(c), corresponds to the configuration in which the multielectron quantum dot is not significantly exchange coupled to the double quantum dot (i.e. $J \simeq 0$). Indeed, here we observe the line that corresponds to the crossing between the states $\ket{S^L; \uparrow}$ and $\ket{\uuu}$.  (The state of the multielectron quantum dot is irrelevant in this regime, and we label it to be $\ket{\uparrow}$, anticipating the following analysis.)

For the intermediate values of the $\varepsilon$ we observe that the line features form a very different pattern than in case of the three electron triple quantum dot (Subsection.~\ref{subsection:TQD}). The line indicating, so far, the position of the crossing between $\ket{S^L; \uparrow}$ and $\ket{\uuu}$ [white triangle in Fig.~\ref{many_charge_states:fig7}(c)] states converges to $B=0$. Meanwhile, a second line emerges and shifts towards larger values of the external magnetic field $B$ (grey square). For more positive values of $\varepsilon$, which correspond to (1,0,$2N \! - \! 2$) occupancy, we observe that the newly appeared line does not diverge to large magnetic field. Instead, it returns towards $B=0$ (green circle). At the point where the line reaches $B=0$ we observe two new lines. The position of one of these new lines is virtually independent on $B$ (pink diamond) while another diverges to large magnetic field for increasing values of $\varepsilon$ (black triangle).

To explain this peculiar behavior, we introduce a modification of the simple microscopic model of Eq.~\ref{eq:spin0Ham} that we used to describe the even-occupancy spin-0 case.  We now consider the gate voltage of the multielectron quantum dot to be tuned such that the ground state is odd-occupied, and is effectively described by a single unpaired spin with a low energy orbital above an effective vacuum state of paired singlets. Generalizing beyond Eq.~\ref{eq:spin0Ham}, we include \emph{two} orbitals of the multielectron quantum dot, indicated with subscript $R1$ and $R2$, and the spin correlation term $\xi$ of Eq.~\ref{eq:JB}. The Hamiltonian of the system (illustrated in Fig.~\ref{many_charge_states:fig6}) is given by
\begin{align}
	\hat{H}_{\text{spin-1/2}} &= \sum_{i=L,M,R1,R2} E_i \hat{n}_i \nonumber \\
	&+ \frac{\xi}{2} \sum_{\sigma,\sigma'} \hat{a}^\dag_{R1,\sigma}\hat{a}^\dag_{R2,\sigma'}\hat{a}_{R1,\sigma'} \hat{a}_{R2,\sigma} \nonumber \\
    &- t_{DD} \sum\limits_{\sigma=\uparrow,\downarrow}  (\hat{a}_{L,\sigma}^\dag \hat{a}_{M,\sigma} + \hat{a}_{M,\sigma}^\dag \hat{a}_{L,\sigma}) \nonumber \\
	&- t_1\sum\limits_{\sigma=\uparrow,\downarrow}  (\hat{a}_{M,\sigma}^\dag \hat{a}_{R1,\sigma} + \hat{a}_{R1,\sigma}^\dag \hat{a}_{M,\sigma}) \nonumber \\
    &- t_2\sum\limits_{\sigma=\uparrow,\downarrow}  (\hat{a}_{M,\sigma}^\dag \hat{a}_{R2,\sigma} + \hat{a}_{R2,\sigma}^\dag \hat{a}_{M,\sigma})\,.
    \label{eq:spin12Ham}
\end{align}
The first term of this equation is the diagonal in terms of the spin occupancy numbers and captures both the charging and Coulomb energies of the dots. The second term, proportional to $\xi$, captures the spin correlation energy: it is a term that favors a $S=1$ triplet configuration when both levels $R1$ and $R2$ are occupied. The remaining terms proportional to $t_{DD}$, $t_1$, and $t_2$ describes tunnel coupling within the double quantum dot, between the middle dot $M$ and $R1$, and between $M$ and $R2$, respectively.

In Fig.~\ref{many_charge_states:fig7}(b) we present the energy diagram of the double quantum dot coupled to spin-1/2 multielectron quantum dot obtained from the 3-electron spectrum of Eq.~\ref{eq:spin12Ham}, using $\Delta E - \xi >0$ and $t_2/t_1 > \sqrt{2}$ (see discussion in subsection~\ref{subsection:phase_diagram}). The left part, for which $J=0$ matches exactly the three-electron triple quantum dot case [Fig.~\ref{many_charge_states:fig5}(b)], and the line feature corresponds to the crossing between $\ket{S^L; \uparrow}$ and $\ket{\uuu}$ states (white triangle). In contrast, for the intermediate values of $\varepsilon$ the singlet-like state $\ket{S^L; \uparrow}$ continuously changes into triplet-like state $\propto \ket{\uparrow; T_0^R} - \sqrt{2} \ket{\downarrow; T_+^R}$. The change from triplet-like to singlet-like character is a cause for the convergence of the line feature to $B=0$. Simultaneously the triplet-like state $\propto \ket{T_0^L; \uparrow} - \sqrt{2} \ket{T_+^L; \downarrow}$ shifts towards higher energies and continuously changes into $\ket{\uparrow; S^R}$. The crossing between this state and the states with the total spin projection $S_z=-1/2$ on the direction of the external magnetic field results in the emergence of the new line in Fig.~\ref{many_charge_states:fig7}(c) (grey square). This interpretation leads to the conclusion that for intermediate values of $\varepsilon$ a triplet configuration consisting of the single electron spin and the spin 1/2 in the multielectron quantum dot has \emph{lower} energy than the singlet.

For larger values of $\varepsilon$, the energy of the singlet-like $\ket{\uparrow; S^R}$ state decreases and ultimately becomes lower than the energy of the triplet-like state. This implies that the ground state of the multielectron quantum dot in $2N \! - \! 2$ occupancy has spin 0, consistent with the evidence presented in Sec.~\ref{many_charge_states:0}. The value of $\varepsilon$ for which the singlet-like and triplet-like states are degenerate corresponds to the crossing of the three line features at $B=0$ [Fig.~\ref{many_charge_states:fig7}(c)].  Specifically, the three lines correspond to the crossings of $\ket{\uparrow; S^R}$ with triplet-like states of different spin projections $S_z$ on the direction of the external magnetic field. The left, middle and the right line correspond to crossing with states having, respectively, $S_z=1/2, -1/2$ and $-3/2$, as indicated by gray sqaure, green circle and pink diamond in Fig.~\ref{many_charge_states:fig7}(b),(c).

\begin{figure}[tb]
	\centering
	\includegraphics[width=0.57\textwidth]{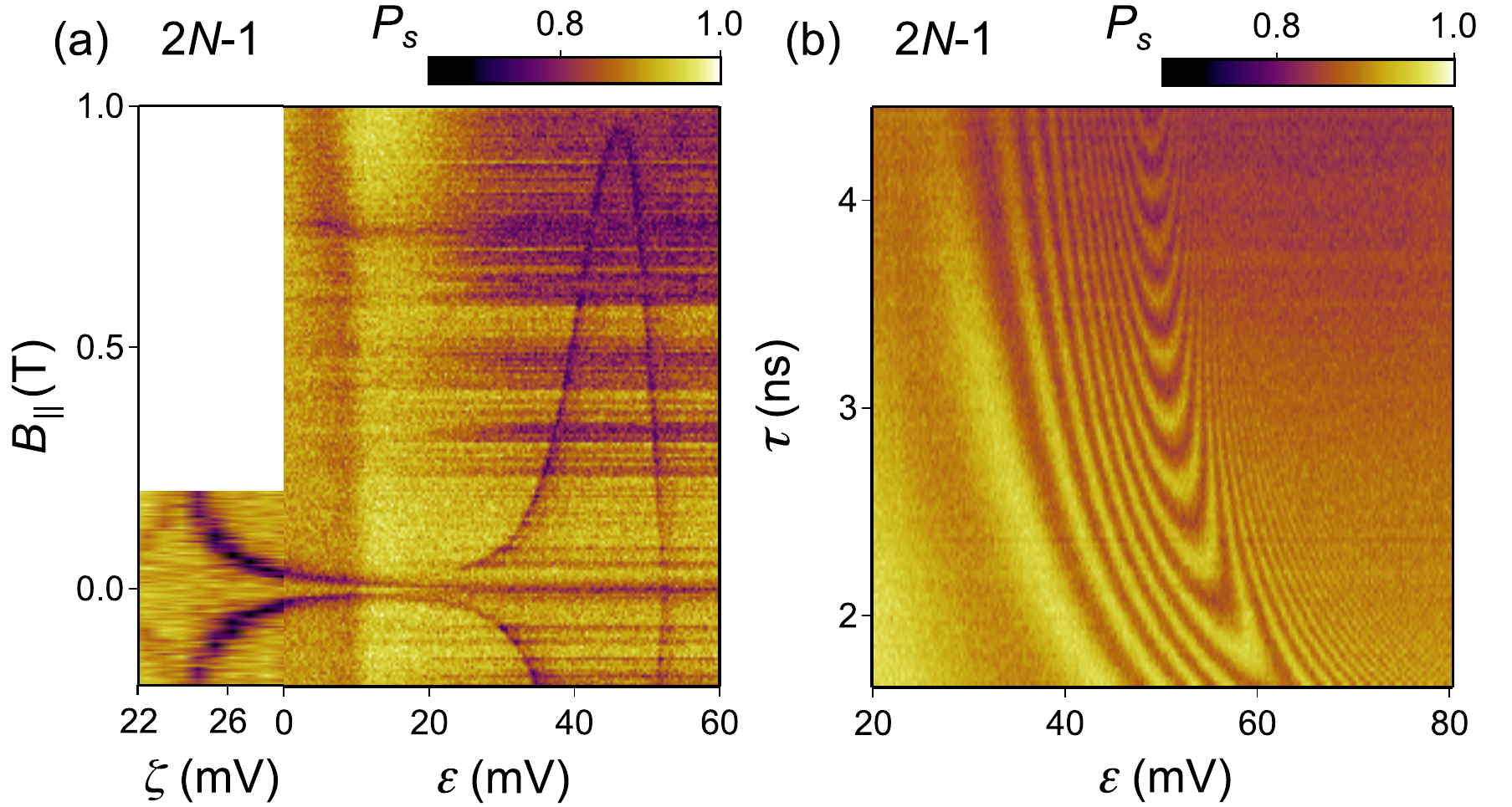}
	\caption[Leakage spectroscopy and exchange oscillations in a two-electron double quantum dot coupled to $(2N \! - \! 1)$-occupied multielectron quantum dot]{Leakage spectroscopy (a) and time-resolved exchange oscillations measurement (b) for the multielectron dot occupied by $2N \! - \! 1$ electrons.
	}
	\label{many_charge_states:fig8}
\end{figure}

An identical analysis of the leakage spectroscopy measurements for the $2N \! - \! 1$ occupancy of the multielectron quantum [Fig.~\ref{many_charge_states:fig8}(a)] dot yields the same conclusion. In particular it indicates that the ground state of the $2N$ occupancy should have spin 0, in agreement with the evidence presented in Sec.~\ref{many_charge_states:0}.

Our microscopic model also allows us to reconstruct the leakage spectrum, as described in Sec.~\ref{many_charge_states:0}.  Our reconstruction, which qualitatively reproduces all of the essential features of the leakage spectroscopy measurement, is shown in Fig.~\ref{many_charge_states:fig7}(e).

This interpretation of the energy diagram from the leakage spectroscopy measurements implies that the exchange interaction strength reaches an extremal value as a function of $\varepsilon$.  This extremum in exchange energy should result in a maximum in the exchange oscillations frequency. In Figs.~\ref{many_charge_states:fig7}(d) and \ref{many_charge_states:fig8}(b) we present the time-resolved measurement of the exchange oscillations for, respectively, $2N \! - \! 3$ and $2N \! - \! 1$ occupancy of the multielectron dot. 
The exchange energy extracted from the leakage spectroscopy pattern for $2N \! - \! 3$ occupancy is presented in Fig.~\ref{many_charge_states:fig7}(f).
 We observe the maximum in the oscillations frequency for the value of $\varepsilon$ that matches exactly the extreme position of the line features in the leakage spectroscopy measurement [Figs.~\ref{many_charge_states:fig7}(c) and \ref{many_charge_states:fig8}(a)]. This behavior confirms that the exchange interaction strength has an extremum, and that it changes sign at the charge transition; however these measurements do not yield information about the sign of the exchange interaction strength. In principle that information could be obtained from the dependence of the oscillations visibility on $\varepsilon$, but the visibility can be affected by the finite waveform rise time and readout infidelity, therefore we restrain ourselves from such analysis.

To summarize, for $2N \! - \! 3$ and $2N \! - \! 1$ occupancies of the multielectron quantum dot we observe a negative exchange interaction (i.e., triplet-preferring) with the single electron as long as the electron resides at the neighboring dot, i.e. in the charge configurations (1,1,$2N \! - \! 3$) or  (1,1,$2N \! - \! 1$).  The exchange interaction then becomes positive (i.e. singlet-preferring) when the electron is transferred onto the multielectron quantum dot, i.e. in (1,0,$2N \! - \! 2$) or (1,0,$2N$) electron configuration, leading to the formation of the spin 0 ground state for the even occupancy of the multielectron quantum dot.

\subsection{Case study: Negative exchange in the multielectron dot}
\label{subsection:negative-J}

\begin{figure}[tb]
	\centering
	\includegraphics[width=0.57\textwidth]{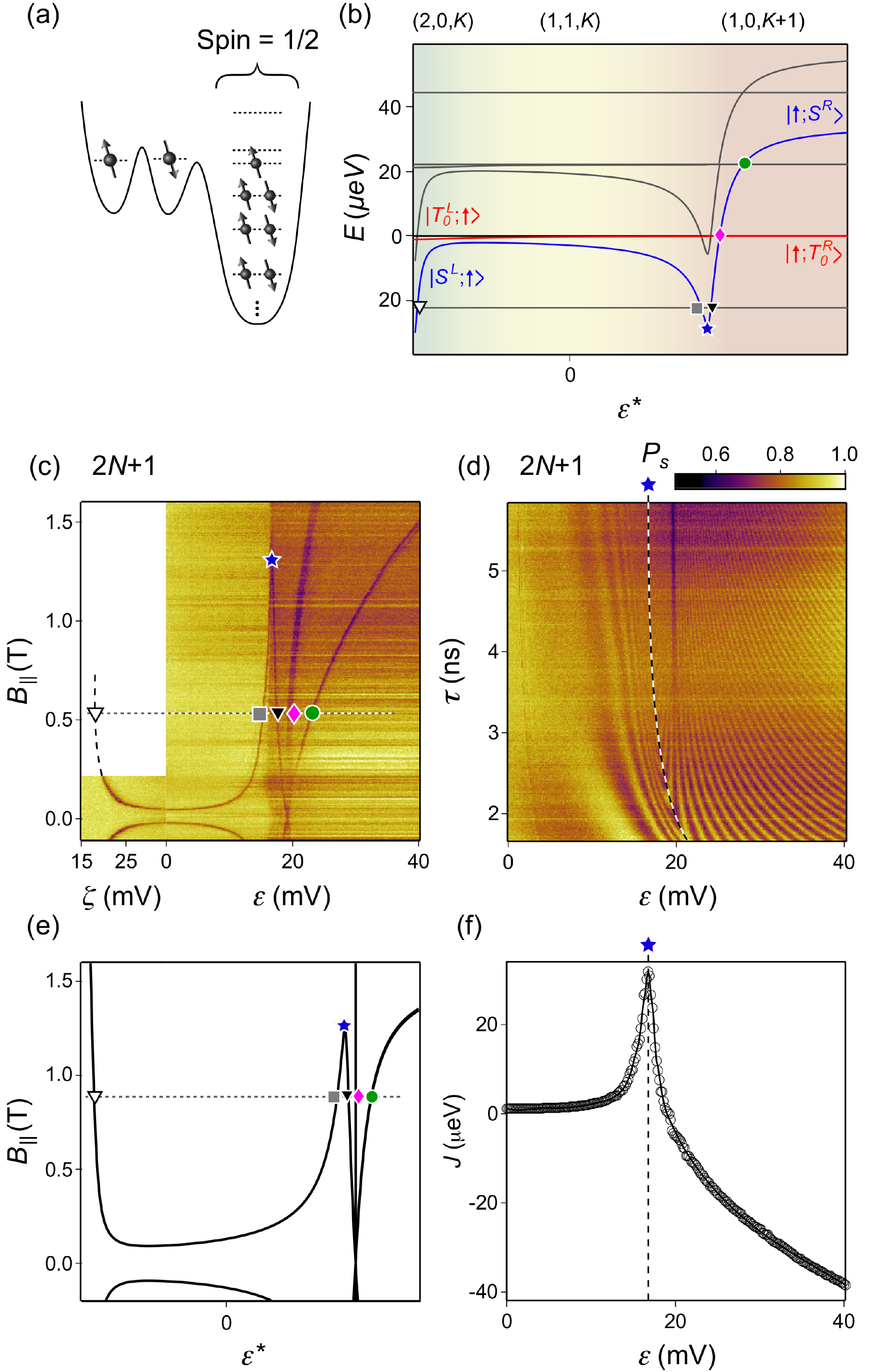}
	\caption[Leakage spectroscopy and exchange oscillations in a two-electron double quantum dot coupled to $(2N \! + \! 1)$-occupied multielectron quantum dot]{
	(a) Schematics of the odd-occupied multielectron quantum dot with spin 1/2 ground state tunnel coupled to the two electron double quantum dot.
	(b) The inferred energy diagram at the transition between (2,0,$2N \! + \! 1$), (1,1,$2N \! + \! 1$) and (1,0,$2N \! + \! 2$) electronic configurations, for a finite magnetic field. The markers indicate the crossings revealed by the leakage spectroscopy measurement presented in panel (c).
	(d) Time resolved measurement of the exchange oscillations between the $2N \! + \! 1$-occupied spin-1/2 multielectron quantum dot and the neighboring electron.
	(e) Reconstruction of the leakage spectrum using the simple microscopic model.
	(f) Dependence of the exchange energy extracted from the pattern of exchange oscillations in panel (d).
	}
	\label{many_charge_states:fig9}
\end{figure}

Next we focus on is $2N \! + \! 1$ multielectron dot occupancy. Similarly to the occupancies $2N \! - \! 3$ and  $2N \! - \! 1$ (Subsection~\ref{subsection:positive-J}) in this case there is a single unpaired electron on the highest occupied orbital of the multielectron dot [Fig.~\ref{many_charge_states:fig9}(a)]. However, the leakage spectroscopy measurement [Fig.~\ref{many_charge_states:fig9}(c)] implies that the exchange interaction with the single neighboring electron is qualitatively different.

In contrast to the previous cases, the line indicating the crossing between singlet-like state $\ket{S^L \uparrow}$ and the fully polarized $\ket{\uuu}$ (white triangle) state does not converge to $B=0$. Instead, the line feature reverses to high magnetic fields for intermediate values of $\varepsilon$, i.e., in (1,1,$2N \! + \! 1$) charge state (grey square). This behavior indicates that as long as the single electron resides on the neighboring dot the singlet-like state $\ket{\uparrow; S^R}$ has lower energy than the triplet-like states. Nevertheless, beyond at the transition to (1,0,$2N \! + \! 2$) the line feature does come back to $B=0$, and the set of the three lines appear as well.

The energy diagram corresponding to this case is presented in Fig.~\ref{many_charge_states:fig9}(b). Indeed, using $\Delta E - \xi <0$ and $t_2/t_1<\sqrt{2}$ we obtain an energy diagram revealing the positive (singlet-preferring) exchange interaction in (1,1,$2N \! + \! 1$) electron configuration and negative (triplet-preferring) in (1,0,$2N \! + \! 2$) configuration. In this energy diagram we can identify all crossings corresponding to the line features in Fig.~\ref{many_charge_states:fig9}(c). In particular the three lines correspond to the crossings between $\ket{\uparrow S^R}$ and the triplet-like states having, respectively, $S_z=-3/2$ (left line, pink diamond), $-1/2$ (middle, green circle) and $1/2$ (right, grey rectangle).

Following the logic from Subsection~\ref{subsection:positive-J} we come to the conclusion that the ground state of the $2N \! + \! 2$ occupancy multielectron dot must have a spin 1. Indeed, in Sec.~\ref{many_charge_states:1} we present the evidence for the exchange interaction between $2N \! + \! 2$ occupancy multielectron quantum dot and the neighboring spin, consistent with a spin-1 ground state.

In Fig.~\ref{many_charge_states:fig9}(d) we present the time resolved exchange oscillations measurement. Similarly to the $2N \! - \! 3$ and $2N \! - \! 1$ occupancies we find the that the oscillations reach locally the maximum frequency for the value of $\varepsilon$ that correspond to the extreme position of the line feature in the leakage spectroscopy measurement. Additionally we observe the emergence of the line for the value of $\varepsilon$ that results in the degeneracy between $\ket{\uparrow; S^R}$ and $\propto \ket{\uparrow; T_0^R} - \sqrt{2} \ket{\downarrow; T_+^R}$ states.

More extensive study of the charge occupancy $2N \! + \! 1$, presented in this subsection, including tunability of the exchange profile, is a topic of Ref.~\cite{Negative-J} (Ch.~\ref{ch:negative-j}).

\subsection{Extreme occupancies of the multielectron quantum dot $2N \! - \! 5$ and $2N \! + \! 3$}

\begin{figure}[tb]
	\centering
	\includegraphics[width=0.57\textwidth]{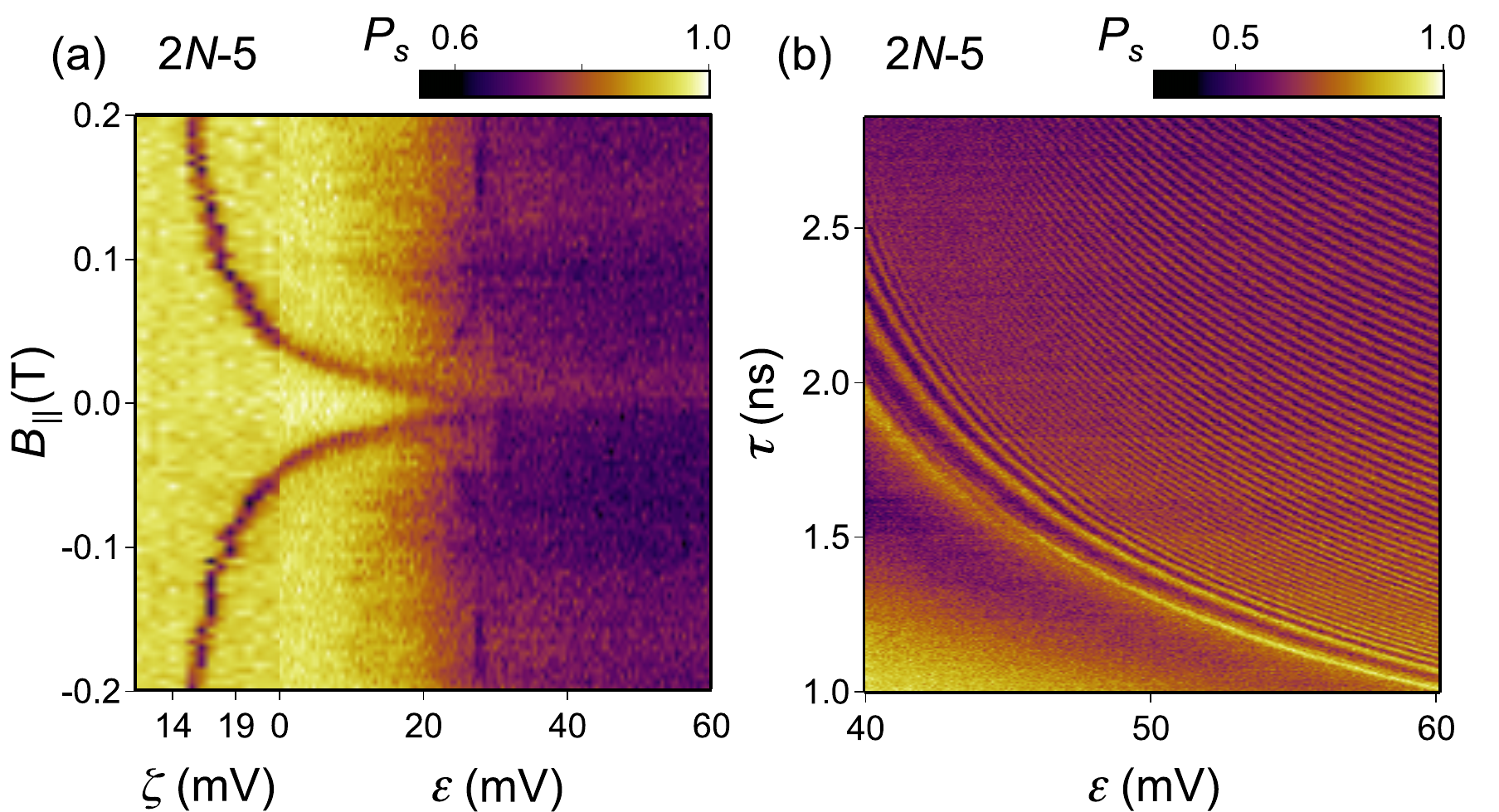}
	\caption[Leakage spectroscopy and time-resolved exchange oscillations measurement for the multielectron dot occupied by $2N \! - \! 5$ electrons]{Leakage spectroscopy (a) and time-resolved exchange oscillations measurement (b) for the multielectron dot occupied by $2N \! - \! 5$ electrons.
	}
	\label{many_charge_states:fig10}
	\vspace{10pt}
%
	\includegraphics[width=0.57\textwidth]{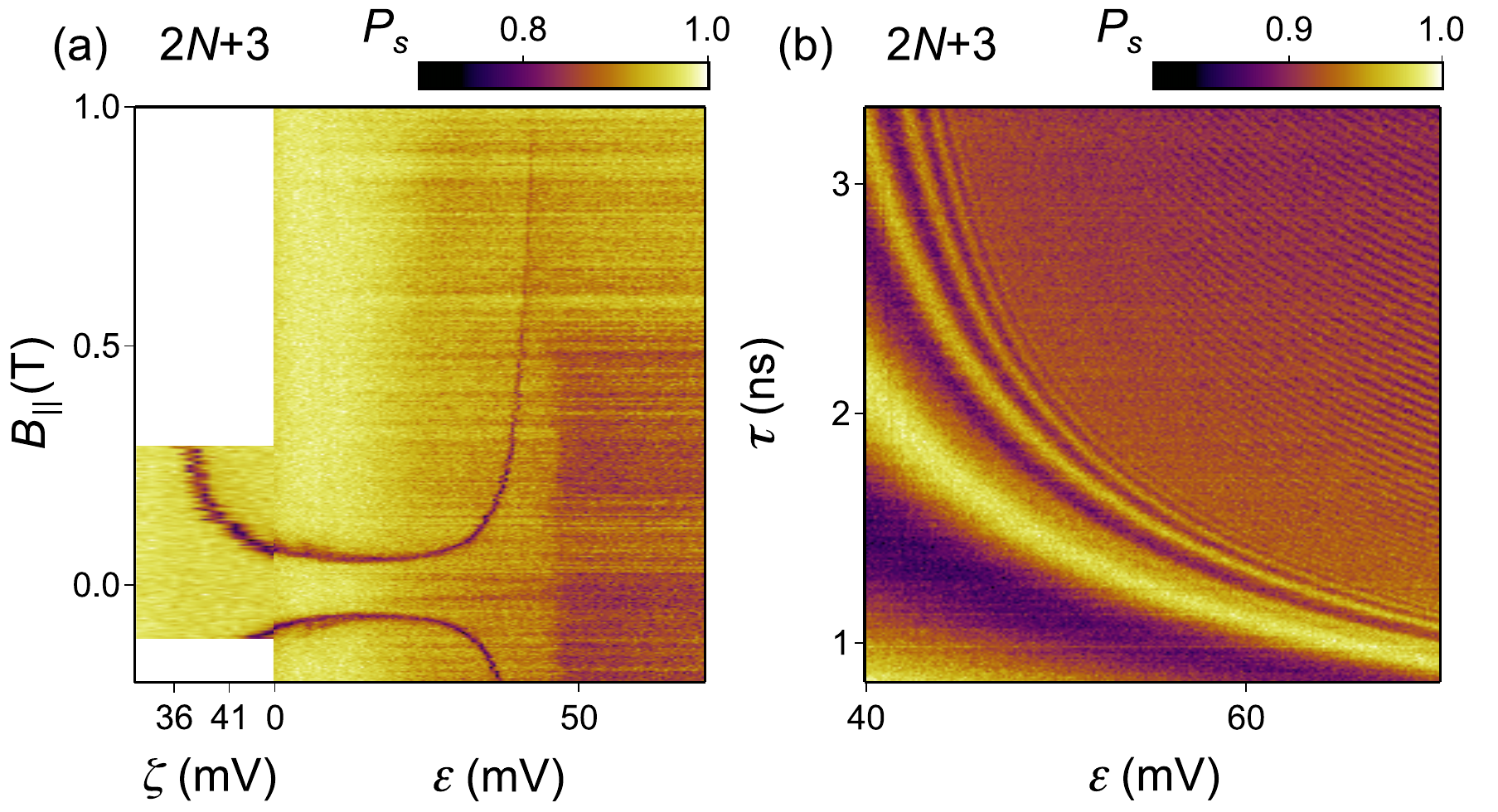}
	\caption[Leakage spectroscopy and time-resolved exchange oscillations measurement for the multielectron dot occupied by $2N \! + \! 3$ electrons]
	{Leakage spectroscopy (a) and time-resolved exchange oscillations measurement (b) for the multielectron dot occupied by $2N \! + \! 3$ electrons.
	}
	\label{many_charge_states:fig11}
\end{figure}

We now present the results of the leakage spectroscopy and exchange oscillation measurements for the $2N \! - \! 5$ and $2N \! + \! 3$ occupancies of the multielectron dot charge. These are the most extreme studied occupancies, which is linked to the limited tunability of the tunnel coupling between the middle and multielectron dot. As a result, the data provides consistent evidence for the presence of a spin on the multielectron quantum dot but the quality is not sufficient to perform the full analysis analogous to Subsections~\ref{subsection:positive-J} and \ref{subsection:negative-J}.

In Fig.~\ref{many_charge_states:fig10}(a) we show the leakage spectroscopy for $2N \! - \! 5$ occupancy of the multielectron quantum dot. Initially the line feature converges towards zero, indicating the decrease of the exchange interaction within the two-electron double quantum dot. However, at $\varepsilon \approx 20$~mV it appears to split. Beyond the apparent splitting, one of the lines converges completely to $B=0$ while the second one diverges. Finally, we suspect that the diverging line returns and crosses zero at about $\varepsilon = 30$~mV. This suggests that one of the two possible scenarios takes place: either the exchange interaction strength between the single electron and the spin-1/2 multielectron dot has a negative sign for small wavefunction overlap (in the (1,1,$2N \! - \! 5$) charge configuration) and positive for large wavefunction overlap (in (1,0,$2N \! - \! 4$) occupancy), or else the opposite behavior is the case. The observation of the spin-0 ground state for $2N \! - \! 4$ occupancy supports the first hypothesis.

Notably, the exchange oscillations presented in Fig.~\ref{many_charge_states:fig10}(b) do not reveal a local maximum in the oscillations frequency. This indicates there is no extremum in the exchange interaction strength, which contradicts our hypothesis that the line feature crosses $B=0$ in Fig.~\ref{many_charge_states:fig10}(a). Regardless, the presence of the exchange oscillations is evidence that the $2N \! - \! 5$ occupancy multielectron quantum dot has a spinful ground state.

The leakage spectroscopy performed for the other extreme occupancy, $2N \! + \! 3$, presented in Fig.~\ref{many_charge_states:fig11}(a), reveals the pattern characteristic for the three-electron triple quantum dot (Subsection~\ref{subsection:TQD}). This suggests that the multielectron quantum dot with this occupancy behaves as an ordinary spin-1/2 with no remarkable phenomena occurring at the $(1,1,2N \! + \! 3)$ to $(1,0,2N \! + \! 4)$ charge transition. However we can not fully exclude the possibility that the exchange interaction changes sign, which may be hard to detect when the tunnel coupling is too large~\cite{Negative-J}.  In addition, the exchange oscillations pattern does not allow us to make a definite statement about the presence of the extremum in the exchange interaction strength due to increased dephasing at the strongly pinched-off interdot charge transition~\cite{Dial2013} [Fig.~\ref{many_charge_states:fig11}(b)].

\subsection{The classification of the observed spin-1/2 ground state characteristics within a Hubbard model}
\label{subsection:phase_diagram}

\begin{figure}[tb]
	\centering
	\includegraphics[width=0.57\textwidth]{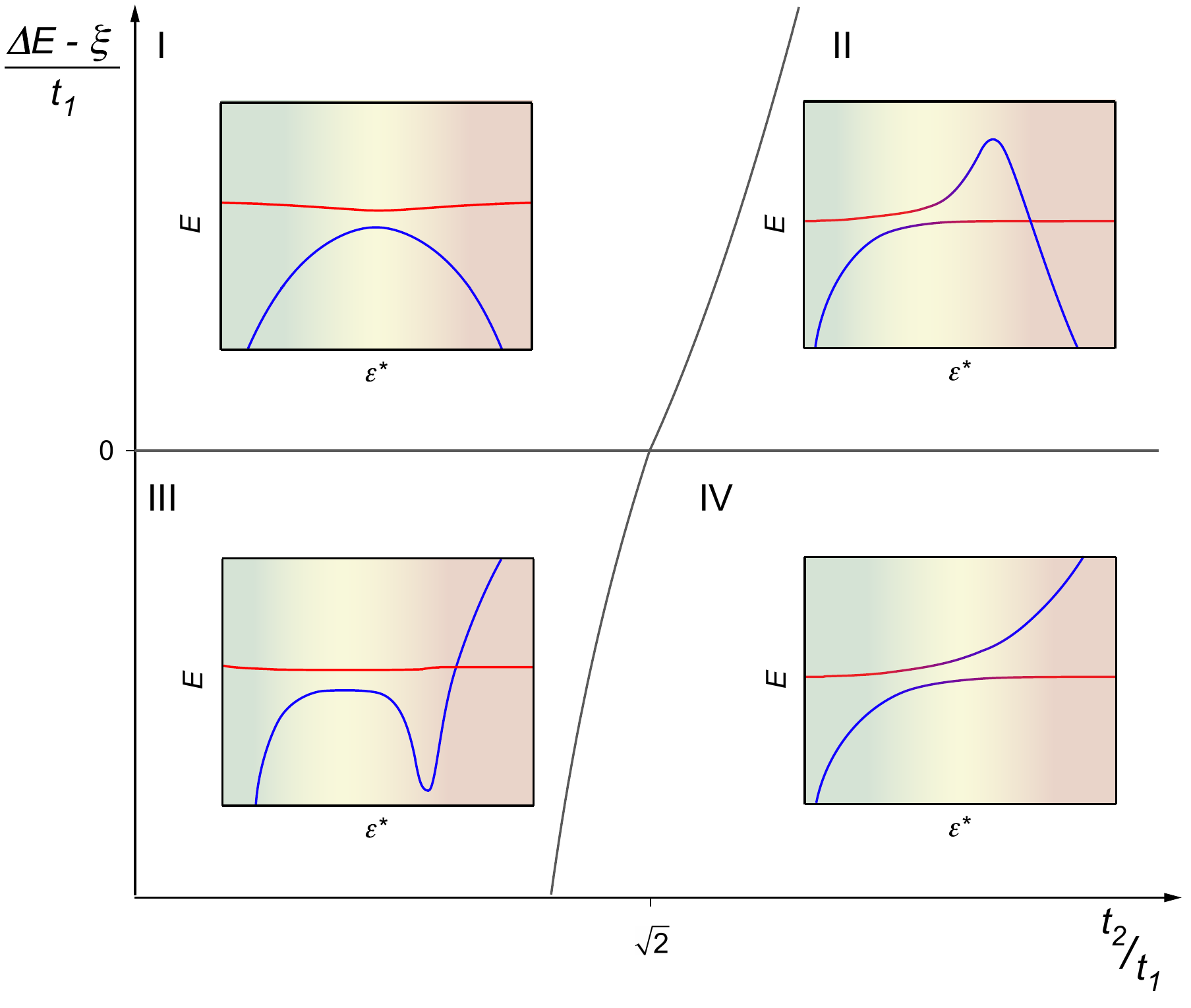}
	\caption[Illustration of qualitatively different exchange profiles arising from the interplay between the level spacing in the multielectron quantum dot $\Delta E$, spin correlation energy $\xi$ and tunnel couplings]{
	Illustration of qualitatively different exchange profiles arising from the interplay between the level spacing in the multielectron quantum dot $\Delta E$, spin correlation energy $\xi$ and tunnel couplings between a single-electron dot and two lowest orbitals of the multielectron quantum dot $t_{1/2}$. Colored lines in the insets I-IV represent the energies of the three-spin states with $S=1/2$, $S_z=-1/2$ as a function of detuning $\varepsilon^*$.
	}
	\label{many_charge_states:fig12}
\end{figure}

The effective exchange coupling depends on many free parameters in our phenomenological model, but we find that its general behaviour falls into four main regimes as shown schematically in Fig.~\ref{many_charge_states:fig12}.  We analyse these regimes of the spectrum in terms of two dimensionless quantities.  The first is $(\Delta E - \xi)/t_1$, which we note can be positive or negative.  When positive, i.e., $\Delta E > \xi$, the energy separation of the two relevant single-particle levels 1 and 2 in the multielectron dot is larger than the spin correlation energy, and the doubly-occupied multielectron dot will ultimately favour a singlet configuration energetically once detuning has crossed the charge transition point.  When this quantity is negative, i.e., $\Delta E < \xi$, the spin correlation energy is larger than the splitting, and a double-occupied triplet configuration is energetically preferred past the charge transition.  Thus, the parameter $(\Delta E - \xi)/t_1$ will determine the spin state of the multielectron dot (singlet or triplet) when detuning into the multielectron dot is very large.

The other parameter we use to describe the spectrum is the ratio $t_2/t_1$.  When $t_2/t_1 \lesssim \sqrt{2}$, tunnel coupling from the middle dot to the lower level of the multielectron dot is dominant, whereas $t_2/t_1 \gtrsim \sqrt{2}$ describes a situation where the tunnelling to the higher level of the multielectron dot is stronger.  The regime $t_2/t_1 \gtrsim \sqrt{2}$ results in an exchange interaction that favours a triplet configuration in the multielectron dot for small detuning, and $t_2/t_1 \lesssim \sqrt{2}$ favours a singlet configuration.

As a function of these two dimensionless quantities, we obtain the four regimes shown in Fig.~\ref{many_charge_states:fig12}.  In regime I, the behaviour is qualitatively the same as for a triple quantum dot; the higher level in the multielectron quantum dot is largely uninvolved.  In regime II, the stronger coupling to the second level of the multielectron dot means that a triplet configuration is preferred for small detuning, but becomes singlet-preferring as the detuning is increased past the charge transition.  In such a regime, we expect exchange coupling to be negative for small detuning, and then change sign and become positive for large detuning, as seen with the $2N-3$ and $2N-1$ occupancies.  Regime III is the opposite; the coupling to the lower level of the multielectron dot dominates, preferring a singlet configuration for small detuning, but for larger detuning the spin correlation energy is dominant and favours a triplet configuration.  This regime is represented by the $2N+1$ occupancy, where exchange coupling is negative past the charge transition into the multielectron dot.  Finally, in Regime IV, the dominant coupling to the higher level of the jellybean together with the strong spin correlation energy ensures that a triplet configuration is energetically preferred throughout.  

\section{Spin 1}
\label{many_charge_states:1}

\begin{figure}[tb]
	\centering
	\includegraphics[width=0.57\textwidth]{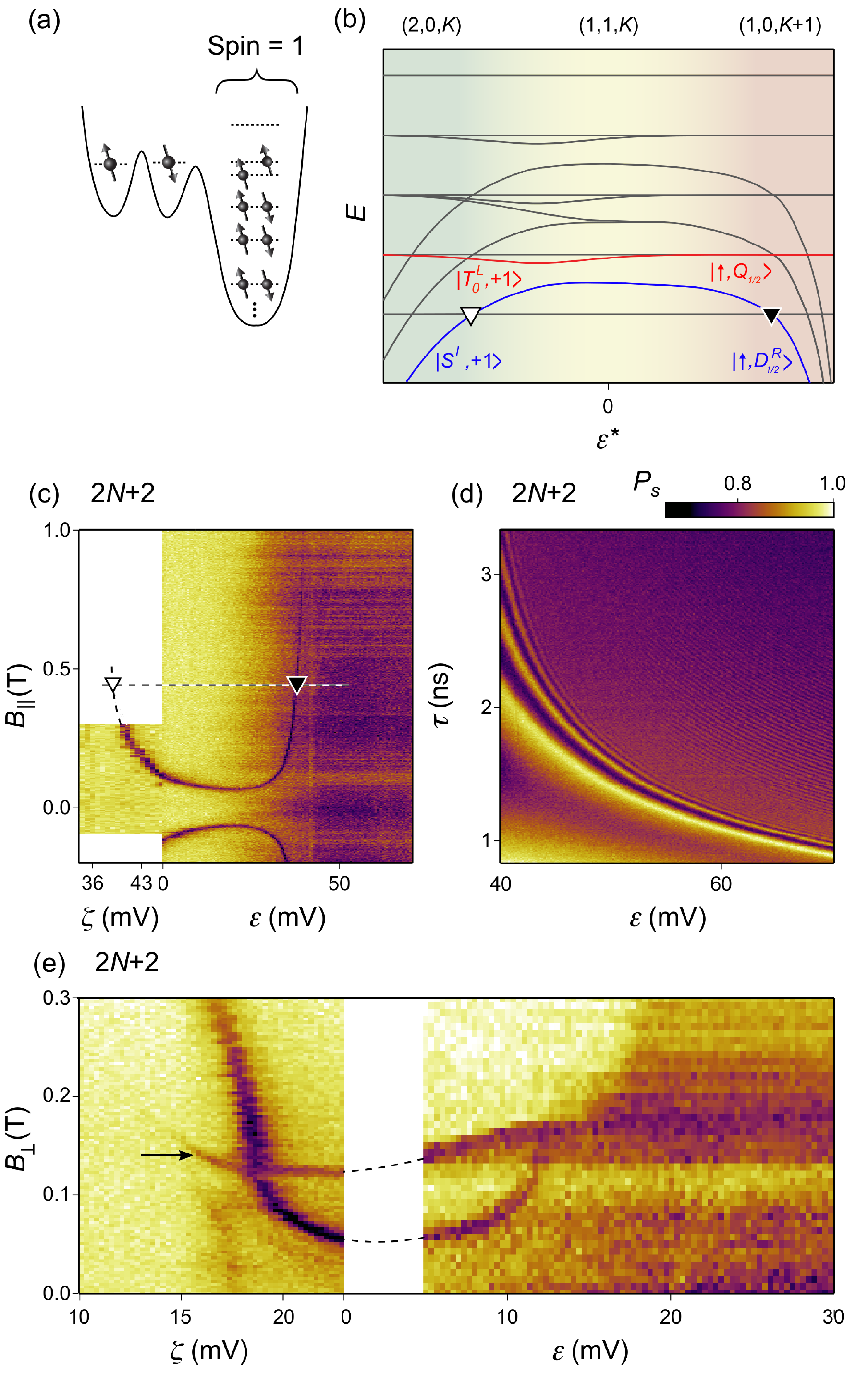}
	\caption[Leakage spectroscopy and exchange oscillations in a two-electron double quantum dot coupled to $(2N \! + \! 2)$-occupied multielectron quantum dot]{
	(a) Schematics of the odd-occupied multielectron quantum dot with spin 1 ground state tunnel coupled to the two electron double quantum dot.
	(b) The inferred energy diagram at the transition between (2,0,$2N \! + \! 2$), (1,1,$2N \! + \! 2$) and (1,0,$2N \! + \! 3$) electronic configurations, for a finite magnetic field. The markers indicate the crossings revealed by the leakage spectroscopy measurement presented in panel (c).
	(d) Time resolved measurement of the exchange oscillations between the $2N \! + \! 2$-occupied spin-1/2 multielectron quantum dot and the neighboring electron.
	(e) Leakage spectroscopy measurement in out-of-plane magnetic field
	}
	\label{many_charge_states:fig13}
\end{figure}

Our final case concerns the $2N \! + \! 2$ multielectron quantum dot occupancy.  Our study of the $2N \! + \! 1$ occupancy (Subsection~\ref{subsection:negative-J}) suggests that the addition of an electron to the spin-1/2 ground state results in the triplet configuration having lower energy than the singlet. This implies that the ground state of the multielectron dot with $2N \! + \! 2$ occupancy should have spin 1 [Fig.~\ref{many_charge_states:fig13}(a)].

Indeed, in the leakage spectroscopy measurement [Fig.~\ref{many_charge_states:fig13}(c)] we observe a pattern that is more similar to the three-electron triple quantum dot case [Fig.~\ref{many_charge_states:fig5}(c)] than to other even occupancies [Figs.~\ref{many_charge_states:fig3}(c) and \ref{many_charge_states:fig4}]. It is an unambiguous evidence for presence of the non-zero spin in the multielectron quantum dot, spin 1 in this case. We note that the line feature diverges to large $B$ for increasing $\varepsilon$ indicating the positive sign of the exchange interaction, i.e. preferring the low-spin state. Also the measurement of the exchange oscillations shows the presence of the exchange interaction [Fig.~\ref{many_charge_states:fig13}(d)]. These observations lead us to the following conclusions.  First, the multielectron dot in $2N \! + \! 2$ occupancy carries a spin.  Second, the exchange interaction with the neighboring spin has a positive sign, and therefore the addition of the electron will lead to reduction of the ground state spin.

We can relate all the observed features to the schematic energy diagram presenting spin states of the two-electron double quantum dot coupled to spin-1. In the left side of the diagram spin-1 is decoupled from the double quantum dot and the eigenstates are the tensor product of double dot states and spin-1 states with different spin projections on the direction of the magnetic field $\ket{0/\pm 1}$. The three double-dot singlet-like states correspond in the diagram to the three lines diverging towards small energies.

On the contrary, in the right side, the eigenstates are the tensor product of spin-1/2 states and strongly coupled spin-1/2 and spin-1. Among the latter six states there are four quadruplet states $\ket{Q_{\pm3/2}}$, $\ket{Q_{\pm1/2}}$ with a total spin 3/2, and two dublet states $\ket{D_{\pm1/2}}$ with the total spin of 1/2 (the subscript indicates the spin projection on the magnetic field direction). Due to positive exchange interaction the dublet states diverge towards small energy.

We conjecture that in the experiment we initialize the triple dot in the $\ket{S^L;+1}$ state (similarly to $\ket{S^L;\uparrow}$ for spin-1/2 multielectron dot). As we change detuning this eigenstate continuously changes into $\ket{\uparrow;D^R_{1/2}}$ [bottom red line in Fig.~\ref{many_charge_states:fig13}(b)] which results in the exchange oscillations in Fig.~\ref{many_charge_states:fig13}(c). Meanwhile, the line features in the leakage spectroscopy correspond to the crossing of this red-coloured state with a fully polarized $\ket{T_+;+1} \equiv \ket{\uparrow; Q_{3/2}}$ state [black and white triangles in Fig.~\ref{many_charge_states:fig13}(b),(c)].

Finally, we present an outcome of the leakage spectroscopy measurement in the out-of-plane magnetic field $B_\perp$ [Fig.~\ref{many_charge_states:fig13}(e)]. Curiously, in this case we observe an additional line (pointed by the black arrow). At the boundary between $(2,0,2N \! + \! 2)$ and $(1,1,2N \! + \! 2)$ the line indicating DD $S$-$T_+$ crossing and the new line do seem not to interact with each other. On the other hand, at the transition between $(1,1,2N \! + \! 2)$ and $(1,0,2N \! + \! 3)$ the conventional line ends at the crossing point. We speculate that the additional line appears because the multielectron-dot orbitals are strongly affected by the out-of-plane magnetic field, which leads to the change of the ground state spin from 1 ($B_\perp \lesssim 180$~mT) to 0 ($B_\perp \gtrsim 180$~mT). Indeed, this would explain lack of interaction between the lines in regime where the double quantum dot is essentially decoupled from the multielectron dot. Moreover, this is consistent with a disappearance of the line at $\varepsilon \approx13$~mV, since we know that this feature is not present when the multielectron quantum dot has a spin-0 ground state case (Sec.~\ref{many_charge_states:0}).

\section{Summary and outlook}
\label{many_charge_states:summary}

To summarize, we apply the methods developed for spin qubit manipulation to study a mesoscopic, multielectron quantum dot. This enables us to detect the electron parity of the dot and study in detail the interaction with the single neighboring electron. We discover a counterintuitive exchange profile between a single electron and an odd-occupied multielectron quantum dot and observe that the exchange interaction rapidly varies. In particular the exchange interaction changes sign as a result of a few-milivolt change of gate voltages. We also explain this observation using simple Hubbard model and classify possible exchange profiles. 

The most important conclusion of this work that can be readily exploited is that the multielectron quantum dot is perfectly suitable as a mediator of the exchange interaction. Indeed we demonstrate that in the follow up experiment~\cite{JB-mediated_exchange}. However several other findings may find application in spin qubits as well. First, at the position of the discovered extrema in the exchange profile, the exchange oscillations have reduced sensitivity to charge noise~\cite{Martins2016,Reed2016}, which can be used to increase the gates fidelity. Second, access to both signs of the exchange lifts constraint on the dynamically decoupled gates~\cite{Wang2012,Wang2014}. Third, since the large quantum dot is characterized by reduced level spacing, it may be possible to exploit charge-noise insensitive singlet-triplet splitting in the regime where the two electrons occupy the multielectron quantum dot, in a manner analogous to the quantum dot hybrid qubit~\cite{Kim2014,Cao2016}. Finally, larger size of the multielectron quantum dot implies reduction of the Overhauser field experienced by the electrons, since its wavefunction overlaps with larger number of nuclei, and therefore reduced dephasing~\cite{Taylor2007}.

Nevertheless, several questions concerning the multielectron quantum dots remain unanswered. First of all, the distribution of the level spacings and the strength of the spin correlation energy were not studied here. In particular the dependence of these two characteristics on the dot size is of both fundamental and practical importance. Another curiosity is that for all three spin-1/2 ground states, for which we performed the full analysis, we observe the extrema in the exchange strength even though they are characterized by different sign of the exchange strength. This may be a hint of the correlation between the level spacing and the ratio of the tunnel couplings.

Finally, for the first time we perform the leakage spectroscopy and exchange oscillations measurement to study the spectrum of a not understood object. We propose that the same principle could be applied to study numerous other systems. Several examples are quantum dot coupled to the quantum hall or fractional quantum hall edge states~\cite{Yang2015,Kiyama2015}, or to the hybrid super-semiconducting quantum dot such as Majorana islands~\cite{Deng2016,Gharavi2016}. Even more brave generalization of this technique could involve study of the exchange interaction between a quantum dot located at the tip of the scanning probe and any surface structure.

\section*{Author contributions}
S.F., G.C.G. and M.J.M. grew the heterostructure. P.D.N. fabricated the device. F.M., F.K.M., F.K. and P.D.N. prepared the experimental setup. F.K.M. and F.M. performed the experiment. T.S., S.D.B. and A.C.D. developed the theoretical model and performed the simulations. F.K.M., S.D.B., F.M., F.K., A.C.D., T.S. and C.M.M. analysed data and prepared the manuscript.

\chapterimage{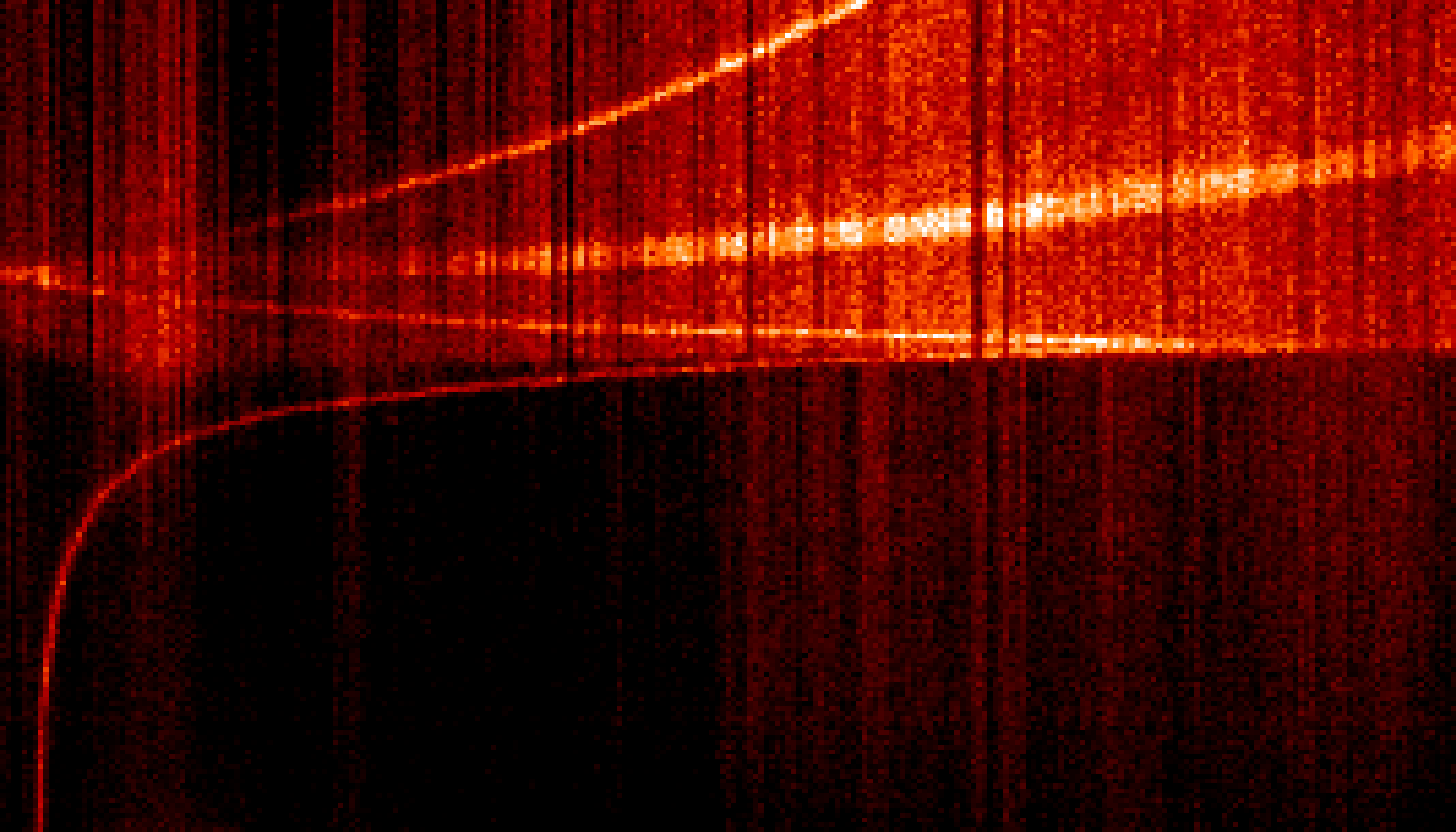}
\chapter[Negative exchange interaction in a multielectron quantum dot]{\protect\parbox{0.9\textwidth}{Negative exchange interaction \\ in a multielectron quantum dot}}
\label{ch:negative-j}

{\let\thefootnote \relax\footnote{This chapter is adapted from the manuscript in preparation.}}
\addtocounter{footnote}{-1}

\begin{center}
Frederico Martins$^{1,*}$, Filip K. Malinowski$^{1,*}$, Peter D. Nissen$^{1}$, \\
Saeed Fallahi$^{2}$, Geoffrey C. Gardner$^{2,3}$, Michael J. Manfra$^{4,5}$, \\
Charles M. Marcus$^{6}$, Ferdinand Kuemmeth$^{1}$
\end{center}

\begin{center}
	\scriptsize
	$^{1}$ Center for Quantum Devices, Niels Bohr Institute, University of Copenhagen, 2100 Copenhagen, Denmark\\
	$^{2}$ Department of Physics and Astronomy, Birck Nanotechnology Center, Purdue University, West Lafayette, Indiana 47907, USA \\
	$^{3}$ School of Materials Engineering and School of Electrical and Computer Engineering, \\ Purdue University, West Lafayette, Indiana 47907, USA \\
	$^{4}$ Department of Physics and Astronomy, Birck Nanotechnology Center, and Station Q Purdue, \\ Purdue University, West Lafayette, Indiana 47907, USA \\
	$^{5}$ School of Materials Engineering, Purdue University, West Lafayette, Indiana 47907, USA \\
	$^{6}$ Center for Quantum Devices and Station Q Copenhagen, Niels Bohr Institute, \\ University of Copenhagen, 2100 Copenhagen, Denmark \\
	$^{*}$ These authors contributed equally to this work
\end{center}

\begin{center}
\begin{tcolorbox}[width=0.8\textwidth, breakable, size=minimal, colback=white]
	\small
	By operating a one-electron quantum dot (fabricated between a multielectron dot and a one-electron reference dot) as a spectroscopic probe, we study the spin properties of a gate-controlled multielectron GaAs quantum dot at the transition between odd and even occupation number.  We observe that the multielectron groundstate transitions from spin-1/2-like to singlet-like to triplet-like as we increase the detuning towards the next higher charge state. The sign reversal in the inferred exchange energy is robust and already occurs in the absence of an external magnetic field, while the exchange strength is tunable by gate voltages and in-plane magnetic fields. Complementing spin leakage spectroscopy data, the inspection of coherent multielectron spin exchange oscillations provides further evidence for the sign reversal and, inferentially, for the importance of non-trivial multielectron spin exchange correlations.
\end{tcolorbox}
\end{center}

\section{Introduction}

Semiconducting quantum dots with individual unpaired electronic spins offer a compact platform for quantum computation~\cite{Kloeffel2013,Awschalom2013}.
They provide submicron-scale two-level systems that can be operated as qubits~\cite{Loss1998,Petta2010,Bluhm2011,Nowack2007,Maune2012,Malinowski2017} 
and coupled to each other via direct exchange or direct capacitive interaction. In these approaches, the essential role of nearest-neighbor interactions in larger and larger arrays of one-electron quantum dots~\cite{Shulman2012,Veldhorst2015,Nichol2017,Ito2016,Zajac2016} poses technological challenges to upscaling, due to the density of electrodes that define and control these quantum circuits. 
This issue has stimulated efforts to study long-range coupling of spin qubits either by electrical dipole-dipole interaction~\cite{Shulman2012,Nichol2017,Trifunovic2012} or via superconducting microwave cavities~\cite{Mi2016,Srinivasa2016, Russ2015}.
However, these approaches involve the charge degree of freedom, which makes the qubit susceptible to electrical noise~\cite{Coish2005,Taylor2007,Dial2013,Martins2016}. Recent work~\cite{Medford2013a, Malinowski2017b} indicates that the effective noise needs to be reduced significantly before long-range two-qubit gates with high fidelity can be reached~\cite{Srinivasa2016,Russ2016}.
Alternatively, symmetric exchange pulses can be implemented that perform fast, charge-insensitive gates~\cite{Weiss2012,Bertrand2015,Martins2016,Reed2016}. 
Even though the exchange interaction is intrinsically short-ranged, its range can be increased by means of a quantum mediator~\cite{Baart2016,Braakman2013}. 
In particular, using a large multielectron quantum dot as an exchange mediator has the potential to do both: provide fast spin interaction~\cite{Mehl2014,Srinivasa2015} and alleviate spatial control line crowding. To avoid entanglement with internal degrees of freedom of the mediator, recent theory~\cite{Mehl2014,Srinivasa2015} motivates the use of a multielectron quantum dot with a spinless ground state and a level spacing sufficiently large to suppress unwanted excitations by gate voltage pulses.

\begin{figure}[t]
	\begin{center}
	\includegraphics[width=0.6\textwidth]{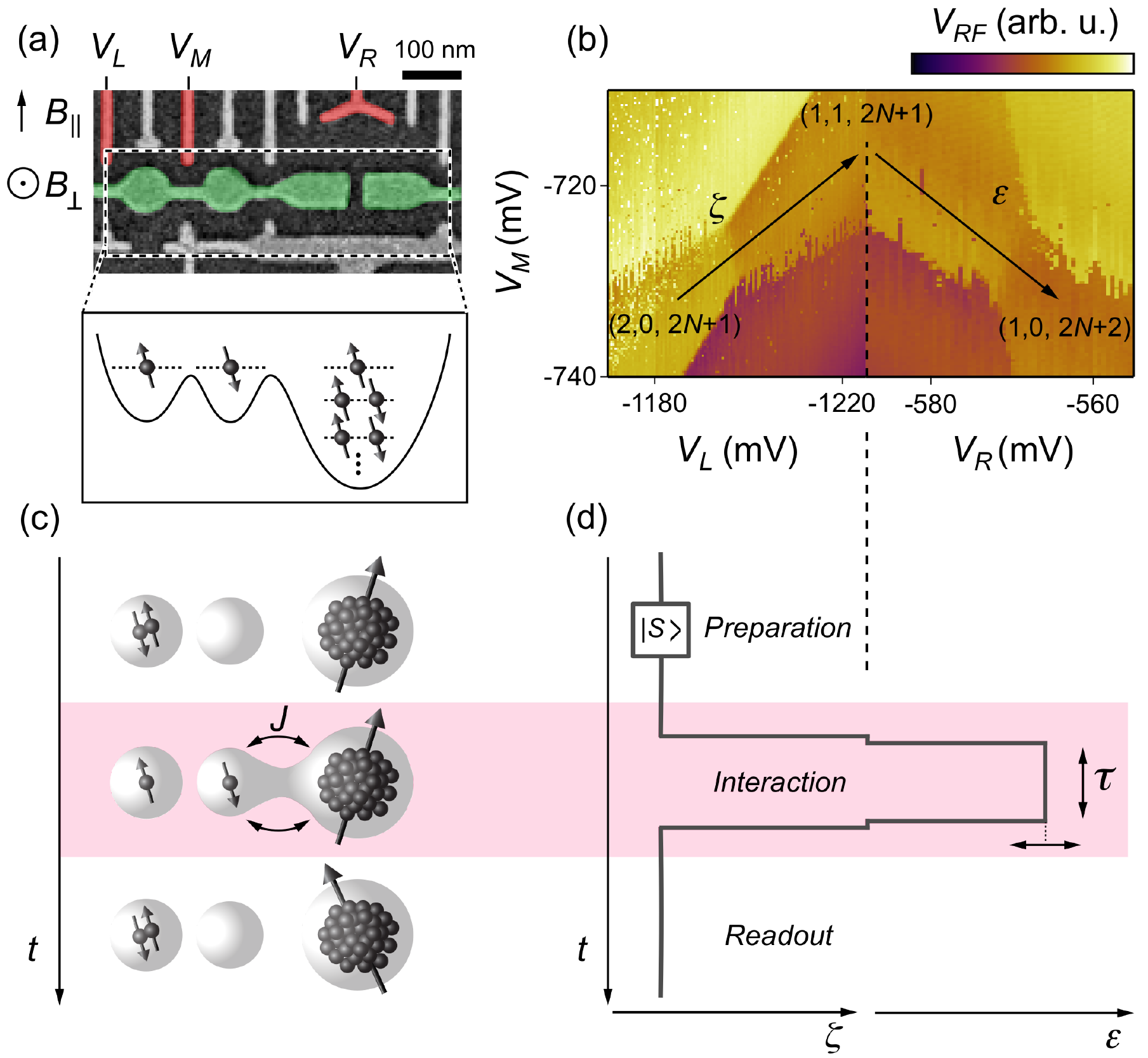}
	\caption[the device consisting of a two-electron double quantum dot next to a multielectron quantum dot]{
	(a) Electron micrograph of the device consisting of a two-electron double quantum dot next to a multielectron quantum dot. The accumulation gate (colored in green) is operated at positive voltage. Remaining gates deplete the underlying two-dimensional electron gas. Gates $V_{L}$, $V_{M}$, and  $V_{R}$, highlighted in red, are connected to high-bandwidth lines. A proximal charge sensor (not shown) coupled to a radio frequency circuit allows fast measurements. The direction of the magnetic field $B_\parallel$ and $B_\perp$ is indicated.
	(b) Charge diagrams indicating the electron occupation of the triple quantum dot as function of $V_{L}$, $V_{M}$, and  $V_{R}$. Arrows indicate $\zeta$ and $\epsilon$ axes in a gate voltage space.
	(c) Concept of the experiment. Two electrons are initialized in a singlet state in the left quantum dot. Thereafter one of the electrons is moved to the middle dot and interacts with the multielectron quantum dot through exchange interaction $J$. At the end, readout is attained by performing spin-to-charge conversion for two-electron spin states in the double quantum dot.
	(d) Implementation of the pulse sequence in respect of the gate-voltage parameters $\zeta$ and $\varepsilon$.
	}
	\label{negative-j:fig1}
	\end{center}
\end{figure}

In this Letter, we investigate a GaAs multielectron quantum dot and show that its spin properties make it suitable for use as a coherent spin mediator. The experiment involves a chain of three quantum dots that can be detuned relative to each other using top-gate voltage pulses. 
The central one-electron dot serves as a probe: its spin can be tunnel coupled either to the left one-electron dot (serving as a reference spin for initialization and readout), or to a large dot on the right, thereby probing its multielectron spin states. 
We focus on a particular odd occupancy of the multielectron dot, $2N$+1, characterized by an effective spin 1/2, and establish that the exchange coupling between the central probe spin and the multielectron spin depends strongly and non-monotonically on the detuning of relevant gate voltages.  
Remarkably, this exchange coupling becomes negative, i.e. triplet-preferring, as the central electron is detuned further into the right dot. We therefore infer a spin-1 ground state for $2N$+2 occupation, even in the absence of an applied magnetic field. 
Besides fundamental implications for the role of non-trivial interactions within a multielectron dot, presented elsewhere for a large range of MED occupations~\cite{Malinowski2017}, our finding has practical applications. 
For example, the nonmonotonicity of the exchange profile results in a sweetspot, whereas its sign reversal removes a long-standing constraint for the construction of compact dynamically corrected exchange gates~\cite{Wang2012,Wang2014}.

\section{Sample}

The three quantum dots were fabricated in a GaAs/Al$_{0.3}$Ga$_{0.7}$As heterostructure hosting a two-dimensional electron gas with a bulk density $n$~=~$2.5\times10^{15}$~m$^{-2}$ and a mobility  $\mu$~=~230~m$^{2}$/Vs, located 57 nm below the wafer surface.
The confining potential and dot occupancy is voltage-tuned by Ti/Au metallic gates deposited on a 10~nm thin HfO$_2$ gate dielectric. 
Figure~\ref{negative-j:fig1}(a) shows the two accumulation gates (colored in green) surrounded by various depletion gates, and a schematic cut through the resulting triple-well potential. Gates labeled $V_{L}$, $V_{M}$, and  $V_{R}$ (colored in red) are connected to high-bandwidth coaxial lines and allow application of nanosecond-scale voltage pulses. 
An adjacent quantum dot (not shown) serves as a fast charge sensor, i.e. changes in its conductance change the amplitude ($V_{RF}$) of a reflected rf carrier~\cite{Barthel2010}.
All measurements were conducted in a dilution refrigerator with mixing chamber temperature below 30~mK. 

The device can be viewed as a two-electron double quantum dot (DQD) tunnel-coupled to a multielectron dot (MED) with an estimated number of electrons between 50 and 100, based on $n$ and the area of the multielectron dot.
By measuring $V_{RF}$ as a function of voltages $V_{L}$, $V_{M}$ and  $V_{R}$ we can map out the dots' occupancies in the vicinity of the charge states (2,0,$2N$+1), (1,1,$2N$+1) and (1,0,$2N$+2). Here, the numbers correspond to electron occupation in the left dot, central dot and the MED, respectively. 
The resulting charge diagram in Fig.~\ref{negative-j:fig1}(b) allows the definition of two detuning axes in gate-voltage space, $\zeta$ and $\varepsilon$, such that a reduction of $\zeta$ pushes the central electron into the left dot, whereas an increase in $\varepsilon$ pushes it to the MED (cf. arrows).

The MED spin states are probed by the pulse sequence illustrated in Fig.~\ref{negative-j:fig1}(c,d).
First, two electrons in a singlet state are prepared in the left dot, by pulsing to the (2,0,$2N$+1) charge state. 
Then a $\zeta$ pulse to the (1,1,$2N$+1) state effectively turns off intra-DQD exchange interactions while maintaining the two-electron spin state. 
The next step probes the interaction between the central electron and the MED in the vicinity of the charge transition between (1,1,$2N$+1) and (1,0,$2N$+2). 
This is done by pulsing $\varepsilon$, i.e. by temporarily applying a negative voltage pulse to $V_M$ and a positive voltage pulse to $V_R$.
After an interaction time $\tau$ we return to (1,1,$2N$+1) and immediately reduce $\zeta$ for single-shot reflectometry readout~\cite{Barthel2009}: If $V_{RF}$ indicates a (2,0,$2N$+1) charge state, we assign a singlet outcome, whereas (1,1,$2N$+1) indicates that a spin interaction with the MED has occured, and we count it as a non-singlet outcome. The fraction of singlet outcomes when repeating typically 1024 identical pulse sequences is denoted by $P_S$. 

\begin{figure*}[t]
	\centering
	\includegraphics[width=\textwidth]{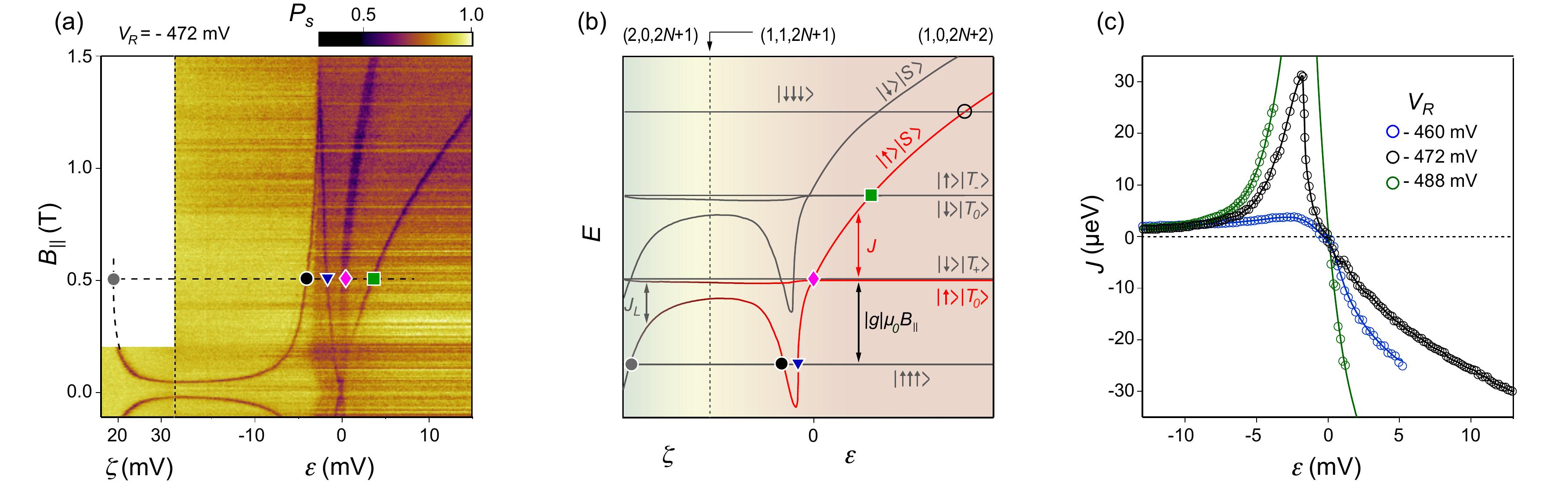}
	\caption[Leakage spectroscopy and inferred energy diagram]
	{
	(a) $P_S$ as a function of $\zeta$, $\varepsilon$ and $B_\parallel$ for a fixed, long interaction time $\tau=150$~ns. 
		(b) Corresponding energy diagram of the spin states of a Heisenberg model, as a function of $\zeta$, $\varepsilon$ for a fixed $B_\parallel$. 
	States highlighted in red witness the interaction between the central probe spin and the effective MED spin, which combine into a singlet-like state, $\ket{\uparrow}\ket{S}$, that is above a triplet-like state, $\ket{\uparrow}\ket{T_0}$, for sufficiently large $\varepsilon$ (negative $J$).  
	The charge character of the groundstate transitions from (2,0,2$N$+1) via (1,1,2$N$+1) to (1,0,2$N$+2) as indicated by the background shading. 
	The sign reversal of $J$ happens at $\varepsilon=0$.
	The Zeeman shift $|g| \mu_B B_\parallel$ and crossings with other states leading to spin leakage features in (a) are indicated (see main text). 
	Leakage from $\ket{\uparrow}\ket{S}$ to the fully polarized $\ket{\downarrow\downarrow\downarrow}$ state (empty circle) is not observed in (a), likely because weak Overhauser gradients or spin-orbit coupling do not allow changes in spin projection by 2. 
	 The $S$-$T_0$ leakage feature in (a) (magenta diamond) is not field independent as predicted by the model, likely due to orbital coupling of $B_\parallel$ to MED states in combination with a small misalignment of the sample.
	(c) Experimental exchange profiles for different operating points (distortions of the confining potential), identified by $V_{R}$ during the readout step (symbols). Black circles are extracted from (a). Solid lines are guides to the eye.
}
\label{negative-j:fig2}
\end{figure*}

\section{Leakage spectroscopy and exchange oscillations}

Leakage spectroscopy is performed by choosing  $\tau$ sufficiently long to detect incoherent spin mixing between the central electron and MED states.  
Figure~\ref{negative-j:fig2}(a) shows $P_S(\varepsilon, B_\parallel)$, where $\varepsilon$ is the detuning voltage during the interaction step and  $B_\parallel$ is the applied in-plane magnetic field. To make connection to the conventional two-electron DQD regime we also plot $P_S(\zeta, B_\parallel)$, acquired by replacing the composite $\zeta$-$\varepsilon$ pulse in Fig.~\ref{negative-j:fig1}(d) by a pure  $\zeta$ pulse. 
Spin leakage is clearly observed as a sharp suppression of $P_S$ for particular detuning values, with a non-trivial magnetic field  dependence for $\varepsilon>-5$~mV. To understand this spectrum we note that all features below $\varepsilon\approx-5$~mV are well explained by mixing with fully polarized spin states, consistent with previous spin leakage experiments: The $\varepsilon$ dependence (``spin funnel'') is analogous to mixing between singlet and $T_+\equiv\ket{\uparrow\uparrow}$ in two-electron DQDs~\cite{Petta2005,Maune2012,Veldhorst2015}, whereas the $\zeta$ dependence is analogues to mixing between a singlet-like state and $\ket{\uparrow\uparrow\uparrow}$ in three-electron triple quantum dots~\cite{Medford2013a,Medford2013}. 
(Here, each arrow indicates the spin state within one quantum dot.)
The characteristic dependence on $B_\parallel$ arises from the linear Zeeman shift of fully polarized spin states~\cite{Laird2010,Taylor2013}.

This identification confirms odd multielectron occupation, i.e. (1,1,$2N$+1), with effective spin 1/2. It also allows us to extract the exchange interaction ($J$) between the central spin and the effective MED spin from the ordinate of the leakage feature (black dot), using  $J = |g| \mu_B B_\parallel$, where $g = -0.44$ is the electronic $g$-factor for GaAs and $\mu_B$ is the Bohr magneton. 
Towards higher detuning, $\varepsilon>-5$~mV, an overall drop in the background of $P_S$ indicates that the MED ground state transitions into (1,0,$2N$+2), approximately concurrent with the sharp leakage feature (black dot) reaching a maximum before turning towards $B_\parallel=0$ (blue triangle). At $\varepsilon=0$ two additonal leakage features appear at $B_\parallel=0$. We interpret this maximum as a maximum in in the exchange profile, $J(\varepsilon)$, and associate the crossing at $B_\parallel=0$ with a sign reversal of $J(\varepsilon)$.

To infer the spin spectrum, we impose the observed exchange profile $J(\zeta,\varepsilon)$ on a Heisenberg model of three spin-1/2 orbitals~\footnote{The specific detuning dependence of $J$ within a Hubbard model is explained in Ref.~\cite{many_JB_charge_states} (chapter~\ref{ch:many_charge_states}), and remains phenomenological within this Letter.}. For simplicity we ignore orbital coupling to $B_\parallel$ and inspect spin Zeeman effects only.
The resulting energy diagram, sketched in Fig.~\ref{negative-j:fig2}(b) for finite $B_\parallel$, allows us to identify all characteristic leakage features.
On the left side of the energy diagram only intra-DQD exchange is significant ($J_{L}$), and the eigenstates are the tensor products of a DQD spin state and a MED ``spectator" spin. For example, the grey dot marks the crossing between $\ket{S}\ket{\uparrow}$ and $\ket{T_+}\ket{\uparrow}$, and relates the ``spin funnel'' in (a) to $J_{L}(\zeta)$.
Analogously, on the right side of the energy diagram, the left dot is decoupled and hosts the spectator spin, while the central spin interacts with the effective MED spin. Here, field dependent crossings map out the positive (black and blue marker) and negative (green marker) regime of $J(\varepsilon)$ (cf. crossings of $\ket{\uparrow}\ket{S}$ with $\ket{\uparrow\uparrow\uparrow}$, $\ket{\uparrow}\ket{T_0}$ and $\ket{\downarrow}\ket{T_+}$). 
At these crossings rapid mixing due to uncontrolled Overhauser gradients is expected to occur, changing electronic spin projections by 1 on a timescale of $T_2^*\approx10$ ns \cite{Petta2010}. 

\begin{figure}[t]
	\begin{center}
	\includegraphics[width=0.6\textwidth]{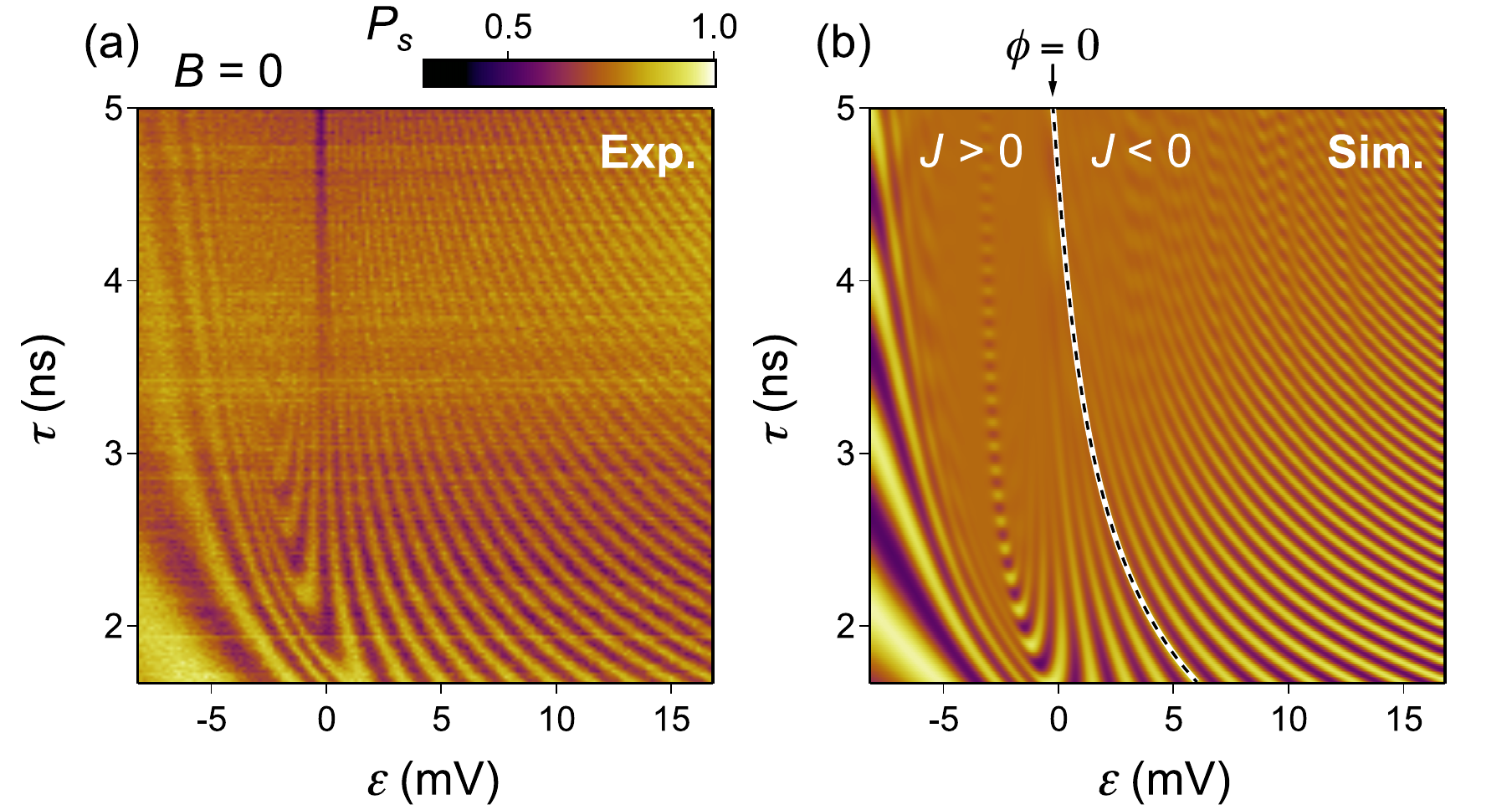}
	\caption[Exchange oscillations between a single electron and a multielectron quantum dot]
	{
	(a) Exchange oscillations in $P_S$ as a function of $\varepsilon$ and exchange time $\tau$, in the vicinity of (1,1,$2N$+1) and (1,0,$2N$+2) charge transition. External magnetic field is zero, and DC tuning voltages are the same as in Fig.~\ref{negative-j:fig2}(a) ($V_{R}$~=~472~mV).
	(b) Simulation of the exchange oscillations using $J(\varepsilon)$ from Fig.~\ref{negative-j:fig2}(c). Simulation assumes a Gaussian $\varepsilon$ low frequency noise with a standard deviation of 0.18 mV, and a rise time of the experimental instrumentation of 0.8~ns. Dashed line indicates zero phase accumulation and divides the area where $J$ is positive and negative. For large $\tau$ a dark feature appears at $\varepsilon = 0$ in (a) but not in (b). We associate it with leakage out of the simulated subspace, corresponding to the crossing in Fig.~\ref{negative-j:fig2}(a,b) indicated by a green square. 
}
	\label{negative-j:fig3}
	\end{center}
\end{figure}

In contrast to three-electron triple dots~\cite{Medford2013a,Medford2013}, where $J$ is always positive, we observe that $\ket{\uparrow}\ket{S}$ and $\ket{\uparrow}\ket{T_0}$ cross each other at $\varepsilon=0$.
This implies that the exchange interaction between the single and multielectron quantum dot changes sign from positive to negative, i.e. it is singlet-preferring for small hybridization and becomes triplet-preferring once the central electron has transferred to the multielectron dot. Next, we test for robustness and gate-tunability of this effect.
In Fig.~\ref{negative-j:fig2}(c) we plot $J(\varepsilon)$ extracted from Fig.~\ref{negative-j:fig2}(a) (black symbols), and compare it to two exchange profiles (green and blue symbols) measured by distorting the confining potential while preserving the charge configuration of the triple dot system (cf. Fig.~\ref{negative-j:figS1}).
In all cases $J(\varepsilon)$ shows the same behavior, namely a maximum and sign reversal at the position of the charge transition, and a negative sign in the (1,0,$2N$+2) configuration.
This interpretation implies that the $2N$+2 charge state of the multielectron dot has a total spin of 1 at zero magnetic field, which is further confirmed by studying the MED behavior over multiple charge states~\cite{Malinowski2017}.

Direct evidence for the sign reversal in $J$ (without the need for a magnetic field) can be obtained from time-domain measurements. To this end, we induce coherent exchange oscillations between central and MED spin by significantly reducing (and varying) the interaction time $\tau$. The observed pattern of $P_S(\varepsilon,\tau)$, shown in Fig.~\ref{negative-j:fig3}(a) for the same DC tuning parameters as in Fig.~\ref{negative-j:fig2}(a), differ from analogous oscillations of the exchange-only qubit \cite{Laird2010,Medford2013}. 
Namely, the appearance of a chevron-like pattern indicates the presence of a local maximum in $J(\varepsilon)$. 
Following contours of equal phase ($\phi$) around this ``sweet spot'', we note that $\phi(\tau)$ has opposite sign for large and small $\varepsilon$, implying a sign reversal in $J(\varepsilon)$. 
To show consistency between time-domain and leakage spectroscopy results, we perform numerical simulations of the exchange oscillations using the experimentally measured exchange profile presented in Fig.~\ref{negative-j:fig2}(c). The simulation is limited to the Hilbert space spanned by $\ket{\uparrow}\ket{S}$ and $\ket{\uparrow}\ket{T_0}$ (indicated with red in Fig.~\ref{negative-j:fig2}(b)) and includes a quasistatic Gaussian noise in $\varepsilon$ with standard deviation $\sigma_{\epsilon}$~=~0.18~mV~\cite{Martins2016,Barnes2016} and a rise time of our instrumentation of 0.8~ns. The simulation reproduces a chevron pattern (Fig.~\ref{negative-j:fig3}(b)), whereas simulations using $J(\varepsilon)=|J(\varepsilon)|$ produce a qualitatively different pattern (not shown). Therefore, the contour $\phi=0$ does indeed separate regions with $J>0$ from regions with $J<0$.

\begin{figure}[t]
	\begin{center}
	\includegraphics[width=0.6\textwidth]{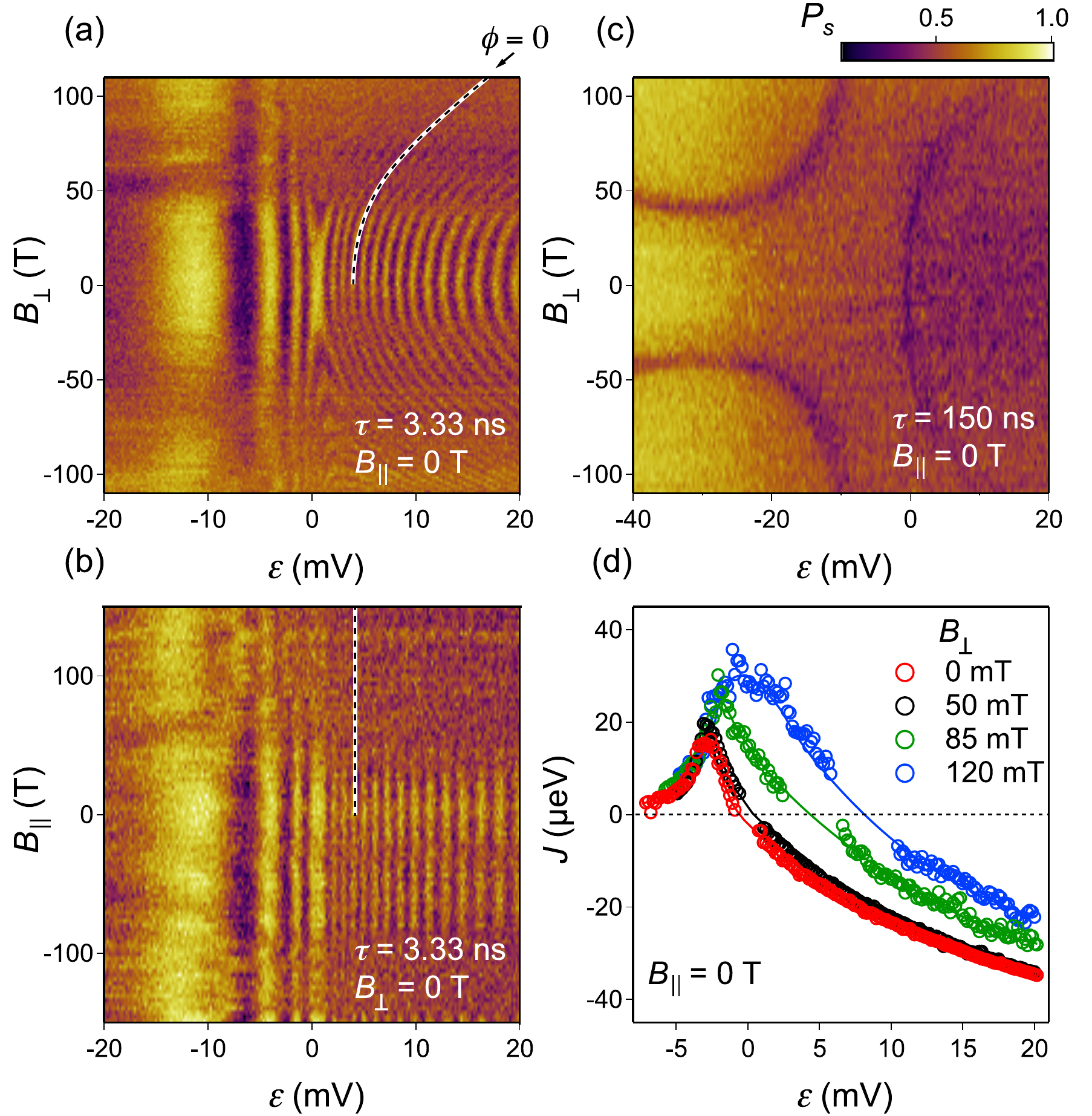}
	\caption[Exchange oscillations and leakage spectroscopy as a function of orbital magnetic field]{
	(a) Exchange oscillations as a function of orbital field $B_\perp$ and $\varepsilon$ for fixed exchange time $\tau$~=~3.33~ns and fixed $B_\parallel$~=~0~T. 
	(b) Exchange oscillations as a function of $B_\parallel$ and $\varepsilon$ for fixed exchange time $\tau$~=~3.33~ns and fixed $B_\perp$~=~0~T.
	(c) Same as (a) but in the leakage spectroscopy regime ($\tau$~=~150~ns). Features of reduced $P_S$ correspond to mixing between $\ket{\uparrow}\ket{S}$ and various other states (cf. horizontal cut of Fig.~\ref{negative-j:fig2}(a) at $B_\parallel=0$). A small deviation between the $J=0$ feature in (c) and the $\phi=0$ contour in (a) is likely due the drastically different choices of $\tau$ in combination with finite-rise-time effects of our instrumentation. 
	(d) Exchange profiles $J(\varepsilon)$ for different values of $B_\perp$.
	}
	\label{negative-j:fig4}
	\end{center}
\end{figure}

\section{Effect of orbital field}

Finally, we study the effects of applied magnetic fields on the exchange profile. 
Figure~\ref{negative-j:fig4}(a) presents $P_S$ as a function of $\varepsilon$ and out-of-plane magnetic field $B_\perp$, while keeping $B_\parallel$~=~0 and $\tau$~=~3.33~ns fixed. 
In such a plot, contours correspond to constant $J$ in the $\varepsilon$-$B_\perp$ plane, and their curvature indicates that out-of-plane magnetic fields move the sign reversal of $J$ towards higher detuning (cf. $\phi=0$ contour, marked by a dashed line). 
For comparison, within the same range, $B_\parallel$ has no observable influence on the pattern of the exchange oscillations (Fig.~\ref{negative-j:fig4}(b)). 
By choosing $\tau$ longer than the coherence time we obtain the $B_\perp$-dependence of the leakage spectrum (Fig.~\ref{negative-j:fig4}(c), using $\tau$~=~150~ns). 
The two leakage features appearing for negative values of $\varepsilon$ correspond to mixing between $\ket{\uparrow}\ket{S}$ and the fully polarized $\ket{\uparrow\uparrow\uparrow}$. The leakage feature appearing for positive values of $\varepsilon$ indicates $J=0$ and resolves into three lines at higher fields (cf. Fig.~\ref{negative-j:fig2}). 

\begin{figure}[t]
\begin{center}
\includegraphics[width=110 mm]{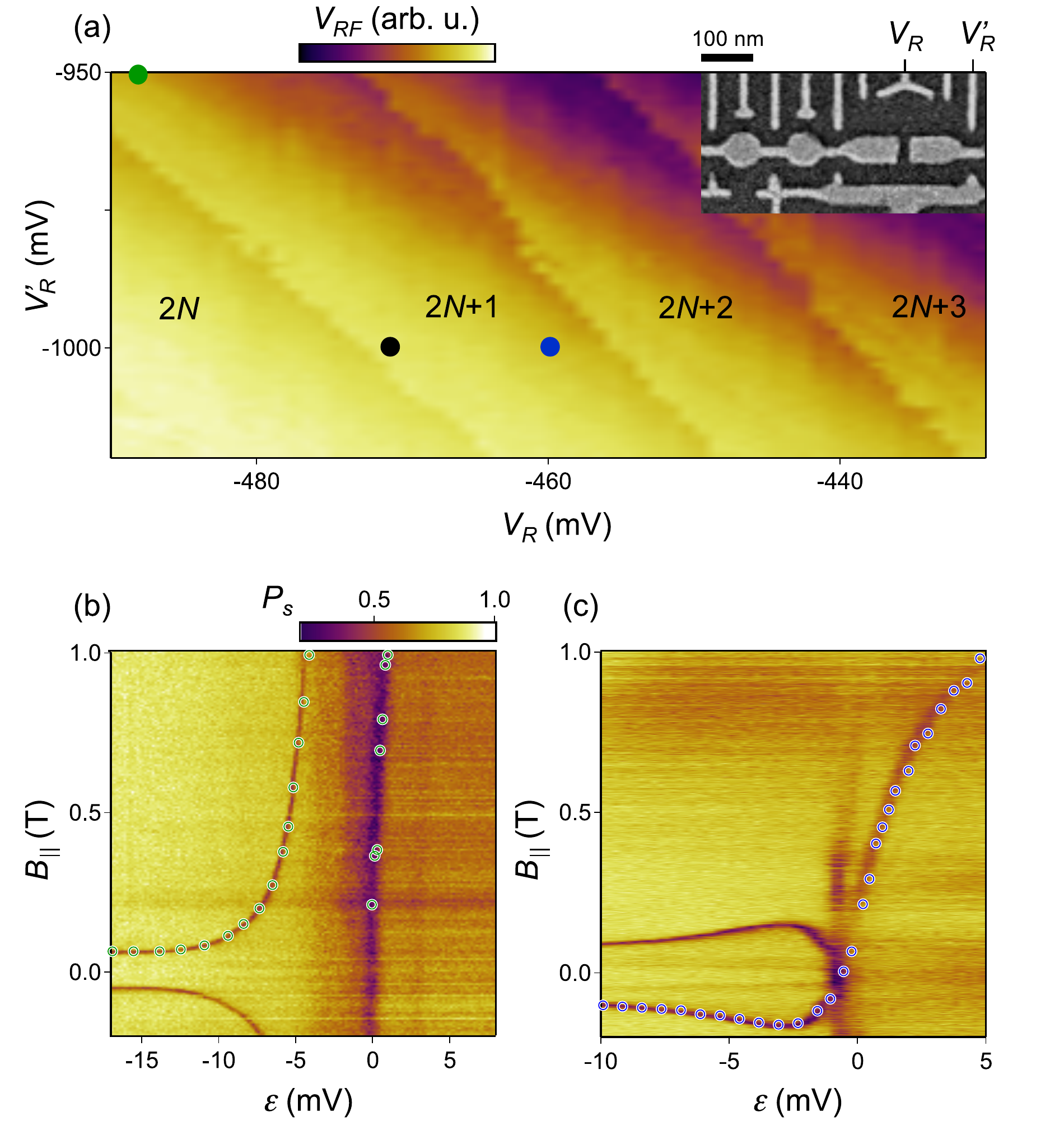}
\caption[Leakage spectroscopy for different tuning of the multielectron dot in tha same occupancy]{
(a) Charge diagram indicating the electron occupation in the multielectron dot as function of gates $V_{R}$ and  $V_{R}'$, as defined in the micrograph shown in the inset. Dots indicate the DC values of $V_{R}$ and $V_{R}'$ at which spectroscopy of the exchange energy have been performed.
(b) Probability of detecting a singlet, $P_{s}$, as a function of $\varepsilon$ and $B_{||}$ for  a exchange time $\tau=150$~ns, $V_{R}$~=~-490~mV and $V_{R}'$~=~-950~mV.
(c) Same as (b) for $V_{R}$~=~-460~mV and $V_{R}'$~=~-1000~mV.
Data corresponding to the black dot is shown in the main article in Fig.~\ref{negative-j:fig2}(a).
Exchange profile, $J$, extracted from these two spectroscopies is shown in Fig.~\ref{negative-j:fig2}(c).
}
\label{negative-j:figS1}
\vspace{10pt}
%
\includegraphics[width=150 mm]{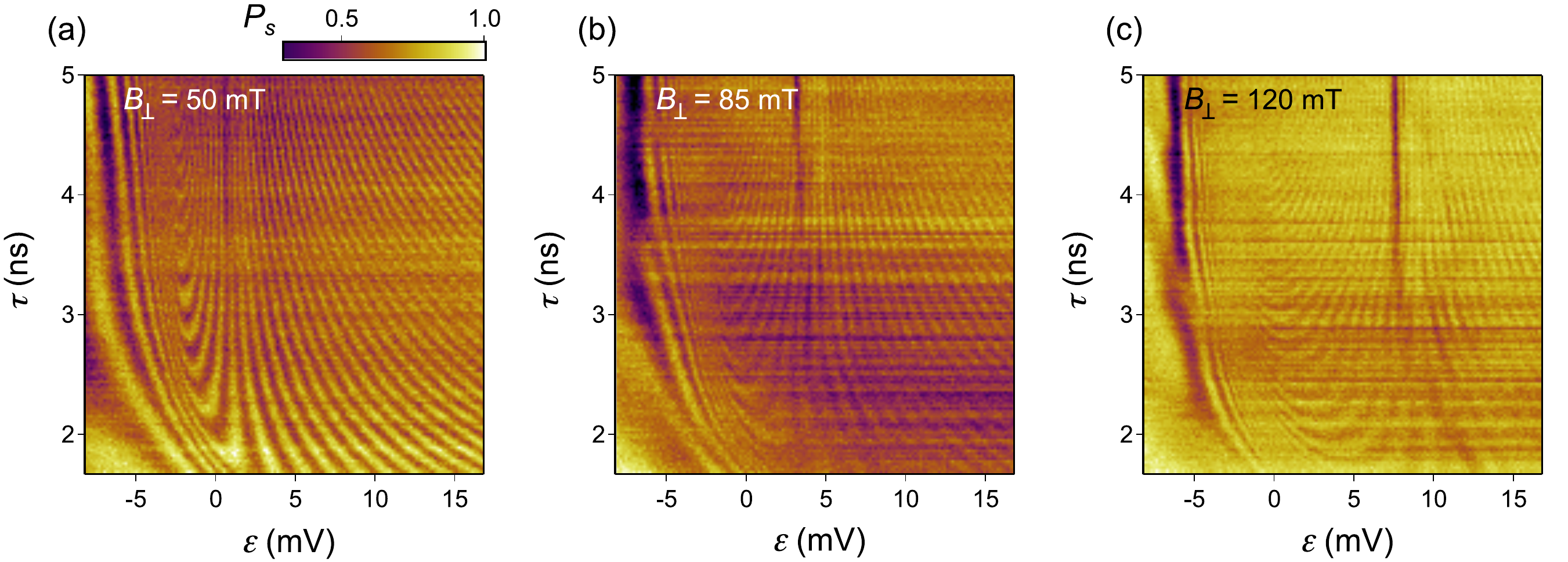}
\caption[Time-resolved exchange oscillations in orbital magnetic field]{ 
Exchange oscillations as a function of $\varepsilon$ for a exchange time $\tau$, in the vicinity of the (1,1,$N$)-(1,0,$N$+1) charge transition for various values of $B_\perp$. 
$J$ profiles extracted from these oscillations are plotted in  Fig.~\ref{negative-j:fig4}(d) in the main text.
(a) $B_\perp =50$~mT; (b) $B_\perp =85$~mT;  (c) $B_\perp =120$~mT.
}
\label{negative-j:figS2}
\end{center}
\end{figure}

Exchange profiles $J(\varepsilon)$ for $B_\perp$ =  0, 50, 85 and 120 mT were extracted from $P_S(\varepsilon,\tau)$ maps obtained for the same tuning voltages as in Fig.~\ref{negative-j:fig4}(a) (Fig.~\ref{negative-j:figS2}). Their $B_\perp$-dependence shown in Fig.~\ref{negative-j:fig4}(d) corroborates again the sensitivity of the exchange profile to the underlying electronic orbitals, and establishes an electrical  sweet spot in $J(\varepsilon)$ that can be precisely tuned by $B_\perp$. 

\section{Conclusion}

In summary, we have investigated experimentally the exchange interaction between a two-electron double quantum dot and a multielectron quantum dot, by complementing incoherent spin leakage measurements with time-resolved coherent exchange oscillations at various tuning voltages and magnetic field configurations. 
We find that the multielectron dot with odd occupation number behaves as a spin-1/2 object that gives rise to a non-monotonic exchange coupling to the neighboring dot. By changing the relative dot detuning voltage by a few millivolt the sign of the exchange interaction can be tuned from positive to negative (also at zero magnetic field), indicating the presence of non-trivial electron-electron interactions. Finally, we show that the exchange profile can be tuned by either changing the gate potentials or applying an out-of-plane orbital magnetic field, giving rise to a tunable electrical sweet spot that might benefit the implementation of high-fidelity exchange gates~\cite{Martins2016,Reed2016} in long-distance quantum mediators. 

\section*{Acknowledgements}

We thank Stephen Bartlett, Andrew Doherty and Thomas Smith for helpful discussions.  This work was supported by LPS-MPO-CMTC, the EC FP7-ICT project SiSPIN no. 323841, the Army Research Office and the Danish National Research Foundation.

\section*{Author contributions}
S.F., G.C.G. and M.J.M. grew the heterostructure. P.D.N. fabricated the device. F.M., F.K.M., F.K. and P.D.N. prepared the experimental setup. F.K.M. and F.M. performed the experiment. F.K.M. performed the simulations. F.M., F.K.M., F.K. and C.M.M. analysed data and prepared the manuscript.

\section{Supplement: Extracting $J(\varepsilon)$ from exchange oscillations}

Exchange profiles $J(\varepsilon)$ plotted in Fig.~\ref{negative-j:fig4}(d) in the main text were obtained from Fig.~\ref{negative-j:fig3}(a) and Figs.~\ref{negative-j:fig2}(a-c) for $B_\perp $~=~0, 50, 85 and 120~mT, respectively.
For each value of $\varepsilon$, the frequency of the exchange oscillations $J$ is obtained in two steps. 
First, we calculate the Fast Fourier transform of $P_\mathrm{s}(\tau)$ and find the frequency bin with the largest weight. 
Then we use this frequency as an initial guess for fitting a damped sine wave of frequency $J$ to $P_\mathrm{s}(\tau)$, with a decaying amplitude of the form exp$\left(-\tau/T_\mathrm{R}\right)$. 
Values of $J(\varepsilon)$ extrated by this method are plotted as circles in Fig.~\ref{negative-j:fig3}(d).



\chapterimage{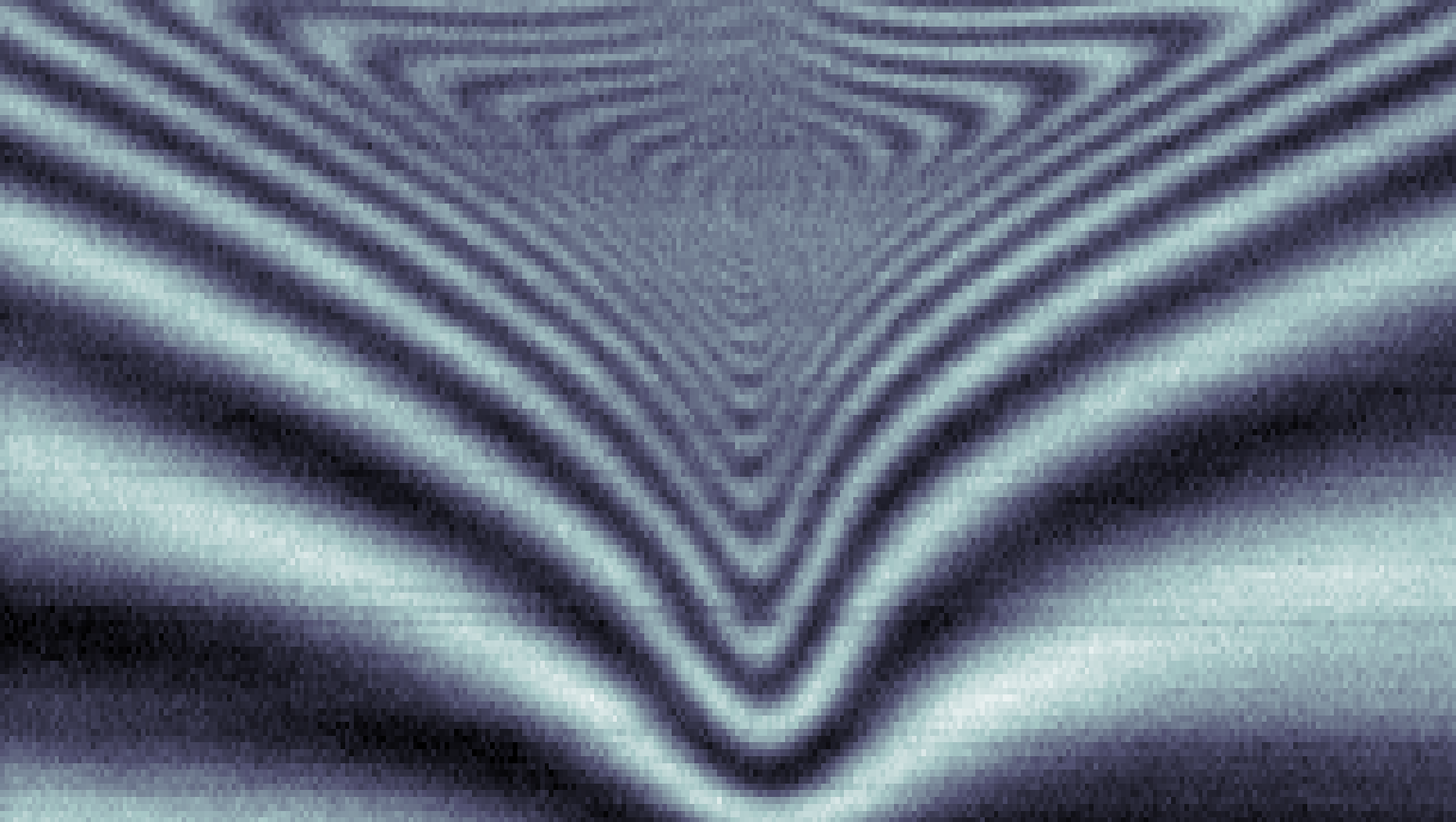}
\chapter[Fast coherent spin-exchange via a multielectron quantum mediator]{\protect\parbox{0.9\textwidth}{Fast coherent spin-exchange via \\ a multielectron quantum mediator}}
\chaptermark{Fast coherent spin-exchange via a multielectron...}
\label{ch:jb-mediated}

{\let\thefootnote \relax\footnote{This chapter and chapter \ref{ch:supp_jb-mediated} are adapted from the manuscript in preparation.}}
\addtocounter{footnote}{-1}

\begin{center}
Filip K. Malinowski$^{1,*}$, Frederico Martins$^{1,*}$, Thomas Smith$^{2}$, Stephen D. Bartlett$^{2}$, \\
Andrew C. Doherty$^{2}$, Peter D. Nissen$^{1}$, Saeed Fallahi$^{3}$, Geoffrey C. Gardner$^{4}$, \\
Michael J. Manfra$^{3,4}$, Charles M. Marcus$^{1}$, Ferdinand Kuemmeth$^{1}$
\end{center}

\begin{center}
	\scriptsize
	$^{1}$ Center for Quantum Devices, Niels Bohr Institute, University of Copenhagen, 2100 Copenhagen, Denmark\\
	$^{2}$ Centre for Engineered Quantum Systems, School of Physics, The University of Sydney, Sydney NSW 2006, Australia  \\
	$^{3}$ Department of Physics and Astronomy, Birck Nanotechnology Center, and Station Q Purdue, \\
	Purdue University, West Lafayette, Indiana 47907, USA\\
	$^{4}$ School of Materials Engineering, Purdue University, West Lafayette, Indiana 47907, USA\\
	$^{*}$ These authors contributed equally to this work
\end{center}

\section{Introduction}

In a scalable quantum processor, the coupling used to perform two-qubit gates must have fulfill three key requirements: it should be fast, coherent and long-range.
Indeed, the presence of such coupling underlies the success of the superconducting~\cite{Riste2013,Kelly2015} and trapped-ion qubits~\cite{Monz2016}. 
The spin qubit community is also focused on finding such coupling scheme. 
However approaches based on superconducting qubits designs~\cite{McDermott2005,DiCarlo2009}, i.e. capacitative coupling~\cite{Shulman2012,Nichol2017} and cavity-mediated interaction~\cite{Mi2016,Stockklauser2017}, tend to be troublesome. 
That is because they involve a charge degree of freedom~\cite{Taylor2005a,Burkard2006}, while spin qubits lack the charge-noise protection~\cite{Taylor2007,Dial2013} inherent to transmons~\cite{Koch2007}. 
Alternatively, exchange coupling can be introduced in a charge-noise insensitive manner~\cite{Martins2016,Reed2016}. 
Moreover, the intrinsically short range of the exchange interaction can be extended by means of a quantum mediator~\cite{Baart2017,Mehl2014a,Srinivasa2015}. 
Here we show that a multielectron quantum dot works extraordinarily well as such mediator~\cite{Srinivasa2015}. 
We show that the exchange interaction mediated by a multielectron quantum dot can be controlled up to a few-gigahertz regime. 
Moreover, many-body effects present in the multielectron quantum dot give rise to the emergence of the extremum in the exchange interaction strength characterized by low charge-noise sensitivity~\cite{Negative-J,many_JB_charge_states}.
We conclude that the exchange interaction mediated by a multielectron dot fulfills all requirements of the scalable coupling mechanism and sketch a clear path for scaling the coherent quantum dot systems.

\begin{figure}[t]
	\centering
	\includegraphics[width=0.6\textwidth]{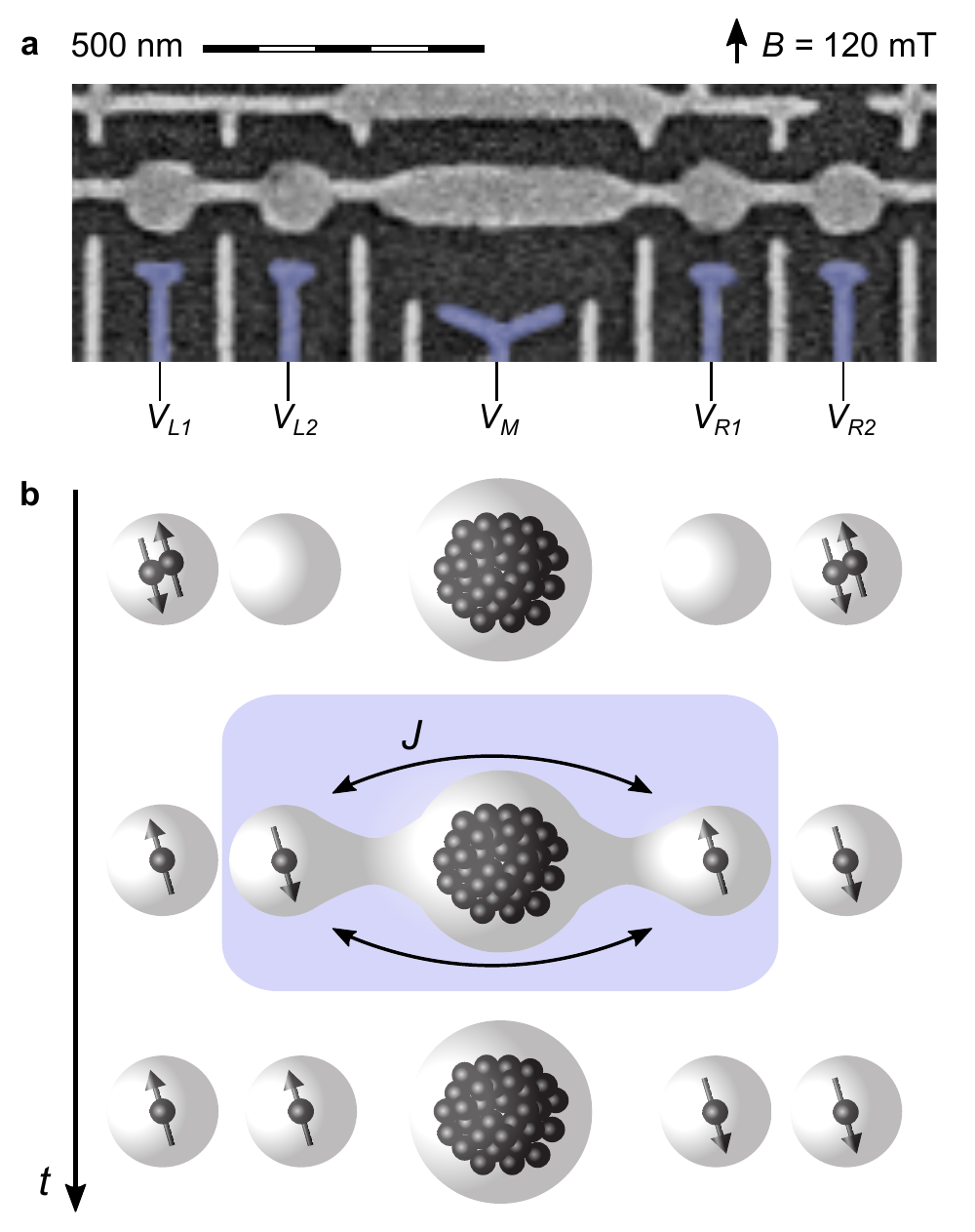}
	\caption[The concept of the experiment]{
	{\bf The concept of the experiment.}
	(a) SEM micrograph of the measured device. Single-electron quantum dots are located below the circular gates, while a multielectron dot is located below the central, elongated gate. Nanosecond voltage pulses applied to the blue-colored gates are used to control the position and the interaction between single electrons.
	(b) Scheme of the experiment. First, pairs of electrons in the left and right double quantum dot are initialized in the singlet state $\ket{S_{L/R}}$ in the outer dots. Than, single electrons are moved to the inner dots, thereby turning off the exchange interaction with the reference spins, which remain in the outer dots. Next, the exchange between the electrons located in the inner quantum dots is induced, by applying a positive voltage pulse on the gate $V_M$, and negative voltage pulse on the remaining gates. The exchange interaction causes flip-flops between electronic spins on the inner dots. As a result a spin state of the electrons within a double dots become one of the triplet states (see text for details). Change of the spin states of both double quantum dots is detected in parallel by means of spin-to-charge conversion followed by the measurement of the two nearby radio-frequency sensor quantum dots.
	}
	\label{jb-mediated:fig1}
\end{figure}

\section{The device and concept of the experiment}

We implement long-range exchange coupling mediated by a multielectron quantum dot in a quintuple quantum dot array (Fig.~\ref{jb-mediated:fig1}(a)). The quantum dots are defined in a 57~nm deep GaAs two-dimensional electron gas, covered by 10 nm Hf$_2$O insulating oxide layer, by means of electrostatic gates deposited on top of the heterostructure. The middle dot has an even number of electrons~\cite{many_JB_charge_states}, between 50 and 100 as estimated from the lithographic size of the device and the density of the 2-dimensional electron gas. The double quantum dots (DQDs) located on both sides are tuned to a single-electron regime.

The exchange interaction is studied by means of a sequence sub-microsecond voltage pulses applied to the blue-colored gates in Fig.~\ref{jb-mediated:fig1}(a). The pulse sequence (Supplementary Section~\ref{supp_jb-mediated:definitions}) realizes the following procedure (Fig.~\ref{jb-mediated:fig1}(b)). First, pairs of electron in the DQDs are pushed to the outmost quantum dots, where they relax. This initializes each of the DQDs in a spin singlet state $\ket{S^{L/R}}=(\ket{\uparrow\downarrow}-\ket{\downarrow\uparrow})/\sqrt{2}$ where the arrows indicate the spin state of the two electrons and subscript $L/R$ indicates left and right DQD. Than, the electrons are rapidly separated, so that each of the small dots is occupied by a single electron. This pulse turns off the exchange interaction between electrons within each DQD. In the third stage the positive voltage pulses on gate $V_M$ and negative on all other gates induces the exchange between the electrons located on the inner pair of small dots (to which we will refer as ``inner dots''). The induced exchange interaction causes flip-flops between those electronic spins. This leads to the correlated decrease of probability that electron pairs in both DQDs are in a singlet state. After the interaction time $\tau$ the exchange-inducing pulse is switched-off. Finally, we employ the spin-to-charge conversion~\cite{Johnson2005a} to readout the relative spin of the electron pairs in each DQD. The reflectometry readout of the two nearby quantum-dot-based charge sensors~\citep{Barthel2010a} allows us to distinguish between singlet and triplet states within each DQD independently and with single-shot fidelity.

\begin{figure}[t]
	\centering
	\includegraphics[width=\textwidth]{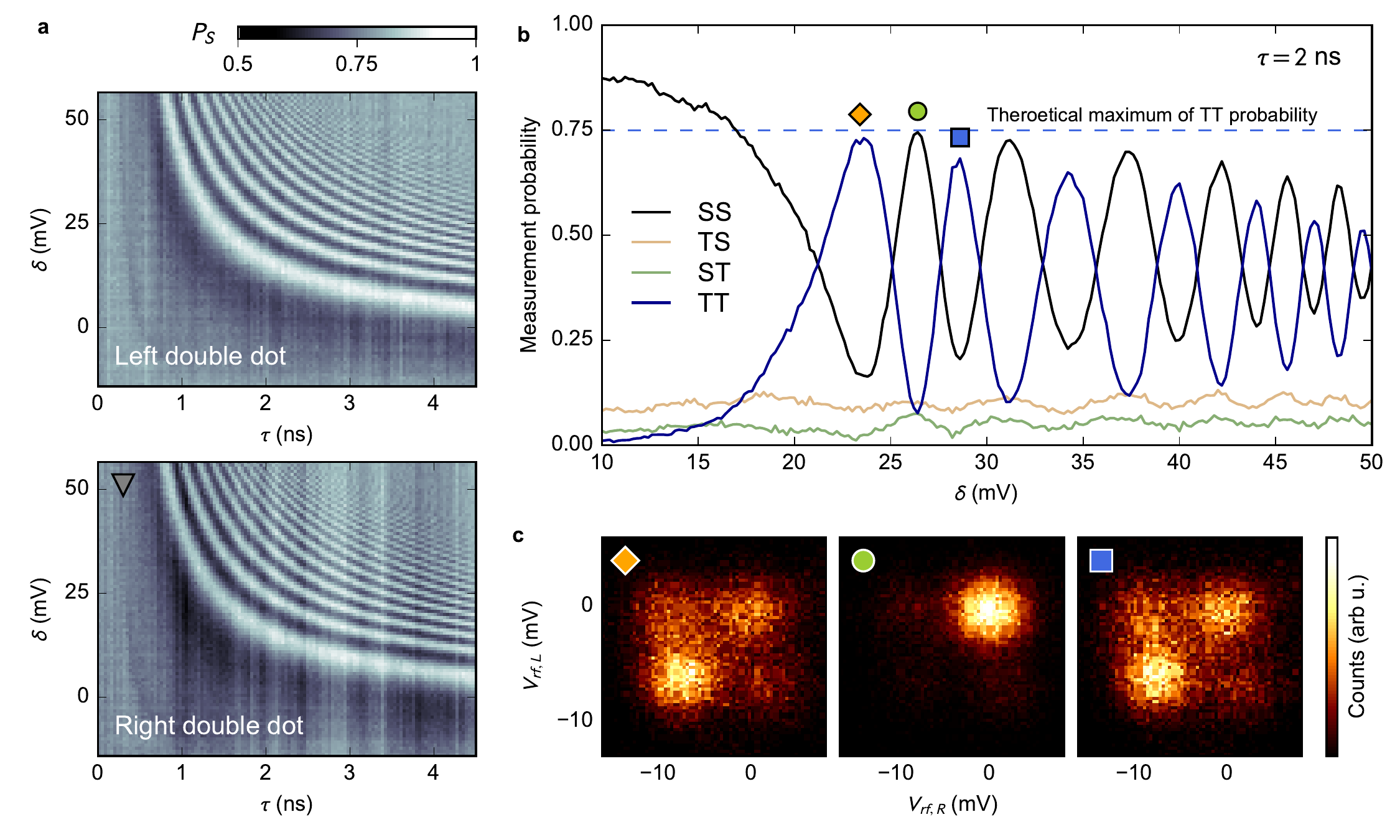}
	\caption[Evidence of the exchange interaction between distant electronic spins]{
	{\bf Evidence of the exchange interaction between distant electronic spins}
	(a) Probability of singlet measurement on both double quantum dots, as a function of the interaction time $\tau$ and the amplitude of the pulse on gate $V_M$. Data presented in both panels is acquired simultaneously.
	(b) Joint probabilities of the final spin states, as a function of the exchange-inducing pulse amplitude, for fixed interaction time $\tau=2$~ns.
	(c) Histograms of the single-shot outcomes for several pulse amplitudes, indicated with colored markers in panel (b).
	}
	\label{jb-mediated:fig2}
\end{figure}

\section{Evidence for the exchange coupling}

The result of such pulse sequence is shown in Fig.~\ref{jb-mediated:fig2}(a). In the two panels, we plot the singlet return probability $P_S$ of each DQD as a function of duration $\tau$ and voltage of the pulse amplitude (Supplementary Section~\ref{supp_jb-mediated:definitions}).
These oscillations are the result of the exchange-driven flip-flops of the electrons located on the inner small quantum dots. 
Noteworthy, the frequency of the oscillations decreases with small $V_M$ values, corresponding to the multielectron dot level detuned very far from the inner dots levels. 
Inversely, for a large positive pulse on gate $V_M$ the oscillations become faster as the multielectron dot level gets into resonance and, ultimately, shifts below the single particle levels of the inner dots (which will follow from the analysis presented later). 
The bending of the oscillatory pattern for short interaction times is a result of 0.8~ns rise time of the voltage pulses. Complementary evidence for the presence of the exchange interaction between the two electrons located on the inner quantum dots is provided by the leakage spectroscopy measurement, mapping the position of the crossings between various spin states (Supplementary Section~\ref{supp_jb-mediated:leakage}).

Next, we confirm the correlation between single-shot outcomes. 
In this measurement we fix interaction time $\tau = 2$~ns and vary the amplitude of the exchange-inducing pulse (see Supplementary Section~\ref{supp_jb-mediated:definitions}). 
The joint probabilities of singlet and triplet readout outcomes in the two DQDs are presented in Fig.~\ref{jb-mediated:fig2}(b). The obtained oscillations of singlet and triplet outcomes are correlated, and we observe the constant background of the anticorrelated counts.
Histograms of the recorded single-shot readouts for several exchange pulse amplitudes are presented in Fig.~\ref{jb-mediated:fig2}(c). The joint probabilities were estimated from the histograms by fitting the quadruple Gaussian, followed by a correction for the relaxation during the measurement time (see Supplementary Section~\ref{supp_jb-mediated:decay}).

The oscillations observed here are a result of the precession between the initialized $\ket{S_L}\ket{S_R}$ state and the fully entangled $\tfrac{1}{2}(\ket{S^L}\ket{S^R}-\ket{T_0^L}\ket{T_0^R}+\ket{T_+^L}\ket{T_-^R}+\ket{T_-^L}\ket{T_+^R})$ state. 
Here the two kets indicates the state of the left and right DQD, respectively. The spin triplet states are labeled according to the standard convention, $\ket{T_0}=(\ket{\uparrow\downarrow}+\ket{\downarrow\uparrow})/\sqrt{2}$, $\ket{T_+}=\ket{\uparrow\uparrow}$, $\ket{T_-}=\ket{\downarrow\downarrow}$. 
As a consequence, we expect fully correlated readouts and oscillations visibility of 75\%. 
Fig.~\ref{jb-mediated:fig2}(b) shows that both predictions are almost fulfilled. 
We attribute the observed background of anticorrelated outcomes, which decreases the oscillations visibility, to the interaction between the electrons located on the outer dots with the nuclear bath, as well as finite bandwidth of the voltage pulses. 

\begin{figure}[t]
	\centering
	\includegraphics[width=0.8\textwidth]{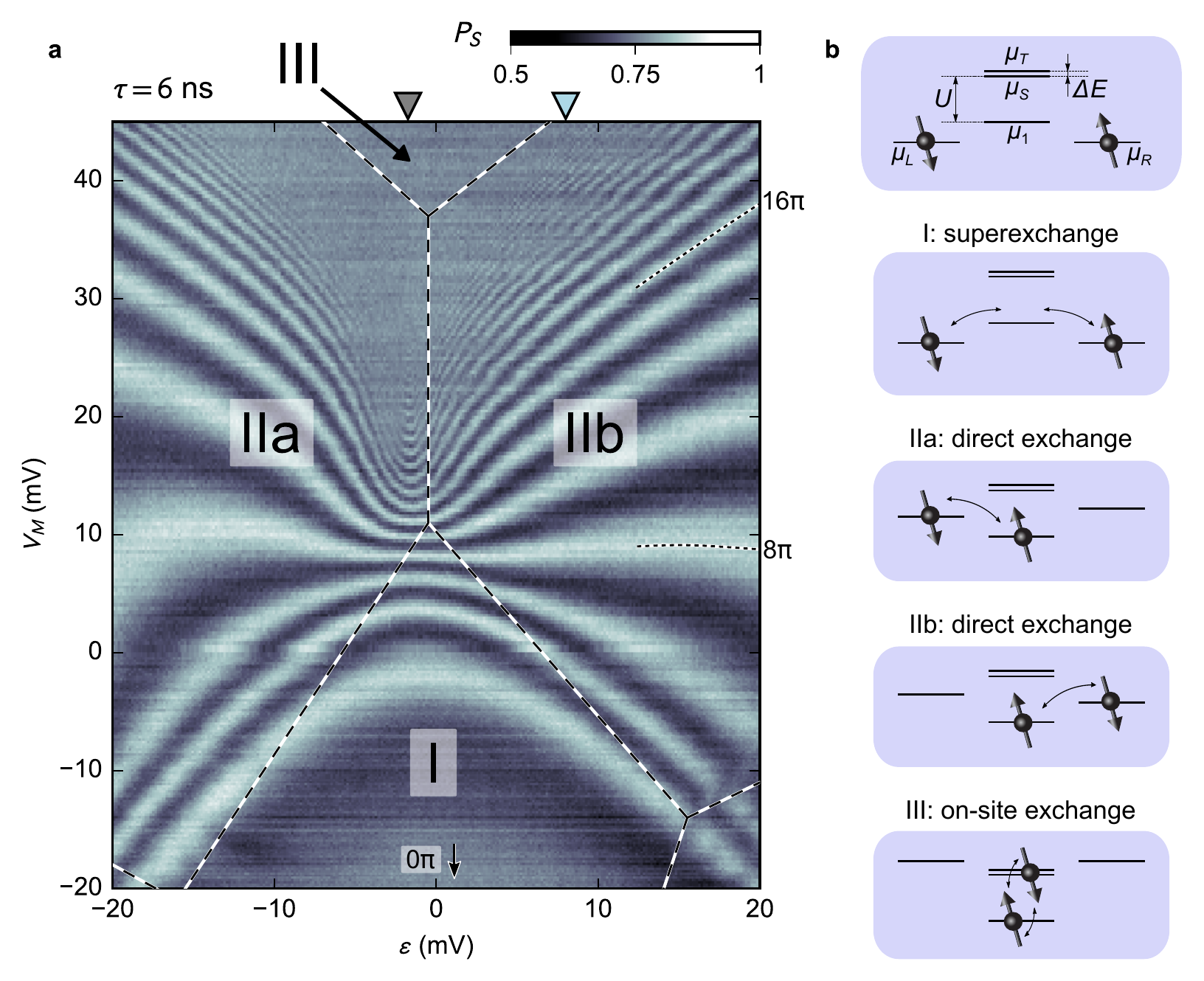}
	\caption[The fingerprint of the three regimes of the exchange interaction]{
	{\bf The fingerprint of the three regimes of the exchange interaction.}
	(a) Probability of the singlet outcome in the right double quantum dot, as a function of the detuning between the inner quantum dots $\varepsilon$, and voltage on the multielectron dot gate $V_M$, for fixed interaction time $\tau=6$~ns. Colored triangles indicate detuning corresponding to the data presented in Fig.~\ref{jb-mediated:fig2}a and \ref{jb-mediated:fig4}a. Dashed lines indicate independently measured positions of charge transitions (Sec.~\ref{supp_jb-mediated:charge}).
	(b) Illustration of the alignment of the chemical potentials of the dots in different interaction regimes. $\mu_{L/R}$ indicates chemical potential of the single-electron in the left/right dot. $\mu_1$ is the chemical potential of the single electron added to the multielectron dot. $\mu_{S/T}$ is the chemical potential of the multielectron dot with two electrons added in the singlet/triplet configuration. $U$ and $\Delta$ indicate charging energy and the level spacing of the multielectron dot.
	}
	\label{jb-mediated:fig3}
\end{figure}

\section{Three regimes of the exchange interaction}

Thereafter, we explore different regimes of the exchange interaction mediated by the multielectron quantum dot. 
For that purpose we define a new gate voltage parameter $\varepsilon=(V_{L2}-V_{R1})/\sqrt{2} + C$ (where $C$ is a constant; Supplementary Section~\ref{supp_jb-mediated:definitions}), which controls the detuning between the single particle levels of the two inner dots. In Fig.~\ref{jb-mediated:fig3}(a), we fix the interaction time to $\tau=6$~ns and map out the oscillations as a function of $\varepsilon$ and $V_M$. 
In this figure fringes correspond to the lines at which the acquired phase is fixed (as indicated by the annotations). 
Since the interaction time $\tau$ is fixed, these fringes correspond also to lines of constant exchange, up to distortions due to the voltage pulse bandwidth.
 
For negative $V_M$ the exchange interaction is the weakest, and it gradually increases for larger $V_M$.
The exchange interaction strength increases most rapidly for $\varepsilon \approx 0$ where, for $V_M\gtrsim$~20~mV, the fringes are blurred due to decoherence and aliasing. 
For large $| \varepsilon |$ the exchange increases more slowly, but ultimately the oscillations become impossible to resolve as well.
We also observe that the obtained pattern is almost symmetric with respect to $\varepsilon$.

To understand the pattern we measure the charge distribution during the interaction (Supplementary Section~\ref{supp_jb-mediated:charge}). The extracted positions of the charge transitions are indicated in Fig.~\ref{jb-mediated:fig3}(a) by the dashed lines while the deduced positions of the electrons are illustrated in Fig~\ref{jb-mediated:fig3}(b). In the region labeled I the inner dots remain singly occupied, and the multielectron dot remains in the initial charge state. In this case the virtual occupation of the multielectron quantum dot mediates the exchange interaction, as explained in detail in Ref.~\cite{Srinivasa2015}.

In the region IIa and IIb one of the electrons is moved to the multielectron dot and the effective many-body spin 1/2 exchange interacts directly with the second electronic spin. The symmetry between these two configurations gives raise to the symmetry of the oscillations pattern with respect to $\varepsilon=0$. The slight asymmetry arises form the inequality between the tunneling of the electrons from the left and the right inner dot to the central multielectron dot.

Finally, in region III the chemical potential of the multielectron dot is decreased so much that both of the electrons tunnel on the multielectron dot, and interact while being on the same site. Depending on the relative spin, either both electrons occupy the same, lowest orbital, or the lowest and the second lowest orbital. 
The energy difference between these levels is a limit to the strength of the exchange interaction, which is given by the two mesoscopic parameters: the level spacing $\Delta E$ and the difference of the Coulomb repulsion between electrons occupying the same or two different orbitals~\cite{many_JB_charge_states,Kurland2000,Folk2001}.

\begin{figure}[t]
	\centering
	\includegraphics[width=0.6\textwidth]{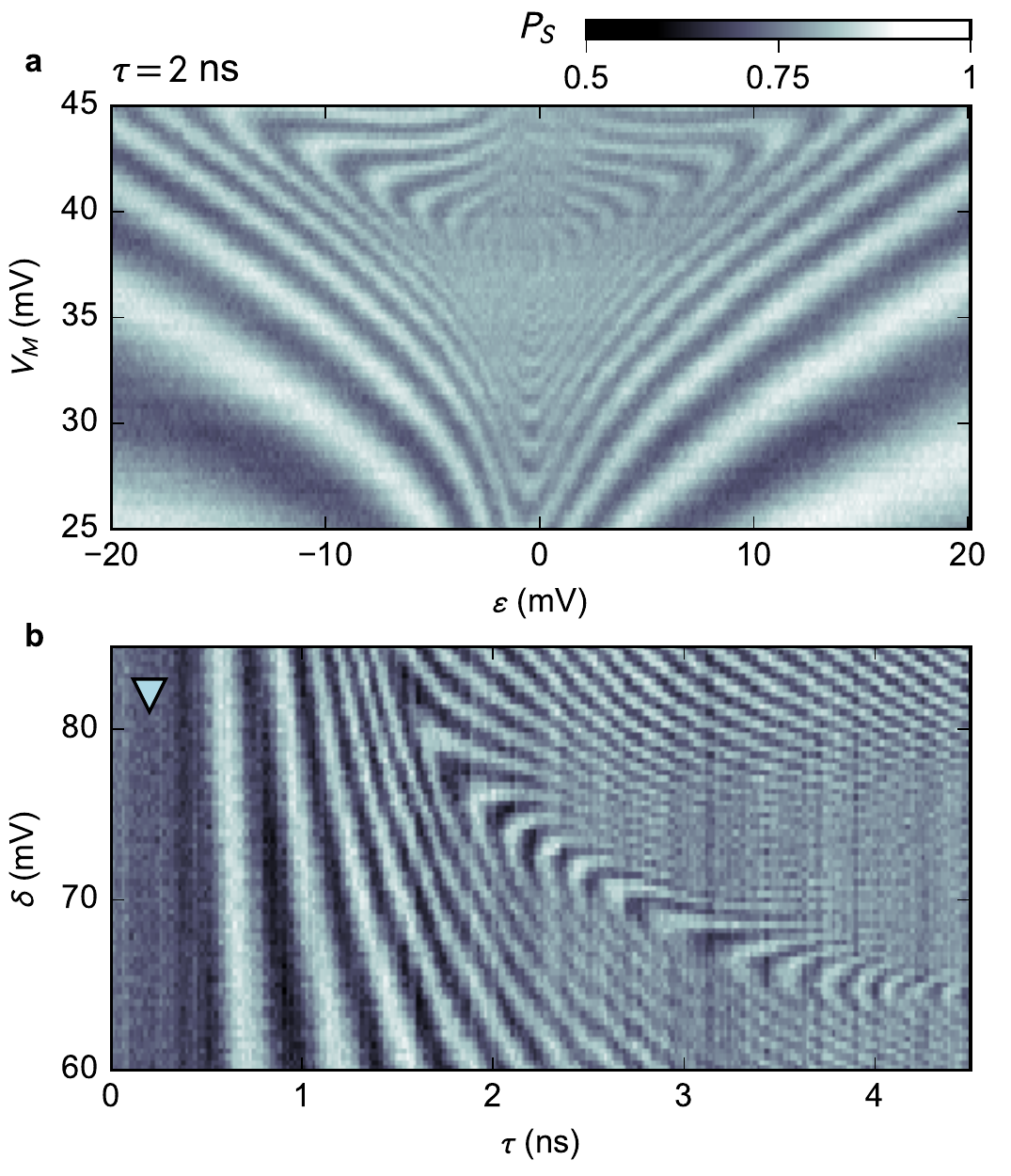}
	\caption[The extremum in the exchange energy at the border of the direct and on-site regimes of the exchange interaction]{
	{\bf The extremum in the exchange energy at the border of the direct and on-site regimes of the exchange interaction.}
	(a) Exchange oscillations for short, fixed interaction time $\tau=2$~ns as a function of the detuning between the inner quantum dots $\varepsilon$, and voltage on the multielectron dot gate $V_M$. A chevron pattern, related to the exchange sweet-spot is emerging at the transition between direct and on-site regimes. Dashed lines indicate independently measured positions of the charge transitions.
	(b) Time dependence of the exchange oscillations. Oscillations have particularly high coherence at the position of the chevron where the natural dynamical decoupling occurs. Shift of the oscillations fringes up for small $\tau$ is a result of the limited bandwidth of the instrumentation.
	}
	\label{jb-mediated:fig4}
\end{figure}

\section{Extremum in the exchange interaction strength}

For the interaction time of $\tau=6$~ns we cannot resolve the pattern in on-site exchange interaction regime. For that purpose we decrease the interaction time to $\tau=2$~ns and measure a pattern of oscillations as a function of the inner quantum dots $\varepsilon$ and voltage on the gate $V_M$ (Fig.~\ref{jb-mediated:fig4}(a)). At the transition between direct and on-site exchange interaction regimes we observe a peculiar series of arcs. These indicate that, instead of increasing monotonously, exchange strength as a function of $V_M$ goes through a maximum.

In Fig.~\ref{jb-mediated:fig4}(b), we present time-dependent exchange oscillations measured in a asymmetric configuration, i.e. for $\varepsilon=8$~mV (indicated with a blue triangle in Fig.~\ref{jb-mediated:fig3}(a)). In this data we observe a chevron pattern indicating the extremum in the exchange energy~\cite{Negative-J,many_JB_charge_states} with respect to $V_M$. The presence of a maximum is a consequence of the small level spacing of the multielectron quantum dot and particular ratio of tunneling to the lowest unoccupied orbitals. The specific conditions for the maximum to occur are discussed in Ref.~\cite{many_JB_charge_states}, where we study the interaction between spin-1/2 multielectron quantum dot and a single electron. The maximum observed here has the same origin, with the only difference being that the effective spin 1/2 is formed for a brief moment by moving one of the electrons into the multielectron dot.

Moreover, we observe two characteristic positions at which the oscillations fidelity is particularly high. First, for large values of $V_M$, the exchange energy is given only by the mesoscopic parameters of the dot, which are virtually insensitive to the small disturbances. This noise insensitive region is identical to the one noted in Ref.~\cite{Dial2013} and one exploited by the hybrid spin qubit realized in the three-electron double quantum dot~\cite{Kim2014,Cao2016}. Second high-fidelity position is located at the chevron pattern. Here, the maximum in the exchange strength provides insensitivity to the detuning between the energy of the electron residing on the multielectron and one residing on the inner dot. In the presented configuration of the device the oscillation frequency at both noise-insensitive points is too high to perform rotations by small angles. However by decreasing tunnel couplings between the single electron and multielectron quantum dots the latter low-frequency point can be tuned down to a manageable frequency~\cite{Negative-J} of approximately 1~GHz.

\section{Perspective for use of the multielectron quantum dots}

The natural next step after this demonstration is to use a multielectron quantum dot of larger dimensions. 
This will enable to define multiple single-electron quantum dots around the multielectron dot, and perform coherent operation on arbitrary pair of electrons. 
Increase of the coupler size has additional advantage of reducing the on-site exchange energy which would enable performing high-fidelity, small-angle rotations. Another challenge is to implement this coupling scheme in a silicon nanostructure, to reduce decoherence effects due to interaction with nuclear spins. Third, demonstrated manipulation of the electrons involving the multielectron quantum dot provides a strong evidence that it is possible to coherently shuttle the electrons through the multielectron quantum dot. Combination of this three achievements will open a path for scaling the quantum dot based systems.

To conclude, we demonstrate that the multielectron quantum dot can serve as an exchange interaction mediator between distant electronic spins. We show that the interaction can be induced in three different ways, by keeping the electronic spins on distant quantum dots or by transferring one or both of them onto a multielectron dot. We show two regions in the gate voltage space at which the qubits susceptibility to the charge noise during the interaction is reduced. Finally, we sketch the plan for transforming this proof-of-principle experiment into a scalable system of quantum dots.

\section*{Author contributions}
S.F., G.C.G. and M.J.M. grew the heterostructure. P.D.N. fabricated the device. F.M., F.K.M., F.K. and P.D.N. prepared the experimental setup. F.K.M. and F.M. performed the experiment. T.S., S.D.B. and A.C.D. developed the theoretical model and performed the simulations\footnote{Simulations will be included in the final version of the manuscript.}. F.K.M., S.D.B., F.M., F.K., A.C.D., T.S. and C.M.M. analysed data and prepared the manuscript.

\chapterimage{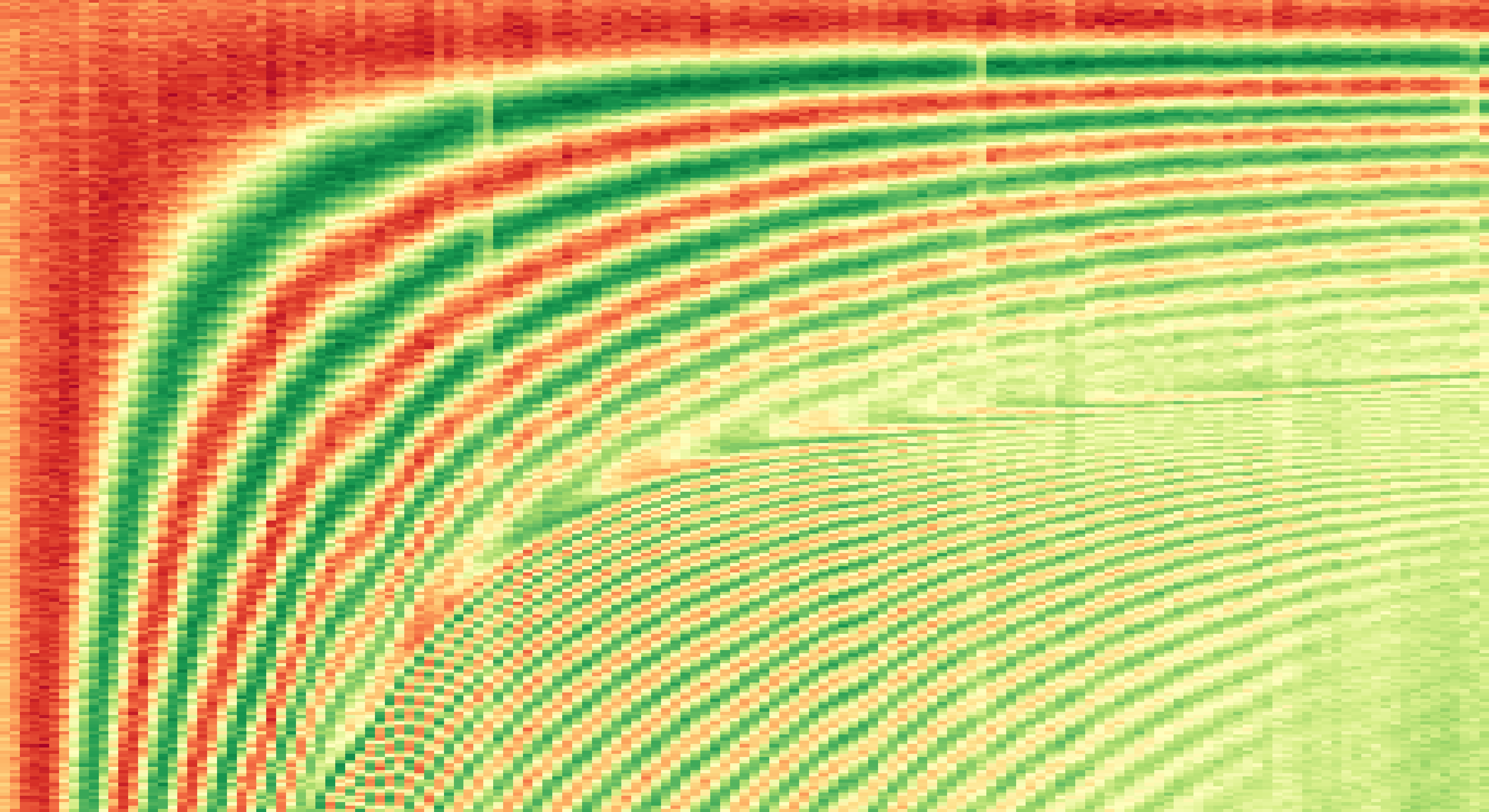}
\chapter[Supplementary Information for ``Fast coherent spin-exchange via a multielectron quantum mediator"]{\protect\parbox{0.9\textwidth}{Supplementary Information for \\ ``Fast coherent spin-exchange via \\ a multielectron quantum mediator''}}
\chaptermark{Supplementary Information for ``Fast coherent spin-exchange...''}
\label{ch:supp_jb-mediated}

\begin{center}
\begin{tcolorbox}[width=0.8\textwidth, breakable, size=minimal, colback=white]
	\small
	This supplementary information discusses the following topics:
	\begin{enumerate}
		\item[\ref{supp_jb-mediated:definitions}] Definitions of the varied  gate-voltage parameters
		\item[\ref{supp_jb-mediated:subns}] Achieving a subnanosecond temporal resolution of the exchange pulses
		\item[\ref{supp_jb-mediated:leakage}] Mapping the position of the crossings between spin states with different total spin projection $\hat{S}_z$
		\item[\ref{supp_jb-mediated:probabilities}] Calculation of the joint probabilities
		\item[\ref{supp_jb-mediated:charge}] Measurement of the charge distribution during interaction between distant electrons
	\end{enumerate}
\end{tcolorbox}
\end{center}

\section{Definitions of the varied gate-voltage parameters}
\label{supp_jb-mediated:definitions}

In the quintuple quantum dot fully-tuned up to the experiment showing exchange interaction mediated by the multielecron quantum dot it is impossible to measure the full five-dot charge diagram. This is because the co-tunneling process from the leads to the multielectron quantum dot is strongly suppressed, since the electron must co-tunnel through one of the two double quantum dots. For that reason the set up of the pulse sequences is performed in steps, starting with the choice of the readout configuration.

The DC tuning of the quintuple quantum dot corresponding to the readout position is therefore an origin of the gate-voltage space in which the pulses are performed. The rough configuration of the readout point is done based of the ``partial'' charge diagrams in which the plunger gates controlling the occupations of the two DQDs are sweeped. Fine tuning of the readout configuration is done my running one of the pulse sequences and maximization of the signal amplitude with small adjustments of the DC gate voltages. Unless stated explicitly, voltages on all gates are measured \emph{relative} to the readout configuration.

\begin{figure}[t]
	\centering
	\includegraphics[width=\textwidth]{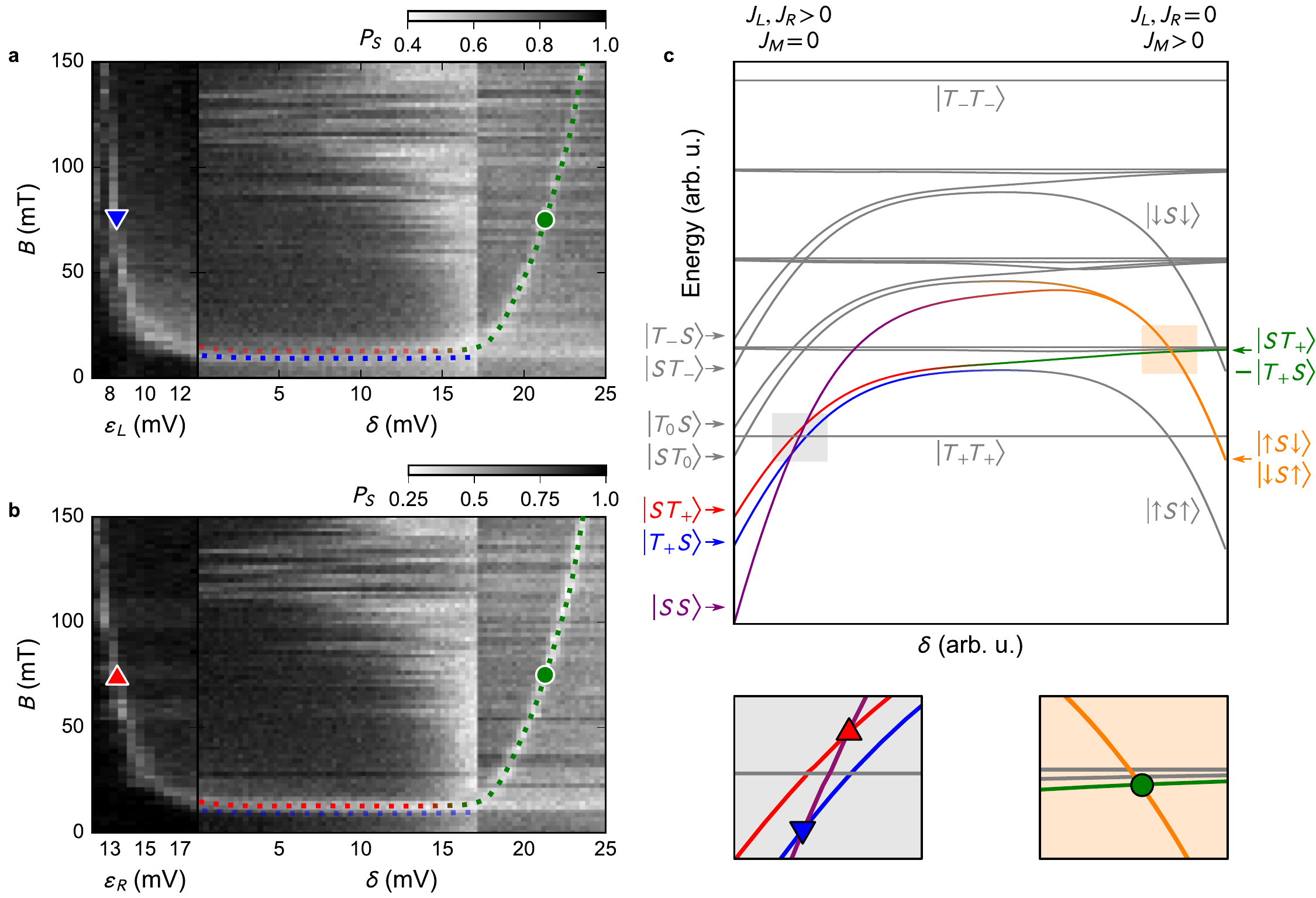}
	\caption[The leakage spectroscopy two exchange-coupled double quantum dots]{
	{\bf The leakage spectroscopy two exchange-coupled double quantum dots}
	Leakage spectroscopy measurement performed simultaneously for the left (a) and the right (b) double quantum dot.
	(c) Schematic energy diagram of the two exchange-coupled double quantum dots, for finite in-plane magnetic field. In the left only the exchange interaction within the left and right double quantum dot ($J_{L/R}$) is non-zero. In the right only exchange mediated my the multielectron quantum dot ($J_M$) is non-zero. Markers indicate the crossing that are detected in the leakage spectroscopy measurement.
	}
	\label{supp_jb-mediated:figS1}
\end{figure}

As a first step of the sequence tune-up we perform simultaneously the leakage spectroscopy~\cite{Negative-J,many_JB_charge_states} (so-called ``spin funnel'' measurement~\cite{Petta2005,Maune2012}) of both double quantum dots. In this measurement we simultaneously vary parameters $\varepsilon_L = (V_{L2}-V_{L1})/\sqrt{2}$ and $\varepsilon_R = (V_{R1}-V_{R2})/\sqrt{2}$ (the gates to which the voltages are applied are shown in Fig.~\ref{jb-mediated:fig1}(a) in the main text). The example of the obtained result is shown in the left panels of Fig.~\ref{supp_jb-mediated:figS1}(a,b). This measurement we use to set the separation point, in which both double quantum dots are in (1,1) charge configuration and the exchange interaction strength is minimized. As a separation point we choose a value of $\varepsilon_i$ for which the line features lies between $B=10$ and $20$~mT. In case of Fig.~\ref{supp_jb-mediated:figS1}(a,b) the separation point is situated at $\varepsilon_L^S=13$~mV and $\varepsilon_R^S=18$~mV.

The separation point (indicated with a superscript $S$) serves as a reference for the pulse inducing the multielectron-dot-mediated exchange interaction. From the separation point we define a new pulse direction~$\delta$
\begin{equation}
	\left(
	\begin{array}{c}
		V_{L1} \\
		V_{L2} \\
		V_{M} \\
		V_{R1} \\
		V_{R2}
	\end{array}
	\right) = \left(
	\begin{array}{c}
		-3 \\
		-2 \\
		3 \\
		-2 \\
		-3
	\end{array} 
	\right)
	\times \frac{\delta}{\sqrt{35}} +
	\left(
	\begin{array}{c}
		V_{L1}^S \\
		V_{L2}^S \\
		0 \\
		V_{R1}^S \\
		V_{R2}^S
	\end{array}
	\right),
	\label{supp_jb-mediated:parameters1}
\end{equation}
where normalization factor $\sqrt{35}$ was chosen to ensure that change of $\delta$ by one was corresponding to distance 1 in the gate voltage space with a Cartesian metric.

\begin{table}[tb]
	\caption{
	List of dataset obtained in slightly different DC configuration.}
	\centering
	\vspace{3pt}
	\label{tunings}
	\begin{tabular}{cccc}
	\hline \hline
	& \textbf{DC configuration} & \textbf{Figures} & \\ \hline
	& 1 & \ref{jb-mediated:fig2}(a), \ref{jb-mediated:fig3}, \ref{jb-mediated:fig4}(a,b), \ref{supp_jb-mediated:figS3} & \\ 
	& 2 & \ref{jb-mediated:fig2}(b,c), \ref{supp_jb-mediated:figS2} & \\ 
	& 3 & \ref{supp_jb-mediated:figS1} & \\ \hline \hline
	\end{tabular} 
\end{table}

On the $\delta$ axis we choose the final reference point (indicated with superscript $0$). With respect to that point we define a variable $\varepsilon = \left[ (V_{R1}-V_{R1}^0) - (V_{L2}-V_{L2}^0) \right]/\sqrt{2}$ which corresponds to detuning between the two inner single-electron quantum dots.

The data presented in this article was acquired in several slightly different DC gate voltage tuning (Table~\ref{tunings}), however the principle of the pulse tune-up applies to all of the datasets. In between dataset there was no significant retuning of the quintuple dot array, and therefore the tunnel couplings can be considered unchanged throughout the entire experiment, while definitions of the reference points (indicated with superscripts $S$ and $0$) as well as $\delta$ axis were changed.

\section{Achieving a subnanosecond temporal resolution of the exchange pulses}
\label{supp_jb-mediated:subns}

To achieve a subnanosecond resolution of the exchange pulse we use combined signal of the two arbitrary waveform generator channels to pulse voltage on the multielectron-dot plunger gate $V_M$. We set the two channels to output a square waveform of the same duration and amplitude, but opposite polarity, and combine them using the inverted power splitter. Rising slope of the pulses is set to the beginning of the intended exchange pulse, while the falling slope happens at the beginning of the spin initialization procedure. The duration of the subnanosecond voltage pulse is then adjusted with a skew between the two channels of the arbitrary waveform generator. Importantly, this method allows to overcome the limitations of the temporal resolution, but is constrained by the 0.8~ns pulse risetime in our setup, which leads to distortion effects in Figs.~2, 3 and 4 in the main text.

\let\mysectionmark\sectionmark
\renewcommand\sectionmark[1]{}
\section{Mapping the position of the crossings between spin states with different total spin projection $\hat{S}_z$}
\let\sectionmark\mysectionmark
\sectionmark{Mapping the position of the crossings between spin states...}
\label{supp_jb-mediated:leakage}

One of the methods for detection and quantification of the exchange interaction is the leakage spectroscopy. In the double dot case it can be used to locate the position of the crossing between the singlet $\ket{S}$ and the fully polarized triplet $\ket{T_{+/-}}$ state (the sign of the electronic g-factor defines which of the triplet states is used), and results in the characteristic funnel shape~\cite{Petta2005,Maune2012}. In more complex case of the triple quantum dot the position of the analogous crossing, which depends on the value of the external magnetic field, enables the reconstruction of the exchange profile~\cite{Gaudreau2011,many_JB_charge_states,Negative-J}. Here we employ the same technique for our case of the two double quantum dots, coupled with multielectron dot mediator.

The sequence of the applied voltage pulses is the same as in the time-resolved study of the exchange interaction mediated my the multielectron quantum dot, with the only exception being that the interaction time is set to $\tau = 150$~ns which is much longer that dephasing time due to Overhauser field and the charge noise. The left panels of Fig.~\ref{supp_jb-mediated:figS1}(a,b) present the position of the $S$-$T_+$ crossing for the two DQDs, acquired simultaneously, identical to the conventional ``spin funnel''~\cite{Petta2005,Maune2012}. The overlap of the two line-features is only apparent, as the horizontal axes correspond to different gate-voltage parameters (detunings of the left $\varepsilon_L$ and the right $\varepsilon_R$ DQD).

The right panels of Fig.~\ref{supp_jb-mediated:figS1}(a,b) present the result in the regime where the long-range exchange turns on. In this case the horizontal axis is shared. We observe that the in the central part the lines indicating the level crossing does not overlap, as underlined with blue and red dotted lines. Meanwhile, in the rightmost part, the line indicating crossing perfectly overlaps (green dotted line) and diverges towards the large field. In the intermediate region a signature of the anticrossing can be observed as well.

\begin{figure}[t]
	\centering
	\includegraphics[width=\textwidth]{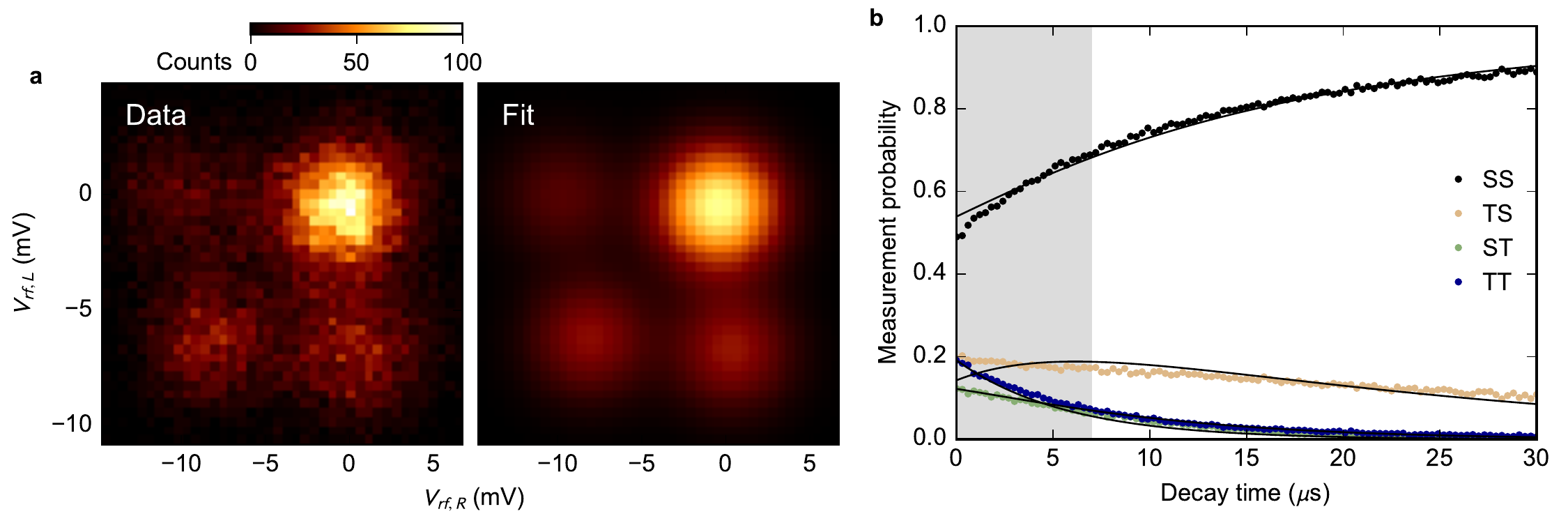}
	\caption[Joint probability estimation]{
	{\bf Joint probability estimation}
	(a) The joint histogram of the single-shot readout (left) and quadrupole Gaussian fit (right).
	(b) Decay of the triplet states in the measurement configuration. Experimentally measured decay (dots) is fitted by the minimal model of the two independent decay rates for the two double quantum dots (lines).
	}
	\label{supp_jb-mediated:figS2}
\end{figure}

This agrees with a Heisenberg model of the 4 exchange-coupled 1/2 spins arranged in a linear array (multielectron dot is neglected in the model, and only the exchange interaction it mediates is considered). Energy diagram with arbitrary dependence of the three exchange interactions reveals all of the relevant features (Fig.~\ref{supp_jb-mediated:figS1}(c)). In the left side of the diagram only exchange interaction within DQDs is nonzero. This allows us to identify the crossings detected in the leakage spectroscopy measurement. On the contrary, in the right only the exchange interaction mediated by the multielectron quantum dot is nonzero.

The lines in the left part of Fig.~\ref{supp_jb-mediated:figS1}(a,b) correspond to the $S$-$T_+$ crossing on the left and the right DQD. In the two-DQDs pictures these are $\ket{SS}$-$\ket{T_+ S}$ and $\ket{SS}$-$\ket{S T_+}$ crossings indicated with, respectively, blue and red triangle. In the middle part of the energy diagram and leakage spectroscopy data $\ket{T_+ S}$ and $\ket{S T_+}$ states start to hybridize due to the exchange mediated by the multielectron dot. As a results only one of the line features continues (as indicated with the red to green dotted line transition), while the other one ends (blue dotted line) and continues slightly shifted (green dashed line). This indicates the position at which $\ket{T_+ S}$ and $\ket{S T_+}$ are no more the eigenstates, but their superposition $(\ket{T_+ S} - \ket{S T_+})/\sqrt{2}$ is (the green line in \ref{supp_jb-mediated:figS1}(b)). At this stage also $\ket{SS}$ state is no longer an eigenstate while $\ket{\uparrow S \downarrow}$ and $\ket{\downarrow S \uparrow}$ are (the orange line).

\section{Calculation of the joint probabilities}
\label{supp_jb-mediated:probabilities}

The joint probabilities, presented in Fig.~\ref{jb-mediated:fig2}(b), are calculated based of the histograms of single shot outcomes for each pulse amplitude (presented in the Supplementary Video 1). In the first step we fit the 2-dimensional quadruple Gaussian to the histogram including multiple pulse amplitudes. From this fit we obtain the position of the four peaks (8 parameters) and their widths (2 parameters, we use different distribution widths for measurements performed with different sensors, and set them to be the same for all 4 Gaussian). The data and the fit are presented in Fig.~\ref{supp_jb-mediated:figS2}(a). Having fixed the positions and widths we fit the amplitude of the Gaussians to outcomes histograms for each voltage pulse amplitude separately. The normalized amplitudes of Gaussians yield the joint probabilities, uncorrected for the decay during the measurement.

\begin{figure}[t]
	\centering
	\includegraphics[width=0.5\textwidth]{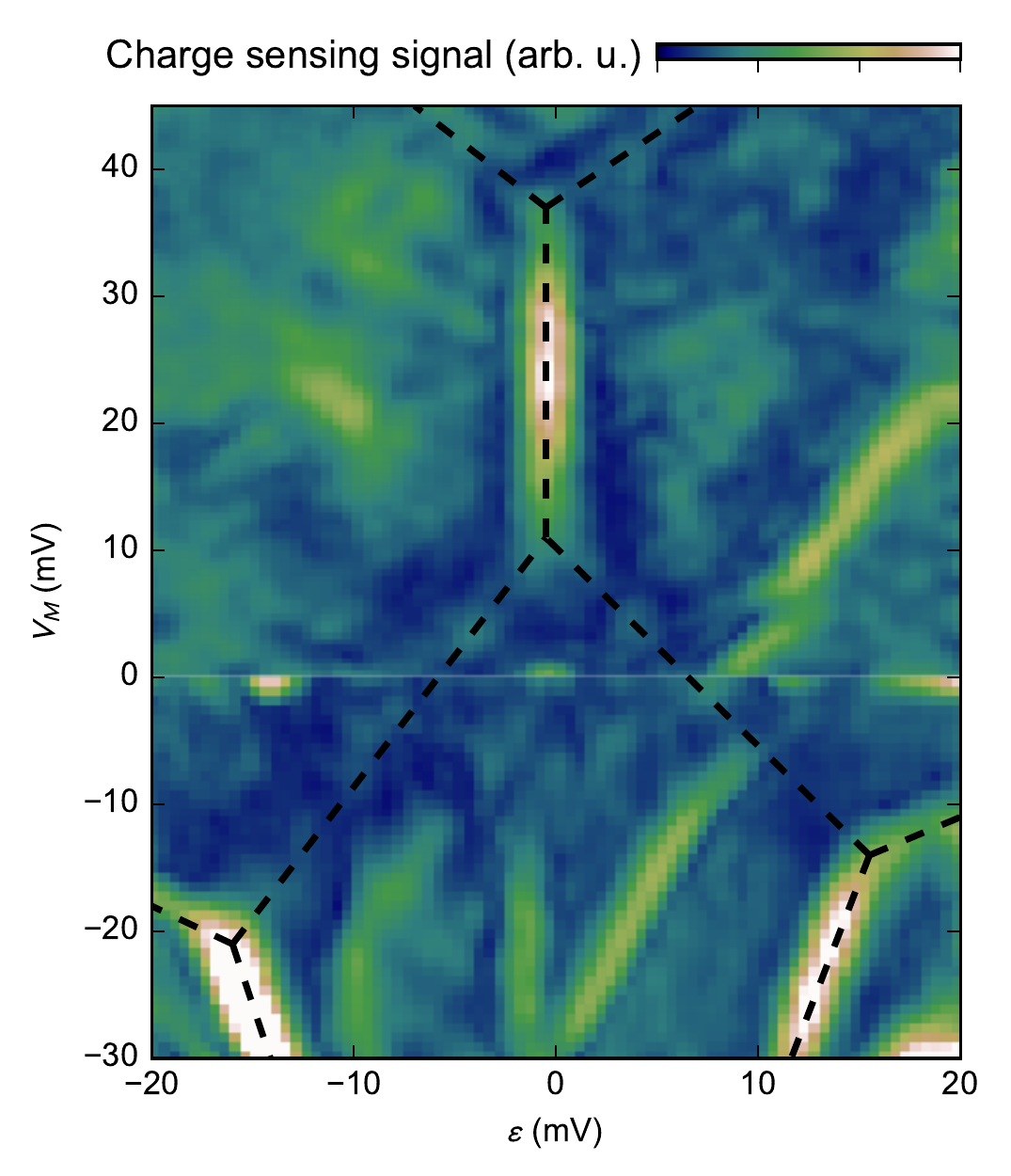}
	\caption[Processed diagram of the charge distribution during the interaction]
	{
	Processed diagram of the charge distribution during the interaction mediated by the multielectron quantum dot. Dashed lines indicate the extracted positions of the charge transitions.
	}
	\label{supp_jb-mediated:figS3}
\end{figure}

To correct for the decay of the two-electron states in the DQDs we fix the amplitude of the exchange-inducing pulse at the value that yields significant number of counts for all 4 possible outcomes and introduce the waiting time in the readout configuration, before we perform the actual measurement. Fig.~\ref{supp_jb-mediated:figS2}(b) presents the obtained decay curves. Next we introduce the minimal model of the decay, in which the triplet  states decay to singlet independently of the state of the other DQD, and with the rates different for both DQDs. This model is not physically accurate for several reasons -- the state of one DQD may affect another and decay rates are expected to differ for different triplet states. Yet in absence of the insight into full relaxation dynamics we decide to limit ourselves to the simplest possible scenario.

Having fitted the decay rates for both DQDs (see Fig.~\ref{supp_jb-mediated:figS2}(b)) we can reverse the relation between measured probabilities and the real probabilities:
\begin{equation}
	\vec{p}_\mathrm{meas} = \frac{1}{T_R} \int\limits_0^{T_R} M(t) \vec{p}_\mathrm{real} \mathrm{d}t
	\label{supp_jb-mediated:decay}
\end{equation}
where $\vec{p}_\mathrm{meas/real}$ are the vectors of possible real/measured outcome probabilities, $M(t)$ is the introduces the decay during the waiting time $t$ and $T_R$ is the total readout time of 7~$\mu$s (as indicated with the gray-haded region in Fig.~\ref{supp_jb-mediated:figS2}(b)). The integration is performed to include decay that occurs \emph{during} the readout time. Application of the numerically inversed relation~\eqref{supp_jb-mediated:decay} yields the calculated joint probability of the four states.

\let\mysectionmark\sectionmark
\renewcommand\sectionmark[1]{}
\section{Measurement of the charge distribution during interaction between distant electrons}
\let\sectionmark\mysectionmark
\sectionmark{Measurement of the charge distribution during interaction...}
\label{supp_jb-mediated:charge}

To independently confirm the position of the electrons during the interaction step we extending the interaction time to 4~$\mu$s, while maintaining the remainder of the pulse sequence unchanged. During this time we perform a measurement using both charge sensors. This is repeated for several settings of the charge sensor since the sensor is sensitive only when it is set to the slope of the sensor quantum dot Coulomb peak. We perform the numerical derivative of each data set, than apply blur by convolving with them Gaussian kernel ($\sigma=1.5$ pixel) and take the absolute value to take advantage of the sensitivity on both slopes of the Coulomb peaks. Finally we sum the obtained data sets with various weights. The processed data obtained in this way is presented in Fig.~\ref{supp_jb-mediated:figS3}. The inferred charge transitions are indicated with dashed black lines. The features indicating the electron transfer from one of the inner dots to the multielectron dot are very weak, due to large tunnel couplings, which were necessary to observe the superexchange. The two region in the bottom left and right of the Fig.~\ref{supp_jb-mediated:figS3} correspond to, respectively, (1,1,$2N \! + \! 1$,1,0) and (0,1,$2N \! + \! 1$,1,1) charge configurations of the quintuple quantum dot.

Except for the indicated charge transitions one can observed additional features which have no counterparts in the data set presenting exchange oscillations (Fig.~\ref{jb-mediated:fig3}(a)). This is the case because the exchange oscillations occur in the metastable electron configuration of the quintuple dot array. As long as the interaction time is much shorter than the relaxation rates it is irrelevant for the spin manipulations. However, as we increase the interaction time to perform the measurement of the charge distribution the relaxation occurs for the significant fraction of the pulse repetitions.


\chapterimage{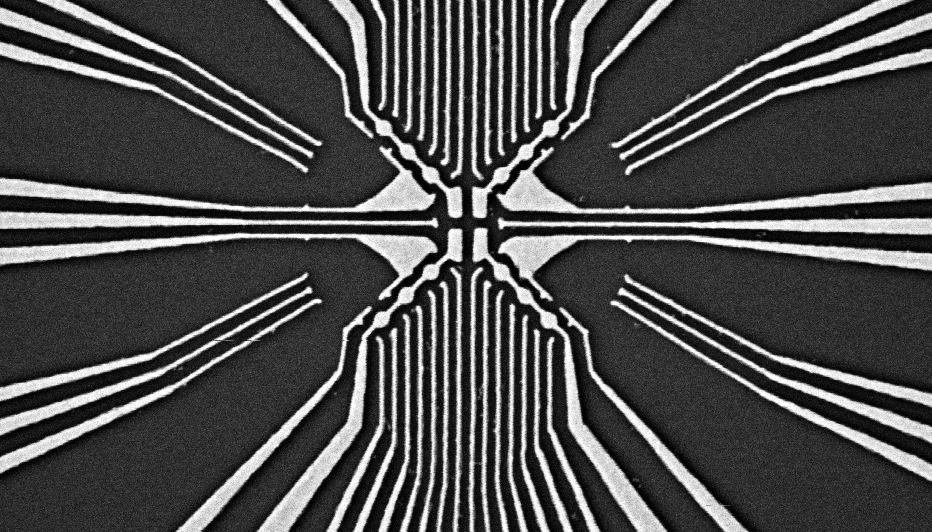}
\chapter[Outlook]{\protect\parbox{0.9\textwidth}{Outlook}}
\label{ch:outlook}

The study of the multielectron quantum dot, and  demonstration of the long range exchange interaction opens several paths for further research, both basic and towards scaling of the spin qubit systems. In this final chapter of the thesis I contrast the long range exchange coupling with other currently studied coupling mechanisms. Then I propose several follow-up experiments towards scaling of the quantum dot systems involving the multielectron-dot--mediated exchange interaction.

\section{Coupling mechanisms for spin qubits}

The exchange interaction mediated by the multielectron quantum dot, demonstrated in chapter~\ref{ch:jb-mediated}, is only one among several mechanisms proposed to realize the two qubit gates. Our approach is motivated by the original proposal of Daniel Loss and David DiVincenzo~\cite{Loss1998}, in which a direct exchange is a basis for two qubit gates. This idea provided the potential for high fidelity operations~\cite{Martins2016,Reed2016}. It is also relatively simple to realize in small quantum dot arrays but it does not take into account the technical limitations related to fabrication of large arrays of closely spaced quantum dots. Nevertheless, this approach can be used in small segments of the large quantum dot arrays, that would communicate with each other by means of one of the other mechanisms I describe below.

One possibility to provide communication between small segments of the processor is to physically shuttle the electrons while maintaining their coherence~\cite{Taylor2005a}. The demonstration of this concept was realized in arrays of up to four quantum dots~\cite{Baart2016,Fujita2017}, where an electron was moved by sequentially adjusting the gate voltages dots to transfer electrons between the neighboring dots. The obtained results provide the evidence that neither relaxation nor dephasing is enhanced during shuttling of the electrons. On the other hand, using an array of small dots to shuttle electrons does not simplify the device fabrication. Another possibility is to create a narrow, empty channel between only two metallic gates and use a surface acoustic wave to carry the electron through. Proof of principle for this approach was presented in Refs.~\cite{McNeil2011,Bertrand2016}, although high reliability of classical information transport, and preservation of coherence are yet to be established.

Meanwhile, a relatively high fidelity of the two qubit gates was demonstrated with capacitively coupled S-T$_0$ qubits~\cite{Shulman2012,Nichol2017}. This method, that can be generalized to other multidot qubits, employs the difference in the charge distribution between singlet and triplet states at the transition between (2,0) and (1,1) dot occupancies~\cite{Taylor2005a}. One can view this interaction as a small change in the detuning $\varepsilon$ of one double quantum dot, conditioned on the state of the other qubit (and vice versa)~\cite{Ward2016,VanWeperen2011}. The challenge in capacitive coupling is that it is relatively short range\footnote{Dipole-dipole interaction scales as $1/r^4$, with the distance $r$.}, especially considering the screening by the dot-defining metallic gates. This constraint can be lifted by means of the floating gates, extending between the two qubits~\cite{Trifunovic2012}. Another, more fundamental problem is that using the charge degree of freedom introduces a susceptibility to the charge noise. Therefore the reduction of the noise\footnote{As opposed to suppression by means of qubit symmetrization. Symmetrization minimizes the difference in charge distribution between the spin states but suppresses the coupling as well. It is worth noting that switching between symmetric and tilted operation of the qubits provides a mean to switch the capacitative coupling on and off.} is necessary if capacitative coupling was to be used in large quantum dot arrays.

Similar limitation constraints the usage of superconducting microwave cavities~\cite{Burkard2006,Srinivasa2016}. For this proposal the potential payoff is the increase in the coupling range up to a millimeter scale (given by the wavelength of the photon propagating in the superconducting co-planar waveguide). Up to date the first step towards cavity-mediated coupling, i.e. the strong coupling between a charge qubit and the cavity photon, was demonstrated by three groups~\cite{Mi2016,Stockklauser2017,Bruhat2017}. In all cases the key to obtain these results was the reduction of the charge noise contribution, combined with maximization of the photon electric field at the dots site. 

Finally, the exchange coupling mediated by the multielectron quantum dot can be used to perform exchange gates between relatively distant spins. The maximum size of the quantum dot that can mediate such interaction is crucial to ultimately assess whether this mechanism is suitable for scalable dot arrays. At the same time the electrons could be shuttled between distant locations through the multielectron quantum dot. The ability to perform both, the long-range coupling and the shuttling of physical qubits, would be unprecedented in the field of solid state quantum computing.

To summarize, given the current state of the art it is virtually impossible to predict which of the coupling mechanisms will be the most successful. Direct exchange interaction suffers from the gate crowding, which may be less limiting with the progress of the nanofabrication techniques. Capacitative coupling provides a mean to create dense arrays, while circumventing the extreme crowding, but requires reduction of the charge noise by several orders of magnitude to achieve the fault-tolerant fidelities. A similar restriction also applies to coupling via the superconducting resonator, which, on the other hand, can ultimately resolve the issue of the dense qubit packing. Finally, multielectron quantum dots can be employed to provide long-range communication between quantum dots, provided that reasonably large quantum dots can host a spin-0 ground state spaced from the first excited state by sufficiently large level spacing.

\let\mysectionmark\sectionmark
\renewcommand\sectionmark[1]{}
\section{Next steps towards scaling of quantum dot arrays with exchange coupling}
\let\sectionmark\mysectionmark
\sectionmark{Next steps towards scaling of quantum dot arrays...}
\label{outlook:steps}

Scaling the array of long-range exchange coupled quantum dots is a challenging task. Yet the demonstration of the multielectron-dot--mediated exchange interaction enables to set several goals that can be achieved in the nearest futures.

\begin{figure}[tb]
\begin{center}
\includegraphics[width=0.9\textwidth]{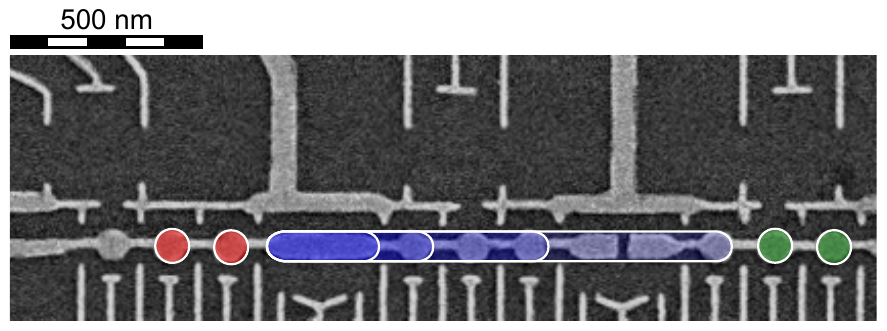}
\caption[Proposal for the study of the multielectron quantum dot with varying area in the working device]{
Proposal for the study of the multielectron quantum dot with varying area in the working device. The multielectron dot size can be easily increased by adjoining the neighboring small dots as indicated with the blue-colored ellipses. Due to the large distance between the gates defining a distant edge, and the multielectron quantum dot and the two-electron double quantum dot (red circles) a wide range of multielectron dot sizes can be explored with little effort. If the largest dot turns out to have sufficiently large level spacing the long range coupling can be readily demonstrated by defining a second double quantum dot in the position indicated by green circles.
}
\label{outlook:different_sizes}
\end{center}
\end{figure}

First, it is essential to demonstrate the long-range exchange in a device fabricated in a material with sparse spinful nuclei, such as Si/SiGe quantum wells or a MOS structure; or in a structure where the spin carriers are more weakly coupled to the nuclear spins, such as GaAs hole quantum wells. The reason is that a short inhomogeneous dephasing time for an electronic spin establishes an ultimate limit to the fidelity of the spin manipulations in GaAs. Stabilization of the Overhauser field via dynamical nuclear polarization in combination with dynamically decoupled gates will delay the moment of transition to a different system, but the amount of the overhead involved will become limiting sooner rather than later. Nevertheless, multiple nearest objectives can be realized with GaAs electron quantum dots, in parallel to progress in the reliability of fabrication of other structures.

For example it is crucial to experimentally establish the dependence of the mean level spacing with the area of the multielectron quantum dot. In particular it is important to establish the maximum dot size that reliably provides a level spacing larger than the energy corresponding to the typical electron temperature ($kT = 8.6$~$\mu$eV for $T=100$~mK, which corresponds to $f = kT/h = 2.1$~GHz). This gives a bound on the maximum distance between the exchange-coupled spins. A ballpark estimate can be obtained from the model of non-interacting electrons in a hard box yields $\langle \Delta E\rangle = \pi /m^* A$. This yields a maximum dot area of 0.36~$\mu$m$^2$ for electrons in GaAs (factor of 12 relative to size of dot studied in Ch.~\ref{ch:jb-mediated}), 0.12~$\mu$m$^2$ for electrons in Si and 0.055~$\mu$m$^2$ for heavy holes in GaAs. However the interaction effects and unknown depletion area around the metallic gates may result in significant deviations from this prediction. Moreover, the spin correlation energy can reduce the spacing between the spin-0 ground state and spin-1 excited state. The outcome may point towards using either elongated dot-buses rather than circular multielectron quantum dots.

Studying if the mean level spacing on the multielectron dot area can be performed on the device that was employed in the experiments presented in this part (Fig.~\ref{outlook:different_sizes}). The size of the multielectron quantum dot can be increased by adjoining the neighbouring regions, intended originally to host single electron dots. This proposal is very simple to realize experimentally, due to very weak capacitative coupling between the gates defining the farthest edge of the multielectron dot and the double quantum dot which can serves for spin initialization and readout. The device shown in Fig.~\ref{outlook:different_sizes} enables to also demonstrate the coupling via an elongated multielectron quantum dot, provided the first experiment yields a promising results.

\begin{figure}[tb]
\begin{center}
\includegraphics[width=0.55\textwidth]{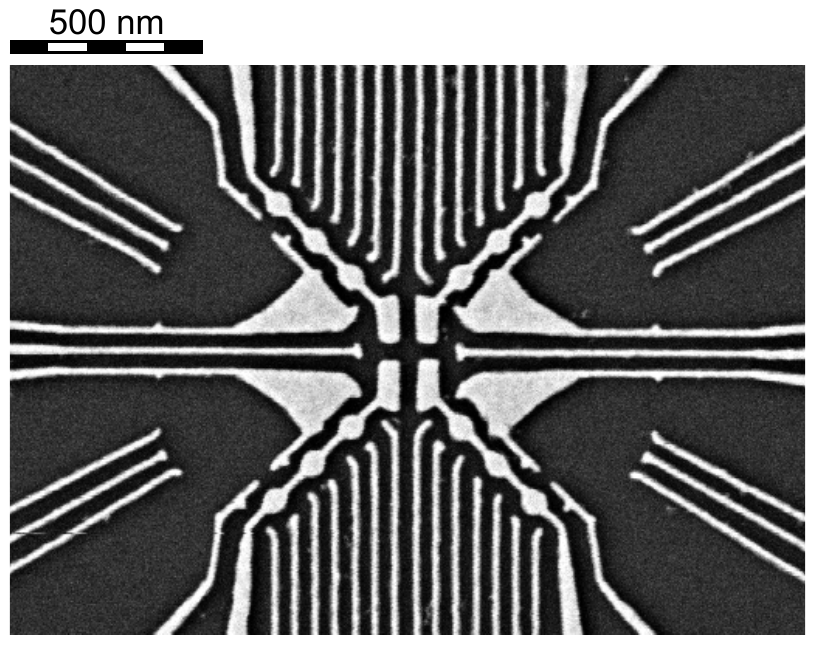}
\caption[Scanning electron micrograph of the test exposure of the 4-qubit device]{
Scanning electron micrograph of the test exposure of the 4-qubit device. The four single, double or triple quantum dots can be directly tunnel-coupled to the multielectron quantum dot with an area of approximately 0.08~$\mu$m$^2$. Each of the qubits is equipped with the individual sensor dot enabling simultaneous, independent spin readouts.
}
\label{outlook:malina}
\end{center}
\end{figure}

Knowing the maximum dimensions of the coupling dot, one can attempt to couple multiple qubits via the same mediator. This would increase the connectivity within a quantum dot array, potentially reducing number of operations required to realize a quantum algorithms. An example of device geometry suitable for such demonstration is presented in Fig.~\ref{outlook:malina}. This design combines the dimensions of both single-electron and sensor dots that proved to be easily tunable, with the multielectron quantum dot that has an area sufficiently small to provide level spacing much larger than $kT$.

The conditions required for the long distance exchange coupling of spins are simultaneously the conditions for the coherent transport of spin through the multielectron quantum dot. As demonstrated in Sec.~\ref{many_charge_states:0} a single \emph{excess} electron preserves coherence while residing on the multielectron quantum dot, enabling the measurement of the Overhauser field gradient. There is no obstacle to transfer it afterwards to a different quantum dot. Such experiment can be conducted in both device geometries presented in Figs.~\ref{outlook:different_sizes} and \ref{outlook:malina}. In particular, the electron transfer between red and green-colored dots in Fig.~\ref{outlook:different_sizes} can be realized either through one, extremely elongated dot, or through two multielectron dots of half the size.

These numerous possibilities, most of which can be demonstrated in available, working and tested devices provide a compelling argument for using the multielectron quantum dots. In particular the possibility to demonstrate the long range coupling and long range coherent transfer of physical qubits via the same mediator is unique to this proposal.

\part{Appendix}
\label{part:experiment}
\appendix

\chapterimage{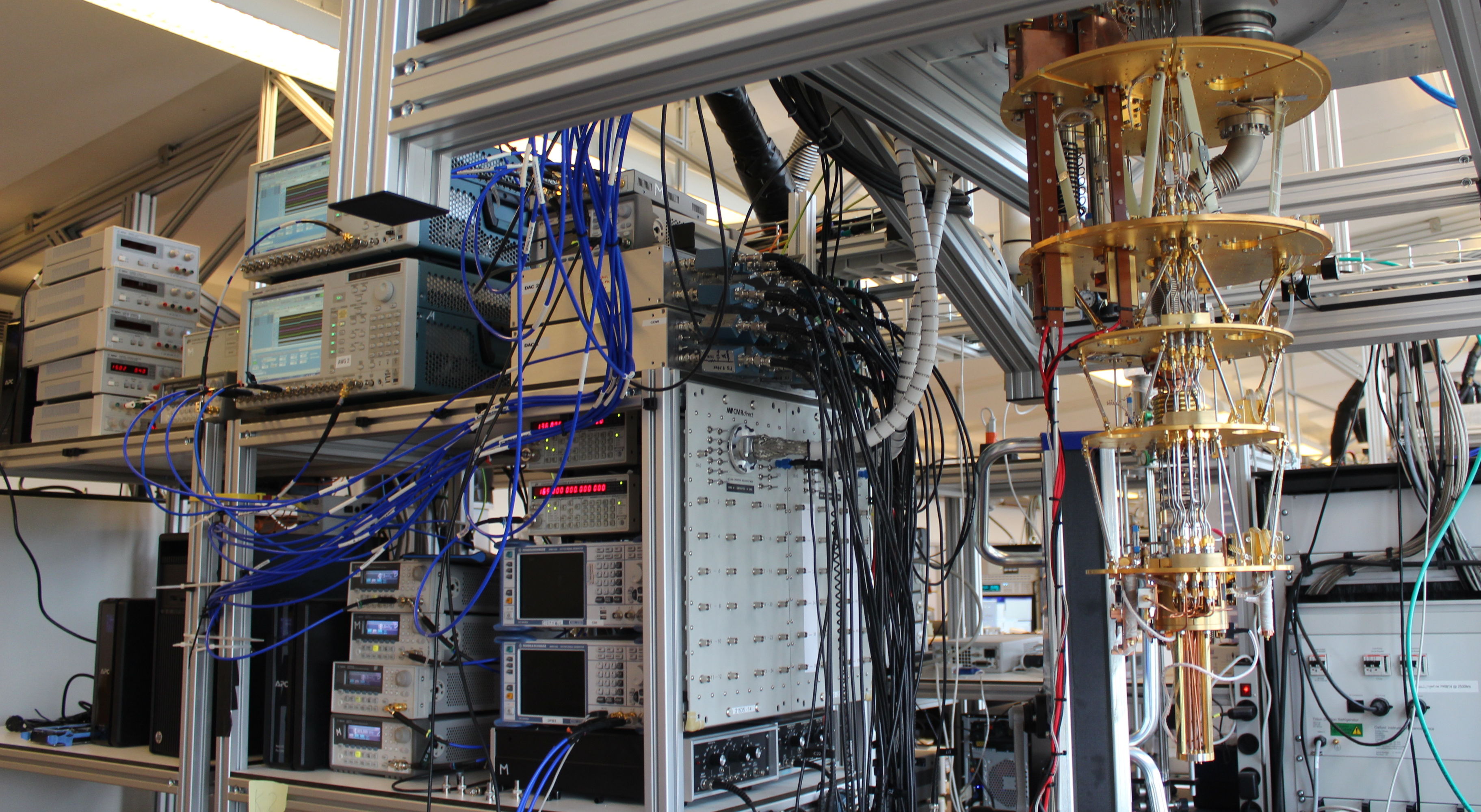}
\chapter[Setup]{\protect\parbox{0.9\textwidth}{Setup}}
\label{ch:setup}

In this chapter you'll find a description of all the electronics in the experimental:
\begin{itemize}
	\item the DC part which allows to shape the quantum dot potential and perform basic transport measurements (\ref{setup:DC});
	\item the HF part that is responsible for applying nano- to milisecond voltage pulses (\ref{setup:HF,RF});
	\item the reflectometry setup and demodulation circuit (\ref{setup:HF,RF}).
\end{itemize}
In further sections I will focus on certain extensions of the setup, that are used to:
\begin{itemize}
	\item combine HF pulses and RF bursts (\ref{setup:RF});
	\item synchronise multiple AWGs (\ref{setup:sync});
	\item readout two sensors simultaneously (\ref{setup:miltiplex}).
\end{itemize}

\section{DC/LF circuit}
\label{setup:DC}

\begin{figure}[bt]
	\centering
	\includegraphics[width=0.7\textwidth]{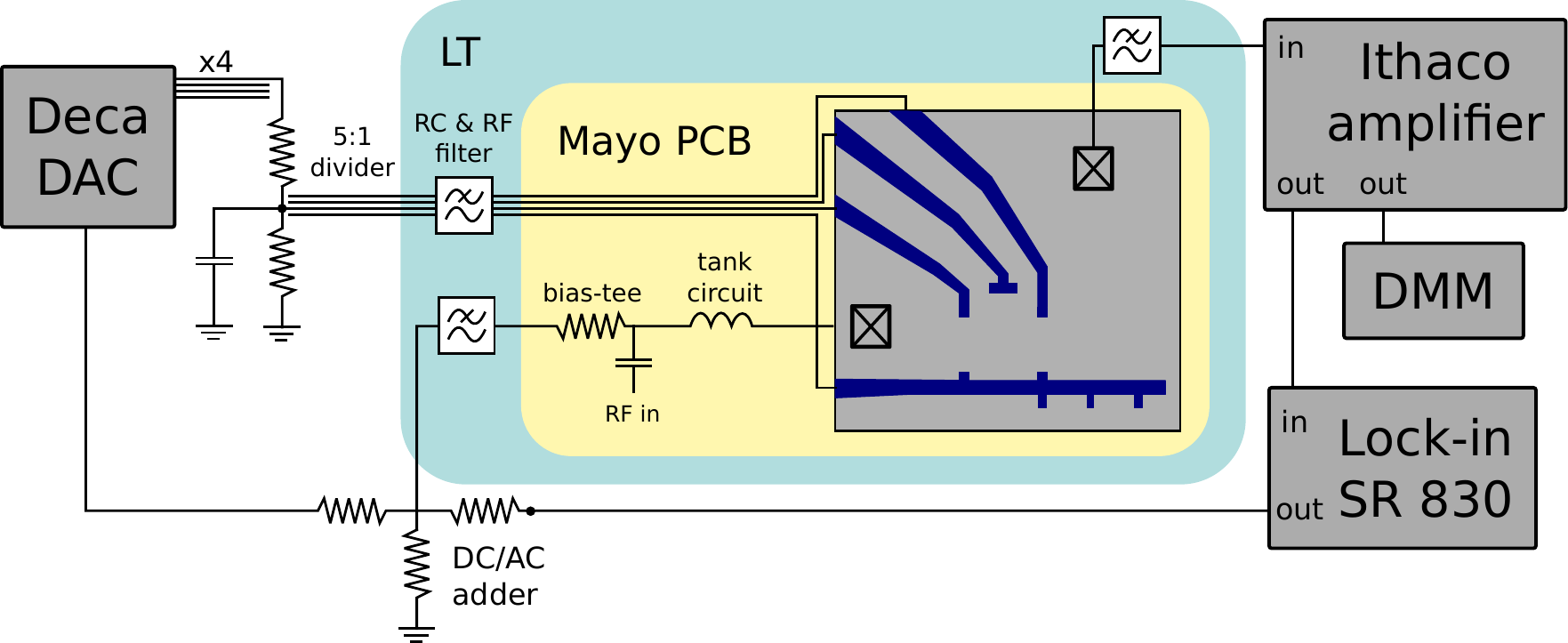}
	\caption[Scheme of DC/LF circuit]{Scheme of DC/LF circuit. Parts of the circuit drawn in a pale blue region are located at mixing chamber plate or on coldfinger, while yellow box shows elements mounted on the Mayo PCB.}
	\label{setup:DC_scheme}
\end{figure}

A low frequency circuit consists of two parts -- for gate control and transport measurements (Fig.~\ref{setup:DC_scheme}). The gate voltages are controlled with homemade DACs (called DecaDACs) to provide stable voltage on gates forming quantum dots. DecaDACs were built by Jim MacArthur from Harvard electronics shop. Each of these has 20 16-bit channels that can operate in ranges 0-10, -10-10 and -10-0 V. To increase resolution of 30 $\mu$V voltage on each channel was divided (1:5). The lines are brought down to the mixing chamber using 24 line constantan looms which are thermalized at every stage of the fridge. At mixing chamber the lines are RC and RF filtered to minimize instrumentation and Johnson's noise. RC filter consists of 80 MHz low-pass filter (Mini-Circuits LFCN-80), two 2 k$\Omega$ resistors and two 2.7 nF capacitors. Cut-off frequency of this filter is approximately 30 kHz. The RF filter consists of three components: 80 MHz low pass filter (Mini-Circuits LFCN-80), 1.45 GHz low pass filter (Mini-Circuits LFCN-1450) and 5 GHz low pass filter (Mini-Circuits LFCN-5000). DC lines that are combined with HF lines (not shown on a scheme) go through additional bias tee, which I will describe in section~\ref{setup:HF,RF}.

Part of the circuit used for transport measurements consists of DecaDAC (a source of DC bias) and SR830 lock (an AC source). Their signals is combined and divided in the adder (with ratios 1:$\sim$400 for DC and 1:$\sim$18000 for AC). Than line goes down the fridge through a set of filters (identical to ones on gate-controlling lines) and is connected to the source ohmic of the sample. Drain ohmic is connected to a low noise Ithaco 1211 current preamplifier. Its two outputs are connected to Agilent 34401A DMM and lock-in SR830 to measure, respectively, the current and conductance. Some ohmic lines are combined on the sample board with reflectometry lines via bias-tees and go through tank circuits. I will describe these parts of the circuit later~(Sec.~\ref{setup:HF,RF}).

\section{HF and reflectometry circuit}
\label{setup:HF,RF}

The primary purpose of the high-frequency (HF) circuit is to apply pulses to the gates with nanosecond temporal resolution and to perform measurements on a microsecond timescale. Additionally, we use them to apply a sawtooth wave of 2 kHz frequency. By synchronising measurements with period of the sawtooth we can speed up measurements of charge diagrams. I will describe this technique in chapter \ref{ch:RT} and now focus only on the experimental setup.

\begin{figure}[bt]
	\centering
	\includegraphics[width=1\textwidth]{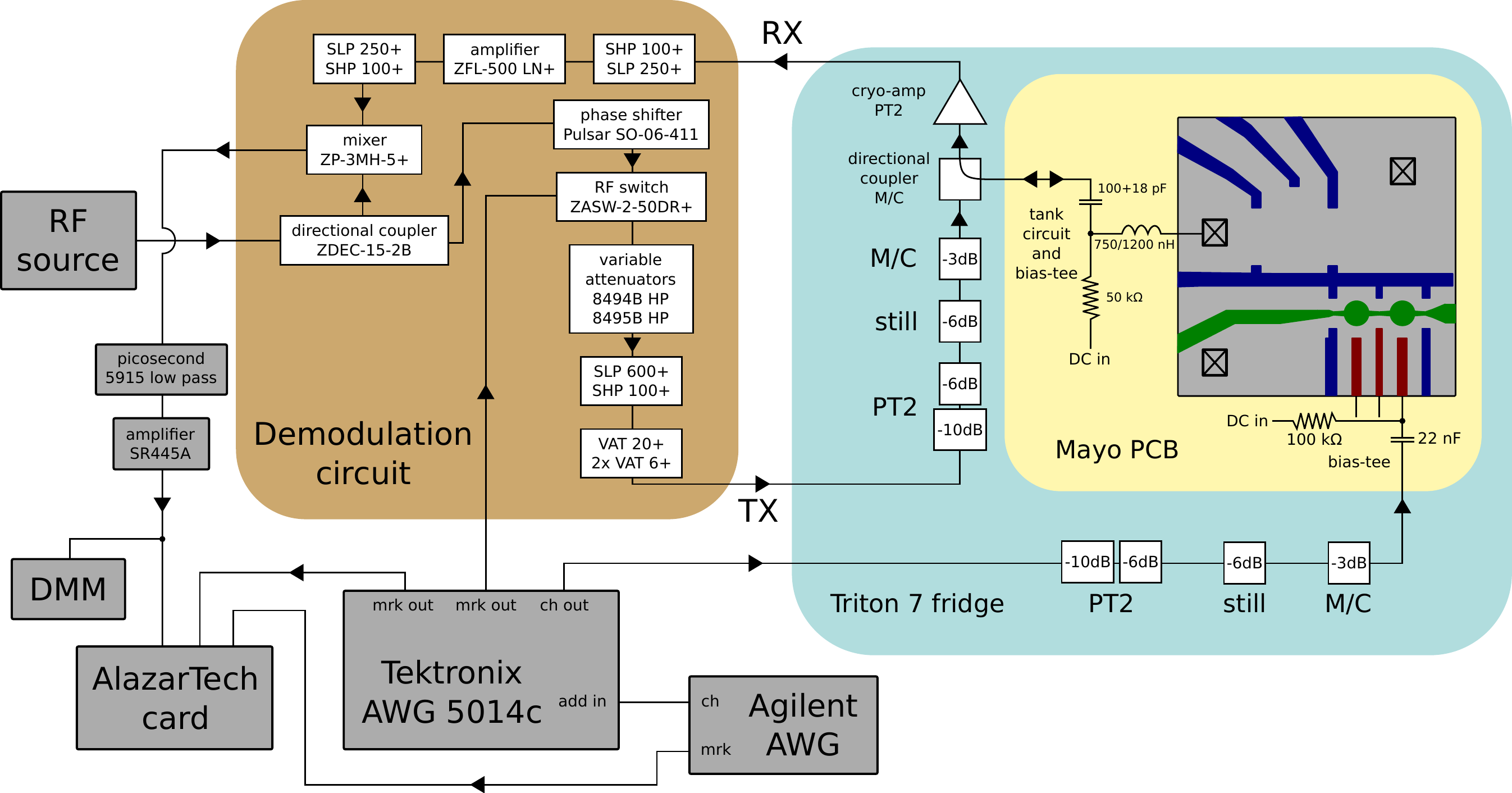}
	\caption[Scheme of HF/RF circuit]{Scheme of HF/RF circuit. Parts of the circuit drawn in a pale blue region are located inside the fridge, while yellow box shows elements mounted on the Mayo PCB. Brown box symbolizes a demodulation circuit and other elements located on the fridge. Symbols M/C, still and PT2 indicate at which level of the fridge relevant component is located.}
	\label{setup:HF_scheme}
\end{figure}

The heart of the circuit is an arbitrary waveform generator (AWG) Tektronix 5014c. As shown in figure~\ref{setup:HF_scheme}, it sends the pulses, via attenuators, down the fridge. At the Mayo PCB high frequency pulses are combined with DC signal using bias-tees ($R=100$ k$\Omega$, $C=22$ nF). Combined signal arrives to the sample. To synchronise pulses with the measurement we connect one of the marker channels to the AlazarTech PCI digitalizer (ATS 9440), which is used for reflectometry measurements.

To perform reflectometry measurements we need to send the RF carrier signal down the fridge, collect reflected signal, amplify, demodulate and deliver to the Alazar card. The RF source is Stanford Research SG384 signal generator, operating at resonant frequency of the tank circuit (see below), typically between 100 and 250 MHz. RF signal is divided in directional coupler. Most of the signal (optimally 13 dBm) goes directly to the mixer and will serve as a reference. Smaller part of the signal goes through the phase shifter (needed to adjust phase of reflected signal relative to the reference signal) and RF switch (which we use to turn the RF carrier on only during the measurements). Than the signal goes through a set of filters and variable attenuators, which allows to control RF power at the sample. At the bottom of the fridge RF carrier arrives to the next directional coupler. Small part of the signal goes through, is combined with DC signal and hits the resonant circuit.

The resonant circuit consists of the inductor (typically 700 to 1200 nH), parasitic capacitance to ground and resistance ($\sim$1~pF), and resistance of the sensor quantum dot ($1/R = \sigma \sim$0.2 $e^2/h$). When properly tuned the resistance of the quantum dot changes depending on the charge state of the qubit. In turn, this resistance modifies the impedance of the resonant circuit and therefore the amplitude of the reflected signal. For detailed description of principles of readout using reflectometry I recommend Christian Barthel's thesis~\cite{Barthel2010b}.

The reflected signal goes back to the directional coupler, now to the low-loss input. At the pulse tube plate 2 the signal is amplified with the CITLF1 cryo-amplifier (co-called Weinreb amplifier). Outside of the fridge the signal goes through a set of filters, additional amplifier and arrives at the mixer. The demodulated signal is once more filtered and amplified and sent to the DMM (for monitoring purposes) and to the Alazar PCI digitalizer.

The setup has one more element: AWG Agilent 33250A that is used to generate sawtooth wave. This sawtooth wave can be overlaid with the fast pulses using ``add input'' option of the Tektronix AWG. Synchronization of the sawtooth with the Alazar card allows to use for the acquisition of the charge diagrams~(chapter~\ref{ch:RT}).

\section{Adding RF to HF pulses}
\label{setup:RF}

\begin{figure}[bt]
	\centering
	\includegraphics[width=0.8\textwidth]{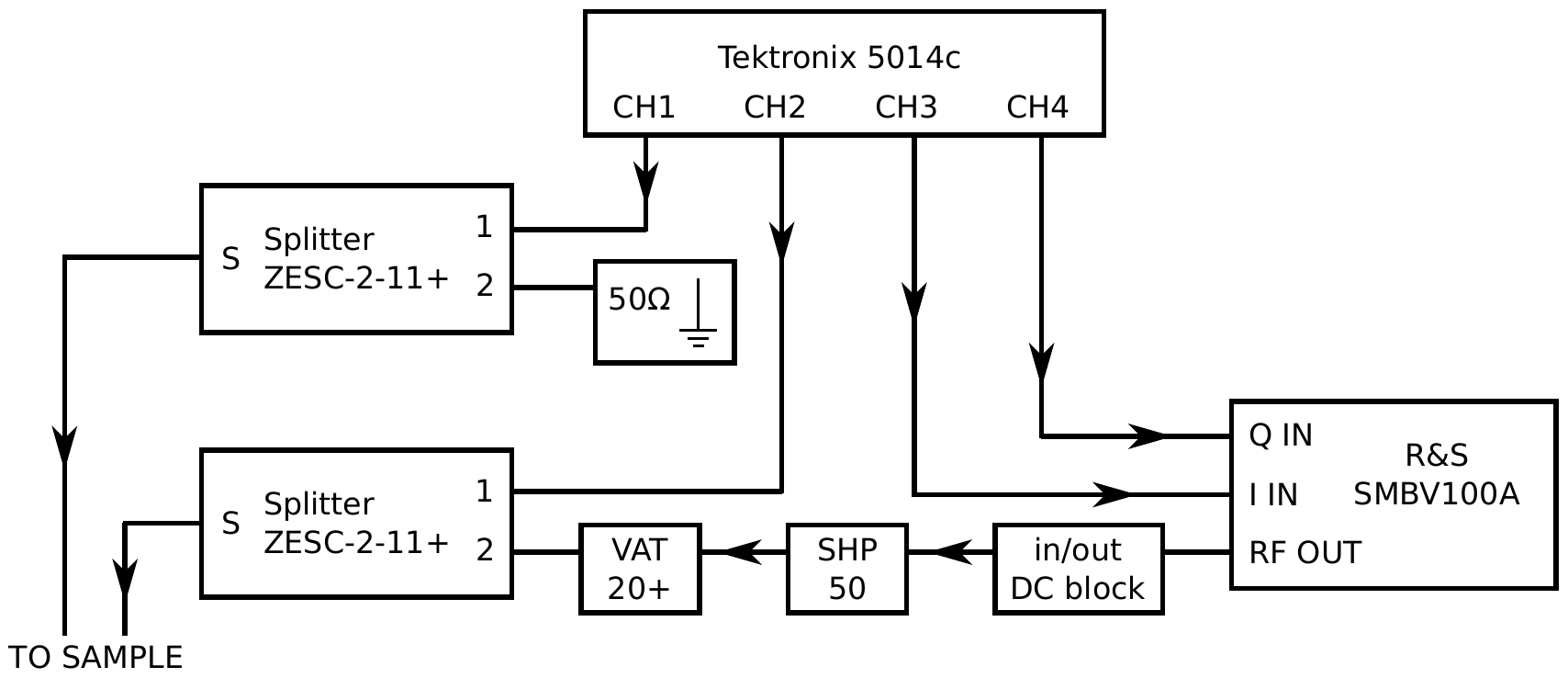}
	\caption[Setup for applying modulated RF pulses]{Setup for applying modulated RF pulses.}
	\label{setup:RFscheme}
\end{figure}

Operation of a spin qubits such as resonant-exchange or Loss-DiVincenzo requires application of RF voltage pulses as well as pulses of arbitrary shape on the same gate. For that purpose we built the following extension of the setup (Fig.~\ref{setup:RFscheme}).

Two of the channels of Tektronix 5014c AWG were used to apply pulses on gates. They both are connected to one of the outputs of ZESC-2-11+ power splitter. In case of the channel 1 the other output of the splitter is 50 $\Omega$ terminated.

Another two channels of the AWG are used to modulate in-phase and quadrature of the RF signal generated by a vector source R\&S SMBV100A. The output of the RF goes through a in/out DC block which breaks a ground loop, a high-pass filter and a 20 dB attenuator. In the end RF is combined with the signal from 2nd AWG channel on with the power splitter.

Observant reader would notice that in this setup the RF signal will necessarily be delayed relative to the signal from the AWG. This delay is of the order of tens of nanoseconds and needs to be calibrated out. Correction in done software by delaying pulses on channels 1 and 2 relative to the channels 3 and 4.

\begin{figure}[t]
	\centering
	\includegraphics[width=0.65\textwidth]{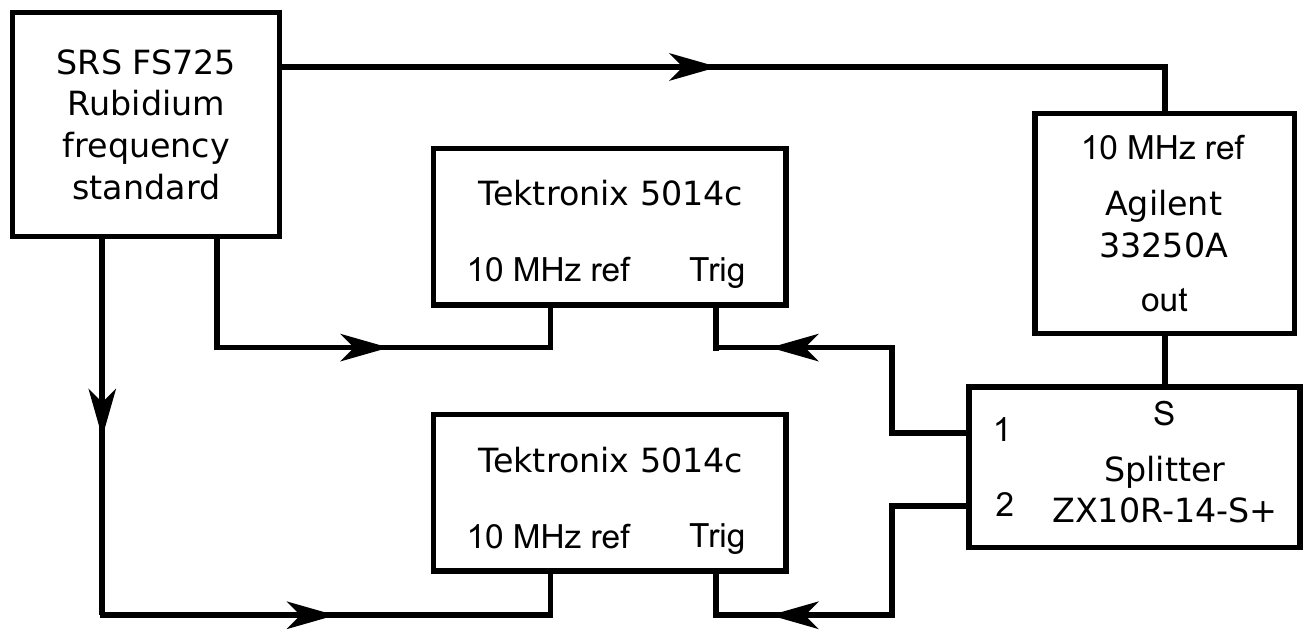}
	\caption[Setup for synchronization of two AWGs]{Setup for synchronization of two AWGs}
	\label{setup:syncsetup}
\end{figure}

\section{Synchronisation of multiple AWGs}
\label{setup:sync}

Most of the multi-qubit experiments require more than 4 fast signals, and therefore synchronisation of multiple AWGs. For synchronisation one needs to provide AWGs with a common 10 MHz reference and a common external trigger. In our experiment we used SRS FS725 Rubidium frequency standard to generate 10 MHz reference and Agilent 33250A AWG to generate a square wave that triggered two Tektronix 5014c AWGs.

However, if you do just that, you will find out that part of the time the AWGs are one clock cycle off. To eliminate that you need to use the cables of the same length for triggering. Moreover if AWGs are supposed to output the sequence, it's good to run the trigger continuously so it would trigger both AWGs at the beginning of \emph{each} pulse, not only at the beginning of each sequence. Importantly, the period of the triggering square wave needs to be an integer multiple of AWG clock period. This means also that it also has to use the same 10 MHz reference. Otherwise AWGs might be one clock cycle off most of the time.

\section{Simultaneous readout of two sensors}
\label{setup:miltiplex}

\begin{figure}[bt]
	\centering
	\includegraphics[width=0.9\textwidth]{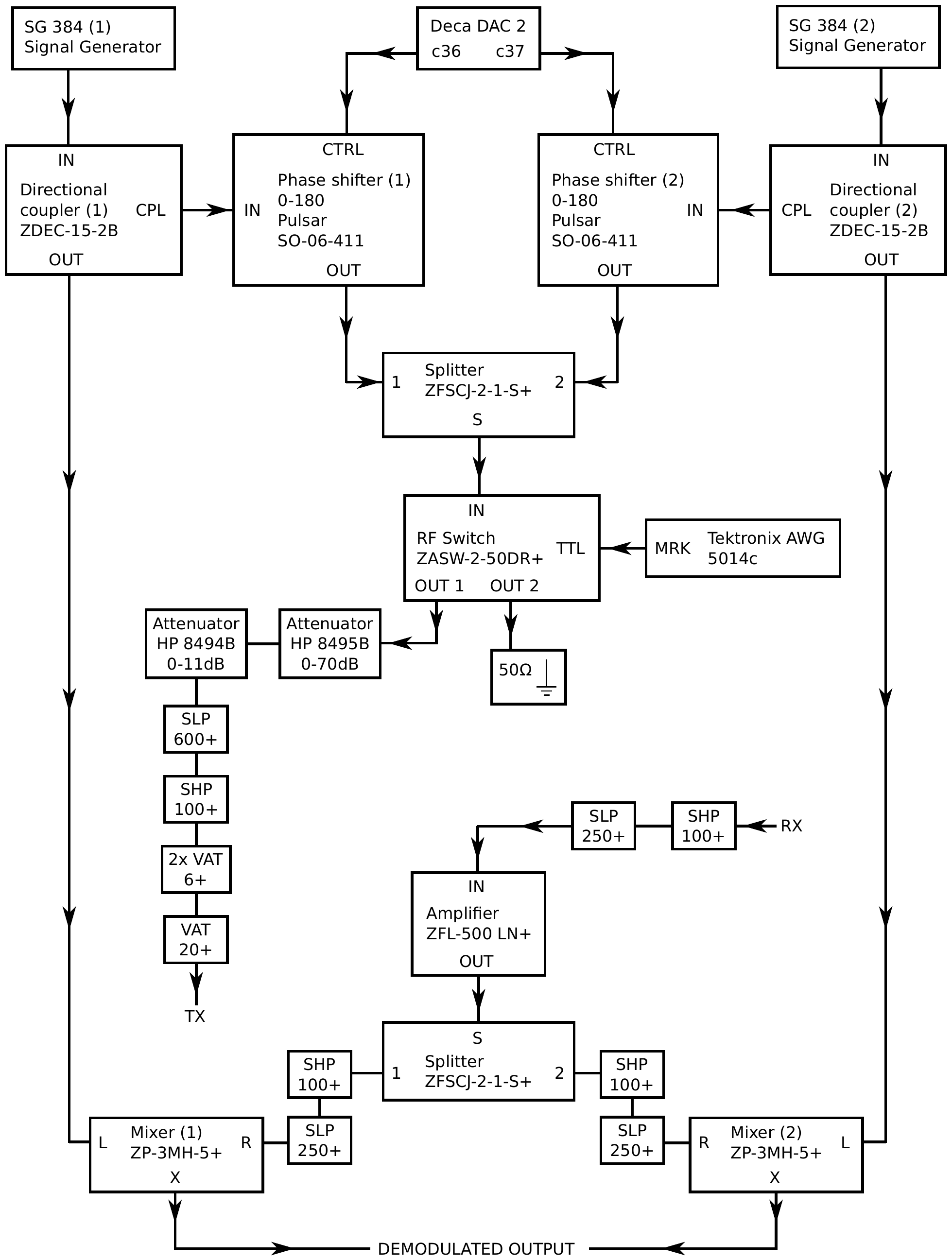}
	\caption[Circuit for reflectometry multiplexing]{Circuit for reflectometry multiplexing. The left and the middle part, as well as the middle and the right part form a demodulation circuit identical to on in Fig.~\ref{setup:HF_scheme}. The signal is combined end splitted with a pair of power splitters.}
	\label{setup:multischeme}
\end{figure}

The final improvement to the setup was an extension of the demodulation circuit (Fig.~\ref{setup:multischeme}) to enable simultaneous measurement of two sensor dots embedded in resonant circuit of distinct resonant frequencies. To large extent this circuit is a doubled circuit from Fig.~\ref{setup:HF_scheme}.

Two signal generators produce two RF tones. Signal from each of them goes through a directional coupler to the mixer. The other outputs of the directional couplers are connected through the phase shifters to outputs of the power splitter where the two frequencies are combined. The combined RF carrier goes through the RF switch, common for both frequencies, and a set of filters and attenuators to the fridge.

The reflected signal is splitted at the output of the fridge and demodulated with two reference signals in two separate mixers. In this way we get two DC outputs that can be connected to DMMs or Alazar card.

\chapterimage{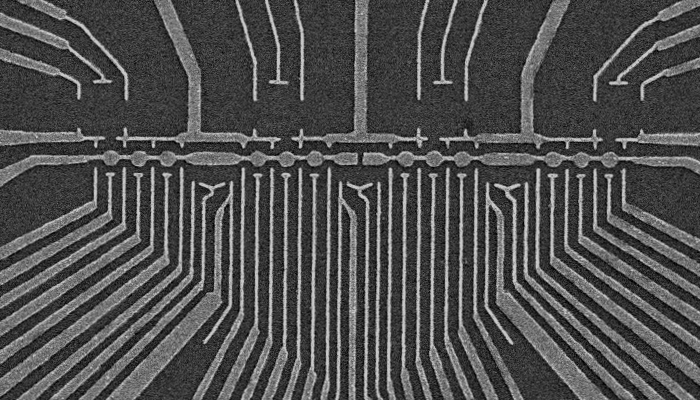}
\chapter[Fabrication recipe]{\protect\parbox{0.9\textwidth}{Fabrication recipe}}
\label{ch:samples}

%
%
%

\begin{multicols}{2}

\subsection*{Mesa}
\begin{itemize}
	\item 3 solvent clean, N$_2$ dry
	\item pre-bake 5 min at 185$^\circ$C
	\item cool down 30 s on glass slide
	\item spin AR 300-80 ($\sim$15 nm)
	\begin{itemize}
		\item 10 s at 500rpm, 500rpm/s
		\item 60 s at 4000rpm, 4000rpm/s
	\end{itemize}
	\item Bake 1 min at 185$^\circ$C
	\item spin EL9 ($\sim$315 nm)
	\begin{itemize}
		\item 10 s at 500 rpm, 500 rpm/s
		\item 60 s at 4000 rpm, 4000 rpm/s
	\end{itemize}
	\item bake 3 min at 185$^\circ$C
	\item expose in Elionix:
	\begin{itemize}
		\item electron energy 100 keV
		\item write field 600 $\mu$m
		\item 20.000 dots
		\item pitch 3
		\item beam current 40-60 nA
		\item aperture 250 $\mu$m
		\item Beamer base dose: 295 $\mu$C/cm$^2$ (0.4 $\mu$s/dot for 60 nA)
	\end{itemize}
	\item develop the resist
	\begin{itemize}
		\item MIBK:IPA 1:3, 90 s
		\item rinse in IPA 15 s
	\end{itemize}
	\item O$_2$ plasma ash for 60-120s ($\sim$12 nm/min)
	\item prepare 1:8:240 H$_2$SO$_4$:H$_2$O$_2$:H$_2$O
	\item measure etch rate on a dumy GaAs chip
	\begin{itemize}	
		\item etch a dummy chip for 60 s
		\item wash in mQ H$_2$O for 20 s
		\item sonicate in Acetone for 5 min and in IPA for 2 min,
		\item N$_2$ dry
		\item measure mesa height
		\item estimate etch rate ($\sim 2$ nm/s)
	\end{itemize}
	\item etch the chip
	\begin{itemize}	
		\item etch $\sim$ 20 nm below 2DEG
		\item wash in mQ H$_2$O for 20 s
		\item sonicate in Acetone for 5 min and in IPA for 2 min,
		\item N$_2$ dry
		\item measure mesa height
		\item estimate etch rate
	\end{itemize}
	\item warm the chip in IPA/acetone for 1-2 hours
	\item O$_2$ plasma ash
\end{itemize}

\subsection*{Ohmics}
\begin{itemize}
	\item 3 solvent clean, N$_2$ dry
	\item bake 5 min 185$^\circ$C
	\item cool down 30s on glass slide
	\item spin EL9 ($\sim$320 nm)
	\begin{itemize}
		\item 10 s at 500 rpm, 500 rpm/s
		\item 60 s at 4000 rpm, 4000 rpm/s
	\end{itemize}
	\item bake 3 min at 185$^\circ$C,
	\item cool down for 60s
	\item spin 4\% PMMA ($\sim$200 nm)
	\begin{itemize}
		\item 10 s 500 rpm, 500 rpm/s
		\item 60 s at 4000 rpm, 4000 rpm/s
	\end{itemize}
	\item bake 3 min at 185$^\circ$C
	\item expose in Elionix:
	\begin{itemize}
		\item electron energy 100 keV
		\item  write field 600 $\mu$m
		\item 20.000 dots
		\item aperture 250 $\mu$m
		\item beam current 40-60 nA
		\item dose 700 $\mu$C/cm$^2$
	\end{itemize}
	\item develop in MIBK:IPA 1:3 60s
	\item rinse in IPA 10 s
	\item O$_2$ plasma ash for 120 s ($\sim$12 nm/min)
	\item load the sample into AJA
	\item ash with Ar plasma for 120 s
	\item deposit (for approx. 60 nm depth):
	\begin{itemize}
		\item 43 nm Ge
		\item 30 nm Pt
		\item 87 nm Au
	\end{itemize}
	\item lift off in 85$^\circ$C NMP for 2+ hours
	\item anneal in rapid thermal annealer for 2 min at 425$^\circ$C
\end{itemize}

\subsection*{Alignment marks}
\begin{itemize}
	\item 3-solvent clean, N$_2$ dry
	\item O$_2$ plasma ash for 30 s
	\item pre-bake 4 min at 185$^\circ$C
	\item cool down 30 s
	\item spin 4\% PMMA
	\begin{itemize}
		\item 10 s at 500 rpm, 500 rpm/s
		\item 60 s at 4000 rpm, 4000 rpm/s
	\end{itemize}
	\item bake 3 min at 185$^\circ$C
	\item expose in Elionix
	\begin{itemize}
		\item electron energy 100 keV
		\item dose 800-900 $\mu$C/cm$^2$
		\item write field 150 $\mu$m
		\item 60.000 dots
		\item beam current 1 nA
		\item aperture 60 $\mu$m
	\end{itemize}
	\item develop in MIBK:IPA for 60 s
	\item rinse in IPA for 15 sec, N$_2$ dry
	\item O$_2$ plasma ash 60 s ($\sim$12 nm/min)
	\item load the sample into AJA
	\item deposit:
	\begin{itemize}
		\item 10 nm Ti
		\item 60 nm Au
	\end{itemize}
	\item lift off in hot acetone/NMP for 2+ hours
\end{itemize}

\subsection*{HfO$_2$}
\begin{itemize}
	\item 3 solvent clean, no sonication, N$_2$ dry
	\item pre-bake 5 min at 185$^\circ$C, cool down 30 s
	\item spin 4\% PMMA
	\begin{itemize}
		\item 10 s at 500 rpm, 500 rpm/s
		\item 60 s at 4000 rpm, 4000 rpm/s
	\end{itemize} 
	\item bake 5 min at 185$^\circ$Ce
	\item expose in Elionix
	\begin{itemize}
		\item electron energy 100 keV
		\item dose 700 $\mu$C/cm$^2$
		\item write field 600 $\mu$m
		\item 20.000 dots
		\item aperture 250 $\mu$m
		\item beam current 40 nA
	\end{itemize}
	\item develop in MIBK:IPA 1:3 for 60 s
	\item rinse in IPA for 15 s
	\item O$_2$ plasma ash for 60 s
	\item deposit 10 nm HfO$_2$
	\begin{itemize}
		\item chamber heater 130$^\circ$C
		\item wall heater 130$^\circ$C
		\item N$_2$ flow 20 SCCM
		\item pulse HfO$_2$ 0.3 sec
		\item purge 90 sec
		\item pulse H$_2$O 0.03 sec
		\item cycles 80
		\item expected thickness $\sim$10 nm
	\end{itemize}
	\item scratch edge to expose resist
	\item lift off in hot NMP for 5 min with $5\times 2$ sec sonication bursts
	\item put in 80$^\circ$C NMP for 2+ hours
	\item $5\times 2$ sec sonication
	\item wet observe and repeat NMP and sonication if necessary
\end{itemize}

\subsection*{Fine gates}
\begin{itemize}
	\item 3 solvent clean, N$_2$ dry
	\item pre-bake 5 min at 185$^\circ$C, cool down 30 s
	\item spin 2\% PMMA
	\begin{itemize}
		\item 10 s at 500 rpm, 500 rpm/s
		\item 60 s at 4000 rpm, 4000 rpm/s
	\end{itemize} 
	\item bake 15 min at 185$^\circ$C
	\begin{itemize}
		\item electron energy 100 keV
		\item base dose 1240 $\mu$C/cm$^2$
		\item PEC done with beamer
		\item write field 150 $\mu$m
		\item 60.000 dots
		\item beam current BC 300 pA
		\item aperture 40 $\mu$m
	\end{itemize}
	\item develop in IPA:H$_2$O 7:3 for 2.5 min at -5$^\circ$C
	\item deposit:
	\begin{itemize}
		\item 10 nm Ti
		\item 60 nm Au
	\end{itemize}
	\item lift off in hot acetone/NMP for 2+ hours
	\item wet observe and repeat NMP and soft sonication if necessary
\end{itemize}

\subsection*{Outer connectors}
\begin{itemize}
	\item 3 solvent clean, N$_2$ dry
	\item pre-bake 5 min at 185$^\circ$C, cool down 30 s
	\item spin 9\% Co+polymer
	\begin{itemize}
		\item 10 s at 500 rpm, 500 rpm/s
		\item 60 s at 4000 rpm, 4000 rpm/s
	\end{itemize}
	\item bake 3 min at 185$^\circ$C
	\item cool down for 30 s
	\item spin 4\% CSAR
	\begin{itemize}
		\item 10 s at 500 rpm, 500 rpm/s
		\item 60 s at 4000 rpm, 4000 rpm/s
	\end{itemize}
	\item bake 3 min at 185$^\circ$C
	\item expose in Elionix
	\begin{itemize}
		\item electron energy 100 keV
		\item write field 600 $\mu$m
		\item 20.000 dots
		\item pitch 3
		\item beam current BC 40-60 nA
		\item aperture 250 $\mu$m
		\item CON-file prepared w Beamer \\
			Dwell times are set for each element assuming 45-46 nA and given base dose. Adjust for different beam current by applying dose factor	
		\item Beamer base dose 190 $\mu$C/cm$^2$ (0.34 $\mu$s/dot for 45-46 nA)
	\end{itemize}
	\item develop, 45 sec AR-600-546
	\item 5 sec O-xylene
	\item 10 sec IPA rinse
	\item N$_2$ dry
	\item O$_2$ plasma ash for 60 sec ($\sim$12 nm/min)
	\item load the sample into AJA
	\item deposit (total of ($1.2\times$ mesa height)
	\begin{itemize}
		\item 10 nm Ti
		\item 110 nm Au
	\end{itemize}
	\item lift off in 55$^\circ$C acetone for 2+ hours
\end{itemize}

\end{multicols}

\chapterimage{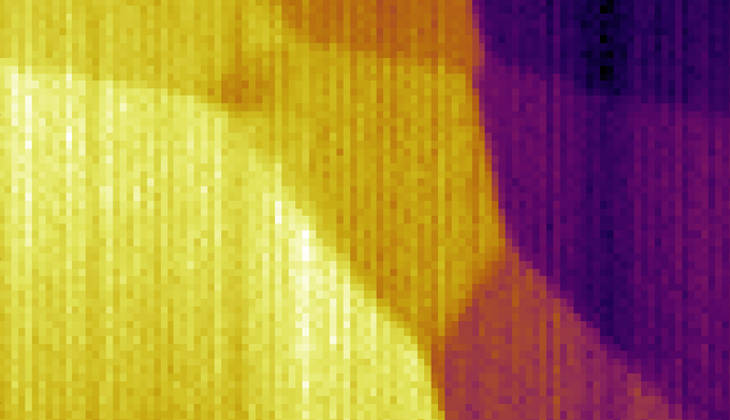}
\chapter[Real-time measurements of charge diagrams]{\protect\parbox{0.9\textwidth}{Real-time measurements\\ of charge diagrams}}
\label{ch:RT}

During the measurements of the PDN 7c sample we realized that reflectometry measurements are sufficiently sensitive, to acquire charge diagrams of reasonable quality with averaging time of 50 $\mu$s per pixel (0.5 s for 100$\times$100 image). On the other we knew that measuring such a charge diagram took us typically 20 s. The difference between these times is is a consequence of sending multiple commands to the DecaDAC to step DC voltage on the gates. In the following I will describe how we used DAC sweeps and sawtooth waves synchronised with the Alazar card to take advantage of high sensitivity, and realized real-time measurements of charge diagrams with up to 5 fps.

The technique was to the large extend based on legacy code developed by Jim Medford. It was implemented in IgorPro, in the experiment file which kept evolving since times of Alex Johnson, Jason Petta and first coherent operations on single electrons~\cite{Petta2005,Johnson2005a}.

While developing the code we learned that similar technique was developed in Jason Petta group~\cite{Stehlik2015}. Their readout technique was based on measurement of transmission of the cavity coupled to nanowire double quantum dot. Usage of higher carrier frequencies ($\sim$8 GHz) allowed to use Josephson parametric amplifier, which boosted signal to noise ratio and let them cut down measurement time of a single charge diagram to 20 ms. The noteworthy difference between the two techniques is that cavity-based readout relies on the tunneling rates being comparable or larger that the RF excitation frequency, which limits the sensitivity to more ``pinched-off'' charge transitions. In case of the SET-based charge sensing this constraint weaker, and is given by the sweeping rate of the gate voltages ($\sim$5~$\mu$s per pixel in our case, corresponding to $\sim$200~kHz tunneling rate).

\begin{figure}
	\centering
	\includegraphics[width=0.6\textwidth]{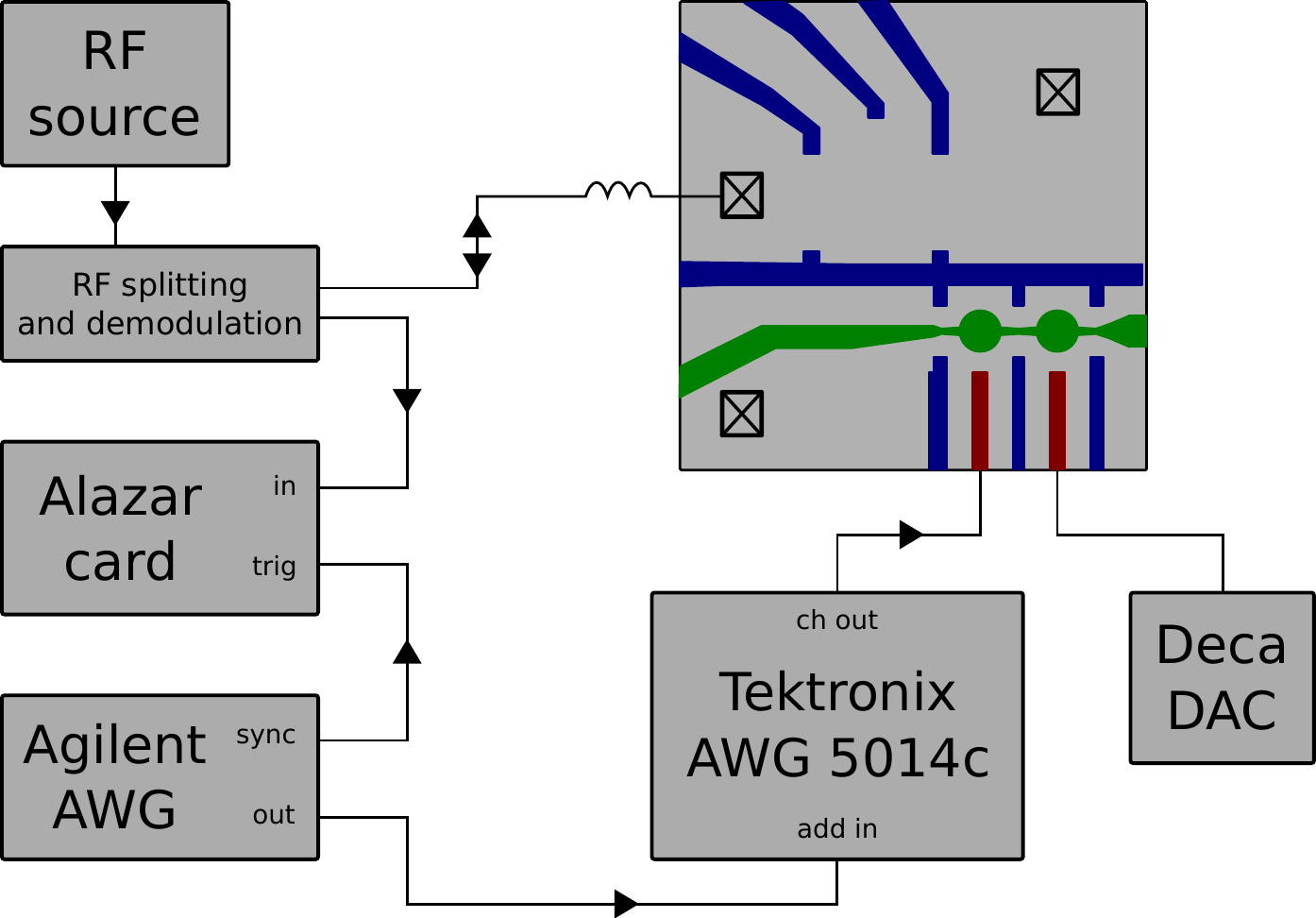}
	\caption[Scheme of a setup for fast acquisition of charge diagrams]{Scheme of a setup for fast acquisition of charge diagrams. All elements that are not necessary for the explanation of the principle of operation are omitted.}
	\label{RT:setup_scheme}
\end{figure}

\section{The principle of operation}
\label{RT:principle}

To perform real-time measurements of charge diagrams you need two voltage sources capable of sweeping voltage or applying a sawtooth wave. We used DecaDAC to sweep voltage on one of the gates with variable speed and Agilent 33250A AWG to apply fixed-frequency sawtooth wave on another (Fig.~\ref{RT:setup_scheme}). The signal from Agilent AWG was transmitted through Tektronix AWG which can add a high frequency waveform to the sawtooth.

\begin{figure}
	\centering
	\includegraphics[width=\textwidth]{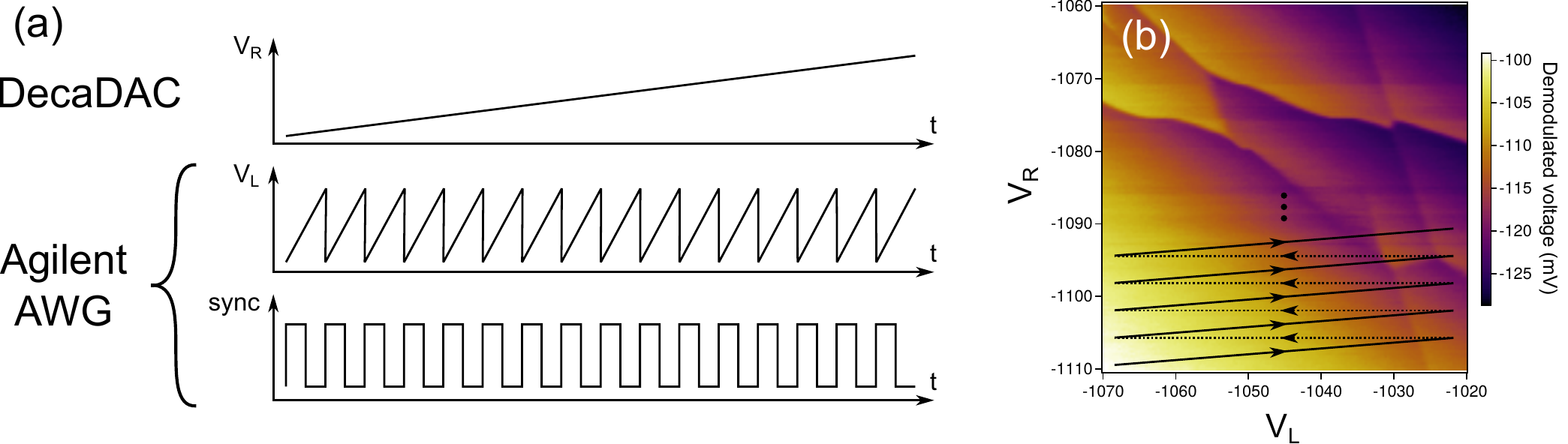}
	\caption[Sweep and sawtooth wave applied for real-time charge diagram measurements]{(a) Sweep and sawtooth wave applied with DAC and Agilent AWG for real-time charge diagram measurements. Sync of the Agilent AWG is used for triggering the Alazar card. (b) Schematic path followed by gate voltages. Solid lines indicate sweep, while dotted lines indicate voltage jump.}
	\label{RT:waveforms}
\end{figure}

As show in Fig.~\ref{RT:waveforms}(a) DecaDAC sweeps voltage only once while acquiring a single charge diagram, while Agilent AWG is continuously applying 1903 Hz sawtooth wave. (Because of timescales separation these devices don't need to be synchronised.) This results in a zig-zag path in gate voltage space that covers the entire mapped area (Fig.~\ref{RT:waveforms}(b))

To acquire the charge diagram Alazar card is triggered at the beginning of each sawtooth sweep and collects data for\footnote{Exact frequeny is irrelevant. It needs to be significantly higher than the cutoff frequency of the bias-tees wich add high-frequency and DC voltage. On the other hand it should be as low as possible to minimize artifacts which appear when the sawtooth frequency is comparable to the tunneling rates.} 1/(1903 Hz) $\approx$ 0.5 ms. Sample speed throughout each of the traces is set to 100 MS/s, which corresponds to $\sim$50000 points per sawtooth period. Subsequent points are then averaged together to obtain specified horizontal resolution. Few of such traces are then averaged together, typically 5-20 in the live-view mode and 50-100 for high quality diagrams. The averaged trace corresponds to a single horizontal line in the final charge diagram. This procedure is repeated to obtain demanded vertical resolution. Total acquisition time of $50\times50$ charge diagram with 10 traces averaged together is therefore {1/1903~Hz~$\times$~10~$\times$~50~=~0.26~s}. The precalculated total acquisition time is used to calculate the DecaDAC voltahe sweeping rate.

\section{Code}
\label{RT:code}

{\em If you are not using the legacy code you can safely skip this section.}

Here I will describe \verb'chargeDiagramTraceAlazar()' function which is a workhorse of real-time charge diagram acquisition. The code is written in IgorPro. Knowledge of Igor will be helpful, but hopefully not necessary to understand the code. I will clarify however, that mysterious ``waves'' are simply arrays of floating point numbers.

For completeness -- the code described here does not adjust frequency and amplitude of the sawtooth wave, although this feature is implemented in GUI.

\vspace{3pt}\hrule\vspace{5pt}

Function \verb'chargeDiagramTraceAlazar' takes the following arguments:
\begin{itemize}
	\item \verb'waveNom' -- string that will be a base name of waves storing the output;
	\item \verb'DACchan' -- integer specifying swept DecaDAC channel;
	\item \verb'from/to' -- starting/final voltage for DAC sweep;
	\item \verb'points_slow' -- resolution of the charge diagram along axis swept by DAC;
	\item \verb'avg_slow' -- number of sawtooth ramps averaged together;
	\item \verb'points_fast' -- resolution of the charge diagram along axis swept by Agilent AWG;
	\item \verb'channels' -- string of letters specifying read Alazar channels, separated by semicolons;
	\item \verb'noDisp' -- variable specifying whether to open new window displaying charge diagram,
\end{itemize}
and does not return anything. But it does create (or overwrite) a wave(s) \verb'waveNom#', where \verb'#' is a letter indicating the channel of the Alazar card.
\begin{verbatim}
function chargeDiagramTraceAlazar(waveNom,DACchan,from,to,points_slow,
                               avg_slow,points_fast,channels[,noDisp])
    string waveNom, channels
    variable DACchan, from, to, points_slow, avg_slow, points_fast,
                                                          noDisp, copy
\end{verbatim}

\vspace{3pt}\hrule\vspace{5pt}

By default don't pop up a new window.
\begin{verbatim}
    if(paramisdefault(noDisp)||noDisp == 1)
        noDisp = 1
    else   
        noDisp = 0
    endif
\end{verbatim}

\vspace{3pt}\hrule\vspace{5pt}

Read the global variable which tells whether we're in single shot readout mode. If so, reconfigure Alazar card settings. Most importantly \verb'cmode()' function chooses the Alazar channel that is used for triggering the measurement and sets sampling rate.
\begin{verbatim}
    NVAR AlazarFast = root:AlazarFast
    if(AlazarFast != 0)
        cmode()
    endif
\end{verbatim}

\vspace{3pt}\hrule\vspace{5pt}

Load information about sampling rate of the Alazar and frequency of the sawtooth from global variables. We used sampling rate of 100 MSa/s (maximum rate of ATS9440 card is 125 MSa/s) and sawtooth frequency of 1903 Hz.
\begin{verbatim}
    NVAR alazarClock = root:alazarClock                       // 1.0e8
    NVAR rampFreq = root:rampFreq                             // 1903
\end{verbatim}

\vspace{3pt}\hrule\vspace{5pt}

Calculate how many samples can be taken during a single sweep of a sawtooth wave. 5 $\mu$s is subtracted from the sweep time to provide a waiting time between end of one measurement and beginning of another. Then the number of samples is rounded down to the nearest multiple of 16, which is required by the Alazar card. For different models of Alazar cards number of samples might need to be set to multiple of different power of 2.
\begin{verbatim}
    variable pointsPerRamp = alazarClock/rampFreq-5e-6*alazarClock // 52048.6
    variable mLength = (floor(pointsPerRamp/16))*16                // 52048
\end{verbatim}

\vspace{3pt}\hrule\vspace{5pt}

Calculate sweeping rate for DecaDAC (in mV/s) according to the sweep range, averaging and sawtooth wave frequency
\begin{verbatim}
    variable speed = abs(from-to)/points_slow/avg_slow*rampFreq
\end{verbatim}

\vspace{3pt}\hrule\vspace{5pt}

Explanation of the next step requires basic understanding of operation of XOPs that control the Alazar card. They arrange and manipulate the date in form of 3-dimensional array. The data acquired within  window that follows each trigger is arranged along the first dimension (points). Along the second dimension (records) I will arrange data from \verb'avg_slow' number of subsequent sweeps of a sawtooth wave. These sweeps will be averaged together to form a single line on 2D charge diagram. Along the third dimension (buffers) I will arrange traces that will make subsequent lines on the charge diagram.
\begin{verbatim}
    variable buffers = points_slow
    variable records = avg_slow
\end{verbatim}

\vspace{3pt}\hrule\vspace{5pt}

Set values of variables controlling which of the Alazar channels to measure
\begin{verbatim}
    variable a,b,c,d
    if(FindListItem("A",channels)>=0)
        a = 1
    endif
    if(FindListItem("B",channels)>=0)
        b = 1
    endif
    if(FindListItem("C",channels)>=0)
        c = 1
    endif
    if(FindListItem("D",channels)>=0)
        d = 1
    endif
\end{verbatim}

\vspace{3pt}\hrule\vspace{5pt}

Set a DAC channel to the initial value and wait a moment to make sure that the command was applied. Than start sweeping the DAC channel. Factor of 1.02 and additional waiting time were chosen experimentally to calibrate out software delays and minimize distortions. Notice that except for this calibration there is no synchronisation between DAC and Agilent AWG.
\begin{verbatim}
    setval("c"+num2str(DACchan),from)
    wait(0.03)
    rampDAC(DACchan,to,speed*1.02,read=0)
    wait(90*0.0005)
\end{verbatim}

\vspace{3pt}\hrule\vspace{5pt}

After the beginning of the DAC sweep run the external operating procedure which acquires data with the Alazar card. \verb'baseName="test"' specifies base wave name with the data that will be created in the memory. \verb'id={1,1}' specifies identifier of the Alazar card. \verb'cd={a,b,c,d}' specifies measured Alazar channels. \verb'sample={mLength,records,buffers}' defines dimensionality of the array of data points described above. \verb'pointsAvg=mLength/points_fast' tells how many of the subsequent samples should be averaged together. This will specify resolution of the charge diagram. \verb'dim=2' specifies that data array should be averaged along 2nd dimension.
\begin{verbatim}
    ATSreadWaveNPT_AVG/q baseName="test",id={1,1}, cd={a,b,c,d},
                               sample={mLength,records,buffers},
                               pointsAvg=mLength/points_fast, dim=2
\end{verbatim}

\vspace{3pt}\hrule\vspace{5pt}

Copy the data from a \verb'test' wave to the wave specified by user. Than convert volts to milivolts. Factor of \verb'buffers' is introduced to correct for the bug in XOP.
\begin{verbatim}
    variable i,j
    for(i=1;i<=4;i+=1)
        if((i==1&&a==0)||(i==2&&b==0)||(i==3&&c==0)||(i==4&&d==0))
            continue
        endif
        string chan = num2char(i+64)
        duplicate /o $("test_avg"+chan),$(waveNom+chan)
        wave outWave =$(waveNom+chan)
        outWave *=1000*buffers
    endfor
\end{verbatim}

\vspace{3pt}\hrule\vspace{5pt}

Create a new window in which charge diagram will be displayed (if user set \verb'nodisp != 1').
\begin{verbatim}
    if(noDisp != 1)
        j=1
        for(i=1;i<=4;i+=1)
            chan = num2char(i+64)
            if(FindListItem(chan,channels)>=0)
                showwaves(waveNom+chan)
                positionwindow(j)
                j += 1
            endif
        endfor
    endif
end
\end{verbatim}

End of function.

\vspace{3pt}\hrule\vspace{5pt}

The rest of convenient features is hidden behind the GUI, but the heart of the code is as simple as this.

\section{GUI and additional features}

\begin{figure}[t]
\begin{center}
\includegraphics[width=\textwidth]{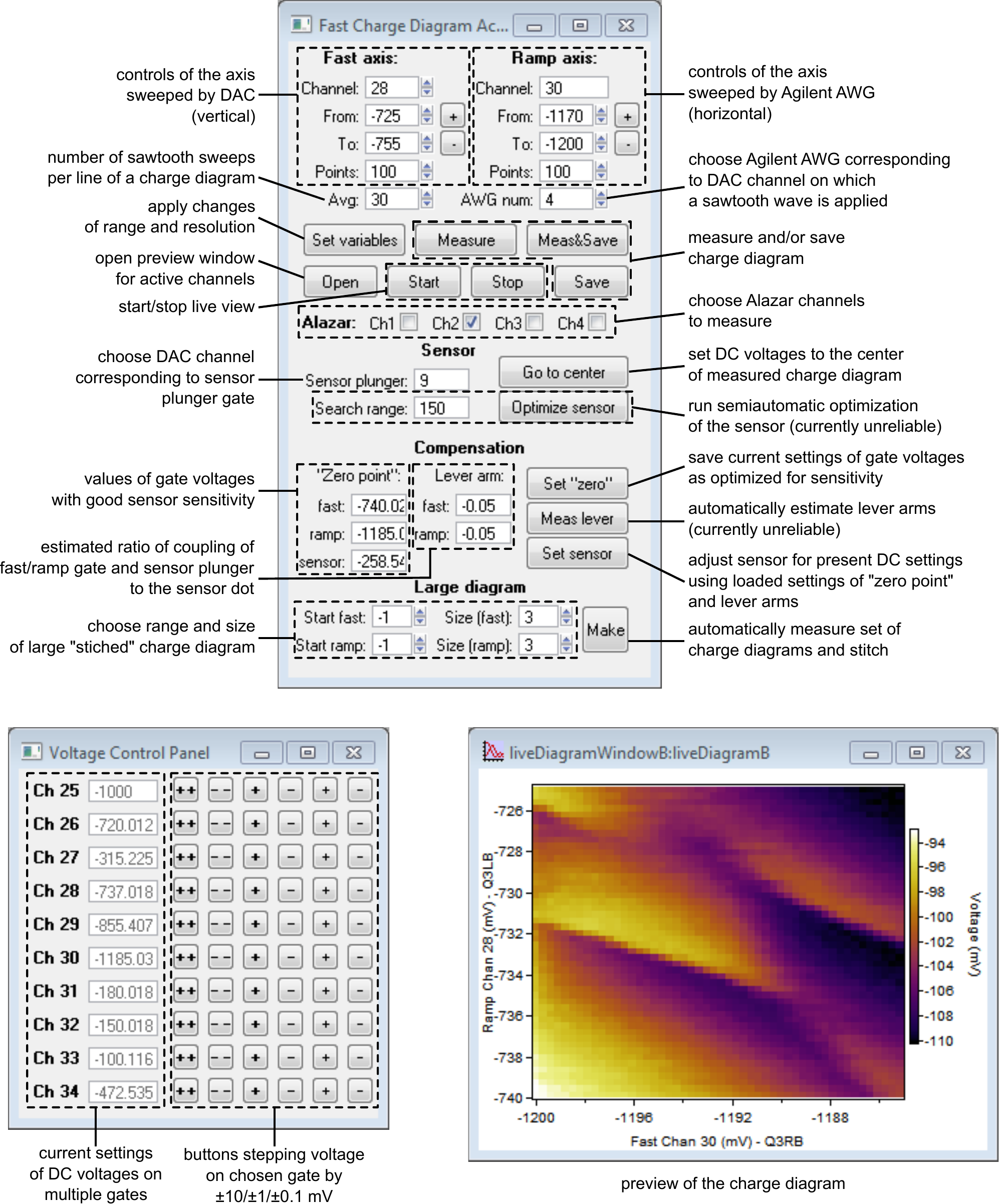}
\caption[Graphical user interface for control of real-time view of a charge diagram]{
Graphical user interface for control of real-time view of a charge diagram
}
\label{RT:GUI}
\end{center}
\end{figure}

As the acquisition of the charge diagrams becomes much faster than typing commands into command line, we added a graphical interface (Fig.~\ref{RT:GUI}). It provides a convenient way of setting the range, resolution and averaging of the scan. Also it allows to choose the measurement of multiple Alazar channels therefore possibility of measuring multiple sensors without adding to the measurement time. Separate panel adjusts to step voltage by $\pm$0.1, 1 or 10 mV on a chosen gate to modify the charge diagram.

Fast tuning requires continuous adjustments of the sensor dot, since sensitivity is quickly lost as voltages on the gates are changed by tens or hundreds of milivolts. These adjustments can be done in a few ways. In practice we found it is most convenient to manually adjust sensor plunger voltage with additional panel.

\begin{figure}[t]
\begin{center}
\includegraphics[width=0.8\textwidth]{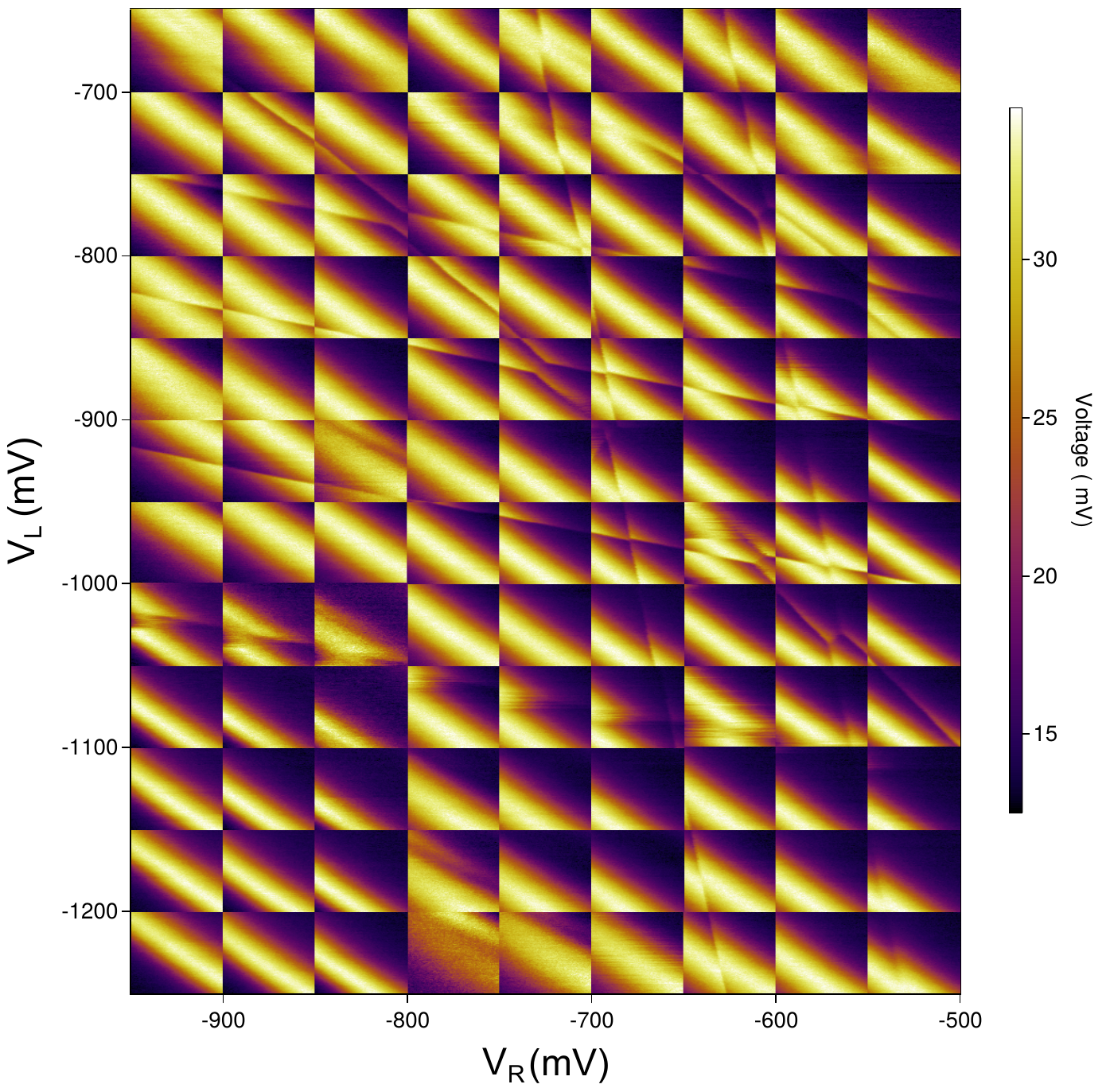}
\caption[Example of a triple quantum dot charge diagram stitched from multiple 50$\times$50 mV (100$\times$100 px) patches]{
Example of a triple quantum dot charge diagram stitched from multiple 50$\times$50 mV (100$\times$100 px) patches.
}
\label{RT:stitched}
\end{center}
\end{figure}

Finally, very often there is need to acquire a large diagram extending over a range of a few hundred milivolts. However we found that increasing of the scan range to much more than 50 mV is unpractical for several reasons. First, the sensor is sensitive only in a narrow range of voltages. Second, a large amplitude of the sawtooth wave causes heating of the electron gas, likely due to energy dissipation at the attenuators mounted at the mixing chamber. Instead we use the approach of automatic stitching of multiple charge diagrams (Fig.~\ref{RT:stitched}). Before taking each of the diagrams, sensor plunger voltage is approximately readjusted according to estimated coupling of various gates to the sensor dot. These can be easily set manually after a few trials and errors.

\section{Artifacts in charge diagrams due to fast sweeping}

Fast measurements of charge diagrams come with certain drawbacks. The biggest one becomes apparent when tuning multiple-dot structures and minimizing tunnel coupling to the leads. In such cas tunneling rates $\Gamma$ between the dots and the leads become comparable or smaller than frequency of the sawtooth wave ($1903$ Hz). This results in distortion, blurring and shifting of the hcharge transitions corresponding to barriers with small $\Gamma$.

\begin{figure}[t]
\begin{center}
\includegraphics[width=\textwidth]{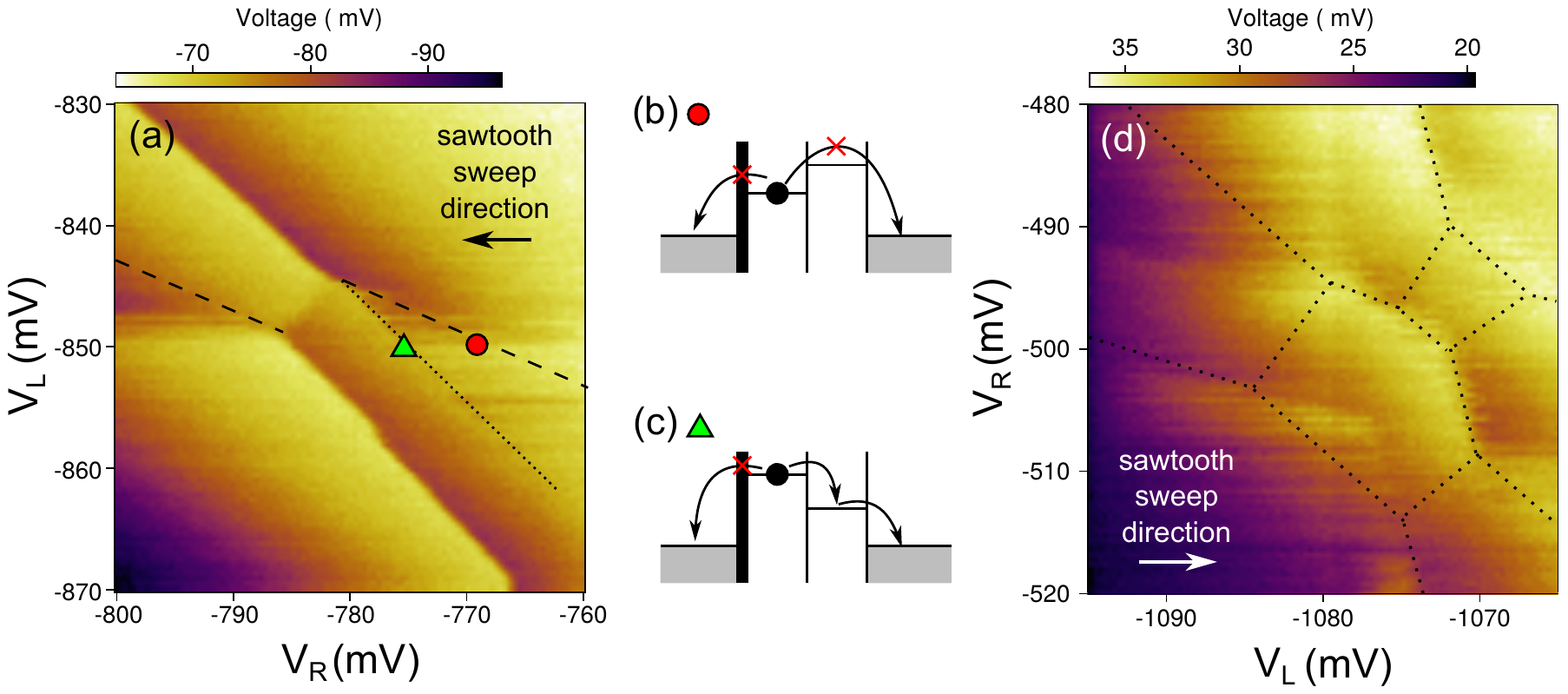}
\caption[Transition blurring and distortion in real-time measurements of charge diagrams]{
Transition blurring and distortion in real-time measurements of charge diagrams. (a) In this double dot diagram lines indicates with dashed lines are gone, because tunneling rate through corresponding barrier is smaller than frequency of the sawtooth wave. (b) Low tunneling rate to the lead prevents the double quantum dot from reaching the ground state. (c) Ground state is reached only when sequential tunneling to the lead via the right dot is avaliable. (d) Triple dot charge diagram, very strongly distorted due to slow tunneling rates. Dotted lines indicate positions of the real charge transitions. This degree of distortion is not unusual at the advanced stage of tuning, when the qubits are already working, and one is pinching off the barriers to the leads to maximize relaxation times.
}
\label{RT:artifact}
\end{center}
\end{figure}

Typical example of such artifact is presented in Fig.~\ref{RT:artifact}(a). In this case the barrier between the left quantum dot and the lead (dashed line) disappears, and corresponding change of the dolt occupation happens at different voltage (dotted line). Ther reason for this is the following. As the voltage is swept from positive the negative values the chemical potential of the left dot moves above the chemical potential of the leads (dashed line; Fig.~\ref{RT:artifact}(b)). However the tunneling does not occur until efficient path for the electron is energetically available. Ground state is reached only when arrangement of chemical potentials makes sequential tunneling through the right quantum dot possible (dotted lines; Fig.~\ref{RT:artifact}(c)).

Fortunately, trained eye can usually easily recognize the artifacts and carry on tuning (for extreme case of the blurring, see Fig.~\ref{RT:artifact}(d)). Moreover, presence of the artifacts does not imply that tuned qubit will be lousy, on the contrary -- quantum dot systems well isolated from electron bath in the leads tend to have longer relaxation times and be better behaved, as the exchange of electron (or spin) with the leads becomes dramatically suppressed.

\section{Applying control pulses while acquiring charge diagrams}

\begin{figure}[t]
	\centering
	\includegraphics[width=\textwidth]{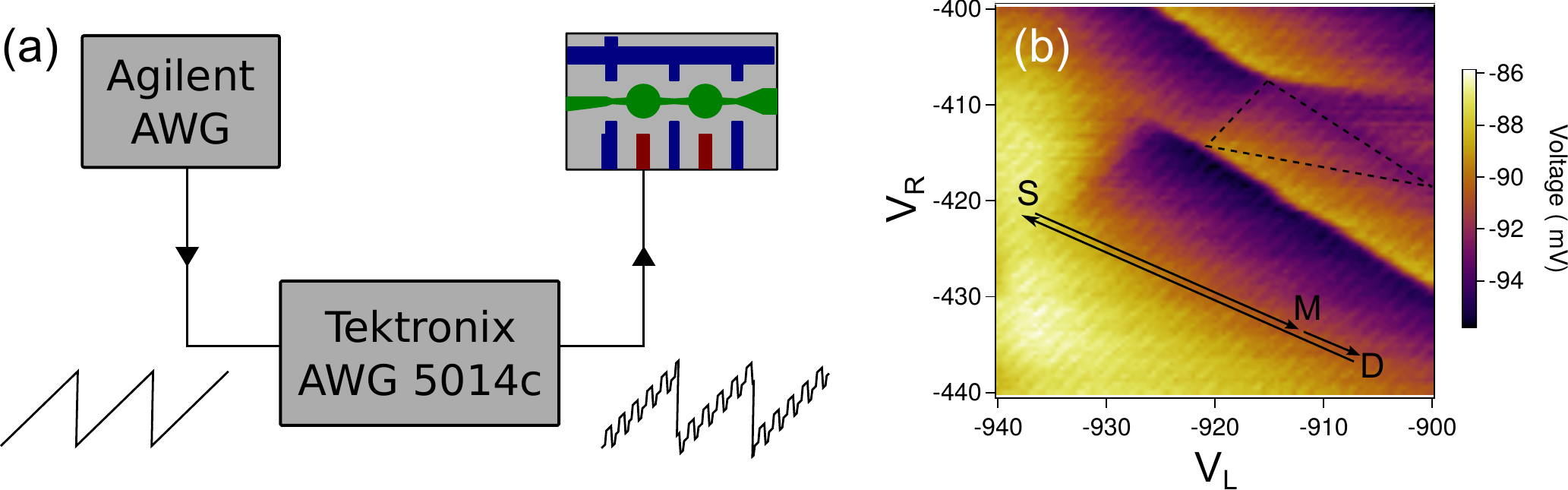}
	\caption[Charge diagram obtained while pulsing on the gates]{
	(a) Scheme of generation of sawtooth wave with fast pulses added on top using Tektronix AWG. (b) Example of charge diagram obtained while pulsing on the gates. Arrows indicate pulse shape in gate voltage space. This pulse has three steps: {\bf S}eparation -- that randomizes spin state of two electrons, {\bf M}easurement -- at which we attempt to read a spin state via spin-to-charge conversion and {\bf D} -- compensation point that sets average voltage during the pulse to zero. Triangle indicated with dashed lines indicates region where signature of a spin blockade is expected.
	}
	\label{RT:pulsing}
\end{figure}

Sometimes in the experiments there is a need to search for certain spots with long relaxation time within the charge diagram. Typically the goal is to find a spin blockade or a good measurement point within a spin-blocked triangle~(Fig.~\ref{RT:pulsing}(b)). In such case there is a need to measure a charge diagram where each pixel represents the value of the measurement performed at this point at the end of certain sequence of pulses.

The convenient hack is to apply a desired pulse using the AWG while acquiring a charge diagram (Fig.~\ref{RT:pulsing}(a)), with a real-time technique or in a standard way. If the RF carrier is switched on only when voltages are set to the measurement point the regions with long relaxation time should show up in the charge diagram. The signal acquired throughout the rest of the pulse cycle contributes only to the background noise. Increased averaging allows to compensate for this background.

There are four additional notes that have to be made on this topic. First, to obtain charge diagrams of good quality the measurement time should be long ,relative to the total pulse length. I recommend to keep it above 50\%.

Second, the application of the fast pulses results in additional artifacts. They appear whenever any of the steps the pulse goes across a transition with small tunneling rate. Therefore you need to pay additional attention and make sure that the designed pulse does not cross such charge transition whenever measurement is performed within interesting region.

Third, if the sawtooth wave period is close to the integer number of pulse periods a strong stripes might appear on the charge diagram, that will dominate the entire measurement (weak diagonal stripes can be seen in Fig.~\ref{RT:pulsing}(b)). In such case one needs to change the length of the pulse a little bit (a few hundred nanoseconds is usually enough).

Fourth, the average voltage within a singleperiod of pulse needs to be 0. This is because DC voltage drop across the bias tees on the sample board, that may cause heating of the fridge. It is convenient to set voltage on Tektronix AWG to 0 at the measurement point. Otherwise you will observe apparent shifts of the charge diagram, which might be tedious to calibrate out.

\chapterimage{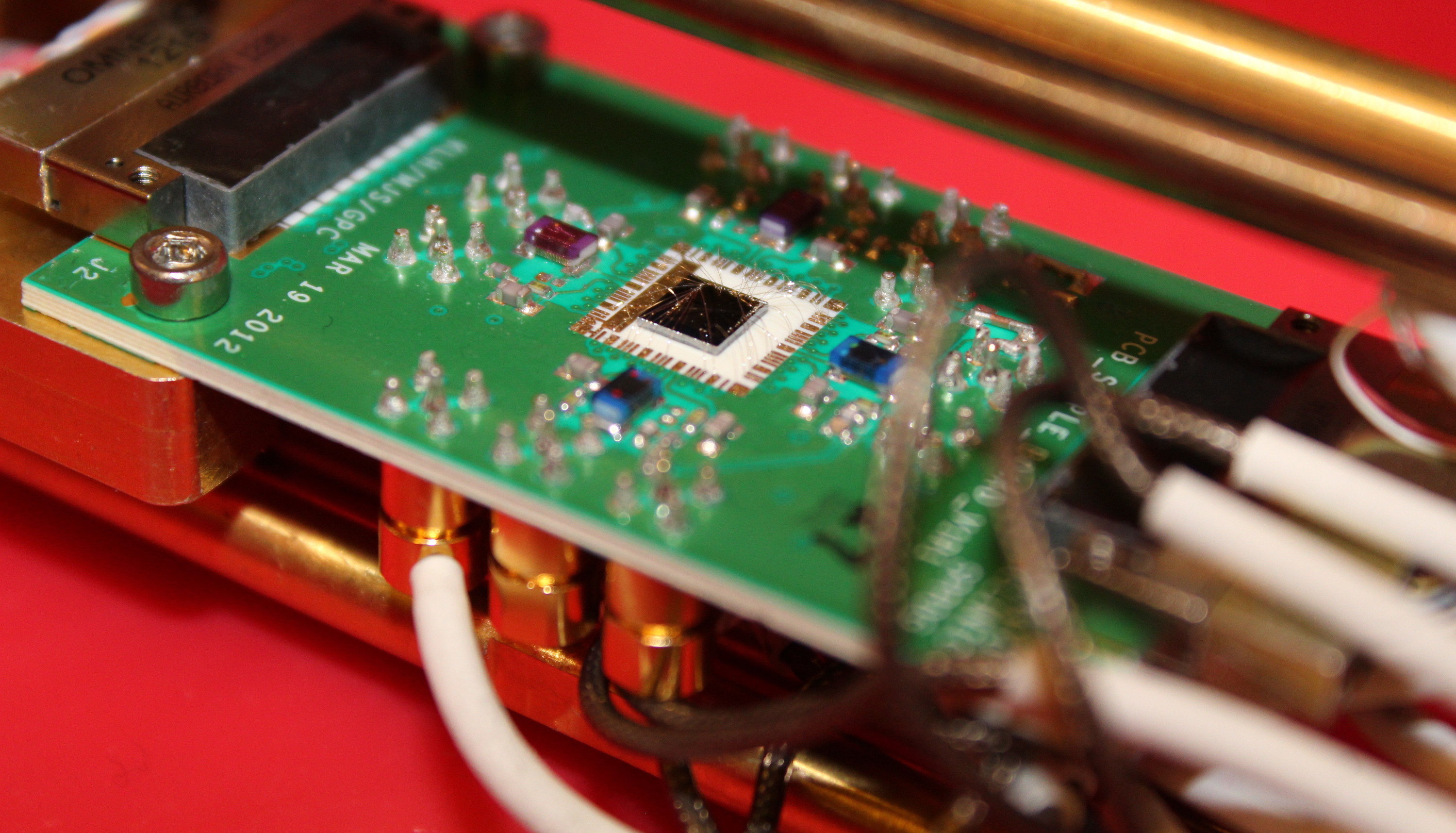}
\chapter[Remote control of multiple arbitrary waveform generators]{\protect\parbox{0.9\textwidth}{Remote control of multiple \\ arbitrary waveform generators}}
\label{ch:AWGs}

\begin{center}
\begin{minipage}{0.8\textwidth}
	\small
	\emph{"Documentation is like sex: When it's good, it's amazing, and when it's bad it's still better than nothing." - Eli Levinson-Falk}
	
	\begin{flushright}
	\begin{minipage}{0.85\textwidth}
		Edward Laird citing Eli Levinson-Falk in the beginning of the Igor procedure file containing drivers for Tektronix 5014c AWG to justify the state of documentation.
	\end{minipage}
	\end{flushright}
\end{minipage}
\end{center}

In this chapter you will find instructions how to use Igor Pro procedures I rewrote to control multiple Tektronix 5014c arbitrary waveform generators (based on older code by Edward Laird, Christian Barthel, Jim Medford and others). In section~\ref{AWGs:guide} I will introduce the features available through graphical interface, introduce most useful commands and point to some important places in the code. In the section~\ref{AWGs:code} I will dive deeper into the structure of the code, explaining the features that are less obvious but need to be remembered.

\section{User's guide}
\label{AWGs:guide}

\subsection{Initialization}

The code controlling the AWGs (excluding measurement procedures) is contained in three procedure files:
\begin{itemize}
	\item \verb'MultipleAWG5014.ipf' -- contains definitions of GUI and some high level functions
	\item \verb'MultipleAWG5014_PulseProcedures.ipf' -- contains most of the procedures for uploading pulses, sequences, compensation of high-pass filters etc.
	\item \verb'LowLevelAWG5014Control.ipf' -- low-level driver for communication with AWGs, adopted from Christian Barthel's code to communicate with multiple AWGs.
\end{itemize}
Compiling them requires cleaning the experiment file from the old legacy code for control of Tektronix 510, 720 and single 5014. You might also need to create two symbolic paths for saving pulses and sequences (Sec.~\ref{AWGs:sec:pulsepanel}).

Once the code is compiling initialize the GUI by calling
\begin{verbatim}
	initAWG(n)
\end{verbatim}
where \verb'n' is the number of AWGs you will be using. This will create several panels that I will describe in a second. For now you care about the main panel (titled \verb'AWG5014panel'). Set the IP addresses of the AWGs and the port number (you can find or change it on AWG in \verb'system\GPIB/LAN configuration' menu).

Depending on the Igor code version you need then to press \verb'Initialize VISA' or call \verb'initVISA()' first. If no errors are thrown the communication to the AWGs is now established.

However before you proceed you need to know one thing. The values of the parameters you see now in the GUI (which I will describe in a second) are the values stored locally. If you connected AWG for the first time or restarted it they do not match the real settings until you change them through GUI. For this reason I recommend to control AWG only via Igor interface, to make sure the values stored in Igor match values that are really set.

\subsection{Main panel}

\begin{figure}
\begin{center}
\includegraphics[width=\textwidth]{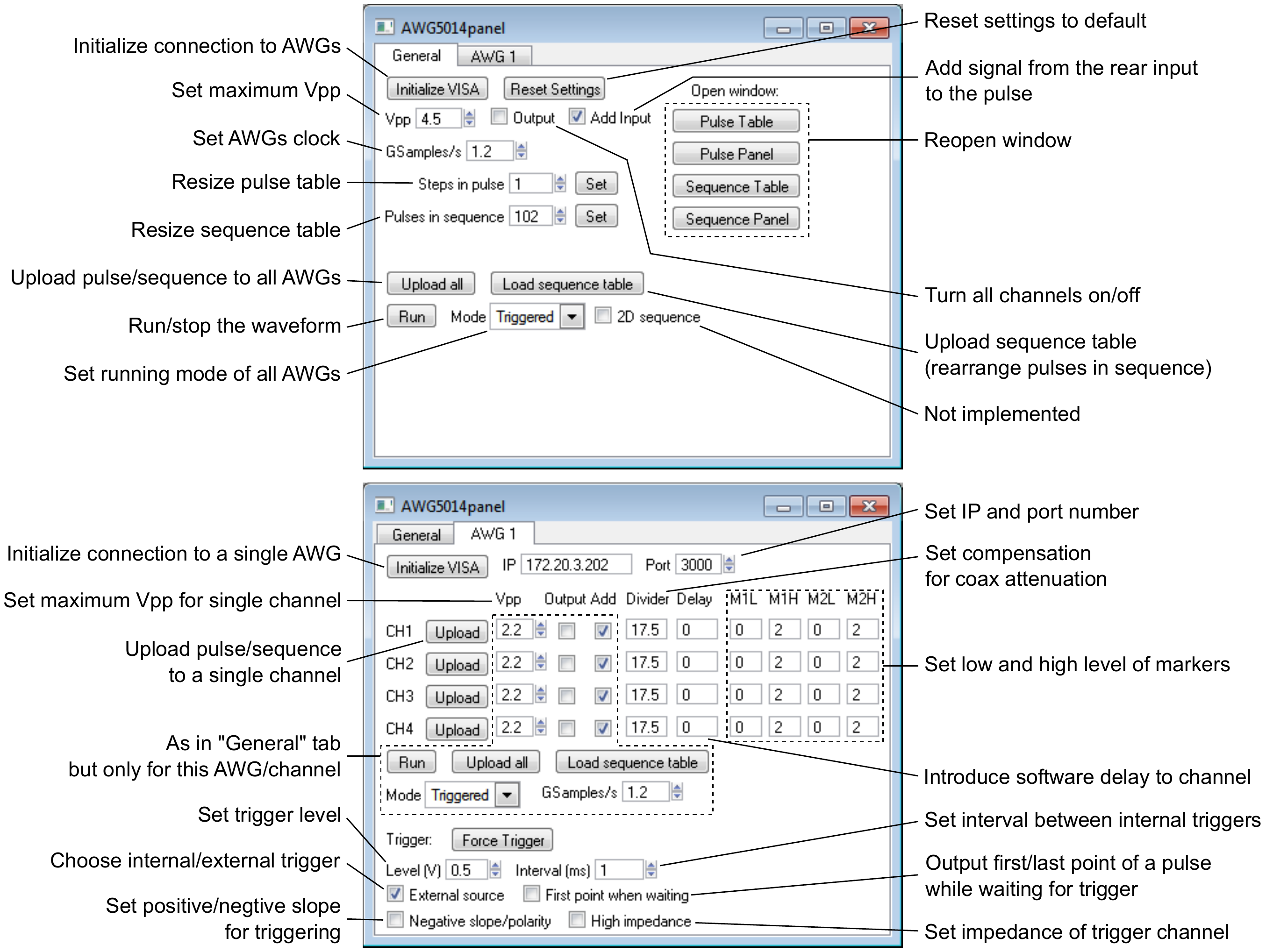}
\caption[Main panel for control of multiple Tektronix 5014c AWGs]{
Main panel for control of multiple Tektronix 5014c AWGs
}
\label{AWGs:main_panel}
\end{center}
\end{figure}

The main panel (Fig. \ref{AWGs:main_panel}) provides easy access to main settings of each AWG and is mostly self-explanatory. \verb'General' tab allows to control all of the AWGs simultaneously. In particular from here you can resize pulse and sequence table, and upload designed pulse (mode \verb'Continuous' or \verb'Triggered') or sequence (mode \verb'Sequence'). If you wish to rearrange pulses in the sequence you can modify sequence table without the need to upload all pulses (section~\ref{AWGs:sequence}). You can also turn all outputs on/off and start running the waveform\footnote{Note: Running the waveform and turning outputs on first time after uploading the sequence can take a few seconds. This might cause errors when you run the first measurement after uploading the sequence}. \verb'Add input' controls whether signal supplied to the rear input will be added to the waveform (this is useful for fast acquisition of charge diagrams; chapter \ref{ch:RT}). Buttons on the right allow to reopen remaining windows and tables.

Other tabs allow to control each AWG separately. Here you can reupload pulse or sequence to single channel (only if length of the pulse and sequence table was not changed) and control trigger and marker channels.

Should you need to reopen the main panel simply call:
\begin{verbatim}
	AWG5014panel()
	buildAWGpanelTabs(n)
	setAWGtab("whatever",0)
\end{verbatim}
where \verb'n' is number of AWGs you use.

\subsection{Uploading pulses}
\label{AWGs:pulse}

\begin{figure}[tb]
\begin{center}
\includegraphics[width=.8\textwidth]{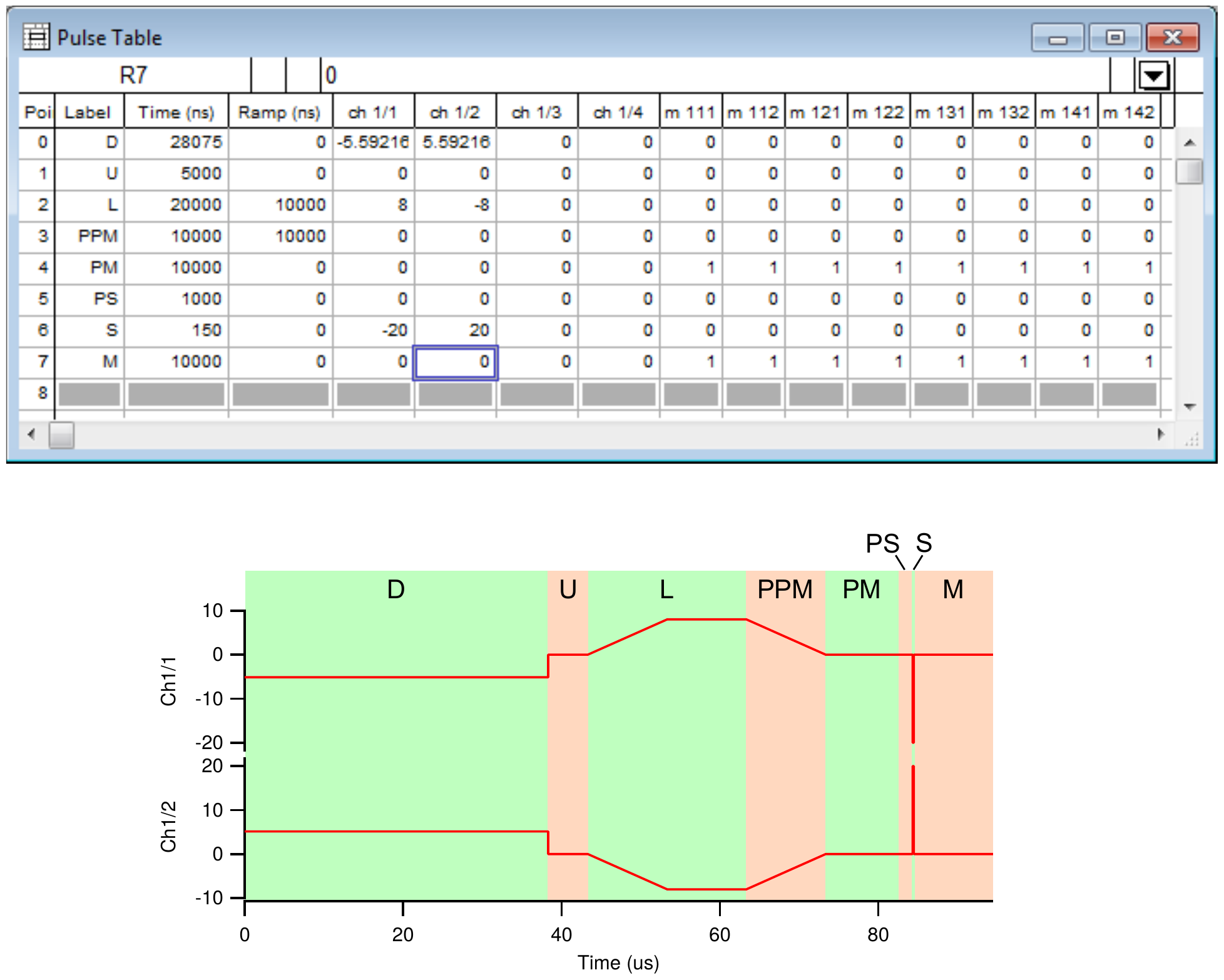}
\caption[Pulse table and pulse based on the table]{
(top) Pulse table for a single AWG and (bottom) pulse created based on the table (only channels 1 and 2 are shown).
}
\label{AWGs:pulsetable}
\end{center}
\end{figure}

Once we familiarized ourselves with the main panel it is time to upload and run our first pulse. Let's take a look at the pulse table and corresponding pulse shape (Fig.~\ref{AWGs:pulsetable}). The rows of the table correspond to segments of the pulse. Each of the segments you can label with a few character string. Segments labeled \verb'D' and \verb'M' play a special role (which I'll mention later) and therefore you should avoid using labels starting with D or M for other segments. Each segment lasts for time specified in nanoseconds in next column. Be aware that inserting time that is not a multiple of AWG clock cycle may sometimes cause troubles. Columns labeled \verb'ch #/$' indicate voltage (in \emph{milivolts}) that will be outputted during this segment on channel \verb'$' of AWG number \verb'#'\footnote{The indices always go in a descending hierarchy: AWG number first, channel number second, marker number third (if needed)}. If the \verb'Ramp' is nonzero it indicates the length of the sweep to the specified voltage.

The example of the pulse created with a pulse table is on the bottom of the Fig.~\ref{AWGs:pulsetable}. It consists of a few steps, each labeled according to the pulse table. In particular notice that step \verb'L' (time 20000, ramp 10000) consists of 10 $\mu$s sweep from 0 to $\pm$8 mV and 10 $\mu$s waiting time at $\pm$8 mV, so ramp time is included in the total segment time. Next segment (\verb'PPM') shows that even more clearly -- entire segment is a sweep.

Final columns titled \verb'm#$%' set the marker \verb'%' of channel \verb'$', AWG number \verb'#' to low (value 0) or high (value 1). In this example marker is turned on during \verb'PM' (reference measurement) and \verb'M' (measurement) segments to turn on RF carrier and trigger the Alazar card (so in fact only two markers are used)\footnote{Side note: notice that there is additional 1 $\mu$s waiting time introduced between {\verb"PM"} and {\verb"M"}. This is because the Alazar card needs to have clear slope to trigger on. So if marker is switched off for just 150 ns there is a risk that trigger will be missed.}.

Once you prepared pulse table set mode of the AWGs to \verb'Continuous' or \verb'Triggered' and press \verb'Upload all'.

\subsection{Correcting for attenuation and delays}

By now you might have noticed two parameters in the GUI that do not have counterparts in the settings of AWG, that is \verb'Divider' and \verb'Delay'. These are software corrections for attenuation and (large) difference in length of the coax lines.

The divider is factor that will multiply the value which is set in the pulsetable (\emph{not} in dB). The goal here is to choose this factor to be equal to the attenuation of the fast lines.  Once you do that, the voltage specified in the pulse table will correspond to the actual voltage on the sample. To estimate the divider factor upload small (let's say mV, period 10-100 $\mu$s) square pulse on one of the gates and run it while acquiring a charge diagram. All charge transitions should become doubled, and the distance between two images will correspond to the actual amplitude of the square pulse on the gate. Now insert divider, upload square pulse again and repeat until you reach satisfying accuracy. In our example (Figs.~\ref{AWGs:main_panel} and \ref{AWGs:pulsetable}) we see that during segment \verb'S' we want to apply $-20$ mV on gate connected to ch 1/1. However estimated divider id 17.5. This means that the voltage on the output of the AWG will be set to $17.5\times (-20$ mV$) = -350$ mV.

The delay parameter can be used to compensate for large (>1 ns) differences in the length of the fast lines. In practice difference between length of coaxes is much smaller and should be corrected by adjusting the skew (directly on AWG, not added to GUI). However it is necessary if you use some of the AWG channels to pulse on the gates and other to modulate the RF. If delay on any of the channels is nonzero the software will insert additional segment in the beginning/end of each pulse, when the voltage is 0. And so if length of the cables delays modulated RF signal by 20 ns you need to insert +20 in the corresponding window. Notice that the delay is introduced by adding additional waiting time in the beginning and/or at the end of each pulse therefore you need to design a pulse in such a way, that this will not introduce artifacts in the measurements.

In fact there is one more in-software correction: correction for high-pass filtering of the fast lines. This feature is not implemented in the interface -- the cutoff frequency is hardcoded but can be modified in \verb'compilePulseTable()' function (Subsec.~\ref{AWGs:compile}).

\subsection{Special role of {\bf D} and {\bf M} segments in compensation of high-pass filtering}

Speaking of proper design of pulse and high-pass filtering it is time to discuss convention considering pulse shapes and segments labeled \verb'D' and \verb'M'. What you would probably like is for the volage on gates to be equal to DC set by DAC whenever you set 0 in the pulsetable. Therefore you must be aware of high-pass filtering (by the bias-tees). Effectively it will shift the average voltage throughout the pulse to 0 (that is to DC value set by DAC) and cause heating effects. On the other it is very often convenient to keep the measurement point \verb'M' at values set by DAC, and therefore all AWG channels output 0 during this segment. To solve this issue it is convenient to introduce additional segment \verb'D' (choice of letter D is arbitrary and is hardcoded in the software). It's duration is set to fixed percentage of the pulse length (set in the code, as I will describe in subsection \ref{AWGs:correctionD}) and amplitude is adjusted to set the average voltage throughout the pulse to 0. By convention segment \verb'D' in the beginning of the pulse, just after the measurement \verb'M' and before initialization. The point is for to reinitialize the system after the \verb'D' pulse so it would not affect the experiment.

There is one more issue related to high-pass filtering, which is related to the fact that while AWG is not running or waiting for trigger (while the outputs are on) it does output the voltage. This will be the voltage equal to the first/last point in the sequence (depending on whether \verb'First point when waiting' check box is checked in the GUI, Fig.~\ref{AWGs:main_panel} (bottom)). Within the convention in which \verb'D' point is first and \verb'M' point is last (and all voltages are equal 0) the checkbox need to remain unchecked, which will make AWG output 0 while it's waiting to run.

\subsection{Modifying a pulse from the command line}
\label{AWGs:modifying}

In the code I define a special function that can be used to modify pulses from the command line
\begin{verbatim}
	setpulseparameter(idStr, value)
\end{verbatim}
It's syntax is analogous to \verb'setval' function. String \verb'idStr' specifies parameter that will be modified and \verb'value' is it's new value. There are a few fundamental predefined \verb'idStr' parameters, which begin with letters \verb't', \verb'v', \verb'r' and \verb'm'. Their function is the following:
\begin{itemize}
	\item \verb'"t<label>"' sets time of the segment \verb'<label>' to \verb'value' nanoseconds
	\item \verb'"v#$<label>"' sets voltage on AWG \verb'#', channel \verb'$' during the segment \verb'<label>' to \verb'value' milivolts
	\item \verb'"r<label>"' sets ramp time of the segment \verb'<label>' to \verb'value' nanoseconds
	\item \verb'"m#$%<label>"' sets marker \verb'%' on AWG \verb'#', channel \verb'$' during the segment \verb'<label>' to \verb'value'
\end{itemize}
There is also one special \verb'idStr="mAll"' which sets all markers to \verb'value' during entire pulse.

Now you might want to specify your own parameter to modify, which is a complex combination of parameters above. To do that you need to get into the code: find definition of \verb'setpulseparameter()' function and conditional tree starting with the line:
\begin{verbatim}
	if(stringmatch(idStr[0],"t")||stringmatch(idStr[0],"v")||
	                    stringmatch(idStr[0],"r")||stringmatch(idStr[0],"m"))
\end{verbatim}
You can add you entry in this tree. For example you can add such commands:
\begin{verbatim}
	elseif(stringmatch(idStr[0,2],"det"))
	    setpulseparameter("v11"+idStr[3,sEnd], -value)
	    setpulseparameter("v12"+idStr[3,sEnd], value)
\end{verbatim}
This will add possibility of calling, for example
\begin{verbatim}
	setpulseparameter("detS",20)
\end{verbatim}
from the command line. This will set the voltage on channel 1 of AWG 1 during segment \verb'S' to $-20$ mV and voltage on channel 2 of AWG 1 during segment \verb'S' to $+20$ mV (that is: it will apply such changes to the pulsetable).

Syntax introduced here will be relevant later, for definitions of pulse sequences.

\subsection{Saving and loading pulses}
\label{AWGs:sec:pulsepanel}

\begin{figure}[tb]
\begin{center}
\includegraphics[width=.8\textwidth]{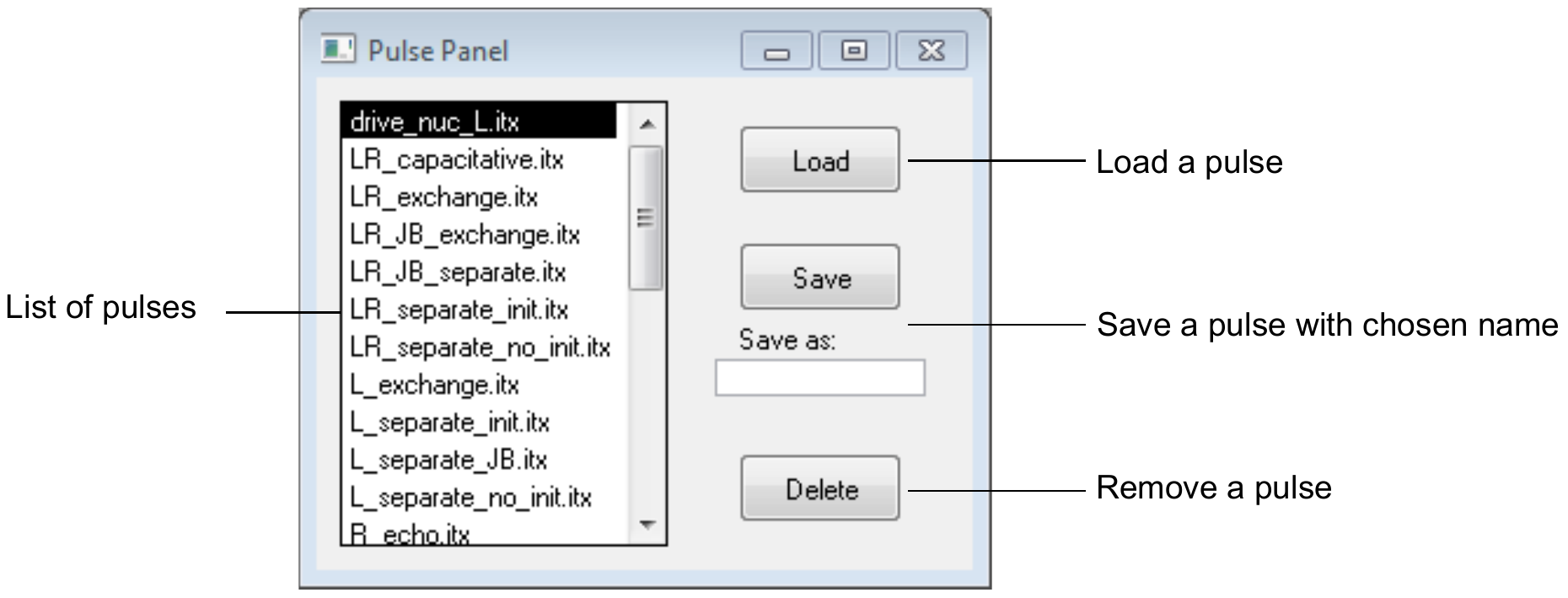}
\caption[Panel for saving and loading the pulses]{
Panel for saving and loading the pulses
}
\label{AWGs:pulsepanel}
\end{center}
\end{figure}

Once you define a fancy pulsetable you will probably like to save it and be able to use at later point in the experiment. For that you can use a \verb'Pulse panel' (Fig.~\ref{AWGs:pulsepanel}; it can be opened from the main tab on AWG panel~\ref{AWGs:main_panel}). It's quite self explanatory. Just remember that \verb'Delete' button removes highlighted pulse without a warning.

Technical detail is that each column of the pulsetable is stored as a wave, and all waves corresponding to a single pulse are saved in a single Igor text file (\verb'.itx').

Pulses will be saved in the symbolic path \verb'pulses', which I suggest to locate in \\ \verb'<experiment_file_location>\pulses\'. Panel identical to this one can be used to save sequences in symbolic path \verb'sequences' (suggested location: \verb'<experiment_file_location>\sequences\'.)

\subsection{Sequences}
\label{AWGs:sequence}

More complex function of the AWG is application of a sequence of pulses. First a set of waveforms is uploaded to AWG, then they are arranged in certain order. Once this is done the AWG can execute the sequence.

\begin{figure}[tb]
\begin{center}
\includegraphics[width=.65\textwidth]{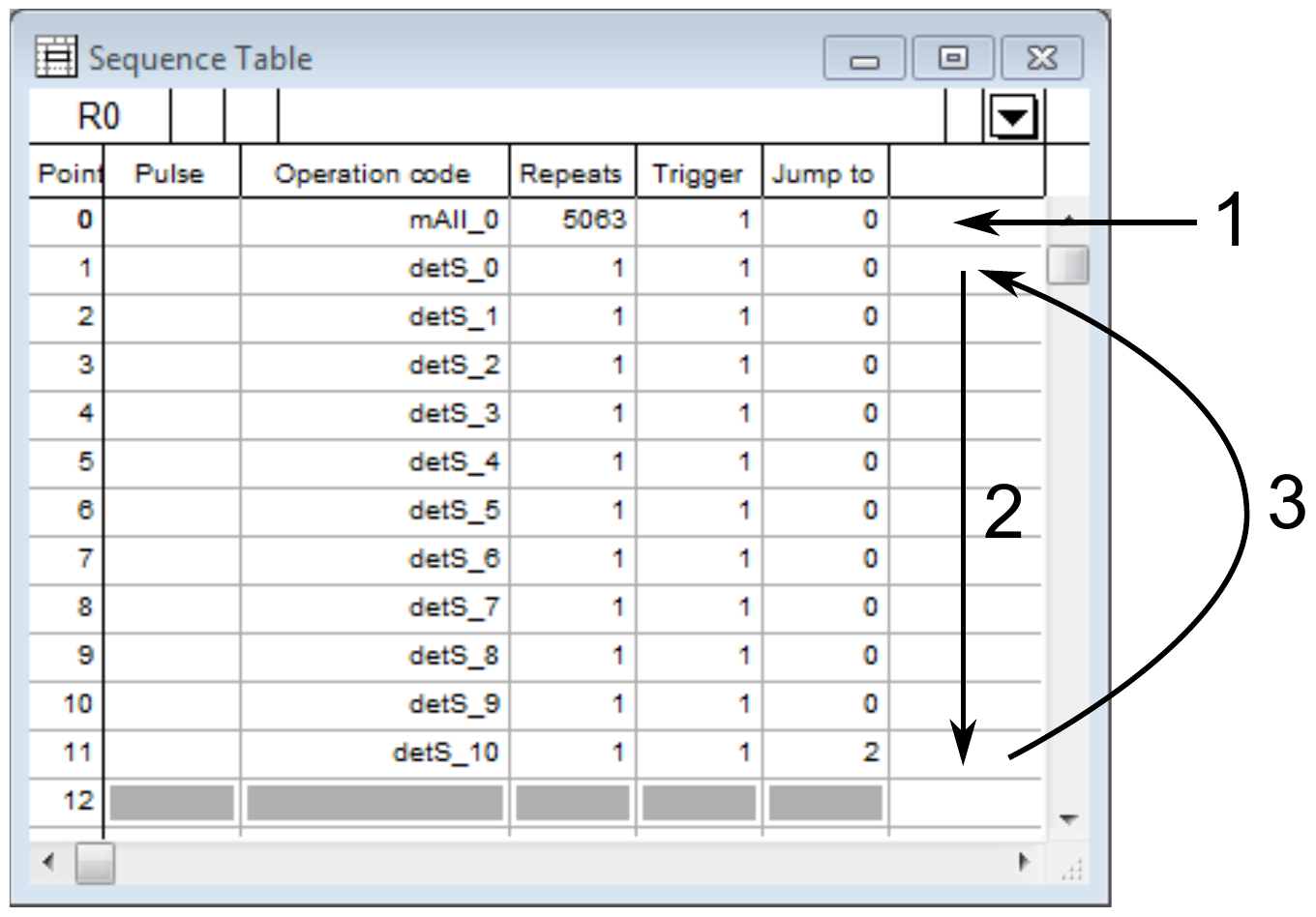}
\caption[Sequence table]{
Sequence table. 1) All markers are turned off and pulse runs 5063 times. 2) Eleven pulses with different values of detS are outputted. 3) AWG returns to the second line of the sequencetable and runs the eleven pulses all over again.}
\label{AWGs:sequencetable}
\end{center}
\end{figure}

To create a sequence we need to take at the final element of the GUI -- \verb'Sequencetable' (Fig.~\ref{AWGs:sequencetable}). This table specifies how pulses will be arranged in a sequence and whether they should be modified in any way.

First column -- \verb'Pulse' -- gives the name of the pulse. This name corresponds to the name of the file in the pulse panel. In case this field is left empty (which is usually the case) currently modified pulsetable is used.

Second column specifies what modifications should be done to the pulse. The operation code is a string that has a form of \verb'<idStr>_<value>'. When the sequence is compiled the software will take specified pulse and call \verb'setpulseparameter("<idstr>",<value>)'.

Third columns will specifies how many times the pulse should be repeated before moving on to executing the next line of the \verb'sequencetable'.

Fourth column tells whether the AWG should wait for the trigger before outputting this pulse.

Final columns specifies to which line in a sequence table to move after executing this line of a \verb'sequencetable'. This is slightly confusing because Igor labels rows of the \verb'sequencetable' from 0, while AWG starts with 1. The convention is this: value 0 means -- execute next line of a \verb'sequencetable'. Any number $n$ larger than 0 means -- execute line $n$, which is labeled in Igor as $n-1$. 

Knowing this conventions you can define sequence manually. However this is impractical, because usually sequences in spin-qubit experiments (perhaps also other) look like one specified in \verb'sequencetable' in Fig.~\ref{AWGs:sequencetable}. First a single pulse runs over and over again while communication with other devices takes place. This pulse has all markers turned off, not to trigger any measurement. Than series of pulses is applied, which differ only by value of one parameter. In this case detuning during separation point (\verb'detS'). The \verb'Go to' value of the last pulse in the sequence points is set to 2 so AWG stays in a loop, outputting the pulses with various \verb'detS' over and over again.

Such \verb'sequencetable' can be created using the function
\begin{verbatim}
	buildsequencetable("<pulse_name>","<idStr>",from,to,steps)
\end{verbatim}
which has similar structure to \verb'do1d()' function. Here \verb'<pulse_name>' is the basic pulse -- this name will appear everywhere in the first column of the \verb'sequencetable'. \verb'idStr' indicates the parameter which will be modified in this sequence. \verb'from' and \verb'to' are limits within which \verb'idStr' will be changed and \verb'steps' specifies how many different values should it take (in fact there will be \verb'steps'+1 values).

For example the \verb'sequencetable' in Fig.~\ref{AWGs:sequencetable} was created by calling
\begin{verbatim}
	buildsequencetable("","detS",0,10,10)
\end{verbatim}

You can make more subtle adjustments, for example whether to wait for trigger before outputting each pulse, how many times to repeat the first pulse etc. by modifying the code (which will be described in second part of this chapter).

\section{Code}
\label{AWGs:code}

\subsection{Basic structure}

\begin{figure}[tb]
	\centering
	\includegraphics[width=\textwidth]{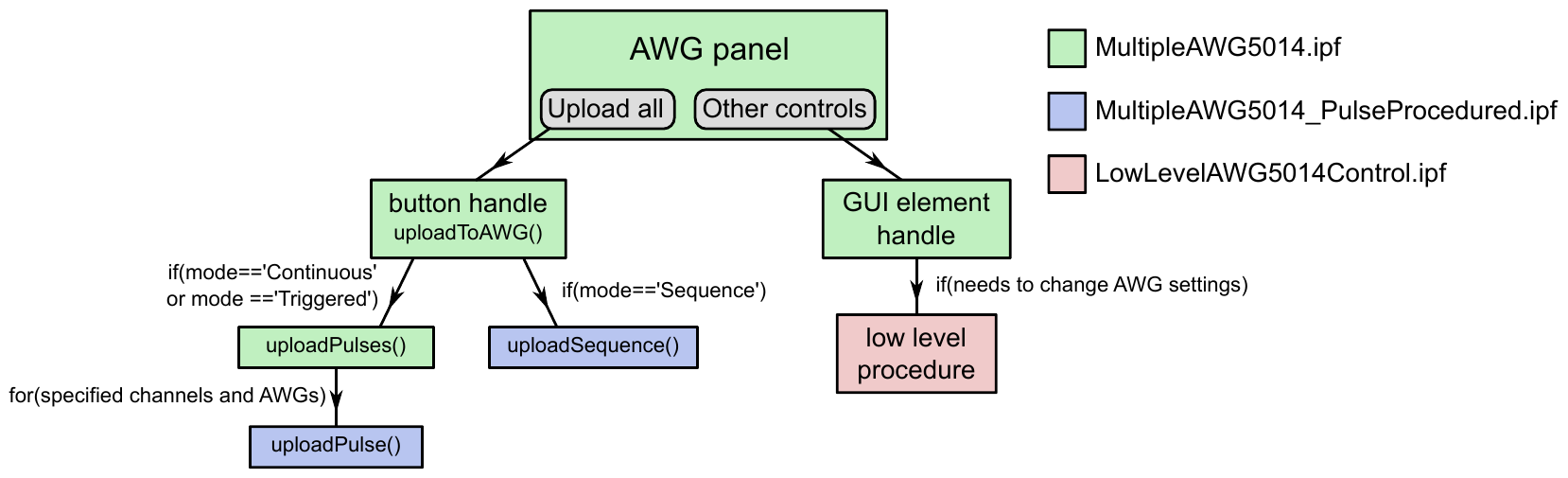}
	\caption[Basic structure of the code]{Basic structure of the code}
	\label{AWGs:main_tree}
\end{figure}

The code was not written systematically however it has quite logical internal structure. In this section I will describe this structure and show certain key points, where crucial procedures and parameters are defined.

The basic structure of the code, with GUI as a root, is shown in Fig.~\ref{AWGs:main_tree}. The most important part of the code is \verb'uploadToAWG()' function which handles clicks of the \verb'Upload all' button and calls subroutines responsible for compilation and uploading of pulses (Subsec.~\ref{AWGs:uploadPulse}) and sequences (Subsec.~\ref{AWGs:uploadSequence}). I'll describe those procedures in detail throughout this section.

Except for that GUI has several elements that instantly change settings of AWG. The functions that handle relevant events call low level functions communicating with AWGs and change local variables that store information about the setup of AWG. These functions are simple and therefore I'll not describe them here. However I remind -- because information of the AWG setup is stored locally, in certain situations the settings presented on the GUI may differ from actual settings of AWGs.

In addition to this basic functionality there are a few additional loosely connected procedures for building a sequence table, visualization of the pulse shape etc. They are very handy, and I'll describe them in the end of the chapter.

\subsection{Function: uploadPulse(AWGNum, chan[, quiet])}
\label{AWGs:uploadPulse}

\begin{figure}[tb]
	\centering
	\includegraphics[width=\textwidth]{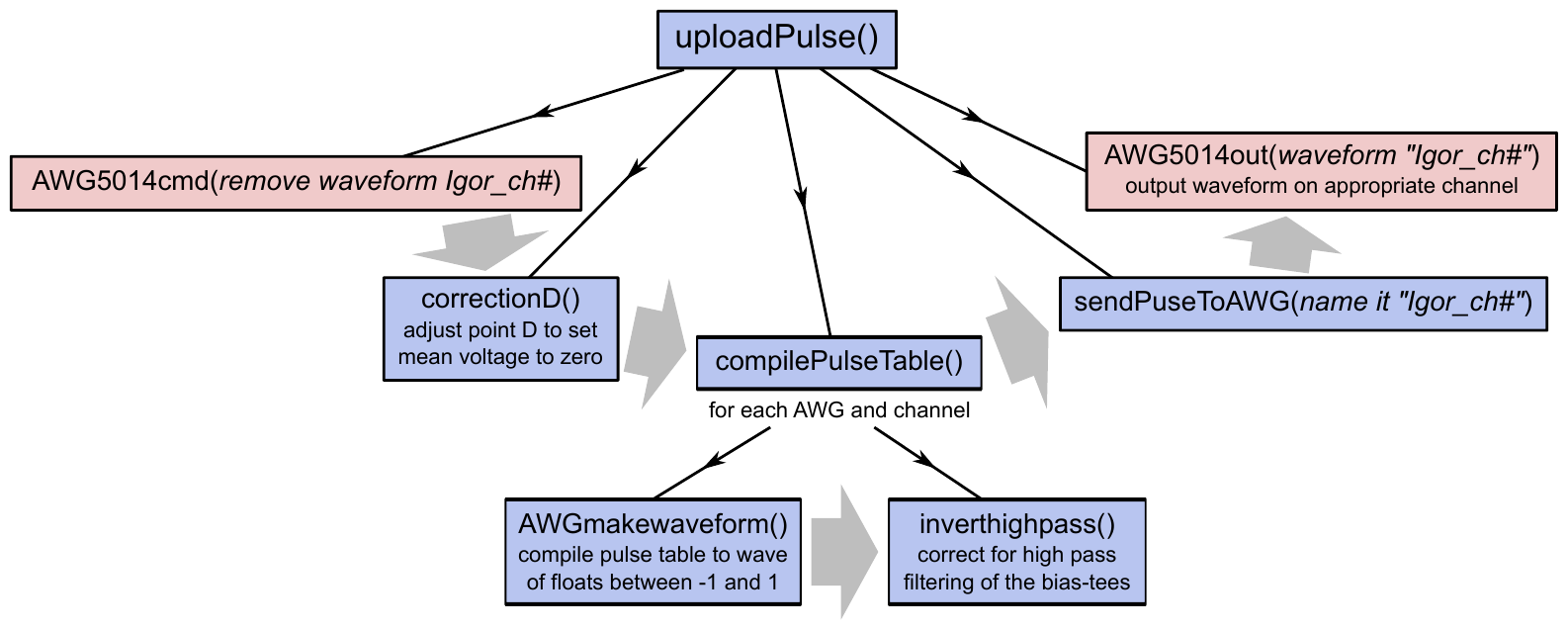}
	\caption[Structure of the code for uploading pulses]{Structure of the code for uploading pulses. Black lines indicates hierarchy. Gray arrows indicate order in which functions are called.}
	\label{AWGs:pulse_tree}
\end{figure}

The purpose of this function is to upload a pulse to a single channel in \verb'Continuous' or \verb'Triggered' mode. It performs the following steps (Fig.~\ref{AWGs:pulse_tree}):

\paragraph{} It removes old waveform from the AWG. Importantly it removes \emph{only} waveform that was previously outputted on this channel in \verb'Continuous' or \verb'Triggered' mode. This means that whatever sequence was uploaded before remains stored on the AWG and can be ''activated'' by switching back to \verb'Sequence' mode. As you'll find out by yourself, it allows to save a lot of time on uploading of the sequence.

\paragraph{} Time and voltages during \verb'D' point in the active pulsetable are adjusted by calling \\ \verb'correctionD(background=0)' (Subsec.~\ref{AWGs:correctionD}).

\paragraph{} It compiles the pulse to wave of floating numbers between $-1$ and $1$, in which each cell corresponds to a single clock cycle of the AWG. The relevant function is called \\ \verb'compilePulseTable(AWGNum, chan, background=0)' (Subsec.~\ref{AWGs:compile}).

\paragraph{} It turns the wave into binary and sends ready waveform to the proper AWG naming it \verb'Igor_ch$' where \verb'$' is a channel number by calling \verb'sendPulseToAWG(AWGNum, chanm "Igor")' (Subsec.~\ref{AWGs:send}). 

\paragraph{} The AWG is commanded to output the waveform on the designated channel.

\subsection{Function: uploadSequence(AWGNum, chan[, quiet])}
\label{AWGs:uploadSequence}

\begin{figure}[tb]
	\centering
	\includegraphics[width=\textwidth]{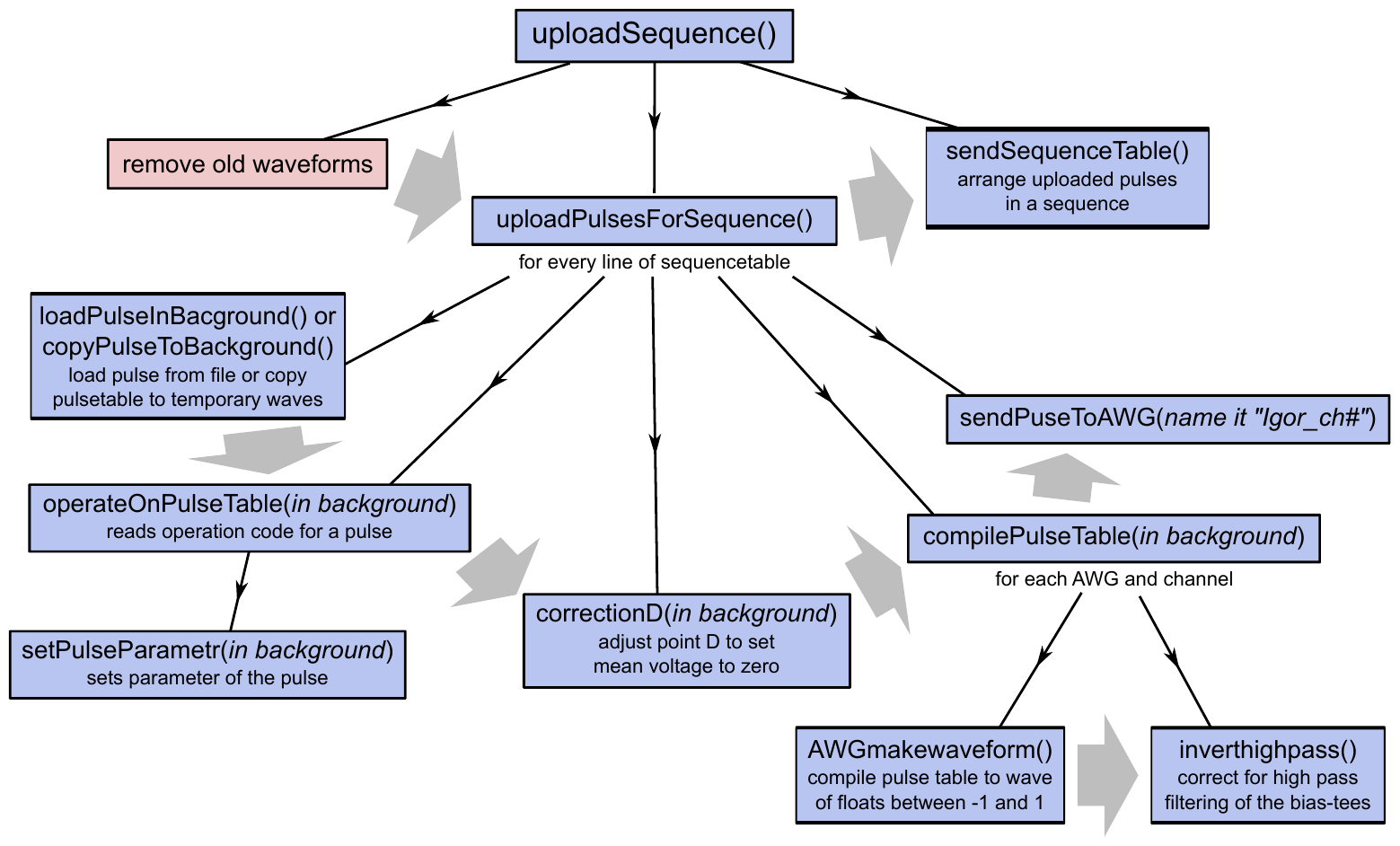}
	\caption[Structure of the code for uploading sequences]{Structure of the code for uploading sequences. Black lines indicates hierarchy. Gray arrows indicate order in which functions are called.}
	\label{AWGs:sequence_tree}
\end{figure}

This function is more branched than one for uploading pulses (Fig.~\ref{AWGs:sequence_tree}) and consists of the following parts:

\paragraph{} The function removes old waveforms from AWGs. If pulses are uploaded to all channels nothing is left behind. This is because communication with AWG is relatively slow, so browsing a list of waveforms and removing them one by one would be very inefficient.

\paragraph{} All pulses the sequence consists of are uploaded to the memory of the AWG with function \verb'uploadPulsesForSequence(AWGNum, chan)' (Subsec.~\ref{AWGs:uploadForSequence}).

\paragraph{} Sequencetable is converted to the series of commands which are sent to AWGs by calling \verb'sendSequenceTable(AWGNum)' (Subsec.~\ref{AWGs:sendSequenceTable}). The commands arrange waveforms uploaded to AWG in sequence.

\subsection{Function: correctionD([background])}
\label{AWGs:correctionD}

The purpose of function is to adjust time and voltage during the pulse segment \verb'D' to set the mean voltage during the pulse to zero. The time of segment \verb'D' in set to be a fixed fraction of the elength of the entire pulse by a set of commands:

\begin{verbatim}
	// set time of D point to 0 and calculate total time
	setPulseParameter("tD",0,background=background)
	variable totTime = sum(AWGt)
	
	// set initial time of D pulse to ##% of pulse time
	NVAR AWGfreq = root:malina:AWGfreq
	variable timeD = round(totTime/2*AWGfreq)/AWGfreq
	setPulseParameter("tD",timeD,background=background)
\end{verbatim}

In this case line \verb'variable timeD = round(totTime/2*AWGfreq)/AWGfreq' (factor 2) will set the segment \verb'D' to last 1/3 of the total length of the pulse. If you wish to adjust the length of the \verb'D' point here is where you should make modifications. The purpose of the multiplication by \verb'AWGfreq', rounding and division is to guarantee that the time of the segment \verb'D' will be a multiple of the AWG clock cycle.

\subsection{Function: compilePulseTable(AWGNum, chan[, background])}
\label{AWGs:compile}

This function main function is to pass parameters of a pulse (just single channel) from pulsetable to \verb'AWGmakewaveform' function. However it also adds segments in the beginning and/or at the of the pulse to introduce delay specified by the user in the GUI.

\paragraph{} First, the waves containing columns of pulse table are copied to locally created waves. To these two segments are added, in the beginning and in the end. In these segments all voltages and markers are set to, and their duration is adjusted based on delays for \emph{all} channels.

\paragraph{} Waves storing the pulse table are passed to \verb'AWGmakewaveform()' function (Subsec.~\ref{AWGs:makeWaveform}), which creates in the memory two waves, one for waveform (normalized to $\pm$1 based on \verb'Vpp' set for a channel) and second for markers (points have values 0, 1, 2 or 3, to represent 2 bits -- for 1st and 2nd marker).

\paragraph{} Functions for high- and low-pass correction are called\footnote{I was not using correction for low-pass filtering, so I am not aware whether it's functioning is correct}. Here the time constants of the filters are hardcoded. This relevant piece of code goes like this (Subsec.~\ref{AWGs:inverthighpass}):
\begin{verbatim}
	// invert high-pass and low-pass filters
	variable highpass = 2.7e6 // time constant of the high-pass filter in ns
	variable lowpass = 0
	if(highpass != 0)                                       
	    inverthighpass(highpass*1e-9, igorPulse, "arbinv")   
	    duplicate/o $"arbinv" $("root:malina:pulses:Igor_AWG_"+num2str(AWGNum)+
		                                                     "_ch"+num2str(chan))
	    KillWaves $"arbinv"
	endif
	if(lowpass != 0)
	    invertLowpass(lowpass, igorPulse, "arbinv")
	    duplicate/o $"arbinv" $("root:malina:pulses:Igor_AWG_"+num2str(AWGNum)+
		                                                     "_ch"+num2str(chan))
	    KillWaves $"arbinv"
	endif
\end{verbatim}

\subsection{Function: AWGmakewaveform\\(AWGNum, chan, arbstr, markstr, tw, rw, hw, mw1, mw2)}
\label{AWGs:makeWaveform}

This function waveform (normalized to $\pm$1 based on \verb'Vpp' set for a channel) and marker-wave (points have values 0, 1, 2 or 3, to represent 2 bits -- for 1st and 2nd marker). \verb'AWGNum' and \verb'chan' need to be specified, since \verb'Vpp' and \verb'divider' can be specified differently for each channel. \verb'arbstr' and \verb'markstr' are names of the waves that will be created in the memory. \verb'tw', \verb'rw', \verb'hw', \verb'mw1', \verb'mw2' are the waves with the parameters read from the pulsetable.

\verb'AWGmakewaveform()' function performs the following steps:

\paragraph{} Reads \verb'AWGFreq', \verb'Vpp' and \verb'divider' from the global variables.

\paragraph{} Calculates number of clock cycles for each segment and ramp.

\paragraph{} Calculates whether the pulse has the minimum length of 256 points. If not, repeats the same pulse many tames in a single waveform.

\paragraph{} Calculates voltage and 2-bit value of the marker wave for each clock cycle.

\paragraph{} Normalizes the voltage according to the \verb'Vpp'.

\subsection{Function: inverthighpass(tau, inputPulse, finalPulseName)}
\label{AWGs:inverthighpass}

The purpose of the \verb'inverthighpass()' function is to correct the pulse shape, for the high pass filter on the bias-tee. Essentially it applies a transformation:
\begin{equation}
	V(t_i) \mapsto V(t_i) + \frac{V(t_i) - V_\mathrm{mean}}{\tau / \Delta t}.
\end{equation}
Here $V(t_i)$ is voltage during $i$-th clock cycle, $\Delta t= t_{i+1}-t_i$, $\tau$ is the time constant of the high pass filter and $V_\mathrm{mean}$ is the mean voltage during the pulse.

Under assumption that $\Delta t \ll \tau$ this reverses the filtering perfectly. Typically $\Delta t \approx 1$~ns and $\tau \approx 1$~ms so this condition is easily fulfilled. 

\subsection{Function: sendPulseToAWG(AWGNum, chan, name)}
\label{AWGs:send}

This function reads compiled waveform for single channel and markerwave and passes it to \verb'awg5014sendPatternToWList()' for uploading. Here also the name of the uploaded waveform file is defined.

\subsection{Function: awg5014sendPatternToWList(AWGNum, filename, arbwave, markwave)}
\label{AWGs:sendpattern}

Low-level \verb'awg5014sendPatternToWList()' function converts the waveform and markerwave to the binary format, which can be efficiently sent to the AWG. More precisely each Voltage and marker value at each step is converted to \emph{two} 8-bit integers (14 bits for voltage and 2 bits for markers) according to the following recipe.
\begin{verbatim}
	binarywave[2*i] = round((arbwave[i]+1)*(2^13))&255
	                     //first 8 bits of 14-bit representation of arbwave
	binarywave[2*j+1] =
	          (round((arbwave[i]+1)*(2^13))&(2^8+2^9+2^10+2^11+2^12+2^13))/255
	          +(markWave[i]&1)*2^6+(markWave[i]&2)/2*2^7
	                     //remaining 6 arbwave and 2 bits of markerwave
\end{verbatim}

Than new waveform is created on AWG and \verb'binarywave' is sent using \verb'AWGVisaWriteBinary()' function.

\subsection{Function: uploadPulsesForSequence(AWGNum, chan[, sequenceName])}
\label{AWGs:uploadForSequence}

This function essentially executes procedure identical to \verb'uploadPulse()' function in a loop for every line of the sequence table.

In every iteration it looks at the row of the sequence table, copies an appropriate pulse table to the temporary waves, applies an operation that modifies the pulse table according to the codeword (Sec.~\ref{AWGs:modifying}), applies \verb'correctionD()', compiles with \verb'compilePulseTable()' and uploads with \verb'sendPulseToAWG()' under a name created based on the codeword.

In case the name of the base pulse does not match any of the saved pulse tables it searches through the list of the saved sequences and calls itself with the subsequence name as an optional argument. This enables to iteratively upload all of the pulses for the subsequence.

\subsection{Function: sendSequenceTable(AWGNum)}
\label{AWGs:sendSequenceTable}

Once all of the pulses of a sequence are uploaded with \verb'uploadPulsesForSequence()' function they are arranged in a sequence by calling this function. Essentially it loops through the rows in a sequence table and sends a series of binary commands to AWG specifying the waveform name, number of repetitions etc. for each element in a sequence. To slightly increase the speed commands ares first memorized in a strings of $\sim100$ commands which are sent with \verb'AWGVisaWriteBinary()'.

\subsection{Function: showpulse(channels)}
\label{AWGs:showpulse}

\verb'Showpulse()' function allows to plot waveforms defined in a pulsetable. For example calling \verb'showpulse("11;12")' will plot waveforms for channels 1 and 2 of AWG 1. This is how the plot in Fig.~\ref{AWGs:pulsetable} was created (except for background shading and segment labels).

\subsection{Function: annotateshape(xcent, ycent, xchan, ychan)}
\label{AWGs:annotateshape}

This function visualises the pulse shape as arrows on a charge diagram. Parameters \verb'xchan' and \verb'ychan' (which are strings consisting of two digits) choose which AWG channels should be visualised on, respectively, horizontal and vertical axis. Parameters \verb'xchan' and \verb'ychan' indicate DC voltages on a charge diagram, which will correspond to 0 voltage on corresponding AWG channel. Special case is \verb'xchan=0' or \verb'ychan=0' which will result in visualisation of the pulse in the center of the charge diagram.

\bookmarksetup{startatroot}

\startcontents
\cleardoublepage
\phantomsection
\markboth{\sffamily Bibliography}{\sffamily Bibliography}
\addcontentsline{toc}{chapter}{Bibliography}
\chapterimage{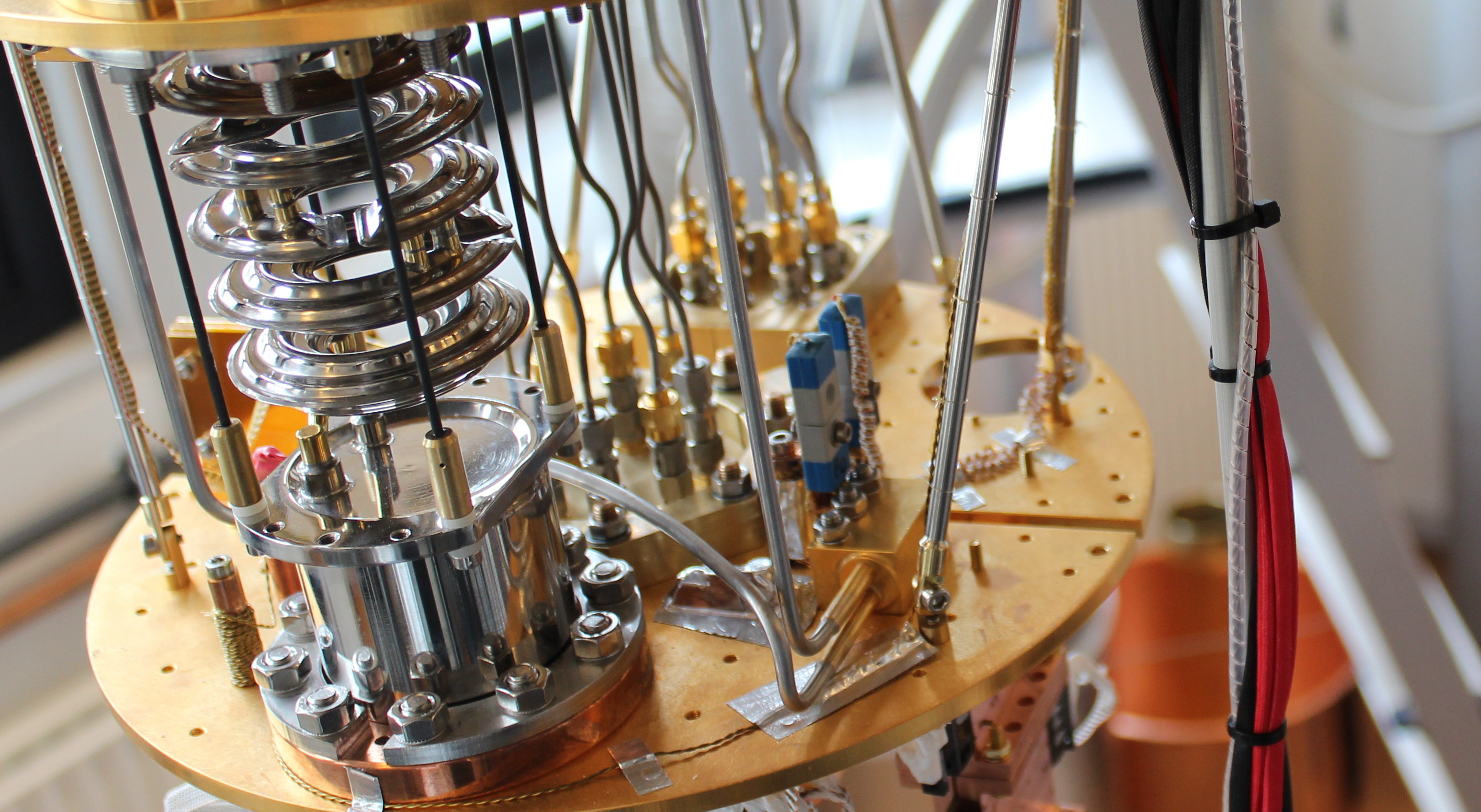}
\bibliography{all_bib,temp_bib}{}
\bibliographystyle{naturemag}

\end{document}